# Applied Causal Inference Powered by ML and AI

Victor Chernozhukov[*] Christian Hansen[†] Nathan Kallus[‡]
Martin Spindler[§] Vasilis Syrgkanis[¶]

March 4, 2024

[*] MIT
[†] Chicago Booth
[‡] Cornell University
[§] Hamburg University
[¶] Stanford University

# Contents

# Notation

| | |
|---|---|
| $:=$ | assignment or definition |
| $\equiv$ | equivalence |
| $\mapsto$ | "maps to" in the definition of a function |
| $X, Y, Z, ...$ | random variables (includes vectors); |
| | for noise vectors, we use $\epsilon, \epsilon_X, \epsilon_j, ...$ |
| $X'$ | transpose of a vector; |
| $\mathsf{x}$ | value of a random variable $X$ |
| P | probability measure |
| $P_X$ | probability distribution of $X$ |
| $P_{Y\|X}$ | conditional law of $Y$ given $X$ |
| $\mathsf{p}$ | density (either probability mass function |
| | or probability density function) |
| $\mathsf{p}_X$ | density of $P_X$ |
| $\mathsf{p}(x)$ | density of $P_X$ evaluated at the point $x$ |
| $\int \mathsf{p}(x)dx$ | integral with respect to the base measure |
| | (Lebesgue for probability density and counting measure for pmf) |
| $\mathsf{p}(y\|x)$ | (conditional) density of $P_{Y\|X=x}$ evaluated at $y$ |
| $z_{\mathsf{a}}$ | the $\mathsf{a}-$quantile of the standard normal distribution |
| $\{X_i\}_{i=1}^n$ | $= (X_1, X_2, ..., X_n)$ |
| | typically an iid. sample of size $n$ with distribution $P_X$ |
| | $X_{j1}, ..., X_{jn}$ when referring to $j^{th}$ component of $X$ |
| $\mathrm{E}[X]$ | expectation of $X$ |
| $\mathrm{E}[Y\|X]$ | conditional expectation of $Y$ given $X$ |
| $\mathbb{E}_n[f(Y, X)]$ | empirical expectation (e.g. $\mathbb{E}_n[f(Y, X)] := \frac{1}{n}\sum_{i=1}^n f(Y_i, X_i)$) |
| $\mathrm{Var}(X)$ | variance of $X$ |
| $\mathbb{V}_n[g(W)]$ | empirical variance |
| | (e.g. $\mathbb{V}_n[g(W)] = \mathbb{E}_n[g(W)g(W)'] - \mathbb{E}_n[g(W)]\mathbb{E}_n[g(W)]'$) |
| $\mathrm{Cov}(X, Y)$ | covariance of $X, Y$ |
| $X \perp Y$ | orthogonality of $X, Y$, i.e. $\mathrm{E}(XY') = 0$ |
| $X \perp\!\!\!\perp Y$ | independence of $X, Y$ |
| $X \perp\!\!\!\perp Y \mid Z$ | conditional independence of $X, Y$ given $Z$ |
| $P_{Y(x)} = P_{Y:do(X=x)}$ | intervention distribution (can be indexed by M) |
| $P_{Y(x)\|X} = P_{Y\|X:fix_Y(X=x)}$ | counterfactual distribution |
| $\mathsf{G}$ | directed graph |
| $\mathrm{pa}_{\mathsf{G}}(X)$, $\mathrm{de}_{\mathsf{G}}(X)$ , $\mathrm{an}_{\mathsf{G}}(X)$ | parents, descendants, and ancestors of node $X$ |
| | in graph $\mathsf{G}$ |
| $\mathbb{R}^p$ | the $p$-dimensional euclidean space |
| $\|\|x\|\|_1 := \sum_{j=1}^p \|x_j\|$ | the $\ell_1$-norm in $\mathbb{R}^p$ |
| $\|\|x\|\| \equiv \|\|x\|\|_2 := \sqrt{\sum_{j=1}^p x_j^2}$ | the $\ell_2$-norm in $\mathbb{R}^p$ |
| $\|\|x\|\|_\infty := \max_{j=1}^p \|x_j\|$ | the $\ell_\infty$-norm in $\mathbb{R}^p$ |
| $\|\|A\|\| := \sup_{x \in \mathbb{R}^p \setminus 0} \frac{x'Ax}{x'x}$ | the operator norm (maximum eigenvalue) of a matrix $A$ |

# Preface

This book aims to provide a working introduction to the emerging fusion of modern statistical inference – aka machine learning (ML) or artificial intelligence (AI) – and causal inference methods. The book is aimed at upper level undergraduates and master's-level students as well as doctoral students focusing on applied empirical research. A sufficient background for the core material is one semester of introductory econometrics and one semester of machine learning. We hope the book is also useful to empirical researchers looking to apply modern methods in their work.

The book provides an overview of key ideas in both predictive inference and causal inference and shows how predictive tools are key ingredients to answering many causal questions. We use the term predictive inference to refer to settings where prediction or description is the main goal such that models and estimates do not need a causal interpretation. ML/AI tools are largely designed to answer predictive inference questions, and we provide a high-level overview of popular ML/AI methods (such as Lasso, random forests, and deep neural networks, among others) to provide background for readers less familiar with these methods.

On the causal inference side, we introduce foundational ideas that provide the underpinning to attaching causal interpretations to statistical estimates. We discuss these ideas using the language of potential outcomes, directed acyclical graphs (DAGs), and structural causal models (SCMs). We view the language of potential outcomes, DAGs, and SCMs as complementary. We recognize that readers coming from different backgrounds may be more familiar or disposed to one of potential outcomes, DAGs, or SCMs, but we strongly believe that individuals interested in causal inference should be familiar with each of these frameworks. We find that they all offer useful insights and being able to communicate using each framework allows one to communicate with audiences interested in understanding causality coming from many different backgrounds.

The book has two main sections: Core Material and Advanced Topics. The Core Material provides the main content of the book. After concluding the Core Material, a reader should have an idea of the key ideas underlying both predictive and causal inference and how to wed these ideas to learn canonical objects in causal inference settings. The Core Material is made up of chapters that move between predictive inference and causal inference, typically by first introducing tools developed for predictive inference and then showing how these tools can be used as inputs to answering causal inference questions. The Advanced Topics then provide extensions of the Core Material to settings with more complicated causal structures, such as instrumental variables models, to settings where understanding heterogeneity in causal effects is the goal, and to specific popular settings in empirical work such as Difference-in-Differences.

Within sections, blocks marked with ⋆ require more substantial preparation in mathematical statistics. We recommend that the reader looking to apply machine learning methods in their work skim or pass them on their first reading and return to them at their leisure.

Short lists of references and study problems are included after each chapter to offer the reader opportunities to investigate further and consolidate their knowledge.

We would like to also acknowledge the tremendous and exceptional help and expertise provided by Philipp Bach, Wenxuan Guo, Andy Haupt, Shunzhuang Huang, David Hughes, Jannis Kück, Malte Kurz, Sven Klassen, Oliver Schacht, Sophie Sun, Vira Semenova, Gulin Tuzcuoglu, Suhas Vijaykumar, John Walker, Thomas Wiemann, Justin Young, and Dake Zhang with both writing and developing supporting Notebooks in R and Python. We are also grateful to Alexander Quispe and Anzony Quispe for developing a Bookdown version of the notebooks and providing other complementary topics and great examples.

*Chernozhukov, Hansen, Kallus, Spindler & Syrgkanis*

# Sneak Peek: Powering Causal Inference with ML and AI

# 0

A primary question we will be concerned with in this book is: What is the *causal effect* of an action on an outcome? For example, we may want to know what the effect of setting a product's price is on the volume of its sales.[1] To consider this question we scraped data on 9,212 toy cars from Amazon.com. Figure 0.1 shows a log-log-scale scatter plot of the 30-day average price at which each was offered and the reciprocal of its sales rank, a publicly available surrogate for sales volume.[2] We let $D$ denote the log of the price and $Y$ the negative log of the sales rank of a toy car randomly drawn from the population of toy cars sold on Amazon.com. We will use this example to preview the book's chapters and how they come together to enable the reader to power applied causal inference on modern datasets using ML and AI.

1: This effect may be referred to as the price *elasticity* of demand for the product.

2: Were the reader to do such an analysis using internal company data they would use actual sales volumes.

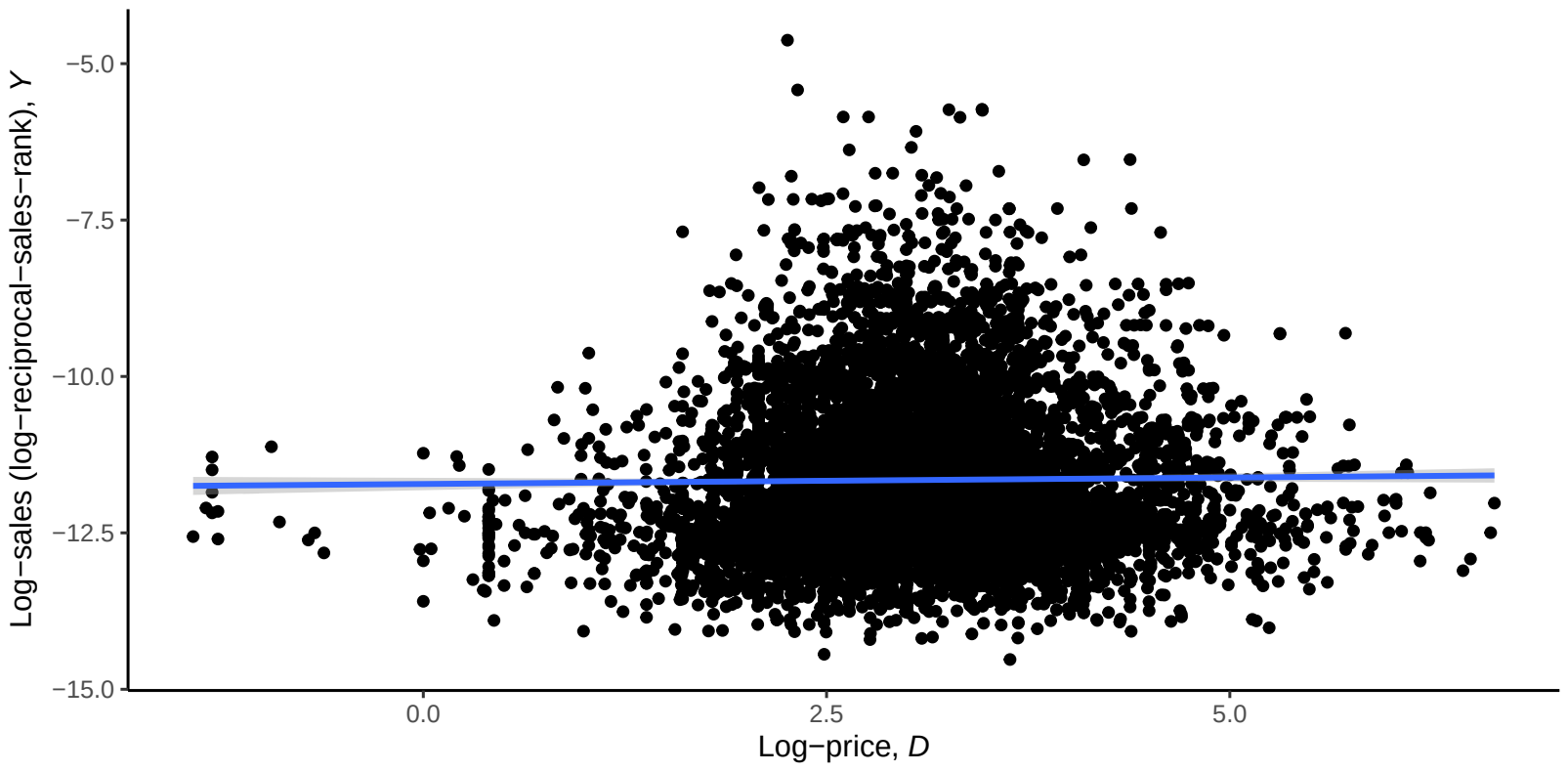

**Figure 0.1:** Log-prices and log-reciprocal-sales-rank of 9,212 toy cars on Amazon.com along with a linear fit.

In Chapter 1, we present linear regression by ordinary least squares (OLS), which can help us understand the relationship between these two variables. Here it suggests that a unit increase in $D$ is associated with anything between a $-0.008$ and a $0.050$ unit change in $Y$ on average over toy cars; that is, $(-0.008, 0.050)$ is the 95% confidence interval on the slope of the best linear predictor. In words, it suggests one cannot rule out small negative or even slightly positive association between price and sales. It would be incorrect, however, to infer that arbitrarily increasing the price on any one toy car would cause almost no effect on its sales volume, or even increase it.

Instead, economic theory would suggest that the unobserved *potential* log-sales, $Y(d)$, of any *one* toy car should in fact *decrease* as the log-price that one sets, $d$, increases. In Chapter 2, we present this notion of potential outcomes and study inference on their averages when actions are *randomized* (or, *exogenous*). For example, we may be interested in the average sales if price were set to a certain level. Unlike the randomized controlled trial (RCT) setting discussed in that chapter, here prices are *not* actually set at random; that is, prices are *endogenous*. Thus, the reason we may see no or a slightly positive association is *confounding* factors that affect both the potential sales at any one price and the particular price that is set. For example, whether a toy car is produced by a brand name or incorporates characters from a popular TV show might increase sales at any one price as well as lead the seller to choose a higher price, whether in anticipation of higher demand or because of higher production or licensing costs.

We formalize this notion of confounding in Chapter 5 and consider causal inference on averages of potential outcomes when one observes *all* confounding variables, $W$. In Chapter 6, we go on to consider a *linear* structural equation,

$$Y(d) = \alpha d + U, \qquad (0.0.1)$$

which posits that, on average, log-sales at any one log-price is a linear function of the log-price, aside from the idiosyncrasies $U$ of each one toy car. Within this structural equation, we interpret $\alpha$ as the causal effect of $d$ on $Y$; that is, the effect of a change in $d$ on $Y$ produced by intervening in the system to change $d$ while holding all other determinants of sales constant. This causal effect is generally not recovered from regression of observed $Y$ on observed price, $D$, as observed price is set in the market and plausibly related to unobserved factors $U$.

In our simple linear structural equation, the assumption that $W$ accounts for all confounding leads us to conclude that we have

$$Y = \alpha D + g(W) + \varepsilon, \quad \mathrm{E}[\varepsilon \mid D, W] = 0 \qquad (0.0.2)$$

for some function $g(W)$. Thus, after all of our causal modeling and assumptions, what remains is inference on a coefficient in a possibly complex regression model of $Y$ on $D$ and $W$, all of them *observed* variables. That is, under our causal modeling and assumptions, making statistical inferences (such as constructing estimates and confidence intervals) on $\alpha$ in Eq. (0.0.2) from data on $(Y, D, W)$ would be *causal inference*. (0.0.1) is the simplest of structural equations – to understand more complex structures we consider *systems* of equations, and in Chapter 7 even *nonlinear* structural equations.

To explain how we power such causal inference with ML and AI, let us now return to the question of *what is* $W$ *in the first place*? There are many features we can observe about each toy car on Amazon.com in addition to its price and sales: all the text on the product page such as name and description, the product subcategory (beyond being a toy), the brand, the color, and the dimensions and weight of both the item and its packaging. What features can we make use of, and how?

Classical methods, like OLS, (Chapter 1) allow us to conduct inference on $\alpha$ when Eq. (0.0.2) is a linear regression with moderately high dimensions, that is, when $W$ is a $p$-dimensional random vector, $g(W) = \beta_1 + \beta_2' W$, and $p$ is much smaller than the number of observations we have (here, 9,212). Letting $g(W) = \beta_1 + \beta_2' W$ in Eq. (0.0.2) we obtain a *linear model*. There are 243 product subcategories for our toy cars. Consider identifying each with a number in $1, \ldots, 243$ and letting $W$ be a 243-dimensional vector with a 1 in the index corresponding to the product's subcategory and 0 elsewhere. OLS regression of $Y$ on $D$ and this particular $W$ explains 7.5% of $Y$'s variance (as measured by adjusted $R^2$) and gives a 95% confidence interval on $\alpha$ of $(-0.026, 0.036)$. These results are not very different from what we inferred in the observed association between $Y$ and $D$ without adjusting for any confounding effects, but at least the upper bound is smaller – we indeed do not believe a positive effect is realistic.

Perhaps we need to control for more confounding effects than just subcategory membership. However, even without departing from linearity, OLS no longer provides reliable inference if we include too many features in $W$. Letting $g(W) = \beta_1 + \beta_2' W$ in Eq. (0.0.2) with a high-dimensional $W$, that is, where $d$ is comparable to or bigger than the number of observations, we obtain a *linear model with high-dimensional controls*. In Chapter 3, we present more advanced ML methods than OLS: predictive inference in high dimensions using regularized linear regression. The use of regularized linear regression may improve prediction relative to OLS but introduces biases that imperil inference on coefficients. In Chapter 4, we show how to remedy this bias when making inferences on any one coefficient. In the context of causal inference, this setup allows us to potentially handle very many confounders, and the hope is that we can then more reliably justify having accounted for *all* confounders. In a nutshell, in the setting of Eq. (0.0.2), if we take $\tilde{Y}$ and $\tilde{D}$ to be the *residuals* from a modern high-dimensional linear regression of $Y$ on $(1, W)$ and of $D$ on $(1, W)$, respectively, then OLS regression of $\tilde{Y}$ on $\tilde{D}$ yields valid inference on $\alpha$ even when $W$ is high-dimensional.

Consider letting $W$ be a 11546 dimensional vector including not only the indicator of subcategory but also the item's physical dimensions, transformed by log and expanded up to third power of the logarithms, missingness indicators, the interaction of these dimension features with subcategory, the indicator of brand (among 1827 brands). In this case, $p$ is greater than the number of observations $n$. Using the methods we present in Chapter 4 to leverage this high-dimensional $W$ in this particular set up, we obtain a 95% confidence interval on $\alpha$ of $(-0.10, -0.029)$. The confidence interval including only negative values is in concordance with the intuition that intervening to increase price would decrease demand. At the same time, we may still worry that a linear model is too restrictive, in essence allowing us only to control linearly for pre-specified confounders.[3]

3: One may include *pre-specified transformations* of confounders as well as discussed in Chapter 1.

In Chapter 9, we present nonlinear ML methods for regression: trees, ensembles, and neural nets. Compared to predicting log-price and log-sales with LASSO, using these methods (with a 2083-dimensional feature vector omitting the expansions and interactions needed for linear models) increases the $R^2$ by 25-53% and 89-189% (evaluated using 5-fold cross-validated $R^2$). Clearly these methods offer significant predictive improvements in this dataset. However, such nonlinear methods have no clear parameter to extract, no coefficient to inspect. While making excellent predictions, it is not immediately clear how to use them to make valid statistical inferences on finite-dimensional parameters, like average effects. We tackle that question in Chapter 10. Letting $g(W)$ be an arbitrary nonlinear function in Eq. (0.0.2) gives rise to what is called the *partially linear model*, which strikes a nice balance between structure and flexibility: the causal-effect part of the model is simple and interpretable – for each unit increase in action we get $\alpha$ increase in outcome – while the confounding part, which we have no interest in interpreting, can be almost-arbitrarily complex.[4] In the setting of Eq. (0.0.2), it turns out we can keep the method of residual-on-residual OLS inference, but using residuals from advanced *nonlinear* regressions, as long as we fit these regressions on parts of the data that exclude where we use them to make predictions and produce the residuals. This is *double machine learning* or *debiased machine learning* or *double/debiased machine learning*[5] for the partially linear model. Using DML together with gradient-boosted-tree regression to make inferences on the price elasticity $\alpha$ in this example yields a confidence interval of $(-0.139, -0.074)$, suggesting an effect whose direction agrees

4: Luckily, even if the partially linear assumption fails, estimates still reflect some average of the *causal* effects of increasing *all* prices by a small amount, provided we have accounted for all confounding effects in in $W$. See Remarks 10.2.2 and 10.3.3.

5: We will use these terms interchangeably and abbreviate them with *DML*.

even more strongly with our intuition, which can be attributed to these more powerful predictive methods being able to better account and correct for the confounding effects that pushed the apparent direction upward.

It is still unclear, however, whether the numeric features we observe can reliably capture all of the confounding effects – if they cannot, then no regression, no matter how flexible, can help. This problem – getting the right data to enable causal inference – is a common challenge when dealing with observational data. It is in using all the available data, where modern AI along with the tools we develop in this book come together to uniquely enable powerful causal inferences using modern observational data sets. Modern data sets are rich, containing far more than just numeric features. This data set, for example, contains text on each product – descriptions that capture many important features about each product that are not clearly tabulated but must be inferred by reading the text. Luckily, modern AI has made great inroads in recent years in machine cognition of text, images, videos, and other rich data.

In Chapter 11, we discuss how these powerful tools can be used in concert with DML. BERT is a large language model leveraging a deep learning architecture known as *transformers* and achieving impressive performance on natural-language-processing benchmarks. Using neural-net-based predictive models for log-price and log-sales built on top of BERT results in a 12-37% and 4-59% increase in cross-validated $R^2$, respectively, relative to the nonlinear models using only numeric features in the data. The non-numeric features in the data therefore seem to account for more than the baseline numeric factors of products in predicting price and sales. Using DML for the partially linear model together with these models that use the non-numeric features, we are able to make causal inferences that account for confounding factors reflected in the rich text on the product page for each toy car. Proceeding in this way, as we explain in greater detail in Chapter 11, we obtain a confidence interval on $\alpha$ of $(-0.21, -0.13)$. That we get a more negative estimate here again suggests that there were residual confounding effects inducing a spurious positive relationship between price and sales that we could only have controlled for and counteracted by using AI to account for the rich text data.

While it is relatively easy to validate predictive models' performance by using held-out test sets and cross-validation, it is difficult – impossible, even – to definitively validate a causal effect, as it will inevitably rest on fundamentally untestable assumptions. Nonetheless, we can have greater confidence in estimates that correctly and fully leverage the available data and do not rely on unnecessary parametric assumptions. Estimates based on DML on top of AI allow us to do just that. We can use rich data without imposing strong functional form restrictions and importantly can do so without imperiling guarantees on valid statistical inference. The Core material outlines the basic ideas and provides fundamental results for using DML with AI learners to estimate and do inference for low-dimensional causal effects.

The Advanced Topics section includes chapters that expand upon the basic material from the Core chapters. In the Core material, we discuss more complex structures than the partially linear model introduced in this preview, but do inference essentially only when all relevant variables are observed. In Chapter 12, we present alternative ways to identify causal effects when we do not believe we observe all confounders – techniques such as sensitivity analysis, instrumental variables, and proxy controls, and we provide methods for causal inference in such settings in Chapter 13. These tools allow us to have confidence in causal estimates that leverage special structure like instruments or proxies without additionally making unnecessary parametric assumptions and with the ability to leverage rich data using powerful AI. In many examples, one may wish to understand heterogeneity in causal effects such as how causal effects differ across observed predictors. Chapter 14 covers DML inference on quantities that characterize this heterogeneity, and Chapter 15 goes beyond inference on low-dimensional causal parameters and discusses learning heterogeneous causal effects from rich individual-level data and even personalizing treatments based on such data. Finally, we consider application of DML in conjunction with two popular methods for identifying causal effects – difference-in-differences and regression discontinuity designs – in Chapter 16 and Chapter 17 respectively.

After studying the book, the reader should also be able to understand and employ DML in many other applications that are not explicitly covered. In the toy car example we focused on sales, but sales may not reflect demand when we reach the limits of on-hand inventory, something known as right-censoring. Censoring is an example of data coarsening, and mathematically it is not too dissimilar from the missingness of potential outcomes for actions not taken. Similarly, we may want to look at distributional effects beyond averages, like effects on the quantiles of sales. DML can often be applied to these problems and there is active research on applying it to ever more intricate problems.

There are also topics beyond our scope. We started by saying we focus on the causal effect of an action on an outcome – a broader yet much more challenging question is, among multiple variables, discovering which have causal effects on which. While we *do* discuss the use of directed acyclic graphs in Chapter 7 and Chapter 8, we only use them to represent assumed structure and only briefly mention how one might try to learn causal structure directly from data, which is the subject of *causal discovery*.

Our aim is rather focused: present the building blocks of predictive inference and of causal inference and illustrate their effective and correct use in concert in a way that allows readers to employ them in real, practical settings. The book interweaves the two kinds of inference, with many real-data examples with code notebooks. We hope the outcome is that we reach an endpoint where the reader is ready to power causal inferences with ML and AI and be able to draw valid, reliable inferences in practice using rich modern data.

# Core Material

# Predictive Inference with Linear Regression in Moderately High Dimensions

# 1

"Infer: to form an opinion or guess that something is true because of the information that you have."

– Cambridge Dictionary [1].

Least squares, and particularly its application to linear regression, is one of the most widely used statistical methods. It is an intuitive tool for predictive inference and for establishing association. The method of least squares was introduced in the 1800s by L. Legendre and C.F. Gauss. Here we review properties of least squares estimation of linear models in moderately high-dimensional problems, focusing on its use in predictive inference and for establishing association. This treatment provides a starting point for our subsequent review of modern statistical (machine) learning methods, which will relax our assumption on dimensionality as well as consider nonlinear models.

## 1.1 Foundation of Linear Regression

### Regression and the Best Linear Prediction Problem

We consider a scalar random variable $Y$, an outcome of interest, and a $p$-vector of covariates

$$X = (X_1, \ldots, X_p)'.$$

We assume that a constant of 1 is included as the first component in $X$; that is, $X_1 = 1$.

For theoretical purposes, we first consider linear regression in the population. Working in the population means that we have access to unlimited amounts of data to compute population moments – such as $\mathrm{E}[Y]$, $\mathrm{E}[YX]$, and $\mathrm{E}[XX']$ – and that we can define "ideal" quantities. After defining these ideal quantities, we then turn to estimation with real data, which we will take to be a sample of observations drawn from the population.

Our first goal is to construct the best linear prediction rule for $Y$ using $X$. That is, the predicted value of $Y$ given $X$ will be of the linear form:

$$\sum_{j=1}^{p} \beta_j X_j = \beta' X, \text{ for } \beta = (\beta_1, \ldots, \beta_p)',$$

where $\beta$'s are called the regression parameters or coefficients.

We define $\beta$ as any solution to the *Best Linear Prediction (BLP) Problem*,

$$\min_{b \in \mathbb{R}^p} \mathrm{E}\left[(Y - b'X)^2\right],$$

where we minimize the Expected or Mean Squared Error (MSE) for predicting $Y$ using the linear rule

$$b'X = \sum_{j=1}^{p} b_j X_j, \quad b = (b_1, \ldots, b_p)'.$$

The solution to this optimization problem, $\beta' X$, is called the *Best Linear Predictor* (BLP) of $Y$ using $X$. This jargon refers to the fact that $\beta' X$ is the best, according to MSE, linear prediction rule for $Y$ among all possible linear prediction rules.

We can compute $\beta$ by solving the first order conditions for the BLP problem:

$$\mathrm{E}\left[(Y - \beta' X)X\right] = 0.$$

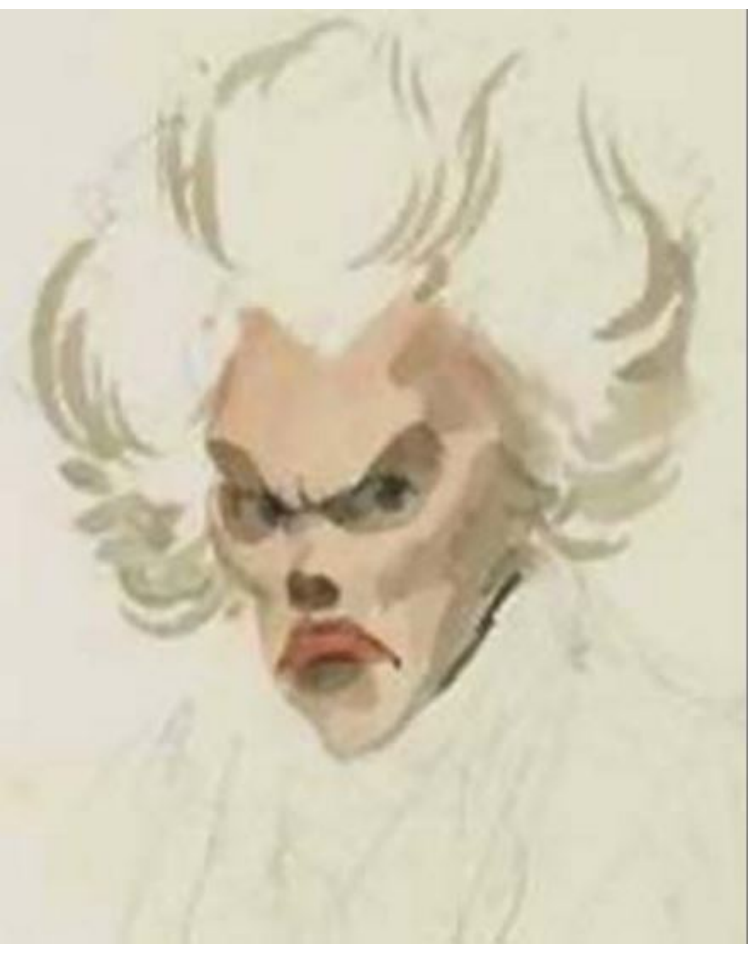

**Figure 1.1:** The only known portrait of Legendre (a friendly caricature) by Julien Léopold Boilly. Source: Wikipedia. The hairstyle is amazing.

These equations are also referred to as the Normal Equations and are obtained by setting the derivative of the objective function $b \mapsto \mathrm{E}\left[(Y - b'X)^2\right]$ with respect to $b$ equal to zero. Thus, any solution to the BLP problems satisfies the Normal Equations.

Defining the regression error or residual as

$$\varepsilon := (Y - \beta'X),$$

we can write the Normal Equations as

$$\mathrm{E}[\varepsilon X] = 0, \quad \text{or equivalently} \quad \varepsilon \perp X.$$

Note that we use $\perp$ to denote orthogonality between random variables, and $\perp\!\!\!\perp$ to denote full statistical independence.. That is, for random variables $U$ and $V$, $U \perp V$ means $\mathrm{E}[UV] = 0$. Further, if $U$ is a *centered random variable*, then $U \perp\!\!\!\perp V$ implies $U \perp V$, but the reverse implication is not true in general. Indeed, let $U \sim N(0,1)$ and $V = U^2 - 1$, then $U \perp V$ but $U \not\!\perp\!\!\!\perp V$.

Therefore, the BLP problem provides a simple decomposition of $Y$:

$$Y = \beta'X + \varepsilon, \quad \varepsilon \perp X,$$

where $\beta'X$ is the part of $Y$ that can be linearly predicted or explained with $X$, and $\varepsilon$ is whatever remains – the so-called unexplained or residual part of $Y$.

## Best Linear Approximation Property

The normal equation $\mathrm{E}[(Y - \beta'X)X] = 0$ implies by the law of iterated expectations that

$$\mathrm{E}[(\mathrm{E}[Y \mid X] - \beta'X)X] = 0.$$

Therefore, the BLP of $Y$ is also the BLP for the conditional expectation of $Y$ given $X$. This observation is important and motivates the use of various transformations of regressors to form $X$.

## From Best Linear Predictor to Best Predictor

Here we explain the use of constructed features or regressors. If $W$ are "raw" regressors/features, *technical (constructed) regressors* are of the form

$$X = T(W) = (T_1(W), ..., T_p(W))',$$

where the set of transformations $T(W)$ is sometimes called the *dictionary* of transformations. Example transformations include polynomials, interactions between variables, and applying functions such as the logarithm or exponential. In the wage analysis reported below, for example, we use quadratic and cubic transformations of experience, as well as interactions (products) of these regressors with education and geographic indicators.

The main motivation for the use of constructed regressors is to build *more flexible and potentially better* prediction rules. The potential for improved prediction arises because we are using prediction rules $\beta' X = \beta' T(W)$ that are *nonlinear* in the original raw regressors $W$ and may thus capture more complex patterns that exist in the data. Conveniently, the prediction rule $\beta' X$ is still linear with respect to the parameters, $\beta$, and with respect to the constructed regressors $X = T(W)$.

In the population, the *best predictor* of $Y$ given $W$ is

$$g(W) = \mathrm{E}[Y \mid W],$$

the *conditional expectation* of $Y$ given $W$. The *conditional expectation function* $g(W)$ is also called the *regression function* of $Y$ on $W$. Specifically, the conditional expectation function $g(W)$ solves the best prediction problem[1]

$$\min_{m(W)} \mathrm{E}\left[(Y - m(W))^2\right].$$

Here we minimize the MSE among all prediction rules $m(W)$ (linear or nonlinear in $W$).

As the conditional expectation solves the same problem as the best linear prediction rule among a larger class of candidate rules, the conditional expectation generally provides better predictions than the best linear prediction rule.[2]

By using $\beta' T(W)$, we are implicitly approximating the best predictor $g(W) = \mathrm{E}[Y|W]$. Indeed, for any parameter $b$,

$$\mathrm{E}\left[(Y - b'T(W))^2\right] = \mathrm{E}\left[(g(W) - b'T(W))^2\right] + \mathrm{E}\left[(Y - g(W))^2\right],$$

That is, the mean squared prediction error is equal to the mean squared approximation error of $b'T(W)$ to $g(W)$ plus a constant that does not depend on $b$. Therefore, minimizing the mean squared prediction error is the same as minimizing the mean squared approximation error. Thus, the BLP $\beta' T(W)$ is the *Best Linear Approximation* (BLA) to the best predictor, which is the regression function $g(W)$. Finally, as the dictionary of transformations $T(W)$ becomes richer, the quality of the approximation of the BLA $\beta' T(W)$ to the best predictor $g(W)$ improves.

1: This result follows by rewriting the objective function as

$$\min_{m(W)} \mathrm{E}[\mathrm{E}[(Y - m(W))^2 \mid W]],$$

noting that it is equivalent to

$$\mathrm{E}[\min_{\mu \in \mathbb{R}} \mathrm{E}[(Y - \mu)^2 \mid W]],$$

and deriving the first order conditions for the inner minimization: $E(Y \mid W) - \mu = 0$.

2: Unless the conditional expectation function turns out to be linear, in which case the conditional expectation and best linear prediction rule coincide.

**Example 1.1.1** (Approximating a Smooth Function with a Polynomial Dictionary) Suppose $W \sim U(0,1)$ where $U$ denotes the continuous uniform distribution, and

$$g(W) = \exp(4 \cdot W).$$

We use

$$T(W) = \underbrace{(1, W, W^2, \dots, W^{p-1})'}_{p \text{ terms}}$$

to form the BLA/BLP, $\beta' T(W)$. Figure 1.2 provides a sequence of panels that illustrate the approximation properties of the BLA/BLP corresponding to $p$ =2, 3, and 4:

- With $p = 2$ we get a linear in $W$ approximation to $g(W)$. As the figure shows, the quality of this approximation is poor.
- With $p = 3$ we get a quadratic-in-$W$ approximation to $g(W)$. Here, the approximation quality is markedly improved relative to $p = 2$ though approximation errors are still clearly visible.
- With $p = 4$ we get a cubic-in-$W$ approximation to $g(W)$, and the quality of approximation appears to be excellent.

This simple example highlights the motivation for using nonlinear transformations of raw regressors in linear regression analysis. What this example does not yet reveal are the *statistical* challenges of dealing with higher and higher dimension $p$ when learning from a finite sample.

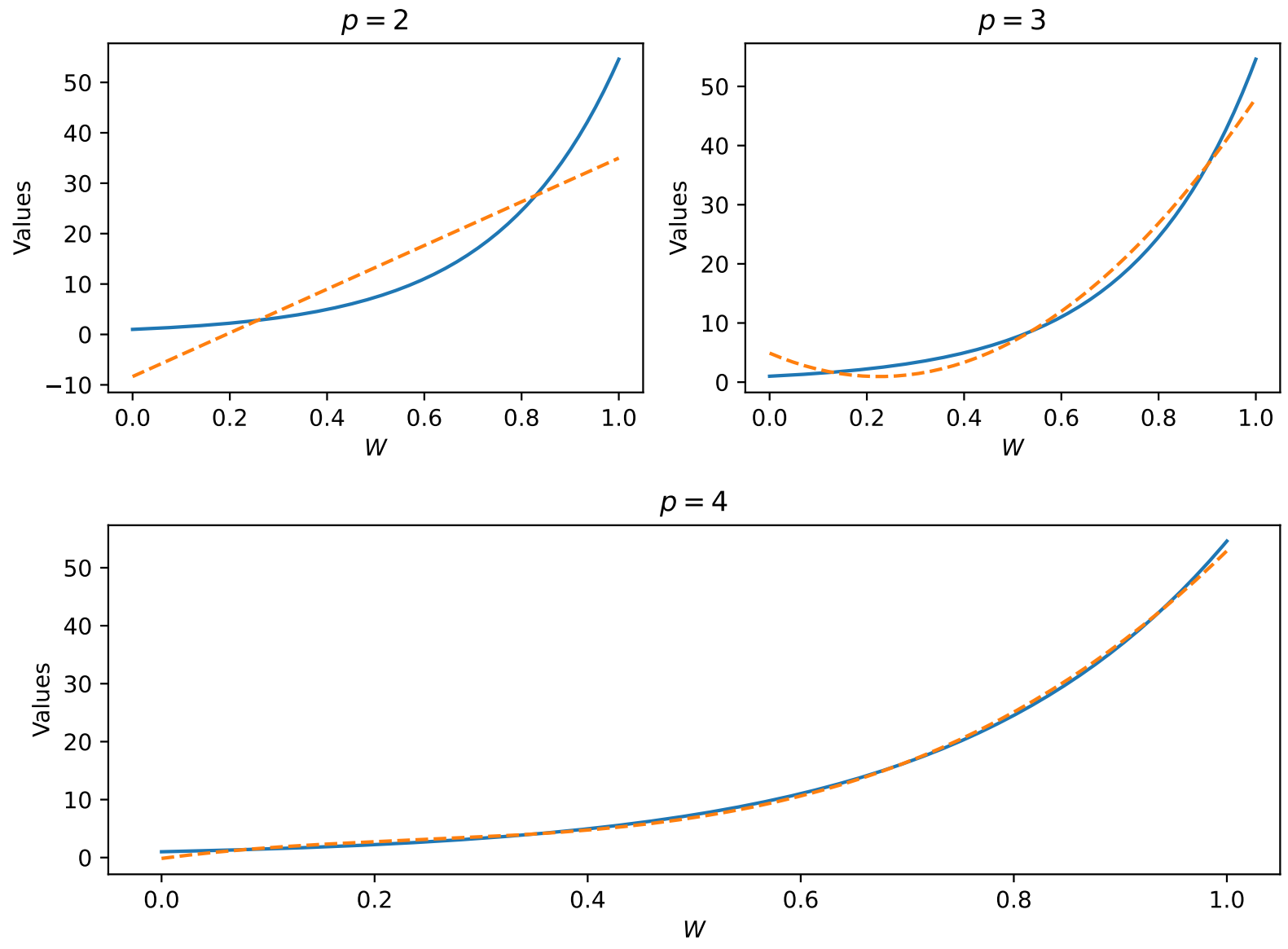

**Figure 1.2:** Refinements of Approximation to Regression Function $g(W)$ by using polynomials of $W$.

There are many ways of generating flexible approximations, which are studied by approximation theory and nonparametric statistical learning theory.[3]

3: See, e.g., Tsybakov [2]. We will also consider nonlinear approximations using trees and neural networks in Chapter 9.

When we have multiple variables, we may generate transformations of each of the variables and employ interactions – products involving these terms. As a simple concrete example, consider a case with two raw regressors, $W_1$ and $W_2$. We could build polynomials of second order in each of the raw regressors – $(1, W_1, W_1^2)$, $(1, W_2, W_2^2)$. We may then collect these variables along with the interaction in the raw regressors, $W_1W_2$ in a vector

$$(1, W_1, W_2, W_1^2, W_2^2, W_1W_2)$$

for use in a regression model. There are, of course, many other possibilities such as considering higher order polynomial terms, e.g. $W_1^3$; higher order interactions, e.g. $W_1^2W_2$; and other nonlinear transformations, e.g. $\log(W_1)$.

## 1.2 Statistical Properties of Least Squares

### The Best Linear Prediction Problem in Finite Samples

In practice, the researcher does not have access to the entire population, but observes only a sample

$$\{(Y_i, X_i)\}_{i=1}^n = ((Y_1, X_1), ..., (Y_n, X_n)).$$

We assume that this sample is a random sample from the distribution of $(Y, X)$. Formally, this condition means that the observations were obtained as realizations of independently and identically distributed (iid) copies of the random variable $(Y, X)$. By treating the observations as iid, we are modeling the data as independent random draws with replacement from a population. Other possible models include sampling without replacement from a finite population, stratified sampling, observations of a process over time, and other schemes or scenarios that induce dependence between the data points. For the most part, we focus on the iid model throughout this book.

We construct the best in-sample linear prediction rule for $Y$ using $X$ analogously to the population case by replacing theoretical expected values, E, with empirical averages, $\mathbb{E}_n$. Specifically,

$\mathbb{E}_n$ abbreviates the notation $\frac{1}{n}\sum_{i=1}^n$. For example,

$$\mathbb{E}_n[f(Y, X)] := \frac{1}{n}\sum_{i=1}^n f(Y_i, X_i).$$

given $X$, our predicted value of $Y$ will be

$$\sum_{j=1}^{p} \hat{\beta}_j X_j = \hat{\beta}' X, \text{ for } \hat{\beta} = (\hat{\beta}_1, ..., \hat{\beta}_p)',$$

where $\hat{\beta}$ is any solution to the *Best Linear Prediction Problem in the Sample*, also known as Ordinary Least Squares (OLS):

$$\min_{b \in \mathbb{R}^p} \mathbb{E}_n[(Y - b'X)^2].$$

We often use the hat decoration $\hat{}$ for quantities that depend on the sample. For example, $\beta$ denotes the BLP in the population, while $\hat{\beta}$ is the BLP in the sample.

That is, $\hat{\beta}$ minimizes the sample MSE for predicting $Y$ using the linear rule $b'X$. The $\hat{\beta}$'s are called the sample regression coefficients.

We can compute $\hat{\beta}$ as any solution to the Sample Normal Equations,

$$\mathbb{E}_n[X(Y - X'\hat{\beta})] = 0,$$

which are obtained as the first order conditions to the Best Linear Prediction Problem in the Sample. Further, defining the residuals (or, in-sample regression errors) as

$$\hat{\varepsilon}_i := (Y_i - \hat{\beta}' X_i),$$

we obtain the decomposition

$$Y_i = X_i'\hat{\beta} + \hat{\varepsilon}_i, \quad \mathbb{E}_n[X\hat{\varepsilon}] = 0,$$

where $X_i'\hat{\beta}$ is the predicted or explained part of $Y_i$, and $\hat{\varepsilon}_i$ is the unexplained or residual part.

## Properties of Sample Linear Regression

The best linear prediction rule in the population is $\beta' X$, and a key question is whether $\hat{\beta}' X$ estimates (that is, approximates using data) $\beta' X$ well.

The best linear prediction rule is also the best linear rule for predicting future values of $Y$ given a new draw $X$, when new $(Y, X)$ are sampled from the same population. Therefore, if we can approximate the best linear prediction rule in the population, we can also approximate the best linear prediction rule for predicting outcomes given future $X$'s sampled from the population.

The fundamental statistical issue is that we are trying to estimate $p$ parameters, $\beta_1, ..., \beta_p$, without imposing any assumptions on these parameters. Intuitively, to estimate each parameter well, we need many observations per parameter. This intuition suggests that $n/p$ should be large, or, equivalently that $p/n$ should be small, in order for estimation error to be small. The following result captures this intuition more formally.

**Theorem 1.2.1** (Approximation of BLP by OLS) *Under regularity conditions,*[a]

$$\sqrt{\mathrm{E}_X[(\beta'X - \hat{\beta}'X)^2]} = \sqrt{(\hat{\beta} - \beta)'\mathrm{E}_X[XX'](\hat{\beta} - \beta)}$$

$$\leq const_{\mathrm{P}} \cdot \sqrt{\mathrm{E}\varepsilon^2}\sqrt{\frac{p}{n}},$$

*where* $\mathrm{E}_X$ *is the expectation with respect to* $X$ *alone, the inequality holds with probability approaching 1 as* $n \longrightarrow \infty$*, and* $const_{\mathrm{P}}$ *is a constant that depends on the distribution of* $(Y, X)$*.*

[a] See Notes (Section 1.5) for references.

Theorem 1.2.1 says that, for nearly all realizations of data, the sample linear regression is close to the population linear regression if $n$ is large and $p$ is much smaller than $n$:

$$\sqrt{\mathrm{E}_X[(\beta'X - \hat{\beta}'X)^2]} \approx 0.$$

In other words, under our requirement of $p/n$ small, the sample BLP approximates the population BLP well.

Given indexed random variables (vectors, elements) $A_n$ and $B_n$ in a metric space equipped with metric $d$, the notation $A_n \approx B_n$ means that the distance between $A_n$ and $B_n$ concentrates around 0 – formally, that $\lim_{n\to\infty} \mathrm{P}(d(A_n, B_n) \leq \varepsilon) = 1$ for each $\varepsilon > 0$.

## Analysis of Variance

Analysis of variance involves the decomposition of the variation of $Y$ into explained and unexplained parts. Explained variation is a measure of the predictive performance of a model. This decomposition can be conducted both in the population and in the sample.

The main idea is to use the previous decomposition of $Y$,

$$Y = \beta'X + \varepsilon, \quad \mathrm{E}[\varepsilon X] = 0,$$

to decompose the variation in $Y$ into the sum of *explained variation* and *residual variation*:

$$\mathrm{E}[Y^2] = \mathrm{E}[(\beta'X)^2] + \mathrm{E}[\varepsilon^2].$$

The quantity

$$\mathrm{MSE}_{pop} = \mathrm{E}[\varepsilon^2]$$

is the population MSE. The ratio of the explained variation to the total variation is the population $R^2$:

$$R^2_{pop} := \frac{\mathrm{E}[(\beta' X)^2]}{\mathrm{E}[Y^2]} = 1 - \frac{\mathrm{E}[\varepsilon^2]}{\mathrm{E}[Y^2]} \in [0,1].$$

That is, $R^2_{pop}$ is the proportion of variation of $Y$ explained by the BLP.

**Remark 1.2.1** The "standard" definition of $R^2$ assumes that either we work with a centered $Y$, that is, we recenter $Y$ such that $\mathrm{E}[Y] = 0$. (However, our definition above does not require this property).*centered random variable*

The decomposition of the variance in the sample proceeds analogously. Using the representation

$$Y_i = \hat{\beta}' X_i + \hat{\varepsilon}_i$$

and the orthogonality condition $\mathbb{E}_n[X\hat{\varepsilon}] = 0$ provided by the sample Normal Equations, we obtain the decomposition

$$\mathbb{E}_n[Y^2] = \mathbb{E}_n[(\hat{\beta}' X)^2] + \mathbb{E}_n[\hat{\varepsilon}^2].$$

Thus, we can define the sample MSE,

$$\mathrm{MSE}_{sample} = \mathbb{E}_n[\hat{\varepsilon}^2],$$

and the sample $R^2$,

$$R^2_{sample} := \frac{\mathbb{E}_n[(\hat{\beta}' X)^2]}{\mathbb{E}_n[Y^2]} = 1 - \frac{\mathbb{E}_n[\hat{\varepsilon}^2]}{\mathbb{E}_n[Y^2]} \in [0,1].$$

By the law of large numbers and Theorem 1.2.1, when $p/n$ is small, we have the following approximations:

$$\mathbb{E}_n[Y^2] \approx \mathrm{E}[Y^2], \quad \mathbb{E}_n[(\hat{\beta}' X)^2] \approx \mathrm{E}[(\beta' X)^2], \quad \mathbb{E}_n[\hat{\varepsilon}^2] \approx \mathrm{E}[\varepsilon^2].$$

Thus, when $p/n$ is small and $n$ is large, the sample fit measures are good approximations to population fit measures:

$$\mathrm{MSE}_{sample} \approx \mathrm{MSE}_{pop} \text{ and } R^2_{sample} \approx R^2_{pop}.$$

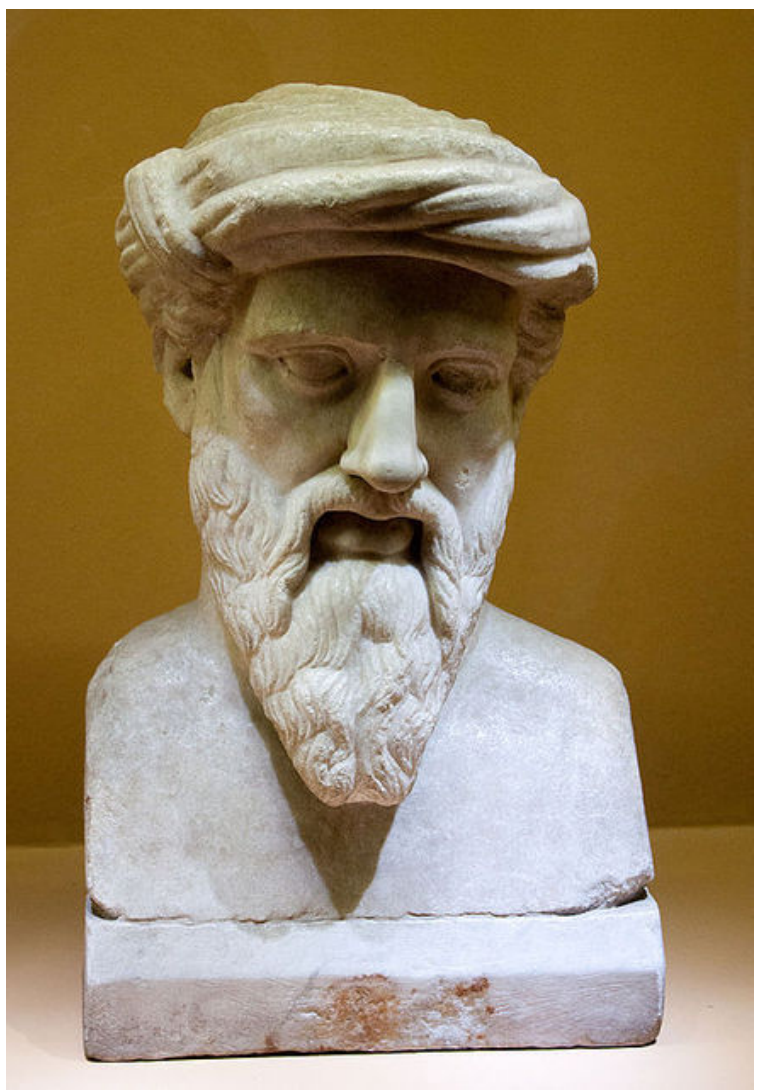

**Figure 1.3:** Pythagoras of Samos invented least squares and analysis of variance for the case of $n = 2$ and $p \leq 2$ around 570 BC. He was therefore the first known machine learner.

## Overfitting: What Happens When $p/n$ Is Not Small

When $p/n$ is not small, the picture about predictive performance of the in-sample BLP becomes inaccurate and possibly misleading. In this setting, the in-sample linear predictor can be substantially different from the population BLP.

Consider an extreme example where $p = n$ and all variables in $X$ are linearly independent. In this case, we have

$$\mathrm{MSE}_{sample} = 0 \text{ and } R^2_{sample} = 1$$

no matter what $\mathrm{MSE}_{pop}$ and $R^2_{pop}$ are. E.g. we could have $R^2_{sample} = 1$ even if $R^2_{pop} = 0$. Therefore, here we have an extreme example of *overfitting*, where the in-sample predictive performance overstates the out-of-sample predictive performance of the linear model. The following example illustrates less extreme cases.

The Linear Model Overfitting R Notebook and the Linear Model Overfitting Python Notebook contain code for the numerical experiment.

**Example 1.2.1** (Overfitting Example) Suppose $X \sim N(0, I_p)$ and $Y \sim N(0, 1)$ are statistically independent. It follows that the best linear predictor of $Y$ is $\beta' X = 0$ and that $R^2_{pop} = 0$.

- If $p = n$, then the typical $R^2_{sample}$ is $1 \gg 0$.
- If $p = n/2$, then the typical $R^2_{sample}$ is about $.5 \gg 0$.
- If $p = n/20$, then the typical $R^2_{sample}$ is about $.05 > 0$.

These results can be deduced by simulation or analytically.

Provided $p < n$, better measures of out-of-sample predictive ability are the "adjusted" $R^2$ and MSE:[4]

$$\mathrm{MSE}_{adjusted} = \frac{n}{n-p}\mathbb{E}_n[\hat{\varepsilon}^2], \quad R^2_{adjusted} := 1 - \frac{n}{n-p}\frac{\mathbb{E}_n[\hat{\varepsilon}^2]}{\mathbb{E}_n[Y^2]}.$$

4: The adjustment factor $\frac{n}{n-p}$ is derived in a homogeneous model, so that $\mathrm{E}[\mathrm{MSE}_{adjusted}] = \mathrm{MSE}_{pop}$, see e.g., p. 8 in [3] for the derivation.

The adjustment by $\frac{n}{n-p}$ corrects for overfitting and provides a more accurate assessment of predictive ability of the linear model in Example 1.2.1 and more generally under the assumption of homogeneous $\varepsilon$. The intuition is that models with many parameters increase the in-sample fit and potentially cause overfitting. Hence, the number of parameters is incorporated in the definition of $\mathrm{MSE}_{adjusted}$ and $R^2_{adjusted}$ in an attempt to account for this phenomenon.

## Measuring Predictive Ability by Sample Splitting

How should we measure the predictive ability of the linear model (or other nonlinear models that we will discuss) more reliably, even in cases when $p/n$ is not small?

> A general way to measure predictive performance is to perform *data splitting*. The idea can be summarized in two parts:
>
> 1. Use a random part of a dataset, called the training sample, for estimating/training the prediction rule.
> 2. Use the other part, called the testing sample, to evaluate the quality of the prediction rule, recording out-of-sample mean squared error and $R^2$.

Generally, a predictive model is trained on a sample and the real test of its predictive ability happens when "new, unseen" observations arrive. With new observations in hand, we learn how far off our predictions are, when compared to the realized values. By partitioning the data set into two parts, we preserve an "unseen" set of observations on which to test our model, mimicking this process of ex-post performance assessment.[5]

5: If the "test set" is used many times to evaluate models, it becomes a "validation" set. The term "test set" is often reserved for the final evaluations of very few models.

The data splitting procedure can be described more formally as follows:

> **Generic Evaluation of Prediction Rules by Sample-Splitting**
>
> 1. Randomly partition the data into training and testing samples. Suppose we use $n$ observations for training and $m$ for testing/validation.
> 2. Use the training sample to compute a prediction rule $\hat{f}(X)$. For example, $\hat{f}(X) = \hat{\beta}'X$ in the linear model.
> 3. Let $\mathcal{J}$ denote the indexes of the observations in the test sample. Then the out-of-sample/test mean squared error is
>
> $$\text{MSE}_{test} = \frac{1}{m}\sum_{k\in\mathcal{J}}(Y_k - \hat{f}(X_k))^2,$$
>
> and the out-of-sample/test $R^2$ is
>
> $$R^2_{test} = 1 - \frac{\text{MSE}_{test}}{\frac{1}{m}\sum_{k\in\mathcal{J}} Y_k^2}.$$

In Section 3.B, we consider a more data-efficient evaluation procedure called cross-validation where test data are reused for training. In brief, we split the data into even parts, for each part we repeat the evaluation procedure taking that part to be the "test" sample, and finally we average the values of $\text{MSE}_{test}$ that we computed in each round.

There is an important variation on the sample splitting procedure, called *stratified splitting* that provides guarantees that the training and test samples are similar.[6] In large samples, training and test samples will be similar by virtue of the laws of large numbers, but similarity is not guaranteed in moderate-sized samples. For more discussion, please see this blog on Data Splitting [4].

6: For example, we can make sure that the proportions of college-graduates and non-college-graduates are the same in both training and test samples. These issues are important in moderate-sized samples.

## 1.3 Inference about Predictive Effects or Association

Here we examine inference on *predictive effects*, which describe how our (population best linear) predictions change if the value of a regressor changes by a unit, while the other regressors remain unchanged.

Specifically, we partition the vector of regressors $X$ into two components:

$$X = (D, W')',$$

where $D$ represents the "target" regressor of interest, and $W$ represents the other regressors, sometimes called the controls. We can therefore write

$$Y = \underbrace{\beta_1 D + \beta_2' W}_{\text{predicted value}} + \underbrace{\varepsilon}_{\text{error}}, \tag{1.3.1}$$

and ask the question:

> How does the predicted value of $Y$ change if $D$ increases by a unit while $W$ remains unchanged?

Note that this question is purely about the properties of the prediction rule and generally has nothing to do with causality.

The answer is the predicted value of $Y$ changes by

$$\beta_1.$$

**Example 1.3.1** (Wage Differences) In the analysis of wages, which we will discuss later in more detail, an interesting question can be formulated as:

- "What is the difference in predicted wages between female and non-female workers with the same job-relevant characteristics?"

Let $D$ represent the female indicator and $W$ represent experience, educational, occupational, and geographic characteristics. The answer to the question is then the population regression coefficient

$$\beta_1$$

corresponding to $D$.

## Understanding $\beta_1$ via "Partialling-Out"

"Partialling-out" is an important tool that provides conceptual understanding of the regression coefficient $\beta_1$.

In the *population*, we define the partialling-out operation as a procedure that takes a random variable $V$ and creates the "residualized" error variable $\tilde{V}$ by subtracting the part of $V$ that is linearly predicted by $W$:

$$\tilde{V} = V - \gamma_{VW}' W, \quad \gamma_{VW} \in \arg\min_{\gamma} \mathrm{E}\left[(V - \gamma' W)^2\right].$$

When $V$ is a vector, we apply the operation to each component. It can be shown that the partialling-out operation is linear in the sense that[7]

$$Y = \nu V + \mu U \implies \tilde{Y} = \nu \tilde{V} + \mu \tilde{U}.$$

7: Verify this as a reading exercise. Use the definition of the BLP decompositions of $U$ and $V$ with respect to regressors $W$, to derive a BLP decomposition of $Y$ with respect to $W$.

Formally, this operation is well defined on the space of random variables with finite second moments.

We apply the partialling-out operation to both sides of our regression equation $Y = \beta_1 D + \beta_2' W + \varepsilon$ to get

$$\tilde{Y} = \beta_1 \tilde{D} + \beta_2' \tilde{W} + \tilde{\varepsilon},$$

which simplifies to the decomposition:

$$\tilde{Y} = \beta_1 \tilde{D} + \varepsilon, \quad \mathrm{E}\left[\varepsilon \tilde{D}\right] = 0. \tag{1.3.2}$$

Decomposition (1.3.2) follows because partialling-out eliminates $\beta_2' W$, since $\tilde{W} = 0$, and leaves $\varepsilon$ untouched, $\tilde{\varepsilon} = \varepsilon$, since $\varepsilon$ is linearly unpredictable by $X$ and therefore by $W$. Moreover, $\mathrm{E}[\varepsilon \tilde{D}] = 0$ since $\tilde{D}$ is a linear function of $X = (D, W')'$ and $\varepsilon$ is orthogonal to $X$ and therefore to any linear function of $X$.

The decomposition (1.3.2) implies that $\mathrm{E}\varepsilon\tilde{D} = 0$ are the Normal Equations for the population regression of $\tilde{Y}$ on $\tilde{D}$. Therefore, we just rediscovered the following result.

**Theorem 1.3.1** (Frisch-Waugh-Lovell, FWL [5],[6],[7]) *The population linear regression coefficient $\beta_1$ can be recovered from the population linear regression of $\tilde{Y}$ on $\tilde{D}$:*

$$\beta_1 = \arg\min_{b_1} \mathrm{E}[(\tilde{Y} - b_1\tilde{D})^2] = (\mathrm{E}[\tilde{D}^2])^{-1}\mathrm{E}[\tilde{D}\tilde{Y}],$$

*where we assume $D$ cannot be perfectly predicted by $W$, i.e., $\mathrm{E}[\tilde{D}^2] > 0$, so $\beta_1$ is uniquely defined.*

In other words, $\beta_1$ can be interpreted as a (univariate) linear regression coefficient in the linear regression of *residualized* $Y$ on *residualized* $D$, where the residuals are defined by partialling-out the linear effect of $W$ from $Y$ and $D$.

Technically, these are regression *errors*, not residuals, as we are here working with the population, whereas residuals refer to errors to the sample regression fit. However, we will not adhere strictly to this distinction as it will be convenient to apply analogous logic to partialling-out in the population and the sample.

When we work with the *sample*, we simply mimic the partialling-out operation in the population in the sample. In what follows, we assume $p/n$ is small, so sample linear regression provides high-quality partialling-out. By the FWL Theorem applied to the sample instead of in the population, the sample linear regression of $Y$ on $D$ and $W$ gives us the estimator $\hat{\beta}_1$ which is identical to the estimator obtained via sample partialling-out.

It is useful to give the formula for $\hat{\beta}_1$ in terms of sample partialling-out:

$$\hat{\beta}_1 = \arg\min_{b} \mathbb{E}_n[(\check{Y} - b\check{D})^2] = (\mathbb{E}_n[\check{D}^2])^{-1}\mathbb{E}_n[\check{D}\check{Y}], \quad (1.3.3)$$

where $\check{V}_i$ is the residual left after predicting $V_i$ with controls $W_i$ in the sample and we assume $\mathbb{E}_n[\check{D}^2] > 0$. That is,

$$\check{V}_i = V_i - \hat{\gamma}_{VW}' W_i, \quad \hat{\gamma}_{VW} \in \arg\min_{\gamma} \mathbb{E}_n[(V - \gamma' W)^2].$$

From Theorem 1.2.1, we know that using sample linear regression for partialling-out will provide high-quality estimates of the residuals when $p/n$ is small. When $p/n$ is not small, using sample linear regression for partialling-out won't be such a good idea and an alternative is to use penalized regression or dimension reduction. We will cover this in Chapter 3, but we can definitely try it out in the empirical example that concludes this chapter before we even attempt to understand it.

Why not?

## Adaptive Inference

We next consider the large sample properties of the estimator $\hat{\beta}_1$.

**Theorem 1.3.2** (Adaptive Inference) *Under regularity conditions and if $p/n \approx 0$, the estimation error in $\check{D}_i$ and $\check{Y}_i$ has no first order effect on the stochastic behavior of $\hat{\beta}_1$. Namely,*

$$\sqrt{n}(\hat{\beta}_1 - \beta_1) \approx \sqrt{n}\mathbb{E}_n[\tilde{D}\varepsilon]/\mathbb{E}_n[\tilde{D}^2] \quad (1.3.4)$$

*and consequently,*

$$\sqrt{n}(\hat{\beta}_1 - \beta_1) \overset{a}{\sim} N(0, \mathsf{V})$$

*where*

$$\mathsf{V} = (\mathrm{E}[\tilde{D}^2])^{-1}\mathrm{E}[\tilde{D}^2\varepsilon^2](\mathrm{E}[\tilde{D}^2])^{-1}.$$

We can equivalently write

$$\hat{\beta}_1 \overset{a}{\sim} N(\beta_1, \mathsf{V}/n).$$

That is, $\hat{\beta}_1$ is approximately normally distributed with mean $\beta_1$ and variance $\mathsf{V}/n$. Thus, $\hat{\beta}_1$ concentrates in a $\sqrt{\mathsf{V}/n}$- neighborhood of $\beta_1$ with deviations controlled by the normal law.

The notation $A_n \overset{a}{\sim} N(0, \mathsf{V})$ reads as $A_n$ is approximately distributed as $N(0, \mathsf{V})$. Approximate distribution formally means that $\sup_{R\in\mathcal{R}} |\mathrm{P}(A_n \in R) - \mathrm{P}(N(0, \mathsf{V}) \in R)| \approx 0$, where $\mathcal{R}$ is the collection of rectangular sets (intervals for the case of $A_n$ being a scalar random variable).

The first result in Theorem 1.3.2, equation (1.3.4), states the estimator minus the estimand is an approximate centered average. The remaining properties stated in the theorem then follow from the central limit theorem.

The *adaptivity* refers to the fact that estimation of residuals $\check{D}$ has a negligible impact on the large sample behavior of the OLS estimator – the approximate behavior is the same as if we had used true residuals $\tilde{D}$ instead. This adaptivity property will be derived later as a consequence of a more general phenomenon which we shall call *Neyman orthogonality*.[8]

8: We'll defer the formal defintion of Neyman orthogonality for a bit. See Section 4.3.

The estimated standard error of $\hat{\beta}_1$ is $\sqrt{\hat{\mathsf{V}}/n}$, where $\hat{\mathsf{V}}$ is any estimator of $\mathsf{V}$ based on the plug-in principle such that $\hat{\mathsf{V}} \approx V$. The standard estimator for independent data is called the Eicker-Huber-White robust variance estimator ([8], [9],[10], [11]):

$$\hat{\mathsf{V}} = (\mathbb{E}_n[\check{D}^2])^{-1}\mathbb{E}_n[\check{D}^2\hat{\varepsilon}^2](\mathbb{E}_n[\check{D}^2])^{-1}.$$

This standard error estimator formally works when $p/n \approx 0$, but fails in settings where $p/n$ is not small; see, e.g., [12].

Consider the set, called the $(1-\mathsf{a})\%$ *confidence interval*,

$$[\hat{l}, \hat{u}] := \left[\hat{\beta}_1 - z_{1-\mathsf{a}/2}\sqrt{\hat{\mathsf{V}}/n}, \hat{\beta}_1 + z_{1-\mathsf{a}/2}\sqrt{\hat{\mathsf{V}}/n}\right],$$

where $z_{1-\mathsf{a}/2}$ denotes the $(1-\mathsf{a}/2)$−quantile of the standard normal distribution. For example, the 95% confidence interval is given by

$$\left[\hat{\beta}_1 - 1.96\sqrt{\hat{\mathsf{V}}/n}, \hat{\beta}_1 + 1.96\sqrt{\hat{\mathsf{V}}/n}\right].$$

If we were to envision drawing samples of size $n$ repeatedly from the same population, a $(1-\mathsf{a}) \times 100\%$ confidence interval would contain $\beta_1$ in approximately $(1-\mathsf{a}) \times 100\%$ of the drawn samples:

$$\mathrm{P}(\beta_1 \in [\hat{l}, \hat{u}]) \approx 1 - \mathsf{a}.$$

In other words, aside from "atypical" samples, which occur with probability smaller than $\approx \mathsf{a}$, the confidence interval contains the population value of the best linear predictor coefficient $\beta_1$. Note that what is random in the coverage event "$\beta_1 \in [\hat{l}, \hat{u}]$" is the confidence interval $[\hat{l}, \hat{u}]$, which depends on the specific sample. The population quantity $\beta_1$ is fixed over draws of samples since the population is unchanged.

An alternative is to use $\tilde{\mathsf{V}} = \frac{n}{n-p}(\mathbb{E}_n[\check{D}^2])^{-1}\mathbb{E}_n[\widehat{\varepsilon}^2]$ instead of $\hat{V}$ and the $(1-\mathsf{a}/2)$−quantile of the Student's t-distribution with $n-p$ degrees of freedom instead of $z_{1-\mathsf{a}/2}$. Under these choices, $\mathrm{P}(\beta_1 \in [\hat{l}, \hat{u}]) = 1-\mathsf{a}$ if $\varepsilon \perp\!\!\!\perp X$ and $\varepsilon$ is normal. Normality here is only required for exact coverage. However, coverage may fail to be even approximately $1-\mathsf{a}$ when $\tilde{\mathsf{V}}$ is used and $\varepsilon \perp\!\!\!\perp X$ does not hold. We thus prefer the *robust* variance estimator $\hat{\mathsf{V}}$ because it ensures $\mathrm{P}(\beta_1 \in [\hat{l}, \hat{u}]) \approx 1-\mathsf{a}$ without relying on $\varepsilon \perp\!\!\!\perp X$, known as homoskedasticity. We do sometimes use Student's t quantiles because they converge to standard normal quantiles from above as $n-p$ grows and thus maintain approximate confidence interval coverage.

## 1.4 Application: Wage Prediction and Gaps

In labor economics, an important question is what determines the wage of workers. Interest in this question goes back to the work of Jacob Mincer (see [13]). While determining the factors that lead to a worker's wage is a causal question, we can begin to investigate it from a predictive perspective. We aim to answer two main questions:

- The Prediction Question: How can we use job-relevant characteristics, such as education and experience, to best predict wages?
- The Predictive Effect or Association Question: What is the difference in predicted wages between male and female workers with the same job-relevant characteristics?

We illustrate using data from the 2015 March Supplement of the U.S. Current Population Survey (CPS 2015). As outcome, $Y$, we use the log hourly wage, and we let $X$ denote various

We here use the binary distinction between "male" and "female" workers only because it is thus recorded as a binary variable in the CPS 2015 data, taking only these two values and denoted by "sex." It is, nonetheless, self-reported. We will investigate differences in wages in the two groups defined by this variable. This may be thought to correspond to what is often referred to as a "gender wage gap."

This example also serves as an important reminder that data is simply speech organized into tables and, as such, can encode specific worldviews, subjective definitions, and biases, even when reflecting external observations. Data, including variable names and variable values, should therefore not be taken as objective truth simply because of its dry, tidy form and should instead be understood critically within the context of its collection.

characteristics of workers. We focus on a (sub) sample of single (never married) workers, which is of size $n = 5,150$. Table 1.1 provides mean characteristics of some key variables.

| | Sample Mean |
|---|---|
| Log Wage | 2.97 |
| Female | 0.44 |
| Some High School | 0.02 |
| High School Graduate | 0.24 |
| Some College | 0.28 |
| College Graduate | 0.32 |
| Advanced Degree | 0.14 |
| Experience | 13.76 |

**Table 1.1:** Descriptive statistics for sample of never married workers.

We will estimate a linear predictive (regression) model for log hourly wage using job-relevant characteristics

$$Y = \beta' X + \varepsilon, \quad \varepsilon \perp X,$$

assess the quality of the empirical prediction rule $\hat{\beta}' X$ using out-of-sample prediction performance, and analyze if there is a gap (i.e., difference) in pay for male and female workers (i.e. analyze the so-called "*gender wage gap*"). Any such gap may partly reflect discrimination in the labor market. We will discuss the potential to learn about discrimination in more detail in Chapter 6.

## Prediction of Wages

Predicting Wages R Notebook and Predicting Wages Python Notebook contain the predictive exercise for wages.

Our goal here is to predict (log) wages using various characteristics of workers, and assess the predictive performance of two linear models using adjusted MSE and $R^2$ and out-of-sample MSE and $R^2$.

We employ three different specifications for prediction:

- In the **Basic Model** $X$ consists of a set of raw regressors (e.g. sex, experience, education indicators, occupation and industry indicators, and regional indicators), for a total of $p = 51$ regressors. Our basic specification is inspired by the famous Mincer equation from labor economics; see, e.g., [13] for a review.
- In the **Flexible Model**, $X$ consists of all raw regressors from the basic model as well as *technical regressors*, which

are transformations of the raw regressors, namely, polynomials in experience (experience$^2$, experience$^3$, and experience$^4$) and additional two-way interactions of the polynomials in experience with all other raw regressors except for sex. An example of a regressor created through a two-way interaction is *experience* times the indicator of having a *college degree*. In total, we have $p = 246$ regressors.

| | p | $R^2_{sample}$ | $MSE_{sample}$ | $R^2_{adj}$ | $MSE_{adj}$ |
|---|---|---|---|---|---|
| **basic** | 51 | 0.30 | 0.23 | 0.30 | 0.23 |
| **flexible** | 246 | 0.35 | 0.22 | 0.31 | 0.23 |
| **flexible** Lasso | 246 | 0.32 | 0.23 | 0.31 | 0.23 |

**Table 1.2:** Assessment of predictive performance with in-sample $R^2$ and $MSE$.

To enable both in- and out-of-sample performance evaluation. We start by randomly selecting 80% of the observations as the training sample and keep the other 20% for use as a test sample.

Table 1.2 shows measures of predictive performance in the training data. That is, the table reports predictive performance on the same data that were used to estimate the model parameters. The flexible regression model performs slightly better than the basic model (higher $R^2_{adj}$ and lower $MSE_{adj}$). Note also that the discrepancy between the unadjusted and adjusted measures is not large, which is expected given that

$$p/n \text{ is small.}$$

We report results for evaluating the prediction rules in the test data in Table 1.3. That is, the table reports predictive performance on *new* data that were not used to estimate the models.

| | $MSE_{test}$ | $R^2_{test}$ |
|---|---|---|
| **basic** | 0.197 | 0.328 |
| **flexible** | 0.206 | 0.296 |
| **flexible** Lasso | 0.200 | 0.317 |

**Table 1.3:** Assessment of predictive performance on a 20% validation sample.

Based on this exercise, it appears that the basic regression model works slightly better than the flexible regression at predicting log wages for new observations. That is, we see that the test (out-of-sample) $MSE$ and $R^2$ for the basic regression model are respectively slightly lower and higher than those of the flexible regression model, indicating slightly superior out-of-sample predictive performance. This behavior is different from that obtained when looking at the within sample fit statistics reported in Table 1.2.

Tables 1.2 and 1.3 also provides the test $MSE$ of the flexible model that has been estimated via Lasso regression. Lasso (*least absolute shrinkage and selection operator*) is a penalized regression method that can be used to reduce the complexity of a regression model when the ratio $p/n$ is not small. We introduce this method in Chapter 3, but this does not prevent us from trying it here even though it may appear as a black box at this point. The out-of-sample $MSE$ can be computed for any other black-box prediction method as well. In this example, this method performs similarly to the basic and flexible regression models estimated using OLS. This finding is not surprising given the modest dimensionality and similarity between the performance of the two OLS-estimated models.

Finally, to highlight the potential of estimating the linear model via OLS to overfit, we consider one more model.

- In the **Extra Flexible Model**, $X$ consists sex and all two way interactions between experience, experience$^2$, experience$^3$, experience$^4$, and all other raw regressors except for sex. In total, we have $p = 979$ regressors in this specification.

| | OLS | Lasso |
|---|---|---|
| $MSE_{sample}$ | 0.178 | 0.210 |
| $MSE_{adj}$ | 0.235 | 0.223 |
| $MSE_{test}$ | 0.250 | 0.199 |
| $R^2_{sample}$ | 0.467 | 0.368 |
| $R^2_{adj}$ | 0.345 | 0.331 |
| $R^2_{test}$ | 0.148 | 0.322 |

**Table 1.4:** Assessment of predictive performance in the extra flexible model with $p = 979$ regressors.

We report measures of predictive performance in the training and test data from OLS and Lasso estimates of our "extra flexible" model in Table 1.4. Here, we see that the model estimated by OLS appears to be overfitting. The in-sample statistics substantially overstate predictive performance relative to the performance we see in the test data. For example, the $R^2$ and adjusted $R^2$ in the training data are 0.467 and 0.345, both of which substantially overstate the $R^2$ obtained in the test data, 0.148. We also see that the performance on the test data for the extra flexible model is substantially worse than the performance of the much simpler basic and flexible models. That is, it looks like the OLS estimates of the extra flexible model have specialized to fitting aspects of the training data that do not generalize to the test data and lead to a deterioration in predictive performance relative to the simpler models.

The performance of the Lasso contrasts sharply with this behavior. We see that the in-sample and out-of-sample predictive performance measures for the Lasso based estimates of the extra flexible model are similar to each other. They are also similar to the performance of the simpler models. It seems that Lasso is finding a competitive predictive model without overfitting even in the extra flexible model. We will return to this behavior in Chapter 3 where we will show that Lasso and related methods are able to find good prediction rules in even extremely high-dimensional settings, where for example $p \gg n$, where OLS breaks down theoretically and in practice.

## Wage Gap

Wage Gaps R Notebook and Wage Gaps Python Notebook contain the code for this section.

An important question is whether there is a gap (i.e., difference) in predicted wages between male and female workers with the same job-relevant characteristics. To answer this question, we estimate the log-linear regression model:

$$Y = \beta_1 D + \beta_2' W + \varepsilon, \quad (1.4.1)$$

where $Y$ is log-wage, $D$ is the indicator of being female (1 if female and 0 otherwise) and the $W$'s are other determinants of wages. $W$ includes education, polynomials in experience, region, and occupation and industry indicators plus all two-way interactions of polynomial in experience with region, occupation, and industry indicators.

| | All | Male | Female |
|---|---|---|---|
| Log Wage | 2.9708 | 2.9878 | 2.9495 |
| Less then High School | 0.0233 | 0.0318 | 0.0127 |
| High School Graduate | 0.2439 | 0.2943 | 0.1809 |
| Some College | 0.2781 | 0.2733 | 0.2840 |
| College Graduate | 0.3177 | 0.2940 | 0.3473 |
| Advanced Degree | 0.1371 | 0.1066 | 0.1752 |
| Experience | 13.7606 | 13.7840 | 13.7313 |

**Table 1.5:** Empirical means for the groups defined by the sex variable for never-married workers.

As we have log-transformed wages, we are analyzing the relative difference in pay for male and female workers. Table 1.5 tabulates mean characteristics given sex. It shows that the difference in average log-wage between never married male and never married female workers is equal to 0.038 with male workers earning more. Thus, in this group, male average wage is about 3.8% higher than female average wage.[9] We also observe that never married female workers are relatively more educated than never married male workers.

9: This interpretation relies on the approximation $log(a) - log(b) \approx (a-b)/b$, which is accurate whenever $(a-b)/b$ is small and $b > 0$.

Table 1.6 summarizes the regression results. Overall, we see that the unconditional wage gap of size 3.8% for female workers increases to about 7% after controlling for worker characteristics. This means we would predict a female worker's wage to be about 7% less per hour on average than the wage of a male worker who had the same experience, education, geographical region and occupation.

The partialling-out approach provides a numerically identical estimate for the coefficient $\beta_1$ ($\beta_1 \approx 7\%$), numerically confirming the FWL theorem. Using Lasso for partialling-out (*p-out w/Lasso*) gives similar results to using OLS. This similarity is expected here, since

$$p/n \text{ is small,}$$

and partialling out by least squares should work well.

| | Estimate | Std. Error |
|---|---|---|
| reg without controls | −0.038 | 0.016 |
| reg with controls | −0.070 | 0.015 |
| partial out reg w/ controls | −0.070 | 0.015 |
| Double Lasso (p-out w/ Lasso) | −0.072 | 0.015 |

**Table 1.6:** Estimated conditional wage gaps for never married workers.

> To sum up, our estimate of the conditional wage gap for never-married workers using OLS is about −7% and the 95% confidence interval is about [−10%, −4%].

One way to understand the estimate with controls (−0.070) is as the part of the total gap (−0.038) that cannot be explained by differences in group characteristics. Namely, take Eq. (1.4.1) and average it in the male and female groups to obtain:

$\mathbb{E}_n[\cdot \mid D = d]$ abbreviates $\mathbb{E}_n$ for the subsample of the data where $D = d$, for $d = 0, 1$.

$$\underbrace{\mathbb{E}_n[Y \mid D=1] - \mathbb{E}_n[Y \mid D=0]}_{-0.038} = \underbrace{\hat{\beta}_1}_{-0.070} + \underbrace{\hat{\beta}_2'(\mathbb{E}_n[W \mid D=1] - \mathbb{E}_n[W \mid D=0])}_{0.031}.$$

Here, 0.031 is the difference in log wages we predict based on differences in characteristics. That is, based on observed characteristics $W$ and slopes $\hat{\beta}_2$, we would predict a higher average log wage for female workers than for male workers. This positive difference based on characteristics is counteracted by a negative difference of −0.070 that is *unexplained* by the characteristics and attributable to the difference in the sex variable alone, holding characteristics fixed.

Note these numerical values are rounded so the numbers under the braces across the equation above do not exactly add up.

One missing part in this interpretation is that the model Eq. (1.4.1) does not consider the possible interaction of sex and the characteristics in the prediction of log wage. We can augment Eq. (1.4.1) to account for this, resulting in the interactive log-linear regression model:

$$Y = \beta_1 D + \beta_2' W + \beta_3' W D + \varepsilon.$$

Fitting this new model provides an alternative decomposition:

Such a decomposition that accounts for an *interaction* term is known as a *Oaxaca-Blinder decomposition* introduced in [14] and [15].

$$\underbrace{\mathbb{E}_n[Y \mid D=1] - \mathbb{E}_n[Y \mid D=0]}_{-0.038} = \underbrace{\hat{\beta}_1}_{-2.320} + \underbrace{\hat{\beta}_2'(\mathbb{E}_n[W \mid D=1] - \mathbb{E}_n[W \mid D=0])}_{0.002} + \underbrace{\hat{\beta}_3'\mathbb{E}_n[W \mid D=1]}_{2.280}.$$

Here, 0.002 is the difference attributed to differences in group characteristics. Next, 2.280 is the difference attributed to the different predictive effect of the characteristics in the two groups captured by the coefficients on the interaction terms, $\beta_3$. The difference in predictive effects were previously not considered in the model without interactions. Finally, −2.320 is the remaining difference that remains unexplained by either difference in characteristics *or* their different predictive effect in the two groups.

In order to wrap up and provide a stylized illustration of the impact of dimensionality $p$ on inference, we revisit the extra-flexible model from the previous section which used $p = 979$ controls within a subset of $n = 1000$ of the original observations. This setting gives us $p/n \approx 1$, so the usual theory for estimating linear model coefficients by OLS no longer applies. [16] provide more refined results for OLS estimates of regression coefficients in the case $p/n \to C < 1$. They find that OLS estimates of single coefficients can be consistent in this regime and provide an estimator of the asymptotic variance that is consistent when $p/n < 1/2$ as long as additional regularity conditions hold. They also find that the usual Eicker-Huber-White robust variance estimator is not consistent in this regime but that the jackknife variance estimator, while not consistent, is conservative.

We report estimates of the conditional wage gap within this setup in Table 1.7. We report point estimates from OLS applied to the full set of variables and provide both the Eicker-Huber-White standard error (HC0) and the jackknife standard error (HC3).[10] These are provided for illustration, but we note that HC0 is known to be inconsistent and to behave very poorly, in the sense of generally being far too small, in the high-dimensional setting. HC3 is more reliable, but one should also be skeptical given that $p/n \approx 1$ in this example. Finally, we also report point estimate and standard error for the Double Lasso procedure which is consistent, asymptotically, normal and has estimable standard errors under structure outlined in Chapter 4 even when $p \gg n$. For now, we can think of it is a point of comparison.

10: The Eicker-Huber-White variance estimator is often referred to as "HC0" and the jackknife as "HC3."

| | Estimate | HC0 | HC3 |
|---|---|---|---|
| regression | -0.067 | 0.039 | 0.073 |
| Double Lasso (p-out w/ Lasso) | -0.054 | 0.034 | 0.034 |

**Table 1.7:** The estimated conditional wage gaps for never married workers with approximately 1000 controls in a sample of 1000 observations.

Comparing to the case with the full data set, we see that point estimates are not wildly different but that standard errors are larger. Part of the standard error difference is predicted simply by the difference in sample sizes. Specifically, $\sqrt{5150/1000} \approx 2.27$, so we would expect standard errors to be 2.27 times bigger with $n = 1000$ observations than with $n = 5150$. This inflation holds almost exactly for the Double Lasso.

More interestingly, now that $p/n \not\approx 0$, we start seeing substantial differences in standard errors between unregularized partialling out (regression) and partialling out with Lasso (aka Double Lasso). While we don't want to take the OLS standard errors too seriously – we know the Huber-Eicker-White standard error does not work in this setting and are also suspect of the jackknife here – the comparison between the OLS and Double Lasso standard errors and comparison to the full sample results are revealing. Compared to the full sample results, the jackknife standard error (HC3) is much larger than would be expected simply due to the decrease in the sample size in this example. The difference from this expectation (partially) reflects the impact of dimensionality on the OLS estimate of the regression coefficient. The Double Lasso seems to be roughly insensitive to the dimensionality of the control variables and scales exactly as one would expect given the difference in sample size.

The punchline of this final example is that OLS is no longer adaptive in the "$p/n$ not small" regime. The lack of adaptivity means that conventional properties of OLS may not hold and that other procedures may be highly preferable to OLS.

## Notebooks

- Predicting Wages R Notebook and Predicting Wages Python Notebook contain a simple predictive exercise for wages. We will return to this dataset and prediction problem repeatedly in future chapters, re-estimating it using a broad range of ML estimators and providing a means of comparing their performance.
- Wage Gaps R Notebook and Wage Gaps Python Notebook contain a simple analysis of wage gaps.
- The Linear Model Overfitting R Notebook and the Linear Model Overfitting Python Notebook contain a set of simple simulations that show how measures of fit perform in a high $p/n$ setting.

## 1.5 Notes

Least squares were invented by Legendre ([17]) and Gauss ([18]) around 1800. Frisch, Waugh, and Lovell ([5],[6],[7]) discovered the partialling-out interpretation of the least squares coefficients in the 1930s. The asymptotic theory mentioned in the note is more recent and has been developed since early work of Huber in the 70s on $m$-estimators (estimators that minimize objective functions that correspond empirical averages of losses) under moderately high dimensions; see e.g. [19] and the textbook [20].

For a good, concise treatment of classical least squares, see for example, Chapter 1 in Amemiya's classical graduate econometrics text [3]; Bruce Hansen's new textbook [21] is an excellent up-to-date reference.

Regularity conditions under which Theorem 1.2.1 and Theorem 1.3.2 hold under $p \to \infty$ and $p/n \to 0$ asymptotics can be found in [22] and [16]. The results of the latter reference allow for $p/n \to c < 1$, which introduces an additional asymptotic variance term when $c > 0$; the case with $c = 0$ recovers Theorem 1.3.2. See also review [23] for some recent understanding of properties of least squares estimators.

## Study Questions

Modern notebooks, including Jupyter Notebooks, R Markdown, and Quarto offer a simple way to integrate code cells and explanations (text and formulas) in a single notebook. This allows the user to execute code in discretized chunks for clarity and ease of debugging as well as to better provide commentary on what the code is doing. See the Notebooks section above for examples.

1. Write a notebook (R, Python, etc.) where you briefly explain the idea of sample splitting to evaluate the performance of prediction rules to a fellow student, and show how to use it on the wage data. The explanation should be clear and concise (one paragraph suffices) so that a fellow student can understand. You can take our notebooks as a starting point, but provide a bit more explanation and modify them by exploring different specifications of the models (or looking at an interesting subset of the data or even other data – for example, the data you use for your research or thesis work).

2. Write a notebook (R, Python, etc), where you carry out a wage gap analysis, focusing on the subset of college-educated workers. The analysis should be analogous to what we've presented – explaining "partialling out," generating point estimates and standard errors – but don't hesitate to experiment and explain more. Exploring other data-sets or similar questions, e.g. wage gaps by race, is always welcome.

3. The half-serious link to Pythagoras was serious in its half. Consider sample linear regression with $n = 2$ and just one regressor, so that $Y_i = \hat{\beta} X_i + \hat{\varepsilon}_i$ for $i = 1, 2$, where $\hat{\beta}$ is the ordinary least squares estimator, a scalar quantity in this case. Let $\mathrm{Y} = (Y_1, Y_2)'$ , $\mathrm{X} = (X_1, X_2)'$, $\hat{\varepsilon} = (\hat{\varepsilon}_1, \hat{\varepsilon}_2)'$, and let $\hat{\mathrm{Y}} = \hat{\beta}\mathrm{X}$. Find the connection between the decomposition $\mathrm{Y}'\mathrm{Y}/n = \hat{\mathrm{Y}}'\hat{\mathrm{Y}}/n + \hat{\varepsilon}'\hat{\varepsilon}/n$ and the Pythagorean theorem. Find the geometric interpretation for $\hat{Y}$, and write the explicit formula for $\hat{\beta}$ in this case. If you get stuck, google the "geometric interpretation of least squares."

## 1.A Central Limit Theorem

### Univariate

Consider the scaled sum $W = \sum_{i=1}^{n} X_i/\sqrt{n}$ of independent and identically distributed variables $X_i$ such that $\mathrm{E}[X] = 0$ and $\mathrm{Var}(X) = 1$. The classical CLT states that $W$ is approximately Gaussian provided that none of the summands are too large,

namely

$$\sup_{x\in\mathbb{R}} |\mathrm{P}(W \le x) - \mathrm{P}(N(0,1) \le x)| \approx 0.$$

This result is reassuring, but the theorem does not inform us how small the error is in a given setting.

The Berry-Esseen theorem provides a quantitative characterization of the error.

**Theorem 1.A.1** (Berry-Esseen's Central Limit Theorem)

$$\sup_{x\in\mathbb{R}} |\mathrm{P}(W \le x) - \mathrm{P}(N(0,1) \le x)| \le K\mathrm{E}[|X|^3]/\sqrt{n},$$

*for a numerical constant* $K < .5$.

The result asserts that the Gaussian approximation error rate declines like $1/\sqrt{n}$. It also states that given $n$, the approximation quality improves as the third absolute moment $\mathrm{E}[|X|^3]$ decreases. This results gives a good guide regarding when the Gaussian approximation gives accurate results.[11] Of course, one can also check the approximation quality via simulation experiments that mimic the practical situation.

11: Consider, for instance, the case when $X_i$ are centered and standardized Bernoulli random variables with success probability $p$, i.e., $X_i = \frac{Z_i - p}{\sqrt{p(1-p)}}$ and $Z_i$ is Bernoulli with success probability $p$. The error in the Berry-Esseen theorem, in this case, becomes $\approx 1/\sqrt{p(1-p)n}$. Thus, the error in the Gaussian approximation is guaranteed to be small by the Berry-Esseen theorem only if $p(1-p)n$ is large. Thus, for extreme probabilities, where either success or failure events are extremely rare for the given sample size, i.e., when $p \cdot n$ or $(1-p) \cdot n$ is small, the use of the Gaussian approximation is not advisable.

## Multivariate

Later in the book, we will use multivariate central limit theorems as well. To this end, we are going to state the following more general result due to [24], which refines earlier results by [25] and [26].

Let $\mathcal{I}$ be a countable set (either finite or infinite) and let $X_i, i \in \mathcal{I}$, be independent $\mathbb{R}^d$-valued random vectors. Assume that $\mathrm{E}[X_i] = 0$ for all $i$ and that $\sum_{i\in\mathcal{I}} \mathrm{Var}(X_i) = I_d$. It is well known that in this case, the sum $W := \sum_{i\in\mathcal{I}} X_i$ exists almost surely and that $\mathrm{E}W = 0$ and $\mathrm{Var}(W) = I_d$.

**Theorem 1.A.2** (Multivariate CLT; [24]) *For* $X_i$ *and* $W$ *as above and all measurable convex sets* $A \subseteq \mathbb{R}^d$, *we have*

$$|\mathrm{P}(W \in A) - \mathrm{P}(N(0, I_d) \in A)| \le \left(42d^{1/4} + 16\right) \sum_{i\in I} \mathrm{E}\left[\|X_i\|^3\right].$$

# Causal Inference via Randomized Experiments

# 2

> "Let us divide them in halfes, let us cast lots, that one half of them may fall to my share, and the other to yours; I will cure them without bloodletting and sensible evacuation; but do you do as ye know [. . . ] we shall see how many Funerals both of us shall have."
>
> – Jan Baptist van Helmont [1].

In this chapter we begin discussion of causal inference by focusing on Randomized Control Trials (RCTs). In a randomized control trial, units are randomly divided into those that receive a treatment and those that receive no treatment. Under randomization and other assumptions, the difference in average outcomes between the treated and untreated groups is an average treatment (causal) effect (ATE). By considering pre-treatment covariates, we can improve the precision of the ATE estimate, explore heterogeneity across subgroups, or both. We describe methods for doing so and apply them to several RCTs. We introduce causal diagrams as a means of visualizing RCTs and their underlying causal assumptions. We conclude by outlining some limitations of RCTs.

## 2.1 Potential Outcomes Framework and Average Treatment Effects

In this section, we discuss the potential outcomes framework for analyzing causality and treatment effects. It offers an elegant way to formalize counterfactuals as a mathematical concept.

We begin by introducing the two *latent* (unobserved) variables

$$Y(1) \text{ and } Y(0).$$

They represent the potential or counterfactual random outcomes for an observational unit when the unit is subject to treatment (treatment state $d = 1$) or no treatment (control or untreated state $d = 0$) [2]. In an economic context, the treatment might be a training program or a policy intervention, and the outcome might be an individual's wage or employment status. In what follows, it is also useful to introduce the potential response or structural function:

$$d \mapsto Y(d),$$

which maps the potential treatment state $d \in \{0, 1\}$ to the random potential outcome $Y(d)$.

For simplicity, we do not consider multivalued or continuous treatments.

In this formulation, we have dependence of the potential outcome $Y(d)(\omega)$ on the underlying state of the world $\omega$. In our formalization, $\omega$ will represent randomness across observational units and from any other sources.[1]

1: Recall that a random variable $V$ is a mapping $\omega \mapsto V(\omega)$ from the underlying state of the world $\omega \in \Omega$ to the real line (or other metric space) such that we can assign a probability law to it.

The quantities $Y(1)$ and $Y(0)$ are "counterfactual" because they can't be simultaneously observed. That is, we generally do not have identical replicas of the observational units that are simultaneously subject to both treatment and control. [3] calls the inability to observe an individual simultaneously under treatment and control "the fundamental problem of causal inference". The inability to observed each individual's treatment and control outcome means that causal inference shares many features with "missing data"problems, see, e.g. [4].

The individual treatment effect is

$$Y(1) - Y(0).$$

This effect will vary across individuals as well as with other sources of randomness encoded in $\omega$. As mentioned above, only one of the two terms is actually observed, and hence it is generally infeasible to uncover the individual treatment effect.[2] However, we can hope to estimate averages and the distribution

2: As an example, we could uncover individual treatment effects if we had identical twins that could be put in treatment and control groups, and we believed that the only difference in outcomes between these twins is induced by treatment – that is, $\omega$ only depends on genetic makeup. Such an example seems unrealistic at best.

of $Y(d)$ at the population level to compute quantities such as the average treatment effect (ATE):

$$\delta = \mathrm{E}[Y(1) - Y(0)] = \mathrm{E}[Y(1)] - \mathrm{E}[Y(0)].$$

Let $D$ denote the actual *assigned treatment*, a random variable, which takes a value of 1 if the observational unit participated in the treatment and 0 otherwise.

**Assumption 2.1.1** (Consistency) *We observe*

$$Y := Y(D).$$

For example, if treatment ($D = 1$) corresponded to completion of a job training program and control ($D = 0$) corresponded to not completing the program, Assumption 2.1.1 says that the observed wage outcome is equal to $Y(1)$ for a given person if she has completed the program (has $D = 1$) and is equal to $Y(0)$ if this person has not completed the training program (has $D = 0$). Assumption 2.1.1 seems almost tautological, but it importantly rules out hidden variation in treatment. That is, it requires that the treatment and control states are well-defined and clearly aligned with the observed treatment status, $D$.

**Assumption 2.1.2** (No Interference) *Potential outcomes for any observational unit depend only on the treatment status of that unit and not on the treatment status of any other unit.*

Assumption 2.1.2 has implicitly been captured in our definition of potential outcomes, $Y(d)$, which give the outcome of each unit *when the unit* is subject to treatment state $d$. This formulation rules out scenarios where the treatment given to one unit may impact the outcome of a different unit. Such spillovers could occur, for example, on social networks where treating an individual could impact all of that individual's friends. Some forms of spillovers are readily accommodated by expanding the definition of treatment and correspondingly adjusting definition of potential outcomes,[3] but treating these extensions is beyond the scope of this book.[4]

Assumptions 2.1.1 and 2.1.2 encapsulate what is often referred to as the Stable Unit-Treatment Value Assumption (SUTVA); see, e.g. Imbens and Rubin [10].

The following analytical example may help gain better understanding of the potential outcomes framework.

3: For example, consider a case where each individual has two friends. We could define potential outcomes allowing for spillovers as $Y(d_0, d_1, d_2)$ where $d_0$ denotes the treatment state of an individual, $d_1$ denotes the treatment state of the individual's friend 1, and $d_2$ denotes the treatment state of the individual's friend 2.

4: For further reading we refer, among many others, to [5], [6], [7], [8] and [9].

**Example 2.1.1** [Analytical Example] Consider the following model

$$\begin{aligned} Y(1) &:= \theta_1 + \epsilon_1 \\ Y(0) &:= \theta_0 + \epsilon_0 \\ D &:= 1(\nu > 0), \\ Y &:= Y(D), \end{aligned}$$

where $\theta_0$ and $\theta_1$ are constants, and $(\epsilon_0, \epsilon_1, \nu)$ are jointly normal random stochastic disturbances with mean 0 and covariance matrix $\Sigma$. Here, $\nu$ represents factors that influence selection into the treatment state. In this example $\mathrm{E}[Y(1)] = \theta_1$, $\mathrm{E}[Y(0)] = \theta_0$, and the ATE is $\delta = \theta_1 - \theta_0$. Importantly, only $D$ and $Y$ are observed.

Under Assumption 2.1.1, population data directly provide the conditional averages

$$\mathrm{E}[Y \mid D = d] = \mathrm{E}[Y(d) \mid D = d], \text{ for } d \in \{0, 1\}.$$

The difference of the two averages gives us the average predictive effect (APE) of treatment status on the outcome:

$$\pi = \mathrm{E}[Y \mid D = 1] - \mathrm{E}[Y \mid D = 0].$$

It measures the association of the treatment status with the outcome.

While the APE is identified – meaning computable from the population data – it may seem surprising (or not at all) that the APE in general does not agree with the ATE $\delta$:

$$\delta \neq \pi. \tag{2.1.1}$$

The difference between the APE and ATE is generally said to be due to *selection bias*. The meaning of selection bias is clarified through the following example, and clarified theoretically below.

**Example 2.1.2** (Selection Bias in Observational Data) Suppose we want to study the impact of smoking marijuana on life longevity. Suppose that smoking marijuana has no causal effect on life longevity:

$$Y = Y(0) = Y(1),$$

so that

$$\delta = \mathrm{E}[Y(1)] - \mathrm{E}[Y(0)] = 0.$$

However, the observed smoking behavior, $D$, is not assigned in an experimental study. Suppose that the behavior determining $D$ is associated with poor health choices such as drinking alcohol, which are known to cause shorter life expectancy, so that $\mathrm{E}[Y \mid D = 1] < \mathrm{E}[Y \mid D = 0]$. In this case, we have negative a predictive effect:

$$\pi = \mathrm{E}[Y \mid D = 1] - \mathrm{E}[Y \mid D = 0] < 0 = \delta,$$

which differs from the true causal effect $\delta = 0$.

To sum up, in the smoking example, the chosen "treatment" variable $D$ is potentially negatively associated with the potential health outcome, inducing the selection bias – the difference between the predictive effect and the causal effect.

**Example 2.1.3** (Analytical Version of the Smoking Example) To capture dependence between $Y(d)$ and $\nu$ in the smoking context analytically, we can go back to Example 2.1.1, and make variables $\epsilon_d$ and $\nu$ be negatively associated:

$$\mathrm{E}[\epsilon_d \nu] < 0.$$

The negative association between the $\epsilon_d$ and $\nu$ then results in the observed smoking status, $D$, being negatively associated with the potential outcomes $Y(d)$. Specifically, we have

$$\mathrm{E}[Y|D = 1] < \mathrm{E}[Y|D = 0],$$

which can be verified through additional analytical calculations or via simulation experiments (a homework).

It is useful to emphasize the main reason for having selection bias is that

$$\mathrm{E}[Y(d)|D = 1] \neq \mathrm{E}[Y(d)]$$

whenever $D$ is not independent of $Y(d)$. If $D$ and $Y(d)$ were independent,

$$\mathrm{E}[Y(d)|D = 1] = \mathrm{E}[Y(d)]$$

would hold since in this case $D$ is uninformative about the potential outcome and drops out from the conditional expectation.

To sum up, the problem with observational studies like our contrived Example 2.1.2 is that the "treatment" variable $D$

is determined by individual behaviors which may be linked to potential outcomes. This linkage generates selection bias - the disagreement between APE and ATE. There are many ways of addressing selection bias, one of which is through an experiment, where we randomly assign the treatment to the units.

## Random Assignment/Randomized Controlled Trials

A way to clearly remove selection bias is through random assignment of treatment.

**Assumption 2.1.3** (Random Assignment/Exogeneity) *Suppose that treatment status is randomly assigned. Namely, $D$ is statistically independent of each potential outcome $Y(d)$ for $d \in \{0, 1\}$, which is denoted as*

$$D \perp\!\!\!\perp Y(d)$$

*and* $0 < \mathrm{P}(D = 1) < 1$.

This assumption states that the treatment assignment mechanism is purely random, and ensures that there are units in treatment and in control.

**Example 2.1.4** (Analytical Example Continued) In the analytical example 2.1.1, Assumption 2.1.3 is satisfied if the stochastic shock $\nu$ determining $D$ is independent of stochastic shocks $\epsilon_0$ and $\epsilon_1$ determining $Y(1)$ and $Y(0)$, i.e.

$$\nu \perp\!\!\!\perp (\epsilon_0, \epsilon_1).$$

A key result is that selection bias is removed under Assumption 2.1.3 which allows us to learn summaries of causal effects.

**Theorem 2.1.1** (Randomization Removes Selection Bias) *Under Assumption 2.1.3, the average outcome in treatment group $d$ recovers the average potential outcome under the treatment status $d$:*

$$\mathrm{E}[Y \mid D = d] = \mathrm{E}[Y(d) \mid D = d] = \mathrm{E}[Y(d)],$$

*for each $d \in \{0, 1\}$. Hence the average predictive effect and average treatment effect coincide:*

$$\begin{aligned} \pi &:= \mathrm{E}[Y \mid D = 1] - \mathrm{E}[Y \mid D = 0] \\ &= \mathrm{E}[Y(1)] - \mathrm{E}[Y(0)] =: \delta. \end{aligned}$$

Assumption 2.1.3 is often not plausible for observational data. In a *randomized controlled trial* (RCT)[5] , the aim is to ensure the plausibility of Assumption 2.1.3 by direct random assignment of treatment $D$. That is, subjects are randomly assigned a treatment state $D$ by the experimenter without regard to any of their characteristics. Because the random assignment of the treatment is unrelated to all subject characteristics by construction, well-executed RCTs guarantee that Assumption 2.1.3 is satisfied. Because of this property, many consider RCTs as the gold standard in causal inference, and RCTs are routinely employed in a variety of important settings.[6] Examples include evaluating the efficacy of medical treatment, vaccinations, training programs, marketing campaigns, and other kinds of interventions.

5: Synonyms are experiments and A/B tests.

6: Of course, RCTs must be correctly done to guarantee Assumption 2.1.3. For example, RCTs where experimental protocols are not followed continue to suffer from selection bias. There are also examples, *quasi-experiments*, where we may believe that Assumption 2.1.3 is plausible that do not correspond to explicit designed experiments.

**Example 2.1.5** (No Selection Bias in Experimental Data) Suppose that in the smoking example (Example 2.1.2), we worked with data where smoking or non-smoking was generated by perfectly enforced random assignment. In this case, we would have agreement between average predictive and treatment effects: $\pi = \delta$. While it is difficult to imagine a long-run RCT where study participants could be forced to smoke or not smoke marijuana (we discuss such limitations as well as ethical considerations in Section 2.4), RCTs are routinely employed in a variety of other important settings.

## Statistical Inference with Two Sample Means

Inference is based on the independent sample $\{(Y_i, D_i)\}_{i=1}^n$ obtained from an RCT, where index $i$ denotes the observational unit. We assume that each $(Y_i, D_i)$ has the same distribution as $(Y, D)$. Estimation of the two means $\theta_d = \mathrm{E}[Y \mid D = d]$ for $d = 0$ and $d = 1$ can be done by considering two group means

$$\hat{\theta}_d = \frac{\mathbb{E}_n[Y1(D = d)]}{\mathbb{E}_n[1(D = d)]}.$$

The two means example can also be treated as a special case of linear regression,[7] but we find it instructive to work out the details directly for the two group means. We provide these details in Section 2.A.

7: Indeed, we can regress $Y$ on $D$ and $1 - D$; that is, estimate the model $Y = \theta_1 D + \theta_0(1-D) + U$. We can then apply the inferential machinery developed in the previous chapter.

Under mild regularity conditions, we have that

$$\sqrt{n}\begin{pmatrix} \hat{\theta}_0 - \theta_0 \\ \hat{\theta}_1 - \theta_1 \end{pmatrix} \overset{a}{\sim} N(0, \mathrm{V}),$$

where

$$V = \begin{pmatrix} \frac{\text{Var}(Y|1(D=0))}{P(D=0)} & 0 \\ 0 & \frac{\text{Var}(Y|1(D=1))}{P(D=1)} \end{pmatrix}$$

so that $\hat{\delta} = \hat{\theta}_1 - \hat{\theta}_0$ obeys

$$\sqrt{n}(\hat{\delta} - \delta) \overset{a}{\sim} N(0, V_{11} + V_{22}).$$

To use this result in practice, variance components are usually estimated using the *plug-in principle*, which amounts to using the sample analogues of the expressions above.

Sometimes we are interested in relative effectiveness of treatment effects (for example, vaccine efficiency):

$$f(\theta) = (\theta_1 - \theta_0)/\theta_0 = \delta/\theta_0.$$

Relative effectiveness can be estimated by $\hat{\delta}/\hat{\theta}_0 = f(\hat{\theta})$, where $\hat{\theta} = \{\hat{\theta}_d\}_{d\in\{0,1\}}$ and $\theta = \{\theta_d\}_{d\in\{0,1\}}$, with approximate distribution obtained using the *delta method*:

$$\sqrt{n}(f(\hat{\theta}) - f(\theta)) \approx G'\sqrt{n}(\hat{\theta} - \theta) \overset{a}{\sim} N(0, G'VG),$$

where $G = \nabla f(\theta)$, $\hat{\theta} = (\hat{\theta}_0, \hat{\theta}_1)'$, $\theta = (\theta_0, \theta_1)$.[8] .

8: The approximation follows from application of the first order Taylor expansion and continuity of the derivative $\nabla f$ at $\theta$.

## Pfizer/BioNTech Covid Vaccine RCT

Pfizer/BNTX was the first vaccine approved for emergency use in the EU and US to reduce the risk of Covid-19 disease. See the Food and Drug Administration (FDA) briefing for details about the RCT and the summary data. Volunteers were randomly assigned to receive either a treatment (2-dose vaccination) or a placebo, without knowing which they received, and the doctors making the diagnoses did not know whether a given volunteer received a vaccination or not. In other words, the trial was a double-blind randomized control trial. The results of the study are presented in the following table.

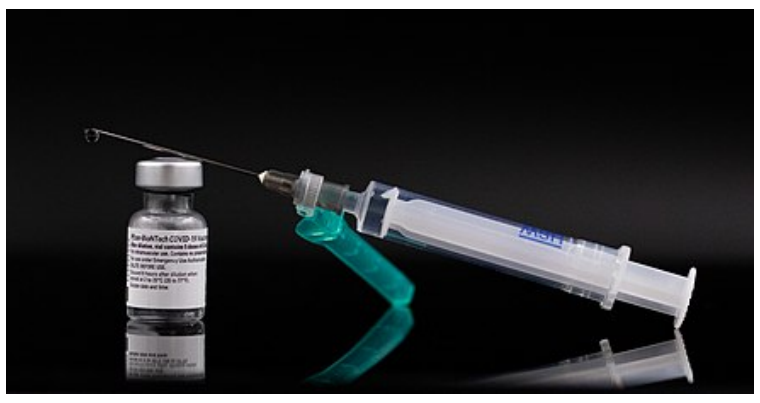

**Figure 2.1:** Tozinameran (Pfizer-BioNTech Covid-19 vaccine); Image Source: Wikipedia / Arne Müseler

Vaccination RCT R Notebook and Vaccination RCT Python Notebook contain the analysis of the Pfizer-BioNTech Covid-19 Vaccine RCTs.

| Efficacy Endpoint Subgroup | BNT162b2 N[a]=19965 Cases n1[b] Surveillance Time[c] (n2[d]) | Placebo N[a]=20172 Cases n1[b] Surveillance Time[c] (n2[d]) | Vaccine Efficacy % (95% CI)[e] |
|---|---|---|---|
| Overall | 9<br>2.332 (18559) | 169<br>2.345 (18708) | 94.6 (89.6, 97.6) |
| Age group (years) | | | |
| 16 to 17 | 0<br>0.003 (58) | 1<br>0.003 (61) | 100.0 (-3969.9, 100.0) |
| 18 to 64 | 8<br>1.799 (14443) | 149<br>1.811 (14566) | 94.6 (89.1, 97.7) |
| 65 to 74 | 1<br>0.424 (3239) | 14<br>0.423 (3255) | 92.9 (53.2, 99.8) |
| ≥75 | 0<br>0.106 (805) | 5<br>0.109 (812) | 100.0 (-12.1, 100.0) |

**Figure 2.2:** The aggregate data from the Pfizer RCT; source: FDA briefing.

We see that the rate of Covid-19 infection was relatively low at the time. Specifically, the treatment group saw 9 Covid-19 cases per 19,965, while the control group saw 169 cases per 20,172.

The estimated average treatment effect is about

$$-792.7 \text{ cases per } 100{,}000,$$

and the 95% confidence band is[9]

$$[-922, -664].$$

Under Assumptions 2.1.3 and 2.2.1 the confidence band suggests that the Covid-19 vaccine caused a reduction in the risk of contracting Covid-19.

We also compute the Vaccine Efficacy metric, which according to [11], refers to the following measure:

$$\mathsf{VE} = \frac{\text{Risk for Unvaccinated - Risk for Vaccinated}}{\text{Risk for Unvaccinated}}.$$

It describes the relative reduction in risk caused by vaccination. Estimating the VE is simple as we can plug-in the estimated group means. We can compute standard errors using the delta method or by simulation. We obtain that the overall vaccine efficacy is 94.6%, replicating the results shown in Figure 2.2. Our 95% confidence interval for VE, based on the normal approximation, is

$$[90.9\%, 98.2\%],$$

which differs only slightly from the FDA briefing table.[10]

9: In this example, we don't need the underlying individual data to evaluate the effectiveness of the vaccine because the potential outcomes are Bernoulli random variables with mean $\mathrm{E}[Y(d)]$ and variance $\mathrm{Var}(Y(d)) = \mathrm{E}[Y(d)(1 - \mathrm{E}Y(d))]$.

10: The analysis in the FDA table is based on the inversion of exact binomial tests, the Cornfield procedure.

**Remark 2.1.1** We notice that the confidence intervals for the VE for the two age groups of seniors are very wide, so to increase precision we pool them together and calculate the effectiveness of the vaccine for the two groups that are 65 or older. The resulting VE estimate is 95% and the two-sided confidence interval based on the normal approximation is

$$[82\%, 106\%]$$

A more refined approach is possible, based on the inversion of exact binomial ratio Cornfield tests [12], which we report in Vaccination RCT R Notebook and Vaccination RCT Python Notebook. This approach, using Vaccination RCT R Notebook, yields a confidence interval of

$$[69\%, 99\%].$$

The reason is that the accumulated counts of binomials are too few for the Gaussian approximations to provide a high-quality approximation, so the exact binomial ratio test inversion delivers a more accurate confidence interval.

## 2.2 Pre-treatment Covariates and Heterogeneity

Sometimes we also have additional *pre-treatment* or *pre-determined* covariates $W$. We might be interested in either using these covariates to estimate average effects more precisely or to describe heterogeneity of the treatment effects. For example, we might be interested in the impact of a treatment across age or income groups.

For this purpose, we consider conditional average treatment effects (CATE):

$$\delta(W) = \mathrm{E}[Y(1) \mid W] - \mathrm{E}[Y(0) \mid W],$$

which compare the average potential outcomes conditional on a set of covariates $W$.

We can directly learn the conditional predictive effects (CAPE),

$$\pi(W) = \mathrm{E}[Y \mid D = 1, W] - \mathrm{E}[Y \mid D = 0, W],$$

from population data. However, these CAPE will generally not agree with the CATE. One assumption that will be sufficient for the CAPE and CATE to agree is having treatment assigned randomly and independently of covariates. As before, the use of RCTs help ensure the plausibility of this assumption.

**Assumption 2.2.1** (Random Assignment Independent of Co-

variates) *Suppose that treatment status is randomly assigned. Namely, $D$ is statistically independent of both the potential outcomes and a set of pre-determined covariates:*

$$D \perp\!\!\!\perp (Y(0), Y(1), W),$$

*and* $0 < \mathrm{P}(D = 1) < 1$.

This assumption spells out that, if we plan to use covariates in the analysis, randomization has to be made with respect to these covariates as well. In practice, it is often tempting to use post-treatment covariates, but the use of such variables runs the danger of violating Assumption 2.2.1. In the extreme case, conditioning on the post-treatment observed outcome $Y$, we find that $\pi(Y) = 0$, even when there is a treatment effect. In a less extreme case, conditioning on post-treatment variables related to the outcome can "control-away" part of the effect, diminishing estimates.

A common scenario where accidentally using a post-treatment covariate may occur is when researchers encounter missing data from imperfect data collection in following-up with control and treated units to collect demographic information. When we drop observations with missing data, we implicitly condition on a post-treatment variable (missingness) which can cause violations of Assumption 2.2.1.

The desire to assess randomization with respect to covariates motivates the following diagnostic procedure.

For random variables $A$ and $B$, $A \sim B$ denotes that $A$ and $B$ have the same distribution.

**Testing Covariance Balance.** The random assignment assumption induces covariate balance. Namely, the distribution of covariates should be the same under both treatment and control:

$$W|D = 1 \sim W|D = 0,$$

and, equivalently,

$$D|W \sim D.$$

A useful implication is that $D$ is not predictable by $W$:

$$\mathrm{E}[D \mid W] = \mathrm{E}[D].$$

This latter conditions is testable using regression tools. It amounts to saying that the $R^2$ of a regression of $D$ on $W$ is 0.

Under Assumption 2.2.1, Theorem 2.1.1 continues to hold, but we now have a stronger result.

**Theorem 2.2.1** (Randomization with Covariates) *Under Assumption 2.2.1, the expected value of $Y$ conditional on treatment status $D = d$ and covariates $W$ coincides with the expected value of potential outcome $Y(d)$ conditional on covariates $W$:*

$$\mathrm{E}[Y \mid D = d, W] = \mathrm{E}[Y(d) \mid D = d, W] = \mathrm{E}[Y(d)|W],$$

*for each $d$. Hence the conditional predictive and average treatment effects agree:*

$$\pi(W) = \delta(W).$$

## Regression and Statistical Inference for ATEs

Empirical researchers often base statistical inference on the ATE using the classical additive linear regression model, where covariates enter additively in the model. This approach has some good practical properties and often empirically leads to improvements in precision over the simple two-means approach, though this precision improvement is not guaranteed. Another approach that we will emphasize is the interactive regression approach, where de-meaned covariates are also interacted with the base treatment. Including interactions of de-meaned covariates with the treatment always improves precision, and it also allows us to discover treatment effect heterogeneity.

## Classical Additive Approach: Improving Precision Under Linearity

We begin explaining the classical additive approach. Here, to simplify the exposition, we make the strong assumption that the conditional expectation function is exactly linear:

$$\mathrm{E}[Y \mid D, W] = D\alpha + \beta' X, \tag{2.2.1}$$

where $X = (1, W)$ contains an intercept and pre-treatment covariates $W$. This setup is clearly restrictive, but the statistical inference result will be valid without this assumption.[11] Later in the book, we will consider fully nonlinear models.

We assume that covariates are centered:[12]

$$\mathrm{E}[W] = 0.$$

By Assumption 2.2.1, there is covariate balance:

$$\mathrm{E}[W \mid D = 1] = \mathrm{E}[W \mid D = 0].$$

11: See Section 2.B for details.

12: Theoretically, this is implemented by redefining $W := W - \mathrm{E}[W]$. In estimation, this is implemented by redefining $W_i := W_i - \mathbb{E}_n[W]$.

Using centered covariates implies that

$$\mathrm{E}[Y(0)] = \mathrm{E}[\mathrm{E}[Y \mid D = 0, X]] = \beta_1$$

$$\mathrm{E}[Y(1)] = \mathrm{E}[\mathrm{E}[Y \mid D = 1, X]] = \beta_1 + \alpha.$$

That is, the average outcome in the untreated state is $\beta_1$, and the average treatment effect $\delta = \mathrm{E}[Y(1)] - \mathrm{E}[Y(0)]$ equals $\alpha$.

Equation (2.2.1) implies that

$$Y = D\alpha + \beta' X + \epsilon, \quad \epsilon \perp (D, X), \tag{2.2.2}$$

implying that $\alpha$ coincides with the coefficient in the BLP of $Y$ on $D$ and $X$. In fact, even if we don't assume the model (2.2.1), we still have that $\alpha = \delta$. That is, the projection coefficient $\alpha$ recovers the ATE $\delta$ without the linearity assumption as we detail in Section 2.B. Furthermore the statistical inference result stated below will hold without requiring linear conditional expectation functions as it is simply a statement about inference on the BLP.

We are interested in statistical inference on the ATE and Relative ATE[13]

$$\alpha \quad \text{and} \quad \alpha/\beta_1.$$

13: Relative ATE is often called *lift* in business applications.

Under regularity conditions, application of the OLS theory from Chapter 1 gives us

$$\begin{pmatrix} \sqrt{n}(\hat{\alpha} - \alpha) \\ \sqrt{n}(\hat{\beta}_1 - \beta_1) \end{pmatrix} \overset{\mathrm{a}}{\sim} N(0, \mathsf{V}),$$

where covariance matrix $\mathsf{V}$ has components:

$$\mathsf{V}_{11} = \frac{\mathrm{E}[\epsilon^2 \tilde{D}^2]}{(\mathrm{E}[\tilde{D}^2])^2}, \quad \mathsf{V}_{22} = \frac{\mathrm{E}[\epsilon^2 \tilde{1}^2]}{(\mathrm{E}[\tilde{1}^2])^2}, \quad \mathsf{V}_{12} = \mathsf{V}_{21} = \frac{\mathrm{E}[\epsilon^2 \tilde{D}\tilde{1}]}{\mathrm{E}[\tilde{1}^2]\mathrm{E}[\tilde{D}^2]},$$

where $\tilde{D} = D - \mathrm{E}[D]$ is the residual after partialling out $X$ from $D$ linearly and $\tilde{1} := (1 - D)$ is the residual after partialling out $D$ and $W$ from 1.

We also obtain the approximate normality for the Relative ATE using the delta method:

$$\sqrt{n}(\hat{\alpha}/\hat{\beta}_1 - \alpha/\beta_1) \overset{\mathrm{a}}{\sim} N(0, G'\mathsf{V}G),$$

where

$$G = [1/\beta_1, -\alpha/\beta_1^2]'.$$

**Improvement in Precision under Linearity**

Now we explain the role of covariates in potentially delivering improvements in precision of estimating the ATE. The underlying idea is that of "denoising." This improvement, however, hinges on the linear model (2.2.1). In the next section, we will obtain improvement without linearity assumptions.

We consider what happens when we do not include covariates in the regression. In this case, the OLS estimator $\bar{\alpha}$ estimates the projection coefficient $\alpha$ in the BLP using $(1, D)$ alone:[14]

$$Y = \alpha D + \beta_1 + U, \quad \mathrm{E}[U] = \mathrm{E}[UD] = 0,$$

where the noise

$$U = \beta'(X - \mathrm{E}[X]) + \epsilon$$

contains the part of $Y$ that is linearly predicted by $X$, $\beta'(X - \mathrm{E}[X]) = \beta' X - \beta_1$. We then have that $\bar{\alpha}$ obeys

$$\sqrt{n}(\bar{\alpha} - \alpha) \overset{a}{\sim} N(0, \bar{\mathsf{V}}_{11}), \quad \bar{\mathsf{V}}_{11} = \frac{\mathrm{E}[U^2 \tilde{D}^2]}{(\mathrm{E}[\tilde{D}^2])^2}.$$

Under the linear model (2.2.1), it follows that

$$\mathsf{V}_{11} \leq \bar{\mathsf{V}}_{11},$$

with the inequality being strict ("<") if $\mathrm{Var}(\beta' X) > 0$.[15] That is, under (2.2.1), using pre-determined covariates improves the precision of estimating the ATE $\alpha$.

However, this improvement theoretically hinges on the correctness of the additive linear model. Statistical inference on the ATE based on the the normal approximation provided above remains valid without this assumption as long as robust standard errors are used.[16] However, the precision can be either higher or lower than that of the classical two-sample approach without covariates. That is, without (2.2.1), $\mathsf{V}_{11}$ and $\bar{\mathsf{V}}_{11}$ are not generally comparable.

**Remark 2.2.1** While the inferential result we derived is robust with respect to the linearity assumption on the CEF, the improvement in precision itself is **not** guaranteed in general and hinges on the validity of the linearity assumption. We provide simulation examples where controlling for predetermined covariates linearly lowers the precision (increases robust standard errors) in Covariates in RCT R Notebook and

14: Here $U = Y - \alpha D - \beta_1$ obeys

$$\begin{aligned} \mathrm{E}[U \mid D = d] &= \mathrm{E}[Y(d) - \alpha d - \beta_1 \mid D = d] \\ &= \mathrm{E}[Y(d) - \alpha d - \beta_1] = 0, \end{aligned}$$

invoking random assignment and the definition of $\alpha$ and $\beta_1$.

15: Verify this as a reading exercise.

16: We always use robust variance formulas throughout the book. However, the default inferential algorithms in R and Python often report the classical Student's formulas as variances, which critically rely on the linearity assumption.

Covariates in RCT Python Notebook.

### The Interactive Approach: Always Improves Precision and Discovers Heterogeneity

Covariates in RCT R Notebook and Covariates in RCT Python Notebook explore the use of covariates to both improve precision and learn about heterogeneity via a simulation experiment.

We can also consider estimation of CATE through the lens of an interactive linear regression model, which interacts treatment indicator $D$ with regressors $X$ constructed from original raw regressors $W$. Including these interactions respects the logic of approximating the conditional expectation of $Y$ given $D$ and raw regressors using linear functional forms. To simplify exposition, we first assume that the interactive model is exactly correct for the CEF:

$$\mathrm{E}[Y \mid D, W] = \alpha' XD + \beta' X. \tag{2.2.3}$$

In Section 2.C, we explain how this approach works without this assumption.

As before, we assume

$$X = (1, W')', \quad \mathrm{E}[W] = 0,$$

which can be achieved in practice by recentering. Here, we recover CATE via

$$\begin{aligned}\delta(W) &= \mathrm{E}[Y(1) \mid W] - \mathrm{E}[Y(0) \mid W] \\ &= \mathrm{E}[Y \mid D = 1, W] - \mathrm{E}[Y \mid D = 0, W] = \alpha' X.\end{aligned}$$

Using that $\mathrm{E}W = 0$, the ATE is then

$$\delta = \mathrm{E}[\delta(W)] = \mathrm{E}[\alpha' X] = \alpha_1,$$

where $\alpha_1$ is the first component of $\alpha$. The function $\alpha_2' W$, where $\alpha_2$ is the vector all elements of $\alpha$ excluding $\alpha_1$, therefore describes the deviation of CATE away from the ATE.

We can verify that $\alpha$ is the coefficient of the linear projection equation:

$$Y = \alpha' DX + \beta' X + \epsilon, \quad \epsilon \perp (X, DX).$$

Therefore, we can treat

$$\bar{D} := DX$$

as a vector of technical treatments[17] and invoke the "partialling

17: A technical treatment refers to any variable obtained as a transformation of the original treatment variable.

out" approach for inference on components of $\alpha$. The variance formulas are given in Section 2.C.

**Remark 2.2.2** (Improvement in Precision Guarantee) Unlike the previous approach, the "interactive" approach always delivers improvements in precision for estimating $\delta$, even if the linearity in (2.2.3) does not hold; this was demonstrated by Lin [13]. Section 2.C explains this point in detail and provides a deeper dive into the properties of the interactive approach without assuming correct linear specification of the CEF.

## Reemployment Bonus RCT

Here we re-analyze the Pennsylvania re-employment bonus experiment [14], which was conducted in the 1980s by the U.S. Department of Labor to test the incentive effects of alternative compensation schemes for unemployment insurance (UI). In these experiments, UI claimants were randomly assigned either to a control group or one of five treatment groups. We focus our discussion on treatment group 4. In the control group the current rules of the UI applied. Individuals in the treatment groups were offered a cash bonus if they found a job within some pre-specified period of time (qualification period), provided that the job was retained for a specified duration; see the Penn Data Codebook for further details on the data.

Reemployment Bonus RCT R Notebook and Reemployment Bonus RCT Python Notebook explore the use of covariates to improve precision and learn about heterogeneity in a Reemployment Bonus RCT.

We consider the

- classical 2-sample approach, no adjustment (CL)
- classical linear regression adjustment (CRA)
- interactive regression adjustment (IRA)
- interactive regression adjustment with double lasso (partialling out by lasso) (IRA-DL)

We use the last approach in the spirit of exploration and experimentation. We describe the last approach and establish its validity in Chapter 4.

Estimates of the ATE on (log) unemployment duration and corresponding estimated standard errors are given in Table 2.1.

| | CL | CRA | IRA | IRA-DL |
|---|---|---|---|---|
| Estimate | -0.0855 | -0.0797 | -0.0755 | -0.0789 |
| Std. Error | 0.0359 | 0.0356 | 0.0356 | 0.0356 |

**Table 2.1:** Estimates of the ATE of the reemployment bonus on log unemployment duration..

The different estimators deliver fairly similar point estimates suggesting that treatment group 4 experiences an average decrease in unemployment duration of around 8%. The three regression estimators deliver estimates that are slightly more precise (have lower standard errors) than the simple difference in means estimator.

We also see that the regression estimators offer slightly lower estimates of the ATE than the difference in means estimator. These differences likely occur due to minor imbalances in the treatment allocation: People older than 54 tended to receive the treatment more than other groups of qualified UI claimants during the later period of the experiment. Loosely speaking, the regression estimators try to correct for this imbalance by "partialling out" the effect of this oversampling and averaging over differences net of these "imbalancing" effects. We will explain how regression adjustment corrects for imbalances in Chapter 5.

See Reemployment Bonus RCT R Notebook and Reemployment Bonus RCT Python Notebook for the results from the balance check.

## 2.3 Drawing RCTs via Causal Diagrams

RCTs can be visualized using causal diagrams. These enable us to simply and clearly show the causal assumptions that underpin our model for retrieving treatment effects. Causal diagrams were introduced as early as 1920s by Sewall and Philip Wright ([15],[16]) and emerged as a fully formal tool due to the work of Judea Pearl and James H. Robins ([17], [18]).

In causal diagrams, random variables are denoted by nodes; and arrows between nodes represent causal effects. In our RCT set-up, we have that the assigned treatment variable causes outcome variable $Y$, and the pre-treatment variables $W$ also cause the outcome variable $Y$, but they don't cause the treatment assignment $D$. This causal diagram is illustrated in Figure 2.3 below.

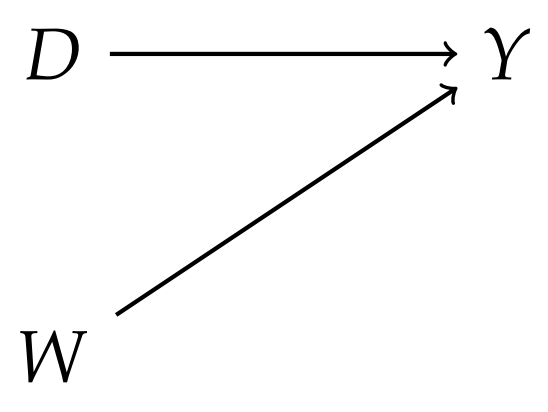

**Figure 2.3:** Causal Diagram for a RCT

Figure 2.4 depicts a version of the diagram that also includes potential outcomes as nodes.

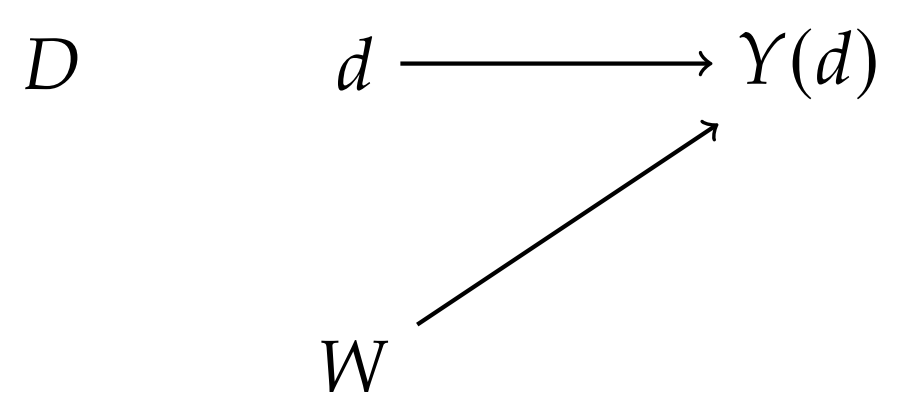

**Figure 2.4:** A Causal Diagram for the RCT Research Design

In Figure 2.4, we show the potential outcomes $Y(d)$ as a single node. The pre-treatment covariates affect this node, which is represented by the arrow from $W$ to the $Y(d)$ node. The assigned treatment variable $D$ is independent of the node $Y(d)$, which is shown by the absence of an arrow connecting the two nodes. The arrow from $d$ to $Y(d)$ shows the causal dependency of $Y(d)$ on the deterministic node $d$. The assigned treatment $D$ is also shown to be independent of the node $W$. The potential outcome process $d \mapsto Y(d)$ and treatment assignment jointly determine the realized outcome variable $Y$ via the assignment $Y := Y(D)$.

We further develop the use of these concepts and the use of causal diagrams as a formal tool in Chapter 7 and Chapter 8.

## 2.4 The Limitations of RCTs

Here, we briefly outline some of the primary limitations of RCTs. We first consider threats to identification, outlining settings in which the stable unit treatment value assumption (SUTVA), an important assumption that underpins causal inference in an RCT setting, is unlikely to hold, and the implications for inference. We then address ethical and practical concerns in RCT implementation and generalizability.

### Externalities, Stability, and Equilibrium Effects

The traditional formulation of Rubin's causal model relies on SUTVA as described in Section 2.1. Part of SUTVA is the requirement that the potential outcomes of one unit should be unaffected by the assignment of treatments to other units [19]. In the following, we consider some cases where this assumption might not hold.

In a vaccine example, this assumption holds if treatment and control populations are "small" (infinitesimal) subpopulations of the entire general population. Our methods measure the average vaccine effects in these settings. However, if we vaccinate a sufficiently large percentage of people, reaching herd immunity, the outcomes for the control group would be essentially the same as outcomes for the treated. SUTVA therefore would not hold.[18]

18: Because SUTVA does not hold in the vaccination context, it is customary to use relative measures of impact like "vaccine efficiency" because they may be a somewhat more stable measure when generalizing from "small" treated subpopulations to a "large" treated population.

In economics, we refer to such spillover effects as externalities or, in some contexts, as general equilibrium effects. For example, there is a positive externality created by people who take the vaccine (and people that don't take vaccine "free ride," once the vaccination level is high enough). Consider another example. We might want to study the earning effect of getting a college degree versus not having a college degree. If treatment will target a relatively small subpopulation of people, there likely won't be any large general equilibrium wage effects. On the other hand, if the treatment will target a large subpopulation, the equilibrium wage will likely adjust (the college wage premium might decrease, for example). In another example, the outcomes for one individual in large-scale training programs may be affected by the number of people trained to perform the same job.

## Ethical, Practical, and Generalizability Concerns

Many RCTs are infeasible because implementing them would be unethical. The general ethical principles and guidelines for research involving human subjects are set out in the 1978 Belmont report ([20]). The key ethical principles are "Respect for persons," "Beneficence," and "Justice." Human subject trials are subject to regulation by an institutional review board, which determines whether the trial is ethical with reference to these guiding principles, or whether it should be prevented from registering.

For example, we previously considered a hypothetical RCT where individuals are assigned to a smoking treatment group. The trial would violate the principle of "beneficence" as the researcher might be causing physical harm to study participants by assigning them to smoking. Thus, RCTs are rarely a feasible means of retrieving the causal effects of harmful interventions as they tend to be unethical.

RCTs may also face practical issues. They can be prohibitively expensive when the treatment is costly, data collection costs are high, or the sample size required for adequate power is high. These issues make it difficult to implement long-term RCTs and find evidence on the long-term effects of interventions, particularly because they are more likely to suffer from attrition.

It may also be politically infeasible for policymakers to enforce randomization of receipt of a desirable treatment.

Even in the best case, where an RCT is successfully implemented and we are confident in our retrieved average treatment effect, it may be difficult to generalize (or extrapolate) the result of an RCT in a specific context to a general finding. This difficulty might be because local conditions or implementation capacity materially differ between where interventions are staged or because the scale of the intervention is important.

## Notebooks

- Vaccination RCT R Notebook and Vaccination RCT Python Notebook contain the analysis of vaccination examples.
- Covariates in RCT R Notebook and Covariates in RCT Python Notebook explore the use of covariates to improve precision and learn about heterogeneity via a simulation experiment.
- Reemployment Bonus RCT R Notebook and Reemployment Bonus RCT Python Notebook explore the use of covariates to improve precision and learn about heterogeneity in a Reemployment Bonus RCT.

## Notes

RCTs have a profound influence on business, economics and science more generally. For example, RCTs are routinely used to study the efficacy of drugs and efficacy of various programs in labor and development economics, among other subfields of economics. The FDA moved to RCTs as the gold standard of proving that treatments work in 1970s-80s. In the tech industry and marketing, RCTs are also called "A/B Tests" and are now widely used. Many major tech companies have their own experimental platforms to carry out thousands of experiments.[19]

The expansion of the use of experimentation in economics is associated with the work of Richard Thaler, the recipient of the 2017 Alfred Nobel Memorial Prize in Economics;[20] Abhijit Banerjee, Esther Duflo, and Michael Kremer, the recipients of the 2019 Alfred Nobel Memorial Prize in Economics;[21] and John List, among many others.

19: See, for example, ExP platform at Microsoft and the WebLab platform at Amazon.

20: "for his contributions to behavioural economics." Source: NobelPrize.org

21: "for their experimental approach to alleviating global poverty." Source: NobelPrize.org

We touched upon very basic ideas here. The basic random design is just one of many possible randomized designs that allow us to uncover causal effects. For an in-depth analysis of design of experiments, please see lecture notes by Art Owen ([21]). For standard RCTs and causal analysis more generally, see the book by Imbens and Rubin [10]. Duflo et al. [22] is another good overview of the use of RCTs with a focus on development economics applications. For real examples of how RCTs are done and designed in practice, see, for example, the FDA registry of RCTs, the American Economic Association for a registry of RCTs in economics, or the The Poverty Action Lab.

## Study Questions

1. Set-up a simulation experiment that illustrates the contrived smoking example, following the analytical example we've presented in the text. Illustrate the difference between estimates obtained via an RCT (smoking generated independently of potential outcomes) and an observational study (smoking choice is correlated with potential outcomes).

2. Sketch out the proof of the large sample properties of the two means estimator.

3. Study the notebook on vaccinations RCTs. Try to replicate the results in the FDA briefing table for each age 18-64 (exact replication is not required). Explain your calculations.

4. Study the notebook on the reemployment example. Experiment with putting even more flexible controls (e.g. use extra interactions of some controls). Report your findings.

5. Work and experiment with the Covariates in RCT notebook. Explain the main points being made.

6. Skim over the information on the Pfizer RCT design briefing. Write down one paragraph summarizing the study design.

7. Skim over one of the RCTs registered with AEA RCT Registry. Write down one paragraph summarizing the study design.

8. Think of some RCTs where stability (SUTVA) is likely to hold and some RCTs where it likely does not.

9. Explain why we can't learn individual treatment effects by first putting a unit in treatment and then putting the individual in control second (or the other way around). A hint is to think of all sources of randomness represented by $\omega$. Would the situation be different if you had a time machine?

## 2.A Approximate Distribution of the Two Sample Means

To demonstrate the result in the text, we note that

$$\hat{\theta}_d - \theta_d = \frac{\mathbb{E}_n[(Y(d) - \mathrm{E}Y(d))1(D = d)]}{\mathbb{E}_n[1(D = d)]}$$

for $d \in \{0, 1\}$ because we can re-write the population group average as

$$\theta_d = \mathrm{E}[Y(d)] = \mathrm{E}[Y(d)]\frac{\mathbb{E}_n[1(D = d)]}{\mathbb{E}_n[1(D = d)]}.$$

Hence, for each $d \in \{0, 1\}$,

$$\sqrt{n}(\hat{\theta}_d - \theta_d) = \sqrt{n}\frac{\mathbb{E}_n[(Y(d) - \mathrm{E}Y(d))1(D = d)]}{\mathbb{E}_n[1(D = d)]}.$$

By the law of large numbers, $\mathbb{E}_n[1(D = d)] \approx \mathrm{P}(D = d)$; so we have the approximation

$$\sqrt{n}\{\hat{\theta}_d - \theta_d\}_{d\in\{0,1\}} \approx \sqrt{n}\frac{\mathbb{E}_n[(Y(d) - \mathrm{E}Y(d))1(D = d)]}{\mathrm{P}(D = d)}.$$

Note that the terms being averaged are

$$\frac{(Y_i(d) - \mathrm{E}[Y(d)])1(D_i = d)}{\mathrm{P}(D = d)}.$$

These terms have zero mean[22] and variance

$$\frac{\mathrm{E}[(Y(d) - \mathrm{E}[Y(d)])^2 1(D = d)^2]}{\mathrm{P}(D = d)^2} = \frac{\mathrm{Var}(Y \mid 1(D = d) = 1)}{\mathrm{P}(D = d)}.$$

22: Why? Hint: Use the law of iterated expectations.

Also note the zero covariance:

$$\mathrm{E}\left[\frac{(Y(1)-\mathrm{E}[Y(1)])1(D=1)}{\mathrm{P}(D=1)}\frac{(Y(0)-\mathrm{E}[Y(0)])1(D=0)}{\mathrm{P}(D=0)}\right]=0.$$

The application of the central limit theorem then yields the claimed result.

## 2.B Statistical Properties of the Classical Additive Approach$^\star$

Here we analyze statistical inference on ATE using OLS and adjusting for $X = (1, W)$, without making the linearity assumptions we made in Section 2.2.

We consider the linear projection equation in the population:

$$Y = D\alpha + X'\beta + \epsilon, \quad \epsilon \perp (D, X).$$

Here, we have that $D$ and $X = (1, W)$ with $\mathrm{E}[W] = 0$, so that $\beta' X = \beta_1 + \beta_2' W$. Moreover, we have that $D \perp W$ in the RCT setting.

First, we'd like to verify that $\alpha = \mathrm{E}[Y(1)] - \mathrm{E}[Y(0)]$ and $\beta_1 = \mathrm{E}[Y(0)]$. For $U := \beta_2' W + \epsilon$, we can write

$$Y = D\alpha + \beta_1 + U, \quad U \perp (1, D).$$

$U \perp (1, D)$ holds because $(1, D) \perp (W, \epsilon)$ using that $\mathrm{E}[W] = 0$ and that $D \perp (W, \epsilon)$. Therefore, $D\alpha + \beta_1$ coincides with the population projection of $Y$ onto $(1, D)$. Hence, the projection coefficients are the same as those obtained by the 2-sample approach in the population. Therefore, $\beta_1 = \mathrm{E}[Y(0)]$ and $\alpha = \mathrm{E}[Y(1)] - \mathrm{E}[Y(0)]$.

Second, we'd like to explain the details of the approximate normality for the estimators of sample OLS coefficients $\hat{\beta}_1$. The OLS theory of the first chapter implies that the OLS estimator $\hat{\alpha}$ obeys

$$\sqrt{n}(\hat{\alpha} - \alpha) \approx \sqrt{n}\frac{\mathbb{E}_n[\epsilon \tilde{D}]}{\mathbb{E}_n[\tilde{D}^2]} \overset{a}{\sim} N(0, \mathsf{V}_{11}),$$

where $\tilde{D} = D - \mathrm{E}[D]$ is the residual after partialling out $X$ from $D$ linearly,[23] and

$$\mathsf{V}_{11} = \frac{\mathrm{E}[\epsilon^2 \tilde{D}^2]}{(\mathrm{E}[\tilde{D}^2])^2}.$$

23: Derive that $\tilde{D} = D - \mathrm{E}[D]$ from Assumption 2.2.1.

Applying the same theory for $\beta_1$ (the intercept coefficient), yields[24]

$$\sqrt{n}(\hat{\beta}_1 - \beta_1) \approx \sqrt{n}\frac{\mathbb{E}_n[\epsilon\tilde{1}]}{\mathbb{E}_n[\tilde{1}^2]} \overset{a}{\sim} N(0, \mathsf{V}_{22}),$$

where $\tilde{1} := (1 - D)$ is the residual after partialling out $D$ and $X$ from 1 and

$$\mathsf{V}_{22} = \frac{\mathrm{E}[\epsilon^2\tilde{1}^2]}{(\mathrm{E}[\tilde{1}^2])^2}.$$

We can also establish that the estimators are jointly approximately normal with covariance

$$\mathsf{V}_{12} = \frac{\mathrm{E}[\epsilon^2\tilde{D}\tilde{1}]}{\mathrm{E}[\tilde{1}^2]\mathrm{E}[\tilde{D}^2]}.$$

24: To explain the derivation, note that by partialling out $D$ and $W$ (recall that $X = (1, W)$) from 1 and $Y$, we obtain

$$\tilde{Y} = \beta_1\tilde{1} + \epsilon; \quad \tilde{1} := (1 - D).$$

The projection of 1 on $D$ and $W$ is given by $D$ since $D$ is binary and we've assumed $\mathrm{E}[W] = 0$.

## 2.C Statistical Properties of the Interactive Regression Approach★

Here we analyze the estimation of the ATE using OLS and adjusting for $(W, DW)$ without making any linearity assumptions on the potential outcomes as we did in Section 2.2. We essentially show that the interactive model can be viewed as estimating the BLP of each of the two potential outcomes $Y(0)$ and $Y(1)$. Using this fact one can then easily argue that the variance of the OLS estimate of the effect using the interactive model can only be lower than the variance of the unadjusted OLS estimate.

Letting $X = (1, W)$ be an intercept and the pre-treatment covariates $W$, let us write the BLP of each of $Y(0)$ and $Y(1)$ using $X$ as

$$Y(d) = \beta_d' X + \varepsilon_d, \quad \varepsilon_d \perp X, \quad d = 0, 1. \tag{2.C.1}$$

Under Assumption 2.2.1, (2.C.1) coincides with the BLP of $Y$ using $X$ in the $D = d$ population. Letting $\varepsilon = D\varepsilon_1 + (1 - D)\varepsilon_0$, we thus have

$$Y = \beta_d' X + \varepsilon, \quad \mathrm{E}[\varepsilon X \mid D = d] = 0, \quad d = 0, 1. \tag{2.C.2}$$

The BLPs in each of the two populations, $D = 0$ and $D = 1$, can be combined across the populations to state the BLP of $Y$ using $(X, DX)$ marginally:

$$Y = \beta_0' X + \beta_\delta' XD + \varepsilon, \quad \varepsilon \perp (X, DX), \tag{2.C.3}$$

where $\beta_\delta = \beta_1 - \beta_0$.[25] Such a linear rule is called *interactive* because it includes the interaction (meaning, product) of $D$ and $W$ as a regressor, in addition to $D$ and $W$.

25: Note that (2.C.1) and (2.C.2) imply $\mathrm{E}[\varepsilon D X] = 0$ and $\mathrm{E}[\varepsilon X] = 0$ and thus that $\varepsilon \perp (X, DX)$.

We assume that covariates are centered:

$$\mathrm{E}[W] = 0.$$

Since $X$ contains an intercept, $\varepsilon_d \perp X$ implies $\mathrm{E}[\varepsilon_d] = 0$. Together with centered covariates, we find that

$$\mathrm{E}[Y(d)] = \mathrm{E}[\beta_d' X + \varepsilon_d] = \beta_{d,1}.$$

This means that the ATE coincides with the coefficient on $D$ in the BLP of $Y$ using $(X, DX)$. That is, $\beta_{\delta,1} = \delta$.

We are often interested in the ATE and Relative ATE

$$\delta \quad \text{and} \quad \delta/\mathrm{E}[Y(0)].$$

If we use OLS to estimate the BLP of $Y$ using $(X, DX)$, then an application of the OLS theory in the previous chapter gives us that, under regularity conditions,

$$\begin{pmatrix} \sqrt{n}(\hat\beta_{\delta,1} - \delta) \\ \sqrt{n}(\hat\beta_{0,1} - \mathrm{E}[Y(0)]) \end{pmatrix} \overset{\mathrm{a}}{\sim} N(0, \mathsf{V}),$$

where covariance matrix $\mathsf{V}$ has components:

$$\mathsf{V}_{11} = \frac{\mathrm{E}[\epsilon^2 \tilde D^2]}{(\mathrm{E}[\tilde D^2])^2}, \quad \mathsf{V}_{22} = \frac{\mathrm{E}[\epsilon^2 \tilde 1^2]}{(\mathrm{E}[\tilde 1^2])^2}, \quad \mathsf{V}_{12} = \mathsf{V}_{21} = \frac{\mathrm{E}[\epsilon^2 \tilde D \tilde 1]}{\mathrm{E}[\tilde 1^2]\mathrm{E}[\tilde D^2]},$$

where $\tilde D = D - \mathrm{E}[D]$ is the residual after partialling out linearly $(1, W, DW)$ from $D$ and $\tilde 1 := (1 - D)$ is the residual after partialling out $(D, W, DW)$ from 1.[26]

26: The derivation follows identical steps as that in Section 2.B with the only exception that when defining $\tilde D$ we need to partial out $(1, W, DW)$ from $D$ and when defining $\tilde 1$ we need to partial out $(D, W, DW)$ from 1. However, since $\mathrm{E}[W] = \mathrm{E}[DW] = 0$, the two residuals take the same form of $D - \mathrm{E}[D]$ and $1 - D$ correspondingly.

We can then obtain the approximate normality for the Relative ATE using the delta method:

$$\sqrt{n}(\hat\beta_{\delta,1}/\hat\beta_{0,1} - \delta/\mathrm{E}[Y(0)]) \overset{\mathrm{a}}{\sim} N(0, G'\mathsf{V}G),$$

where

$$G = [1/\mathrm{E}[Y(0)], -\delta/(\mathrm{E}[Y(0)])^2]'.$$

We can rewrite (2.C.3) as

$$Y = \beta_{0,1} + D\beta_{\delta,1} + U, \quad U = \beta_{0,2}' W + \beta_{\delta,2}' WD + \varepsilon.$$

From $\varepsilon \perp (X, D, DX)$, $\mathrm{E}[W] = 0$, and Assumption 2.2.1, we obtain that $U \perp (1, D)$, meaning that $\beta_{0,1} + D\beta_{\delta,1}$ is the BLP of $Y$ using $(1, D)$. We can therefore estimate the ATE as the coefficient on $D$ *either* in the OLS of $Y$ on $(1, D)$ *or* in the OLS of $Y$ on $(X, DX)$. The former exactly coincides with the unadjusted estimator $\hat{\delta}$ from Section 2.1, which obeys

$$\sqrt{n}(\hat{\delta} - \delta) \overset{a}{\sim} N(0, \bar{\mathsf{V}}_{11}), \quad \bar{\mathsf{V}}_{11} = \frac{\mathrm{E}[U^2 \tilde{D}^2]}{(\mathrm{E}[\tilde{D}^2])^2}.$$

Since $\epsilon$ satisfies the BLP conditions for each of the treatment populations, i.e. $\mathrm{E}[\varepsilon W \mid D = d] = 0$, it then follows that

$$\mathsf{V}_{11} \leq \bar{\mathsf{V}}_{11}.$$

Moreover, the inequality is strict if $\mathrm{Var}(\beta'_{0,2} W) > 0$ or $\mathrm{Var}(\beta'_{1,2} W) > 0$.[27] That is, pre-determined covariates improve the precision of estimating the ATE $\delta$, when using the interactive model, without any linearity assumptions on the CEF.

27: Verify this as a reading exercise.

# 3 Predictive Inference via Modern High-Dimensional Linear Regression

"Il semble que la perfection soit atteinte non quand il n'y a plus rien à ajouter, mais quand il n'y a plus rien à retrancher."
(It seems perfection is attained not when there is no longer anything to add, but when there is no longer anything to take away.)

– Antoine de Saint-Exupéry [1].

Here we discuss the use of penalized regressions for constructing predictions in high-dimensional settings, particularly when $p > n$. We first motivate the high-dimensional setting as arising both from having a high-dimensional regressor set and from constructing technical regressors from raw regressors. We then discuss Lasso, which penalizes the size of the model by the sum of the absolute value of its coefficients. We conclude with an overview of other penalized regression methods.

## 3.1 Linear Regression with High-Dimensional Covariates

### The Framework

We consider a regression model

$$Y = \beta' X + \epsilon, \quad \epsilon \perp X,$$

where $\beta' X$ is the population best linear predictor of $Y$ using $X$, or simply the population linear regression function. The vector $X = (X_j)_{j=1}^p$ is $p$-dimensional. That is, there are $p$ regressors, and

$p$ is large, possibly much larger than $n$.

This case where $p$ is large relative to the sample size is what we call a *high-dimensional* setting. High-dimensional settings arise when

- data have large dimensional features (i.e. many covariates are available for use as regressors),
- we construct many technical regressors[1] from raw regressors, or
- both.

1: Recall, a *technical regressor* is any variable obtained as a transformation of a basic regressor.

Examples of datasets where many covariates are available and potential corresponding exemplary applications include

- country characteristics in cross-country wealth analysis,
- housing characteristics in house pricing/appraisal analysis,
- individual health information in electronic health records and claims data, and
- product characteristics at the point of purchase in demand analysis.

Another source of high-dimensionality is the use of constructed features or regressors of the form

$$X = T(W) = (T_1(W), ..., T_p(W))',$$

where $W$ denotes original raw regressors. As we discussed in Chapter 1, the set of transformations $T(W)$ is sometimes called the *dictionary* of transformations. Example transformations include polynomials, splines, interactions between variables, and applying functions such as the logarithm or exponential. In the wage analysis in Chapter 1, for example, we used quadratic and cubic transformations of experience, as well as interactions (products) of these regressors with education and geographic indicators. Recall that the main motivation for the use of constructed regressors is to build *more flexible and potentially better* prediction rules.

The potential for improved prediction arises because we are using prediction rules $\beta'X = \beta'T(W)$ that are *nonlinear* in the original raw regressors $W$ and may thus capture more complex patterns that exist in the data. Conveniently, the prediction rule $\beta'X$ is still linear with respect to the parameters, $\beta$, and with respect to the constructed regressors $X = T(W)$, so inherits much from the previous discussion of linear regression provided in Chapter 1.

In summary, we have provided two motivations for using high-dimensional regressors in prediction:

- The first motivation is that modern datasets have high-dimensional features that can be used as regressors.
- The second motivation is that we can use nonlinear transformations of features or raw regressors and their interactions to form constructed regressors. Using transformations allows us to better approximate the best prediction rule – the conditional expectation of the outcome given raw regressors.

## Lasso

Recall that we are considering a regression model

$$Y = \beta'X + \epsilon = \sum_{j=1}^{p} \beta_j X_j + \epsilon, \quad \epsilon \perp X \tag{3.1.1}$$

where $p$ is possibly much larger than $n$.

Classical linear regression or least squares fails in these high-dimensional settings because it *overfits* in finite samples. Intuitively, overfitting refers to using patterns that are idiosyncratic to a specific dataset and do not generalize out of sample. That is, it corresponds to using a prediction rule that is overly complex in that it uses patterns that help explain a given dataset, increasing in-sample measures of fit, but are not present in different data even if the data are drawn from the same population, potentially harming out-of-sample prediction performance.

The potential for classical linear regression estimated with least squares to overfit is especially apparent when $p \geq n$. In this case,

conventional least squares will perfectly fit the data regardless of the value of $\beta$ as long as the covariate matrix is rank $n$.[2] We therefore make some assumptions and modify the regression method to deal with cases where $p$ is large.

2: Recall that we illustrated the problem with overfitting in Section 1.2.

An intuitive starting point is the assumption of *approximate sparsity*. Under approximate sparsity, there is a small group of regressors with relatively large coefficients whose use alone suffices to approximate the BLP $\beta'X$ well. The rest of the regressors are assumed to have relatively small coefficients and contribute little to the approximation of the BLP.

An example of approximate sparsity is captured by regression coefficients of the form[3]

3: The notation $\propto$ reads as "proportional to."

$$\beta_j \propto 1/j^2, \quad j = 1, \ldots, p.$$

Here, the first few coefficients capture almost all the explanatory power of the full vector of coefficients as shown in Figure 3.1.

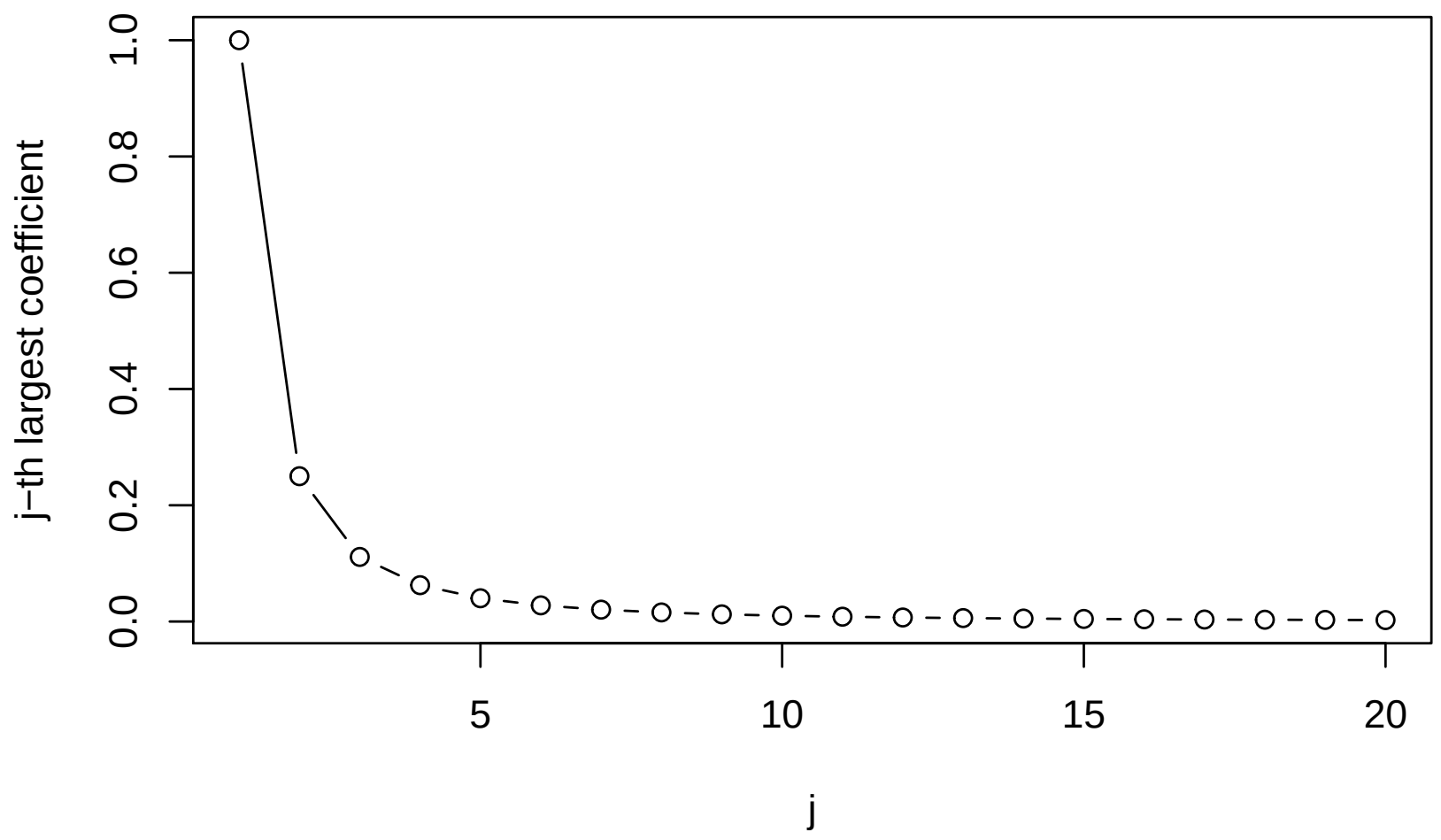

**Figure 3.1:** Example of regression coefficients, $\beta_j = 1/j^2$ that satisfy approximate sparsity.

Next, we define approximate sparsity formally.

**Definition 3.1.1** ***Approximate sparsity:*** *The sorted absolute values of the coefficients decay quickly. Specifically, the $j^{th}$ largest coefficient (in absolute value) denoted by $|\beta|_{(j)}$ obeys*

$$|\beta|_{(j)} \leq Aj^{-a}, \quad a > 1/2, \tag{3.1.2}$$

*for each $j$, where the constants $a$ and $A$ do not depend on the sample size $n$.*

For estimation purposes, we have a random sample $\{(Y_i, X_i)\}_{i=1}^n$. We seek to construct a good linear predictor $\hat{\beta}'X$, which works well when $p/n$ is not small.

Before defining the Lasso problem, it is important to note that we are treating all variables as centered and thus do not include an intercept in the model. In practice, this construction means that, for raw variables $Y^*$ and $X^*$, we start by defining demeaned versions of these variables $Y = Y^* - \mathbb{E}_n[Y^*]$ and $X = X^* - \mathbb{E}_n[X^*]$ for use in estimation of model parameters.[4] We note that the centered model (3.1.1) is equivalent to starting with the model

$$Y^* = \alpha + \beta' X^* + \epsilon \quad \epsilon \perp X^*$$

with intercept $\alpha = \mathrm{E}[Y^*] - \beta'\mathrm{E}[X^*]$. For estimates $\hat{\beta}$ obtained by estimating (3.1.1), we can thus recover an estimate of $\alpha$ as $\hat{\alpha} = \mathbb{E}_n[Y^*] - \hat{\beta}'\mathbb{E}_n[X^*]$.

A centered random variable $U$ has $\mathrm{E}[U] = 0$, and a centered variable $U$ in a sample has $\mathbb{E}_n[U] = 0$.

4: When performing validation exercises, demeaning and any other transformations that depend on features of the data, such as standardization, should be done in both training and test data using the features of the *training* data rather than of the full sample or the test data.

When discussing theoretical properties, we will further assume that regressors are normalized,

$$\mathrm{E}[X_j^2] = 1.$$

We do state the estimation algorithms without assuming this normalization. The combination of centering and normalization – *standardization* – is commonly employed in practice and is done by default in many software packages.

*Lasso* constructs $\hat{\beta}$ as the solution of the following penalized least squares problem:

$$\min_{b \in \mathbb{R}^p} \quad \sum_i (Y_i - b'X_i)^2 + \lambda \cdot \sum_{j=1}^{p} |b_j| \hat{\psi}_j, \tag{3.1.3}$$

which is called the Lasso regression problem. The first term is $n$ times the sample mean squared error, and the second term is called a *penalty term*. The penalty term introduces a cost to the complexity of the prospective model where complexity is captured by the sum of the products of the absolute values of the coefficients $b_j$ with the *penalty loadings* $\hat{\psi}_j$ all multiplied by the *penalty level* $\lambda$.

Rather than work with centered variables, we could equivalently define (3.1.3) with an intercept where the intercept *does not* enter the penalty function. The important thing to keep in mind is that it is rarely appropriate to penalize the intercept.

The penalty loadings are typically set as

$$\hat{\psi}_j = \sqrt{\mathbb{E}_n[X_j^2]}.$$

The use of this penalty ensures invariance of Lasso predictions to rescaling $X_j'$. Note that many software packages implement the Lasso with simple penalty loadings $\hat{\psi}_j = 1$. In such cases, the use of standardized variables produces the same results as using these penalty loadings.

As long as $\lambda > 0$, the introduction of the penalty term in (3.1.3) leads to a prediction rule which is less complex than the rule that would be obtained via solving the unpenalized least squares problem. Specifically, the penalty term in the Lasso problem, $\sum_{j=1}^{p} |b_j|\hat{\psi}_j$, provides a measure of complexity of a regression model in terms of the overall magnitude of the coefficients. When $\lambda$ is positive, minimizing the Lasso problem requires trading off in-sample fit with this measure of complexity. As a result, the overall magnitude of the estimated coefficients, as measured by the penalty term, will be smaller than the overall magnitude of the coefficients absent this penalty. That is, the Lasso solution will have coefficients that are "shrunk" towards 0 relative to the unpenalized least squares problem.[5]

5: This overall shrinkage towards zero relative to the unpenalized problem is sometimes referred to as *shrinkage bias* or *regularization bias*.

One important benefit of introducing the penalty term is that it helps guard against overfitting by introducing a cost to model complexity. Intuitively, overfitting occurs as a model is made increasingly complex in an effort to make improvements to in-sample fit that are small relative to sampling error and could thus correspond to idiosyncrasies of a specific finite sample. The penalty term imposes a cost to complexity which help keep increases to complexity that have small benefit in terms of improving fit from being made. Through careful choice of $\lambda$, we can theoretically guarantee that the Lasso predictor is similar to the optimal predictor, and thus generalizable, even in high-dimensional settings.

A second important feature of Lasso is that it imposes the approximate sparsity condition on the estimated coefficients $\hat{\beta}$. Approximate sparsity is produced because the penalty function in (3.1.3) has a kink at zero which results in the marginal cost of including regressor $X_j$ ($\lambda\hat{\psi}_j > 0$) always being positive when $\lambda > 0$ . Therefore, Lasso includes a regressor $X_j$ with non-zero coefficient only if its marginal predictive ability is higher than this marginal cost threshold. That is, Lasso does *variable selection*: The Lasso solution drops any variable (equivalently sets the variable's coefficient to 0) whose marginal predictive benefit does not exceed the marginal cost of inclusion. We illustrate this variable selection property numerically in Example 3.1.1 below.

It is important to note that Lasso will not generally select the "right" set of variables. Lasso will tend to exclude variables with small, but non-zero population coefficients. Lasso will also tend to fail to select the right variables in settings where the $X$ variables are correlated.[6] That is, one should not conclude that Lasso has selected exactly the variables with non-zero coefficients in the population unless one can rule out variables with small, but non-zero coefficients and ensure that variables are all at most weakly correlated.[7] This failure does not mean that the Lasso predictions are poor quality, but does mean that care should be taken in interpreting the selected variables.

6: For example, consider a scenario where variable $X_1$ has coefficient $\beta_1 = 0$ but is highly correlated to variables $X_2, \ldots, X_k$ that have non-zero coefficients. It is quite plausible that the marginal predictive benefit of including $X_1$ in the model is very high when $X_2, \ldots, X_k$ are not in the model while the marginal predictive benefit of any one of $X_2, \ldots, X_k$ is relatively low. In this case, $X_1$ may enter the Lasso solution with a non-zero coefficient while all of $X_2, .., X_k$ are excluded.

7: This inability to select *exactly* the right regressors is not special to Lasso but shared by all variable selection procedures.

**Example 3.1.1** (Simulation Example) Consider

$$Y = \beta' X + \epsilon, \quad X \sim N(0, I_p), \quad \epsilon \sim N(0, 1),$$

with approximately sparse regression coefficients:

$$\beta_j = 1/j^2, \quad j = 1, \ldots, p$$

and

$$n = 300, \quad p = 1000.$$

Figure 3.2 shows that $\hat{\beta}$ is sparse and is close to $\beta$. We see that Lasso sets most of regression coefficients to zero. It figures out *approximately* the right set of regressors, including only those with the two largest coefficients. Note that Lasso does not, and in fact cannot, select the regressors with non-zero coefficients in this example as all variables have non-zero coefficients.

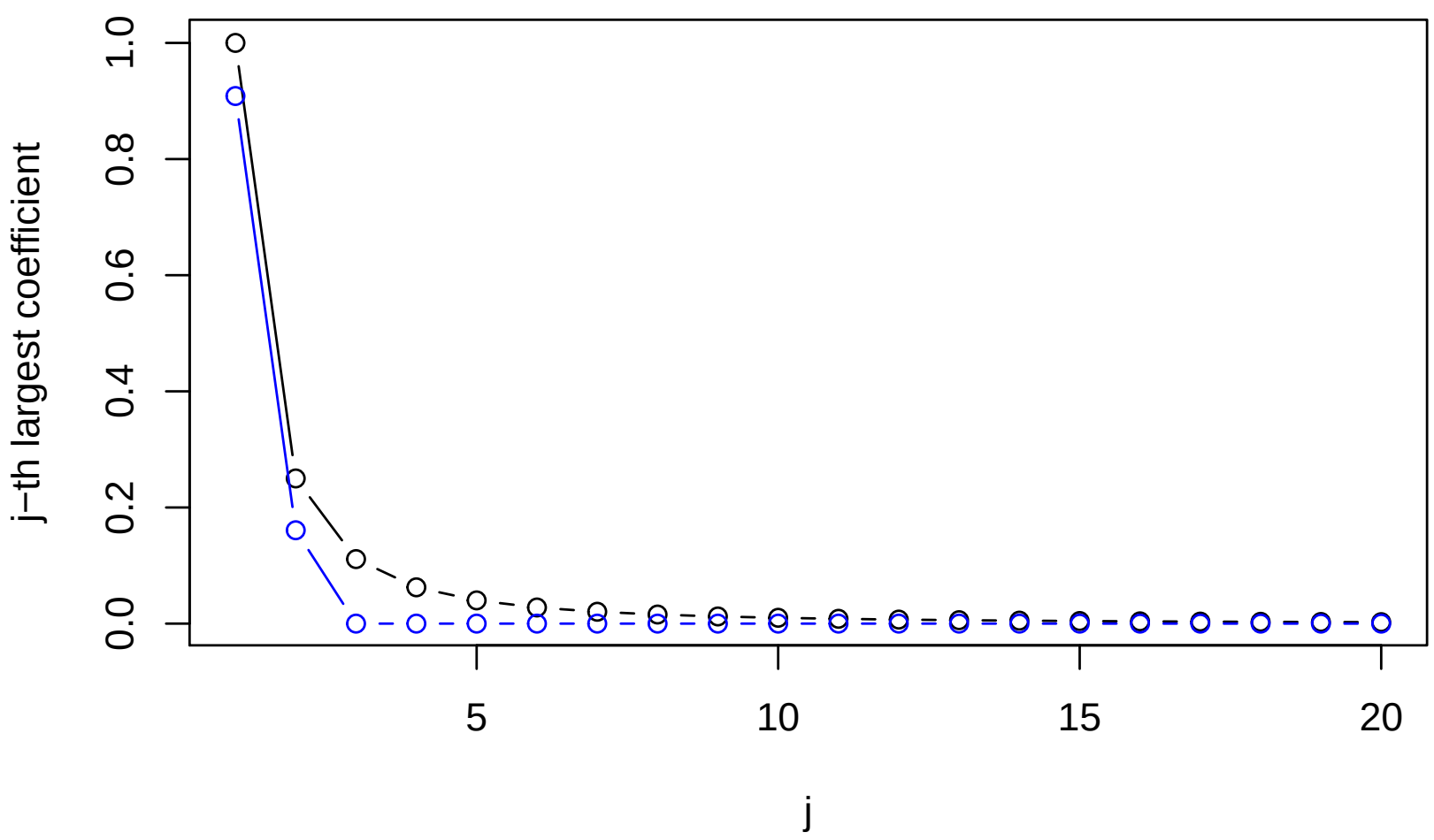

**Figure 3.2:** The true coefficients (black) vs. coefficients estimated by Lasso (blue) in Example 3.1.1.

A crucial point for the two Lasso properties that we have discussed is the choice of the penalization parameter $\lambda$. A theoretically valid choice is[8]

$$\lambda = 2 \cdot c\hat{\sigma}\sqrt{n} z_{1-\mathrm{a}/(2p)} \tag{3.1.4}$$

8: Recall that $z_t$ is such that $\mathrm{P}((N(0,1) \leq z_t) = t$.

where $\hat{\sigma} \approx \sigma = \sqrt{\mathrm{E}[\epsilon^2]}$ is obtained via an iteration method defined in Appendix 3.A, $c > 1$, and $1 - \mathrm{a}$ is a confidence level.[9] We can further simplify the choice using Feller's tail inequality:

$$z_{1-\mathsf{a}/(2p)} \leq \sqrt{2\log(2p/\mathsf{a})},$$

where the inequality becomes sharp as $p \to \infty$.

This penalty level ensures that the Lasso predictor $\hat{\beta}'X$ does not overfit the data and delivers good predictive performance under approximate sparsity ([2, 3]). Another good way to pick the penalty level when building a model for prediction is by cross-validation ([4]).[10]

9: Practical recommendations, based on theory and that seem to work well in practice, are to set $c = 1.1$ and $\mathsf{a} = .05$.

10: Cross-validation is a repeated data-splitting method for choosing penalty parameters for Lasso and for selecting among predictive models more generally. We outline the basic idea of cross-validation in Section 3.B.

## Quick Heuristics for Lasso Properties and Penalty Choice★

Here, we provide a sketch of the mathematics of the Lasso estimator illustrating its variable selection properties and motivating the choice of $\lambda$ in (3.1.4).

Assume $\hat{\psi}_j = 1$ for simplicity. The $j$-th component $\hat{\beta}_j$ of the Lasso estimator $\hat{\beta}$ is set to zero if the marginal predictive benefit of changing $\hat{\beta}_j$ away from zero is smaller than the marginal increase in penalty (see Figure 3.3):

$$\hat{\beta}_j = 0 \text{ if } \left| \frac{\partial}{\partial \hat{\beta}_j} \sum_i (Y_i - \hat{\beta}' X_i)^2 \right| < \lambda.$$

That is,

$$\hat{\beta}_j = 0 \text{ if } |-\hat{S}_j| < \lambda, \quad \hat{S}_j = 2 \sum_i (Y_i - \hat{\beta}' X_i) X_{ji}.$$

We discuss more detailed heuristics for penalty level selection in the appendix, but the rough idea is that the penalty should dominate the noise $S_j = 2\sum_i (Y_i - \beta' X_i) X_{ji}$ in the measurement of the marginal predictive ability. By the high-dimensional central limit theorem ([5]), we have that

$$(S_j)_{j=1}^p \overset{\mathsf{a}}{\sim} 2\sqrt{n}\sigma(\mathcal{N}_j)_{j=1}^p, \quad \mathcal{N}_j \sim N(0,1).$$

Therefore, to guarantee that Lasso sets to zero the any coefficient whose actual value is zero, we would like to choose $\lambda$ to dominate

$$2\sqrt{n}\sigma \max_{j=1,\dots,p} |\mathcal{N}_j|$$

with high probability, say $1 - \mathsf{a}$. Then by the union bound and symmetry of centered normal variables,

$$\mathrm{P}\Big( \max_{j=1,\ldots,p} |\mathcal{N}_j| > z_{1-\mathsf{a}/(2p)} \Big)$$

$$\leq 2 \sum_{j=1}^{p} \mathrm{P}\left( \mathcal{N}_j > z_{1-\mathsf{a}/(2p)} \right)$$

$$= 2p\Big(1 - (1 - \mathsf{a}/(2p))\Big) = \mathsf{a}.$$

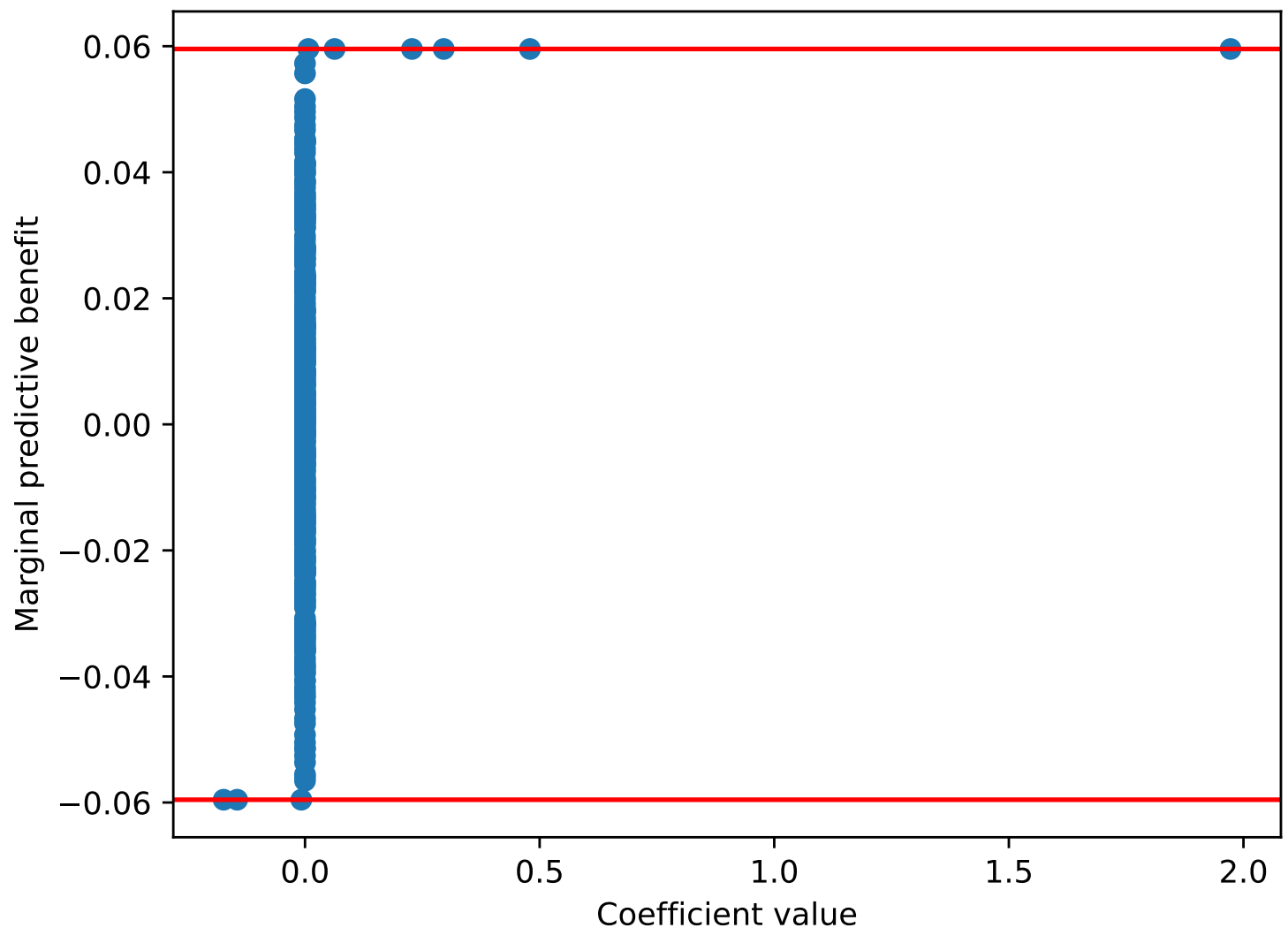

**Figure 3.3:** Example relationship between coefficient value and (signed) marginal predictive value $\hat{S}_j$ at the optimal solution to the Lasso objective. The red lines correspond to $\{-\lambda, \lambda\}$.

The union bound here is crude, but the bound is not very loose. In particular, when the $\mathcal{N}_j$'s are independent, the bound becomes sharp as $p \longrightarrow \infty$. Finally, setting

$$\lambda = 2\sigma\sqrt{n} z_{1-\mathsf{a}/(2p)}$$

we conclude that

$$\mathrm{P}(\max_j |S_j| \leq \lambda) \geq 1 - \mathsf{a},$$

up to a vanishing error. That is, this choice of $\lambda$ guarantees that variables with $\beta_j = 0$ are excluded from the model (have $\hat{\beta}_j = 0$) with high probability.

## OLS Post-Lasso

We can use the Lasso-selected set of regressors, those regressors whose Lasso coefficient estimates are non-zero, to refit the model by least squares. This method is called "least squares post Lasso" or simply *Post-Lasso* ([3]). Compared to Lasso,

Post-Lasso undoes the overall shrinkage toward zero relative to unconstrained least squares from the estimated non-zero coefficients, as we illustrate in Figure 3.1.5 below.[11] Removing this shrinkage towards zero from the non-zero coefficients sometimes delivers improvements in predictive performance.

11: Note that the estimates of the large coefficients are nearly perfect after OLS refitting of the model selected by Lasso in this example.

**Post-Lasso.** We define the Post-Lasso

$$\tilde{\beta} \in \arg\min_{b \in \mathbb{R}^p} \sum_i (Y_i - b'X_i)^2 \text{ such that } b_j = 0 \text{ if } \hat{\beta}_j = 0 \text{ for each } j, \quad (3.1.5)$$

where $\hat{\beta}$ is the Lasso coefficient estimator. The formal properties of the Post-Lasso estimator $\tilde{\beta}$ are similar to those of Lasso $\hat{\beta}$; see Section 3.2.

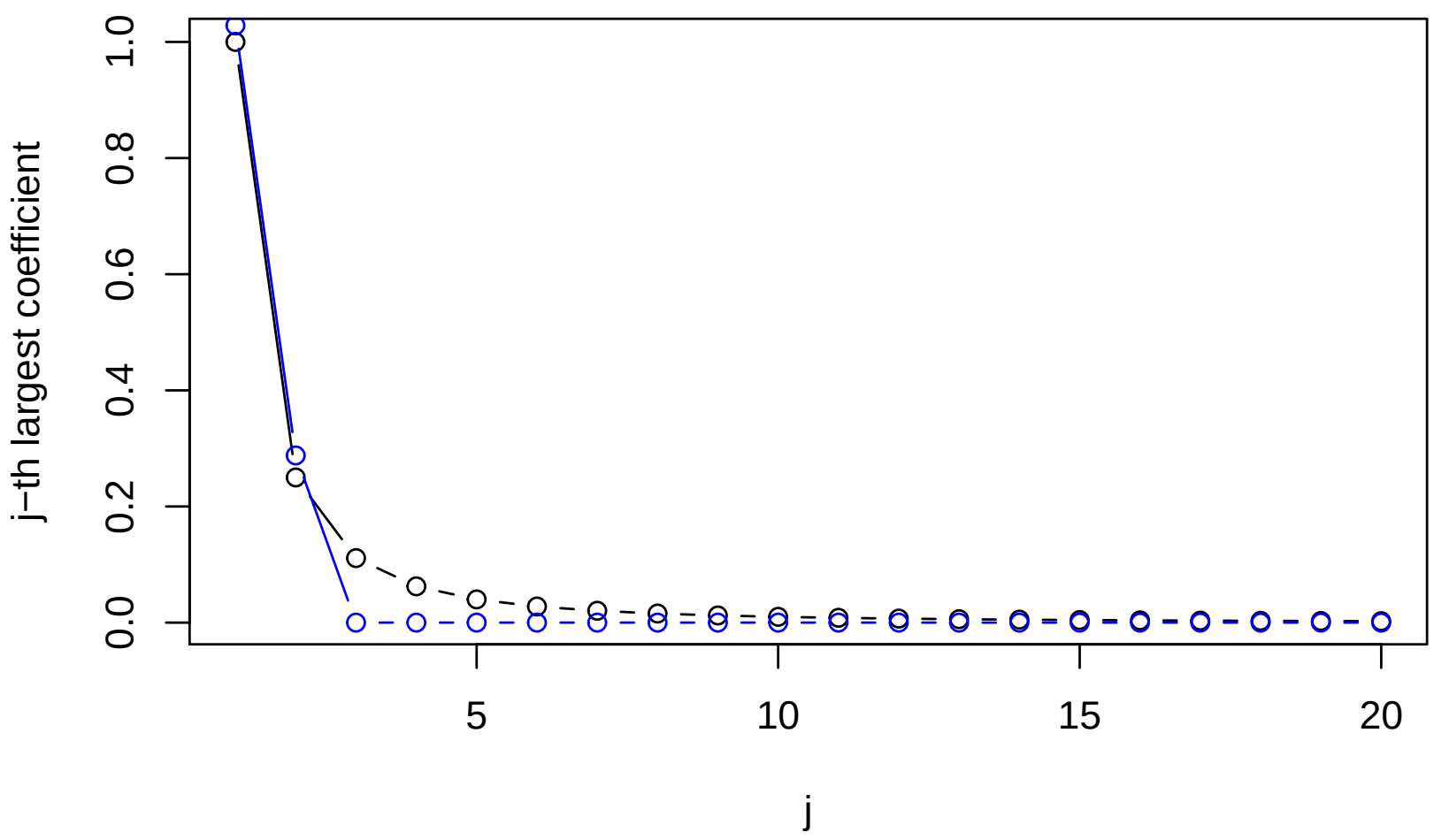

**Figure 3.4:** The true coefficients (black) vs. coefficients estimated by Post-Lasso (blue) in the Example 3.1.1. Post-Lasso tends to remove regularization bias from the estimated non-zero coefficients.

**Remark 3.1.1** (Cross Validation and OLS Post-Lasso) Note that, when using Post-Lasso, one should either use the theoretically justified penalty parameter ([3]) as outlined above or cross-validation for the overall OLS Post-Lasso process. That is, one *should not* apply cross-validation to the Lasso to find a value for $\lambda$ and then use this same value of $\lambda$ with Post-Lasso. Unsurprisingly, using a penalty parameter chosen to optimize cross-validation performance for Lasso tends to lead to poor empirical performance when applied to an entirely different procedure, Post-Lasso.

## 3.2 Predictive Performance of Lasso and Post-Lasso

The best linear prediction rule (out-of-sample) is $\beta' X$. We want to understand the quality of the Lasso prediction rule, $\hat{\beta}' X$. That is,

- Does $\hat{\beta}' X$ provide a good approximation to $\beta' X$?

Recall that with Lasso, we are trying to estimate $p$ parameters $\beta_1, \ldots, \beta_p$, imposing approximate sparsity via penalization. Under approximate sparsity, only a few, say $s$, parameters will be "important." We can call $s$ the *effective dimension*. Lasso approximately figures out which parameters are important to keep. Further, intuitively, to estimate each of the "important" $s$ parameters well, we need many observations for each such parameter. This means that $n/s$ must be large, or, equivalently $s/n$ must be small. Using previous reasoning from least squares theory, we might also conjecture that the key determinant of the rate at which Lasso approximates the best linear predictor is $\sqrt{s/n}$. This conjecture is almost correct.

**Theorem 3.2.1** *Under approximate sparsity as defined in Definition 3.1.1, restricted isometry conditions stated below, choosing $\lambda$ as in* (3.1.4)*, and other regularity conditions stated e.g. in [3, 6], with probability approaching $1 - \alpha$ as $n \to \infty$, the following bound holds:*

$$\sqrt{\mathrm{E}_X \left[ (\beta' X - \hat{\beta}' X)^2 \right]} \leq \text{const} \cdot \sqrt{\mathrm{E}[\epsilon^2]} \sqrt{\frac{s \log(\max\{p, n\})}{n}},$$

*where $\mathrm{E}_X$ denotes expectation with respect to $X$, and the effective dimension is*

$$s = \text{const} \cdot A^{1/a} \cdot n^{\frac{1}{2a}},$$

*where constant $a$ is the speed of decay of the sorted coefficient values in the approximate sparsity definition, Definition 3.1.1. Moreover, the number of regressors selected by Lasso is bounded by*

$$\text{const} \cdot s$$

*with probability approaching $1 - \mathsf{a}$ as $n \to \infty$. The constants* const *are different in different places and may depend on the distribution of $(Y, X)$ and on* $\mathsf{a}$.

The definition of effective dimension stated in this theorem applies, for instance, under the regularity condition that $\max_{j=1}^{p} \|\mathrm{E}[X_j X]\|_1 \leq$ const; i.e. the sum of the absolute values of every row of the covariance matrix $\mathrm{E}[XX']$ is at most a constant. One can also obtain appropriate notions of effective dimension under weaker assumptions on the covariance matrix. For example, one obtains $s \propto n^{1/(2a-1)}$ if $\max_{j=1}^{p} \|\mathrm{E}[X_j X]\|_2 \leq$ const or $s \propto n^{1/(2(a-1))}$ if $\max_{j=1}^{p} \|\mathrm{E}[X_j X]\|_\infty \leq$ const where $\propto$ means "is proportional to."

Therefore, if $s \log(\max\{p, n\})/n$ is small, Lasso and Post-Lasso regression come close to the population regression function/best linear predictor. Relative to our conjectured rate $\sqrt{s/n}$, there is an additional factor $\sqrt{\log(\max\{p, n\})}$ in the bound. This factor captures the price of not knowing *a priori* which of the $p$ regressors are the $s$ important ones. Lasso approximately finds these important predictors, but correspondingly suffers a loss relative to a predictor estimated with knowledge of the best $s$-dimensional model ("oracle estimator"). A theoretical guarantee similar to Theorem 3.2.1 has been established for cross-validated Lasso [4], though with number of selected regressors diverging slowly relative to $s$ rather than achieving $s = \text{const} \cdot s$.

Under approximate sparsity and with appropriate choice of penalty parameters, Lasso and Post-Lasso will approximate the best linear predictor well. Theoretically, they will not overfit the data, and we can thus use the sample and adjusted $R^2$ and $MSE$ to assess out-of-sample predictive performance. Of course, it is always a good idea to verify the out-of-sample predictive performance by using sample splitting.

**Remark 3.2.1** (Exact Sparsity) It is helpful to consider the exactly sparse case, in which there are only $k$ non-zero coefficients bounded by some constant and the rest of the coefficients are exactly zero. In this case, the effective dimension is (up to constants) equal to the number of non-zero coefficients, i.e.

$$s = \text{const} \cdot k.$$

To see this, note that $\beta$ satisfies the approximate sparsity condition with $A = \text{const} \cdot k^a$ for $a \geq 1$, since $\beta_j \leq \text{const} \leq \text{const} \cdot k^a / j^a$ for $j \leq k$ and $\beta_j = 0 \leq \text{const} \cdot k^a / j^a$ for $j > k$. Then $s \leq \text{const} \cdot kn^{1/2a}$, which yields the result as $a \to \infty$.

**On regularity conditions**$^\star$. A sufficient condition under which Theorem 3.2.1 can be established is the restricted isometry condition:

**Definition 3.2.1** (Restricted Isometry) *The following conditions hold:*

$$\textit{Uniformly in } Z \subset X : \dim(Z) \leq L = s \log(n),$$

$$\sup_{\|a\|=1} |a'(\mathbb{E}_n[ZZ'] - \mathrm{E}[ZZ'])a| \approx 0,$$

$$0 < C_1 \leq \inf_{\|a\|=1} a'\mathrm{E}[ZZ']a \leq C_2 < \infty,$$

*where $C_1$ and $C_2$ are constants.*

This condition says that "small groups" of regressors are not collinear and are well-behaved. I.e. we have that subvectors $Z$ of $X$ with dimension $L = s\log(n)$ have empirical Gram matrices $\mathbb{E}_n[ZZ']$ that are close to their population analogues $\mathrm{E}[ZZ']$ in the operator norm and have population covariance matrix $\mathrm{E}[ZZ']$ with eigenvalues bounded away from zero and from above. This condition is simple and intuitive but is stronger than necessary. Results similar to Theorem 3.2.1 have been shown to hold under considerably weaker conditions. The condition $\sup_{\|a\|=1} |a'(\mathbb{E}_n[ZZ'] - \mathrm{E}[ZZ'])a| \approx 0$ has been demonstrated to be valid under various more primitive conditions; see Appendix 3.C.

## 3.3 A Helicopter Tour of Other Penalized Regression Methods for Prediction

Instead of the Lasso penalty, other penalty schemes can be used, leading to different regression estimators with different properties. These estimators are motivated by different structures for the coefficients on the set of regressors in a high-dimensional model. We consider three important settings where coefficient are sparse, dense, or sparse+dense.

We have already seen that sparse coefficient vectors have a small number of relatively large, non-zero coefficients with the rest of the coefficients being close enough to zero to be ignorable. A dense coefficient vector has the vast majority or all coefficients non-zero and of comparable magnitude. A sparse+dense structure has the vast majority of coefficients being non-zero and of similar magnitude along with a small number of relatively large coefficients. Figure 3.5 illustrates each setting.

Throughout this section, we assume that regressors have been centered and normalized to have second empirical moment equal to 1. We thus ignore coefficient specific penalty parameters like the $\hat{\psi}_j$ in the Lasso problem (3.1.3).

We have already outlined Lasso regression, which performs best in an approximately sparse setting. We next consider the Ridge method, which performs best in the dense setting.

**Ridge.** The Ridge method estimates coefficients by penalized least squares, where we minimize the sum of squared prediction error plus the penalty term given by the sum of

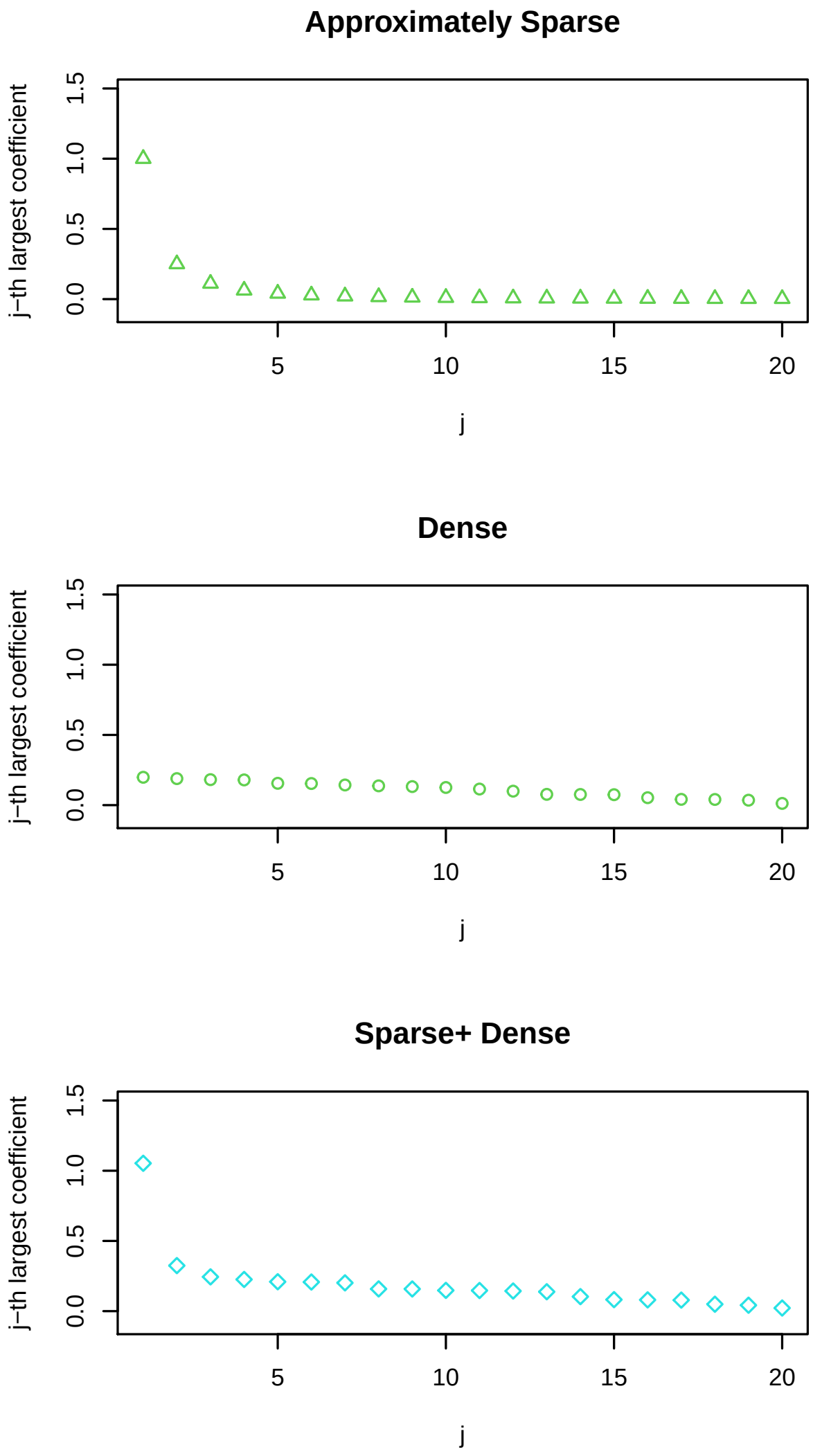

**Figure 3.5:** The Lasso penalty is best suited for approximately sparse models, and the Ridge penalty for models with small dense coefficients. The Elastic Net can be tuned to perform well with either sparse or dense coefficients. The Lava penalty is best suited for models with coefficients generated as the sum of approximately sparse coefficients and small dense coefficients.

the squared values of the coefficients times a penalty level $\lambda$:

$$\hat{\beta}(\lambda) = \arg\min_{b \in \mathbb{R}^p} \sum_{i=1}^{n} (Y_i - b'X_i)^2 + \lambda \sum_{j} b_j^2.$$

Ridge balances the complexity of the model measured by the sum of squared coefficients with the goodness of in-sample fit. In contrast to Lasso, Ridge penalizes the large values of coefficients much more aggressively and small values much less aggressively – indeed, squaring big values makes them even bigger and squaring small numbers makes them even smaller.

Because of the latter property,

- Ridge does not set estimated coefficients to zero and so it does not do variable selection.
- The Ridge predictor $\hat{\beta}'X$ is especially well suited for prediction in "dense" models, where the $\beta_j$'s are all small without necessarily being approximately sparse.
- Ridge regression is also well suited when the matrix $\mathrm{E}[XX']$ is poorly behaved, as measured by the decay of its eigenvalues to zero.

In the dense case, the Ridge predictor can easily outperform the Lasso predictor.

Like Ridge, the Lasso predictor empirically seems to have reasonable prediction performance in the presence of ill-behaved design matrices, although we don't understand its theoretical properties well in this case.

**Remark 3.3.1** (Theoretical Properties of the Ridge Procedure*) For excellent analysis of Ridge properties, see [7], who present the following bound for the fixed (conditional on) $X_1, \ldots X_n$ case holding with high probability:

$$\mathbb{E}_n\left[(\hat{\beta}'X - \beta'X)^2\right] \lesssim \sum_{j=1}^{p} \frac{\lambda^2 \zeta_j \gamma_j^2}{(\zeta_j^2 + \lambda)^2} + \frac{\mathrm{E}[\epsilon^2]}{n} \sum_{j=1}^{p} \left( \frac{\zeta_j^2}{(\zeta_j + \lambda)^2} \right),$$

where $(\zeta_j)_{j=1}^p$ are eigenvalues of $\mathbb{E}_n[XX']$ and $\gamma_j$ are such that $\beta = \sum_{j=1}^p \gamma_j c_j$ with $c_k$ being the eigenvectors of $\mathbb{E}_n[XX']$. The theoretically optimal penalty level can be chosen to minimize the right hand side, though doing so is infeasible as the right hand side depends on $\beta$. In practice, the penalty level is generally chosen by cross-validation. An analogous result holds for bounding $\mathrm{E}_X\left[(\hat{\beta}'X - \beta'X)^2\right]$ in the case of random $X_1, \ldots X_n$; see [7] for the statement.

The first component on the right hand side can be thought of as squared bias, and the second component is mean squared estimation error. Observe that when $\zeta_j = 1$ and $\lambda$ is bounded, the second term is of order $p/n$, which translates to the rate of $\sqrt{p/n}$ after taking the square root. Having the second term go to 0 thus requires $\sqrt{p/n} \to 0$. In contrast, $p$ can be larger than $n$ and the second term can still vanish when eigenvalues decay to zero. In this case, the effective dimension for a given $\lambda$ is

$$d(\lambda) = \sum_{j=1}^{p} \frac{\zeta_j^2}{(\zeta_j + \lambda)^2},$$

and the second term is of order $d(\lambda)/n$. The ratio $d(\lambda)/n$ then determines the rate at which the Ridge predictor approximates the optimal predictor if the square bias term is of smaller order. Of course, it is hard to know that the square bias term is of smaller order than the second term in practice. The squared bias term will also not be of small order when there is a large $\gamma_j$ associated with a large eigenvalue $\zeta_j$.

**Remark 3.3.2** (Connection to Principal Components$^\star$) Ridge regression is closely related to *principal components regression* which regresses an outcome of the first $K$ principal components of the predictor variables $X_i$. Principal components provide mutually orthogonal rotations of the original $X_i$'s that maximize fit to the overall design matrix. Here, we consider a case where we have $p < n$ centered predictor variables that are linearly independent. We let $P_{ki}$ denote the $i^{\text{th}}$ element of the $k^{\text{th}}$ normalized principal component - the principal component divided by it's standard deviation which is given by the $k^{\text{th}}$ largest eigenvalue of $\mathbb{E}_n[XX']$, $\zeta_k$. Under these conditions, the ridge prediction can be expressed as

$$X_i'\hat{\beta} = \sum_{k=1}^{p} P_{ki} \frac{\zeta_k}{\zeta_k + \lambda} \mathbb{E}_n[P_k Y].$$

Note that principal components regression using the first $K$ principal components would produce predictions

$$\hat{y}_i = \sum_{k=1}^{K} P_{ki} \mathbb{E}_n[P_k Y].$$

That is, Ridge and principal components regression are tightly connected. Unlike principal components regression, Ridge regression does not pre-select which principal components to use but instead places less weight on low variance principal components according to $\frac{\zeta_k}{\zeta_k+\lambda}$. We find the implicit use of principal components in ridge to be interesting, but note that we can explicitly use principal components as input variables in all penalized methods and in the more advanced methods that we discuss in Chapter 9. We visit using Principal Component Analysis for feature extraction when we outline feature engineering in Chapter 11. For further discussion, see [8] p. 64-67 or the blog post Ridge vs PCA.

Ridge and Lasso have other useful modifications or hybrids that can perform well in the sparse, dense or sparse + dense settings. One popular modification is the Elastic Net [9] that can perform well in either the sparse or the dense scenario with appropriate tuning.

**Elastic Net.** The Elastic Net method estimates coefficients by penalized least squares with the penalty given by a linear combination of the Lasso and Ridge penalties:

$$\hat{\beta}(\lambda_1, \lambda_2) = \arg\min_{b \in \mathbb{R}^p} \sum_i (Y_i - b'X_i)^2 + \lambda_1 \sum_j b_j^2 + \lambda_2 \sum_j |b_j|.$$

We see that the penalty function has two penalty levels $\lambda_1$ and $\lambda_2$, which are chosen by cross-validation in practice.

- By selecting different values of penalty levels $\lambda_1$ and $\lambda_2$, we have more flexibility with Elastic Net for building a good prediction rule than with just Ridge or Lasso.
- The Elastic Net performs variable selection unless we completely shut down the Lasso penalty by setting $\lambda_2 = 0$.
- With proper tuning, Elastic Net works well in regression models where regression coefficients are either approximately sparse or dense.

See [10] for some theoretical results on Elastic Net.

Another way to combine the Lasso and Ridge penalties is the Lava method, which is intended to work well in sparse+dense settings.

**Lava**. The Lava method ([11], [12]) estimates coefficients by solving the penalized least squares problem:

$$\hat{\beta}(\lambda_1, \lambda_2) = \arg\min_{b: b = \delta + \xi \in \mathbb{R}^p} \sum_i (Y_i - b'X_i)^2$$

$$+ \lambda_1 \sum_j \delta_j^2 + \lambda_2 \sum_j |\xi_j|.$$

Here components of the parameter vector are split into a "dense part" $\delta_j$ and "sparse part" $\xi_j$, where the $\delta_j$'s are penalized like in Ridge, and the $\xi_j$'s are penalized like in Lasso. The minimization program automatically determines the best split into the dense and sparse parts. There are two corresponding penalty levels $\lambda_1$ and $\lambda_2$, which can be chosen by cross-validation in practice.

- Compared to the Elastic Net, the Lava method penalizes large and small coefficients much less aggressively – large coefficients are penalized like Lasso and small coefficients like Ridge. Like Ridge, Lava does not do variable selection.
- Lava is designed to work well in

  "sparse + dense"

  regression models where there are several large coefficients and many small coefficients that do not vanish quickly enough to satisfy approximately sparsity.
- With proper tuning that allows either $\lambda_1$ or $\lambda_2$ to be set to large values, Lava can also work in either "sparse" or "dense" models.

Theoretical guarantees for these methods are given in [11] and [12]. Theoretically and practically, Lava can significantly outperform Lasso, Ridge and Elastic Net in "sparse+dense" models, and, with appropriate tuning, has comparable performance to Lasso in "sparse" models and to Ridge in "dense" models.

## 3.4 Choice of Regression Methods in Practice

How should we select the appropriate penalized regression method? The answer is simple if we are interested in building the best prediction. We can split the data into training and testing sets and simply choose the method that performs the best on the test set. Rigorous theoretical guarantees for this approach have been provided by [13].

We show an example of this approach in R Notebook on ML for Prediction of Wages and Python Notebook on ML for Prediction of Wages which illustrate the use of penalized regression methods for predicting log-wages using CPS 2015 data. We can also use ensemble methods to aggregate prediction methods to get boosts in predictive performance – we describe these aggregation methods in Chapter 9.

## Notebooks

- R Notebook on Penalized Regressions and Python Notebook on Penalized Regressions provide details of implementation of different penalized regression methods and examine their performance for approximating regression functions in a simulation experiment. The simulation experiment includes one case with approximate sparsity, one case with dense coefficients, and another case with both approximately sparse and dense components.

- R Notebook on ML for Prediction of Wages and Python Notebook on ML for Prediction of Wages provide details of implementation of different penalized regression methods and examine their performance for predicting log-wages using CPS 2015 data.

## Notes

Lasso was introduced by Frank and Friedman [14], and its geometric and computational properties were elaborated on by Tibshirani [15], who also gave it its name. The first general theoretical analysis of Lasso was done by Bickel, Ritov, and Tsybakov [2]. Hastie, Tibshirani, and Wainwright [16] provides a good textbook introduction.

There are many variations on the basic Lasso theme, only some of which we mentioned in this chapter. The properties of the Post-Lasso estimator in approximately sparse models (without assuming that Lasso perfectly selects the "right model") were first established in [3]. The properties of Lasso and Post-Lasso don't hinge on the assumption of Gaussian or sub-Gaussian errors, as proven in [6], though such assumptions are often imposed. Fundamentally, the properties of these procedures rely on a high-dimensional central limit theorem ([5]) that allows Gaussian approximations to key average-like quantities.While cross-validation has been frequently used to select the penalty level, validity of this approach for Lasso was only proven recently – [4]. The Lasso has been extended to clustered dependence by [17] and to time series and many time series by [18], with the corresponding package available at this Link.

There is a large literature on Ridge estimation, with the reference [7] providing what seems to be the state of the art. The Lava approach has been proposed and analyzed in [11] and [12]. [12] also discusses applications to problems with latent confounding and, for this reason, refers to Lava as the spectral deconfounder. We discuss other approaches to dealing with latent confounding in Chapter 12 and Chapter 13.

## Study Problems

1. Solve the Lasso optimization problem analytically with only one regressor and interpret the solution.

2. Experiment with the R Notebook on Penalized Regressions, trying out modifications of the Monte-Carlo experiments. As examples, you might change parameters that govern the speed of decay of coefficients to zero, change the error distribution, or alter the structure of dependence among the design variables. Try to explain the results to a fellow student, linking explanations to the theoretical properties of these methods.

3. Experiment with the R Notebook on ML Prediction of Wages. Try to explain the results to a fellow student, linking explanations to the theoretical properties of these methods.

## 3.A Additional Discussion and Results

### Iterative Estimation of $\sigma$

The plug-in choice of $\lambda$ given in equation (3.1.4) requires an estimate of $\sigma$. We can estimate $\sigma$ using the following iterative method. Let $X^0$ be a small set of regressors (a trivial choice is just the intercept, but we may include, for example, the five regressors that are most strongly correlated with the $Y_i$'s). Let $\widehat{\beta}_0$ be the least squares estimator of the coefficients on the covariates associated with $X^0$, and define

$$\hat{\sigma}_0 := \sqrt{\mathbb{E}_n[(Y_i - \hat{\beta}_0' X_i^0)^2]}.$$

Set $k = 0$, and specify a small constant $\nu \geq 0$ as a tolerance level and a constant $K > 1$ as an upper bound on the number of iterations:

We find that $K = 1$ works well in practice.

1. Compute the Lasso estimator $\widehat{\beta}$ based on the penalty level $\lambda$ given in equation (3.1.4) using $\hat{\sigma}_k$.
2. Set $\hat{\sigma}_{k+1} = \sqrt{\mathbb{E}_n[(Y_i - \hat{\beta}' X_i)^2]}$.
3. If $|\hat{\sigma}_{k+1} - \hat{\sigma}_k| \leqslant \nu$ or $k > K$, stop; otherwise set $k \leftarrow k + 1$ and go to (1).

We note that the plug-in choice of $\lambda$ given in equation (3.1.4) relies on assuming homoskedasticity of the BLP residuals, i.e. $\epsilon \perp\!\!\!\perp X$. This independence implies that $\mathrm{E}[\epsilon^2 X_j^2] = \mathrm{E}[\epsilon^2]\mathrm{E}[X_j^2]$. With independent observations where we do not have $\epsilon \perp\!\!\!\perp X$, we should use penalty loadings $\hat{\psi}_j = \sqrt{\mathbb{E}_n[\hat{\epsilon}^2 X_j^2]}$, where $\hat{\epsilon}_i \approx \epsilon_i$ can be estimated in a similar iterative manner as described above. In this case, we would then take $\hat{\sigma} = 1$ in formula (3.1.4) for $\lambda$ (see [6] for more details).

We expect the homoskedastic formula for the penalty provided in (3.1.4) will work well in many cases, especially when random variables $\epsilon, X_j$ are expected to have fast decaying tail probabilities. For example, when fourth moments of $\epsilon, X_j$ are bounded by some constant factor of their second moments, an application of the Cauchy-Schwarz inequality implies that $\mathrm{E}[\epsilon^2 X_j^2] \leq \text{const} \cdot \mathrm{E}[\epsilon^2]\mathrm{E}[X_j^2]$, which is, up to a constant, the simplifying condition implied by homoskedasticity.

## Some Lasso Heuristics via Convex Geometry★

Assume $\hat{\psi}_j = 1$ for each $j$ for simplicity, which amounts to normalizing regressors to have the second empirical moment equal to 1. Consider

$$\widehat{\beta} \in \arg\min_{b \in \mathbb{R}^p} \widehat{Q}(b) + \frac{\lambda}{n}\|b\|_1, \tag{3.A.1}$$

where

$$\widehat{Q}(b) = \mathbb{E}_n[(Y_i - b' X_i)^2].$$

The key quantity in the analysis of (3.A.1) is the score – the gradient of $\widehat{Q}$ at the true value:

$$S = -\nabla \widehat{Q}(\beta_0) = 2\mathbb{E}_n[X\epsilon].$$

The score $S$ is the effective "noise" in the problem that should be dominated by the regularization. However, we would like to make the regularization bias as small as possible. This reasoning suggests choosing the smallest penalty level $\lambda$ that is just large enough to dominate the noise with high probability, say $1 - \mathsf{a}$, which yields

$$\lambda > c\Lambda, \text{ for } \Lambda := n\|S\|_\infty. \tag{3.A.2}$$

Here, $\Lambda$ is the maximal score scaled by $n$, and $c > 1$ is a theoretical constant that guarantees that the score is dominated.

It is useful to mention some simple heuristics for the principle (3.A.2) which arise from considering the simplest case where all of the regressors are irrelevant so that $\beta = 0$. We want our estimator to perform at a near-oracle level in all cases, including this case, but here the oracle estimator $\beta^*$ sets $\beta^* = \beta = 0$. We thus also want $\widehat{\beta} = \beta = 0$ in this case, at least with a high probability, say $1 - \mathsf{a}$. From the subgradient optimality conditions for (3.A.1), we must have

$$-S_j + \lambda/n > 0 \text{ and } S_j + \lambda/n > 0 \text{ for all } 1 \le j \le p \tag{3.A.3}$$

for the Lasso estimator for each coefficient to be exactly 0. We can guarantee (3.A.3) holds by setting the penalty level $\lambda/n$ such that $\lambda > n \max_{1\le j\le p} |S_j| = n\|S\|_\infty$ with probability at least $1 - \mathsf{a}$, which is precisely what the rule (3.A.2) does.

Gaussian approximations to this score motivate the following X-dependent penalty implementation.

**Remark 3.A.1** (Refining Penalty Levels) An X-dependent penalty level can be specified as follows:

$$\lambda = c \cdot 2\hat{\sigma}\Lambda(1 - \mathsf{a}|\{X_i\}_{i=1}^n), \tag{3.A.4}$$

where

$$\Lambda(1 - \mathsf{a}|\{X_i\}_{i=1}^n) \\ = (1 - \mathsf{a}) - \text{quantile of } \; n\|\mathbb{E}_n[Xg/\Psi]\|_\infty \mid \{X_i\}_{i=1}^n,$$

$g_i$ are i.i.d. $N(0, 1)$, and $\Psi = \text{diag}(\hat{\psi}_j)_{j=1}^p$. $\Lambda(1 - \mathsf{a}|\{X_i\}_{i=1}^n)$ can be thus be easily approximated by simulation. The use of normal errors $g_i$ could be motivated by assuming the Gaussian errors $\epsilon_i$ in the model or by appealing to a high-dimensional central limit theorem. We note that by the union bound and Feller's tail inequality,

$$\begin{aligned}\Lambda(1-\mathsf{a}|\{X_i\}_{i=1}^n) &\leq \sqrt{n} z_{1-\mathsf{a}/(2p)} \\ &\leq \sqrt{2n\log(2p/\mathsf{a})}.\end{aligned} \tag{3.A.5}$$

Thus, $\sqrt{2n\log(2p/\mathsf{a})}$ provides a simple upper bound on the penalty level.

Refined penalty levels are important when components of $X_i$ are highly correlated, in which case the X-dependent penalty will be much lower that the bounds given in 3.A.5. Using the lower penalty level can offer both practical and theoretical boosts in performance in such cases.

## Other Variations on Lasso

Here and below we assume that

$$\hat{\psi}_j = 1, \quad j = 1, ..., p$$

to simplify notation. A variant of Lasso, called the *Square-root Lasso* estimator ([19],[20]), is defined as a solution to the following program:

$$\min_{b\in\mathbb{R}^p} \sqrt{\mathbb{E}_n[(Y-b'X)^2]} + \frac{\lambda}{n}\|b\|_1. \tag{3.A.6}$$

Analogously to Lasso, we may set the penalty level as

$$\lambda = c \cdot \widetilde{\Lambda}(1-\mathsf{a}|\{X_i\}_{i=1}^n), \tag{3.A.7}$$

where $c > 1$ and

$$\begin{aligned}&\widetilde{\Lambda}(1-\mathsf{a}|\{X_i\}_{i=1}^n) \\ &\qquad = (1-\mathsf{a}) - \text{quantile of } \; n\|\mathbb{E}_n[Xg]\|_\infty/\sqrt{\mathbb{E}_n[g^2]} \mid \{X_i\}_{i=1}^n,\end{aligned}$$

with $g_i \sim N(0,1)$ independent for $i = 1, \ldots, n$. As with Lasso, there is also a simple asymptotic option for setting the penalty level:

$$\lambda = c \cdot 2\sqrt{n} z_{1-\mathsf{a}/(2p)}. \tag{3.A.8}$$

The main attractive feature of (3.A.6) is that the penalty level $\lambda$ specified above is independent of the value $\sigma$. This estimator has statistical performance that is as good as the iterative or cross-validated Lasso. Moreover, the estimator is a solution to a highly tractable conic programming problem:

$$\min_{t\geq 0, b\in\mathbb{R}^p} t + \frac{\lambda}{n}\|b\|_1 : \sqrt{\mathbb{E}_n[(Y - b'X)^2]} \leq t, \tag{3.A.9}$$

where the criterion function is linear in parameters $t$ and positive and negative components of $b$, while the constraint can be formulated with a second-order cone, informally known as the "ice-cream cone."

There are several other estimators that make use of penalization by the $\ell_1$-norm. A final important case is the *Dantzig selector* estimator [21]. It also relies on $\ell_1$-regularization but exploits the notion that the residuals should be nearly uncorrelated with the covariates. The estimator is defined as a solution to

$$\min_{b\in\mathbb{R}^p} \|b\|_1 \quad : \quad \|\mathbb{E}_n[X(Y - b'X)]\|_\infty \leq \lambda/n. \tag{3.A.10}$$

Again, one may set $\lambda = \sigma\Lambda(1-\mathsf{a}|\{X_i\}_{i=1}^n)$. Here, we focused our discussion on Lasso but virtually all theoretical results carry over to other $\ell_1$-regularized estimators including (3.A.6) and (3.A.10). We also refer to [22] for a feasible Dantzig estimator that combines the square-root Lasso method (3.A.9) with the Dantzig method.

## 3.B Cross-Validation

Cross-validation is a common practical tool that provides a way to choose tuning parameters such as the penalty level in Lasso. The idea of cross-validation is to rely on repeated splitting of the training data to estimate the out-of-sample predictive performance.

**Definition 3.B.1** (Cross-Validation in Words)

- *We partition the data into K blocks called "folds." For example, with $K = 5$, we split the data into 5 non-overlapping blocks.*
- *Leave one block out. Fit a prediction rule on all the other blocks. Predict the outcome observations in the left out block, and record the empirical Mean Squared Prediction Error. Repeat this for each block.*
- *Average the empirical Mean Squared Prediction Errors over blocks.*
- *We do these steps for several or many values of the tuning*

*parameters and choose the value of the tuning parameter that minimizes the Averaged Mean Squared Prediction Error.*

We can also consider many different methods for constructing prediction rules as well. For example, we could try Lasso with many different values of the penalty parameter and Ridge with many different values of the penalty parameter and choose the tuning parameter and method (Lasso or Ridge) that minimizes the cross-validated Mean Squared Prediction Error.

**Definition 3.B.2** (Cross-Validation: Formal Description)

- *Randomly partition the observation indices* $1, \ldots., n$ *into* $K$ *folds* $B_1, \ldots, B_K$.
- *For each* $k = 1, \ldots, K$, *fit a prediction rule denoted by* $\hat{f}^{[k]}(\cdot; \theta)$, *where* $\theta$ *denotes the tuning parameters such as penalty levels and* $\hat{f}^{[k]}$ *depends only on observations with indices not in the fold* $B_k$.
- *For each* $k = 1, \ldots, K$, *the empirical out-of-sample MSE for the block* $B_k$ *is*

$$MSE_k(\theta) = \frac{1}{m_k} \sum_{i \in B_k} (Y_i - \hat{f}^{[k]}(X_i; \theta))^2,$$

  *where* $m_k$ *is the size of the block* $B_k$.
- *Compute the cross-validated MSE as*

$$CV\text{-}MSE(\theta) = \frac{1}{K} \sum_{k=1}^{K} MSE_k(\theta).$$

- *Choose the tuning parameter* $\hat{\theta}$ *as a minimizer of* $CV\text{-}MSE(\theta)$.

**Remark 3.B.1** (On Guarantees of Cross-Validated Predictors) A common step people do in practice is to retrain the predictor $\hat{f}(X)$ on the entire data with the best tuning parameter $\hat{\theta}$ found by cross-validation. Theoretical properties of the resulting cross-validated predictor $\hat{f}(X)$ are only well understood for some high-dimensional problems. E.g., see [4] for results on Lasso with cross-validation.

**Remark 3.B.2** (Guarantees for Pooled Cross-Validated Estimator) On the other hand, there are rigorous theoretical guarantees for the pooled cross-validated predictor:

$$\hat{f}(X) = \frac{1}{K} \sum_{k=1}^{K} \hat{f}^{[k]}(X; \hat{\theta}),$$

which are provided by [23] and [13] who establish that the resulting prediction rule has optimal or near-optimal rates for approximating the best predictor in a given class.

Note that the pooled procedure is different from the default CV procedure implemented in many software packages and used in many applications.

## 3.C Laws of Large Numbers for Large Matrices*

The following results are useful for justifying the restricted isometry condition for empirical Gram matrices $\mathbb{E}_n[XX']$.

Let $s_n, p_n, k_n$ be sequences of positive constants, $\ell_n = \log(n)$, and $C$ a fixed positive constant. Let $(X_i)_{i=1}^n$ be iid. vectors. Denote by $(Z_i)_{i=1}^n$ corresponding subvectors.

Suppose that $\max_{\|a\|=1} \mathrm{E}[(Z'a)^2] \leq C$ for all $Z \subset X$ such that $\dim(Z) \leq s_n \ell_n$ and that one of the following holds:

(a) $X_i$ is a sub-Gaussian, namely

$$\sup_{\|u\|\leq 1} \mathrm{P}(|X_i'u| > t) \leq 2\exp(-t^2/c_2^2)$$

for all $t \geq 0$, and $s_n(\log n)(\log(\max\{p_n, n\}))/n \to 0$;

(b) $X_i$ has bounded components, namely

$$\max_j |X_{ij}| \leq k_n$$

and $k_n^2 s_n \log^2 n \log(s_n \log n) \log(\max\{p_n, n\})/n \to 0$.

Then with probability $1 - \delta_n$

$$\max_{Z\subset X:\dim(Z)\leq s_n\ell_n} \max_{\|a\|=1} |a'(\mathbb{E}_n[ZZ'] - \mathrm{E}[ZZ'])a| \leq \Delta_n,$$

where $(\delta_n, \Delta_n)$ are decreasing sequences and $(\delta_n, \Delta_n) \to 0$.

Under (a) the result follows from Theorem 3.2 in [24] and under (b) the result follows from [25]. These references also imply finite-sample characterization of error bounds $(\delta_n, \Delta_n)$.

## 3.D A Sketch of the Lasso Guarantee Under Exact Sparsity★

Let us assume that the population BLP $\beta_0$ satisfies exact sparsity, i.e. only $s$ out of $p$ coefficients are non-zero. Denote with $A$ the set of non-zero coefficients and with $A^c$ the complement of that set. Since the Lasso minimizes the objective $\hat{Q}(b) + \frac{\lambda}{n}\|b\|_1$ for $\widehat{Q}(b) = \mathbb{E}_n[(Y - b'X)^2]$, we have

$$\hat{Q}(\hat{\beta}) - \hat{Q}(\beta_0) \leq \frac{\lambda}{n}(\|\beta_0\|_1 - \|\hat{\beta}\|_1). \tag{3.D.1}$$

Let $\nu := \hat{\beta} - \beta_0$. Since the objective $\hat{Q}(\beta)$ is convex in $\beta$, we have by an application of the Cauchy-Schwarz inequality that

$$\hat{Q}(\hat{\beta}) - \hat{Q}(\beta_0) \geq \nabla\hat{Q}(\beta_0)'\nu = -S'\nu \geq -\|S\|_\infty \|\nu\|_1$$

for $S = -\nabla\widehat{Q}(\beta_0) = 2\mathbb{E}_n[X\epsilon]$.

We will assume that $\lambda$ is chosen such that we have $\frac{\lambda}{n} \geq 2\|S\|_\infty$ with probability $1 - \mathsf{a}$.[12] We focus then on the good event where the above inequality is satisfied. Then we can combine the above two inequalities:

12: The High-Dimensional CLT bounds tell us that if we set $\lambda \approx \sqrt{n \log(\max\{p/\mathsf{a}, n\})}$, then this inequality holds with probability $1 - \mathsf{a}$.

$$\frac{\lambda}{n}(\|\beta_0\|_1 - \|\hat{\beta}\|_1) \geq -\|S\|_\infty \|\nu\|_1 \geq -\frac{\lambda}{2n}\|\nu\|_1.$$

Hence with with high probability,

$$\hat{\beta} - \beta_0 \in RC = \{\nu : \|\beta_0 + \nu\|_1 \leq \|\beta_0\|_1 + \|\nu\|_1/2\}.$$

Note also that $\nu \in RC$ implies[13]

13: Verify this as a reading exercise.

$$\|\nu_{A^c}\|_1 \leq 3\|\nu_A\|_1 \tag{3.D.2}$$

where $\nu_A$ denotes the entries from $\nu$ in $A$ and $\nu_{A^c}$ denotes the entries of $\nu$ in $A^c$. This inequality roughly states that the error vector $\nu = \hat{\beta} - \beta_0$ is primarily supported on $A$.

We impose the following regularity condition:

$$0 < C_1 \leq \min_{\nu \in RC \setminus 0} \frac{\nu' \mathrm{E}[XX']\nu}{\|\nu\|^2} \leq C_2 < \infty. \tag{3.D.3}$$

The restricted isometry conditions we impose in the text are known to imply this condition.[14]

14: See, e.g. Lemma 10 in [26] for an argument based on [2].

Suppose that we can argue that we have, for any vector $\nu \in$

$RC$,

$$\nu'\mathbb{E}_n[XX']\nu \geq \hat{C}_1 \|\nu\|_2^2 \tag{3.D.4}$$

for some $\hat{C}_1 > 0$ that will be generally be related to $C_1$ and features of the population. (3.D.4) is oftentimes referred to as the empirical Restricted Strong Convexity (RSC) property. We provide an example and the corresponding $\hat{C}_1$ below.

Then, using the fact that $\hat{Q}(\beta)$ is quadratic in $\beta$, we can invoke the exact second order Taylor expansion:

$$\hat{Q}(\hat{\beta}) - \hat{Q}(\beta_0) = S'\nu + \nu'\mathbb{E}_n[XX']\nu \geq -\|S\|_\infty \|\nu\|_1 + \hat{C}_1 \|\nu\|_2^2.$$

When combined with the upper bound from the optimality of $\hat{\beta}$ for the penalized empirical loss and the fact that $\frac{\lambda}{n} \geq 2\|S\|_\infty$, this expansion yields

$$\frac{\lambda}{n}\|\nu\|_1 \geq \hat{Q}(\hat{\beta}) - \hat{Q}(\beta_0) \geq -\frac{\lambda}{2n}\|\nu\|_1 + \hat{C}_1 \|\nu\|_2^2.$$

The second crucial inequality that

$$\|\nu\|_2^2 \leq \frac{3\lambda}{2\hat{C}_1 n}\|\nu\|_1 \tag{3.D.5}$$

then follows.

Finally, note that for any vector $\nu$ that is primarily supported on $A$, the $\ell_2$ and $\ell_1$ norms are within a factor $\approx \sqrt{s}$ of each other:

$$\|\nu\|_1 = \|\nu_A\|_1 + \|\nu_{A^c}\|_1 \leq 4\|\nu_A\|_1 \leq 4\sqrt{s}\|\nu_A\|_2 \leq 4\sqrt{s}\|\nu\|_2$$

where we used the norm inequality, that for an $s$-dimensional vector $v$, we have $\|v\|_1 \leq \sqrt{s}\|v\|_2$. Thus, we can conclude

$$\|\nu\|_2 \leq \frac{6\lambda}{\hat{C}_1 n}\sqrt{s}. \tag{3.D.6}$$

Using the assumption that $\nu'\mathrm{E}[XX']\nu \leq C_2\|\nu\|_2$ for $\nu \in RC$, we get the final bound:

$$\sqrt{\mathrm{E}_X[(X'\hat{\beta} - X'\beta_0)^2]} = \sqrt{\nu'\mathrm{E}[XX']\nu} \leq C_2\|\nu\|_2 \leq \frac{6\lambda\, C_2}{\hat{C}_1 n}\sqrt{s}.$$

It remains to argue the empirical RSC property. Note that if

$$\|\mathbb{E}_n[XX'] - \mathrm{E}[XX']\|_\infty \leq \mu_n$$

with probability approaching 1,[15] then we have

15: Application of the high-dimensional CLT implies that we can take $\mu_n \propto \sqrt{\frac{\log(\max\{p,n\})}{n}}$.

$$
\begin{aligned}
\nu'\mathbb{E}_n[XX']\nu &\geq \nu'\mathrm{E}[XX']\nu - \|\nu\|_1^2 \|\mathbb{E}_n[XX'] - \mathrm{E}[XX']\|_\infty \\
&\geq (C_1 - 16s\mu_n)\|\nu\|_2^2
\end{aligned}
$$

by Condition (3.D.3) and an application of the Hölder inequality. Thus, if $n$ is large enough such that $16s\mu_n \leq \frac{C_1}{2}$, we conclude that the empirical RSC condition holds with $\hat{C}_1 = \frac{C_1}{2}$.

# Statistical Inference on Predictive and Causal Effects in High-Dimensional Linear Regression Models 4

"The partial trend regression method can never, indeed, achieve anything which the individual trend method cannot, because the two methods lead by definition to identically the same results."
(An in-words restatement of the FWL theorem.)

– Ragnar Frisch and Frederick V. Waugh [1].

Here we discuss inference on predictive effects using Double Lasso methods, where we use Lasso (at least) twice to residualize outcomes and a target covariate of interest whose predictive effect we'd like to infer. Double Lasso methods rely on the approximate sparsity of the best linear predictors for the outcome and for the target covariate. The resulting estimator concentrates in a $1/\sqrt{n}$ neighborhood of the true value and is approximately Gaussian, enabling the construction of confidence bands. We explain the low bias property of the Double Lasso method using Neyman orthogonality, and isolate the latter as a critical property for further generalizations.

## 4.1 Introduction

We recall the predictive effect question:[1]

- How does the predicted value of $Y$ change if a regressor $D$ increases by a unit, while other regressors $W$ remain unchanged?

As before, we denote the set of regressors as $X = (D, W)$. In Chapter 1, we discussed how we could use the population regression coefficient corresponding to the variable $D$, denoted $\alpha$, to answer this question. We also discussed how to estimate this effect and construct confidence intervals with regression. Now we turn to estimation and construction of confidence intervals for $\alpha$ in the high-dimensional setting, using the tools we developed in Chapter 3.

1: We discuss assumptions and modeling frameworks under which the predictive effect question has a causal interpretation in detail in Chapter 5 through Chapter 8. Under the framework developed in those chapters, the tools in this chapter offer one approach to performing statistical inference for causal effects. Here, we simply note that we may be interested in providing statistical inference for predictive effects regardless of whether they have a causal interpretation.

Here we focus on using Lasso methods. We can use other penalized methods with the caveat that theoretical guarantees are not available unless we perform additional data splitting. We will discuss the use of data splitting and more general machine learning methods in detail when we introduce "double machine learning" or "debiased machine learning" in Chapter 10.

## 4.2 Inference with Double Lasso

### Inference on One Coefficient

The key to inference will be the application of Frisch-Waugh-Lovell partialling-out. Consider the simple predictive model:

$$Y = \alpha D + \beta' W + \epsilon, \tag{4.2.1}$$

where $D$ is the target regressor and $W$ consists of $p$ controls. After partialling-out $W$,

$$\tilde{Y} = \alpha \tilde{D} + \epsilon, \quad \mathrm{E}[\epsilon \tilde{D}] = 0, \tag{4.2.2}$$

where the variables with tildes are residuals retrieved from taking out the linear effect of $W$ (practically, via linear regression):

$$\tilde{Y} = Y - \gamma_{YW}' W, \quad \gamma_{YW} \in \arg\min_{\gamma \in \mathbb{R}^p} \mathrm{E}[(Y - \gamma' W)^2],$$

$$\tilde{D} = D - \gamma_{DW}' W, \quad \gamma_{DW} \in \arg\min_{\gamma \in \mathbb{R}^p} \mathrm{E}[(D - \gamma' W)^2].$$

$\alpha$ can then be recovered from population linear regression of $\tilde{Y}$ on $\tilde{D}$:

$$\alpha = \arg\min_{a\in\mathbb{R}} \mathrm{E}[(\tilde{Y} - a\tilde{D})^2] = (\mathrm{E}[\tilde{D}^2])^{-1}\mathrm{E}[\tilde{D}\tilde{Y}].$$

Note also that $a = \alpha$ solves the moment equation:

$$\mathrm{E}[(\tilde{Y} - a\tilde{D})\tilde{D}] = 0.$$

We now consider estimation of $\alpha$ in a high-dimensional setting. For estimation purposes, we maintain that we have a random sample $\{(Y_i, X_i)\}_{i=1}^n$ where $X_i = (D_i, W_i)$.

To estimate $\alpha$, we will mimic the partialling-out procedure in the population in the sample. In Chapter 1, where $p/n$ was small, we employed ordinary least squares as the prediction method in the partialling-out steps. We are now considering cases where $p/n$ is not small, and we instead employ Lasso-based methods in the partialling-out steps.

The estimation procedure for a target parameter $\alpha$ in a high-dimensional linear model setting can be summarized as follows:

**The Double Lasso procedure:**

1. We run Lasso regressions of $Y_i$ on $W_i$ and $D_i$ on $W_i$

$$\hat{\gamma}_{YW} = \arg\min_{\gamma\in\mathbb{R}^p} \sum_i (Y_i - \gamma' W_i)^2 + \lambda_1 \sum_j \hat{\psi}_j^Y |\gamma_j|,$$

$$\hat{\gamma}_{DW} = \arg\min_{\gamma\in\mathbb{R}^p} \sum_i (D_i - \gamma' W_i)^2 + \lambda_2 \sum_j \hat{\psi}_j^D |\gamma_j|,$$

   and obtain the resulting residuals:

$$\check{Y}_i = Y_i - \hat{\gamma}_{YW}' W_i,$$

$$\check{D}_i = D_i - \hat{\gamma}_{DW}' W_i.$$

   In place of Lasso, we can use Post-Lasso or other Lasso relatives (the Dantzig selector, square-root Lasso, and others).

2. We run the least squares regression of $\check{Y}_i$ on $\check{D}_i$ to

obtain the estimator $\hat{\alpha}$:

$$\begin{aligned}\hat{\alpha} &= \arg\min_{a \in \mathbb{R}} \mathbb{E}_n[(\check{Y} - a\check{D})^2] \\ &= (\mathbb{E}_n[\check{D}^2])^{-1}\mathbb{E}_n[\check{D}\check{Y}].\end{aligned} \tag{4.2.3}$$

We can use standard results from this regression, ignoring that the input variables were previously estimated, to perform inference about the predictive effect, $\alpha$.

Good performance of the Double Lasso procedure relies on approximate sparsity of the population regression coefficients $\gamma_{YW}$ and $\gamma_{DW}$, with a sufficiently high speed of decrease in the sorted coefficients and on careful choice of the Lasso tuning parameters. For approximate sparsity, we will impose that the sorted coefficients satisfy

$$|\gamma_{YW}|_{(j)} \leq Aj^{-a} \text{ and } \quad |\gamma_{DW}|_{(j)} \leq Aj^{-a}$$

for $a > 1$ and $j = 1, \ldots, p$.[2] Under these sparsity conditions, we can use the plug-in rule outlined in Chapter 3 for choosing $\lambda_1$ and $\lambda_2$. Importantly, using these tuning parameters theoretically guarantees that we produce high quality prediction rules for $D$ and $Y$ while simultaneously avoiding overfitting under approximate sparsity. Absent these guarantees, we cannot theoretically ensure that first step estimation of $\check{D}$ and $\check{Y}$ does not have first-order impacts on the final estimator $\hat{\alpha}$. Practically, we have found that Lasso with penalty parameter selected via cross-validation can perform poorly in simulations in moderately sized samples. We return to this issue in Chapter 10 where we discuss a method that allows the use of complex machine learners, including Lasso and other regularized estimators, and data-driven tuning (e.g. cross-validation).

2: Note that in this case the effective dimension $s$ of the problem is $s \approx A^{1/a}n^{1/2a} \ll n^{1/2}$. Intuitively, the effective number of non-zero coefficients grows slower than $\sqrt{n}$.

The following theorem can be shown for the Double Lasso procedure:

**Theorem 4.2.1** (Adaptive Inference with Double Lasso in High-Dimensional Regression) *Under the stated approximate sparsity, the conditions required for Theorem 3.2.1 (e.g. restricted isometry), and additional regularity conditions, the estimation error in $\check{D}_i$ and $\check{Y}_i$ has no first order effect on $\hat{\alpha}$, and*

$$\sqrt{n}(\hat{\alpha} - \alpha) \approx \sqrt{n}\mathbb{E}_n[\tilde{D}\epsilon]/\mathbb{E}_n[\tilde{D}^2] \overset{a}{\sim} N(0, \mathrm{V}),$$

*where*
$$\mathsf{V} = (\mathrm{E}[\tilde{D}^2])^{-1}\mathrm{E}[\tilde{D}^2\epsilon^2](\mathrm{E}[\tilde{D}^2])^{-1}.$$

The above statement means that $\hat{\alpha}$ concentrates in a $\sqrt{\mathsf{V}/n}$-neighborhood of $\alpha$, with deviations controlled by the normal law. Observe that the approximate behavior of the Double Lasso estimator is the same as the approximate behavior of the least squares estimator in low-dimensional models; see Theorem 1.3.2 in Chapter 1.

Just like in the low-dimensional case, we can use these results to construct a confidence interval for $\alpha$. The standard error of $\hat{\alpha}$ is

$$\sqrt{\hat{\mathsf{V}}/n},$$

where $\hat{\mathsf{V}}$ is a plug-in estimator of $\mathsf{V}$. The result implies, for example, that the interval

$$[\hat{\alpha} \pm 1.96\sqrt{\hat{\mathsf{V}}/n}]$$

covers $\alpha$ about 95% of the time.

## Application to Testing the Convergence Hypothesis

R Notebook on Double Lasso for Growth Convergence and Python Notebook on Double Lasso for Growth Convergence provides code for the convergence hypothesis example.

We provide an empirical example of partialling-out with Lasso to estimate the regression coefficient $\alpha$ in the high-dimensional linear regression model:

$$Y = \alpha D + \beta' W + \epsilon.$$

In this example, we are interested in how economic growth rates ($Y$) are related to the initial wealth levels in each country ($D$) controlling for a country's institutional, educational, and other similar characteristics ($W$).

The relationship is captured by $\alpha$, the "speed of convergence/-divergence," which predicts the speed at which poor countries catch up ($\alpha < 0$) or fall behind ($\alpha > 0$) rich countries, after controlling for $W$. Here, we are interested in understanding if poor countries grow faster than rich countries, controlling for educational and other characteristics. In other words, is the speed of convergence negative: Is $\alpha < 0$?

$\alpha < 0$ corresponds to the Convergence Hypothesis predicted by the Solow growth model. Robert M. Solow is a world-renowned MIT economist who won the Nobel Prize in Economics in 1987.

In our data, the outcome ($Y$) is the realized annual growth rate of a country's wealth (Gross Domestic Product per capita). The target regressor ($D$) is the initial level of the country's wealth. The controls ($W$) include measures of education levels, quality of institutions, trade openness, and political stability in the country. The sample, which is based on the Barro-Lee data set [2], contains 90 countries and about 60 controls. Thus $p \approx 60, n = 90$ and $p/n$ is not small. We expect the least squares method to provide a poor/ noisy estimate of $\alpha$. We expect the method based on partialling-out with Lasso to provide a high-quality estimate of $\alpha$.

| | Estimate | Std. Error | 95% CI |
|---|---|---|---|
| OLS | -0.009 | 0.032 | [-0.073, 0.054] |
| Double Lasso | -0.045 | 0.018 | [-0.080, -0.010] |

**Table 4.1:** Estimates for the convergence coefficient. We report specification robust standard errors with finite sample correction, i.e., "HC1."

Least squares provides a rather noisy estimate of convergence speed, which does not allow drawing strong conclusions about the convergence hypothesis. For example, the 95% confidence interval is wide and includes both positive and negative values. Given that $p/n$ is not small in this example, we should also be highly skeptical of the OLS results and especially the standard error. For example, [3] show that conventional robust standard errors are not even consistent in linear models when $p/n$ is not small. In sharp contrast, Double Lasso provides a precise estimate for which we can obtain theoretically justified inferential statements even though $p/n$ is not close to 0. The Lasso-based point estimate is $-4.5\%$ and the 95% confidence interval for the (annual) convergence rate is $-8\%$ to $-1\%$. This empirical evidence is consistent with the conditional convergence hypothesis.

## 4.3 Why Partialling-out Works: Neyman Orthogonality

### Neyman Orthogonality

In the Double Lasso approach, $\alpha$ is the target parameter and $\eta$ are *nuisance* projection *parameters*[3] with true value

$$\eta^o = (\gamma'_{DW}, \gamma'_{YW})'.$$

As the learned value $\hat{\alpha}$ of $\alpha$ depends on the values of the nuisance parameters, it is useful to explicitly consider the dependence of $\hat{\alpha}$ on the nuisance parameters:

$$\hat{\alpha}(\eta).$$

3: *Nuisance parameters* refer to parameters that must be learned or otherwise adjusted for in order to learn the parameter of interest but are not of direct interest themselves. That is, they are nuisances - we'd like to ignore them if we could.

For the majority of the estimation processes we will describe in this book, we can construct a population analogue

$$\alpha(\eta)$$

of the estimator $\hat{\alpha}(\eta)$, such that the in-sample estimation procedure converges to it, in a formal sense.

For instance, the Double Lasso process constructs the residuals

$$\check{Y}_i(\eta) = Y_i - \eta_1' W_i, \quad \check{D}_i(\eta) = D_i - \eta_2' W_i$$

and then obtains $\hat{\alpha}(\eta)$ as the solution to the empirical estimating equation

$$\widehat{\mathsf{M}}(a, \eta) := \mathbb{E}_n[(\check{Y}(\eta) - a\check{D}(\eta))\check{D}(\eta)] = 0.$$

This process implicitly defines the function $\hat{\alpha}(\eta)$. We can think of the population analog of this process, where we construct the residuals

$$\tilde{Y}(\eta) = Y - \eta_1' W, \quad \tilde{D}(\eta) = D - \eta_2' W$$

and solve the population moment equation

$$\mathsf{M}(a, \eta) := \mathrm{E}[(\tilde{Y}(\eta) - a\tilde{D}(\eta))\tilde{D}(\eta)] = 0, \qquad (4.3.1)$$

which again implicitly defines the function $\alpha(\eta)$.

The main idea of the Double Lasso approach is that, in the population limit, it corresponds to a procedure for learning the target parameter $\alpha$ that is first-order insensitive to local perturbations of the nuisance parameters around their true values, $\eta^o$:

$$\partial_\eta \alpha(\eta^o) = 0. \qquad (4.3.2)$$

Formally, we use $\partial_\eta$ to denote the Gateaux derivative. See Remark 10.4.2 in Chapter 10 for more details.

We will call the local insensitivity of target parameters to nuisance parameters as in (4.3.2) Neyman orthogonality of the estimation process.

Neyman orthogonality is important for providing high-quality estimation and inference, especially in high-dimensional settings. In high-dimensional settings, we use regularization procedures to estimate the nuisance parameters as solutions to suitable prediction problems. The use of regularization generally results in bias, and we may heuristically view using regularized estimates of nuisance parameters as plugging in estimates of these parameters that are close to, but not exactly equal to, the true values of the nuisance parameters $\eta^o$. Neyman orthogonality, which guarantees that the target parameter is locally insensitive to perturbations of the nuisance parameters around their true values, then ensures that this bias does not transmit to the estimation of the target parameter, at least to the first order.

Let us prove the claim $\partial_\eta \alpha(\eta^o) = 0$ for the Double Lasso process. Since the function $\alpha(\eta)$ is implicitly defined as the solution to the equation $\mathsf{M}(a, \eta) = 0$, by the implicit function theorem and letting $\alpha = \alpha(\eta^o)$:

$$\partial_\eta \alpha(\eta^o) = -\partial_a \mathsf{M}(\alpha, \eta^o)^{-1} \partial_\eta \mathsf{M}(\alpha, \eta^o).$$

Here

$$\partial_\eta \mathsf{M}(\alpha, \eta^o)$$

consists of two components

$$\partial_{\eta_1} \mathsf{M}(\alpha, \eta^o) = \mathrm{E}[W \tilde{D}(\eta^o)] = \mathrm{E}[W(D - \gamma'_{DW} W)] = 0$$

and

$$\begin{aligned} \partial_{\eta_2} \mathsf{M}(\alpha, \eta^o) &= -\mathrm{E}[W \tilde{Y}(\eta^o)] + 2\mathrm{E}[\alpha W \tilde{D}(\eta^o)] \\ &= -\mathrm{E}[W(Y - \gamma'_{YW} W)] + 2\mathrm{E}[\alpha W(D - \gamma'_{DW} W)] = 0. \end{aligned}$$

We summarize the discussion as follows:

**Neyman Orthogonality.** The parameter of interest $\alpha$ that depends on nuisance parameters $\eta$ with true value $\eta^o$ is Neyman orthogonal with respect to these parameters if

$$\partial_\eta \alpha(\eta^o) = 0.$$

If the parameter $\alpha$ is defined as a root in $a$ of the equation $\mathsf{M}(a, \eta) = 0$, which depends on the nuisance parameters $\eta$ with true value $\eta^o$, then the equation is Neyman orthogonal if

$$\partial_\eta \mathsf{M}(\alpha, \eta^o) = 0.$$

The principle is applicable to problems outside the high-dimensional linear model problem considered in this chapter.

## What Happens if We Don't Have Neyman Orthogonality?

If we don't have Neyman orthogonality, we should not expect to get high-quality estimates of the target parameters. For example, a seemingly sensible approach that one might consider for statistical inference in the high-dimensional linear model context is as follows:

**(Invalid) Single Selection/Naive Method.**
In this invalid method, one applies Lasso regression of $Y$ on $D$ and $W$ to select relevant covariates $W_Y$, in addition to the covariate of interest, then refits the model by least squares of $Y$ on $D$ and $W_Y$. Inference for the target parameter is then carried out using conventional inference based on the latter regression.

Despite its simplicity and seeming intuitive appeal, the approach outlined above is not a valid approach if the goal is to perform inference on $\alpha$. It is a fine approach if the goal is solely the prediction of the outcome, but it can result in very misleading conclusions about the parameter of interest $\alpha$, as we demonstrate in Example 4.3.1 below.

The naive approach outlined above relies on the moment condition

$$\mathsf{M}(a, b) = \mathrm{E}[(Y - aD - b'W)D] = 0.$$

When $b = \beta$, this moment condition is satisfied by the true value, $a = \alpha$. In this case, t coincides with the classical moment condition for $\alpha$ underlying low-dimensional ordinary least squares which sets prediction errors to be orthogonal to each predictor variable.

However, this moment condition does not exhibit Neyman orthogonality since

$$\partial_b \mathsf{M}(\alpha, \beta) = \mathrm{E}[DW] \neq 0$$

unless $D$ is orthogonal to $W$.[4] Because $\mathsf{M}(a, b)$ is not Neyman orthogonal, the bias and the slower than parametric rate of convergence,

$$\sqrt{s \log(p \vee n)/n},$$

of our estimate of $\beta'W$ will transmit to bias and slower than $\sqrt{n}$ convergence in estimates of $\alpha$ provided by solving the empirical analog of $\mathsf{M}(a, b)$. The "Single Selection" procedure outlined

4: In "pure" RCTs where treatment is assigned independently of everything, $D$'s are orthogonal to $W$, after de-meaning $D$, so Neyman orthogonality automatically holds in this setting.

above exactly provides the solution to this moment condition. Consequently, while this naive procedure provides an estimator of $\alpha$ that will approach the true value in large samples (at a slower than $\sqrt{n}$-rate), the bias of the estimator converges too slowly for standard inference methods to provide reliable inference.

We can set up a simulation experiment to verify that this naive approach provides low-quality estimates for $\alpha$.

**Example 4.3.1** In R Notebook with Experiment on Orthogonal vs Non-Orthogonal Learning and Python Notebook with Experiment on Orthogonal vs Non-Orthogonal Learning, we compare the performance of the naive and orthogonal methods in a computational experiment where $p = n = 100$, $\beta_j = 1/j^2$, $(\gamma_{DW})_j = 1/j^2$, and

$$Y = 1 \cdot D + \beta' W + \varepsilon_Y, \quad W \sim N(0, I), \quad \varepsilon_Y \sim N(0,1)$$

$$D = \gamma'_{DW} W + \tilde{D}, \quad \tilde{D} \sim N(0,1)/4.$$

From the histograms shown in Figure 4.1, we see that the naive estimator is heavily biased, as expected from the lack of Neyman orthogonality in its estimation strategy. We also see that the Double Lasso estimator, which is based on principled partialling-out such that Neyman orthogonality is satisfied, is approximately unbiased and Gaussian.

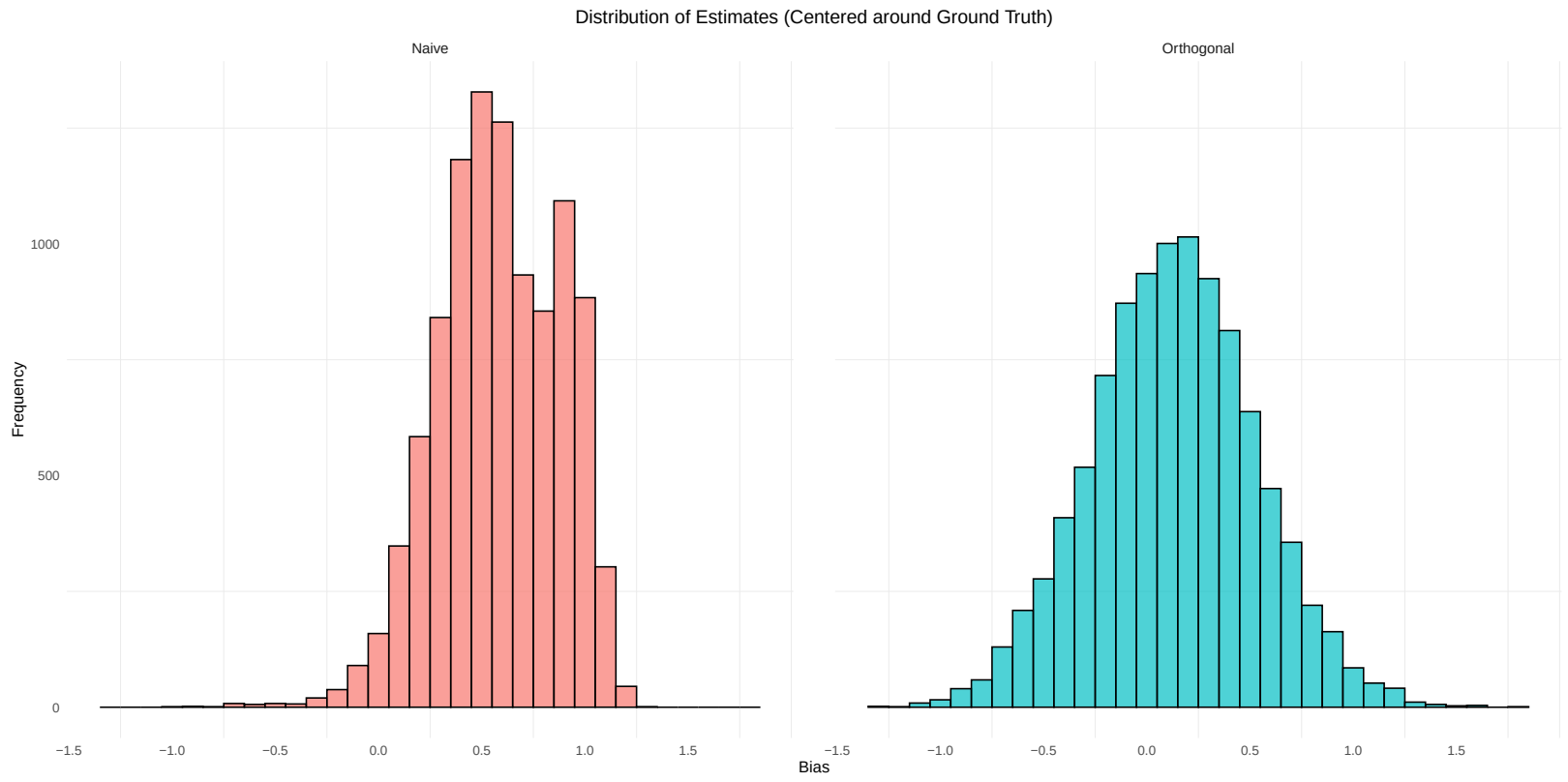

**Figure 4.1: Left Panel:** Simulated distribution of the naive (single-selection) non-orthogonal estimator centered around the true value. **Right Panel:** Simulated distribution of the orthogonal estimator centered around the true value.

The reason that the naive estimator does not perform well is that it only selects controls that are strong predictors of the outcome, thereby omitting weak predictors of the outcome. However, weak predictors of the outcome could still be strong predictors of $D$, in which case dropping these controls results in a strong omitted variable bias. In contrast, the orthogonal approach solves two prediction problems – one to predict $Y$ and another to predict $D$ – and finds controls that are relevant for either. The resulting residuals are therefore approximately "de-confounded."

## 4.4 Inference on Many Coefficients

If we are interested in more than one coefficient, we can repeat the one-by-one Double Lasso procedure for each of the coefficients of interest and obtain valid estimation and inference on each component under regularity conditions.

We consider the model

$$\underbrace{Y}_{\text{Outcome}} = \underbrace{\sum_{\ell=1}^{p_1} \alpha_\ell D_\ell}_{\text{Target Predictors}} + \underbrace{\sum_{j=1}^{p_2} \beta_j \bar{W}_j}_{\text{Controls}} + \epsilon,$$

where we use $D_\ell$ for $\ell = 1, ..., p_1$ to denote the predictors of interest and $\bar{W}_j$ for $j = 1, ..., p_2$ to denote other predictors in the model. Here, both the number of predictors of interest, $p_1$, and the number of additional variables, $p_2$, can both be very large.

There are at least three motivations for considering many coefficients of interest:

- there can be multiple policies whose predictive effect we would like to infer;
- we can be interested in heterogeneous predictive effects across pre-specified groups;
- we can be interested in nonlinear effects of policies.

This setting encompasses examples where we are interested in *heterogeneous effects*, where $D_\ell' s$ are generated as

$$D_\ell = D_0 \bar{X}_\ell, \quad \ell = 1, ..., p_1,$$

where $D_0$ is a base variable of interest – for example, a treatment indicator, a price, or a group indicator – and $(\bar{X}_\ell)_{\ell=1}^{p_1}$ are known transformations of controls $\bar{W}$ – for example, various subgroup indicators.

The setting also encompasses cases where *nonlinear effects* are of interest. For example, we could consider $D_\ell$'s generated as polynomial transformations of a multi-valued base variable, such as a price:

$$D_\ell = D_0^\ell, \quad \ell = 1, ..., p_1.$$

We could further interact these transformations with other variables to study nonlinear heterogeneous effects.

**One by One Double Lasso for Many Target Parameters.** For each $\ell = 1, \ldots, p_1$, we apply the one-by-one Double Lasso procedure for estimation and inference on the coefficient $\alpha_\ell$ in the model

$$Y = \alpha_l D_\ell + \gamma_\ell' W_\ell + \epsilon, \quad W_\ell = ((D_k)'_{k \neq \ell}, \bar{W}')'.$$

Under approximate sparsity conditions, the Double Lasso method provides a high-quality estimate $\hat{\alpha} = (\hat{\alpha}_\ell)_{\ell=1}^{p_1}$ of $\alpha = (\alpha_\ell)_{\ell=1}^{p_1}$ that is approximately Gaussian. We can thus easily construct individual confidence intervals or even joint confidence bands. Under regularity conditions, these results allow for simultaneous inference on $p_1 > n$ coefficients.

**Theorem 4.4.1** (Double Lasso for Many Coefficients) *Under regularity conditions including approximate sparsity as in Definition 3.1.1 with parameters $(A, a)$ with $a > 1$ in all partialling out steps and provided $(\log p_1)^5/n$ is small, we have the adaptivity property,*

$$\sqrt{\log p_1} \max_{\ell \leq p_1} \left| \sqrt{n}(\hat{\alpha}_\ell - \alpha_\ell) - (\mathbb{E}_n[\tilde{D}_\ell^2])^{-1} \sqrt{n} \mathbb{E}_n[\tilde{D}_\ell \epsilon] \right| \approx 0,$$

*and, consequently, the Gaussian approximation*

$$\sqrt{n}(\hat{\alpha} - \alpha) \overset{a}{\sim} N(0, \mathsf{V}),$$

*where*

$$\mathsf{V}_{\ell k} = (\mathrm{E}[\tilde{D}_\ell^2])^{-1} \mathrm{E}[\tilde{D}_\ell \tilde{D}_k \epsilon^2] (\mathrm{E}[\tilde{D}_k^2])^{-1}.$$

Recall that the above distributional approximation formally means that

$$\sup_{R \in \mathcal{R}} \left| \mathrm{P}\left( \sqrt{n}(\hat{\alpha} - \alpha) \in R \right) - \mathrm{P}\left( N(0, \mathsf{V}) \in R \right) \right| \to 0,$$

where $\mathcal{R}$ is a collection of all (hyper) rectangles. The latter result allows the construction of *simultaneous confidence bands* on all target parameters $\alpha_\ell$'s of the form:

$$\widehat{CR} = \times_{\ell=1}^{p_1} \left[ \hat{\alpha}_\ell \pm c \sqrt{\hat{\mathsf{V}}_{\ell\ell}/n} \right],$$

The critical value $c$ in the simultaneous confidence band is

chosen so that

$$\begin{aligned}\mathrm{P}(\alpha \in \widehat{CR}) &= \mathrm{P}\left(\sqrt{n}(\alpha - \hat{\alpha}) \in \sqrt{n}(\hat{CR} - \hat{\alpha})\right)\\ &= \mathrm{P}\left(\sqrt{n}(\alpha_\ell - \hat{\alpha}_\ell) \in [\pm c\hat{\mathrm{V}}_{\ell\ell}^{1/2}]\ \forall\ \ell \in \{1, \ldots, p_1\}\right)\\ &\approx 1 - \mathrm{a}\end{aligned}$$

where $1 - \mathrm{a}$ denotes the confidence level.

The use of a simultaneous confidence band when looking at multiple coefficients allows us to control the probability that even one coefficient from the set we are investigating falls outside of the interval. For instance, a 95% simultaneous confidence band implies that, if we were to repeat the data sampling process many times, then in 95% of these repetitions *all coefficients* would lie within their respective interval.

There is nothing special about 95% here. You could replace all instances with $1 - \mathrm{a}$ if you were interested in $(1 - \mathrm{a})$% confidence statements.

On the contrary, standard 95% confidence intervals for each coefficient – typically referred to as "marginal confidence intervals" – only guarantee that *separately* each coefficient falls in its interval in 95% of the experiments. However, these *success events* for different coefficients can happen on different repetitions. In the worst-case, these success events could be independent random variables with success probability 95%. In this case, the probability that we observe one failure when we look at $p_1$ coefficients could be much larger than 5%; i.e. $1 - \mathrm{P}(\text{no confidence interval failed}) = 1 - (1 - 0.05)^{p_1} \gg 0.05$ and approaches 1 as $p_1$ grows.

These properties mean that marginal confidence intervals are generally inappropriate for judging statistical relevance when multiple coefficients are of interest. For example, if we declare any variable whose marginal 95% confidence interval excludes zero "statistically significant" or a "discovery," the probability that we mistakenly make discoveries – in the sense of claiming a coefficient is not zero when it in fact is – is not 0.05 but potentially substantially larger, e.g. $1 - (1 - 0.05)^{p_1}$ under independence of success events. If instead we report a 95% simultaneous confidence band, this probability of making false discoveries is at most 0.05. Of course, false discovery rate control is only one reason why one might care about the stronger guarantee that a simultaneous confidence band provides. [5] For a survey on simultaneous inference in high dimensions see [6].

5: If one is particularly interested in false discovery rate (FDR) control, then more tailored procedures could potentially be less conservative than the simultaneous confidence band and can be combined with the marginal confidence interval and marginal $p$-value constructions we provide in this book. See e.g. [4]. See also [5] for more on FDR control and the use of multidimensional Gaussian approximations.

**Remark 4.4.1** (Details on critical values) It can be shown that an "ideal" choice of $c$ is

$$c = (1 - \mathsf{a}) - \text{ quantile of } \left\| N\left(0, \mathsf{D}^{-1/2}\mathsf{V}\mathsf{D}^{-1/2}\right)\right\|_\infty ,$$

where $\mathsf{D} = \text{diag}(\mathsf{V})$ is a matrix with variances $(\mathsf{V}_{\ell\ell})_{\ell=1}^{p_1}$ on the diagonal and zeroes off the diagonal. The critical value $c$ can therefore be approximated by simulation plugging in $\mathsf{V} = \hat{\mathsf{V}}$. Please see [6], for example, for more details. Note that $c$ is generally no smaller than the $(1 - \mathsf{a}/2)$-quantile of a $N(0,1)$, so the simultaneous confidence bands are always no smaller than the component-wise confidence bands.

## Discovering Heterogeneity in the Wage Gap Analysis

R Notebook on Double Lasso for the Heterogeneous Wage Gap and Python Notebook on Double Lasso for the Heterogeneous Wage Gap provide code for the wage gap illustration.

We apply the Double Lasso method to analyze heterogeneity of wage gaps using our CPS 2015 data. As in Chapter 1, we use the log hourly wage as the outcome variable. To explore heterogeneity, we interact the female indicator with group indicators capturing education groups (Some High School (shs), High School Graduate (hsg), Some College (scl), College Graduate (clg), Advanced Degree (ad)), region indicators – Midwest (mw), South (so), West (we)) and a fourth degree polynomial in experience (exp1= Experience, exp2= Experience$^2$/100, exp3= Experience$^3$/1000, exp4= Experience$^4$/10000). In total these are 12 target parameters corresponding to the 11 interactive variables and the non-interactive variable that corresponds to the female indicator. All engineered variables used for heterogeneity were de-meaned prior to taking the interaction with sex, while the sex variable was not de-meaned. Hence, the interaction coefficients can be interpreted as "predictive effect modifiers," and the coefficient associated with the non-interactive variable sex as the average predictive effect. As additional variables, we also include all pairwise interactions of the aforementioned variables (excluding sex), as well as one-hot-encodings for occupation and industry sector, providing 990 engineered features. All engineered variables used as controls were also de-meaned prior to estimation.

Table 4.2 provides estimated coefficients, standard errors, pointwise p-values, and the 95% simultaneous confidence band for the coefficients on sex and its interactions with the schooling (shs, hsg, scl, clg, and ad), region (mw, so, and we), and experience (exp1, exp2, exp3, and exp4) variables described above. Rows give variable names with "*" indicating interaction; e.g.

| | Estimate | Std. Error | p-value | Sim. Band lower | Sim. Band upper |
|---|---|---|---|---|---|
| sex | -0.07 | 0.02 | 0.00 | -0.11 | -0.02 |
| sex:shs | -0.20 | 0.11 | 0.07 | -0.53 | 0.14 |
| sex:hsg | 0.01 | 0.05 | 0.80 | -0.14 | 0.16 |
| sex:scl | 0.02 | 0.05 | 0.65 | -0.12 | 0.17 |
| sex:clg | 0.06 | 0.04 | 0.16 | -0.08 | 0.20 |
| sex:mw | -0.11 | 0.04 | 0.01 | -0.23 | 0.01 |
| sex:so | -0.07 | 0.04 | 0.07 | -0.19 | 0.04 |
| sex:we | -0.05 | 0.04 | 0.22 | -0.18 | 0.07 |
| sex:exp1 | 0.02 | 0.01 | 0.01 | -0.00 | 0.04 |
| sex:exp2 | 0.02 | 0.05 | 0.64 | -0.12 | 0.17 |
| sex:exp3 | -0.05 | 0.03 | 0.10 | -0.16 | 0.06 |
| sex:exp4 | -0.01 | 0.00 | 0.00 | -0.01 | -0.00 |

**Table 4.2:** Estimates of Heterogeneous Predictive Effects in the CPS 2015 data. Row labels correspond to variable names as described in the text; e.g. the row "sex*shs" corresponds to the interaction between sex and shs (a dummy for having completed some high school). Estimated coefficients and standard errors are given in the "Estimate" column and "Std. Error" column respectively. The marginal p-value is given in the "p-value" column. The remaining columns "Sim. Band lower" and "Sim. Band upper" provide the lower and upper bounds of the simultaneous confidence band for each variable.

the row sex*shs provides results for the interaction between sex and shs.

Looking coefficient by coefficient, we see evidence that having a college degree increases the predictive effect, i.e. decreases the wage gap, while the largest increase in wage gap occurs for the least educated workers. However, as judged by pointwise p-values, these heterogeneities are not statistically significant at the usual 5% level. We also see that the wage gap is predicted to be larger in the Midwest region, and this is effect is statistically significant at the 5% level based on the marginal p-value. However, care should be taken when looking at pointwise results. The simultaneous confidence regions are relatively wide and include 0 for all coefficients except for the main effect on sex, suggesting that it may be difficult to draw any strong conclusions about heterogeneity of predictive effects in this example.

## 4.5 Other Approaches That Have the Neyman Orthogonality Property

### Double Selection

One way to fix the naive "single selection" approach outlined in Section 3 would be to have "double selection":

**Double Selection**

- find controls $W_Y$ that predict $Y$ as judged by Lasso;
- find controls $W_D$ that predict $D$ as judged by Lasso;
- regress $Y$ on $D$ and the union of controls $W_Y \cup W_D$; proceed with standard inference.

This procedure is approximately equivalent to the partialling out approach, and therefore inherits the orthogonality property. This approach is more conservative compared to single selection, as it makes sure that we have not omitted controls that are strong confounders for $D$. It therefore guards against large omitted variable biases.

## Desparsified Lasso

Yet another procedure that has the orthogonality property and is approximately equivalent to the partialling out approach under suitable conditions is desparsified Lasso.

This approach uses the fact that $a = \alpha$ solves the equation,

$$\mathrm{M}(a, \eta) = \mathrm{E}[(Y - aD - b'W)\tilde{D}(\gamma)] = 0,$$

when $\eta = (b', \gamma')' = \eta^o := (\beta', \gamma'_{DW})'$ for $\gamma_{DW}$ the best linear predictor coefficient from regressing $D$ onto $W$ and

$$\tilde{D}(\gamma) = D - \gamma' W.$$

One can verify that

$$\alpha(\eta) = \left(\mathrm{E}[D\tilde{D}(\gamma)]\right)^{-1} \mathrm{E}\left[(Y - b'W)\tilde{D}(\gamma)\right],$$

and that

$$\alpha = \alpha(\eta^o).$$

Further, the moment condition is Neyman orthogonal – verification of which is left to the reader – which implies that

$$\partial_\eta \alpha(\eta^o) = 0,$$

similarly to the argument for Double Lasso.

**Desparsified Lasso**

- Run a Lasso estimator with suitable choice of $\lambda$ as discussed in Chapter 3 of $Y$ on $D$ and $W$, and save the coefficient estimate $\hat{\beta}$.

- Run a Lasso estimator with suitable choice of $\lambda$ as discussed in Chapter 3 of $D$ on $W$ and save the coefficient estimate $\hat{\gamma}$.
- The estimator $\hat{\alpha}$ is then the solution of the empirical analog of the moment condition above:

$$\mathbb{E}_n[(Y - \hat{\alpha}D - \hat{\beta}'W)\tilde{D}(\hat{\gamma})] = 0,$$

which has the explicit form

$$\hat{\alpha} = \left(\mathbb{E}_n[D\tilde{D}(\hat{\gamma})]\right)^{-1} \mathbb{E}_n\left[(Y - \hat{\beta}'W)\tilde{D}(\hat{\gamma})\right],$$

where $\hat{\beta}$ and $\hat{\gamma}$ are Lasso estimators.

Estimators of this form are referred to in econometrics as "instrumental variable estimators." In purely technical terms, we are using residualized $\tilde{D}$ to "instrument" for $D$.

## Revisiting the Price Elasticity for Toy Cars

Next, we revisit the example from Chapter 0. We are interested in the coefficient $\alpha$ in the high-dimensional linear regression model:

$$Y = \alpha D + \beta' X + \epsilon,$$

where $Y$ is log-reciprocal=sales-rank, $D$ is log-price, and $X = (1, W)$ with product features $W$. We here take $X$ to be the same 11546-dimensional transformed regressors as described in Chapter 0, constructed from product brand, subcategory, and physical dimensions. Here we have $p > n = 9212$, so OLS is underspecified, and even if we consider a specific solution to the normal equations such as the one with the minimum norm, standard errors are unavailable or unreliable. We can still run OLS when we subset the regressors, or equivalently impose that the coefficients on the rest are zero. In Table 4.3 we report the results for such an approach with OLS with three specifications of increasing size: $p = 243$ with only subcategory features (as in Chapter 0), $p = 2069$ after also adding brand features, and $p = 2073$ after also adding log of the physical dimensions features (but without any transformations or interactions). We see that in all cases we cannot exclude 0 from the confidence interval, while the more flexible we make our model (larger $p$), the more negative our estimates and confidence intervals.

Next, we consider estimating elasticities using double lasso, double selection, and desparsified lasso applied to all $p = 11546$

features. In all cases, we pick the regularization parameter by 5-fold cross validation (for the regression of each of $Y$ and $D$). Then we apply the three methods using the lasso models fit or the variables chosen by them. The results are reported in Table 4.3. We see that all three methods result in confidence intervals that are strictly negative, in agreement with the theory that increasing price for any one product decreases its sales.

| | Estimate | Std. Error | 95% CI |
|---|---|---|---|
| OLS ($p = 242$) | 0.005 | 0.016 | [-0.026, 0.036] |
| OLS ($p = 2068$) | -0.003 | 0.021 | [-0.045, 0.039] |
| OLS ($p = 2072$) | -0.033 | 0.022 | [-0.076, 0.010] |
| Double Lasso | -0.064 | 0.018 | [-0.099, -0.029] |
| Double Selection | -0.074 | 0.019 | [-0.111, -0.037] |
| Desparsified Lasso | -0.062 | 0.017 | [-0.096, -0.028] |

**Table 4.3:** Estimates for price elasticity. We report specification robust standard errors with finite sample correction, i.e., "HC1." All non-OLS methods have $p = 11546$.

## Notebooks

- R Notebook with Experiment on Orthogonal vs Non-Orthogonal Learning and Python Notebook with Experiment on Orthogonal vs Non-Orthogonal Learning presents the simulation experiment comparing orthogonal (partialling-out) with non-orthogonal learning (naive method).
- R Notebook with Hard Sparsity on Orthogonal vs Non-Orthogonal Learning and Python Notebook with Hard Sparsity on Orthogonal vs Non-Orthogonal Learning presents an alternative simulation to that shown in the main text comparing orthogonal (partialling-out) with non-orthogonal learning. In this simulation, we consider orthogonal and non-orthogonal learning in a stylized treatment effects simulation.
- R Notebook on Double Lasso for Growth Convergence and Python Notebook on Double Lasso for Growth Convergence presents a Double Lasso analysis of the conditional convergence hypothesis in growth economics.
- R Notebook on Double Lasso for the Heterogeneous Wage Gap and Python Notebook on Double Lasso for the Heterogeneous Wage Gap presents a Double Lasso analysis of the heterogeneous wage gap.

## Notes

We mainly follow the Double Lasso approach developed in [7] and [8], because it is nicely connected to partialling out and will later generalize seamlessly to double machine learning [9]. Desparsified Lasso was developed by [10] and [11]; a closely related approach is the debiased Lasso proposed by [12]. The double selection method was developed by [13] and [14]. Inference on many coefficients using Double Lasso was first developed by [15] and [16]. [17] provide results for Double Lasso with clustered dependence. The Double Lasso and desparsified Lasso approaches have also been extended to time series and many time series by [18]. Both [17] and [18] take into account the temporal dependencies in the data when fitting Lasso and performing inference on the coefficients of interest.

Failure of single selection even when $p$ is small is discussed in simple terms in [14], but the problem was first systematically examined by [19]. A recent paper [20] develops debiasing methods for shape constrained high-dimensional linear regression models.

[6] provide a recent survey on methods for simultaneous inference in high-dimensional settings.

For an in-depth analysis of heterogeneity in the wage gap based on Lasso, we refer to [21].

## Study Problems

1. Experiment with the first notebook, R Notebook with Experiment on Orthogonal vs Non-Orthogonal Learning or Python Notebook with Experiment on Orthogonal vs Non-Orthogonal Learning. Try different models. For example, try different coefficient structures for $\beta$ and $\gamma_{DW}$ and/or different covariance structures for $W$. Provide an explanation to a friend for what each step in the Double Lasso procedure is doing.

2. Explore R Notebook on Double Lasso for Growth Convergence or Python Notebook on Double Lasso for Growth Convergence. Provide an explanation to a friend for what each step in the Double Lasso procedure is doing. Explain the empirical results to a friend. Experiment with making the set of controls more flexible and higher-dimensional by adding nonlinear and/or interaction terms that seem potentially interesting. Comment on how the results differ from the baseline results.

3. Explore R Notebook on Double Lasso for the Heterogeneous Wage Gap and Python Notebook on Double Lasso for the Heterogeneous Wage Gap. Provide an explanation to a friend for what each step in the inference procedure is doing. Explain the empirical results to a friend.

4. Verify that Neyman orthogonality holds for the "de-sparsified" Lasso strategy.

## 4.A High-Dimensional Central Limit Theorems*

Let $X_1, \ldots, X_n$ be independent (but not necessarily identically distributed) random vectors with dimension $p$. Assume that $X_i$'s have mean zero (otherwise, work with $X_i - \mathrm{E}[X_i]$ instead of $X_i$). Consider the scaled sample mean

$$S_n = \frac{1}{\sqrt{n}} \sum_{i=1}^{n} X_i.$$

Let $\bar{\sigma}, \underline{\sigma}$ be given positive constants such that $\underline{\sigma} \leqslant \bar{\sigma}$, and let $B_n \geqslant 1$ be a sequence of constants that may diverge as $n \to \infty$. Let $\Sigma_n = \mathrm{E}[S_n S_n'] = n^{-1} \sum_{i=1}^{n} \mathrm{E}\left[X_i X_i'\right]$. Also, let $\mathcal{R}$ denote the collection of closed rectangles in $\mathbb{R}^p$.

We first present a high-dimensional CLT over the rectangles under a sub-exponential condition on the coordinates. Suppose that the coordinates of $X_1, \ldots, X_n$ are sub-exponential with scale $B_n$, then

$$\sup_{R \in \mathcal{R}} |\mathrm{P}(S_n \in R) - \mathrm{P}(N(0, \Sigma_n) \in R)| \approx 0, \quad (4.A.1)$$

provided that $B_n^2 \log^5(pn)/n \approx 0$. Note that this allows $p$ to be much larger than $n$. It turns out that a similar result applies without sub-exponential conditions, as stated formally below.

To state the results in a finite-sample form, let

$$\delta_{1,n} := \left( \frac{B_n^2 \log^5(pn)}{n} \right)^{1/4} \text{ and } \delta_{2,n}^{[q]} := \sqrt{\frac{B_n^2 (\log(pn))^{3-2/q}}{n^{1-2/q}}},$$

for $q > 2$.

**Theorem 4.A.1** (High-Dimensional CLT, [22]) *Suppose second moments are non-degenerate,* $\min_{j\leqslant p} n^{-1}\sum_{i=1}^{n} \mathrm{E}\left[X_{ji}^2\right] \geqslant \underline{\sigma}^2$*, and fourth moments obey* $\max_{j\leqslant p} n^{-1}\sum_{i=1}^{n} \mathrm{E}\left[X_{ji}^4\right] \leqslant B_n^2\bar{\sigma}^2$.

*(A) If coordinates are subexponential, i.e.,* $\max_{i\leqslant n; j\leqslant p} \mathrm{E}\left[e^{|X_{ji}|/B_n}\right] \leqslant 2$*, then*

$$\sup_{R\in\mathcal{R}} |\mathrm{P}(S_n \in R) - \mathrm{P}(N(0,\Sigma_n) \in R)| \leq C\delta_{1,n},$$

*where* $C$ *is a constant that depends only on* $\underline{\sigma}$ *and* $\bar{\sigma}$.

*(B) If the envelope of the coordinates admits a moment bound* $\max_{i\leqslant n} \mathrm{E}\left[\|X_i\|_\infty^q\right] \leq B_n^q$ *for some* $q > 2$*, then*

$$\sup_{R\in\mathcal{R}} |\mathrm{P}(S_n \in R) - \mathrm{P}(N(0,\Sigma_n) \in R)| \leq C\left(\delta_{1,n} \vee \delta_{2,n}^{[q]}\right)$$

*where* $C$ *is a constant that depends only on* $q, \underline{\sigma}$ *and* $\bar{\sigma}$.

Notably, the above theorem does not impose any restrictions on the correlation structure between the coordinates of the random vectors, so $\Sigma_n$ is permitted to be singular.

As discussed in [23], the assumption of Part (A) is satisfied if, for example, $|X_{ji}| \leqslant B_n$ for all $(i,j)$, but also allows for unbounded coordinates. Part (B) covers the following scenario relevant to regression applications: $X_i = \epsilon_i v_i$ where $\epsilon_i$ is a univariate "error" term while $v_i \in \mathbb{R}^p$ is a vector of fixed "covariates." In this case, $\mathrm{E}\left[\|X_i\|_\infty^q\right] \leqslant \|v_i\|_\infty^q \mathrm{E}\left[|\epsilon_i|^q\right]$, so if the covariates are uniformly bounded and the $q$-th moments of the error terms are bounded, then $B_n = O(1)$. Notably this only requires $\epsilon_i$ to have $q = 2 + \delta$ bounded moments.

Often, statistics of interest are not exactly sample means, but can be well approximated by sample means. For example, the Double Lasso estimator, $\widehat{\alpha} = (\mathbb{E}_n[\check{D}^2])^{-1}\mathbb{E}_n[\check{D}\check{Y}] \approx (\mathrm{E}[\tilde{D}^2])^{-1}\mathbb{E}_n[\tilde{D}\tilde{Y}]$, takes this form. In order to claim a High-Dimensional CLT for such statistics, we need the approximation error to vanish at the rate faster than $1/\sqrt{\log p}$.[6]

6: The requirement that approximation error, denoted $\mathrm{R}_n$, vanishes faster than $1/\sqrt{\log p}$ arises from the fact that the maximum of a Gaussian random vector $N(0,\Sigma)$ concentrates in (i.e., places a probability mass of near 1 to) a $1/\sqrt{\log p}$-neighborhood of its expected value, but not in smaller neighborhoods (anti-concentration). The approximation error $\mathrm{R}_n$ needs to be much smaller than the size of the neighborhood. Otherwise, the probabilistic errors incurred by Gaussian approximation to the distribution of $\hat{S}$ can be as large as 1, meaning that the Gaussian approximation fails.

**Lemma 4.A.2** (High-dimensional CLT for approximate sample mean) . *Suppose that* $S_n$ *obeys (4.A.1), but* $S_n$ *is not directly available. Suppose instead that we have access to* $\widehat{S}_n$ *that approximates* $S_n$ *such that* $\widehat{S}_n = S_n + \mathrm{R}_n$ *with* $\sqrt{\log p}\,\|\mathrm{R}_n\|_\infty \approx 0$. *Assume* $\min_{j\leqslant p} \Sigma_{jj} \geqslant \underline{\sigma}^2$. *Then the same conclusion holds with* $S_n$ *replaced by* $\widehat{S}_n$.

The lemma follows from Nazarov's anticoncentration inequality for Gaussian vectors over rectangles; see [23] for the proof.

# Causal Inference via Conditional Ignorability 5

"compare apples and/to/with apples: to compare things that are very similar."

– Merriam Webster Dictionary [1].

Here we discuss how average causal effects may be identified using regression when treatment is not randomly assigned but instead depends on observed covariates. We discuss the conditional or adjustment method, which relies on comparing the average difference between expected outcomes for treated and untreated units that are comparable (formally, identical) in terms of their characteristics $X$. If treatment is as good as randomly assigned conditional on $X$, then this approach recovers average causal or treatment effects. This key condition is commonly referred to as conditional ignorability, conditional exogeneity, or unconfoundedness.

## 5.1 Introduction

In a cross-country analysis, higher chocolate consumption predicts a higher number of Nobel laureates per capita.

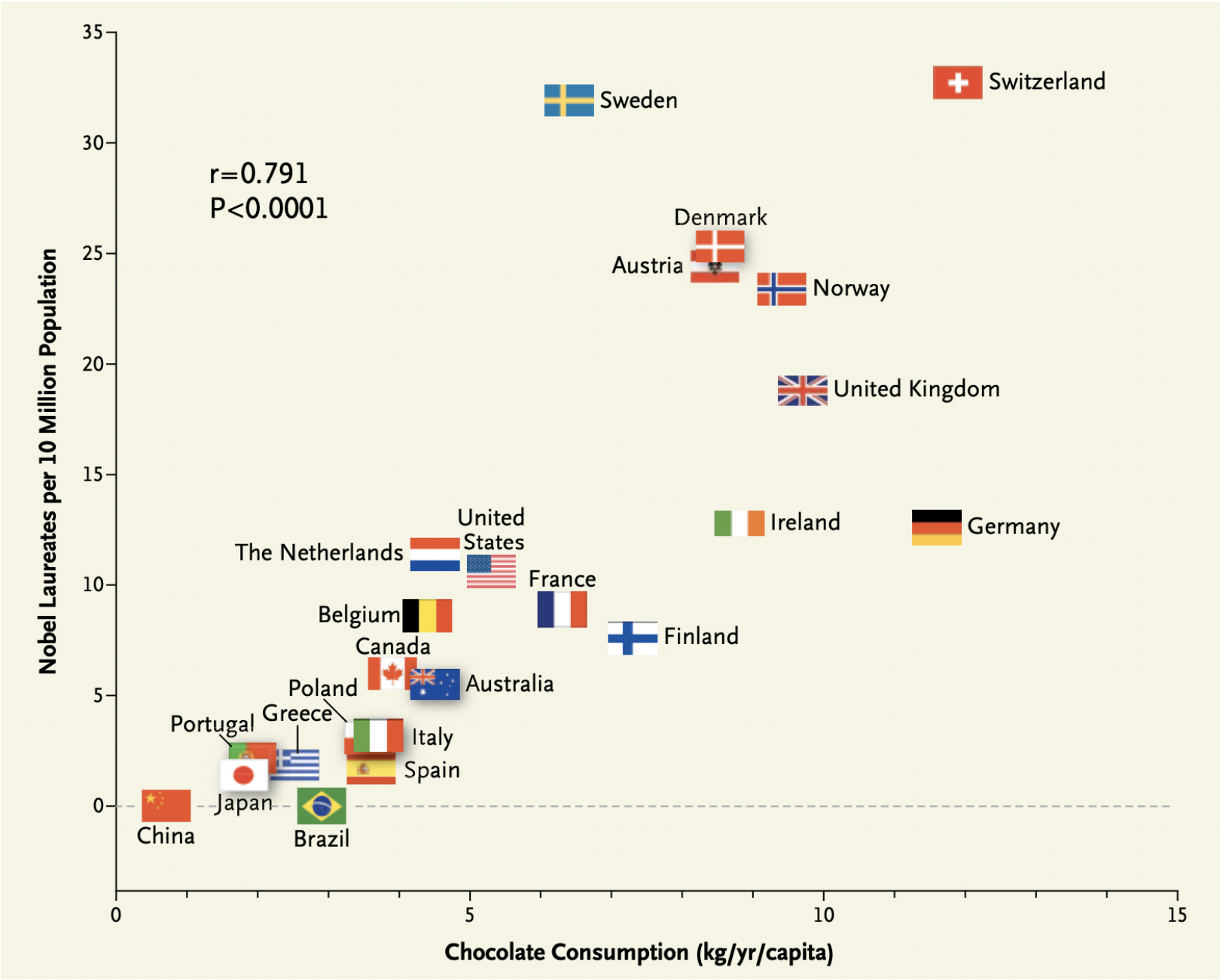

**Figure 5.1:** Source: Franz H. Messerli, "Chocolate Consumption, Cognitive Function, and Nobel Laureates," New England Journal of Medicine. 2012

Is this a reflection of a true causal effect and therefore an actionable insight? If it were, countries could generate more Nobel laureates per capita by making chocolate abundant to everyone. (This wouldn't be a bad thing.) Is this perhaps what Switzerland did? Switzerland has the highest number of Nobel laureates per capita.

Or is there a common cause[1] that creates non-causal association? Perhaps wealthy countries invest more in science and higher wealth causes people to consume luxury goods like chocolate. See for instance plots (D) and (E) in Figure 5.3. Comparative analysis, where we compare nations with identical or similar wealth, would probably reveal that the correlation is not causal.[2]
Probably we should be comparing Switzerland to similar countries in terms of wealth – the "apples-to-apples" comparison, so to speak. This type of analysis is very common in causal

1: We often refer to these common causes as "omitted variables" that give rise to "omitted variable bias."

2: It remains a fundamental empirical problem to confirm this conjecture or disprove this conjecture. The causal channel through which chocolate (and other flavonoids) may affect Nobel production is by documented improvement in the cognitive function.

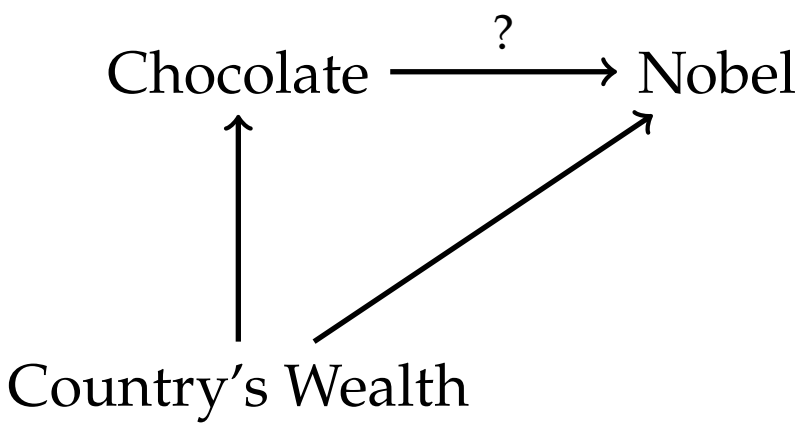

**Figure 5.2:** A Contrived Causal Path Diagram for the Effect of Country's Wealth on Chocolate Consumption and Nobel Prize Production per capita.

inference and is implemented via a set of tools introduced in this chapter.

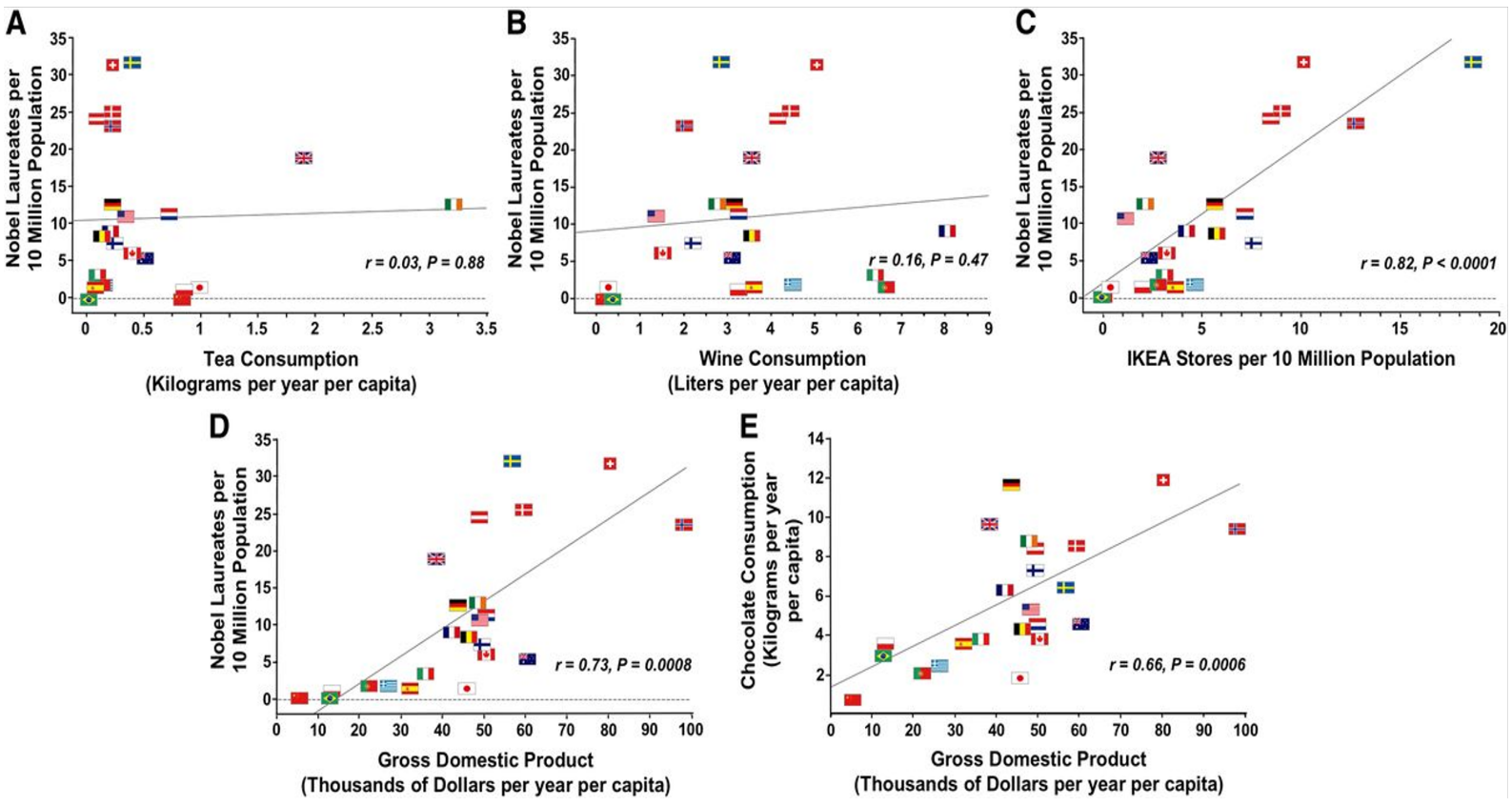

**Figure 5.3:** Source: J Nutr, Volume 143, Issue 6, June 2013, Pages 931–933, "Does Chocolate Consumption Really Boost Nobel Award Chances? The Peril of Over-Interpreting Correlations in Health Studies," ©2013 American Society for Nutrition

In what follows, we work within Rubin's [2] potential outcomes framework, as introduced in Chapter 2. The idea is that if we can think of observed treatment $D$ as generated randomly – independently of potential outcomes – conditional on some pre-treatment variables $X$, then we can learn the average causal (treatment) effects by regression

$$\text{of } Y \text{ on } D \text{ and } X,$$

or, as is often said, by "adjusting" or "controlling" for $X$.

### Notation

Recall that we denote the independence of two random variables (these can include random vectors) $U$ and $V$ as

$$U \perp\!\!\!\perp V.$$

Independence, conditional on a third variable $X$, is denoted by

$$U \perp\!\!\!\perp V \mid X.$$

## 5.2 Potential Outcomes and Ignorability

Recall that we use $Y(d)$ to denote potential outcome in the treatment state $d$, where we consider only the case $d \in \{0, 1\}$ for simplicity. We also recall our example of smoking from Chapter 2. Suppose we want to study the impact of smoking marijuana on life longevity. Suppose that smoking marijuana has no causal/treatment effect on life longevity:

$$Y = Y(0) = Y(1), \text{ so that } \delta = \mathrm{E}[Y(1)] - \mathrm{E}[Y(0)] = 0.$$

However, the observed smoking behavior, $D$, results not from an experimental study, but from observational data in which an individual's smoking decisions are driven by other behavioral choices $X$ (drinking alcohol for example) which cause shorter life longevity. In this case, the predictive effect recovered by regression without adjusting for $X$ does not match the average causal effect

$$\mathrm{E}[Y \mid D = 1] - \mathrm{E}[Y \mid D = 0] < 0 = \delta,$$

because higher $D$ predicts higher $X$, which predicts lower $Y$. This difference between the predictive effect and average causal effect is the result of confounding or *selection bias*.

In this example, conditioning on $X$ can remove the selection bias (see Figure 5.4)

$$\mathrm{E}[\mathrm{E}[Y \mid D = 1, X] - \mathrm{E}[Y \mid D = 0, X]] = \delta,$$

provided that conditional on $X$ variation in $D$ is independent of the potential health outcomes.

The following provides a formal assumption under which we can eliminate the confounding bias by controlling for $X$.[3]

3: The assumption is fundamentally untestable and is an assumption in the purest sense. Given assumed domain knowledge encoded in causal DAGs, we study a systematic way of finding $X$ that satisfy this assumption in subsequent chapters.

**Assumption 5.2.1** (Conditional Ignorability and Consistency) *Ignorability: Suppose that treatment status $D$ is independent of potential outcomes $Y(d)$ conditional on a set of covariates $X$: For each $d$,*

$$D \perp\!\!\!\perp Y(d) \mid X.$$

*Consistency: Suppose that $Y$ is generated as $Y := Y(D)$.*

## Identification by Conditioning

The ignorability assumption[4] says that variation in treatment assignment $D$ is as good as random conditional on $X$. This assumption means that if we look at units with the same value of the covariates, e.g. units with $X = x$, then treatment variation among these observationally identical units, $D \mid X = x$, is indeed produced as if by a formal randomized control trial.

4: You may wonder why the term "ignorability" is used. The distribution of $Y(d)$ depends only on $X$ and not on $D$, so the latter is "ignorable." Note that the conventional name used in econometrics for the ignorability assumption is the *conditional exogeneity* or *conditional independence* assumption.

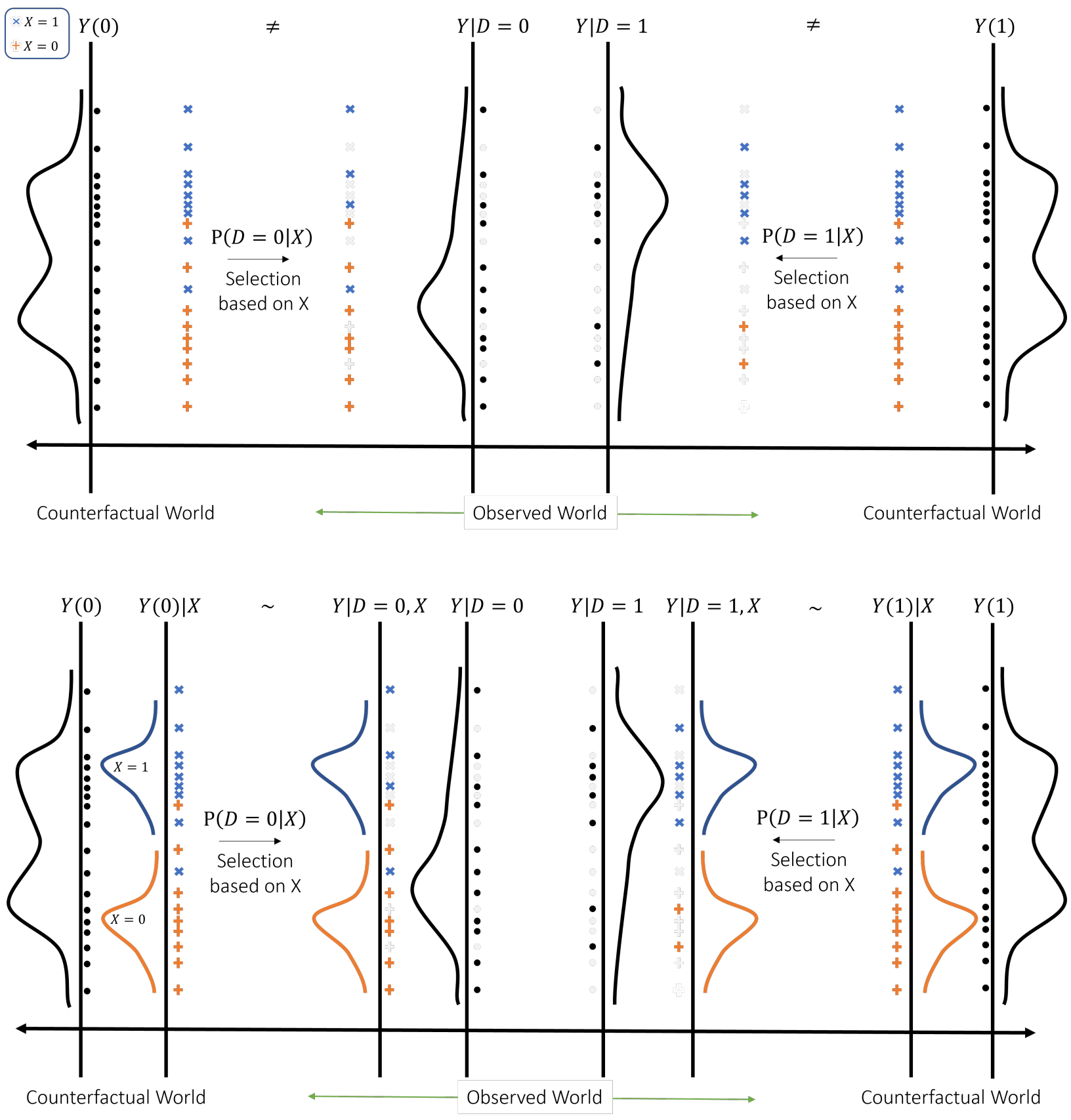

**Figure 5.4:** Pictorial representation of how selection on $X$ can lead to biased observed outcomes between treated and control populations, while conditioning on $X$ removes the selection bias. In this example, the potential outcomes $Y(0)$ and $Y(1)$ have identical distributions shown in the far left and right of the figure. We also have a binary covariate $X$ that is related to treatment probability in the sense that $\mathrm{P}(D = 1|X = 1) > \mathrm{P}(D = 1|X = 0)$ and $\mathrm{P}(D = 0|X = 1) < \mathrm{P}(D = 0|X = 0)$ which leads to selection bias when we do not condition on $X$. This bias is illustrated by the difference in the distribution of (observed) $Y$ given $D = 0$ and $D = 1$ shown in the black curves in the middle of the figure. The bottom panel then shows that selection bias is removed by conditioning on $X$ as the distribution of potential outcomes given $X$ (blue and orange curves under $Y(0)|X$ and $Y(1)|X$) equals the distribution of observed outcomes given $D$ and $X$ (blue and orange curves under $Y|D = 0, X$ and $Y|D = 1, X$).

Therefore, we can learn about the causal effect of $D$ by comparing outcomes across treated and control units who have identical characteristics $X = x$ under the conditional ignorability assumption. The idea of comparing observations who have identical characteristics is the essence of the so-called *conditioning* or *adjustment* strategy to learning causal effects. As conditioning approaches produce a different contrast for every potential value of $X$, we may also wish to average the contrasts at different values of $X$ over the distribution of characteristics to produce a summary measure of the causal effects.

The conditional probability of receiving treatment, *the propensity score*, plays an important role in this approach.

**Assumption 5.2.2** (Overlap/Full Support) *The probability of receiving treatment given $X$, the propensity score*

$$p(X) := \mathrm{P}(D = 1|X),$$

*is non-degenerate:*

$$\mathrm{P}(0 < p(X) < 1) = 1.$$

The overlap assumption requires that there is proper randomization or variation in $D$ at each value $x$ in the support of $X$. Without this condition, there are values $x$ in the support of $X$ where we cannot construct a contrast between treatment and control units. We cannot learn the conditional average treatment effect at these values of $X$ and thus are also unable to learn the unconditional average effect of the treatment.

**Remark 5.2.1** Assumption 5.2.2 is also often called the *full support* condition because it requires

$$\text{support}(D, X) = \{0, 1\} \times \text{support}(X).$$

The following is the most important theoretical result that states that we can recover expectations of potential outcomes from regressions.

**Theorem 5.2.1** (Conditioning on $X$ Removes Selection Bias) *Under Conditional Ignorability and Overlap, the conditional expectation function of observed outcome $Y$ given $D = d$ and $X$ recovers the conditional expectation of the potential outcome $Y(d)$ given $X$:*

$$\mathrm{E}[Y \mid D = d, X] = \mathrm{E}[Y(d) \mid D = d, X] = \mathrm{E}[Y(d) \mid X].$$

To prove Theorem 5.2.1, note that the overlap assumption makes it possible to condition on the events $\{D = 0, X\}$ and $\{D = 1, X\}$ at any value in the support of $X$ and that the second equality holds by ignorability.

Hence, the Conditional Average Predictive Effect (CAPE),

$$\pi(X) = \mathrm{E}[Y \mid D = 1, X] - \mathrm{E}[Y \mid D = 0, X],$$

is equal to the Conditional Average Treatment Effect (CATE),

$$\delta(X) = \mathrm{E}[Y(1) \mid X] - \mathrm{E}[Y(0) \mid X].$$

Thus, the APE and ATE also agree:

$$\delta = \mathrm{E}[\delta(X)] = \mathrm{E}[\pi(X)] = \pi.$$

## Conditional Ignorability via Causal Diagrams

It is possible to illustrate the key ignorability assumption, Assumption 5.2.1, graphically as follows:[5]

5: Note that what we present is just one of many casual diagrams that are compatible with the conditional ignorability condition. There are others, as will become apparent in subsequent chapters.

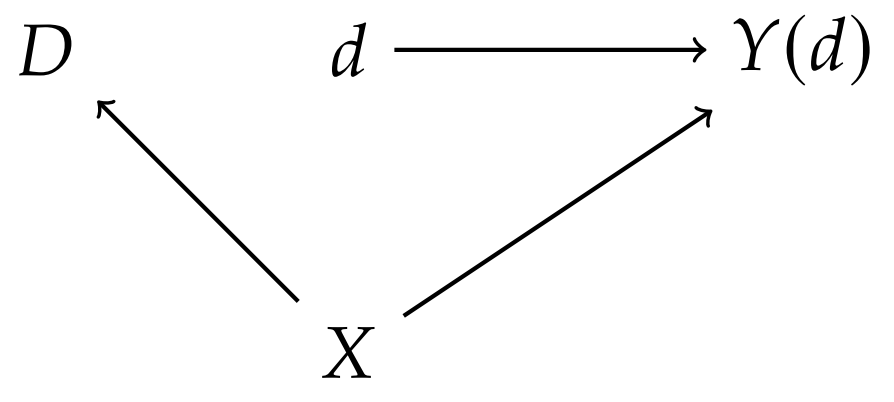

**Figure 5.5:** A Causal Diagram for the Conditional Ignorability Research Design

In this graph, we show the potential outcome $Y(d)$ as a node and the potential treatment status $d$ as another node. The latter node is deterministic. There is an arrow from $d$ to $Y(d)$ indicating the dependency. The pre-treatment covariates $X$ affect both the realized treatment variable $D$ and the potential outcomes $Y(d)$, as shown by the arrow from $X$ to $D$ and from $X$ to $Y(d)$. The assigned treatment variable $D$ is independent of the node $Y(d)$, conditional on $X$. Independence can be derived from the graph by observing the absence of any path between the $D$ and $Y(d)$ nodes other than the path through the variable $X$ upon which we've conditioned. Note that Assumption 5.2.2, the overlap condition, is not illustrated in the graph.

The potential outcome process $d \mapsto Y(d)$ and treatment assignment jointly determine the realized outcome variable $Y$ via the assignment $Y := Y(D)$. This generates the following causal diagram. This graph says that $X$ is generated first. $D$ is then generated, with the distribution of $D$ depending on $X$. Finally, $Y$ is generated, with its distribution depending on both $D$ and $X$. Here, after conditioning on $X$, the statistical dependence (association) between $D$ and $Y$ only reflects the causal channel, $D \rightarrow Y$ allowing us to uncover the ATE, for example.

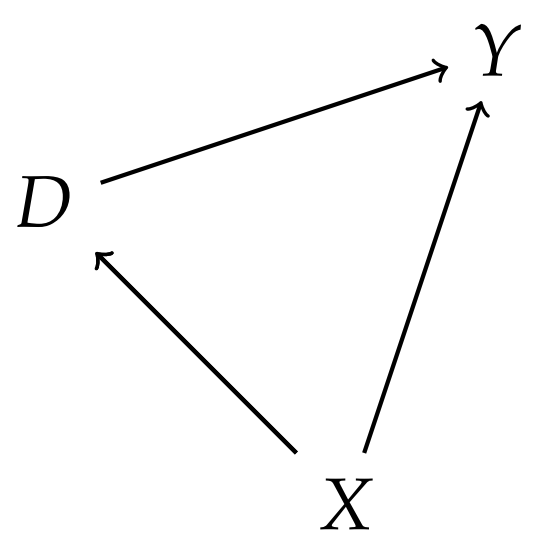

**Figure 5.6:** A Causal Diagram with Conditional Ignorability

## Connections to Linear Regression

The tools from Chapter 1 and Chapter 4 can be used to perform statistical inference on ATEs. We briefly discuss how (high-dimensional) regression can be used to retrieve causal estimates when conditional ignorability holds in this section.

The simplest instance of the problem is when the conditional expectation function of $Y$ given $D$ and $X$ is linear,

$$\mathrm{E}[Y \mid D, X] = \alpha D + \beta' W,$$

which gives a model

$$Y = \alpha D + \beta' W + \epsilon, \quad \mathrm{E}[\epsilon \mid D, X] = 0.$$

Here it is understood that $W$ may include $X$ as well as pre-specified nonlinear transformations of $X$.

In this model, $\alpha$ identifies $\delta$

$$\delta = \alpha$$

under the linearity assumption and ignorability, and our inference tools for $\alpha$ automatically carry over to $\delta$. Note that the linearity assumption and ignorability assumptions imply that treatment effects are homogeneous; that is, $\delta(x) = \delta$ for all $x$ in the support of $X$.

Of course, the assumption of linearity and homogeneous treatment effects is restrictive. A simple way to relax this is to consider interactions. One version of this approach takes all interactions between $W$ and $D$ and assumes

$$\mathrm{E}[Y \mid D, X] = \alpha_1 D + \alpha_2' W D + \beta_1 + \beta_2' W,$$

where we also maintain that we are working with centered covariates: $\mathrm{E}W = 0$.[6]

6: This model is still linear and results for linear models carry over to this case as well.

We then recover the ATE as

$$\delta = \alpha_1$$

and CATE as

$$\delta(X) = \alpha_1 + \alpha_2' W.$$

We can use partialling out methods, such as OLS in the low-dimensional case and Double Lasso (and variants) in the high-dimensional case, to perform inference on $\alpha_1$ and components of $\alpha_2$. We can use these same methods to perform inference over $\beta_1$ and components of $\beta_2$, though these parameters will often not be of interest.

Note, we used this approach in the heterogeneous wage gap example in Chapter 1. The discussion of whether the wage gap analysis has a causal interpretation is given in the next causal inference chapter, Chapter 6.

As demonstrated in Theorem 5.2.1, the ultimate targets are the conditional expectation functions $\mathrm{E}[Y(d)|X]$ if our goal is to learn average causal effects under ignorability. This being our target makes the relevance of considering transformations $W = T(X)$ of $X$ important as we would like to have the linear model provide a good approximation to these conditional expectation functions. See the discussion in "From Best Linear Predictor to Best Predictor" in Chapter 1. If the linear model is misspecified in the sense that it does not approximate the conditional expectation functions well, the estimated causal effects - e.g. $\alpha_1$ in the interactive model - do not necessarily have any causal interpretation. This potential failure is a major reason we consider more flexible, modern machine learning methods.

What about fully nonlinear strategies? We will explore them in Chapter 10.

## 5.3 Identification Using Propensity Scores

The identification by conditioning approach requires being able to accurately model the "outcome process," i.e. the conditional expectation function $\mathrm{E}[Y \mid D, X]$. This conditional expectation function might correspond to a complicated real world process that is hard to model or approximate.

When the outcome process is hard to model, we might have a much better handle on the "treatment selection process," i.e. the propensity score:

$$p(X) = \mathrm{P}(D = 1 \mid X).$$

An alternative approach, known as the Horvitz-Thompson method [3], uses propensity score reweighting to recover aver-ages of potential outcomes. Using the propensity score rather than identification by conditioning on $X$ is a useful empirical strategy when $X$ is high-dimensional and $p(X)$ is available or can be approximated accurately.[7] An example of a setting where the propensity score is known is a *stratified RCT*, which is an experiment where treatment is assigned at random with probability $p(X)$ to individuals with different observed covariates $X$. In this case, the treatment assignment probability $p(X)$ is exactly the propensity score.

7: An interesting example where the propensity score is not known but can be well-approximated is the examination in [4] of the causal effect of attendance at a particular school or group of schools relative to one or more alternative schools (e.g., "elite" vs. "non-elite" schools) in settings where matching algorithms are used to assign students to schools. In this example, we can think of these student assignment mechanisms as $p(X)$.

**Theorem 5.3.1** (Horvitz-Thompson: Propensity Score Reweighting Removes Bias) *Under Conditional Ignorability and Overlap, the conditional expectation of an appropriately reweighted observed outcome $Y$, given $X$, identifies the conditional average of potential outcome $Y(d)$ given $X$:*

$$\mathrm{E}\left[Y\frac{1(D=d)}{\mathrm{P}(D=d|X)} \mid X\right] = \mathrm{E}[Y(d) \mid X]$$

*Then, averaging over $X$ identifies the average potential outcome:*

$$\mathrm{E}\left[Y\frac{1(D=d)}{\mathrm{P}(D=d|X)}\right] = \mathrm{E}[Y(d)]$$

To prove this result, note

$$\begin{aligned}\mathrm{E}\left[Y\frac{1(D=d)}{\mathrm{P}(D=d|X)} \mid X\right] &= \frac{\mathrm{E}[Y1(D=d) \mid X]}{\mathrm{P}(D=d|X)} \\ &= \mathrm{E}[Y(d) \mid X]\frac{\mathrm{E}[1(D=d) \mid X]}{\mathrm{P}(D=d|X)} \\ &= \mathrm{E}[Y(d) \mid X],\end{aligned}$$

where we used conditional ignorability in the second equality.

As a consequence, we can identify average treatment effects by simple averaging of transformed outcomes:

$$\delta = \mathrm{E}[YH], \quad H = \frac{1(D=1)}{\mathrm{P}(D=1|X)} - \frac{1(D=0)}{\mathrm{P}(D=0|X)},$$

where $H$ is called the Horvitz-Thompson transform. Similarly, we can identify conditional average treatment effects as a conditional average of transformed outcomes:

$$\delta(X) = \mathrm{E}[YH \mid X].$$

Note that propensity score reweighting reduces to the difference of means in the control and treatment groups when the propensity score is constant.

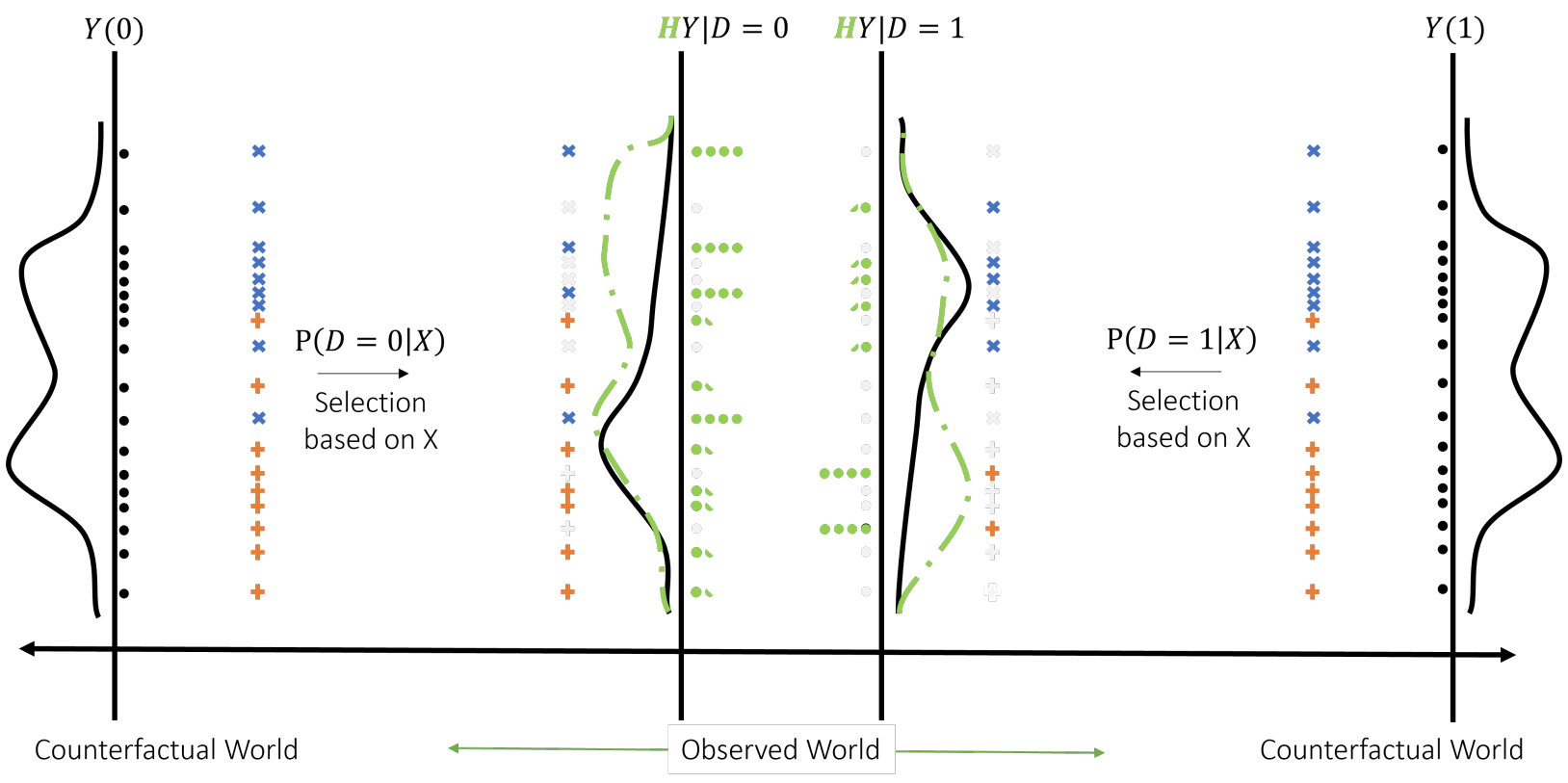

**Figure 5.7:** Pictorial representation of inverse propensity reweighting. As in Figure 5.4, the outer black curves represent the distribution of potential outcomes and the inner black curves represent the distribution of observed $Y$ given only $D$ – showing the selection bias. The green curves represent the observed distribution of $HY$ given $D$ which align and illustrate that the selection bias has been removed.

## Stratified RCTs

In the case where the propensity score $p(X)$ is known, we are essentially back to a classical RCT.

**Definition 5.3.1** (Generalized/Stratified RCT) *If under Assumption 5.2.1, the propensity score $p(X)$ is known, the setting is called a generalized or stratified RCT.*

**Remark 5.3.1** Propensity score reweighting is generally not the most efficient approach to estimating treatment effects from a statistical point of view because it ignores any dependence between the outcomes and controls, $X$, that is not captured by the propensity score. By exploiting dependence between the outcomes and $X$ not captured by the propensity score, more efficient estimation of treatment can occur as using this dependence "de-noises" the outcome. Moreover, estimation based on only propensity score reweighting fails under imbalances that might arise due to imperfect data collection. Later, we will use *both* regression and reweighting as part of "double machine learning" to operationalize efficient statistical inference on treatment effects in fully nonlinear (nonparametric) models.

## Covariate Balance Checks

Given a propensity score $p(X)$, we can check if the RCT is valid (randomization is successful) by performing a *covariate balance*

*check*.. Specifically, conditional ignorability implies that

$$\mathrm{E}[H \mid X] = 0.$$

Thus, if covariates predict $H$, we can conclude that conditional ignorability does not hold. Heuristically, covariates predicting $H$ means that covariates are imbalanced in the sense that, after reweighting by $X$ dependent treatment probability, there are systematic differences in $X$ across treatment and control observations which can be exploited to predict treatment assignment.

In a low-dimensional linear model framework, a covariate balance check can be done by regressing $H$ on $W$, a dictionary of transformations of $X$, and testing if $W$ predicts $H$. $W$ predicting $H$ suggests that the RCTs randomization protocol did not go as planned.

## Connections to Linear Regression

Note that by the Horvitz-Thompson transform characterization of the CATE, $\delta(X) = \mathrm{E}[YH \mid X]$, we can view the conditional average treatment effect as the solution to a prediction problem of predicting the transformed outcome $YH$ from the regressors $X$.

A useful strategy is to consider (potentially high-dimensional) linear regression models where $HY$ is the dependent variable; see, e.g., [5]. Note that if we assume that $\mathrm{E}[Y \mid D, X] = \alpha_1 D + \alpha_2' WD + \beta_1 + \beta_2' W$, where $W$ is a dictionary of transformations of $X$, then we have

$$\mathrm{E}[YH \mid X] = \alpha_1 + \alpha_2' W.$$

Thus, we can simply run a regression of $YH$ on $(1, W')'$. In this regression model, we recover the ATE as

$$\delta = \alpha_1$$

and CATE as

$$\delta(X) = \alpha_1 + \alpha_2' W.$$

We can use partialling out methods, such as Double Lasso, to perform inference on $\alpha_1$ and components of $\alpha_2$. We also discuss estimating CATE using more general machine learning methods in Chapter 14 and Chapter 15.

## 5.4 Conditioning on Propensity Scores*

The fact that conditioning on the right set of controls removes selection bias has long been recognized by researchers employing regression methods. Rosenbaum and Rubin [6] made the much more subtle point that conditioning on only the propensity score

$$p(X) = \mathrm{P}(D = 1 \mid X)$$

also suffices to remove the selection bias.

**Theorem 5.4.1** (Rosenbaum and Rubin: Conditioning on the Propensity Score Removes Selection Bias) *Under Ignorability and Overlap, $D$ is generated independently of $Y(d)$ for each $d$, conditional on the propensity score $p(X)$: For each $d$,*

$$D \perp\!\!\!\perp Y(d) \mid p(X).$$

In other words, conditional on $p(X) = p$, variation in $D$ is as good as randomly assigned. Hence, whenever it suffices to use $X$ for identification by conditioning, it also suffices to use $p(X)$. This fact makes $p(X)$ a "minimal sufficient" statistic, conditioning on which removes selection bias under ignorability.

In scenarios with a known propensity score, we can simply use $p(X)$ as a control in place of the high-dimensional set of characteristics, $X$, and thus bypass a potentially complicated high-dimensional estimation problem. In other words, we can identify the conditional average potential outcome as

$$\mathrm{E}[Y(d) \mid p(X)] = \mathrm{E}[Y \mid D = d, p(X)].$$

Thus, it suffices to learn the CEF $\mathrm{E}[Y \mid D, p(X)]$. We learn good approximations of these CEFs by incorporating polynomials or other transformations of $p(X)$ to make things more flexible and running linear regression methods. Finally, we can also employ nonlinear machine learning methods introduced in Chapter 9 to overcome the limitations of linear models.

After controlling for $p(X)$, we can also consider the use of high-dimensional methods to include other transformations $W$ of the raw variables $X$ in order to improve precision, estimating the more flexible CEF $\mathrm{E}[Y \mid D, p(X), W]$. It is especially advisable to include transformations $W$ that fail the covariate balance checks discussed in Section 5.3. Including $W$ can reduce the selection bias (and, hopefully, set it equal to zero). In the reemployment experiment, for example, we observed that balance did not seem satisfied across age groups. Hence, further controlling for age makes sense and results in modest changes to estimates of the treatment effect. Of course, there is no guarantee that controlling for observed covariates can overcome selection bias in compromised RCTs in general because unobserved covariates may be driving the bias.

**Remark 5.4.1** ("Clever Covariate") Finally, we note that the simple OLS regression of $Y$ on the single constructed regressor

$$\phi(D, X) := \frac{1(D=1)}{\mathrm{P}(D=1|X)} - \frac{1(D=0)}{\mathrm{P}(D=0|X)} = H$$

can be used to estimate the ATE. Specifically, for $\beta$ the coefficient in the model $Y = \beta H + \varepsilon$ with $\varepsilon \perp H$, we have that the ATE is equal to $\mathrm{E}[\beta(\phi(1, X) - \phi(0, X))]$. This result holds even though the CEF function is not given by $\beta H$; see Section 5.B. As such, incorporating the technical regressor $H$ in a linear regression model (without penalization if high-dimensional estimation tools are used) can be a good idea. This approach is referred to as the "clever covariate" approach in the literature [7, 8].

## 5.5 Average Treatment Effect for Groups and on the Treated

In addition to unconditional average treatment effects (ATE) or average treatment effects at specific values of the covariates $X = x$, we may be interested in average effects within specific subpopulations.

A leading example of an interesting subpopulation treatment effect is a group ATE (GATE):

$$\delta_G = \mathrm{E}[Y(1) - Y(0)|G = 1]$$

where $G$ is a group indicator defined in terms of $X$'s. For example, we might be interested in the effects of a training program among younger people, say between 18 and 30 years old ($G = 1(18 \leq \texttt{age} \leq 30)$); among people older than 30 years old (so $G = 1(30 < \texttt{age})$); and differences between these two groups.

We can immediately obtain the GATE using the identification results above and the law of iterated expectations:

$$\begin{aligned}\mathrm{E}[Y(1) - Y(0)|G = 1] \\ &= \mathrm{E}[\mathrm{E}[Y|D = 1, X] - \mathrm{E}[Y|D = 0, X]|G = 1] \\ &= \mathrm{E}[HY|G = 1].\end{aligned}$$

That is, we can identify GATEs either by taking the difference in regression functions or applying propensity score reweighting of outcomes and then averaging over group $G$.

We next consider treatment effects for the subpopulation of treated units, the *average treatment effect on the treated* (ATET):[8]

$$\delta_1 = \mathrm{E}[Y(1) - Y(0) \mid D = 1].$$

8: Rather than ATET, some use the abbreviation AToT or ATT.

For example, consider training completion as a treatment, $D$, and $X$ a vector of pre-treatment variables such that unconfoundedness holds. Consider the question:

- On average, how much more do trainees earn after going through the training program than they would have earned had they not gone through the program?

Note that this question is a counterfactual question as it requires us to compare outcomes for trainees in the treated state, where they receive training, and the unobserved control state, where they did not receive training. The ATET, $\delta_1$, is the parameter that answers such questions about counterfactuals. The ATET is identified by

$$\mathrm{E}[\mathrm{E}[Y|D = 1, X] - \mathrm{E}[Y|D = 0, X] \mid D = 1]$$

similarly to what we had above. It is also possible to bypass the use of $\mathrm{E}[Y|D = 1, X]$ in this case; see Appendix 5.C for more details.

## Study Problems

1. Use one or two paragraphs to explain conditioning and its use in learning treatment effects/causal effects in observational data and randomized trials where treatment probability depends on pre-treatment variables. This discussion should be non-technical as if you were writing an explanation for a smart friend with relatively little exposure to causal modeling.

2. Use one or two paragraphs to explain the propensity score reweighting approach for identification of average

treatment effects. This discussion should be non-technical as if you were writing an explanation for a smart friend with relatively little exposure to causal modeling.

3. Use one or two paragraphs to explain why group ATE and the ATE on the treated may be of interest in empirical work. This discussion should be non-technical as if you were writing an explanation for a smart friend with relatively little exposure to causal modeling.

## 5.A Rosenbaum-Rubin's Result

Recall the propensity score is

$$p(X) := \mathrm{P}(D = 1|X),$$

which is the probability of receiving treatment given $X$. A simple useful intermediate property is the balancing property of the propensity score which states that treatment is independent of $X$ conditional on the propensity score:

$$D \perp\!\!\!\perp X \mid p(X) \quad \Leftrightarrow \quad \mathrm{P}(D = 1|X, p(X)) = \mathrm{P}(D = 1|p(X)).$$

This result follows simply from (i) $\mathrm{P}(D = 1|X, p(X)) = \mathrm{P}(D = 1|X) = p(X)$ and (ii) $\mathrm{P}(D = 1|p(X)) = \mathrm{E}[D = 1|p(X)] = \mathrm{E}[\mathrm{E}[D|X, p(X)]|p(X)] = \mathrm{E}[p(X)|p(X)] = p(X)$. This property underlies covariate balance checks.

We now turn to the theorem of Rosenbaum and Rubin. By Theorem 5.3.1 and the law of iterated expectations, we have that for any function of the form $g(y) = 1(y \leq t), t \in \mathbb{R}$:

$$\begin{aligned}
\mathrm{E}\left[g(Y(1)) \mid p(X)\right] &= \mathrm{E}[\mathrm{E}[g(Y(1))|X, p(X)]|p(X)] \\
&= \mathrm{E}[\mathrm{E}[g(Y(1))|X]|p(X)] \\
&= \mathrm{E}\left[g(Y)\frac{1(D = 1)}{p(X)} \mid p(X)\right] \\
&= \mathrm{E}\left[g(Y)\frac{1(D = 1)}{p(X)} \mid D = 1, p(X)\right] P(D = 1|p(X)) \\
&\quad + \mathrm{E}\left[g(Y)\frac{1(D = 1)}{p(X)} \mid D = 0, p(X)\right] P(D = 0|p(X))
\end{aligned}$$

$$
\begin{aligned}
&= \mathrm{E}[g(Y) \mid D = 1, p(X)] \frac{P(D = 1 \mid p(X))}{p(X)} \\
&= \mathrm{E}[g(Y) \mid D = 1, p(X)] \\
&= \mathrm{E}[g(Y(1)) \mid D = 1, p(X)]
\end{aligned}
$$

where we use $P(D = 1 \mid p(X)) = p(X)$. We can similarly argue for the case of $d = 0$. Thus, the conditional distribution of $Y(1)$ does not depend on $D$, once we condition on $p(X)$, which verifies Theorem 5.4.1.

## 5.B Clever Covariate Regression

Here we show that if we care only about estimating the ATE, then it suffices to learn the BLP of the outcome $Y$ using the single covariate

$$
\phi(D, X) := H = \frac{1(D = 1)}{p(X)} - \frac{1(D = 0)}{1 - p(X)}.
$$

We can then use this BLP model as a proxy for the CEF $\mathrm{E}[Y \mid D, p(X)]$. Specifically, we learn a decomposition $Y = \beta\phi(D, X) + \epsilon, \epsilon \perp \phi(D, X)$ by running OLS of $Y$ on $\phi(D, X)$ and then use $\mathrm{E}[\beta(\phi(1, X) - \phi(0, X))]$ as the ATE. This approach, referred to in the literature as the "clever covariate" approach, was first proposed in [7] and further developed in [8].

Note that the random variable $H$ satisfies

$$
\mathrm{E}[f(D, X)H \mid X] = f(1, X) - f(0, X)
$$

for any function $f(D, X)$.[9] Then, by Theorem 5.3.1 and orthogonality of $\epsilon$ in the BLP decomposition:

9: Verify this as a reading exercise.

$$
\begin{aligned}
\mathrm{E}[Y(1) - Y(0)] = \mathrm{E}[YH] &= \mathrm{E}[\beta\phi(D, X)H] \\
&= \mathrm{E}\left[\beta(\phi(1, X) - \phi(0, X))\right].
\end{aligned}
$$

Note that even though this approach allows us to identify the ATE, it does uncover the CATE $\mathrm{E}[Y(1) - Y(0) \mid X]$. The reason for the failure in learning the CATE is that the residual $\epsilon$ does not necessarily satisfy conditional orthogonality; i.e. we do not have $\mathrm{E}[(Y - \beta\phi(D, X))H \mid X] = 0$.

## 5.C Details of ATET

In observational studies, the ATET is identified under weaker conditions than the ATE because

$$\mathrm{E}[Y(1) \mid D = 1] = \mathrm{E}[Y \mid D = 1],$$

so we only need to identify $\mathrm{E}[Y(0) \mid D = 1]$. We can state the weaker version of the ignorability and overlap conditions as follows:

**Assumption 5.C.1** (Ignorability and Overlap for Treated) *(a) Ignorability. Suppose that the treatment status $D$ is independent of $Y(0)$ conditional on a set of covariates $X$, that is*

$$D \perp\!\!\!\perp Y(0) \mid X.$$

*(b) Weak Overlap. Suppose that the propensity score satisfies:*

$$\mathrm{P}(p(X) < 1) = 1.$$

**Theorem 5.C.1** (Identification of ATET) *Under Assumption 5.C.1,*

$$\delta_1 = \mathrm{E}[Y \mid D = 1] - \mathrm{E}[\mathrm{E}[Y \mid X, D = 0] \mid D = 1].$$

Theorem 5.C.1 follows because, by iterated expectations and ignorability,

$$\begin{aligned} \mathrm{E}[Y(0) \mid D = 1] &= \mathrm{E}[\mathrm{E}[Y(0) \mid D = 1, X] \mid D = 1] \\ &= \mathrm{E}[\mathrm{E}[Y(0) \mid D = 0, X] \mid D = 1] \\ &= \mathrm{E}[\mathrm{E}[Y \mid D = 0, X] \mid D = 1], \end{aligned}$$

where the outer expectation is well-defined because the support of $X$ conditional on $D = 1$ is a subset of the support of $X$ conditional on $D = 0$ by the overlap condition.

The Horvitz-Thompson method can be also used to recover averages of potential outcomes for the treated. Indeed,

$$\frac{\mathrm{E}[DY]}{\mathrm{E}[D]} = \frac{\mathrm{E}[DY(1)]}{\mathrm{E}[D]} = \mathrm{E}[Y(1) \mid D = 1]$$

and

$$
\begin{aligned}
\frac{\mathrm{E}\left[\frac{(1-D)}{1-p(X)}p(X)Y\right]}{\mathrm{E}[D]} &= \frac{\mathrm{E}\left[\frac{p(X)}{1-p(X)}\mathrm{E}[(1-D)Y \mid X]\right]}{\mathrm{E}[D]} \\
&= \frac{\mathrm{E}\left[\frac{p(X)}{1-p(X)}\mathrm{E}[(1-D)Y(0) \mid X]\right]}{\mathrm{E}[D]} \\
&= \frac{\mathrm{E}\left[\frac{p(X)}{1-p(X)}\mathrm{E}[1-D|X]\mathrm{E}[Y(0) \mid X]\right]}{\mathrm{E}[D]} \\
&= \frac{\mathrm{E}[p(X)\mathrm{E}[Y(0) \mid X]]}{\mathrm{E}[D]} \\
&= \frac{\mathrm{E}[\mathrm{E}[D \mid X]\mathrm{E}[Y(0) \mid X]]}{\mathrm{E}[D]} \\
&= \frac{\mathrm{E}[\mathrm{E}[DY(0) \mid X]]}{\mathrm{E}[D]} \\
&= \frac{\mathrm{E}[DY(0)]}{\mathrm{E}[D]} = \mathrm{E}[Y(0) \mid D = 1]
\end{aligned}
$$

where in the second to last step we used that $D \perp\!\!\!\perp Y(0) \mid X$, implies $\mathrm{E}[DY(0) \mid X] = \mathrm{E}[D|X]\mathrm{E}[Y(0) \mid X]$. Hence, we obtain the following result:

**Theorem 5.C.2** (Propensity Score Reweighting for the Treated) *Under Assumption 5.C.1,*

$$\mathrm{E}[Y\bar{H}] = \delta_1, \quad \bar{H} = Hp(X)/\mathrm{E}[D].$$

# Causal Inference via Linear Structural Equations

# 6

"the scientific [. . . ] problem of causality is essentially a problem regarding our way of thinking, not a problem regarding the nature of the exterior world."

– Ragnar Frisch [1].

Here we present the linear structural equation model framework and causal diagrams. The advantage of these models is they are closely related to underlying structural models commonly used in economics and other fields. They allow for transparent derivation of the conditional ignorability assumption from the structure of the model. While linearity is imposed in this chapter, it will be dispensed with in later chapters.

## 6.1 Structural Equation Modelling and Conditional Exogeneity

Basic ideas that appeared in econometrics between the 20s and 40s (P. Wright [2], S. Wright [3], J. Tinbergen [4], T. Haavelmo [5]) provide another take on and language for causality that is closely related to the potential outcomes framework.

### A Simple Triangular Structural Equation Model (TSEM)

We illustrate the basic ideas using a simple model of a household's (say weekly) demand for gasoline, motivated by Hausman and Newey [6].

We start with a log-linear (Cobb-Douglas [7]) model for log-demand $y$ given the log-price $p$

$$y(p) := \delta p,$$

where $\delta$ is the elasticity of demand. Demand is random across households, and we may model this randomness as

$$Y(p) := \delta p + U, \quad \mathrm{E}[U] = 0, \tag{6.1.1}$$

where $U$ is a stochastic shock that describes variation of demand across households (or across time, but assume that we are just looking at a particular time point). We immediately recognize that $Y(p)$ plays the same role as a potential outcome in Rubin's potential outcome model.[1]

1: The subtle difference here is that $U$ does not depend on the index $p$, though we could make $U$ be indexed by $p$ at the cost of more complicated exposition. The distinction drawn is not superficial. Later on, when we discuss models with instruments, the dependence of $U$ on $p$ can create non-trivial problems which are not present in this section.

The stochastic function

$$p \mapsto Y(p)$$

describes a household's log-demand at a given log-price $p$. The expected log-demand at log-price $p$ is given by $\mathrm{E}[Y(p)] = \delta p$. The function encodes various structural causal effects: If we change $p$ from $p_0$ to $p_1$, the expected demand change would be

$$\mathrm{E}[Y(p_1)] - \mathrm{E}[Y(p_0)] = \delta(p_1 - p_0).$$

Model (6.1.1) is very simple, and we may want to introduce covariates to capture other observable factors that may be associated with demand. That is, we may think there are observable parts of the stochastic shock, characterized by $X$, which help us predict household demand. Leading examples are household characteristics. For example, we may think demand is associated with features such as family size, income, number of cars, or geographical location. We can incorporate these features by modelling $U = X'\beta + \epsilon_Y$, where $\epsilon_Y$ is independent of $X$ and has mean zero. Employing this model structure, we can write our augmented model as

$$Y(p) := \delta p + X'\beta + \epsilon_Y, \quad \epsilon_Y \perp\!\!\!\perp X. \tag{6.1.2}$$

Equation (6.1.2) is a structural stochastic model of economic outcomes. This model has nothing to do with regression or a statistical predictive model. Rather, it is a model that provides counterfactual predictions: If log-price is set to $p$, then a household with characteristics $X$ can be predicted to purchase

$$\delta p + X'\beta$$

log-units. Here $p$ is not a random variable – it is an index describing potential values of the price.

Then we ask the question:

- What data $(Y, P, X)$ on quantities, prices, and characteristics should we collect to allow us to estimate the structural parameter $\delta$?

**Assumption 6.1.1** (Conditional Exogeneity) *(i) (Consistency) Suppose the observed variables $(Y, P, X)$ are such that*

$$Y = Y(P)$$

*i.e. the outcome is generated from the structural model, (ii) (Conditional Exogeneity) The observed $P$ is determined outside of the model, independently of $\epsilon_Y$ conditional on $X$:*

$$P \perp\!\!\!\perp \epsilon_Y \mid X \implies P \perp\!\!\!\perp \{Y(p)\}_{p\in\mathbb{R}} \mid X$$

Assumption 6.1.1 is the econometric analog of ignorability.[2] In the context of household demand, this condition requires that $P$ is determined independently of a household's demand shock $\epsilon_Y$, conditional on characteristics $X$. This assumption seems plausible for household level decisions, especially if we include geography in the set of covariates $X$.

2: At a general level, gasoline prices are determined by aggregate supply and demand conditions, with small local geographic adjustments (e.g., gasoline prices in areas with higher prices of land may be higher than in other areas to reflect the higher land costs for gasoline stations). Conditional on being in a given small geographic region, we may think of price fluctuations as independent of household-specific demand shocks.

If the conditional exogeneity condition holds, then

$$Y = Y(P) = \delta P + X'\beta + \epsilon_Y, \quad \epsilon_Y \perp (P, X).$$

This means that the projection parameters of $Y$ on $P$ and $X$ coincide with the structural parameters $\delta$ and $\beta$.

We stress that our parameters $\delta$ and $\beta$ are not defined by regression; they are defined by the model. Under the conditional exogeneity condition, these parameters coincide with the projection parameters.[3]

3: A weaker starting condition than the conditional exogeneity condition for the above result is simply

$$(P, X) \perp \epsilon_Y.$$

That is, the observed $P$ and $X$ are orthogonal to the structural error $\epsilon_Y$.

We might further postulate a structural equation for log-prices:

$$P(x) := x'\nu + \epsilon_P,$$

where $P(x)$ is the stochastic price process indexed by a household characteristics and $\epsilon_P$ describes the centered stochastic price shock. We assume that observed $X$ is independent of price shock $\epsilon_P$,

$$X \perp\!\!\!\perp \epsilon_P.$$

Independence between $\epsilon_P$ and observed $X$ implies that $\nu$ coincides with the projection coefficient of $P$ on $X$.

The price process $P(x)$ captures the belief that prices faced by households may differ depending on household characteristics. Note that this notation allows for only a subset of household characteristics to be systematically related to price; that is, we can have $P(x) = P(x_1)$ for some subvector $x_1$ of $x$. For example, it seems reasonable that households located in different regions would experience different prices, in which case $x_1$ could represent a household's geographic characteristics. Independence of the price shock $\epsilon_P$ from observed $X$ may be plausible if household characteristics are determined well before gasoline prices faced by individual households in any specific time period are set.

Putting the equations together, we have a triangular structural equation model (TSEM):

$$\begin{aligned} Y &:= \delta P + X'\beta + \epsilon_Y, \\ P &:= X'\nu + \epsilon_P, \\ X&, \end{aligned} \tag{6.1.3}$$

where $\epsilon_Y$, $\epsilon_P$, and $X$ are mutually independent (or at least uncorrelated) and determined outside of the model. They are called exogenous variables. $Y$ and $P$ are determined within the model and called the endogenous variables. The structural parameter $\delta$ can be identified by linear regression provided $\text{Var}(\epsilon_P) > 0$, and the structural parameter $\nu$ can be identified by linear regression provided $\text{Var}(X) > 0$.

Under the conditions stated above the parameters of these structural equations coincide with the projection parameters.

**What do we mean by the model being structural?** The term structural means that each of the equations is *assumed* to provide comparative statics and answers to counterfactual questions. Setting the right-hand-side variables to their potential values, we have

$$Y(p, x) := \delta p + x'\beta + \epsilon_Y,$$
$$P(x) := x'\nu + \epsilon_P.$$

The conceptual operation of "setting" or "fixing" the variables is supposed to leave the structure invariant. More generally, the structural parameters are supposed to be invariant to changes in the distribution of exogenous variables – $X, \epsilon_Y, \epsilon_P$ – that have been generated outside of the model. Therefore, we can use these structural parameters to generate counterfactual predictions.

The jargon *comparative statics* refers to the determination of how endogenous variables change in response to changes in exogenous variables. Similarly, *counterfactual questions* coincide with asking how outcomes or endogenous variables change when variables are set to new values with other features of the model remaining fixed; e.g. asking how demand changes when price is set to some new value by a firm with household characteristics, price shocks, and demand shocks unaffected.

## 6.2 Drawing the Model: Causal Diagrams, aka DAGs

Sewall and Philip Wright [2], [3] would have depicted system of equations (6.1.3) graphically as a causal (path) diagram as in Figure 6.1. Observed variables are shown as nodes, causal paths are shown by directed arrows, and the structural (causal) parameters are given by the symbols placed next to the arrows.

The graph represents a structural economic model that can answer causal (comparative statics) questions. For example, the elasticity parameter $\delta$ tells us how household demand will respond to a firm *setting* a new price. Note that a firm setting a new price will not alter household characteristics or the other exogenous features of the model, and thus only the parameter $\delta$ is relevant for answering this question within the model.

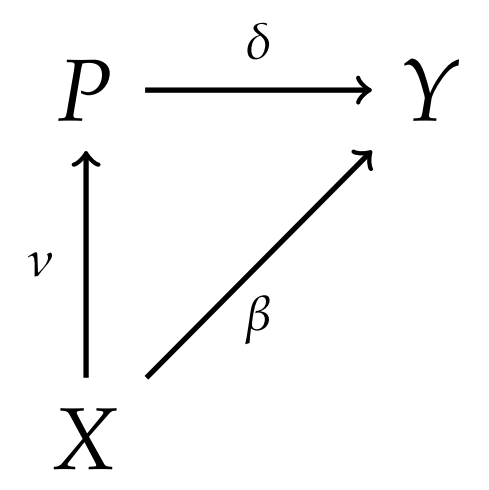

**Figure 6.1:** A simple causal diagram representation of the TSEM for the household gasoline demand example.

We could have expanded the previous graph to include unobserved shocks $\epsilon_P$ and $\epsilon_Y$ as follows:

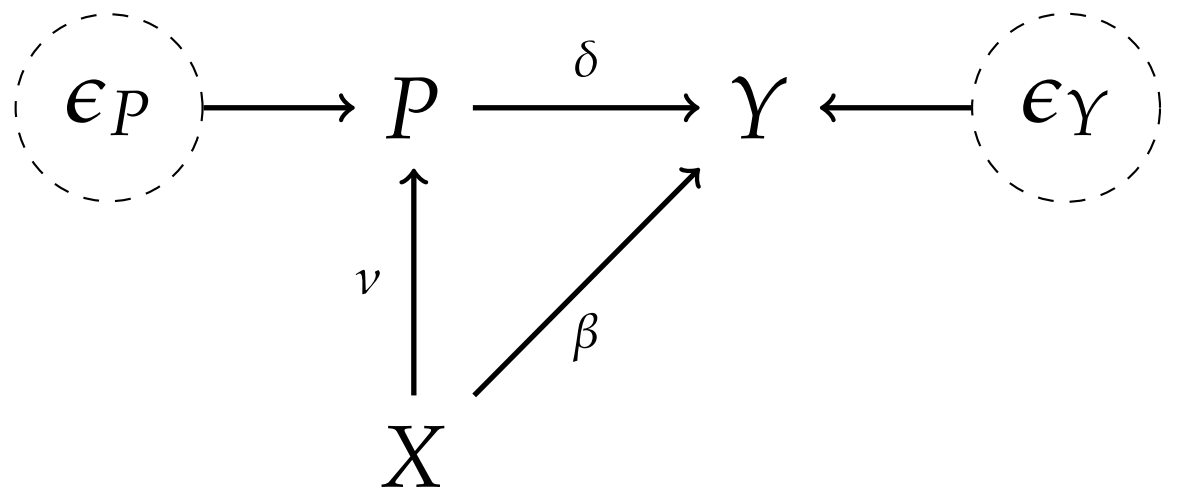

**Figure 6.2:** An expanded causal diagram representation of the TSEM that shows the unobserved shocks $\epsilon_P$ and $\epsilon_Y$ as root nodes.

The graph initiates with the *root nodes* $\epsilon_P$, $X$, and $\epsilon_Y$. The absence of links between the root nodes signifies the orthogonality between the nodes: namely, the absence of correlation. Understanding the orthogonality structure between nodes is an important input into identification of structural parameters via projection. The nodes $X$ and $\epsilon_P$ are *parents* of $P$; the nodes $P, X$, and $\epsilon_Y$ are *parents* of $Y$. The node $Y$ is a *collider* on all paths, because it contains only incoming arrows.

The main effect of interest is $\delta$, which we call the structural causal effect of $P$ on $Y$. This effect is identified after adjusting for $X$. In terms of the graph above, there are two paths connecting $P$ and $Y$:

$$P \longrightarrow Y \text{ and } P \longleftarrow X \longrightarrow Y.$$

The second path is called a *backdoor path* because there is an arrow pointing back to $P$ from $X$. This connection indicates that there is a common cause for $P$ and $Y$. Figuratively speaking, controlling or adjusting for $X$ is said to be like "closing the backdoor path," shutting down the non-casual sources of statistical dependence between $Y$ and $P$.

This visual characterization of the adjustment for $X$ is due to J. Pearl [8] and generalizes to much more complicated graphs. We revisit these ideas throughout subsequent chapters.

How do household characteristics impact our model? $X$ affects $Y$ through two paths:

- the direct effect $\beta$ via $X \to Y$,
- and the indirect effect $\upsilon\delta$ via $X \to P \to Y$.

The indirect effect is said to be "mediated" by $P$. We saw in Section 6.1 that we can identify $\delta$ and $\beta$ from projection of $Y$ on $P$ and $X$, and we can identify $\nu$ by projection of $P$ on $X$. Therefore both the direct and indirect effects are identified.

Mediation structures appeared right at the outset in the Wrights' work [2], [3].

The total effect of $X$ on $Y$ is

$$\nu\delta + \beta,$$

which can be identified in this case by projection of $Y$ on $X$. To verify this, we plug the first equation from the TSEM in (6.1.3) into the second equation producing

$$Y = (\nu\delta + \beta)'X + V; \quad V = \epsilon_Y + \delta\epsilon_P.$$

We see that the composite disturbance $V$ is orthogonal to $X$,

$$V \perp X,$$

and, therefore, $(\nu\delta + \beta)$ coincides with the projection coefficient in the projection of $Y$ on $X$. The latter point can be seen graphically: There are no "backdoor" paths from $X$ to $Y$, so it is not necessary to adjust or control for anything to identify the total effect of $X$ on $Y$.

In fact, while conditioning on $P$ would allow us to identify the direct effect of $X$, $\beta$, it would prevent us from retrieving the total effect $\nu\delta + \beta$. In empirical practice, we may think of conditioning on $P$ as "conditioning on the outcome," as $P$ is determined by its parents, including $X$, so may be thought of as an outcome relative to $X$.

**Remark 6.2.1** (Statistical Identification) Statistical identification typically relies on a combination of orthogonality or conditional independence restrictions and additional conditions – referred to as "rank conditions" in some settings – that ensure there is variation available for learning parameters of interest. For example, we need that $\mathrm{Var}(\epsilon_P) > 0$ if we wish to learn $\delta$ in the TSEM in (6.1.3), and we need overlap for learning ATE as discussed in Chapter 5. Graphical methods provide a tool for representing orthogonality and conditional independence relationships. They typically do not immediately reveal the additional rank-type conditions one would use in establishing statistical point identification. Examining the graphical structure does reveal what causal effects are potentially learnable within the structure, and additional restrictions, such as $\mathrm{Var}(\epsilon_P) > 0$ in the TSEM, can then be deduced. Throughout the remainder of this book, we abstract away from rank-type conditions when discussing graphical models and talk about identifying parameters from the implied orthogonality or conditional independence structure.

To summarize, to learn a causal parameter, we must first define the causal parameter of interest and then carefully consider the choice of what to condition on to learn this effect. These choices are particularly important given the existence of *collider bias*.

Collider Bias R Notebook and Collider Bias Python Notebook provide a simple simulated example of collider bias based on the SEM (6.3.1).

## 6.3 When Conditioning Can Go Wrong: Collider Bias, aka Heckman Selection Bias

Consider the following SEM:

$$\begin{aligned} T &:= \epsilon_T \\ B &:= \epsilon_B \\ C &:= T + B + \epsilon_C \end{aligned} \tag{6.3.1}$$

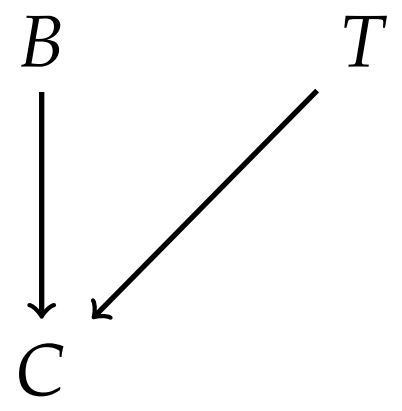

**Figure 6.3:** DAG with a collider representing SEM (6.3.1).

where $\epsilon_T$, $\epsilon_B$, and $\epsilon_C$ are independent $N(0,1)$ shocks. Here the average structural function for $T$, which does not depend on what values $B$ might take, is zero,

$$\mathrm{E}[T] = 0.$$

Regression without conditioning on $C$ correctly identifies that $T$ is not causally impacted by $B$:

$$\mathrm{E}[T \mid B = b] = 0.$$

However, further conditioning on $C$ removes the causal interpretation of the projection coefficient:[4]

$$\mathrm{E}[T \mid B, C] = (C - B)/2; \implies \mathrm{E}[\mathrm{E}[T \mid B = b, C]] = -b/2 < 0.$$

4: Dividing by 2 may seem counter-intuitive, but it is correct. See Collider Bias R Notebook or Collider Bias Python Notebook for detail.

This regression suggests that, controlling for $C$, the predictive effect of $B$ on $T$ is $-1/2$. This predictive effect is not a causal effect.

Collider bias illustrates that conditioning on outcomes may produce the wrong conclusions about causality, so conditioning on outcomes should be always approached with care. In econometrics, collider bias is known as a form of sample selection bias[5] ("conditioning on endogenous variables" or Heckman selection bias [9]).

5: J. Heckman was awarded the Nobel Memorial prize "for his development of theory and methods for analyzing selective samples." Source: Nobelprize.org

**A Serious Digression on Colliders**. Within our toy SEM framework, regression on a collider is clearly the wrong thing to do if one wants to identify the causal effect of $B$ on $T$. However, we do note that regression on a collider can be *very useful* for other predictive tasks.

The following example draws on the discussion given in the "Book of Why" [10] to illustrate collider bias.

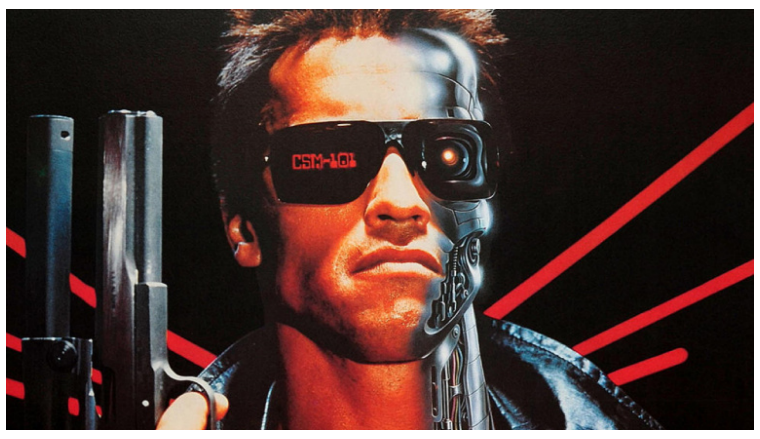

**Figure 6.4:** Our SEM predicts that this actor, A. Terminator, is (essentially) the most talented actor in Hollywood.

**Example 6.3.1** (Structural Model of Hollywood) Suppose that the preceding SEM provides a cartoon depiction of people in Hollywood where $T$ denotes acting talent, $C$ denotes celebrity (i.e. success or popularity), and $B$ denotes bonhomie (i.e. approachability or friendliness). Note that the SEM indicates that more talent and approachability cause more success. Further, for a person to remain in Hollywood, we would expect $C > 0$. As shown above, the causal effect of $B$ on $T$ in this SEM is 0. However, the best linear predictor of $T$ given $B$ conditional on $C > 0$ is

$$\approx .6 - B/4.$$

That is, bonhomie and talent are negatively correlated in Hollywood despite the fact that approachability does not causally impact talent. This correlation is useful for making predictions. For example, the individual depicted in the margin appears quite imposing and not approachable, perhaps with $B = -20$. We would then predict the expected value of his talent to be $t \in [+5.6 \pm 2]$, which is at least 3.6 standard deviations above the average talent of zero in the overall population within our model. From that, we should *predict* that this person is an incredibly talented actor but should not draw any conclusions about causality between $B$ and $T$.

The example illustrates how simple theoretical models are often used in economics. Causal reasoning is made within a simple model, such as the SEM (6.3.1). This reasoning then leads to some testable restrictions, such as negative correlation between $T$ and $B$ conditional on $C > 0$. Even though we may not believe that the stylized model provides a complete model of reality, the implications of the simple model provide some insight into how observed phenomena, such as a negative correlation between $T$ and $B$ conditional on $C > 0$, may arise. Such reversion of the correlation between two variables has been observed empirically in several cases, a prominent one being the birth-weight paradox [11] described below.

**Example 6.3.2** (Birth-weight "paradox" [11]) In a study conducted in 1991 in the US, it was found that infants born to smokers had higher risk of low birth-weight (LBW) and higher risk of infant mortality than infants born to non-smokers. However, when looking at the sub-group of infants with LBW, the comparison is reversed and the risk of infant mortality is lower for infants born to smokers, than for infants born to non-smokers. How is that possible? Does smoking have a positive causal effect on infant mortality conditional on LBW?

A more plausible alternative explanation can be uncovered through the lens of SEMs and Causal Diagrams if one starts to think of competing risks and collider bias. Let's denote with $S$ the smoking indicator, $Y$ the infant death outcome, and $B$ the low birth-weight indicator. We will also denote with $U$ an abstract variable corresponding to the multitude of competing risks that can cause LBW. It is highly plausible that smoking is a risk factor for LBW and also has a direct effect on mortality. Moreover, LBW and the competing risk factors can also have a direct effect on mortality. Putting these factors together leads to the Causal Diagram depicted in Figure 6.5. In this setting, an infant with a smoking parent may be highly likely to have LBW caused by smoking. At the same time, LBW can be much less frequent for non-smoking parents. When we further focus in on the group of infants of non-smoking parents with LBW, it is highly probable that LBW was caused by some other competing risk which can adversely affect mortality. Thus, conditioning on LBW, we could essentially be comparing infants of smoking parents without competing risks to infants of non-smoking parents with competing risks.

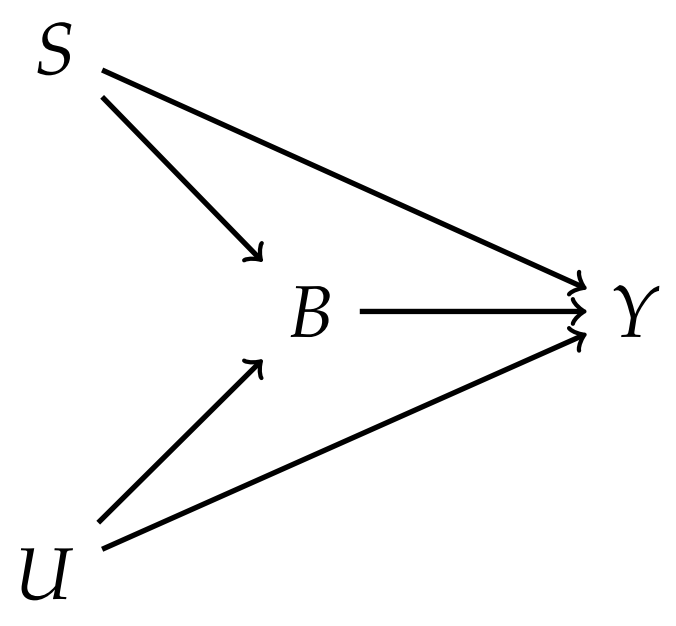

**Figure 6.5:** DAG with a collider representing low birth-weight "paradox" Example 6.3.2.

To illustrate how the unconditional association between $Y$ and $S$ uncovers the true causal effect, while conditioning on $B$ introduces bias and can even reverse the sign of the true effect, let's look at a simple linear SEM that corresponds to the causal diagram depicted in Figure 6.5:

$$\begin{aligned} Y &:= S + B + \kappa U + \epsilon_Y \\ B &:= S + U + \epsilon_B \\ S &:= \epsilon_S \\ U &:= \epsilon_U \end{aligned} \tag{6.3.2}$$

where $\epsilon_Y, \epsilon_B, \epsilon_S$ and $\epsilon_U$ are independent $N(0,1)$ shocks. Note that if we simply project $Y$ on $S$, then we recover the correct positive causal effect of 2, since conditional exogeneity is satisfied. However, when we project $Y$ on $S$ and $B$, we learn a CEF of the form:

$$\begin{aligned} \mathrm{E}[Y \mid S, B] &= S + B + \kappa \mathrm{E}[U \mid S, B] \\ &= S + B + \kappa(B - S)/2 = (1 - \kappa/2)S + (1 + \kappa/2)B. \end{aligned}$$

If the competing risks increase infant mortality a lot, i.e. $\kappa \gg 1$, then this projection recovers an erroneous large negative(!) effect $1 - \kappa/2$ of smoking on mortality.

## 6.4 Wage Gap Analysis and Discrimination

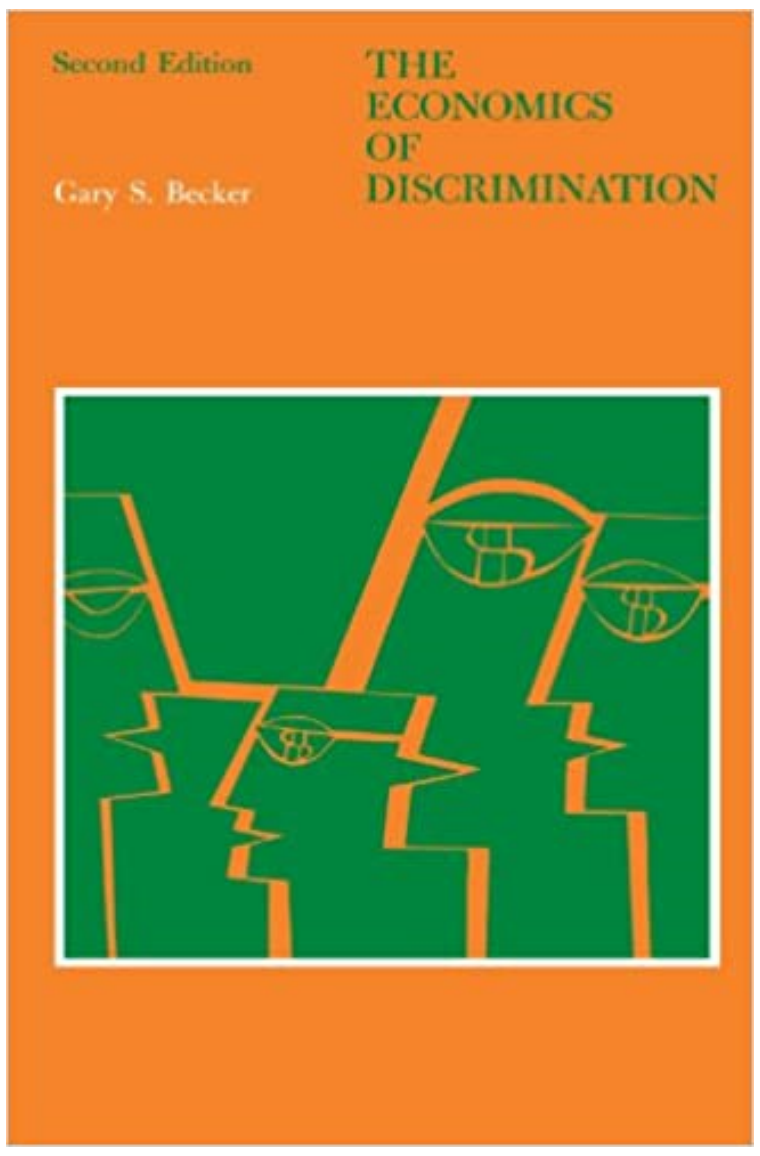

> "The central question in any employment-discrimination case is whether the employer would have taken the same action had the employee been of a different race (age, sex, religion, national origin etc.) and everything else had remained the same." (In Carson versus Bethlehem Steel Corp., 70 FEP Cases 921, 7th Cir. (1996) [12]).

Wage regressions are widely used by labor economists to characterize the wage gap between men and women and to link the wage gap to discrimination; see, e.g., [13] and [14]. Some economists have asserted that it is wrong to study discrimination by doing wage gap regressions, e.g. [15], and that we should instead look at the unconditional difference in outcomes across groups. Their reasoning is based on the argument that key job characteristics – e.g., education and occupation – are determined in response to both a group identity and discrimination and are therefore (intermediate) outcomes. Controlling for these characteristics may then introduce a form of selection bias. Which of these two sets of economists is right?

In what follows, we present a simple SEM in (6.4.1), which postulates that different groups receive equal wages if there are no conditional productivity differences between the groups. We will see that, in this SEM, wage gap regressions do uncover well-defined discrimination effects that occur in wage-setting mechanisms. In contrast, the unconditonal average wage gap uncovers a more complicated causal object, which absorbs discrimination in wage setting, discrimination in human capital and occupational acquisitions, as well as group specific preferences for occupations.

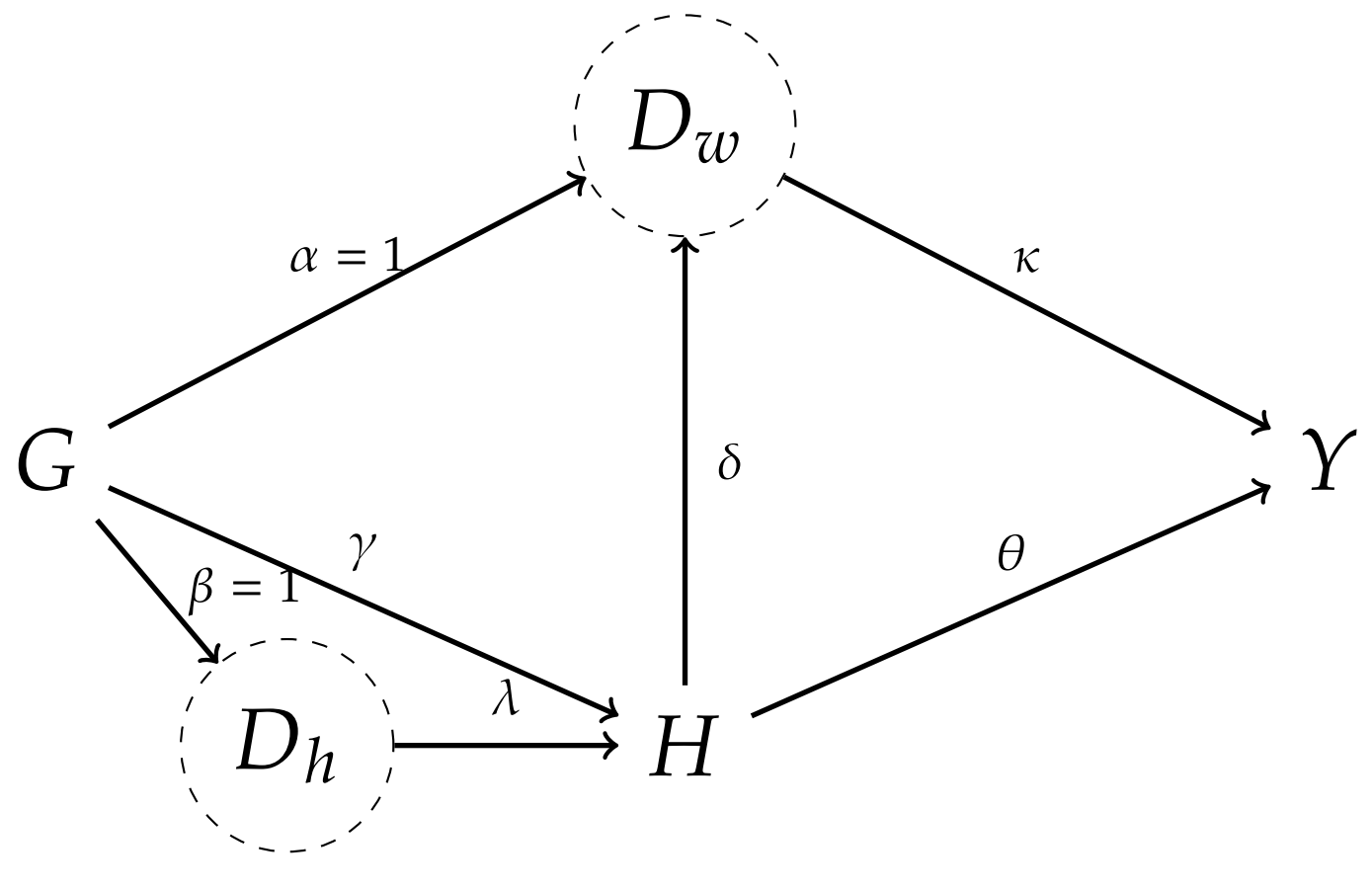

**Figure 6.6:** A Simple Model of Discrimination. Here $G$ denotes a group (e.g., sex), $H$ is human capital, and $Y$ is the wage. $D_w$ denotes unobserved wage discrimination occuring in the work place, and $D_h$ denotes unobserved discrimination that occurs in the accumulation of human capital.

Here we begin with the linear SEM and the equivalent DAG shown in Figure 6.6:

$$\begin{aligned}
&Y := \kappa D_w + \theta H + \epsilon_Y, \\
&D_w := \alpha G + \delta H + \epsilon_{D_w}, \\
&H := \gamma G + \lambda D_h + \epsilon_H, \\
&D_h := \beta G + \epsilon_{D_h}, \\
&G,
\end{aligned} \tag{6.4.1}$$

where the shocks $\epsilon_Y, \epsilon_{D_w}, \epsilon_H, \epsilon_{D_h}$, and $G$ are all mean zero and uncorrelated.

The outcome $Y$ is wage, $G$ is group (e.g., sex), $H$ is human capital (a scalar index that includes labor-relevant characteristics such as education, occupation, etc.),[6] $D_w$ is latent wage discrimination arising in the work-place, and $D_h$ is latent discrimination arising in acquisition of human capital. There could be other observed confounders that we don't show for the sake of simplicity.

6: $H$ can be easily made a vector with a slightly more complicated notation.

The discrimination variables $D_w$ and $D_h$ are latent variables that are important for our model but cannot be directly observed. We maintain throughout that these variables are non-degenerate and related to group identity $G$. Under these assumptions, the scale of these latent variables is non-zero but arbitrary, so we normalize the effect $G \longrightarrow D_w$ to unity, $\alpha = 1$, and the effect $G \longrightarrow D_h$ to unity as well, $\beta = 1$. There is no edge from $G$ to $Y$, reflecting our assumption that there is no systematic group difference in productivity conditional on $H$ and $D_w$. In the absence of productivity differences between workers, economic reasoning suggests that they would be assigned the same wage in a discrimination-free economy [16]. Thus, we would expect $\kappa = 0$ in a discrimination-free economy in the case that $H$ captures all sources of productivity differences between workers.

Within this model, the parameter of interest is then the causal or structural effect of discrimination on wages given by

$$\kappa.$$

If $\kappa \neq 0$, we can conclude that wages are assigned unfairly within the framework of this SEM.

If we observed $D_w$ directly, we could learn the effect of discrimination on wages, $\kappa$, by regression of $Y$ on $D_w$ and $H$. Identification of $\kappa$ from this regression follows from the backdoor criterion discussed in Section 6.2. We don't observe $D_w$ directly, but we postulate that this variable is determined only by $G$, $H$, and a stochastic shock. Dependence on $H$ captures the idea that discrimination may be larger or smaller depending on education level, profession, etc. We return to using this additional structure to learn about $\kappa$ below.

Discrimination may operate through channels other than simple wage differences. For example, in the 1960s, there were relatively few women or African American lawyers, a highly paid occupation. Discrimination that operates through occupational choice or human capital formation is captured by latent variable $D_h$. In our model, $H$, which captures productivity differences between individuals, can be determined as a result of both discrimination and group preferences.[7] The parameter $\gamma$ then captures the effect of group preferences on the formation of $H$, while the effect of discrimination on $H$ is captured by $\lambda$. Since $D_h$ is not observed, there is no way to separately identify these two effects.

7: For example, 90% of firefighters in the US are men, which may reflect a genuine preference for this occupation among men. At the same time, even preference for occupation may be a result of cultural institutions that could themselves be interpreted as discriminatory in broader, cross-cultural, contexts.

It is easy to show, within the model, that the population linear regression of $Y$ on $G$ and $H$ recovers the wage dis-crimination effect,

$$\kappa,$$

and that the linear regression of $H$ on $G$ recovers

$$\gamma + \lambda,$$

the sum of the group preference effect and the human capital discrimination effect; see Appendix 6.A for details. If a further strong assumption is made that there is no group preference effect, $\gamma = 0$, the linear regression of $Y$ on $G$ recovers the total discrimination effect:

$$\kappa + \lambda(\kappa\delta + \theta).$$

**Endogenous Sample Selection**. There is an important issue with our empirical example. We are only able to look at earnings of people who are employed. Thus, we are conditioning on

$$Y > R,$$

where $R$ is the reservation wage. In other words, we are conditioning on the outcome which may cause major selectivity issues: People get employed, and end up in our data, only if the offered wage is higher than some reservation wage. This sample selection on the basis of the outcome can cause major biases in the analysis. The potential for large biases was recognized by James J. Heckman [9] in the 70s and led to the development of the celebrated Heckman selection correction and related methods.

An alternate approach to applying a selection correction in our example is to select a subset $S$ of people who are employed with probability one (or very close to one). For example, one could look at highly educated, unmarried people. Within this subset, we would then have

$$P(Y > R|S) \approx 1.$$

That is, the value of the wage offer, $Y$, is approximately unrelated to whether we observe individual wages for this subset of people. This type of strategy has been employed by Casey Mulligan and Yona Rubinstein [17]. Mulligan and Rubinstein continue to find evidence in favor of the existence of wage gaps in their analysis of a subsample where selection effects are likely small. This finding then suggests that the broad conclusion of the existence of wage gaps is not driven entirely by sample selection issues.

In summary, we have the following observations:

- In general, wage gap regressions just estimate predictive effects or associations.
- When we assume a SEM like the one above holds and there are no endogenous sample selection effects, wage gap regressions estimate wage discrimination effects.
- Unconditional wage gaps generally reflect a combination of different types of discrimination and group preferences and thus do not isolate solely the effects of discrimination.

## Notebooks

- Collider Bias R Notebook and Collider Bias Python Notebook provide a simple simulated example of collider bias, informing our discussion of conditioning on Celebrity in our Structural Model of Hollywood.

## Notes

This chapter presented an approach to causal inference that goes back to the works of Sewall and Philip Wright [2], [3], Tinbergen [4], Haavelmo [5], and others. This tradition lives in modern structural casual models used in econometrics (especially, industrial organization) and in the artificial intelligence community. The latter community, inspired by the foundational work of J. Pearl [8], strongly adopted the use of causal diagrams, known as directed acyclical graphs (DAGs). We continue exploring this approach throughout the remainder of our treatment on causal inference.

## Study Problems

1. Explain collider bias to a friend in simple terms. Use no more than two paragraphs. Illustrate your explanation using a simulation experiment.

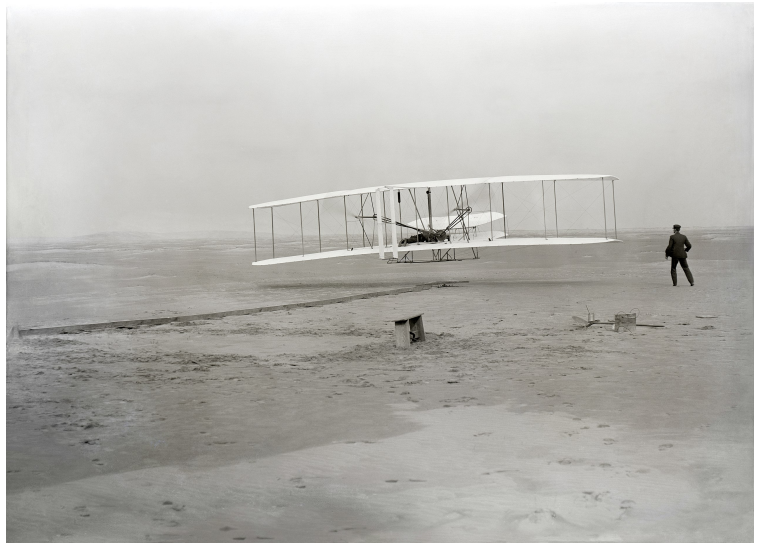

**Figure 6.7:** Early 20th century: The work of Sewall and Philip Wright made it possible for humans to begin to "fly" in the space of causal models. Another family of Wrights made it possible for humans to begin to fly in the air.

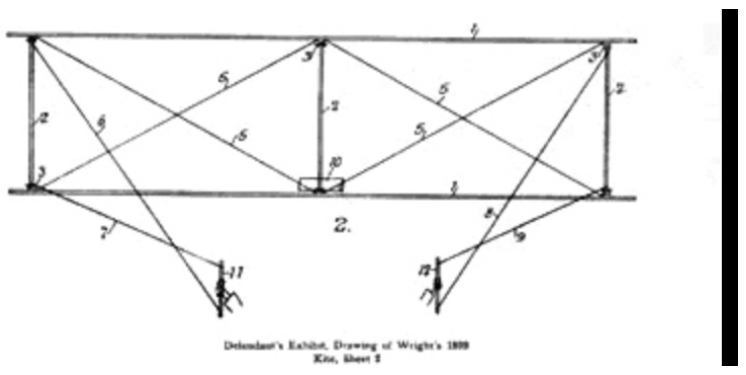

**Figure 6.8:** An early drawing for an airplane appears very much like an early drawing of a DAG.

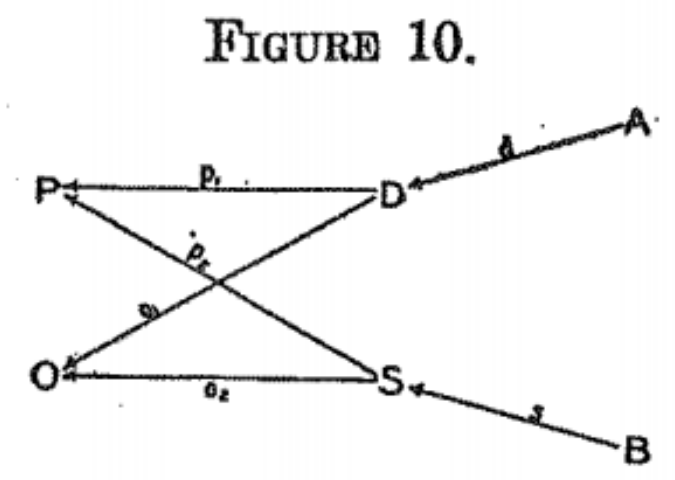

**Figure 6.9:** DAG for Supply-Demand Systems in P. Wright's work in 1928 [2].

2. Empirical: Revisit the group wage gap analysis from Chapter 4, focusing on college-educated workers. Is there a structural/causal interpretation for the estimated wage gap? Is there a group gap in education achievement? Does this group gap in education have a structural/causal interpretation? Some of these questions are open ended and have no simple answers, but it is useful to think about them. (If you have other data sets that might illuminate discrimination in other settings, please use them in place of the wage data set).

3. Free-style exercise: The model for wage discrimination presented in our notes is very stylized and subject to multiple criticisms. For example, it does not deal with promotion and hiring decisions. There are several interesting models of discrimination in hiring, college admissions, and pay. For example, see "The Book of Why"[10] and the Bickel et al. 1975 paper [18] for an analysis of Berkeley undergraduate admissions decisions. Nina Roussile's (2020) [19] paper isolates the ask gap as the central mechanism for the subsequent wage gap. Referring to one such analysis, draw or write down a linear structural causal model that captures the structural idea of the analysis and discuss identification in the model.

## 6.A Details of the Wage Discrimination Analysis

We write out some of the structural equations corresponding to our stylized DAG for discrimination (Figure 6.6):

$$Y := \kappa D_w + \theta H + \epsilon_Y, \quad \epsilon_Y \perp D_w, H, G$$
$$D_w := G + \delta H + \epsilon_{D_w}, \quad \epsilon_{D_w} \perp G, H$$

where the orthogonality relations are implied by the model.

Linear regression analysis would use observable variables only, so we substitute the model for the unobserved $D_w$ in terms of $G$ and $H$ into the equation for $Y$ to obtain

$$Y = \kappa G + (\kappa\delta + \theta)H + U, \quad U := \kappa\epsilon_{D_w} + \epsilon_Y \perp (G, H).$$

The composite error term $U$ is orthogonal to $G$ and $H$. Therefore, regression of $Y$ on $G$ and $H$ learns $\kappa$ and $(\kappa\delta + \theta)$, with our main target being $\kappa$. We can also see that by partialling out $H$,

$$\tilde{Y} = \kappa\tilde{G} + U, \;\; U \perp \tilde{G}.$$

"This is elementary, my dear Watson," said Sherlock Holmes after seeing this.

Thus, $\kappa$ is retrievable only if there is non-zero variation in $\tilde{G}$ after taking out the linear effect of $H$.

Now suppose we want to study discrimination effects in occupational choices, captured by $H$ in our model. We write out the relevant structural equations:

$$H := \gamma G + \lambda D_h + \epsilon_H, \quad \epsilon_H \perp (G, D_h),$$
$$D_h := G + \epsilon_{D_h}, \qquad \epsilon_{D_h} \perp G.$$

Recall that $\gamma$ is the group preference effect and $\lambda$ is the discrimination effect. Since $D_h$ is not directly observed, we substitute it out to arrive at

$$H = (\gamma + \lambda)G + V; \quad V := \gamma\epsilon_{D_h} + \epsilon_H \perp G.$$

Therefore, $\gamma + \lambda$ is the projection coefficient in the projection of $H$ on $G$. Hence, we can identify $\gamma + \lambda$, but we can't identify $\gamma$ and $\lambda$ separately.

Going further, suppose that the group preference effect is zero, so $\gamma = 0$. Then, the previous argument would identify $\lambda$ and we could identify the total discrimination effect arising from two different channels:

$$\kappa + \lambda(\kappa\delta + \theta).$$

from the regression of $Y$ on $G$.

We can assert that the unconditional difference in wages measures discrimination only if the group preference effect in determining $H$ is zero ($\gamma = 0$). Of course, most economists would probably not agree with the assumption that $\gamma = 0$. Empirically, there are large differences in group composition among different professions. These differences likely reflect both discrimination and genuine preferences.

# 7 Causal Inference via Directed Acyclical Graphs and Nonlinear Structural Equation Models

"you are smarter than your data. Data do not understand causes and effects; humans do."

– Judea Pearl [1].

Here we explore a fully nonlinear, nonparametric formulation of causal diagrams and associated structural equation models. These provide a useful tool for thinking about structures underlying causal identification.

## 7.1 Introduction

The purpose of this module is to provide a more formal and general treatment of acyclic nonlinear (and nonparametric) structural equation models (SEMs) and corresponding causal directed acyclic graphs (DAGs). We discuss the concepts and identification results provided by Judea Pearl and his collaborators and by James H. Robins and his collaborators.

In 2011, J. Pearl was awarded the A.M. Turing award, the highest award in the field of Computer Science and Artificial Intelligence: "For fundamental contributions to artificial intelligence through the development of a calculus for probabilistic and causal reasoning." In the Biometrika 1995 article [2], J. Pearl presents his work as a generalization of the SEMs put forward by T.Haavelmo [3] in 1944 and others.

These models and concepts allow us to rigorously define structural causal effects in fully nonlinear models and obtain conditional independence relationships that can be used as inputs to establishing nonparametric identification from the structure of the causal DAGs alone.[1] Structural causal effects are defined as hypothetical effects of interventions in systems of equations. We discuss identification of effects of *do interventions* introduced by Pearl [2] and *fix interventions* introduced by Heckman and Pinto [4] and Robins and Richardson [5].[2] fix interventions induce counterfactual DAGs called SWIGs (Single World Intervention Graphs) and can recover the causal graphs we've seen in previous chapters.

1: We abstract away from rank-type conditions. See Remark 6.2.1.

2: Fix interventions also had appeared as part of do calculus in Pearl [2].

Whether causal effects derived from SEMs approximate policy or treatment effects in the real world depends to a large extent on the degree to which the posited SEM approximates real phenomena. In thinking about the approximation quality of a model, it is important to keep in mind that we will never be able to establish that a model is fully correct using statistical criteria. However, we may be able to reject a given model using formal falsifiability criteria – though not all models are statistically falsifiable – or contextual knowledge. Further, evidence for some causal effects inferred from SEMs can be provided by further use of explicit randomized controlled trials, though the use of experiments is not an option in many cases. Ultimately, contextual knowledge is often crucial for making the case that a given structural model represents real phenomena sufficiently well to produce credible estimates of causal effects when using observational data.

### Notation

Consider a pair of random variables (or equivalently, random vectors) $U$ and $V$ with joint distribution probability (mass) function $\mathsf{p}_{UV}(u,v)$ at generic evaluation points $(u,v)$. We will simply denote $\mathsf{p}_{UV}(u,v)$ by $\mathsf{p}(u,v)$ whenever there is no ambiguity. We will denote the marginal probability (mass) functions by $\mathsf{p}_U(u)$ and $\mathsf{p}_V(v)$, or simply by $\mathsf{p}(u)$ and $\mathsf{p}(v)$. The random variables $U$ and $V$ are independent, which we denote as

$$U \perp\!\!\!\perp V,$$

if and only if the joint probability density (or mass) function $\mathsf{p}(u, v)$ can be factorized as

$$\mathsf{p}(u, v) = \mathsf{p}(u)\,\mathsf{p}(v)$$

or equivalently if and only if

$$\mathrm{E}[g(U)\ell(V)] = \mathrm{E}[g(U)]\mathrm{E}[\ell(V)]$$

for any bounded functions $g$ and $\ell$. This definition of independence implies the ignorability or exclusion results,

$$\mathsf{p}(u \mid v) = \mathsf{p}(u), \quad \mathsf{p}(v \mid u) = \mathsf{p}(v),$$

which follow from Bayes' law:

$$\mathsf{p}(u \mid v) = \frac{\mathsf{p}(u)\,\mathsf{p}(v)}{\mathsf{p}(v)}.$$

Conditional independence is defined similarly by replacing distributions and expectations with their conditional analogs. Appendix 7.B reviews some useful results on conditional independence.

## 7.2 From Causal Diagrams to Causal DAGs: TSEM Example

Formal causal nonlinear DAGs generalize linear parametric models to general nonparametric forms. Recall our previous discussion of a model for a household's log-demand for gasoline (Y), which is a function of log-price (p) and household characteristics (X). We can generalize the simple TSEM to a nonlinear DAG as follows.

**Example 7.2.1** (TSEM) We have a system of triangular structural equations:

$$\begin{aligned} Y &:= f_Y(P, X, \epsilon_Y), \\ P &:= f_P(X, \epsilon_P), \\ X &:= \epsilon_X, \end{aligned} \qquad (7.2.1)$$

where $f$'s are said to be deterministic structural functions and $\epsilon_Y, \epsilon_P, \epsilon_X$ are structural shocks that are independent of each other. The dimension of structural shocks is not restricted. Also, note the independences:

$$\epsilon_Y \perp\!\!\!\perp (P, X), \quad \epsilon_P \perp\!\!\!\perp X.$$

A causal diagram depicting the algebraic relationship defining the TSEM in Example 7.2.1 is shown in Figure 7.1. The absence of edges between nodes encodes the model's independence restrictions. Thus, as before, we can see that we can view graphs as representations of independence relations in statistical models. The graph visually depicts independence restrictions and the propagation of information or structural shocks from root nodes to their children, grandchildren, and so forth.

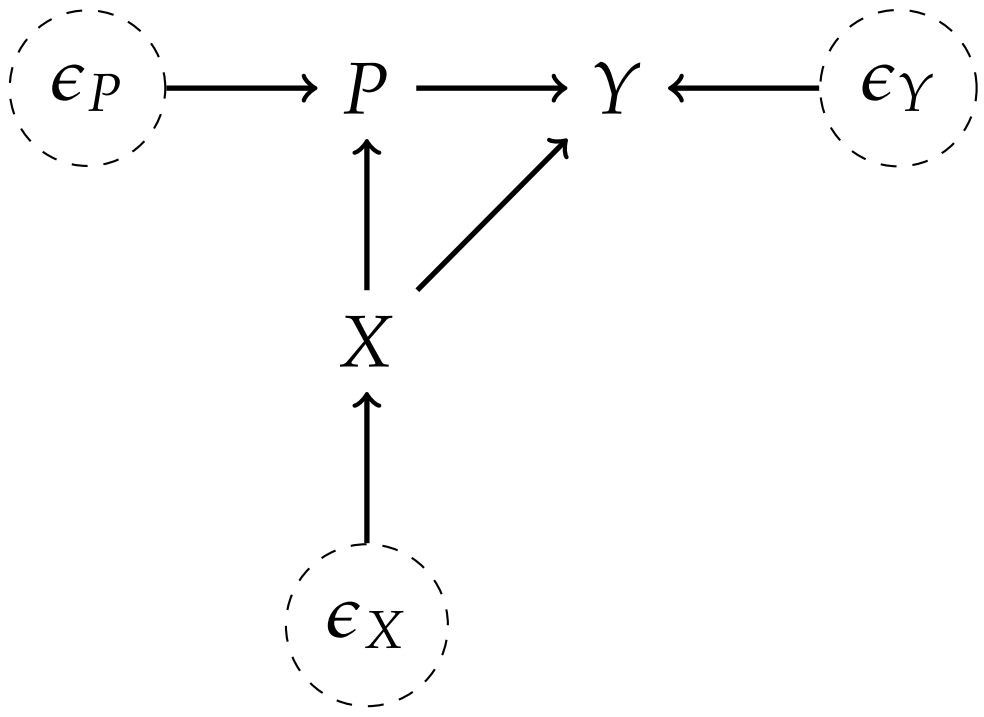

**Figure 7.1:** The causal DAG equivalent to the TSEM in Example 7.2.1.

It is also common to draw graphs based on only observed variables. We can erase the latent root nodes from Figure 7.1 to produce the equivalent diagram illustrated in Figure 7.2.

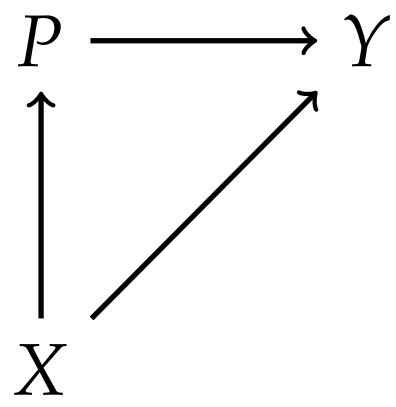

**Figure 7.2:** The causal DAG corresponding to the TSEM in Example 7.2.1 with latent root nodes erased.

The TSEM is purely a statistical model. We can view this model as structural under invariance restriction, following Haavelmo [3].

**Definition 7.2.1** (Structural Form) *When we say that the TSEM is structural, we mean that it is defined by a structure made up of a set of stochastic processes:*

$$\begin{aligned} Y(p,x) &:= f_Y(p,x,\epsilon_Y), \\ P(x) &:= f_P(x,\epsilon_P), \\ X &:= \epsilon_X, \end{aligned}$$

*indexed by $(p,x) \in \mathcal{P} \times \mathcal{X}$, called structural functions or structural potential outcome processes. Moreover,*

- *(Exogeneity) Stochastic shocks $\epsilon_P, \epsilon_X$, and $\epsilon_Y$ are generated as independent variables outside of the model;*
- *(Consistency) The endogenous variables are generated by recursive substituations:*

$$Y := Y(P,X), \quad P := P(X), \quad X := \epsilon_X;$$

- *(Invariance) The structure remains invariant to changes of the distribution of stochastic shocks $\epsilon$.*

The structure will be assumed to be preserved under various interventions as defined below.

While SEMs are statistical models, assumptions akin to those in Definition 7.2.1 endow them with a structural meaning. Structural meaning may be generated by economic or other scientific reasoning. For example, structural functions may correspond to demand functions, supply functions, and expenditure functions, with these notions going back at least to Marshall [6] in the 19$^{\text{th}}$ century.

**Remark 7.2.1** (Link to Potential Outcomes) Consider binary $p \in \{0, 1\}$ for simplicity. Consider potential outcomes, given by the structure:

$$Y(p, X) := g(p, X, \epsilon_Y(p)).$$

We can view potential outcomes through a SEM framework as follows. Let $\epsilon_Y := \{\epsilon_Y(p) : p \in \{0, 1\}\}$, then we have that

$$Y(p, X) = g(p, X, \epsilon_Y(p)) = f_Y(p, X, \epsilon_Y),$$

for

$$f_Y(p, x, e) := 1(p = 0)g(p, x, e(0)) + 1(p = 1)g(p, x, e(1))$$

for the argument $e = \{e(p) : p \in \{0, 1\}\}$. This example emphasizes that the dimensionality of $\epsilon$'s is not restricted in the general framework.

## Identification by Regression

By conditioning on $X = x$ in the graph in Figure 7.1, we obtain the graph shown in Figure 7.3. We can equivalently express the relationship shown in Figure 7.3 in terms of equations as

$$Y(x) = f_Y(P(x), x, \epsilon_Y), \quad \epsilon_Y \perp\!\!\!\perp P(x).$$

If $P(x)$ is non-degenerate, we can further condition on $P(x) = p$ to learn the average structural function

$$\mathrm{E}[f_Y(p, x, \epsilon_Y)]$$

via regressions. We formally record this result as follows.

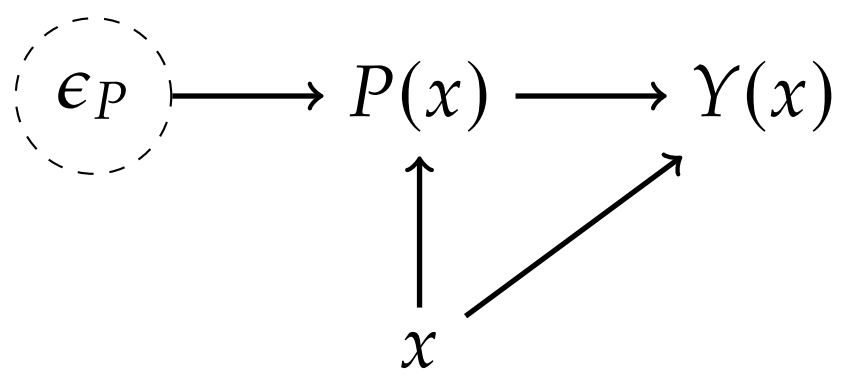

**Figure 7.3:** The graph produced from Figure 7.1 by conditioning on $X = x$. Here $X$ is a parent to both $P$ and $Y$. After conditioning, the remaining source of variation in $P(x)$ is $\epsilon_P$. $\epsilon_P$ is determined exogenously – as if by an experiment – which allows measurement of the causal effect $P(x) \to Y$.

In the TSEM, the conditional average structural function

$$\mathrm{E}[f_Y(p, x, \epsilon_Y)]$$

can be identified by conditioning on $P$ and $X$:

$$\begin{aligned}
&\mathrm{E}[Y|P = p, X = x] \\
&\quad = \mathrm{E}[f_Y(P, X, \epsilon_Y)|P = p, X = x] \\
&\quad = \mathrm{E}[f_Y(p, x, \epsilon_Y)|P = p, X = x] \\
&\quad = \mathrm{E}[f_Y(p, x, \epsilon_Y)]
\end{aligned}$$

provided the event $\{P = p, X = x\}$ is assigned positive density.

This average structural function has the interpretation as the expected outcome when $P$ and $X$ are exogenously set (set outside of the model as if by a policy maker or experiment) to $P = p$ and $X = x$.

Hence, we can use the average structural function to provide counterfactual predictions – predictions for the outcome under exogenous assignment of the policy variable $P$ at fixed values for $X$. Within the TSEM, these counterfactual predictions align with the usual prediction rule $\mathrm{E}[Y|P = p, X = x]$.

If the confounder $X$ is not observed, the causal relationship $P(x) \to Y$ is not identified.

If we can identify the conditional average structural function, we can also identify the conditional average structural causal effect:

$$\begin{aligned}
&\mathrm{E}[f_Y(p_1, x, \epsilon_Y)] - \mathrm{E}[f_Y(p_0, x, \epsilon_Y)] \\
&\qquad = \mathrm{E}[Y|P = p_1, X = x] - \mathrm{E}[Y|P = p_0, X = x].
\end{aligned} \tag{7.2.2}$$

The right hand side of (7.2.2) is a statistical quantity that can clearly be learned from data on $Y$, $P$, and $X$ under reasonable assumptions. The left hand side of (7.2.2) defines a structural quantity of interest: the average effect of exogenously changing $P$ from $p_0$ to $p_1$ at $X = x$.

## Interventions

**Do Interventions.** The do operation $do(P = p)$ or do intervention corresponds to creating the counterfactual graph shown in Figure 7.4. On the graph, we remove $P$ and replace it with a deterministic node $p$. In terms of equations (7.2.1) defining the TSEM, we replace the equation for $P$ with $p$ and then set $P$ equal to $p$ in the first equation. The corresponding counterfactual SEM is

$$\left(\begin{bmatrix} Y \\ P \\ X \end{bmatrix} : \mathrm{do}(P = p)\right) := \begin{bmatrix} f_Y(p, X, \epsilon_Y) \\ p \\ X \end{bmatrix} = \begin{bmatrix} Y(p) \\ p \\ X \end{bmatrix}.$$

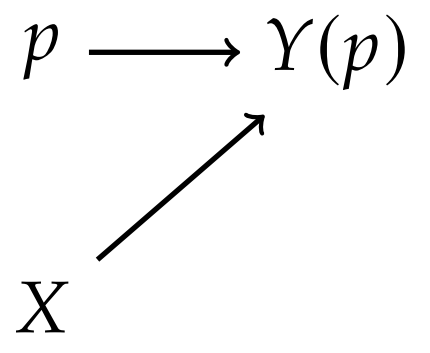

**Figure 7.4:** Causal DAG describing the counterfactual SEM induced by doing $P = p$.

The variables $Y(p)$ and $X$ are the counterfactuals generated by the intervention $do(P = p)$. Note that the intervention keeps $X$ and stochastic shocks $\epsilon_Y$ invariant.

The do operation has been extended to generate other types counterfactuals. For instance, another class of interventions are soft interventions[3] where the intervening variable is set to a value that is a function of its natural value (e.g., increasing a price by 10%). We could represent such interventions by the modified counterfactual SEM:

$$\left(\begin{bmatrix} Y \\ P \\ X \end{bmatrix} : \mathrm{soft}_Y(P, \alpha)\right) := \begin{bmatrix} f_Y(\alpha(P), X, \epsilon_Y) \\ f_P(X, \epsilon_P) \\ X \end{bmatrix} = \begin{bmatrix} Y(\alpha(P)) \\ P \\ X \end{bmatrix}.$$

As an additional general example, we now consider *fix interventions* that induce single-world intervention graphs (SWIGs).[4]

3: The ideas of constructing counterfactuals go back at least to P. Wright's work in 1928 [7], which involved replacing one structural equation with a different equation to define a counterfactual SEM. Specifically, Wright replaced the supply equation with another one reflecting a multiplicative tariff on the price that producers receive. This intervention is a (multiplicative) soft intervention. Building on P. Wright's work, soft interventions have been widely used in empirical economics (e.g., decomposition analysis of wages to study discrimination, carbon and emission taxes in environmental economics and industrial organization). See also [8, 9] for recent theoretical research in the computer science literature, framed in terms of DAGs and nonlinear ASEMs.

4: The fix intervention was introduced in Heckman and Pinto [4], as an extension of the do operation, and SWIGs were developed by Richardson and Robins [5].

**Fix Interventions and SWIGs.** Instead of removing $P$ from the graph in Figure 7.2, we can split it into two nodes – $P$ and a deterministic node $p$ – where all the outgoing arrows from $P$ are removed. The fixed node $p$ then inherits the outgoing arrows from the original $P$.

The corresponding counterfactual SEM is

$$\left(\begin{bmatrix} Y \\ P \\ X \end{bmatrix} : \mathrm{fix}_Y(P = p)\right) := \begin{bmatrix} f_Y(p, X, \epsilon_Y) \\ P \\ X \end{bmatrix} = \begin{bmatrix} Y(p) \\ P \\ X \end{bmatrix}.$$

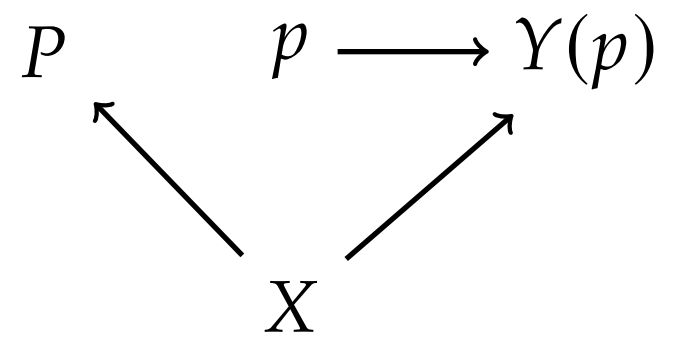

**Figure 7.5:** Causal DAG describing the counterfactual SEM induced by setting $P = p$ in the $Y$ equation in (7.2.1) (formally a SWIG).

The fix intervention merely says that we are setting $P = p$ in the $Y$ equation. Figuratively speaking, it is a "localized do" operation. The variables $Y(p), P$, and $X$ are the counterfactuals generated by this intervention. The intervention does not affect the $P$ and $X$ equations, nor does it affect $\epsilon_Y$ in the $Y$ equation.

The SWIG allows us to immediately see that conditional exogeneity (ignorability) holds:

$$Y(p) \perp\!\!\!\perp P \mid X.$$

Therefore we can identify the counterfactual regression $\mathrm{E}[Y(p) \mid X]$ by the "factual" regression $\mathrm{E}[Y \mid P = p, X]$,

$$\mathrm{E}[Y(p) \mid X] = \mathrm{E}[Y(p) \mid P = p, X] = \mathrm{E}[Y \mid P = p, X],$$

invoking conditional independence and consistency arguments.

The do and fix interventions generate the same counterfactual distribution for $(Y(p), X)$, so the average causal effects of simple interventions coincide in the two approaches. However, the fix intervention creates a triple $(Y(p), X, P)$, which is useful for answering more complicated counterfactual questions.

For example, the counterfactual prediction $\mathrm{E}[Y(0) \mid P = 1]$ tells us what trainees ($P = 1$) would have earned on average, had they not gone through the training program ($p = 0$). In treatment effect analysis, this quantity is crucial for defining the average treatment effects for the treated:

$$\mathrm{E}[Y(1) \mid P = 1] - \mathrm{E}[Y(0) \mid P = 1].$$

Thus, the fix intervention allows us to seamlessly talk about conditional on $P$ counterfactuals:[5]

$$\mathrm{E}\Big[Y(p) \mid P = \bar{p}\Big] := \mathrm{E}\Big[(Y \mid P = \bar{p}) : \mathrm{fix}_Y(P = p)\Big].$$

5: The same statement is formally not true with the do operation in place of the fix operation. Of course, one can also define these conditional counterfactuals by reverting to potential outcomes notation within causal DAGs; see [10].

## 7.3 General Acyclic SEMs and Causal DAGs

We will now turn to generalizing the concepts of the previous section from the TSEM case to general Directed Acyclic Graphs (DAGs) and the corresponding acyclic structural equation models (ASEMs).

## DAGs and Acyclic SEMs via Examples

We now give a sequence of formal definitions, which can be easily understood by looking at just a single example.

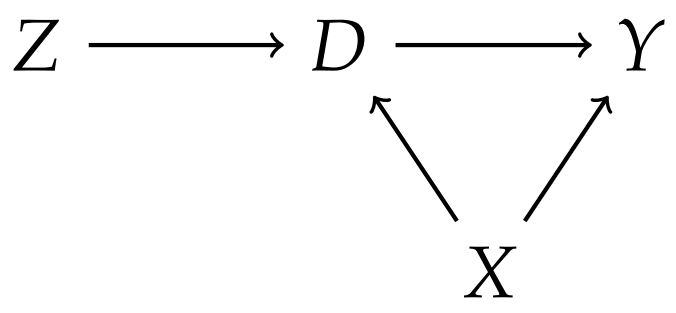

**Figure 7.6:** LS-DAG Example

**Example 7.3.1** (Less Simple DAG (LS-DAG)) A directed acyclic graph (DAG) is a collection of nodes and directed edges with no cycles.

Consider the DAG in Figure 7.6: Here we can *say* that

- $X$ is a parent of its children $D$ and $Y$;
- $D$ and $Y$ are descendants of $Z$;
- There is a directed path from $Z$ to $Y$;
- There are two paths from $Z$ to $X$, but no directed path;
- $D$ is a collider of the path $Z \longrightarrow D \longleftarrow X$;
- $D$ is a noncollider of the path $Z \longrightarrow D \longrightarrow Y$;
- $Y \longleftarrow X \longrightarrow D$ is a backdoor path from $Y$ to $D$.
- There are no cycles (there is no directed path that returns to the same node).

**Example 7.3.2** (ASEM Corresponding to the LS-DAG) A system of triangular structural equations corresponding to Example 7.3.1 is

$$\begin{aligned} Y &:= f_Y(D, X, \epsilon_Y), \\ D &:= f_D(Z, X, \epsilon_D), \\ X &:= \epsilon_X, \\ Z &:= \epsilon_Z, \end{aligned}$$

where $\epsilon_Y$, $\epsilon_X$, $\epsilon_D$, and $\epsilon_Z$ are mutually independent.

Factual distributions in DAG models have a beautiful Markov factorization structure, which allows for a simple representation of the joint distribution of all variables.

**Example 7.3.3** (Factual Law in LS-DAG) Noting the dependences of each variable in the LS-DAG, we can write the joint distribution (density) $\mathrm{p}$ of $Y, D, X, Z$ as

$$\mathrm{p}(y, d, x, z) = \mathrm{p}(y|d, x)\,\mathrm{p}(d|x, z)\,\mathrm{p}(x)\,\mathrm{p}(z).$$

Indeed,

$$\mathrm{p}(y, d, x, z) = \mathrm{p}(y|d, x, z)\,\mathrm{p}(d, x, z),$$

by Bayes' law. Then $\mathrm{p}(y|d, x, z) = \mathrm{p}(y|d, x)$ as the distribution

of $Y$ is independent of $Z$, given its parents $D$ and $X$. Further, $\mathsf{p}(d,x,z) = \mathsf{p}(d|x,z)\,\mathsf{p}(x,z)$, by Bayes' law, and $\mathsf{p}(x,z) = \mathsf{p}(z)\,\mathsf{p}(x)$ by independence.

## General DAGs

The purpose of the rest of this section is to give concise general definitions.

A graph $\mathsf{G}$ is an ordered pair $(V,E)$, where $V = \{1,...,J\}$ is a collection of vertices/nodes and $E$ is a matrix of edges $e_{ij} \in \{0,1\}$ – that is, $E = \{e_{ij} : (i,j) \in V^2\}$.

Given a collection of random variables $X = (X_j)_{j\in V}$, we associate each index $j$ with the name "$X_j$" whenever convenient. If the edge $(i,j)$ is present, namely $e_{ij} = 1$, we read it as

"$X_i \longrightarrow X_j$" or "$X_i$ is an immediate cause of $X_j$."

Consider a strict partial order $<$ on $V$ induced by $E$, where $X_j < X_k$ (we read this as "$X_j$ is determined before $X_k$") means that either $X_j \longrightarrow X_k$ or $X_j \longrightarrow X_{v_1} \longrightarrow ...X_{v_m} \longrightarrow X_k$ is true for some $v_\ell$'s in $V$. A partial ordering of $V$ exists if for each $j$ the statement $X_j < X_j$ is not true. [6] Note that we may interchangeably use random variable names, $X_\ell$, or their indices ,$\ell$, when referring to nodes in the graph.

6: The latter statement means that there are no *cycles*.

**Definition 7.3.1** (DAG) *The graph* $\mathsf{G} = (V,E)$ *is a DAG if the graph has no cycles, that is, if $V$ is partially ordered by the edge structure $E$.*

**Example 7.3.4** (LS-DAG continued) In our example (Example 7.3.1), we had vertices $V = \{1,2,3,4\}$ identified with $Y,D,X,Z$, and the edge set

$$E = \begin{pmatrix} 0 & 0 & 0 & 0 \\ 1 & 0 & 0 & 0 \\ 1 & 1 & 0 & 0 \\ 0 & 1 & 0 & 0 \end{pmatrix}$$

The partial ordering is $X < D, X < Y, Z < D, D < Y$.

**Definition 7.3.2** (Parents, Ancestors, Descendants on a DAG) *The parents of $X_j$ are the set $Pa_j := \{X_k : X_k \longrightarrow X_j\}$. The children* *of $X_j$ are the set $Ch_j := \{X_k : X_j \longrightarrow X_k\}$. The ancestors of $X_j$ are the set $An_j := \{X_k : X_k < X_j\} \cup \{X_j\}$. The descendants of $X_j$ are the set $Ds_j := \{X_k : X_k > X_j\}$.*

**Definition 7.3.3** (Paths and Backdoor Paths on DAGs) *A directed path is a sequence $X_{v_1} \longrightarrow X_{v_2} \longrightarrow ...X_{v_m}$. A non-directed path is a path, where some arrows (but not all) arrows are replaced by $\longleftarrow$. A collider node is a node $X_j$ such that $\longrightarrow X_j \longleftarrow$. A backdoor path from $X_l$ to $X_k$ is an undirected path that starts at $X_l$ and ends with an incoming arrow $\longrightarrow X_k$.*

## From DAGs to ASEMs

Every causal DAG implicitly defines a nonparametric acyclic structural equation model. Thus the two objects are simply different representations or views of the same assumptions on the data generating process and the stochastic potential or counterfactual outcome processes. DAGs are simply a visual depiction of ASEMs and ASEMs are simply a structural equation based expression of DAGs.

**Definition 7.3.4** (ASEM) *The ASEM corresponding to the DAG $\mathsf{G} = (V, E)$ is the collection of random variables $\{X_j\}_{j\in V}$ such that*

$$X_j := f_j(Pa_j, \epsilon_j), \quad j \in V,$$

*where the disturbances $(\epsilon_j)_{j\in V}$ are jointly independent.*

**Definition 7.3.5** (Linear ASEM) *The linear ASEM is an ASEM where the equations are linear:*

$$f_j(Pa_j, \epsilon_j) := f_j' Pa_j + \epsilon_j;$$

*here we identify functions $\{f_j\}$ with coefficient vectors $\{f_j\}$.*

In linear ASEMs we may replace the requirement of independent errors by the weaker requirement of uncorrelated errors.

**Definition 7.3.6** (Structural/Potential Response Processes) *The structural potential response processes for the ASEM corresponding to the DAG $\mathsf{G} = (V, E)$ are given by the structure:*

$$X_j(pa_j) := f_j(pa_j, \epsilon_j), \quad j \in V,$$

*viewed as stochastic processes indexed by the potential parental values $pa_j$.*

**Definition 7.3.7** (Consistency) *The observable variables are generated by drawing $\{\epsilon_j\}_{j\in V}$ and then solving the system of equations for $\{X_j\}_{j\in V}$.*

The stochastic shocks $\{\epsilon_j\}_{j\in V}$ are called exogenous variables, and the variables $\{X_j\}_{j\in V}$ are called endogenous variables. Endogenous variables are determined by the model equations, while exogenous variables are not.

The joint distribution of variables in ASEMs is generally characterized as follows:

**Theorem 7.3.1** (Factual Law via Markovian Factorization) *The general ASEM model, given by $(X_j)_{j\in V}$ with an associated DAG $\mathsf{G}(V, E)$, obeys the following equivalent properties:*

- ***Factorization**: The law admits factorization:*

$$\mathrm{p}(\{x_\ell\}_{\ell\in V}) = \prod_{\ell\in V} \mathrm{p}(x_\ell \mid pa_\ell).$$

- ***Local Markov Property**: All variables are independent of their non-descendants given their parents.*

## Counterfactuals Induced by Interventions

We next discuss counterfactuals generated by interventions. We first consider counterfactuals in the Less Simple DAG example (Example 7.3.1). Note that we use the abbreviation "CF" to denote "counterfactual."

**Example 7.3.5** (CF-ASEM Induced by Do for LS-DAG Example) Consider the ASEM from Example 7.3.1. A counterfactual system induced by $do(D = d)$ is

$$\begin{aligned} &Y(d) := f_Y(X, d, \epsilon_Y),\\ &d,\\ &Z = \epsilon_Z,\\ &X = \epsilon_X, \end{aligned}$$

where $\epsilon_X, \epsilon_Z, \epsilon_Y$ are mutually independent. The corresponding graph, provided in Figure 7.7, is denoted by $\mathsf{G}(d)$.

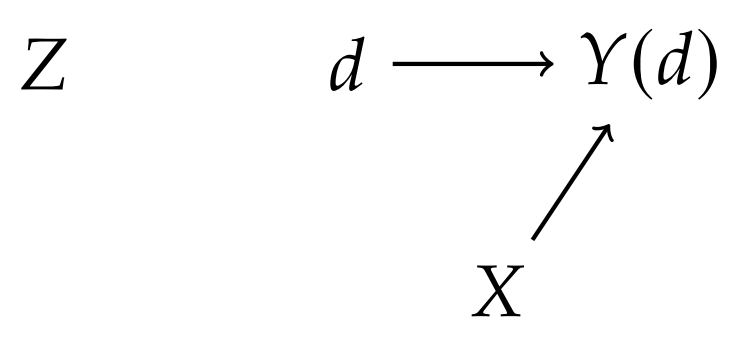

**Figure 7.7:** CF LS-DAG induced by $do(D = d)$ intervention.

**Example 7.3.6** (CF-ASEM Induced by Fix for LSDAG Example) Consider the ASEM from Example 7.3.1. A counterfactual SEM induced by fix($D = d$) takes the following form:

$$\begin{aligned} &Y(d) := f_Y(X, d, \epsilon_Y), \\ &d, \\ &D := f_D(X, Z, \epsilon_D), \\ &Z := \epsilon_Z, \\ &X := \epsilon_X, \end{aligned}$$

where $\epsilon_X$, $\epsilon_Z$, $\epsilon_D$, $\epsilon_Y$ are mutually independent. The corresponding graph, provided in Figure 7.8, is denoted by $\widetilde{\mathsf{G}}(d)$.

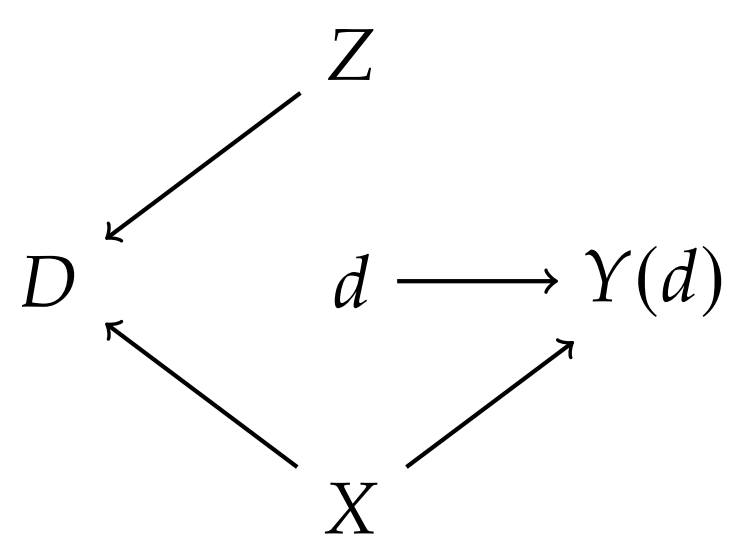

**Figure 7.8:** CF LS-DAG (SWIG) induced by the $fix_Y(D = d)$ intervention.

We now give a more general definition.

**Definition 7.3.8** (Counterfactual ASEM induced by Do Intervention) *The intervention $do(X_j = x_j)$ on an ASEM is said to create the CF-ASEM defined by the modified graph*

$$\mathsf{G}(x_j) = (V, E^*)$$

*and collection of counterfactual variables*

$$(X_k^*)_{k \in V}$$

*where*

- *the edges incoming to the node $j$ are set to zero, namely $e_{ij}^* = 0$ for all $i$,*
- *the remaining edges are preserved, namely $e_{ik}^* = e_{ik}$, for all $i$ and $k \neq j$, and*
- *the counterfactual random variables are defined as*

$$\begin{aligned} X_k^* &:= f_k(Pa_k^*, \epsilon_k), \text{ for } k \neq j. \\ X_j^* &:= x_j \end{aligned}$$

*where $Pa_k^*$ are parents of $X_k^*$ ($k \neq j$) under $E^*$.*

The do intervention modifies the graph $\mathsf{G}$ to $\mathsf{G}(x_j)$ by removing edges. Pearl [10] has described this process as "surgery."[7] We next define the $do$ notation to mean

$$\Big((X_\ell)_{\ell \in V} : \mathrm{do}(x_j)\Big) := (X_\ell^*)_{\ell \in V}.$$

7: This sounds a bit painful.

**Definition 7.3.9** (Counterfactual ASEM induced by Fix Intervention) *The intervention* fix$(X_j = x_j)$ *on an ASEM is said to create the CF-ASEM defined by the modified SWIG*

$$\tilde{\mathsf{G}}(x_j) := (\tilde{V}, \tilde{E}),$$

*and collection of counterfactual variables*

$$(X_k^*)_{k \in V} \cup (X_a^*)$$

*where we split the node $X_j$ into $X_j^* := X_j$ and the new deterministic node $a$*

$$X_a^* := x_j,$$

*where*

- *the node $X_a$ inherits only outgoing edges from $X_j$ and no incoming edges; namely $\tilde{e}_{ai} = e_{ji}$ for all $i$ and $\tilde{e}_{ia} = 0$ for all $i$;*
- *the node $X_j^*$ inherits only incoming edges from $X_j$ and no outgoing edges, namely $\tilde{e}_{ij} = e_{ij}$ for all $i$ and $\tilde{e}_{ji} = 0$ for all $i$;*
- *all the remaining edges are preserved, namely $\tilde{e}_{ik} = e_{ik}$, for all $i$ and $k \neq j$ and $k \neq a$; and*
- *the counterfactual random variables are assigned according to*

$$X_k^* \quad := f_k(Pa_k^*, \epsilon_k), \text{ for } k \neq a,$$

*where $Pa_k^*$ are parents of $X_k^*$ $(k \neq j)$ under $\tilde{E}$.*

Intervention induces new counterfactual distributions for the endogenous variables; see Appendix 7.A for details.

## 7.4 Testable Restrictions and d-Separation

Next we examine the constraints on the data generating process that are implied by a given DAG.

For this we turn to a fundamental theorem in DAGs. We will define the concept of d-separation and prove that d-separation implies conditional independence. This property is typically referred to as a global Markov condition that is implied by the DAG. In order to define this property, we need a few more definitions.

The "d" here denotes "directional" as the direction of arrows in a DAG is important for understanding conditional independence relations; see, e.g., Pearl [10] Chapter 11.

**Definition 7.4.1** (Blocked Paths) *A path $\pi$ is is said to be blocked by a subset of nodes $S$ if and only if*

*(a) $\pi$ contains a chain $i \to m \to j$ or a fork $i \leftarrow m \to j$ such that $m$ is in $S$;*
*(b) Or, $\pi$ contains a collider $i \to m \leftarrow j$, where neither $m$ nor any descendant of $m$ is in $S$.*

*A path that is not blocked is called open.*

In Figure 7.9 the (backdoor) path $Y \leftarrow X \to D$ is blocked by $S = X$.

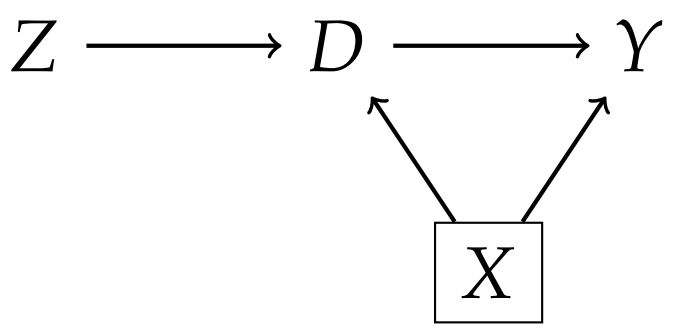

**Figure 7.9:** The path $Y \leftarrow X \to D$ is blocked by conditioning on $X$.

The following definition allows empty sets as conditioning sets.

**Definition 7.4.2** (Opening a Path by Conditioning) *A path containing a collider is opened by conditioning on it or its descendant.*

In Figure 7.10 the path $Y \to C \leftarrow D$ is blocked, but becomes open by conditioning on the collider $S = C$.

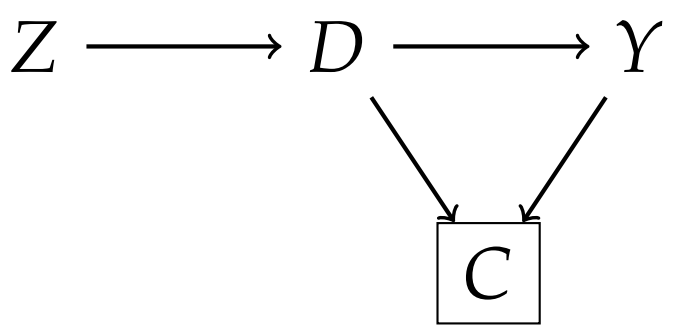

**Figure 7.10:** The path $Y \to C \leftarrow D$ is blocked, but becomes open by conditioning on $C$.

The following defines a key graphical property of DAG, which can be used to deduce key statistical independence restrictions.

**Definition 7.4.3** (d-Separation) *Given a DAG* $\mathsf{G}$, *a set of nodes $S$ d-separates nodes $X$ and $Y$ if nodes in $S$ block all paths between $X$ and $Y$. d-separation is denoted as*

$$(Y \perp\!\!\!\perp_d X \mid S)_{\mathsf{G}}.$$

The following is a fundamental result concerning the conditional independence relations encoded in the graphs.

**Theorem 7.4.1** (Verma and Pearl [11]; Conditional Independence from d-Separation) *d-Separation implies conditional independence:*

- ***Global Markov**: $(Y \perp\!\!\!\perp_d X \mid S)_{\mathsf{G}} \implies Y \perp\!\!\!\perp X \mid S$.*

Figuratively speaking, conditioning on $S$ breaks the information flow between $Y$ and $X$, meaning that $Y$ can't be predicted by $X$, conditional on $S$, and vice versa.

This fundamental result is very intuitive and can be verified directly in simple examples. However, the formal proof is difficult. The reverse implication is not true in general, but is argued to hold "generically" as we discuss in Section 7.5.

**Example 7.4.1** We show a couple of examples illustrating that $d$-separation implies conditional independence:

1. In Figure 7.11, the variables $X$ and $Y$ are d-separated by $S = (Z, U)$, because $S$ blocks all paths between $X$ and $Y$. We also have $Y$ is independent of $X$ conditional on $S$: By the Markov factorization property, $\mathrm{p}(y, x \mid u, z) = \mathrm{p}(y \mid x, z, u)\,\mathrm{p}(x \mid z, u) = \mathrm{p}(y \mid u, z)\,\mathrm{p}(x \mid z, u)$. This equality provides a testable restriction.

2. In Figure 7.12, the variables $X$ and $Y$ are d-separated by $S = Z$, because $S$ blocks all paths between $X$ and $Y$. We also have $Y$ is independent of $X$ conditional on $S$: By the Markov factorization property, $\mathrm{p}(y, x \mid z) = \mathrm{p}(y \mid z)\,\mathrm{p}(x \mid z)$. This equality provides a testable restriction.

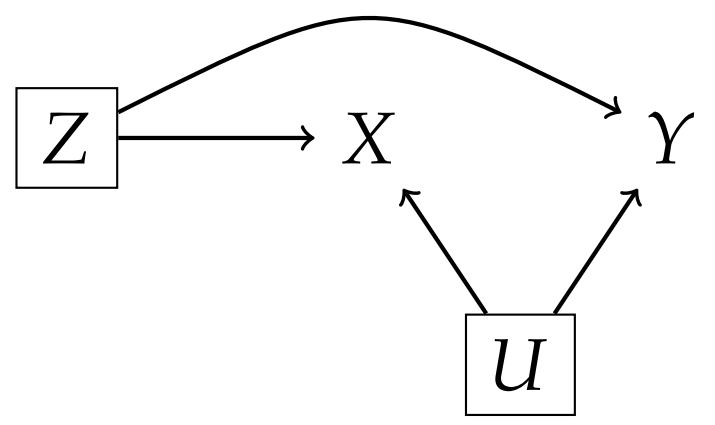

**Figure 7.11:** Example of d-separation.

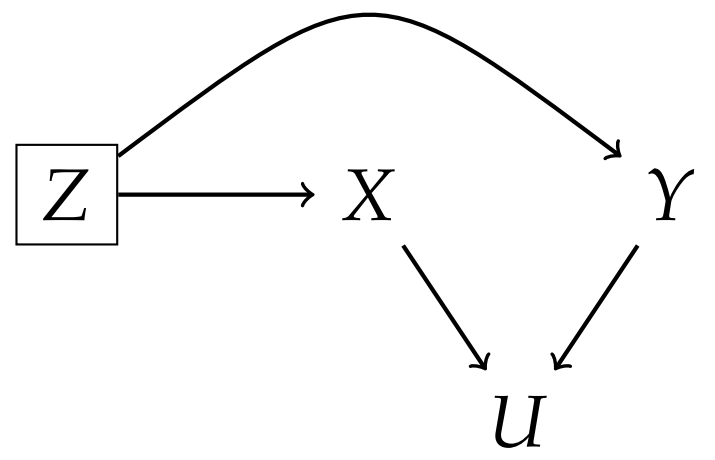

**Figure 7.12:** Example of d-separation.

These testable restrictions are called exclusion restrictions in econometrics because

$$Y \perp\!\!\!\perp X \mid Z \text{ is equivalent to } \mathrm{p}(y \mid x, z) = p(y \mid z), \tag{7.4.1}$$

where the equivalence follows from Bayes' law. In particular,

$$\mathrm{E}[g(Y) \mid X, Z] = \mathrm{E}[g(Y) \mid Z] \tag{7.4.2}$$

for any bounded function $g$ of $Y$. (7.4.2) means that $X$ is excluded from the best predictor of $g(Y)$ using $X$ and $Z$. There are many tests of such restrictions available in the literature.[8] Perhaps one of the reasons for which there are many such tests is that conditional independence testing is formally impossible; see [12]. In practice, the formal impossibility means that any test must be carefully crafted to target specific features within a statistical model as no generic, uniformly valid testing procedure exists.

8: E.g, the reader can search Google Scholar for conditional independence tests, exclusion restrictions tests, or conditional moment tests.

With specific structure provided, conditional independence testing can be relatively straightforward. For example, it reduces to testing hypotheses about linear regression coefficients within a linear ASEM.

**Implementation of Tests in Linear ASEMs.** Consider the hypothesis that $Y$ is independent of $X$, given $Z$. In linear ASEMs, we can test this hypothesis by testing whether the coefficient $\alpha = 0$ in the projection equation

$$Y = \alpha' X + \beta' Z + \epsilon, \epsilon \perp Z.$$

We can perform this test easily with the tools we've developed so far. See R: Dagitty Notebook and Python: Pgmpy Notebook for an example.

Such tests are of course available under structures that are more general than in linear model. For example, [12] exploits debiased ML ideas (introduced in Chapter 4 and further developed in Chapter 10 in this text) to set up testing of exclusion restrictions in some nonlinear models.

**Remark 7.4.1** (Equivalence of Local and Global Markov Properties) The local Markov property, the Markov factorization, and the global Markov property are equivalent (Pearl [10]). Therefore, one can use any of these properties to set up tests of the validity of the Markov structure.

## 7.5 Falsifiability and Causal Discovery*

Here, we provide a brief discussion of whether it is possible to falsify (reject) a causal structure encoded by a DAG with data.

### Equivalence Classes and Falsifiability

**Definition 7.5.1** (Equivalence Classes) *The class of DAGs that induce the same joint distribution of variables is called an equivalence class, and members of an equivalence class may be described as Markov equivalent. DAGs that produce the same joint distribution variables cannot be distinguished from each other.*

Pearl [10] shows that the equivalence class of a DAG is given by reversing any edges such that any such reversal does not destroy existing or create new *v-structures*: converging arrows whose tails are not connected by an edge.

The equivalence classes of a DAG are called PDAGs (partially directed acyclic graphs). We plot them by erasing arrowheads that can be oriented in the opposite direction without adding or removing v-structures. We illustrate PDAGs in Figures 7.13 and 7.14.

Figure 7.13 starts with the triangular structural equation model from Example 7.2.1. Figure (a) is the original DAG implied by the model. To produce the PDAG, shown in (b), we consider reversing each of the arrows from $X$ to $Y$, $X$ to $P$, and $P$ to $Y$. Because each of the nodes is connected, there are no v-structures in the original DAG, and there is similarly no possible reversal that could add a v-structure. As such, the PDAG is simply the original DAG with all arrows removed. In this case, the DAG structure produces no testable implications.

Figure 7.14 starts from a more elaborate DAG than the simple TSEM. We refer to this DAG as "Pearl's Example" because it shows up repeatedly as an illustration in Pearl's work; see, e.g., [2]. Figure (a) is the original DAG defining the model. We produce the PDAG in (b) by considering the reversal of all combinations of arrows connecting the eight nodes. Here, there are only two reversals, changing $Z_2 \longrightarrow X_3$ to $Z_2 \longleftarrow X_3$ and changing $Z_1 \longrightarrow X_1$ to $Z_1 \longleftarrow X_1$, that do not destroy any existing v-structures or create new v-structures. For example, reversing the arrow $Z_2 \longrightarrow X_2$ would destroy the v-structure $Z_2 \longrightarrow X_2$ and $Z_1 \longrightarrow X_2$. As such, the PDAG in (b) is almost identical to the DAG in (a) with the exception that the arrows between $Z_2$ and $X_3$ and between $Z_1$ and $X_1$ have been removed. In this case, the DAG encodes a model which includes exclusion restrictions or testable implications and is potentially falsifiable.

**Remark 7.5.1** (Falsifiability) The edge matrix $E$ of a graph is *triangular* if rows of $E$ can be rearranged to have only 1's below the diagonal. In the absence of any further restrictions, an ASEM with graph $\mathsf{G} = (V, E)$ has testable implications if $E$ is not triangular. If $E$ is triangular, then any law $\mathsf{p}$ of any arbitrary collection of random variables $(X_j)_{j \in V}$ indexed by $V$ can be factorized as

$$\mathsf{p}(\{x\}_{j\in V}) = \prod_{j \in V} \mathsf{p}(x_j \mid pa_j).$$

With population data we have $\mathsf{p}$ and can check if it factorizes according to $V$. If matrix $E$ is triangular, $\mathsf{p}$ always obeys the factorization property. This is to say that there are no exclusion restrictions in the model.

**Example 7.5.1** (TSEM continued) In the TSEM example (Example 7.2.1, we have vertices $V = \{1, 2, 3\}$ identified with

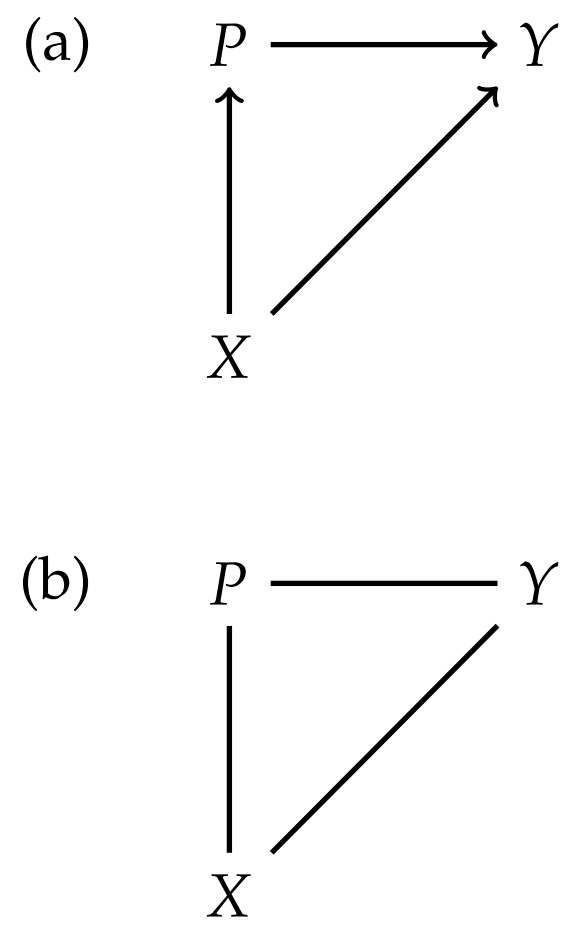

**Figure 7.13:** The original DAG, (a), and the equivalence class or PDAG, (b), for the TSEM example, Example 7.2.1. The undirected edges in the PDAG mean that they can be directed in any direction as long as this does not create a cycle. In empirical analysis directionality must therefore be deduced and assumed from the context.

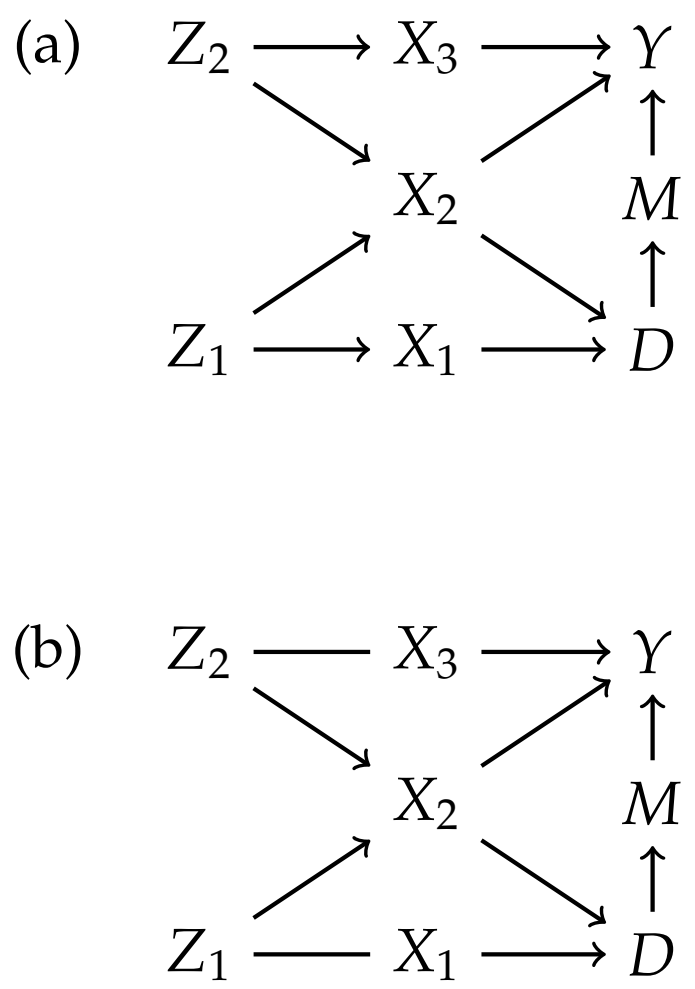

**Figure 7.14:** The original DAG, (a), and the equivalence class or PDAG, (b), for the Pearl's Example. The undirected edges in the PDAG mean that they can be directed in any direction as long as this does not create a cycle. Only two edges can be reoriented here.

$Y, P, X$ and the "triangular" edge set

$$E = \begin{pmatrix} 0 & 0 & 0 \\ 1 & 0 & 0 \\ 1 & 1 & 0 \end{pmatrix}.$$

In the absence of other assumptions, the corresponding TSEM implies no falsifiable restrictions. The equivalence class of the DAG model for this case is generated by rearranging the rows of $E$ in 3! ways, which is equivalent to rearranging the names $(Y, P, X)$ for the nodes.

**Example 7.5.2** (Pearl's Example) The DAG given in Figure 7.14 has vertices $V = \{1, ..., 8\}$ identified with $Y$, $M$, $D$, $X_1$, $X_2$, $X_3$, $Z_1$ ,$Z_2$ and the edge set

$$E = \begin{pmatrix} 0 & 0 & 0 & 0 & 0 & 0 & 0 & 0 \\ 1 & 0 & 0 & 0 & 0 & 0 & 0 & 0 \\ 0 & 1 & 0 & 0 & 0 & 0 & 0 & 0 \\ 0 & 0 & 1 & 0 & 0 & 0 & 0 & 0 \\ 1 & 0 & 1 & 0 & 0 & 0 & 0 & 0 \\ 1 & 0 & 0 & 0 & 0 & 0 & 0 & 0 \\ 0 & 0 & 0 & 1 & 1 & 0 & 0 & 0 \\ 0 & 0 & 0 & 0 & 1 & 1 & 0 & 0 \end{pmatrix}.$$

This edge set cannot be rearranged to have only ones below the diagonal. The DAG in this case has testable implications, and the equivalence class of the DAG model can only involve changing edges between $Z_1$ and $X_2$ and between $Z_2$ and $X_3$.

## Faithfulness and Causal Discovery

Given that DAGs effectively encode conditional independence relations, it is tempting to try to infer conditional independence directly from the data. *Causal discovery* refers to methods that indeed attempt to learn conditional independence relationships from data with one application being attempting to recover causal structures. The possibility of recovering causal structures perfectly from the population data critically relies on the concept of faithfulness.

Recall that d-separation implies conditional independence, but the reverse implication

$$Y \perp\!\!\!\perp X|S \implies (Y \perp\!\!\!\perp_d X|S)_{\mathsf{G}} \tag{7.5.1}$$

is not true in general. If we restrict attention to the set of distributions $p$ of random variables associated with graph G such that implication (7.5.1) holds, we are said to impose the *faithfulness* assumption on p.

**Example 7.5.3** (Unfaithfulness) A trivial example is the DAG

$$X \to Y$$

where

$$Y := \alpha X + \epsilon_Y; \quad X := \epsilon_X;$$

with $\epsilon_X$ and $\epsilon_Y$ independent standard normal variables. Consider $S$ to be the empty set. In this model we have that $Y \perp\!\!\!\perp X$ when $\alpha = 0$, but $Y$ and $X$ are not d-separated in the DAG $X \to Y$. The distribution p of $(Y, X)$ corresponding to $\alpha = 0$ is said to be unfaithful. However, the exceptional point $\alpha = 0$ has a measure 0 on the real line, so this exception is said to be non-generic.

The observation about the simple example above generalizes: If probabilities p themselves are viewed as generated by Nature as a draw from a continuum P, where each p ∈ P factorizes according to G, then the set of models where the reverse implication (7.5.1) does not hold has measure zero. This observation motivates the argument that the faithfulness assumption is a weak requirement; that is, a given p is "very unlikely" to be unfaithful.

**Remark 7.5.2** (Causal Discovery) The use of the faithfulness assumption should allow us to discover the equivalence class of the true DAG from the population distribution p: We can compute all valid conditional independence relations and then discover the equivalence class of DAGs. See, for example, the PC algorithm [13] for an explicit causal discovery algorithm and the review provided in [14]. We can then apply contextual knowledge to further orient the edges of the graph.

Even though the set of unfaithful distributions has measure zero, the neighborhood of this set may not be small in high-dimensional graphs, which creates difficulty in inferring the DAG structure from an estimated version p̂.

**Example 7.5.4** (Unfaithfulness Continued) In the trivial example above, suppose that we have that $\hat{\alpha} = .1$ and $\hat{\alpha} \sim N(\alpha, \sigma^2)$ where $\sigma = .1$. Then we can't be sure whether $\alpha = 0$, $\alpha = .1$, or $\alpha$ equals any other number, though say a 95% confidence interval would have $\alpha$ between $-.1$ and $.3$. Therefore, we can't be sure whether the true model is

$$X \to Y \text{ or } X \quad Y.$$

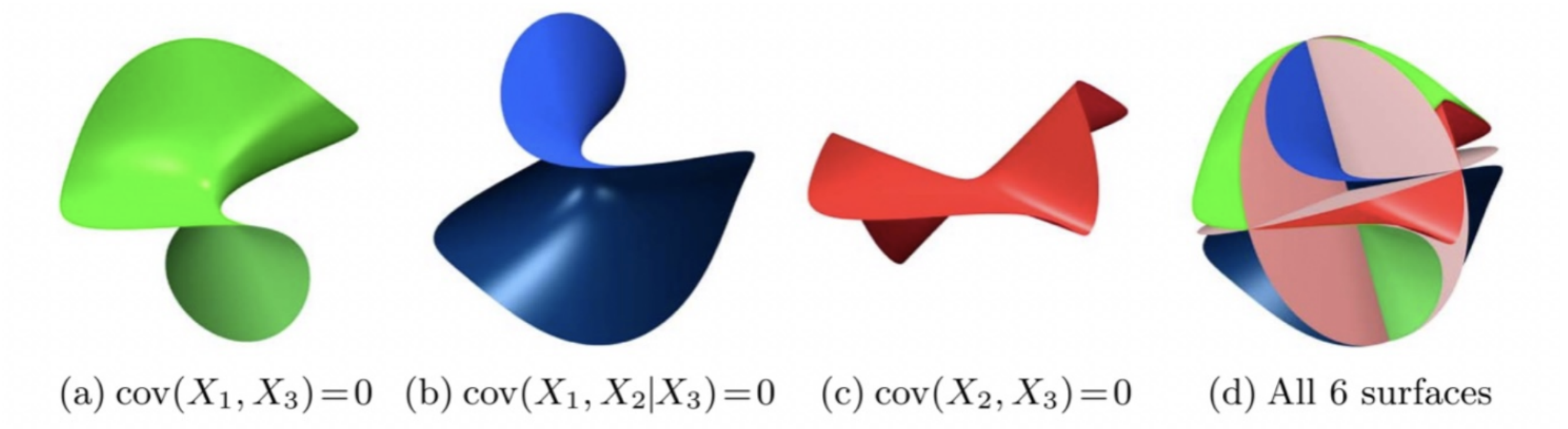

**Figure 7.15:** Uhler et. al [15]: A set of "unfaithful" distributions p in the simple triangular Gaussian SEM/DAG: $X_1 \to X_2, (X_1, X_2) \to X_3$. The set is parameterized in terms of the covariance of $(X_1, X_2, X_3)$. The right panel shows the set of unfaithful distributions, and the three other panels show 3 of 6 components of the set. Each of the cases corresponds to the non-generic case which would make faithfulness fail, leading to discovery of the wrong DAG structure. While the exact setting where faithfulness would fail is non-generic, there are many distributions that are "close" to these unfaithful distributions. This observation means that, in finite samples, we are not able distinguish models that are close to the set of unfaithful distributions from unfaithful distributions and may thus also discover the wrong DAG structure and correspondingly draw incorrect causal conclusions.

Informally speaking, it is impossible to discover the true graph structure in this example when $\alpha \approx 0$. In econometrics jargon, this statement amounts to saying that we can't distinguish exact exclusion restrictions from "approximate" exclusion restrictions.

Thus, it is hard to distinguish exact independence from approximate independence with finite data. In high-dimensional graphs, the possibility that $\hat{p}$ lands in the "near-unfaithful" regions can be substantial, as Uhler et. al.[15]'s analysis shows.

See Uhler et al's [15] figure; reproduced in Figure 7.15.

The observations above motivate a form of sensitivity analysis – e.g., Conley et al. [16] – where one replaces exact exclusion restrictions by approximate exclusion restrictions that can't be distinguished from exact exclusion restrictions and examines the sensitivity of causal effect estimates.

## Notebooks

- R: Dagitty Notebook employs the R package "dagitty" to analyze Pearl's Example (introduced in Figure 7.14) as well as simpler ones. Python: Pgmpy Notebook employs the analogue with Python package "pgmpy" and conducts the same analysis. Both packages automatically list all conditional independence in a DAG; these are obtained by using the graphical d-separation criterion. We then go ahead and test those restrictions assuming a linear ASEM structure. The notebook also illustrates the analysis from the next chapter.

- R: Dosearch Notebook employs the R-package "dosearch" to analyze Pearl's Example (introduced in Figure 7.14). This package automatically finds identification answers to causal queries, allowing us to also answer these types of queries under different data sources, sample selection, and other deviations from the standard framework. Python: Dosearch Notebook does the same thing by loading the R "dosearch" package into Python.

## Additional resources

- Dagitty.Net is an excellent online resource where you can plot and analyze causal DAG models online. It contains many interesting examples of DAGs used in empirical analysis in various fields.

- Causalfusion.Net is another excellent online resource where you can plot and analyze causal DAG models. This resource covers many different deviations from the standard framework.

## Study Problems

The study problems ask learners to analyze Pearl's Example (introduced in Figure 7.14). The provided notebooks are a useful starting point for answering these questions.

Recall that Pearl's Example is structured as follows:

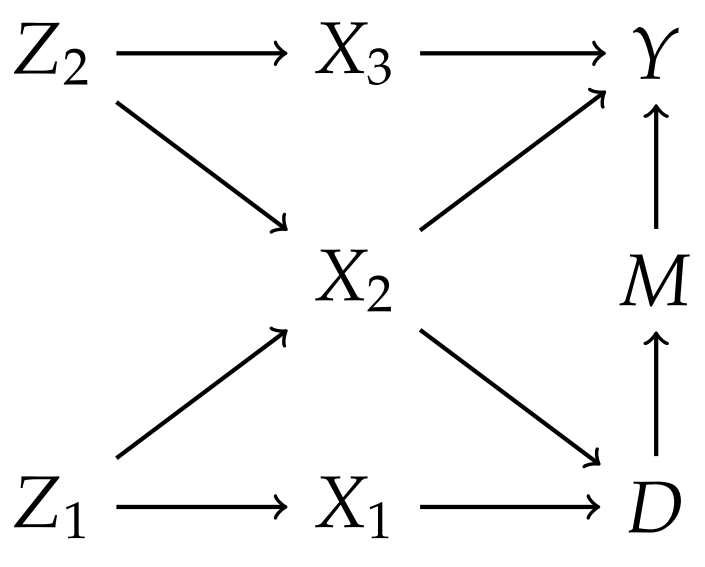

**Figure 7.16:** Pearl's Example

1. Consider Pearl's Example and answer the following questions. The best way to answer this question is to use computational packages (but please explain the principles the package is using).

   a) What are the testable implications of the assumptions embedded in the model? Hint: The testable implications are derived from the d-separation criterion.
   b) Assume that only variables $D$, $Y$, $X_2$ and $M$ are measured, are there any testable implications?
   c) Now assume only $D, Y$, and $X_2$ are measured. Are there any testable implications?
   d) Now assume that all of the variables but $X_2$ (7 in total) are measured. Are there any testable restrictions?
   e) Assume that an alternative model, competing with Model 1, has the same structure, but with the $X_2 \to D$ arrow reversed. What statistical test would distinguish between the two models?

2. Work through the proof that d-separation implies conditional independence in Section 7.C. Supply the steps of the proof that were left as a homework or reading exercise.

## 7.A Counterfactual Distributions★

Interventions induce new counterfactual distributions for endogenous variables. We can readily compute these distributions from the definitions of interventions, as illustrated in the following for the do intervention.

**Example 7.A.1** (Counterfactual Law for Do Intervention in LS-DAG (Example 7.3.1)) We can write the counterfactual distribution of $Y(d), Z, X$ in terms of the factual distribution as

$$\mathsf{p}(y, z, x : \mathrm{do}(d)) = \mathsf{p}(y|d, x)\, \mathsf{p}(z)\, \mathsf{p}(x).$$

Indeed,

$$\mathsf{p}(y, z, x : \mathrm{do}(d)) = \mathsf{p}(y|z, x : do(d))\, \mathsf{p}(z, x : do(d)),$$

by definition and Bayes' law. We also have $\mathsf{p}(y|z, x : do(d)) = \mathsf{p}(y|d, x)$ and $\mathsf{p}(z, x : do(d)) = \mathsf{p}(z, x)$ by the definition of the counterfactual ASEM, and $\mathsf{p}(z, x) = \mathsf{p}(z)\, \mathsf{p}(x)$ by indepen-dence of $Z$ and $X$.

**Theorem 7.A.1** (Counterfactual Law Induced by the Do Intervention) *The induced law* $\mathsf{p}_{X^*}$ *of the counterfactual variables* $X^* = (X^*_\ell)_{\ell \in V \setminus j}$ *induced by* $do(X_j = x_j)$ *can be stated in terms of the factual law as follows:*

$$\mathsf{p}(\{x_\ell\}_{\ell \in V \setminus j} : \mathrm{do}(x_j)) := \mathsf{p}_{X^*}(\{x\}_{\ell \in V \setminus j}) = \prod_{\ell \in V \setminus j} \mathsf{p}(x_\ell \mid pa^*_\ell),$$

*where* $\{x\}_{\ell \in V \setminus j}$ *denotes the point where the density function is evaluated,* $pa^*_j$ *denotes the parental values under the new edge structure, and* $\mathsf{p}$ *denotes the factual law.*

The result follows immediately from the Markov factorization property and the definition of counterfactuals under the do intervention. This characterization is interesting in its own right, because it can be used for identification and inference on the counterfactual laws directly, provided that we are willing to model the distribution of the variables. The use of Bayesian methods can be fruitful for this purpose.

These type of formulas are often called "g-formulas" and first appeared in the work [17] of James Robins in 1986 (using another "tree-based" form of causal graphs).

## 7.B Review of Conditional Independence

The following lemma reviews various ways in which conditional independence can be established.

**Lemma 7.B.1** (Equivalent Forms of Conditional Independence) *Variables $X$ and $Y$ are conditionally independent given $Z$ if and only if one of the following conditions is met:*

1. $\mathrm{p}(x \mid y, z) = \mathrm{p}(x \mid z)$ *if* $\mathrm{p}(y, z) > 0$.
2. $\mathrm{p}(x \mid y, z) = f(x, z)$ *for some function* $f$.
3. $\mathrm{p}(x, y \mid z) = \mathrm{p}(x \mid z)\mathrm{p}(y \mid z)$ *if* $\mathrm{p}(z) > 0$.
4. $\mathrm{p}(x, y \mid z) = f(x, z)g(y, z)$ *for some functions* $f$ *and* $g$.
5. $\mathrm{p}(x, y, z) = \mathrm{p}(x \mid z)\mathrm{p}(y \mid z)\mathrm{p}(z)$ *if* $\mathrm{p}(z) > 0$.
6. $\mathrm{p}(x, y, z) = \mathrm{p}(x, z)\mathrm{p}(y, z)/\mathrm{p}(z)$ *if* $\mathrm{p}(z) > 0$.
7. $\mathrm{p}(x, y, z) = f(x, z)g(y, z)$ *for some functions* $f$ *and* $g$.

As a reading exercise prove the equivalence of (1) and (2), of (1) and (7), and of any other pair.

## 7.C Theoretical Details of d-Separation*

Here we explain why d-separation implies conditional independence.[9]

**Lemma 7.C.1** (Easy Form of d-Separation) *Let* $\mathsf{X}, \mathsf{Y}$, *and* $\mathsf{Z}$ *be three disjoint sets of variables in an ASEM such that their union is an ancestral set, that is, for any* $X \in \mathsf{X} \cup \mathsf{Y} \cup \mathsf{Z}$ *and* $X' < X$ *we have* $X' \in \mathsf{X} \cup \mathsf{Y} \cup \mathsf{Z}$. *If* $\mathsf{Z}$ *d-separates* $\mathsf{X}$ *and* $\mathsf{Y}$, *then*

$$\mathsf{X} \perp\!\!\!\perp \mathsf{Y} \mid \mathsf{Z}.$$

*Proof.* Let $\mathsf{Z}_1$ be the set of nodes in $\mathsf{Z}$ that have parents in $\mathsf{X}$. And let $\mathsf{Z}_2 = \mathsf{Z} \backslash \mathsf{Z}_1$.

Because $\mathsf{Z}$ d-separates $\mathsf{X}$ and $\mathsf{Y}$, we have that (see Figure 7.17):

- For any $W \in \mathsf{X} \cup \mathsf{Z}_1$, $Pa_W \subseteq \mathsf{X} \cup \mathsf{Z}$;[10]
- For any $W \in \mathsf{Y} \cup \mathsf{Z}_2$, $Pa_W \subseteq \mathsf{Y} \cup \mathsf{Z}$.[11]

Let $\mathsf{U}$ denote the set of variables not included in $\mathsf{X}$, $\mathsf{Y}$, or $\mathsf{Z}$. We then obtain a factorization

$$\begin{aligned} \mathrm{p}(\mathsf{x}, \mathsf{z}, \mathsf{y}) &= \int \prod_{W \in \mathsf{U} \cup \mathsf{X} \cup \mathsf{Y} \cup \mathsf{Z}} \mathrm{p}(w \mid Pa_W = pa_W) d\mathsf{u} \\ &= \int \prod_{W \in \mathsf{U}} \mathrm{p}(w \mid Pa_W = pa_W) d\mathsf{u} \\ &\quad \times \prod_{W \in \mathsf{X} \cup \mathsf{Z}_1} \mathrm{p}(w \mid Pa_W = pa_W) \\ &\quad \times \prod_{W \in \mathsf{Z}_2 \cup \mathsf{Y}} \mathrm{p}(w \mid Pa_W = pa_W), \end{aligned}$$

where in the last equality we used the fact that $\mathsf{u}$ does not appear at all in the second and third factors, since $\mathsf{X} \cup \mathsf{Y} \cup \mathsf{Z}$ is ancestral. Moreover, the second factor is a function of $\mathsf{x}$ and $\mathsf{z}$ alone and the third factor is a function of $\mathsf{y}$ and $\mathsf{z}$ alone. The integral is 1 by total probability.[12] It follows that $\mathsf{X} \perp\!\!\!\perp \mathsf{Y} \mid \mathsf{Z}$.[13] □

Now we restate the main claim we'd like to demonstrate, which is that d-separation implies conditional independence.

**Global Markov**. Let $X$ and $Y$ be two variables and $\mathsf{Z}$ be a set of variables that does not contain $X$ or $Y$. If $\mathsf{Z}$ d-separates $X$ and $Y$, then

$$X \perp\!\!\!\perp Y \mid \mathsf{Z}$$

*Proof of Theorem 7.4.1.*

9: We follow the proof sketch presented in Nevin L. Zhang's lecture notes, but rely on ASEMs to simplify some arguments and supply a proof for a key claim.

10: Suppose that any such node has a parent in $\mathsf{Y}$. If it were a node in $\mathsf{X}$, then we get a violation of d-separation. If it were a node in $\mathsf{Z}_1$, then we have that $\mathsf{Z}_1$ has one parent in $\mathsf{X}$ and one parent in $\mathsf{Y}$ and therefore it is a collider that was included in $\mathsf{Z}$, violating d-separation.

11: Suppose that any such node has a parent in $\mathsf{X}$. By the definition of $\mathsf{Z}_1$ it has to be a node in $\mathsf{Y}$. But then we have that a node in $\mathsf{Y}$ has a parent in $\mathsf{X}$, violating d-separation.

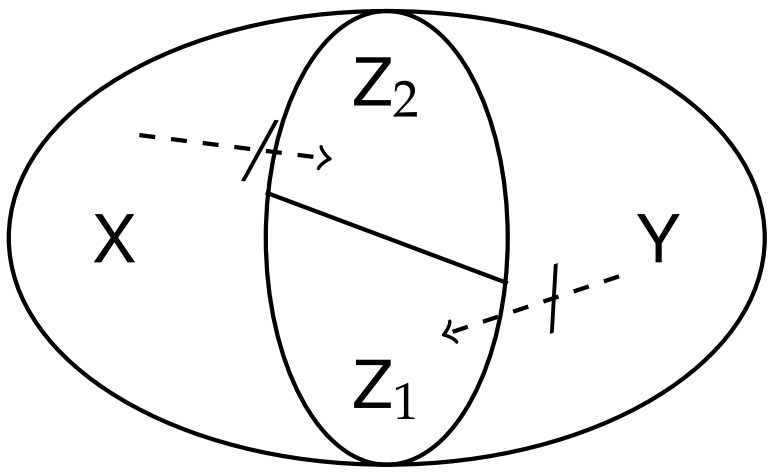

**Figure 7.17:** Pictorial representation of key argument in Lemma 7.C.1.

12: Prove this as a reading exercise by integrating over the variables in $\mathsf{U}$ in reverse order with respect to the DAG ordering.

13: Prove this as a reading exercise, i.e., prove bullet (7) of Lemma 7.B.1.

Let $\mathsf{X}$ be the set of all ancestors of $\{X, Y\} \cup \mathsf{Z}$ that are *not* d-separated from $X$ by $\mathsf{Z}$. Let $\mathsf{Y}$ be the set of all ancestors of $\{X, Y\} \cup \mathsf{Z}$ that are neither in $\mathsf{X}$ nor in $\mathsf{Z}$.

**Key Claim:** The set $\mathsf{Z}$ d-separates the sets $\mathsf{X}$ and $\mathsf{Y}$.

The claim follows from the careful use of the definition of d-separation, and is proven below.

Given the key claim, Lemma 7.C.1 implies that $\mathsf{X} \perp\!\!\!\perp \mathsf{Y} \mid \mathsf{Z}$, since $\mathsf{X} \cup \mathsf{Y} \cup \mathsf{Z}$ is ancestral by its exhaustive construction. This implies that there must exist functions $f(\mathsf{x}, \mathsf{z})$ and $g(\mathsf{z}, \mathsf{y})$ such that

$$p(\mathsf{x}, \mathsf{z}, \mathsf{y}) = f(\mathsf{x}, \mathsf{z}) g(\mathsf{z}, \mathsf{y}).$$

Since $X$ is in $\mathsf{X}$ and $Y$ in $\mathsf{Y}$, the conclusion is reached.[14] □

14: Prove this explicitly, as a reading exercise, by integrating over all variables in $\mathsf{X} \setminus \{X\}$ and $\mathsf{Y} \setminus \{Y\}$ and invoking Lemma 7.B.1.

*Proof of the Key Claim.* Suppose that $\mathsf{Z}$ does not d-separate the sets $\mathsf{X}$ and $\mathsf{Y}$ and that there exists a node $X' \in \mathsf{X}$ which is not d-separated from some node $Y' \in \mathsf{Y}$. Thus, there is an open path $X \text{ - - } X'$,[15] and an open path $X' \text{ - - } Y'$. Consider the concatenation of these two paths. If $X'$ is not a collider on this concatenated path, then the path $X \text{ - - } X' \text{ - - } Y'$ is also open, and therefore $X$ is not d-separated from $Y'$, which is in contradiction with the definition of $\mathsf{X}$ and $\mathsf{Y}$. Thus $X'$ has to be a collider on this concatenated path. Moreover, note that since we are only restricting our analysis to the ancestral set $An_{\{X,Y\} \cup \mathsf{Z}}$, we have that $X'$ must be an ancestor of either $\mathsf{Z}$ or $Y$ or $X$:

15: In this proof, we denote with $U \text{ - - } V$ a path from a node $U$ to a node $V$ and with $U \dashrightarrow V$ a directed path from $U$ to $V$.

If $X'$ is an ancestor of some node in $\mathsf{Z}$ then the path $X \text{ - - } X' \text{ - - } Y'$ is again open, leading to a contradiction with the definition of $\mathsf{X}$ and $\mathsf{Y}$.

If $X'$ is an ancestor of $Y$, then there is a directed path $X' \dashrightarrow Y$. If that path is open, then there is an open path $X \text{ - - } X' \dashrightarrow Y$, violating the fact that $\mathsf{Z}$ was d-separating $X$ from $Y$. For the path to be closed, it must be that some node $Z \in \mathsf{Z}$ is on the path. However, in this case $X'$ is an ancestor of a node in $\mathsf{Z}$, which has already been excluded.

Finally, if $X'$ is an ancestor of $X$, then there exists a directed path $X' \dashrightarrow X$. This path also has to be open, as if a node in $\mathsf{Z}$ existed on that path, then $X'$ would be an ancestor of a node in $\mathsf{Z}$, which has been excluded. However, in this case, we have an open path $Y' \text{ - - } X' \dashrightarrow X$, from $Y'$ to $X$, which violates the definition of $\mathsf{X}$ and $\mathsf{Y}$. □

# 8 Valid Adjustment Sets from DAGs

"if 'good' is taken to mean 'best' fit, it is tempting to include anything in $x$ that helps predict [treatment]"

– Jeffrey Wooldridge [1].

DAGs give us an intuitive approach to take domain knowledge and turn it into an identification strategy. In this section, we focus on identification by conditioning and discuss graphical criteria that lead to the construction of valid adjustment sets for the identification of average causal effects via regression adjustment. We also discuss how graphical criteria can help us differentiate between "good" and "bad" controls.

## 8.1 Valid Adjustment Sets

Consider any variable $D$ of an ASEM as a treatment of interest and any of its descendants $Y$ as an outcome of interest. An adjustment set $S$ is said to be valid for identification of the causal effect of $D$ on $Y$ if the conditional exogeneity/ignorability condition holds

$$Y(d) \perp\!\!\!\perp D \mid S.$$

In what follows, we present an exhaustive (complete) approach for finding valid adjustment sets by using SWIGs.

> We write down the counterfactual SWIG induced by the
>
> $$\text{fix}(D = d)$$
>
> intervention, which operates on all structural equations defining the descendants of $D$ by setting $D = d$ in these equations.
>
> Then, if we have that the potential outcome $Y(d)$ is $d$-separated from the (policy) variable $D$ by a set of variables $S$, conditional exogeneity/ignorability holds:
>
> $$Y(d) \perp\!\!\!\perp D \mid S.$$

Given that conditional exogeneity/ignorability holds, we can identify counterfactual expectations,

$$\mathrm{E}[Y|S = s : \mathrm{do}(d)] := \mathrm{E}[Y(d)|S = s],$$

from expectations of observed variables,

$$\mathrm{E}[Y|S = s, D = d],$$

provided that the positivity condition $p(s, d) > 0$ holds. The agreement between counterfactual and conditional expectations follows because

$$\mathrm{E}[Y(d)|S = s] = \mathrm{E}[Y(d)|D = d, S = s]$$

by exogeneity and

$$\mathrm{E}[Y(d)|D = d, S = s] = \mathrm{E}[Y|D = d, S = s]$$

by consistency.

We can recover unconditional counterfactual means by integration:

$$\mathrm{E}[Y : \mathrm{do}(d)] := \mathrm{E}[Y(d)] = \mathrm{E}[\mathrm{E}[Y|S, D = d]],$$

provided that the positivity condition $p(s, d) > 0$ for each $s$ in the support of $S \mid D = d$ holds.

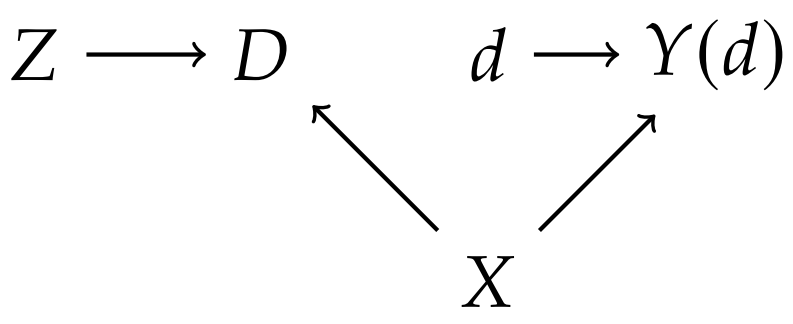

**Figure 8.1:** CF LS-DAG induced by $\mathrm{fix}(D = d)$ intervention.

**Example 8.1.1** (Identification in LS-DAG.) In the SWIG graph in Figure 8.1 corresponding to the LS-DAG model from Example 7.3.1, we see that either $S = X$ or $S = (X, Z)$ d-separates $Y(d)$ from $D$. Therefore, either choice of $S$ provides a valid adjustment set for identifying counterfactual predictions. Here conditioning on $Z$ is not necessary, though we maintain robustness with respect to the presence of a directed edge from $Z$ to $Y$ by including $Z$ in the conditioning set.

We can identify the entire conditional distribution

$$\mathrm{P}(Y(d) \leq t \mid S = s)$$

from the conditional distribution

$$\mathrm{P}(Y \leq t \mid D = d, S = s).$$

We achieve identification of the distribution by replacing $Y$ with $1(Y < t)$ in all previous statements and applying the same arguments for each $t \in \mathbb{R}$. The unconditional distribution of potential outcomes is retrieved by integrating out $S$:

$$\mathrm{P}(Y(d) \leq t) := \mathrm{E}[\mathrm{P}(Y(d) \leq t \mid S)].$$

The following theorem, essentially due to [2], records the discussion formally.

**Theorem 8.1.1** (A Complete Criterion for Identification by Conditioning) *Consider any ASEM with DAG* $\mathsf{G}$. *Let us re-label a policy node* $X_j$ *as* $D$, *and let* $Y$, *an outcome of interest, be any other descendant of* $D$.

*Consider a SWIG DAG* $\tilde{\mathsf{G}}(d)$ *which is induced by the* $\mathrm{fix}(D = d)$ *intervention. Consider any other subset of nodes* $S$ *that appears in both* $\mathsf{G}$ *and* $\tilde{\mathsf{G}}(d)$, *such that*

$$Y(d) \text{ is d-separated from } D \text{ by } S \text{ in } \tilde{\mathsf{G}}(d).$$

▶ *Then the following conditional exogeneity/ignorability holds:*

$$Y(d) \perp\!\!\!\perp D \mid S.$$

▶ *Then*

$$\mathrm{E}[Y(d)|S = s] = \mathrm{E}[Y \mid D = d, S = s]$$

*holds for all s such that* $p(d, s) > 0$.

**Example 8.1.2** (Pearl's Example) Consider the DAG in Figure 8.2, which we introduced as Pearl's Example in Figure 7.14, and the corresponding ASEM, which we don't write out. Here, we are interested in the causal effect $D \to Y$, that is, the effect $d \mapsto Y(d)$. The corresponding SWIG-intervention DAG is shown in Figure 8.3. In this DAG, valid adjustment sets $S$ include

$$\{X_1, X_2\}, \{X_2, X_3\}, \{X_2, Z_2\}, \{X_2, Z_1\},$$

because each d-separates $Y(d)$ and $D$ by blocking all open paths. Conditioning on just $X_2$ won't work, because it blocks the inner backdoor paths from $Y(d)$ to $D$, but opens the outer path on which $X_2$ is a collider. To close this opened path, it suffices to also condition on one of $X_1$, $X_3$, $Z_1$ or $Z_2$.

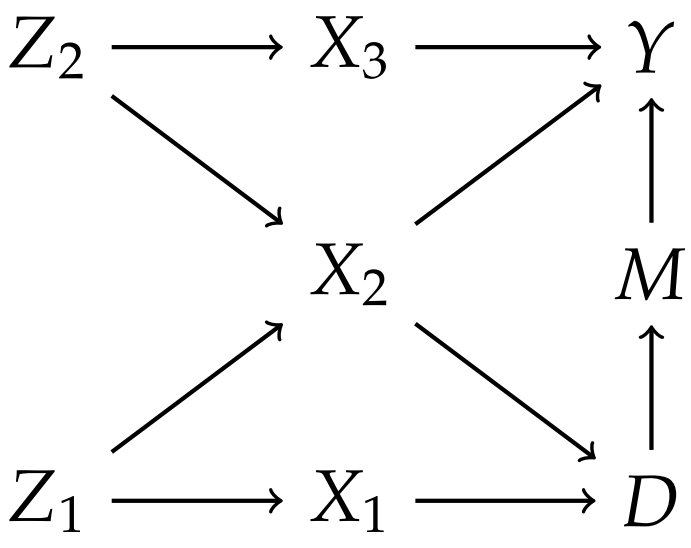

**Figure 8.2:** A DAG in Pearl's Example

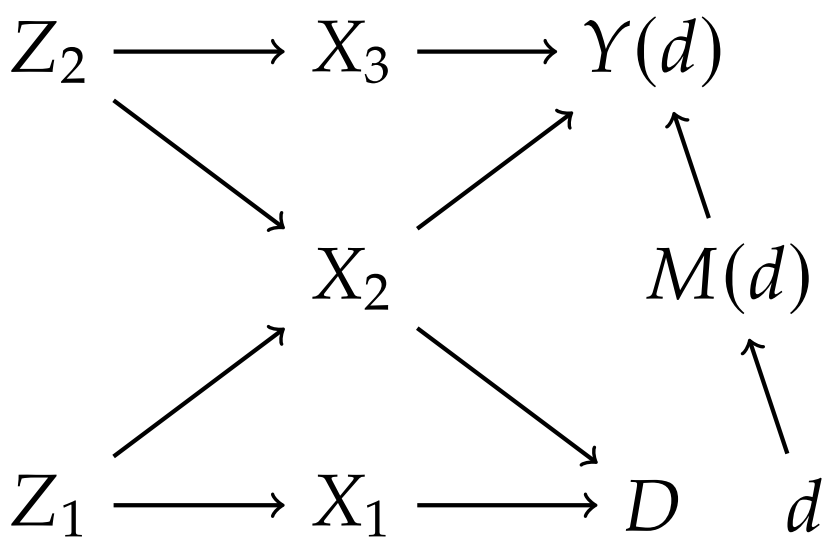

**Figure 8.3:** The DAG induced by the Fix/SWIG intervention fix($D = d$) in Pearl's Example.

## 8.2 Useful Adjustment Strategies

Theorem 8.1.1 provides an exhaustive criterion for finding valid adjustment sets. We now discuss other frequently used strategies for obtaining valid adjustment sets which are strictly less general. Some of these strategies are quite helpful because they are either very simple to apply or can also be used under partial knowledge of the DAG.[1]

1: See [3] for a more detailed discussion of identification by conditioning under limited knowledge of DAGs.

We consider three approaches that allow us to identify the causal effect of D on Y:

- Conditioning on one of **all parents** of $Y$ (that are not descendants of $D$), **all parents** of $D$, or **all parents** of both $D$ and $Y$ is sufficient. This approach provides a valid adjustment set irrespective of the remaining structure of the problem.
- Conditioning using the **backdoor criterion** enables us to find all minimal adjustment sets.

- Conditioning on **all common causes** of $D$ and $Y$ is also sufficient.

## Conditioning on Parents

A very simple strategy is conditioning on one of the parents of $D$, the parents of $Y$, or the parents of both $D$ and $Y$.

**Example 8.2.1** (Pearl's Example Continued) One simple principle is that conditioning on parents of $D$, namely $X_1$ and $X_2$, is sufficient. Alternatively, conditioning on all parents of $Y$ that are non-descendants of $D$, namely $X_2$ and $X_3$, is also sufficient. We should not condition on $M$, because it is a descendant of $D$.

**Corollary 8.2.1** (Adjustment for Parents) *Consider any ASEM. Re-label a policy node $X_j$ as $D$, and let $Y$, an outcome of interest, be any other descendant of $D$.*

- *Let $Z$ be all parents of $D$, and let $A$ be any other set of nodes that are not descendants of $D$. Then $S = (A, Z)$ is a valid adjustment set.*
- *Let $Z$ be the set of all parents of $Y$ that are non-descendants of $D$ and let $A$ be any other set that are not descendants of $D$. Then $S = (A, Z)$ is a valid adjustment set.*

Note that $A$ is allowed to be an empty set. Also note that, in the second case, the additional adjustment set $A$ is redundant, since $\mathrm{p}(y \mid a, z, d) = \mathrm{p}(y \mid z, d)$ in this case.

Adjusting for parents is a very useful strategy, because it only requires knowledge of parents in a DAG without precise knowledge of the remaining graph structure. Conditioning on parents is also behind the propensity score strategies used in many experimental or quasi-experimental empirical analyses. If the propensity score is known, it can be used as a parent of $D$ itself. Finally, conditioning on parents of $Y$ is most useful for attaining maximal statistical efficiency, but may be less robust than conditioning on *both* sets of parents under unforeseen deviations from the given graph structure. See [3] for further detailed discussion of robustness of adjusting for both sets of parents.

## Conditioning by Backdoor Blocking

Pearl [4] developed the following powerful criterion.

**Corollary 8.2.2** (Backdoor Criterion) *Consider any ASEM. Relabel a policy node $X_j$ as $D$, and let $Y$, an outcome of interest, be any other descendant of $D$. The adjustment set $S$ is valid if the backdoor criterion is satisfied: No element of $S$ is a descendant of $D$, and all backdoor paths from $Y$ to $D$ are blocked by $S$.*

In other words, if a collection of random variables $S$ satisfies the backdoor criterion with respect to $(D, Y)$, then conditioning on $S$ identifies the causal effect of $D$ on $Y$. The basic idea is that if we block the backdoor path, we remove all channels of non-causal association between $D$ and $Y$.

**Example 8.2.2** (Pearl's Example Again, using the Backdoor Criterion) The graph in Figure 8.2 has two backdoor paths from $D$ to $Y$: the inner path $D \leftarrow X_2 \rightarrow Y$ and the outer path $D \leftarrow X_1 \leftarrow Z_1 \rightarrow X_2 \leftarrow Z_2 \rightarrow X_3 \rightarrow Y$. Conditioning on just $X_2$ does not allow us to identify the causal effect of $D$ on $Y$ because $X_2$ blocks the inner backdoor path from $Y$ to $D$ but opens the outer path on which $X_2$ is a collider. To close this opened path, it suffices to condition on $X_1$, $X_3$, $Z_1$, or $Z_2$. For example, conditioning sets $S_1 = \{X_1, X_2\}$ or $S_2 = \{X_2, X_3\}$ are valid. Figuring out other valid conditioning sets is left as an exercise. (You can find the answers using the notebook R: Dagitty Notebook or Python: Pgmpy Notebook.) Conditioning on $M$ is obviously not valid – it is a descendant of $D$, an intermediate outcome.

Application of the backdoor criterion can produce all minimal adjustment sets. Relative to the complete strategy formalized in Theorem 8.1.1, we exclude the descendants of $D$ from valid adjustment sets when we focus on backdoor paths. A simple example of a graph where the backdoor criterion does not find all valid adjustment sets is

$$Z \leftarrow D \rightarrow Y.$$

Here conditioning on $Z$ is valid but unnecessary. Conditioning on $Z$ may thus decrease statistical efficiency.[2]

2: We may think that conditioning on $Z$ here could be useful to uncover heterogeneity. However, $Y(d)$ does not depend on $Z$, so conditioning on $Z$ is not useful for describing heterogeneity and can decrease the efficiency of the estimator.

## Conditioning on All Common Causes of $D$ and $Y$

Another simple and widely used adjustment strategy is conditioning on all common causes of the outcome variable of interest and the treatment variable.

**Example 8.2.3** (Pearl's Example Again, using the All Common Causes Criterion) The set of common causes of $D$ and $Y$ is $\{Z_1, Z_2, X_2\}$. This set is a valid adjustment set that differs from the sets found using the parental strategy. We can push the All Common Causes criterion further. For example, we can omit $Z_1$ and $Z_2$ from the DAG, and we can create a new node $X = (X_1, X_2, X_3)$ producing the DAG shown in Figure 8.4. This DAG corresponds to a valid ASEM model where $X$ now represents all common causes of $D$ and $Y$, making it a sufficient adjustment set. This set is bigger than some of the sets found by the previous criteria. It is also tempting to see if the "root common" causes $Z_1$ and $Z_2$ in the original DAG, Figure 8.2, form a valid adjustment set – and they actually do not (why?).

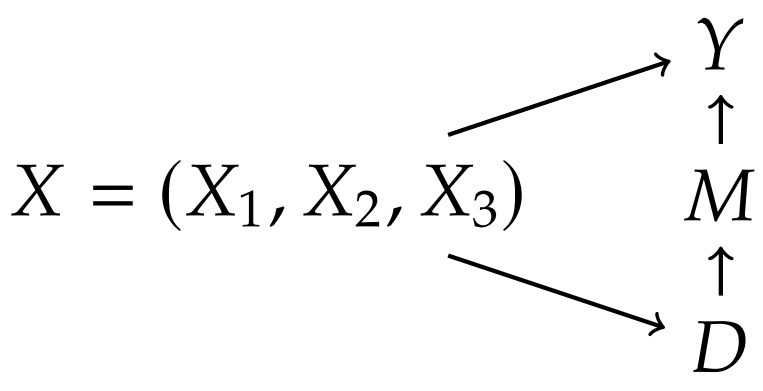

**Figure 8.4:** Reduced DAG for Pearl's Example

Let $\underline{An}_X$ denote the set of strict ancestors of node $X$, where strict means that $X$ is excluded. That is,

$$\underline{An}_X = An_X \setminus X.$$

**Corollary 8.2.3** (Adjustment for All Common Causes) *Consider any ASEM. Re-label a policy node $X_j$ as $D$, and let $Y$, an outcome of interest, be any other descendant of $D$. Let $S$ be the intersection of the strict ancestors of $D$ and $Y$, called the common causes:*

$$S = (\underline{An}_D \cap \underline{An}_Y).$$

*Then $S$ is a valid adjustment set. Furthermore, the set of variables $S'$ that completely mediates the effects of $S$ on $Y$ and $D$ also constitutes a valid adjustment set.*

The strategy above is commonly used in empirical work. However, [3] recommend adjusting for the union $S$ of causes of $Y$ or $D$ (excluding descendants of $D$) in practice as they formally quantify this strategy as the maximally robust strategy under perturbations of a specified DAG structure that preserves $S$. This strategy is useful when we don't know the parents of $Y$ or $D$, but only know that $S$ are their ancestors.

**Corollary 8.2.4** (Adjustment for the Union of Causes) *Consider any ASEM. Re-label a policy node $X_j$ as $D$, and let $Y$, an outcome of interest, be any other descendant of $D$. Let $S$ be the union of the ancestors of $D$ and $Y$ that excludes descendants of $D$ other than $Y$:*

$$S = \underline{An}_D \cup \underline{An}_Y \setminus Ds_D.$$

*Then $S$ is a valid adjustment set.*

**Example 8.2.4** (Pearl's Example Continued) Application of the Union of Causes criterion gives $\{Z_1, Z_2, X_1, X_2, X_3\}$ as a valid adjustment set.

## 8.3 Examples of Good and Bad Controls

We now present a series of simple example DAGs that might arise in empirical research. Within these examples, we discuss what would be good and bad variables to adjust for in each case (aka good and bad controls), when one is interested in estimating the average treatment effect of a treatment $D$ on an outcome $Y$.[3] Similar to the collider bias examples we presented in Section 6.3, we will see how adjusting for some of the observed variables can introduce bias and lead to estimating a parameter that is far from the causal effect of interest. In each case, we will denote the candidate control of interest with $Z$ and will denote unobserved variables with $U$. We depict unobserved variables with a dashed circle in the figures.

3: The content in this section draws heavily from the excellent research paper of Cinelli, Forney and Pearl [5].

We start by analyzing a group of potential control variables that in most empirical applications would correspond to *pre-treatment variables*, i.e. variables whose value was determined prior to the treatment assignment. It is common empirical practice to adjust for as many pre-treatment variables as available in an attempt to ensure that conditional ignorability holds. However, we will see that bias can be introduced by controlling even for pre-treatment variables if one is not careful. Rather than always control for all pre-treatment variables, a better approach is to adjust only for pre-treatment variables that are ancestors of either the treatment, the outcome, or both. If one is willing to believe that identification by conditioning is feasible, then following this approach is a safe strategy.

We then consider the use of *post-treatment variables*, i.e. variables that correspond to quantities whose value is determined after the treatment assignment. We will see that in this case there are relatively few good control cases. In some cases, controlling for post-treatment variables might not hurt and may even improve precision (reduce variance). However, such settings seem unlikely to be common in empirical practice. Hence, as a high-level rule, controlling for post-treatment variables should be avoided when one is interested in estimating causal effects.

Finally, we provide a separate discussion of post-treatment but *pre-outcome variables*, i.e. variables whose value is determined prior to the determination of the value of the outcome of interest. Pre-outcome variables should be included if one is interested in estimating direct effects of the treatment on the outcome while excluding indirect effects. This type of direct effect is referred to as a *controlled direct effect* to distinguish it from other forms of direct effects appearing in mediation analysis. We will see again that one should be careful that the mediation variables that one conditions on are not themselves confounded through unobserved factors even in this case.

## Pre-Treatment Variables or Proxies of Pre-Treatment Variables

**Observed common causes or proxies of common causes.** A common example of a good control that we have discussed so far is an observed common cause, $Z$, of $D$ and $Y$ (Figure 8.5a). Even if the common cause is unobserved, it suffices that we have a proxy control variable that controls all the information flow to either the treatment (complete treatment proxy; Figure 8.5b) or to the outcome (complete outcome proxy; Figure 8.5c). Controlling for such a proxy also blocks the backdoor path $D \leftarrow U \rightarrow Y$. Of course, the proxy blocking the backdoor path only holds if the proxy variable captures *all* the information flow from the unobserved confounder. If, for instance, there are also direct paths from the unobserved variable to the treatment (in the case of a treatment proxy), then controlling for a proxy does not remove confounding bias. In this case, we will see that one can follow more advanced approaches related to proxy controls under additional structure in Chapter 12.

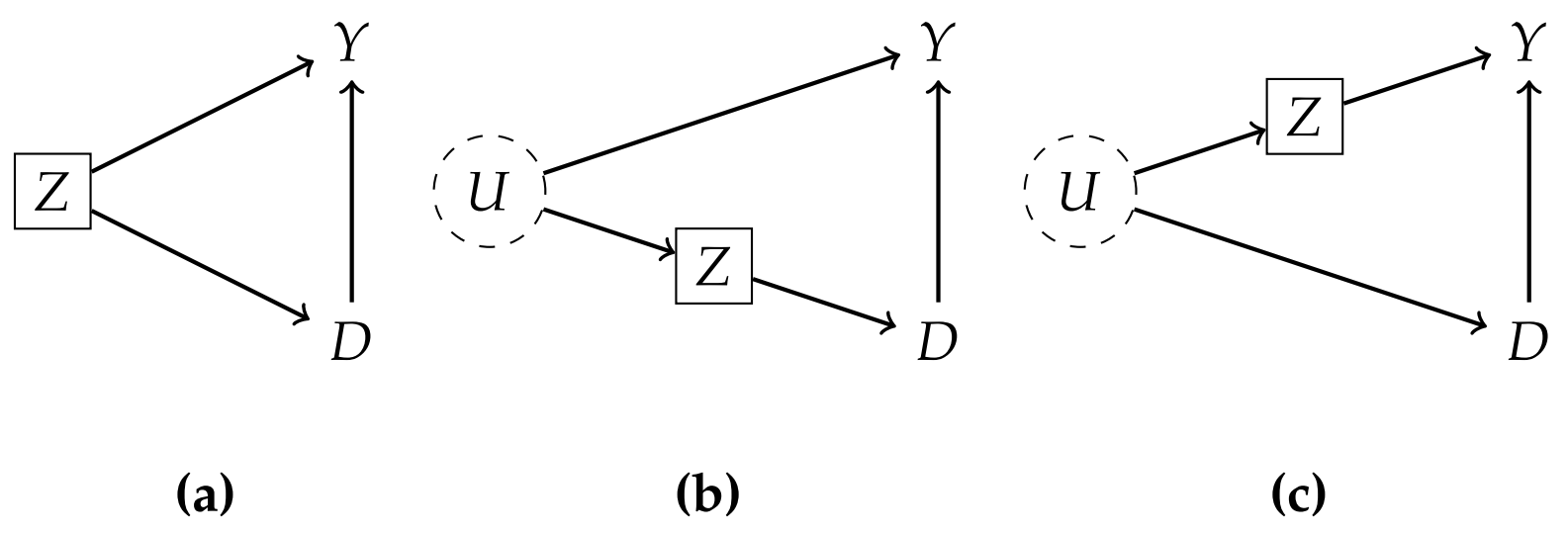

**Figure 8.5:** Good controls: **(a)** observed common cause, **(b)** complete treatment proxy control of unobserved common cause, **(c)** complete outcome proxy control of unobserved common cause.

**Example 8.3.1** (Effect of Multivitamin Consumption on Birth Defects [6]) Suppose we want to estimate the effect of prenatal multivitamin consumption $D$ on birth defects $Y$. One factor that can potentially influence a mother's decision on multivitamin consumption is prior history of birth defects in the family ($Z$); see e.g. [7]. Such prior history is possibly due to unobserved genetic factors $U$ that also have a direct effect on the risk of malformation $Y$; see e.g. [8]. In this case, family medical history $Z$ provides a complete treatment proxy of the unobserved confounder (as in Figure 8.5b) as long as the behavior of a mother is solely driven by the family medical history. Controlling for medical history would thus remove the confounding bias in this scenario.

**Confounded mediators with observed common cause or proxies of unobserved common cause.** It is important to note that confounding occurs even when there exists a common cause $Z$ of the treatment $D$ and some mediator $M$ in a path from $D$ to $Y$ (Figure 8.6a). In such cases, if we don't condition on the common cause of $D$ and $M$, there is an open backdoor path $D \leftarrow Z \rightarrow M \rightarrow Y$. In such cases, $Z$ is a good control as it blocks this backdoor path. Similarly, if a common cause $U$ of $D$ and $M$ is unobserved, but some complete treatment proxy control $Z$ (Figure 8.6b) or some complete outcome proxy control $Z$ (Figure 8.6c) is observed, then it suffices to adjust for this proxy $Z$.

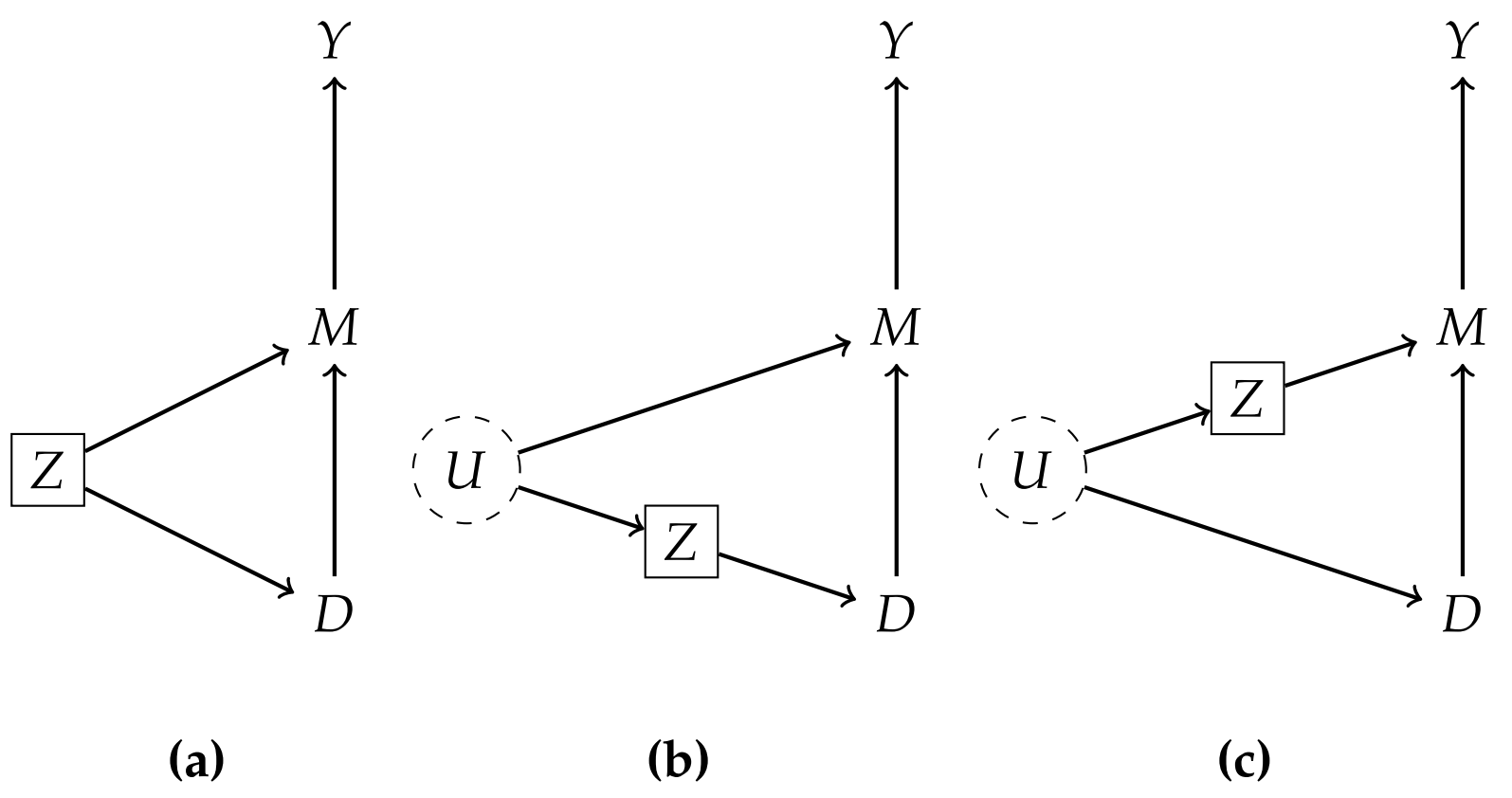

**Figure 8.6:** Good controls: **(a)** confounded mediator with observed common cause, **(b)** confounded mediator, with observed complete treatment proxy control of unobserved common cause, **(c)** confounded mediator with observed complete outcome proxy control of unobserved common cause.

**Causes of only treatment or only outcome.** As stated in Corollary 8.2.4, a conservative empirical practice is to include the union of parents of $D$ and $Y$ in the adjustment set. Including variables that are parents of the outcome (Figure 8.7a) can lead to reduced variance during estimation as explained in Chapter 2 where we discuss including pre-treatment covariates in RCTs. Including variables $Z$ that affect the treatment $D$ but have no causal path to the outcome (Figure 8.7b) is potentially more controversial. Including these variables does not introduce bias. However, their inclusion can be detrimental for precision, as such variables can potentially explain away all of the useful variation in the treatment, leaving little variation for the identification of causal effects.

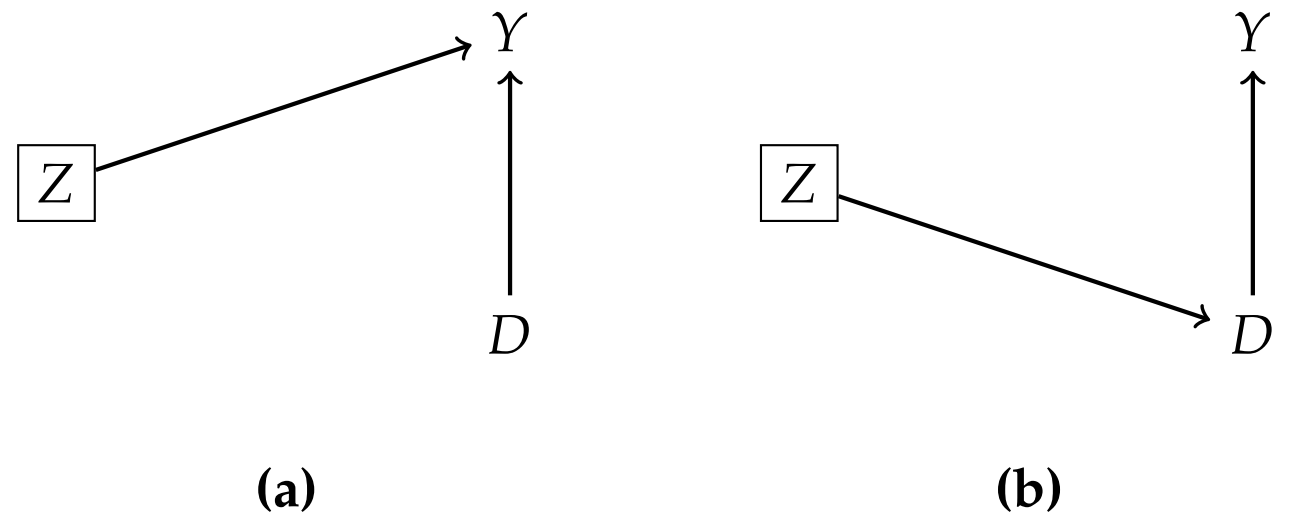

**Figure 8.7:** Neutral controls: **(a)** Outcome-only cause. Can improve precision; decrease variance. **(b)** Treatment-only cause. Can decrease precision; introduce variance.

Even more importantly, when there are unobserved common causes of $D$ and $Y$ as illustrated in Figure 8.8, adjusting for a treatment-only cause, $Z$, can exacerbate the bias stemming from unobserved confounding. Essentially, controlling for $Z$ removes exogenous variation in the treatment $D$ that is useful for identifying the causal effect but leaves the confounded variation - as $Z$ is not related directly to the unobserved confounder $U$. As such, the resulting estimated effect may be essentially driven by the unobserved confounder and thus be heavily biased. For this reason, one should avoid controlling for variables that are *known* to have no causal path to the outcome that does not pass through the treatment. As we will see in Chapter 12, such variables are actually what are referred to as *instruments*. These variables can be thought as useful natural experiments that can be leveraged for causal identification even in the presence of unobserved confounding. However, we will need to use alternative identification arguments and estimation strategies to make use of instruments. We introduced these *instrumental variable* approaches in Chapter 12 and Chapter 13. Importantly, instruments *should not* be used in an identification by adjustment strategy.

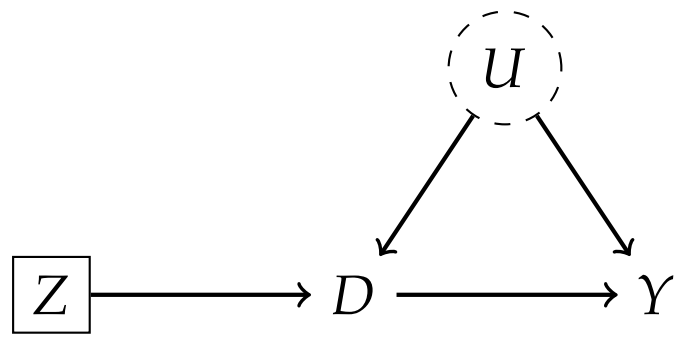

**Figure 8.8:** Bad control. Bias amplification by adjusting for an *instrument*. Treatment-only cause (*instrument*) that can amplify unobserved confounding bias.

**M-bias** The DAG in Figure 8.9, typically referred to in the literature as the M structure, is the source of much debate; see e.g. [9, 10]. If such cases were impossible, the high-level strategy of controlling for all pre-treatment variables when attempting to identify causal effects by conditioning would be an unambiguously safe empirical route resulting in no harm other than potentially increasing variance by including an *instrument*. However, this structure shows that there exist settings where adjusting for a pre-treatment covariate $Z$ can lead to a wrong causal effect, while not adjusting for $Z$ would have yielded the correct causal effect. A better high-level strategy is the one highlighted in the prior sections: If we are willing to assume that identification by conditioning is possible, then we should adjust only for pre-treatment variables that are either an ancestor of the treatment, of the outcome, or of both treatment and outcome.

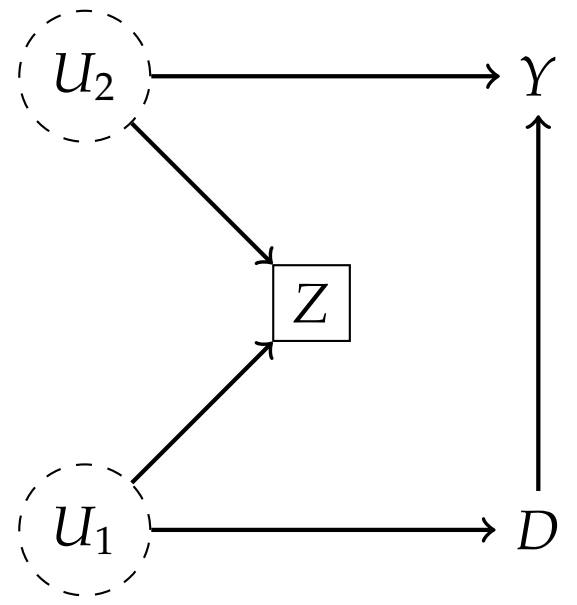

**Figure 8.9:** Bad control. M-Bias. Pre-treatment variable that introduces Heckman selection bias between two uncorrelated unobserved causes.

More concretely, in the M structure graph (Figure 8.9), $D$ and $Y$ are driven by two independent unobserved causal factors $U_1, U_2$. The variable $Z$ is a common outcome of these two unobserved causal factors. When conditioning on $Z$, we introduce collider bias between $U_1, U_2$, making them correlated factors. Conditioning on $Z$ can thus lead to a causal effect estimate that is solely driven by this spurious correlation between $U_1$ and $U_2$, introduced by the collider bias. In graphical terms, adjusting for $Z$ closes the path $D \leftarrow U_1 \rightarrow Z \leftarrow U_2 \rightarrow Y(d)$ in the SWIG DAG $\tilde{G}(d)$ produced by the $\text{fix}(D = d)$ operation. However, there is no open path connecting $D$ to $Y(d)$ when we do not condition on $Z$. Hence, the effect identified by not adjusting for any variable is the correct causal effect within this example structure.

**Example 8.3.2** (Homophily bias in estimating peer effects) A classical example where M-bias arises in empirical work in social sciences is in the estimation of peer effects on social networks [11, 12]. As a concrete example, suppose that we want to understand the spread of civic engagement among friends. Suppose that we look at data that consist of friendship pairs and let $D$ be the level of civic engagement level of one friend at time $t$ and $Y$ the level of civic engagement of the other friend at time $t+1$. Note that when we are estimating the correlation of these two variables, we are implicitly conditioning on the friendship variable $Z$, since we only have data from friendship pairs. Due to homophily, friendship could be driven by the unobserved intrinsic characteristics of each of the two individuals ($U_1$ and $U_2$ in Figure 8.9). It is reasonable to assume that these characteristics are independent as they are determined well before any friendship is formed. Moreover, these qualitative characteristics (e.g. levels of altruism) could very well have a direct effect on each individual's civic engagement. Thus, the estimation of peer effects can be heavily biased due to exactly M-bias.

Homophily refers to the tendency of associate with similar individuals - i.e. similar people tend to become friends.

Finally, note that the M-bias argument is very sensitive to the exact independence of the unobserved factors $U_1, U_2$. In most empirical applications, we expect these unobserved factors that drive the treatment and outcome of interest to be correlated with each other as in Figure 8.10a. In this case, note that even if we don't adjust for $Z$, the calculated effect is biased due to the backdoor path $D \leftarrow U_1 \rightarrow U_2 \rightarrow Y$. Thus, neither adjusting nor not adjusting for $Z$ gives the correct answer.

Moreover, it is not clear whether adjusting for $Z$ increases or decreases the correlation between $U_1$ and $U_2$ and hence exacerbates or ameliorates the confounding bias. Similarly, if $Z$ itself has a direct effect on the outcome (as in Figure 8.10b), on the treatment, or on both (as in Figure 8.10c), then not adjusting for $Z$ opens the backdoor paths $D \leftarrow U_1 \rightarrow Z \rightarrow Y$ and $D \leftarrow Z \rightarrow Y$, correspondingly. Hence, it is not clear that removing the bias induced by these open backdoor paths, by adjusting for $Z$, is more beneficial than the extra M-bias incurred by closing the path $D \leftarrow U_1 \rightarrow Z \leftarrow U_2 \rightarrow Y$. Work of [9, 13] argues that M-bias in many realistic data generating processes is of lower order than confounding bias and therefore argues that one should err on the side of adjusting for pre-treatment covariates even in the potential presence of M-bias. [10] provides a counterpoint, arguing that the strength of the different biases will differ in general and thus careful consideration of the strength of each of the causal paths at play should be done on a case-by-case basis.

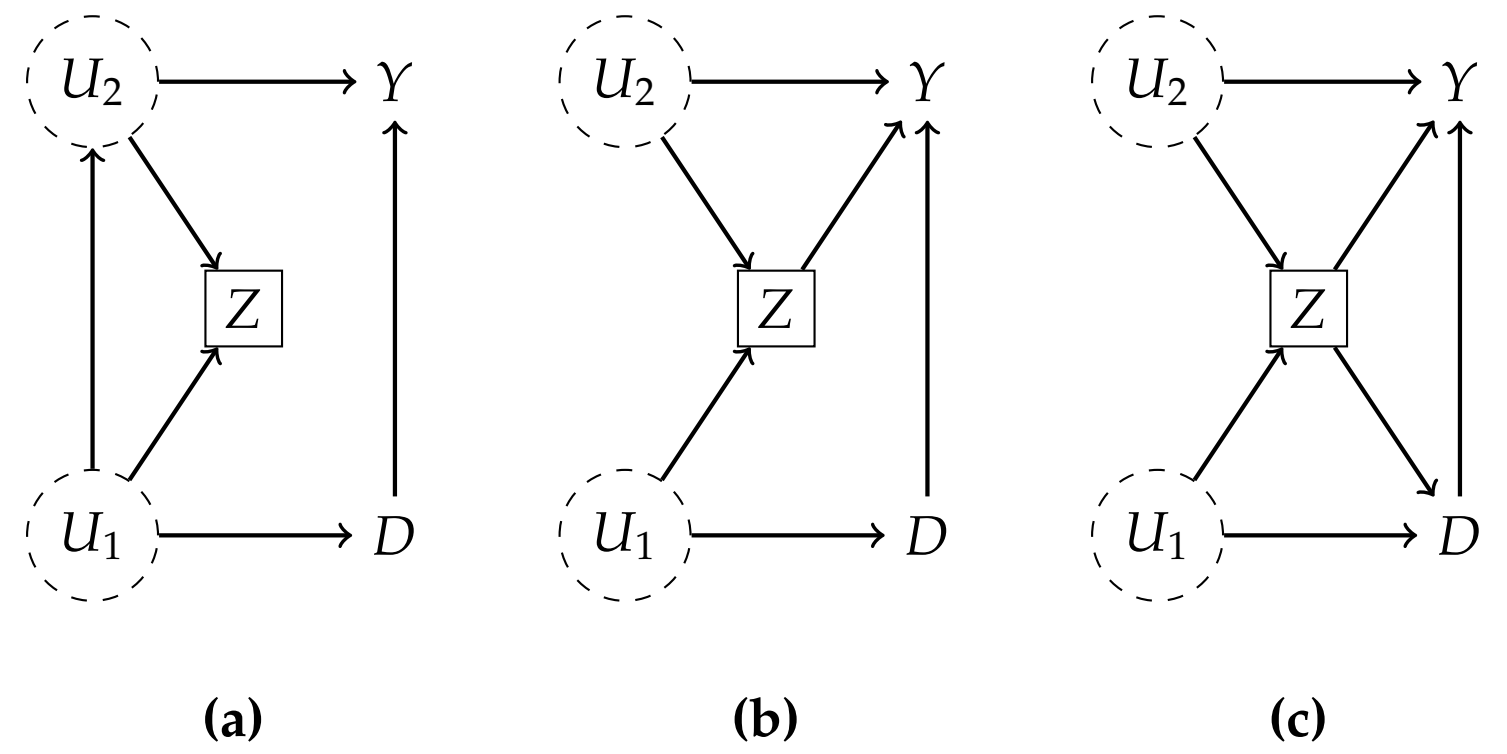

**Figure 8.10:** No perfect control solutions: **(a)** M-bias with correlated unobserved factors. **(b)** M-Bias with confounding. Pre-treatment variable that introduces Heckman selection between two uncorrelated unobserved causes and is a confounder itself. **(c)** Butterfly Bias. M-bias with direct confounding.

## Post-Treatment Variables

Now we turn to adjustment for post-treatment variables. The general message of this section is that explicitly adjusting for post-treatment variables is almost always a bad idea. Importantly, the general message implies that researchers should be careful to avoid implicitly adjusting for post-treatment variables through the way they have structured their observational analysis, data collection, and variable definitions – see e.g. [6] for examples from epidemiology. For instance, when estimating the effect of education on wages using data on *employed* individuals, we are implicitly conditioning on "employment" which is a post-treatment variable and can lead to selection bias.

**Mediation.** A common way a post-treatment variable can lead to bias in identifying the full causal effect of $D$ on $Y$ is if it lies on a causal path from the treatment to the outcome (Figure 8.11a). In this case, the causal influence that flows through that path is blocked and we are only measuring a partial effect. It is important to note, that the causal influence of such a path can be partially blocked even if one conditions on a descendant of the mediator (Figure 8.11b).

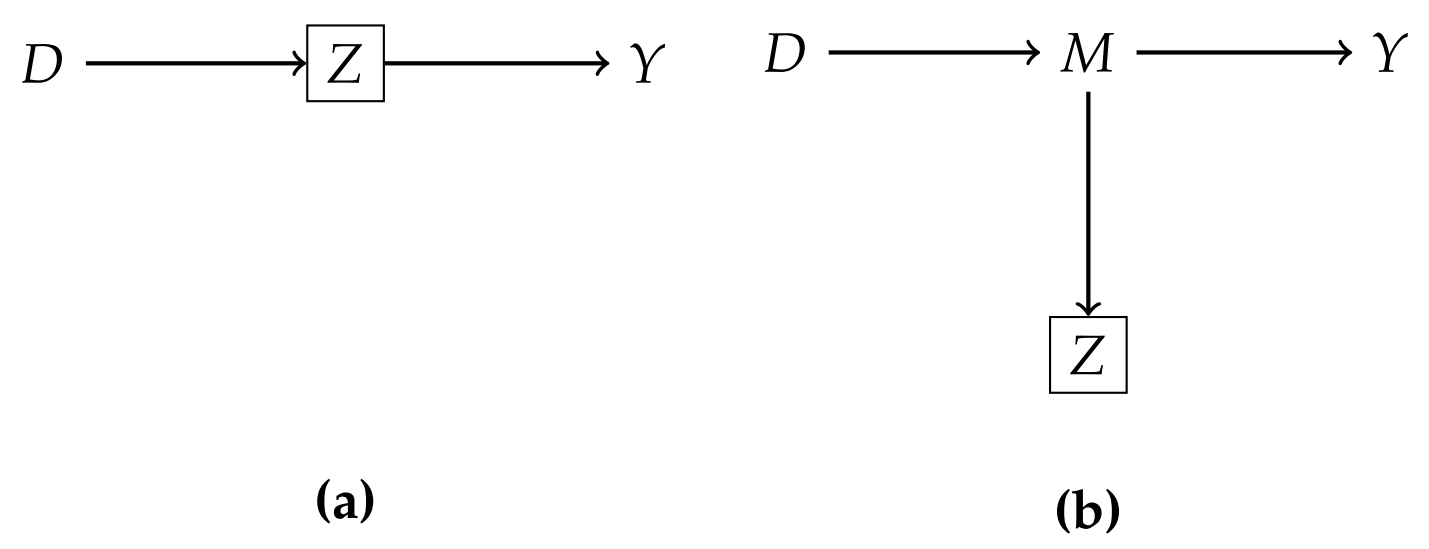

**Figure 8.11:** Bad controls for learning the full direct effect of $D$ on $Y$: **(a)** over-control bias, by controlling for a mediator. **(b)** over-control bias, by controlling for an outcome caused by a mediator.

Interestingly, controlling for an ancestor of a mediator (Figure 8.12) does not impede us from learning the full direct effect of $D$ on $Y$. In this case, the flow through the causal path $D \to M \to Y$ is not blocked by $Z$. For example, d-separation can be easily checked in the SWIG $\tilde{\mathsf{G}}(d)$ produced by fix$(D = d)$.

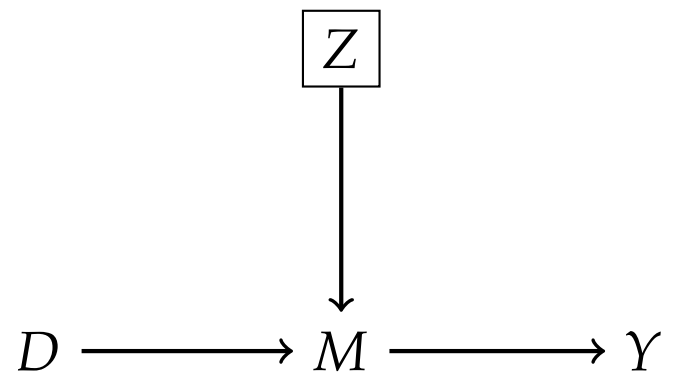

**Figure 8.12:** Neutral control. Cause of a mediator. Can potentially improve precision.

When we are controlling for a post-treatment variable that mediates the effect of the treatment as in Figure 8.11a, we are only capturing direct effects from the treatment to the outcome that do not work through this mediator. This type of direct effect after controlling for mediators is typically referred to as a *controlled direct effect*. Identifying the controlled direct effect is many times a relevant empirical question, in which case controlling for $Z$ is not problematic. However, even when we are interested in the controlled direct effect, we should pay attention to cases where the mediators are themselves confounded through unobserved factors as illustrated in Figure 8.13. In such settings, by controlling for the mediator, we are opening a collider path $D \to Z \leftarrow U \to Y$ which can lead to severe bias, such as calculating non-zero direct effects even when they are zero.

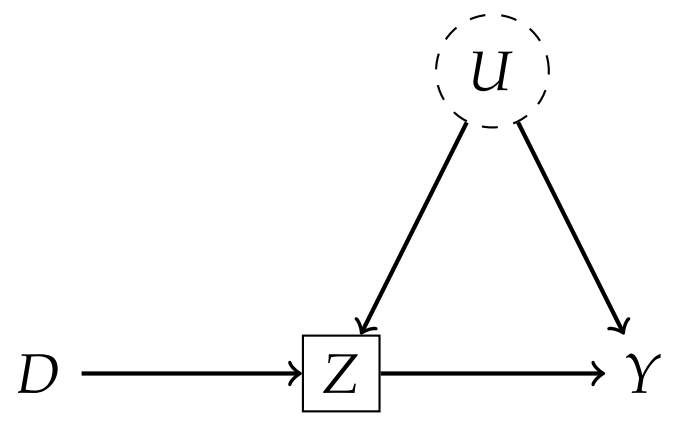

**Figure 8.13:** Bad control even for the *controlled direct effect*. Confounded mediator bias.

**Heckman selection bias** Another common way that post-treatment variables can lead to bias is due to collider bias or Heckman selection, as described in Section 6.3. In this case, conditioning on the post-treatment variable introduces spurious correlations between the treatment variable and some other variable which opens new paths of non-causal influence from the treatment to the outcome. For instance, Figure 8.14a corresponds to the low birthweight paradox we presented in Example 6.3.2. Similarly, Figure 8.14b corresponds to the Hollywood Example Example, Example 6.3.1. Finally, Figure 8.14c arises when we are controlling for an outcome of the outcome as might be produced by recall bias in a case-control study.

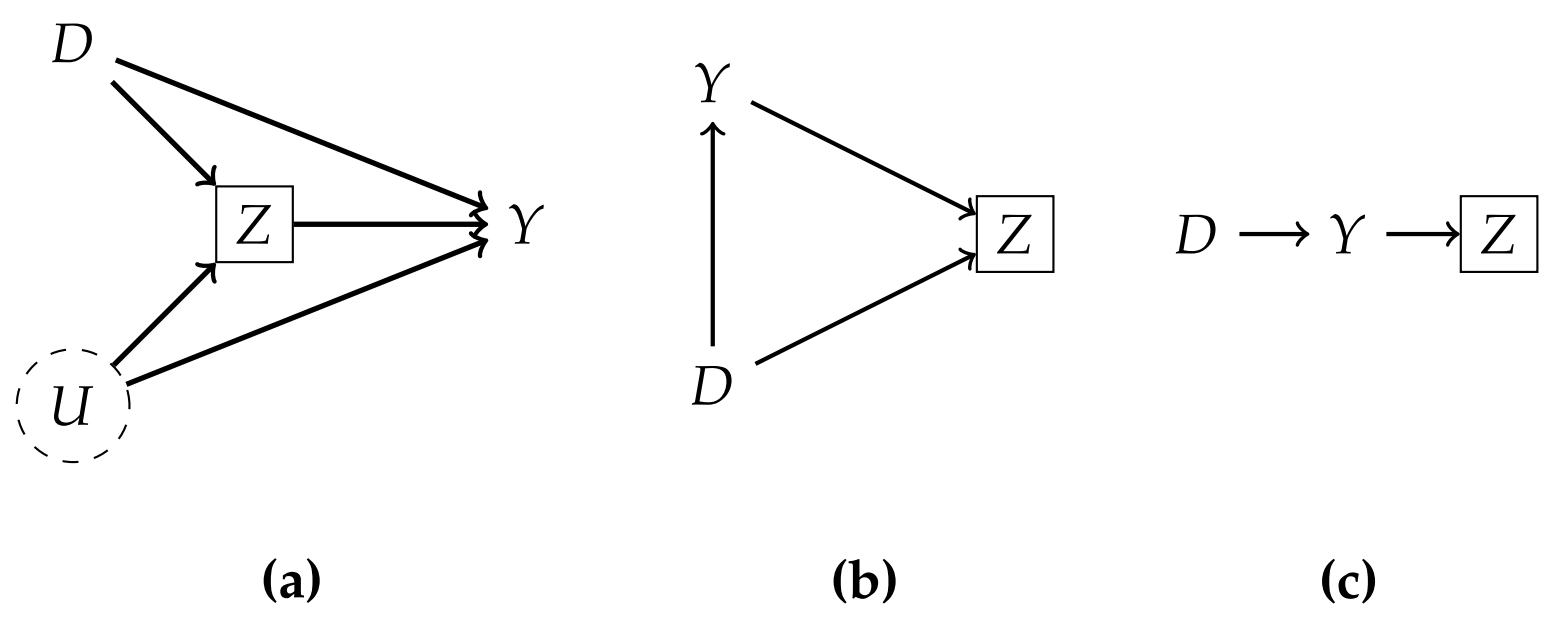

**Figure 8.14:** Bad controls: **(a)** collider stratification bias (e.g. low birth-weight "paradox" example), **(b)** collider stratification bias, **(c)** controlling for an outcome of the outcome of interest.

**Example 8.3.3** (The Industrial Growth Puzzle [14]) In a study conducted during the nineteenth century in the US and Britain, it was found that despite nutrition quality $D$ having improved, the height of men $Y$ decreased. One possible explanation of the results of this study is that the subjects of the study were people who were enlisted in the army or in prison. Both of these variables, enlisted in the army and being in prison, are plausibly determined *after* the outcome variable of height is realized. It might, for example, be that taller men had more civilian opportunities growing up and did not end up enlisting in the army. In this case, looking at a sample of enlistees is implicitly controlling for an outcome of the outcome of interest which could lead to a biased estimate of the effect of nutrition on height.

There are of course some edge cases where controlling for a post-treatment variable $Z$ does not lead to selection bias – e.g. Figure 8.15a and Figure 8.15b. In each of these two cases, the post-treatment variable is not a collider on a path from $D$ to $Y$. However, it is not clear that adjusting for $Z$ improves the analysis in any respect even in these cases, and adjusting for $Z$ could potentially hurt precision.

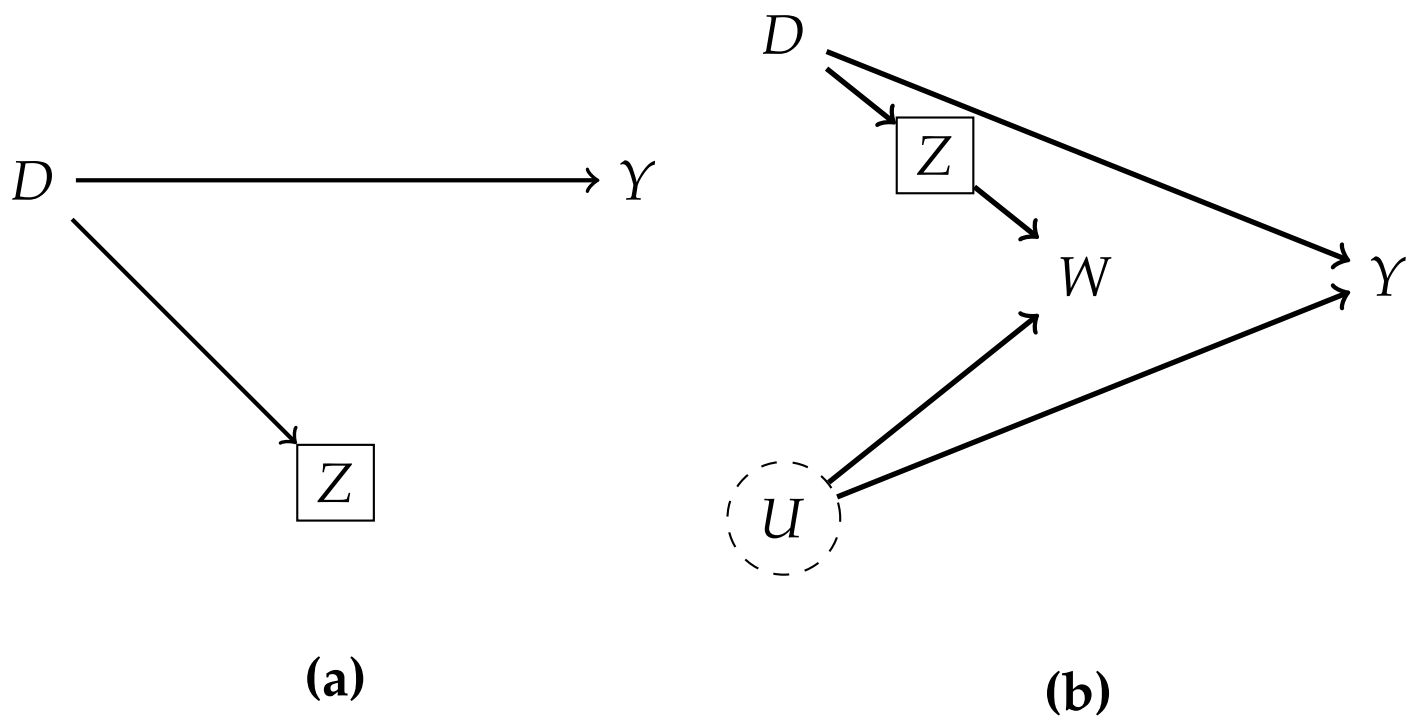

**Figure 8.15:** Neutral controls: **(a)** outcome of the treatment that is unrelated to the outcome of interest, **(b)** outcome of the treatment that does not introduce Heckman selection.

## Notes

Any empirical study that tries to learn the causal effect of $D$ on $Y$ by conditioning on $S$ must have a thought process that justifies this approach. The DAG/ASEM framework is a rigorous representation of such a thought process which enables explicit incorporation of domain knowledge, automatic checking of identifiability, and automatic deduction of testable restrictions. Graphs also provide an effective way of visualizing and communicating models.

## Notebooks

- R: Dagitty Notebook employs the R package "dagitty" to analyze some simple DAGs as well as Pearl's Example. This package automatically finds adjustment sets and also lists testable restrictions in a DAG. Python: Pgmpy Notebook employs the analogue with Python package "pgmpy" and conducts the same analysis.

## Study Problems

The study problems ask learners to continue the analysis of Pearl's Example DAG that we started in the Study Problems to Chapter 7. The provided notebooks are a useful starting point. Recall that Pearl's Example is structured as follows:

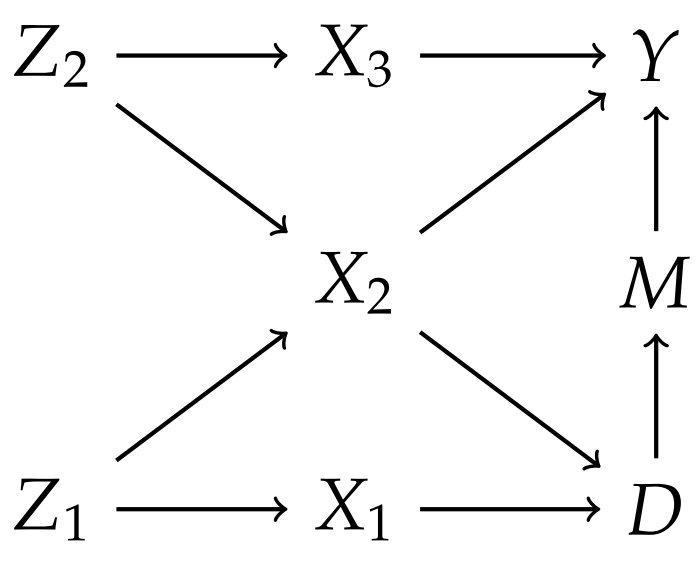

**Figure 8.16:** Pearl's Example

1. For Pearl's Example, write out the parents, non-parents, descendants, and non-descendants of nodes $X_2$ and $M$. List all the backdoor paths between $Y$ and $X_2$. Can you identify the effect of $X_2$ on $Y$ by conditioning?

2. (Front-Door-Criterion) For Pearl's Example, show that we can identify the effect $D \to M$ by conditioning on an empty set and the effect $M \to Y$ by conditioning on $D$. Combining the two results, we can identify the total effect of $D$ on $Y$. Solving this exercise analytically is a nice exercise; you can compare your results against causal identification packages. (Identification via this strategy is known as the Front-Door criterion; see Appendix 8.A.

3. Add an arrow $Z_2 \to Z_1$ in Pearl's Example and figure out how to identify the effect of $D \to Y$ by conditioning, of $D \to M$ by conditioning, and of $M \to Y$ by conditioning. (Note that valid conditioning sets may be empty.) Can you identify the effect of $X_2 \to Y$? If so, how? You may solve this analytically or using a causal identification package.

4. Add an arrow $X_1 \to M$ in Pearl's Example and figure out how to identify the effect of $D \to Y$ by conditioning, of $D \to M$ by conditioning, and of $M \to Y$ by conditioning. Can you identify the effect of $X_2 \to Y$? If so, how? You may solve this analytically or using a causal identification package.

5. Try to ask an instruction-following LLM (such as ChatGPT) about identification and valid adjustment sets, both for the original Pearl's Example as well as the variations in the latter two problems. Can you verify or find mistakes in the response? If you find mistakes, how might they be corrected? When mistakes are pointed out to the LLM, is it able to correct them? For example, you can try starting with the following prompt and make variations on it: "I have a causal graph with nodes Z1, Z2, X1, X2, X3, D, M, Y and edges Z1->X1, Z1->X2, Z2->X2, Z2->X3, X1->D, X2->D, X2->Y, X3->Y, D->M, M->Y. Is the effect of D on Y identified? What are the valid adjustment sets?"

## 8.A Front-Door Criterion via Example

We examine identification in Pearl's Example (Figure 8.2), via the front-door criterion. First note that we can write the potential outcome of interest $Y(d)$ as $Y(M(d))$, since in the SWIG $\tilde{G}(d)$ there is no other path from $d$ to $Y(d)$ other than through $M(d)$.

$$
\begin{aligned}
\mathrm{E}[Y(d)] &= \mathrm{E}[Y(M(d))] \\
&= \int \mathrm{E}[Y(M(d)) \mid M(d) = m]\mathrm{P}(M(d) = m)dm \\
&= \int \mathrm{E}[Y(m) \mid M(d) = m]\mathrm{P}(M(d) = m)dm
\end{aligned}
$$

Suppose that we make a further surgery to the SWIG graph in Figure 8.3 by adding an intervention on the variable $M(d)$, i.e. take the modified SWIG graph induced by intervention fix$(D = d)$ and on that graph make a further intervention fix$(M(d) = m)$. This leads to the new SWIG:

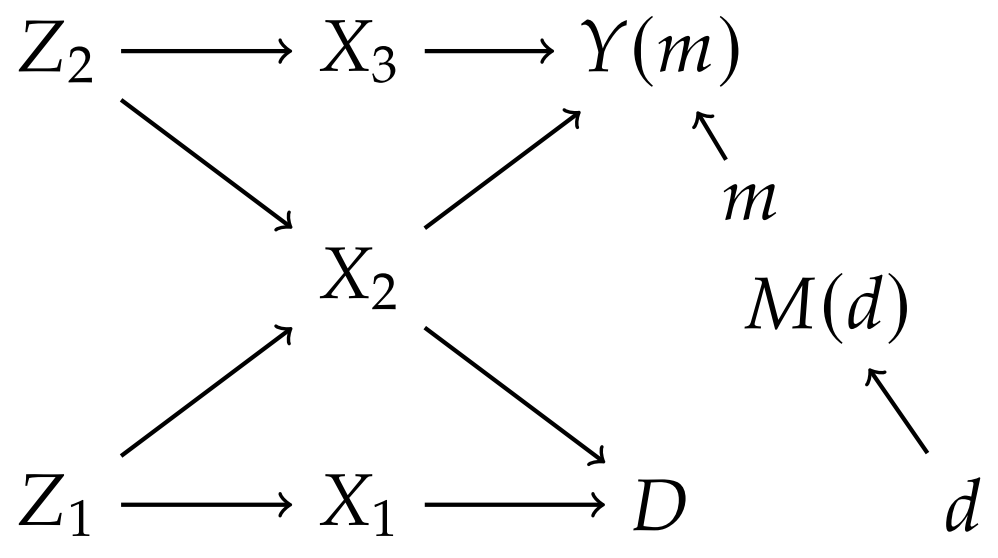

**Figure 8.17:** The DAG induced by a recursive Fix/SWIG intervention fix$(M(d) = m)$ on the SWIG in Figure 8.3.

Note that in this SWIG, we have $Y(m) \perp\!\!\!\perp M(d)$. Thus we have:

$$\mathrm{E}[Y(m) \mid M(d) = m] = \mathrm{E}[Y(m)],$$

leading to the front-door formula:

$$\mathrm{E}[Y(d)] = \int \mathrm{E}[Y(m)]\mathrm{P}(M(d) = m)dm$$

The term $\mathrm{E}[Y(m)]$ is the mean counterfactual response of $Y$ when we intervene on $M$ and $\mathrm{P}(M(d) = m)$ is the probability law of the counterfactual response of $M$ when we intervene on $D$. Both of these interventional quantities can be separately identified via backdoor adjustment. More concretely, $\mathrm{E}[Y(m)] = \mathrm{E}[\mathrm{E}[Y \mid M = m, D]]$, and $\mathrm{P}(M(d) = m) = \mathrm{P}(M = m \mid D = d)$.[4] Note that under linearity assumptions on the CEFs – i.e. $\mathrm{E}[Y \mid M = m, D] = \alpha m + \beta D + c$ and $\mathrm{E}[M \mid D = d] = \gamma d + \delta$ – we get $\mathrm{E}[Y(1) - Y(0)] = \alpha\gamma$.[5] Thus, the average treatment effect $\alpha\gamma$, can be estimated by estimating $\alpha$ via OLS of $Y$ on $M, D$ and $\gamma$ via OLS of $M$ on $D$.

4: See Exercise 2.

5: Prove this as a reading exercise.

# 9 Predictive Inference via Modern Nonlinear Regression

"Nowhere is it written on a stone tablet what kind of model should be used to solve problems involving data."

– Leo Breiman [1].

Here we discuss nonlinear regression methods based on tree models and (deep) neural network models. Tree-based methods include regression trees, random forests, and boosted trees. Regression trees are great for exploration and explainable analytics, while random forests and boosted trees are great predictive tools for structured data and data sets of intermediate size (say, up to several million observations). Neural networks are extremely flexible nonlinear regression methods and are particularly successful for data sets of larger size.

## 9.1 Introduction

We are interested in predicting an outcome $Y$ using raw regressors $Z$, which are $k$-dimensional. The best prediction rule $g(Z)$ under square loss is the conditional expectation function (CEF) of $Y$ given $Z$:

$$g(Z) = \mathrm{E}(Y \mid Z).$$

In previous chapters, we used best linear prediction rules to approximate $g(Z)$ and linear regression or Lasso regression for estimation. Now we consider nonlinear prediction rules to approximate $g(Z)$, focusing on tree-based methods and neural networks.

The use of best prediction rules (CEFs) is not just important for generating good predictions but is crucial for causal inference. Identification of causal parameters such as ATEs via conditioning strategies requires us to work with CEFs rather than with best linear prediction rules. Previously we tried to make best linear prediction rules flexible to try to approximate best prediction rules. Here we explore fully nonlinear strategies.

## 9.2 Regression Trees and Random Forests

### Introduction to Regression Trees

Regression trees are based on partitioning the regressor space (the space where $Z$ takes on values) into a set of rectangles. A simple model is then fit within each rectangle.

The most common approach fits a simple constant model within each rectangle, which corresponds to approximating the unknown function by a "step function." Given a partition into $M$ regions, denoted $R_1, \dots, R_M$ the approximating function when a constant is fit within each rectangle is given by

$$f(z) = \sum_{m=1}^{M} \beta_m 1(z \in R_m),$$

where $\beta_m, m = 1, \dots, M$ denotes a constant for each region and $1(\cdot)$ denotes the indicator function.

Suppose we have $n$ observations $(Z_i, Y_i)$ for $i = 1, \dots, n$. The estimated coefficients for a given partition are obtained by minimizing the in-sample MSE:

$$\hat{\beta} = \arg \min_{b_1,\dots,b_M} \mathbb{E}_n \left( Y - \sum_{m=1}^{M} b_m 1(Z \in R_m) \right)^2 ,$$

which results in

$$\hat{\beta}_m = \text{ average of } Y_i \text{ where } Z_i \in R_m.$$

The regions $R_1, \dots, R_M$ are called nodes, and each node $R_m$ has a predicted value $\hat{\beta}_m$ associated with it.

A nice feature of regression trees is that you get to draw cool pictures, so let's explore their usage graphically in the context of our wage example. In this example, the outcome variable $Y$ is (log) hourly wage; and $Z$ includes experience, geographic, and educational characteristics.

Figure 9.1 illustrates a simple regression tree for the wage data. This tree has a depth of two, meaning that predictions are produced as a sequence of two binary decisions (or partitions of the data). Starting at the top of the tree and working down provides a simple prediction rule for any observation. For example, the predicted wage for a worker without a college degree (`college` = 0) and with less than 14 years of experience (`exper` < 14) is 12 dollars an hour. We obtain this prediction by starting at the top of the tree and taking the left branch because `college` = 0 < .5. We then go left again at the second step because `exper` < 14 and arrive at the predicted value of 12.

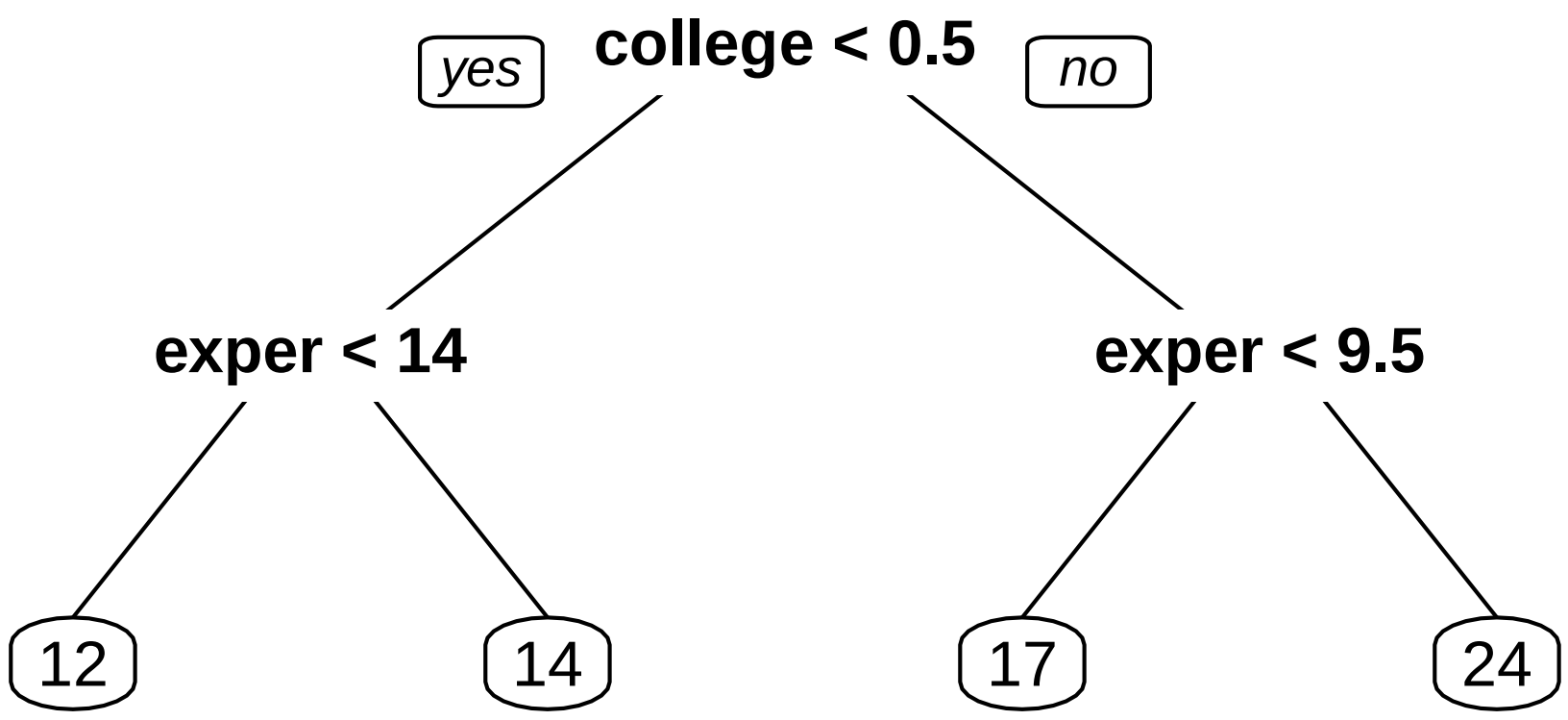

**Figure 9.1:** Regression tree based on wage data. The bottom nodes on the tree provide prediction rules for different subsets of observations. For example, the predicted hourly wage for a college educated worker with 9.5 or more years of experience (a worker with `college` = 1 and `exper` ≥ 9.5) is 24 dollars.

The key feature of trees is that the cut points for the partitions are adaptively chosen based on the data. That is, the splits are not pre-specified but are purely data dependent. So, how did we use the data to grow the tree in Figure 9.1?

To make computation tractable, we use recursive binary partitioning or splitting of the regressor space:

- **Growing the Tree: Level 1.** First, we cut the regressor space into two regions by choosing the regressor and splitting point such that using the prediction rule fit within each region produces the best improvement in the in-sample MSE.[1]

  Applying this procedure in the wage data gives us the depth 1 tree shown Figure 9.2. In this case, the best regressor to split on is the indicator of college degree, that takes values 0 or 1. Here splitting at any point between 0 and 1 provides the same rule, and an often used convention for binary variables is to use the "natural" split point of .5. Applying this split point yields the initial prediction rule: an hourly wage of $20 for college graduates and $13 for others.

1: To be clear, note that, in principle, finding this split point requires trying the partition produced by splitting the data along every possible value of every observed variable. That is, we are neither pre-specifying which variables nor which split points are important in providing a good prediction rule.

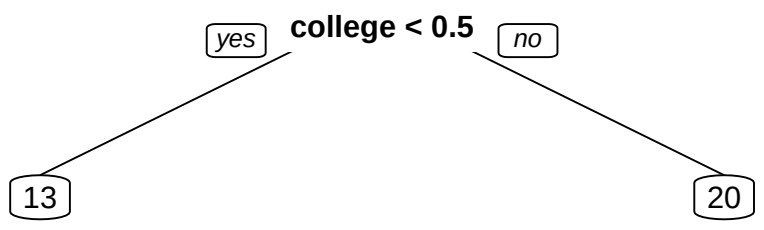

**Figure 9.2:** Depth 1 tree in the wage example

- **Growing the Tree: Level 2.** To grow the tree to depth 2, we then repeat the procedure for choosing the first partition rule within the two regions resulting from the first step. This step will result in a partition of the covariate space into four new regions. It is important to note that the two splits produced at this point may use different variables/splitting points than before. This feature means that the tree algorithm can create "interactions" and "nonlinearities" without requiring input from the user.

  In our example, the regions resulting from applying the first splitting rule correspond to college graduates and non-college graduates). For college graduates, the partitioning rule that minimizes in-sample MSE is to split this group into those with less than 9.5 years of experience and those with 9.5 years or more of experience. We have thus refined the prediction rule for graduates to be $24 an hour if experience is greater than or equal to 9.5 years, and $17 an hour otherwise. For non-graduates the procedure works similarly, though here the in-sample MSE minimizing split is produced by dividing non-graduates into those with less than 14 years of experience and those with 14 years of experience or more.

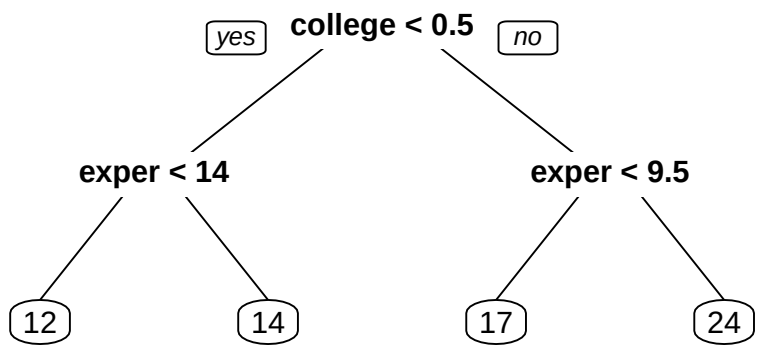

**Figure 9.3:** Depth 2 tree in the wage example

- **Growing the Tree: Higher Levels and Stopping Rule.** To grow deeper trees corresponding to more complex prediction rules, we simply keep repeating. We stop when the desired depth of the tree is reached,[2] or when a prespecified minimal number of observations per region, called minimal node size, is reached.

  In the wage example, we can grow a depth 3 tree by

2: One practical choice of the depth of a tree is to stop just before we get a headache from looking at a complicated tree. This rule is indeed useful if we want to present the tree as a communication device.

repeating the basic procedure within each of the four nodes of the depth 2 tree. The resulting tree is illustrated in Figure 9.4. Here, we see that the indicator for self-reported sex (`female`), high-school graduate indicator (`hsg`), and Southern region indicator (`so`) are the splitting variables chosen in the third level.

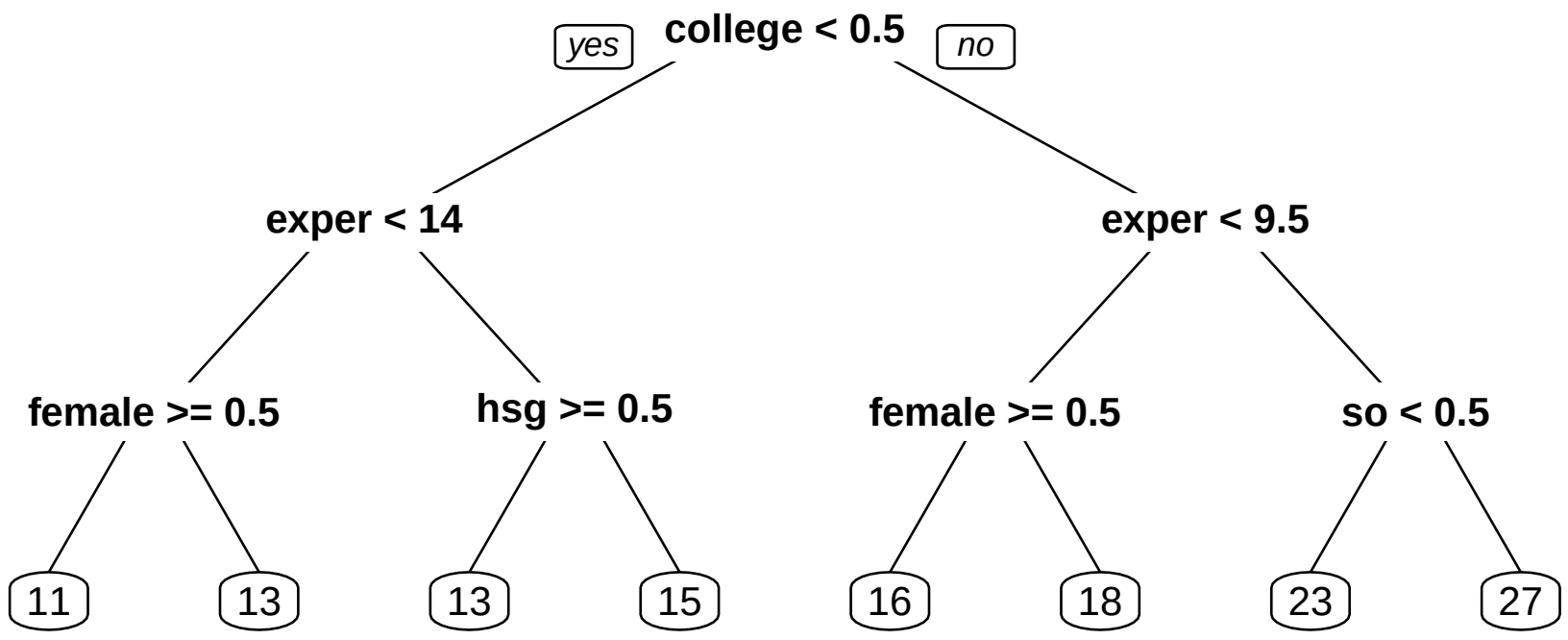

**Figure 9.4:** Depth 3 tree in the wage example. The depth of three was chosen to avoid getting headaches from looking at a more complicated tree.

**Pruning Regression Trees**. We now make several observations.

First, the deeper we grow the tree, the better is our approximation to the regression function $g(Z)$. However, the deeper the tree, the noisier our estimate $\hat{g}(Z)$ becomes, since there are fewer observations per terminal node to estimate the predicted value for this node. From a prediction point of view, we can try to find the right depth or the structure of the tree by a validation exercise such as using a single train/test split or cross-validation. For example, in the wage example, the tree of depth 2 performs better in terms of cross-validated MSE than the tree of depth 3 or 1. The process of cutting down the branches of the tree to improve predictive performance is called "Pruning the Tree."

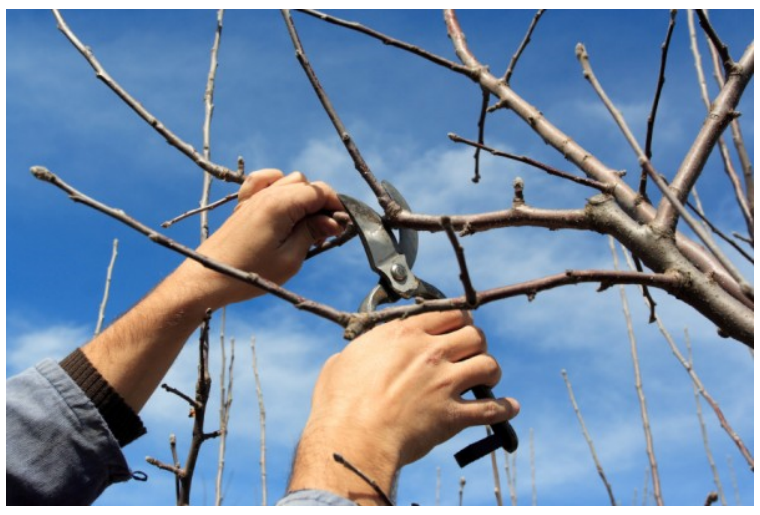

**Figure 9.5:** "To prune a tree." Source: Wikipedia

Often for business analytics and explainability, simple trees like the ones shown are used. If we only care about building good prediction rules, we may build complicated trees and apply pruning to improve predictive performance. A simple penalty for the complexity of the tree is the number of leaves (terminal nodes) times a penalty level, where the penalty level is chosen heuristically; see, e.g, [2]. For example, we can always use a train/test split or cross-validation to settle on a penalty level. There is not a rigorously justified plug-in penalty level for trees like there is for Lasso. Figuring out such a plug-in rule is actually a good research problem.

## Random Forests

In practice, regression trees often do not provide the best predictive performance, because a single regression tree provides a relatively crude approximation to a smooth regression function g(Z). We illustrate the potential poor approximation of regression trees in Figures 9.6 and 9.7. These figures simply illustrate that step functions, which are the outputs of typical regression tree implementations, struggle in approximating smooth functions.

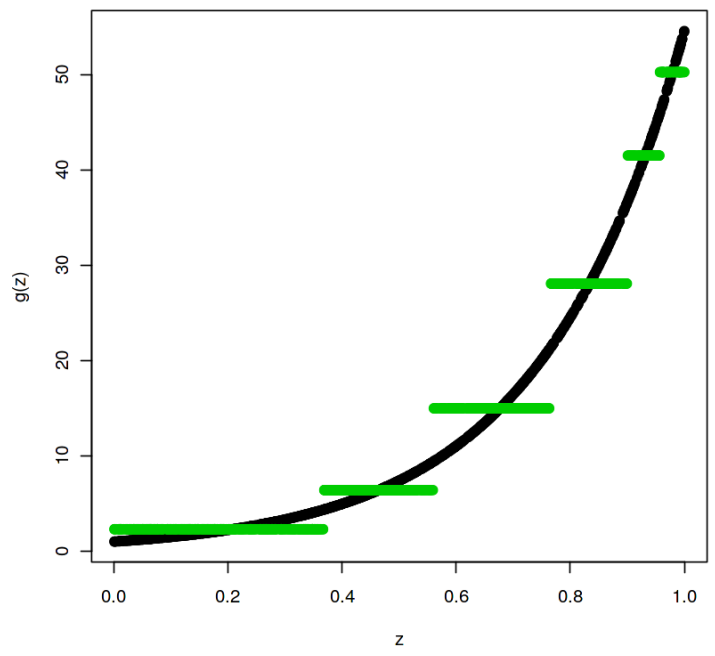

**Figure 9.6:** Approximation of $g(Z) = \exp(4Z)$ by a shallow regression tree in the noiseless case.

A powerful and widely used approach that aims to improve upon simple regression trees is to build a random forest, as proposed by Leo Breiman [3]. The idea of a random forest is to grow many different deep trees that have low approximation error and then average the prediction rules across trees.

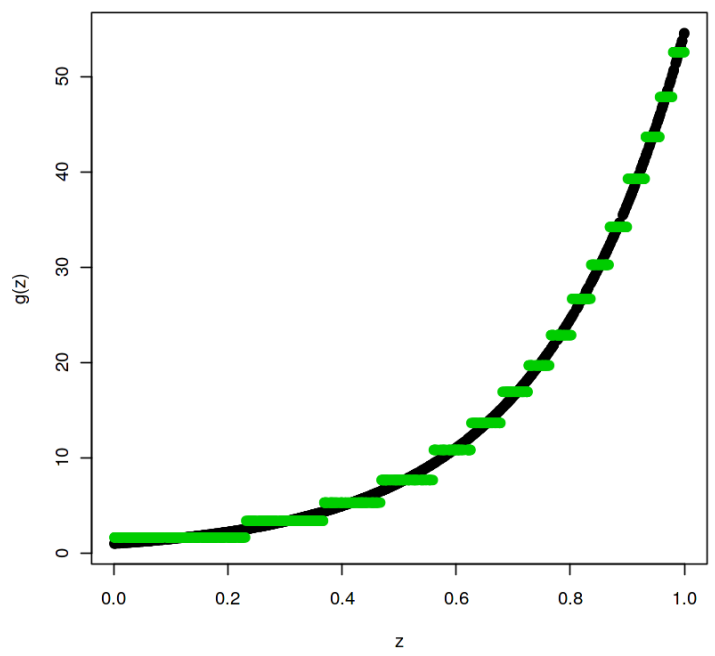

**Figure 9.7:** Approximation of $g(Z) = \exp(4Z)$ by a deep regression tree in the noiseless case.

To produce different trees using only the observed data, the trees going into a random forest are grown from artificial data generated by sampling randomly with replacement from the original data; that is, each tree in a random forest is fit to a *bootstrap sample*.[3] Within the bootstrap samples, trees are grown deep to keep approximation error low. Averaging across the trees produced in the bootstrap samples is then meant to reduce the noisiness of the individual trees. The procedure of averaging noisy prediction rules over bootstrap samples is called Bootstrap Aggregation or Bagging. When the data set is large, we can also rely on fitting trees within *subsamples*[4] instead of using the bootstrap. Using subsamples offers some computational advantages and also simplifies theoretical analysis.

The idea seems very unusual, so let us explain again.

Each bootstrap sample is created by sampling from our data on pairs $(Y_i, Z_i)$ randomly, with replacement. Hence, some observations are drawn multiple times and some aren't redrawn at all. Given a bootstrap sample, indexed by $b$, we build a tree-based prediction rule $\hat{g}_b(Z)$. We repeat the procedure $B$ times in total, and then average the prediction rules that result from each of the bootstrap samples:

$$\hat{g}_{\text{random forest}}(Z) = \frac{1}{B}\sum_{b=1}^{B} \hat{g}_b(Z).$$

The use of the bootstrap here is unusual, yet corresponds to an intuitive idea: If we could have many independent copies of

3: *bootstrap sample*: typically a sample of the same or similar size to the size of the original dataset produced by sampling uniformly from the original data with replacement. Other sampling schemes may also be used, e.g. to accommodate dependence.

4: *subsample*: typically a sample of size much smaller than the original dataset produced by sampling uniformly from the original data without replacement. Other sampling schemes may also be used, e.g. to accomodate dependence.

the data, we could obtain low-bias but potentially very noisy prediction rules in each copy of the data and then average the prediction rules obtained over these copies to reduce the noise. Since we don't have many copies in reality, we rely on the bootstrap to create many quasi-copies of the data. Another feature of this idea is that the cut-points defining partitions for the tree obtained within each bootstrap sample will be different, producing a different step function approximation. Averaging over many step functions with steps at different locations will potentially produce a much smoother approximation to the underlying function. The improved approximation relative to simple trees is illustrated in Figure 9.8.

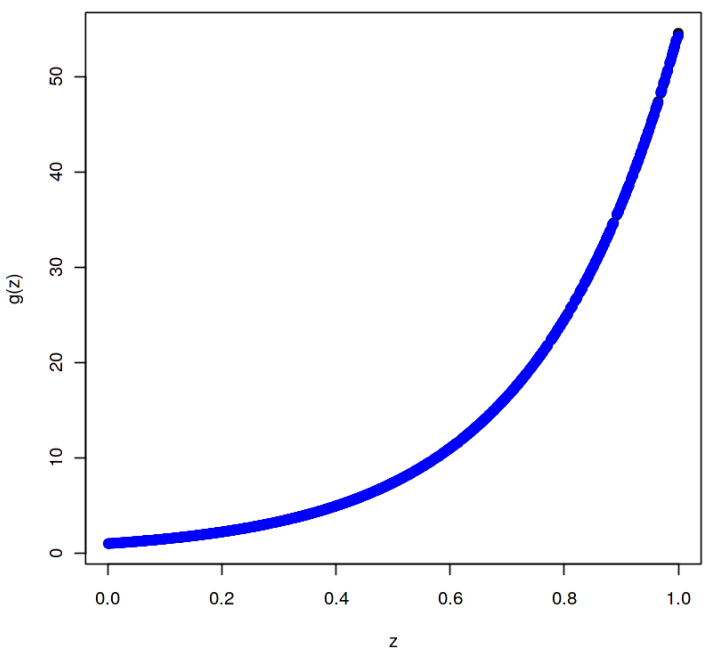

**Figure 9.8:** Approximation of $g(Z) = \exp(4Z)$ by a random forest in the noiseless case.

There are many modifications of the simple version of bootstrap aggregation that we have discussed. The most important modification is the use of additional randomization to "decorrelate" the trees: When we build trees over different bootstrap samples, we also randomize over the variables that trees are allowed to use in forming partitions. This additional layer of randomization encourages trees in different bootstrap samples to have different structure throughout the tree – both near the top and at the bottom – by forcing consideration of distinct sets of variables.

In summary, a random forest is an average of tree based prediction rules (a forest) produced from bootstrap or subsample data (generated randomly).

## Boosted Trees

The idea of boosting is that of recursive fitting: We estimate a simple prediction rule, then take the *residuals*[5] and estimate another simple prediction rule for these residuals. We then take the residuals produced from this new prediction rules and build yet another simple model to predict them. We keep repeating this process until we reach some stopping criterion. The sum of these prediction rules fitted at each step then gives us the overall prediction rule for the outcome.

5: *residuals*: the unexplained part of an outcome we want to predict, after subtracting the prediction from the observed outcome.

Boosting can be applied with any type of base prediction rule. A common use of boosting is with regression trees which leads to *boosted trees*. Boosted trees are built up using shallow trees as the simple prediction rule. Shallow trees are trees with very few levels of depth. By keeping depth low, shallow trees produce low noise prediction rules. However, shallow trees also tend to have high approximation error because they rely on step functions with very few steps to approximate the target regression function. That is, a single shallow regression tree tends to produce a high bias, low variance prediction rule. Boosting then helps alleviate the bias of shallow regression trees. At each step, fitting a model to the residuals from the previous step reduces the approximation error from the previous step. The improved approximation of boosted trees relative to simple trees is illustrated in Figure 9.9.

**The boosting algorithm**

1. Initialize the residuals: $R_i := Y_i, i = 1, ..., n$.
2. For $j = 1, ..., J$
   a) fit a tree-based prediction rule $\hat{g}_j(Z)$ to the data $(Z_i, R_i)_{i=1}^n$;
   b) update the residuals $R_i := R_i - \lambda \hat{g}_j(Z_i)$, where $\lambda$ is called the learning rate.
3. Output the boosted prediction rule:

$$\hat{g}(Z) := \sum_{j=1}^{J} \lambda \hat{g}_j(Z).$$

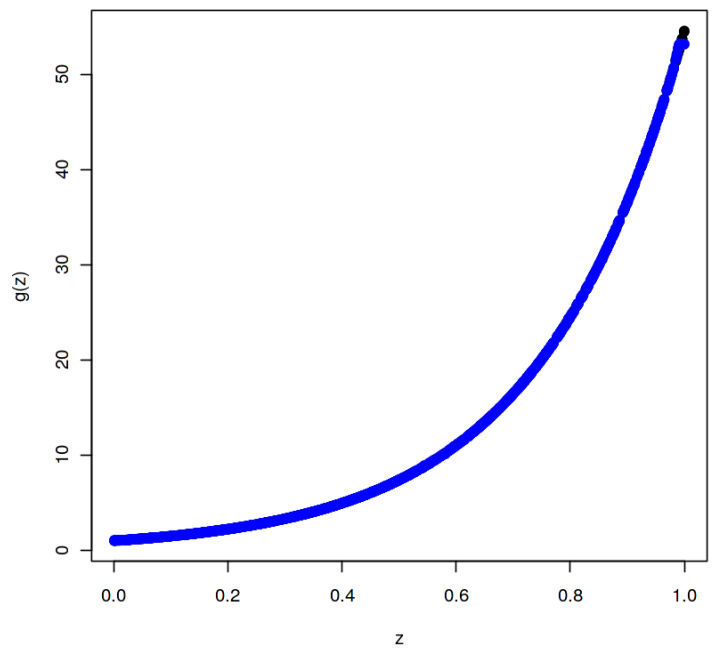

**Figure 9.9:** Approximation of $g(Z) = \exp(4Z)$ by boosted trees in the noiseless case with a sufficient number of steps $J$.

In practice, using boosted trees requires making several choices. One needs to define the tree-based prediction rule used at each step and also choose the number of learning steps, $J$, and the learning rate, $\lambda$. These tuning parameters are typically chosen by cross-validation.[6]

6: We need $0 < \lambda < 1$, and a common default value for $\lambda$ is 0.1. The idea of boosting is to fit simple prediction rules, so one will typically specify the prediction rule by setting the depth of the trees to a small number. For example, at each step, the prediction rule may be a regression tree of depth one (so-called stumps) or depth two. Typically, one will try several small values for depth and again choose among them by cross-validation.

Note that the boosting algorithm is quite general and can be applied to non-tree uses. Note that the number of learning steps for boosting is important across any implementation. Because each step is building a model to predict the unexplained part of the outcome from the previous step, the in-sample prediction errors – the fit to the outcomes used to train the model – must weakly increase with each additional step. If too many iterations are taken, it is thus likely that overfitting will occur, but too few iterations may leave significant bias in the final prediction rule. In practice, the number of iterations is typically chosen by stopping the procedure once there is no marginal improvement to cross-validated MSE. A very popular implementation widely used in industry is xgboost, which has the capability to impose qualitative shape constraints like monotonicity in one or several variables. Other frequently used implementations are lightgbm and catboost.

## 9.3 Neural Nets / Deep Learning

Neural networks are a very powerful tool for modelling non-linear relationships. They rely on many constructed regressors to approximate $g(Z)$, the conditional expectation given the regressors. The method and the name "neural networks" were loosely inspired by the mode of operation of the human brain, and developed by scientists working in Artificial Intelligence. They can be represented by cool graphs and diagrams.

### Basic Ideas

First, we focus on a single layer neural network to introduce the more formal definition of neural nets. The estimated prediction rule will take the form:

$$\hat{g}(Z) := \hat{\beta}' X(\hat{\alpha}) := \sum_{m=1}^{M} \hat{\beta}_m X_m(\hat{\alpha}_m),$$

where the $X_m(\hat{\alpha}_m)$'s are constructed regressors called *neurons*,

$$\alpha = (\alpha_m)_{m=1}^M, \quad \beta = (\beta_m)_{m=1}^M, \quad X(\alpha) = (X_m(\alpha_m))_{m=1}^M.$$

We always take $Z$ to include a constant of 1 as a component and set $X_1(\alpha) = 1$. The remaining neurons are generated as

$$X_m(\alpha_m) = \sigma(\alpha_m' Z), \;\; m = 2, \ldots, M,$$

where $\alpha_m$'s are neuron-specific vectors of parameters called weights, and $\sigma$ is an activation function chosen by the practitioner. Popular activation functions are

- the sigmoid function,

$$\sigma(v) = \frac{1}{1 + e^{-v}},$$

- the rectified linear unit function (ReLU),

$$\sigma(v) = \max(0, v),$$

- the smoothed rectified linear unit function (SReLU),

$$\sigma(v) = \log(1 + \exp(v)),$$

- the leaky rectified linear unit function (Leaky-ReLU),

$$\sigma(v) = \alpha v 1(v < 0) + v 1(v \geq 0)$$

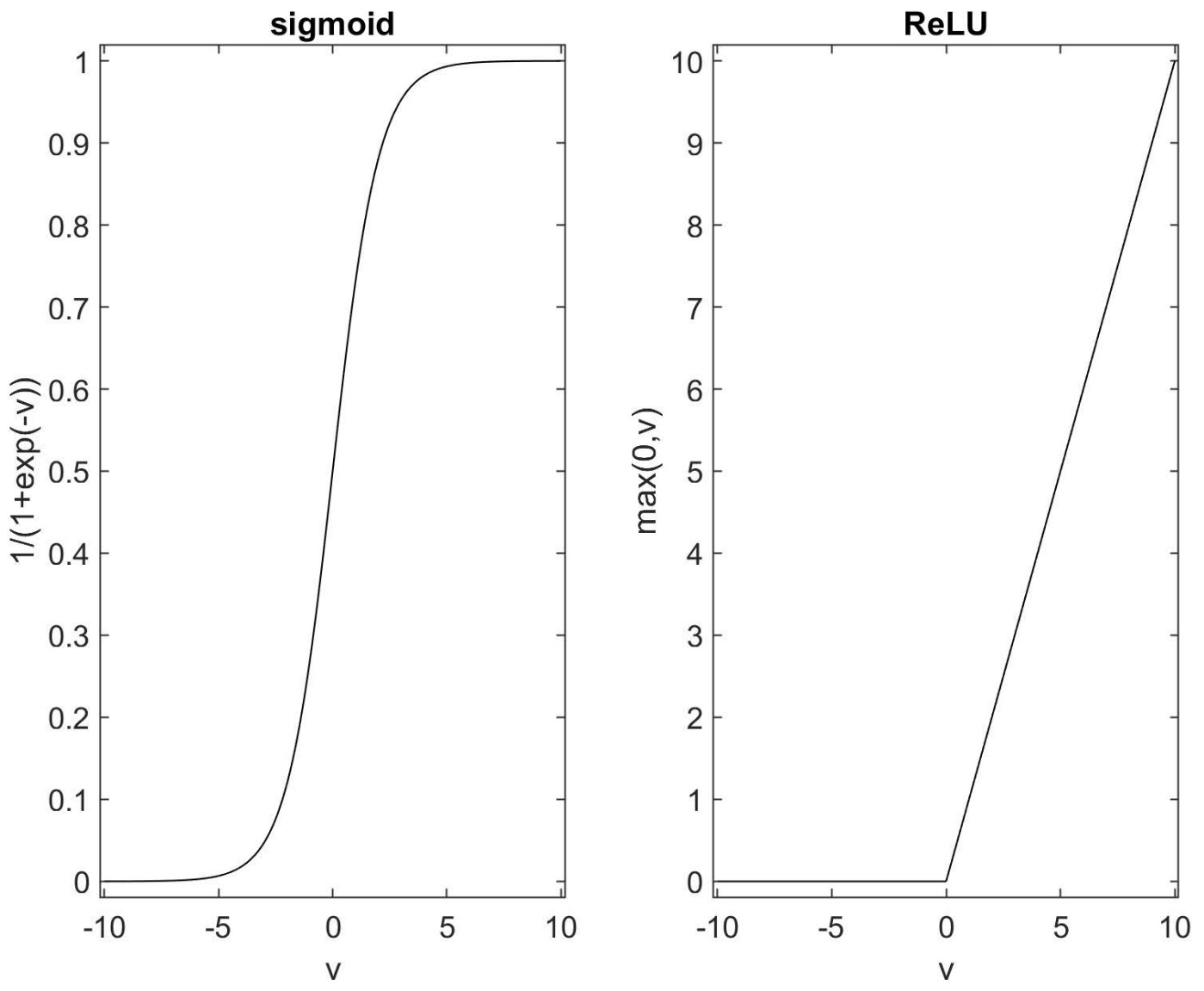

**Figure 9.10:** The sigmoid (logit) and ReLU activation functions

- or the linear function,

$$\sigma(v) = v.$$

The use of nonlinear activation functions is critical for generating high-quality approximations.

The estimators $\{\hat{\alpha}_m\}$ and $\{\hat{\beta}_m\}$, for $m = 1, ..., M$, are obtained as the solution to a penalized nonlinear least squares problem. For example, we could obtain parameter estimates by solving

$$\min_{\{\alpha_m\},\{\beta_m\}} \sum_i \left( Y_i - \sum_{m=1}^{M} \beta_m' X_{im}(\alpha_m) \right)^2 + \text{pen}(\alpha, \beta; \lambda), \quad (9.3.1)$$

where $\text{pen}(\alpha, \beta; \lambda)$ is a penalty function with penalty parameter $\lambda$. Common penalty functions are lasso-type $\ell_1$ penalties,

$$\lambda \left( \sum_m \sum_j |\alpha_{mj}| + \sum_m |\beta_m| \right),$$

and Ridge-type $\ell_2$ penalties,[7]

$$\lambda \left( \sum_m \sum_j (\alpha_{mj})^2 + \sum_m (\beta_m)^2 \right).$$

Neural network estimates are typically computed using stochastic gradient descent (SGD) algorithms. In its simplest version, SGD proceeds as follows: At each step, parameters are updated

7: In many implementations of neural network training the $\ell_2$ penalty is referred to as the "weight decay" parameter; inspired by the fact that the $\ell_2$ penalty adds an extra $-2\lambda w$ term in the gradient calculated at each gradient step of SGD for each parameter $w$, with $w$ being the parameter's current value. Thus it always "decays" the parameter towards zero.

based on the update formula

$$(\alpha, \beta) \leftarrow (\alpha, \beta) - \eta \partial_{\alpha,\beta} \operatorname{Loss}(B; \alpha, \beta)$$

where $B \subset \{1, \ldots, n\}$ is a subset of the samples and the loss is the penalized non-linear least squares objective in Equation (9.3.1) calculated on the subset $B$:

$$\operatorname{Loss}(B; \alpha, \beta) := \sum_{i \in B} \left( Y_i - \sum_{m=1}^{M} \beta_m' X_{im}(\alpha_m) \right)^2 + \operatorname{pen}(\alpha, \beta; \lambda).$$

In other words, every time we take a small step in the direction opposite to an approximate (or stochastic) version of the gradient of the loss that we want to minimize. The gradient designates the direction of parameters towards which the loss increases the most and the opposite is the direction that the loss decreases the most.[8] The magnitude of the step is controlled by the parameter $\eta$, which is many times referred to as the *step-size*.

8: This is typically referred to as the direction of *steepest descent*

In SGD, gradients are computed on subsamples of data (often consisting of a single observation) called batches, and a single cycle through all subsamples is termed an "epoch." By only making use of batches of observations, SGD algorithms are able to scale to massive data sets. Using subsamples of data introduces "stochasticity" relative to using the "full" gradient computed on the entire data. This noise in the computation of gradients also seems to have advantages in helping SGD algorithms avoid local saddle points. There are many fine practical details in terms of efficient computation of gradients for deep neural nets, how updating is done in SGD algorithms in general, and in the application of SGD to learning parameters of deep neural nets.[9]

9: These details are outside of the scope of this monograph. Interested readers might refer to *Deep Learning* by Goodfellow, Bengio, and Courville [4] for a textbook treatment of these issues. A popular method for training neural networks is called Adam; see this Towards Data Science blog for a detailed explanation [5].

The optimization methods employed for learning neural network parameters provide avenues for regularization beyond simply penalizing the size of the coefficients. A popular regularization method is *dropout* regularization where each neuron in a given layer can be set to zero with a given probability – for example, .1 – during parameter update steps. Dropout encourages more robust networks: If a particular neuron is important, the dropout regularization encourages creation of very similar neurons that can replicate the properties of the given neuron. Therefore, dropout regularization can be viewed as a penalty that forces similar weights for groups of neurons.

Another commonly used regularization device used with neural networks is *early stopping*. With early stopping, a measure of out-of-sample prediction accuracy is monitored along with the value of the in-sample objective function (9.3.1). Rather than optimizing until the in-sample objective function is minimized, optimization proceeds until out-of-sample performance appears to start to degrade. By updating parameters based on in-sample fit but stopping based on out-of-sample performance, early stopping helps guard against overfitting.

As can be seen from the preceding paragraphs, using neural networks in practice relies on the choice of many tuning parameters. As there is relatively little theoretical guidance on these choices, tuning parameters are typically chosen using data splitting. An important choice that clearly relates to model flexibility is the number of neurons and neuron layers when considering the deeper networks discussed below. Having more neurons or layers gives us additional flexibility, just like having more constructed regressors provides more flexibility in high-dimensional linear models. Other choices about regularization then interact with the choice of how many neurons and layers to use in preventing overfitting.[10]

10: There has been a flurry of recent research considering the use of very large neural networks with many more parameters than the number of observations that may easily overfit the data. These papers find that such highly overparameterized neural networks tend to find solutions that generalize well.

To visualize the working of a neural network, we rely on a resource called playground.tensorflow.org [7], with which we produced a prediction regression model using a simple single layer neural network model based on two input variables. A screenshot taken after training the model is shown below.

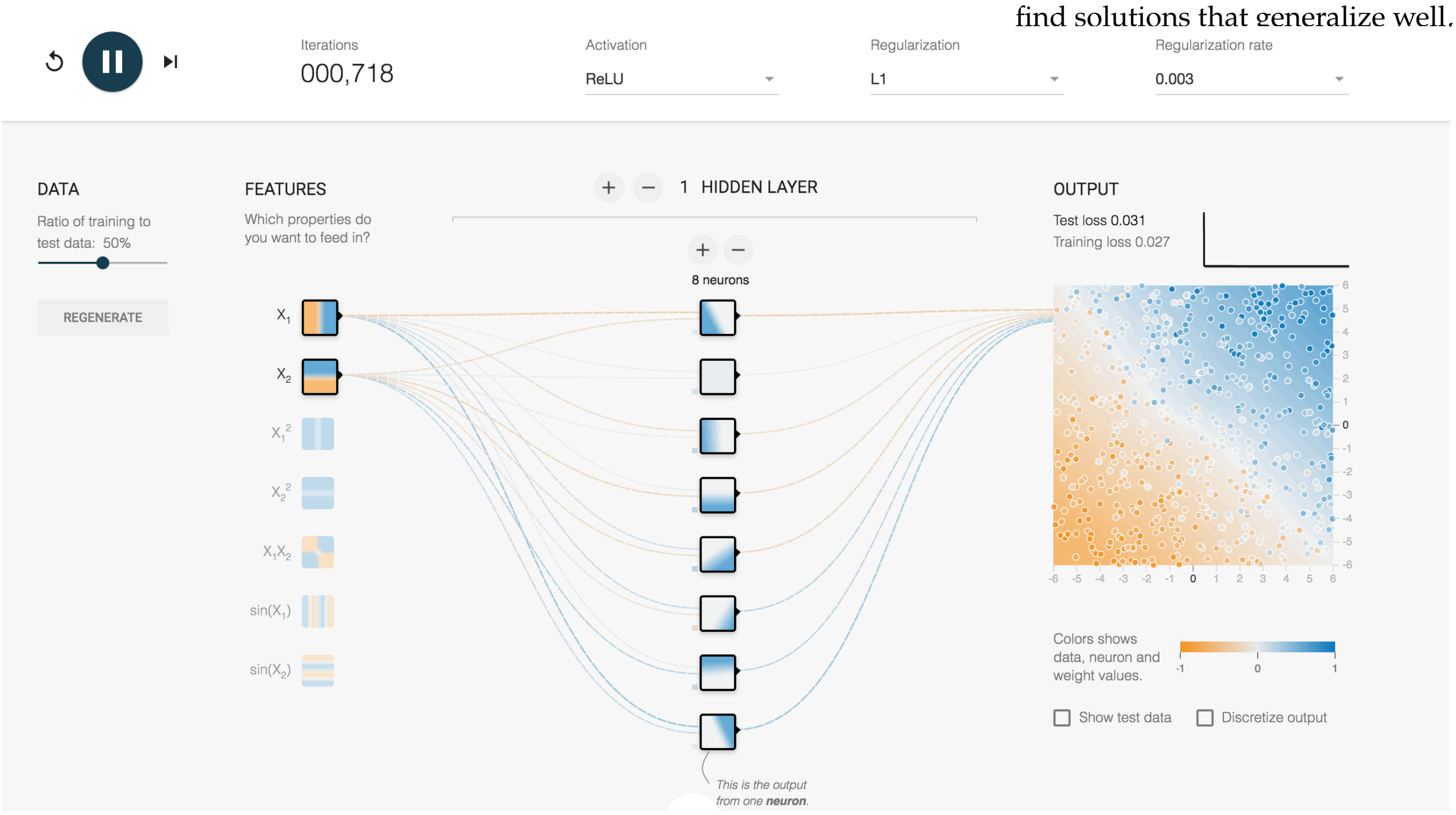

The network depicts the process of taking raw regressors and transforming them into predicted values. In the second column

(labeled "FEATURES"), we see the inputs – our two raw regressors. The third column depicts a "hidden layer" made up of eight neurons.[11] Each neuron is constructed as a (weighted) linear combination of the raw regressors transformed by an activation function. Here we use the ReLU activation function. The neurons are connected to the inputs and the connections represent the $\hat{\alpha}_m$ coefficients. The coloring represents the sign of the coefficients (orange is negative and blue positive) and the width of the connections represents the size of the coefficients.

11: "Hidden" refers to the fact that these layers are typically not reported. However, these layers can be extracted and used as technical regressors for other tasks. We discuss using hidden layers as features in Chapter 11 which deals with feature engineering.

Finally, the neurons are combined linearly to produce the output – the prediction rule. The connections going outwards from the neurons to the output represent the coefficients $\hat{\beta}_m$ of the linear combination of the neurons that produce the final output. The coloring and the width again represent the sign and the size of these coefficients.

The output (prediction) is shown here by the "heat" map in the box on the right. On the horizontal and vertical axes we see the values of the two inputs. The color and its intensity in the "heat" map represent the predicted value.

At the top of the screenshot, we also see that we used "L1" for the type of regularization, which corresponds to using the Lasso type penalty. Here, the penalty level is called the regularization rate and is provided as the last entry in the top line of the screenshot.

In this example, we used a single layer neural network. If we add one or two additional layers of neurons constructed from the previous layer of neurons we get a "deep" network. We illustrate a two-layer network in the following figure.

Prediction methods based on neural networks with several layers of neurons are called "deep learning" methods.

## Deep Neural Networks

Here, we present the structure of a neural network with general depth. Networks with depth greater than one are called deep neural networks (DNN).

For the sake of generality, we consider networks of the multitask form, where we try to predict multiple outputs $Y^t$, $t = 1, ..., T$, where $t$ stands for the "task."[12] A typical scenario is to just have one task, $T = 1$, as in all of our preceding discussion. However, there are many cases where we can use a single DNN to solve multiple tasks.

12: For example, we might be interested in predicting the price of a product using product characteristics across multiple markets or time periods, $t$. In treatment effect analysis, we may build a single neural network to predict both the outcome, $Y$, and the treatment, $D$, using other covariates. We could view this as a multitask learning problem where we are interested in two outputs, $Y^1 = Y$ and $Y^2 = D$.

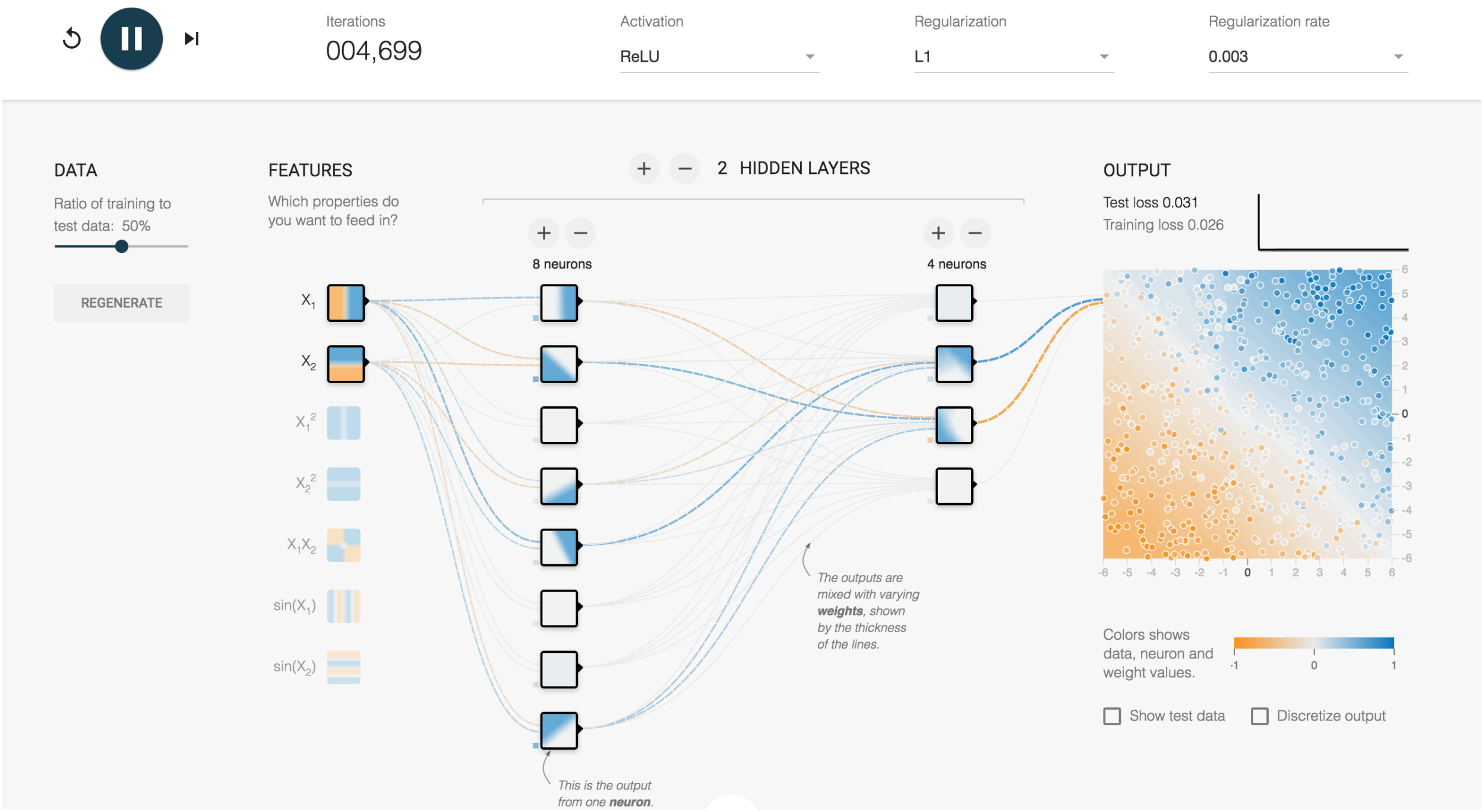

The general nonlinear regression model we work with takes the form

$$Z \overset{f_1}{\longmapsto} H^{(1)} \overset{f_2}{\longmapsto} \quad \dots \overset{f_m}{\longmapsto} H^{(m)} \overset{f_{m+1}}{\longmapsto} \{X^t\}_{t=1}^T, \qquad (9.3.2)$$

where

$$H^{(\ell)} = \{H_k^{(\ell)}\}_{k=1}^{K_\ell}$$

are called neurons, $Z$ is the original input, and the map $f_\ell$ maps one layer of neurons to the next. The maps $f_\ell$ are defined as

$$f_\ell : v \longmapsto \{H_k^{(\ell)}(v)\}_{k=1}^{K_\ell} := (1, \{\sigma_{k,\ell}(v'\alpha_{k,\ell})\}_{k=2}^{K_\ell}), \qquad (9.3.3)$$

where $\sigma_{k,\ell}$ is the activation function that can vary with the layer $\ell$ and across neurons $k$ in a given layer. We always include a constant of 1 as a component of $Z$, and we always designate one of the neurons in each layer up to $m$ to be 1. The final layer, $f_{m+1}$, does not output the constant of 1 as a component:[13]

$$f_{m+1} : v \longmapsto \{X^t(v)\}_{t=1}^T := (\{\sigma_{t,m+1}(v'\alpha_{t,\ell})\}_{t=1}^T). \qquad (9.3.4)$$

13: Common architectures employ activation functions that do not vary with $k$. However, custom architectures, such as ResNet50 discussed in Figure 9.13, can be viewed as having an activation function that depends on $k$, with some neurons linearly activated and some non-linearly.

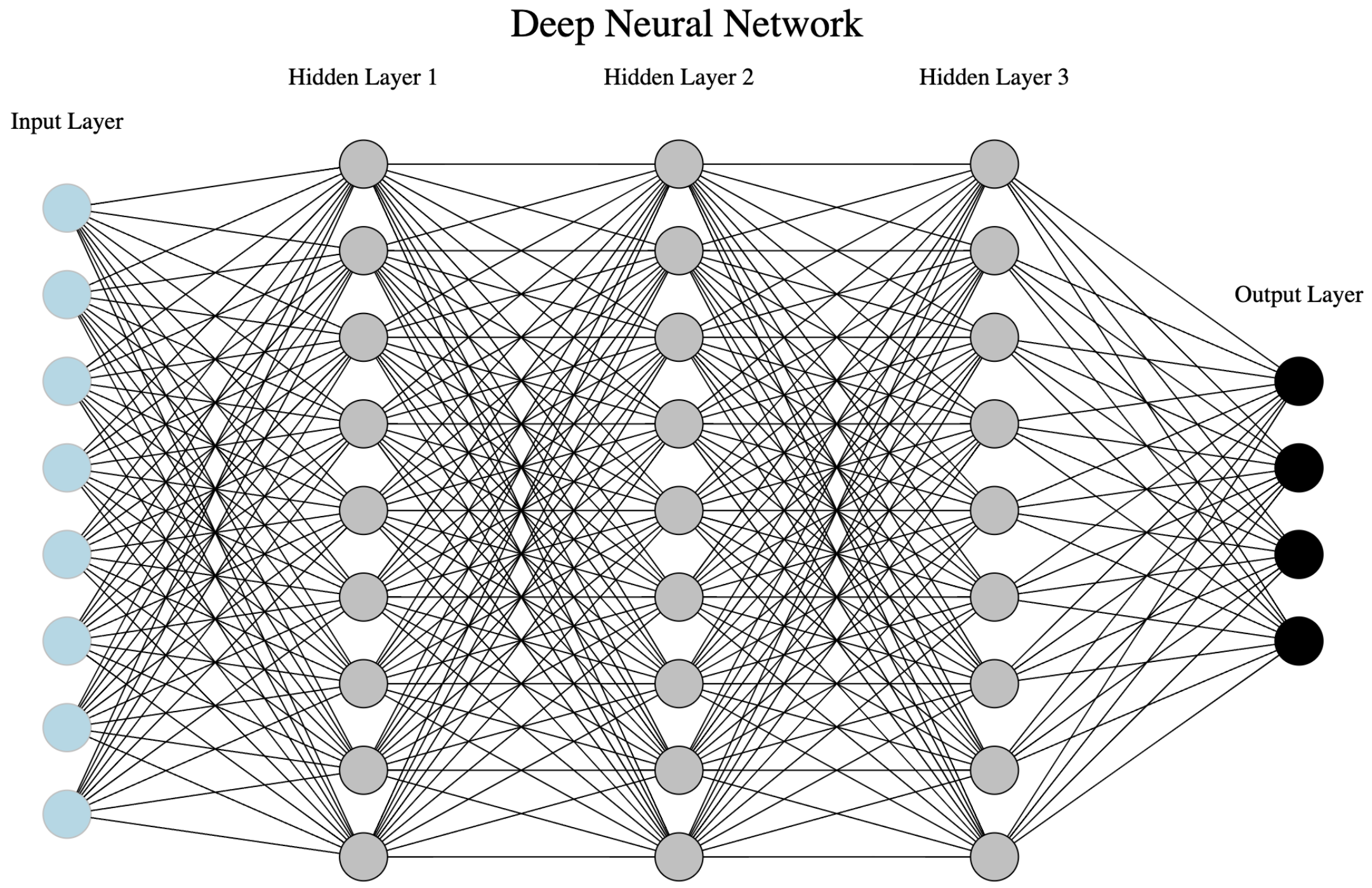

**Figure 9.11:** Standard Architecture of a Deep Neural Network. The input is mapped nonlinearly into the first hidden layer of the neurons. The output of this first mapping is then mapped nonlinearly into the second layer. This process is then repeated $m$ times. The output of the penultimate layer is finally mapped (linearly or nonlinearly) into the output layer, which can have multiple outputs corresponding to different tasks.

The network mapping (9.3.2) consists of repeated composition of nonlinear mappings. This structure has been shown to be an extremely powerful tool for generating flexible functional forms which yields successful approximations in a wide range of empirical problems and is backed by approximation theory. Good approximations can be achieved by both considering sufficiently many neurons and sufficiently many layers (Yarotsky, 2017 [8]; Schmidt-Hieber, 2020 [9]; Farrell et. al, 2021 [10]; Kidger and Lyons, 2020 [11]). In empirical economic examples, it is common to just use a few hidden layers, while much deeper networks are typically used in image processing and text applications.

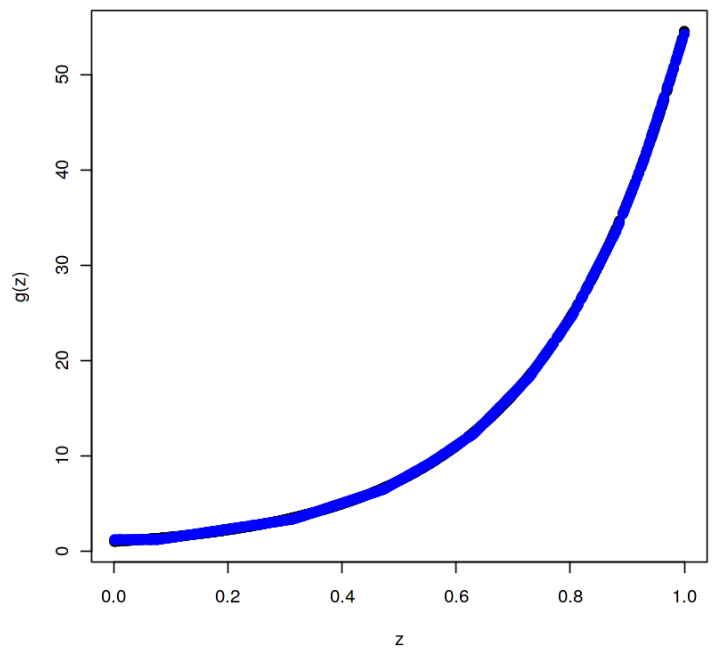

**Figure 9.12:** Approximation of $g(Z) = \exp(4Z)$ by a Neural Network

Similarly to single layer neural networks, the DNN model can be trained by minimizing the loss function

$$\min_{\eta \in \mathcal{N}} \sum_t w_t \sum_i (Y_i^t - X_i^t(\eta))^2 + \text{pen}(\eta; \lambda), \tag{9.3.5}$$

where $\eta$ denotes all of the parameters of the mapping

$$Z_i \mapsto X_i^t(\eta),$$

$w_t$ denotes the weight given to a task $t$, and $\text{pen}(\eta; \lambda)$ is a penalty function with $\lambda$ denoting the penalty level.

## 9.4 Prediction Quality of Modern Nonlinear Regression Methods

As we have already mentioned, the best prediction rule for an outcome $Y$ using features/regressors $Z$ is the function $g(Z)$, equal to the conditional expectation of $Y$ using $Z$:

$$g(Z) = \mathrm{E}[Y \mid Z].$$

Modern nonlinear regression methods, when appropriately tuned and under some regularity conditions, provide estimated prediction rules $\hat{g}(Z)$ that approximate the best prediction rule $g(Z)$ well.

Theoretical work demonstrates that under appropriate regularity conditions and with appropriate choices of tuning parameters, the mean squared approximation error of prediction rules produced by modern nonlinear regression methods is small once the sample size $n$ is sufficiently large, namely,

$$\|\hat{g} - g\|_{L^2(Z)} = \sqrt{\mathrm{E}_Z[(\hat{g}(Z) - g(Z))^2]} \to 0, \quad \text{as } n \to \infty,$$

where $\mathrm{E}_Z$ denotes the expectation taken over $Z$, holding everything else fixed. To deliver these guarantees in high-dimensional settings where the number of features is large, we rely on structured assumptions, such as sparsity in the case of Lasso. Under these conditions we expect that the in-sample MSE and $R^2$ would agree with the out-of-sample MSE and $R^2$.

### Learning Guarantees of DNNs

We say that a function $g : \mathbb{R}^d \to \mathbb{R}$ is $\beta$-smooth if it has $\beta \geq 1$ continuous and uniformly bounded higher-order derivatives.[14] If the regression function $g$ is only known to be $\beta$-smooth, then the best estimator of this function has estimation error, in the worst case, that converges at the rate

$$n^{-\beta/(2\beta+d)},$$

as shown by Charles Stone [12]. When $d$ is not small, this rate of convergence is extremely slow, suggesting that learning a function in $d$ variables is difficult if the dimension $d$ is moderate and the target function is only known to be $\beta$-smooth.[15]

We can achieve better rates of convergence under some kind of structured sparsity or parsimony assumptions as we saw

14: A more general definition allows $\beta$ to be non-integer, but we focus on integer $\beta$ for simplicity.

15: For instance, suppose that $\beta = 1$, i.e. the function is simply assumed to have a uniformly bounded first-order derivative. Moreover, suppose that we have $d = 10$ variables. Then the bound says that if we want an error of $\epsilon = 0.1$, we need $n$ to be such that $n^{-1/12} \approx 0.1$, equivalently we need $n \approx 10^{12} = 1$ trillion samples! If $\beta = 2$, we would only need a petty 10 million samples...

in the rates for high-dimensional linear models in Chapter 3. DNNs are able to take advantage of a nonlinear form of sparsity assumptions that we formulate below following Schmidt-Hieber [9].[16]

16: See also [13] for more recent theoretical developments on provable guarantees for neural networks under sparsity conditions.

**Assumption 9.4.1** (Structured Sparsity of Regression Function) *We assume that $g$ is generated as a composition of $q+1$ vector-valued functions:*

$$g = f_q \circ \ldots \circ f_0$$

*where the i-th function $f_i$*

$$f_i : \mathbb{R}^{d_i} \longrightarrow \mathbb{R}^{d_{i+1}},$$

*has each of its $d_{i+1}$ components $\beta_i$-smooth and depends only on $t_i$ variables, where $t_i$ can be much smaller than $d_i$.*

The rate guarantee will depend on the parsimony/smoothness pairs:

$$(t_i, \beta_i), \quad i = 0, \ldots, q.$$

For example, consider $g : \mathbb{R}^{100} \mapsto \mathbb{R}$,

$$g(x_1, x_2, x_3, x_4, ..., x_{100}) = f_1(f_{01}(x_3), f_{02}(x_2)).$$

Then

$$g_0 = f_1 \circ f_0; \quad d_0 = 100, d_1 = 2; \quad t_0 = 1, t_1 = 2.$$

**Theorem 9.4.1** (Learning Guarantee for DNNs under Approximate Sparsity) *Suppose that (a) the regression function $g$ obeys the structured sparsity assumption (Assumption 9.4.1); (b) the depth of the DNN model is proportional to* $\log n$, *(c) the width of the DNN model is no less than*

$$s \cdot \log n$$

*where $s$ is the effective dimension of the regression function $g$,*

$$s := \max_{i=0,\ldots,q} n^{\frac{t_i}{2\beta_i + t_i}};$$

*and (d) other regularity conditions hold as specified in [9]. Then, there exists a sparse DNN estimator $\hat{g}$ with order $s \log n$ non-zero*

*parameters such that, with probability approaching 1,*

$$\|\hat{g} - g\|_{L^2(Z)} \leq \mathrm{const}_{\mathrm{P}} \sigma \sqrt{\frac{s}{n}}\, \mathrm{polylog}(n),$$

*where* $\mathrm{polylog}(n)$ *is a polynomial in* $\log(n)$, $\sigma^2 = \mathrm{E}[(Y - g(Z))^2]$, *and* $\mathrm{const}_{\mathrm{P}}$ *is a constant that depends on the distribution of the data.*

This fundamental result is due to Schmidt-Hieber [9], where the reader may find the complete statement of regularity conditions and further technical details of the result.

In the example above, despite the high-dimensional setting, $d = 100$, if $f_{01}, f_{02}, f_{11}$ are $\beta$-smooth with $\beta \geq 2$, a sparse DNN is able to achieve the rate (ignoring logs):[17]

$$\sqrt{\frac{s}{n}} = n^{-\beta/(2\beta+2)} \leq n^{-1/3}$$

where the effective dimension is

$$s = n^{\frac{2}{2\beta+2}}.$$

17: Comparing to our earlier numerical example that made no sparsity assumptions, here we see that irrespective of the number of input variables, if we want the error to be $\epsilon = 0.1$, then we need $n \approx 1000$ samples, which is more realistic.

## Learning Guarantees of Trees and Forests

One important property of adaptively built trees is that they are able to identify the relevant dimensions along which the regression function varies. To isolate this type of behavior of trees and forests, we consider a setting where all the regressors are binary, i.e. $Z \in \{0, 1\}^d$. This is without loss of generality for categorical (discrete-valued) regressors, since each level of the regressor can be coded as a binary indicator.[18]

18: Continuous regressors can also be discretized. However, discretization entails some loss of generality, and approximation properties following discretization have not been formally investigated.

Without further assumptions on the regression function $g : \{0, 1\}^d \to \mathbb{R}$, the best convergence rates that one could hope for scale at least at a $\sqrt{2^d/n}$ rate. Even for a moderate number of variables $d$, this rate of convergence can be prohibitively slow.

Adaptively built trees are particularly successful when there is only a small subset $S$, of size $|S| = r$, among the $d$ variables that is relevant. Using this principle, we can formulate a non-parametric analogue of the sparsity assumption that we analyzed in the case of high-dimensional linear regression with Lasso that allows us to improve on the convergence rate obtained without restrictions.

**Assumption 9.4.2** (Nonparametric Sparsity of a Regression Function with Binary Regressors) *We assume that there exists a subset $S$ of size $|S| = r$, such that the function $g$ can be written as a function of only the variables in $S$; i.e. we can write*

$$g(Z) = f(Z_S)$$

*where $Z_S$ is the subvector of $Z$ containing only the coordinates in $S$.*

The assumption can probably be relaxed to "approximate" sparsity.[19]

19: This relaxation has not been formally investigated.

Observe that, unlike the sparsity assumption we made in the case of high-dimensional penalized linear regression, Assumption 9.4.2 imposes no restrictions on the form of the function $f$ that takes as input the relevant variables. Here, under the nonparametric sparsity assumption together with several other regularity conditions, we can prove that the mean squared approximation error of shallow regression trees or "honest" and arbitrarily deep regression forests[20] scales at a

$$\sqrt{2^r \log(d) \log(n)/n}$$

rate. Thus, the convergence rate depends strongly on the sparsity level $r$ while the overall number of regressors $d$ enter only logarithmically. Moreover, even if we knew the relevant variables $S$, we could not hope for a rate faster than $\sqrt{2^r/n}$ since we make no further assumptions on the function $f$. Thus not knowing the relevant set of regressors $S$ adds an extra multiplicative cost on the achievable rate that only grows logarithmically with the number of regressors and the sample size. See [14] for results of similar flavor for variants of regression trees in settings beyond the binary regressor case.

20: An "honest" training approach makes use of subsampling. See Theorem 9.4.3 and the discussion immediately preceding its statement.

**Theorem 9.4.2** (Learning Guarantee for Shallow Regression Trees) *Suppose that (a) the regressors are binary and the outcome variable is bounded; (b) the regression function $g$ obeys Assumption 9.4.2; (c) regularity conditions hold that lower bound the density of the support of the distribution of covariates and upper bound the degree of variance reduction in MSE that can be achieved by features not in $S$ [15]. Then a regression tree estimator $\hat{g}$, where the regression tree is greedily grown based on the MSE criterion up to a depth that is at least $r$ and at most some constant multiple of $r$,*

A greedy algorithm is any algorithm that follows the problem-solving heuristic of making the locally optimal choice at each stage. In our case, a greedily grown tree optimizes over the name of regressor and splitting point that achieve the best one-step improvement in the in-sample MSE at each node.

*satisfies, for* $n \geq \text{const}_P 2^r \log(d/\delta)$*, with probability* $1 - \delta$*,*

$$\|\hat{g} - g\|_{L^2(Z)} \leq \text{const}_P \sigma \sqrt{\frac{2^r \log(d/\delta) \log(n)}{n}},$$

*where* $\sigma^2 = \mathrm{E}[(Y - g(Z))^2]$ *and* $\text{const}_P$ *is a constant that depends on the distribution of the data.*

Capping the depth of the regression tree as in Theorem 9.4.2 helps avoid overfitting, since otherwise we could potentially construct binary trees that achieve zero error on the training data and have large error out-of-sample.

An alternative to avoiding overfitting is to use an ensemble approach based on sub-sampled data. To implement an ensemble approach, we train multiple regression trees, each on a random sub-sample (without replacement) of the original data-set of size $s < n$ and average the predictions of each of these trees. Moreover, to formally argue about the approximation error of such sub-sampled forests, we will require the forests to be trained in an "honest" manner.

In our setting, an honest training approach is as follows: When we train a tree on a sub-sample, we randomly partition the data in half and we use half of the data to find the best splits in a greedy manner, and the other half of the data to construct the estimates at each node of the tree. Such sub-sampled honest forests have been recently popularized by the work of [16]. Subsequent work of [15] showed that honest forests provably adapt to non-parametric sparsity of the regression function.

**Theorem 9.4.3** (Learning Guarantee for Sub-Sampled Honest Forests) *Suppose that (a) the regressors are binary and outcome variable is bounded; (b) the regression function* $g$ *obeys Assumption 9.4.2; (c) regularity conditions hold that lower bound the density of the support of the distribution of covariates and upper bound the degree of variance reduction in MSE that can be achieved by features not in* $S$ *[15]. Then a regression forest estimator* $\hat{g}$*, where each regression tree is built in an honest manner and on a random sub-sample (without replacement) of size* $s = \text{const}_P\, 2^r \log(d/\delta)$ *of the original data, satisfies, for* $n \geq \text{const}_P\, 2^r \log(d/\delta)$ *with probability* $1 - \delta$*,*

$$\|\hat{g} - g\|_{L^2(Z)} \leq \text{const}_P \sigma \sqrt{\frac{2^r \log(d/\delta)\ \text{polylog}(n)}{n}}$$

*where* $\sigma^2 = \mathrm{E}[(Y - g(Z))^2]$ *and* $\text{const}_P$ *is a constant that depends on the distribution of the data and* $\text{polylog}(n)$ *is a polynomial factor* *of* $\log(n)$.

The rate guarantee for Honest Forests in Theorem 9.4.3 is the same as the rate for shallow trees in Theorem 9.4.2. This theory thus does not shed light on why random forests seem to achieve superior predictive performance over simple trees in many applications. Moreover, practical random forest algorithms tend to work well with default tuning choices, whereas the theory requires a careful alignment of the tuning parameters to get good rate guarantees. The regularity conditions also require the explanatory power of the subset of the covariates that are relevant, $S$, to dominate the explanatory power of the irrelevant covariates.[21] This condition on signal strength is a sufficient condition, but it may not be necessary for good performance. That is, there seem to remain substantial gaps in our theoretical understanding of the performance of tree-based algorithms. Further exploring these properties may be an interesting area for further study.

21: Irrelevance here only means that, given the set $S$ of relevant covariates, the other variables do not contribute to the best prediction rule. It does not mean that the irrelevant covariates have no predictive power on their own.

## Trust but Verify

Both tree-based methods and neural networks provide powerful, flexible models that can deliver high-quality approximations of regression functions. However, the high degree of flexibility can lead to overfitting. Therefore, it is always important to verify the performance on test data to make sure that the predictive model being used is actually a good one.

A simple verification procedure is data splitting, which can be performed in the following way:

1. We use a random subset of data for estimating/training the prediction rule.
2. We use the other part of the data to evaluate the quality of the prediction rule, recording out-of-sample mean squared error, $R^2$, or some other desired measure of prediction quality.

Recall that the part of the data used for estimation is called the training sample. The part of the data used for evaluation is called the testing or validation sample. We have a data sample containing observations on outcomes $Y_i$ and features $Z_i$. Suppose we use $n$ observations for training and $m$ for testing/validation. We use the training sample to compute prediction rule $\hat{g}(Z)$. Let $V$ denote the indices of the observations in the test sample. Then the out-of-sample/test mean squared error is

$$\mathrm{MSE}_{test} = \frac{1}{m}\sum_{k\in V}(Y_k - \hat{g}(Z_k))^2.$$

The out-of-sample/test $R^2$ is[22]

$$R^2_{test} = 1 - \frac{\mathrm{MSE}_{test}}{\frac{1}{m}\sum_{k\in V} Y_k^2}.$$

22: In typical empirical applications, these quantities are calculated after de-meaning/centering the outcome.

## A Simple Case Study using Wage Data

We illustrate ideas using a data set of 5150 observations from the March Current Population Survey Supplement 2015. $Y_i$'s are log wages of never-married workers living in the U.S. $Z_i$'s include experience, education, 23 industry and 22 occupation indicators, and some other characteristics. We consider a variety of linear and nonlinear rules for predicting $Y$ with $Z$.

For the linear models, we estimate prediction rules of the form $\hat{g}(Z) = \hat{\beta}'X$ using $X$ generated in two ways:

- (basic model) $X$ consists of the 51 raw regressors in $Z$.
- (flexible model) $X$ consists of 246 variables composed of the 51 raw regressor in $Z$, a fourth order polynomial in experience, and two-way interactions between the polynomial terms in experience and the non-experience variables in $Z$.

We estimate $\hat{\beta}$ by linear regression/least squares and by the following penalized regression methods: Lasso and Post-Lasso with plug-in choice of $\lambda$, cross-validated Lasso, Ridge, and Elastic Net.

For the nonlinear models, we estimate prediction rules of the form $\hat{g}(Z)$ without imposing that $\hat{g}(Z) = \hat{\beta}'X$. That is, we do not assume prediction rules to be linear. We estimate the prediction models by random forests, regression trees, boosted trees, and Neural Networks. We use an implementation of the random forest where, at the step of growing a regression tree, we choose the best variable to split upon among $\sqrt{p} \ll p$ randomly selected variables.

Table 9.1 displays results based upon a single split of data into training and testing sets. It shows the test MSE in column 1, the standard error of the test MSE in column 2, and the test $R^2$ in column 3. We see that the best performing prediction rules are provided by OLS using the raw 51 regressors and Lasso using the basic 51 predictors with penalty parameter selected by cross-validation. The performance of both Elastic Net with the basic set of regressors and boosted trees are also nearly identical to those of the two best methods. Looking at standard errors, we see that the vast majority of methods have test MSE's that are within one standard error of the best test MSE, suggesting relatively little difference in performance across methods.

| | MSE | S.E. | $R^2$ |
|---|---|---|---|
| Least Squares (basic) | 0.229 | 0.016 | 0.282 |
| Least Squares (flexible) | 0.243 | 0.016 | 0.238 |
| Lasso | 0.234 | 0.015 | 0.267 |
| Post-Lasso | 0.233 | 0.015 | 0.271 |
| Lasso (flexible) | 0.235 | 0.015 | 0.265 |
| Post-Lasso (flexible) | 0.236 | 0.016 | 0.261 |
| Cross-Validated Lasso | 0.229 | 0.015 | 0.282 |
| Cross-Validated Ridge | 0.234 | 0.015 | 0.267 |
| Cross-Validated Elastic Net | 0.230 | 0.015 | 0.280 |
| Cross-Validated Lasso (flexible) | 0.232 | 0.015 | 0.275 |
| Cross-Validated Ridge (flexible) | 0.233 | 0.015 | 0.271 |
| Cross-Validated Elastic Net (flexible) | 0.231 | 0.015 | 0.276 |
| Random Forest | 0.233 | 0.015 | 0.270 |
| Boosted Trees | 0.230 | 0.015 | 0.279 |
| Pruned Tree | 0.248 | 0.016 | 0.224 |
| Neural Net | 0.276 | 0.012 | 0.148 |

**Table 9.1:** Prediction Performance for the Test/Validation Sample.

The outliers, in terms of performing relatively poorly, are OLS using the flexible set of covariates as well as the regression tree (Pruned Tree) and the neural net. OLS with the flexible set of predictors uses a relatively large number of variables relative to the sample size and seems likely to be overfit. On the other hand, neither the regression tree nor the neural net is fully tuned. Thus, there may be room to improve the performance of these methods.

## 9.5 Combining Predictions - Aggregation - Ensemble Learning

Given different prediction rules, we can choose either a single method or an aggregation of several methods as our prediction approach. An aggregated prediction is a linear combination of the basic predictors.

In econometrics and statistics, the procedures for combining several methods are called "model averaging" and "aggregation." In machine learning, these terms are relabeled as "ensembles" and "stacking."

Specifically, we consider an aggregated prediction rule of the

form:

$$\tilde{g}(Z) = \sum_{k=1}^{K} \tilde{\alpha}_k \hat{g}_k(Z),$$

where $\hat{g}_k$'s denote basic predictors, potentially including a constant. The basic predictors are computed on the training data.

If the number of prediction rules, $K$, is small, we can figure out the coefficients of the optimal linear combination of the rules, $\tilde{\alpha}_k$, using test data $V$ by simply running least squares of the outcomes in the test data on their associated predicted values:

$$\min_{(\alpha_k)_{k=1}^K} \sum_{i \in V} \left( Y_i - \sum_{k=1}^{K} \alpha_k \hat{g}_k(Z_i) \right)^2 .$$

We wish to emphasize that here we are minimizing the sum of squared prediction errors in the test sample using the prediction rules from the training sample as the regressors. If $K$ is large, we can instead use Lasso for aggregation:

$$\min_{(\alpha_k)_{k=1}^K} \sum_{i \in V} \left( Y_i - \sum_{k=1}^{K} \alpha_k \hat{g}_k(Z_i) \right)^2 + \lambda \sum_{k=1}^{K} |\alpha_k|.$$

### Aggregation Results for the Case Study

We consider the prediction rules based on OLS, Post-Lasso, Elastic Net, Pruned Tree, random forest and boosted trees to build an ensemble method.

| | Weight OLS | Weight Lasso |
|---|---|---|
| Constant | -0.162 | -0.147 |
| Least Squares (basic) | 0.281 | 0.293 |
| Post-Lasso (flexible) | 0.237 | 0.223 |
| CV Elastic Net (flexible) | -0.068 | -0.056 |
| Pruned Tree | -0.140 | 0.000 |
| Random Forest | 0.377 | 0.344 |
| Boosted Trees | 0.367 | 0.245 |

**Table 9.2:** Weights of the ensemble method.

The estimated weights are shown in Table 9.2. The adjusted $R^2$ for the test sample gets improved by about 1%.

### Auto ML Frameworks

There are a variety of new frameworks emerging that do automated search and aggregation of different prediction methods. These automatic aggregation procedures use approaches like the one we outlined above or other heuristics. Example implementations of automatic aggregation methods include H20, AutoML [17], Auto Gluon [18] (which relies on Neural Nets), Auto-Sklearn, Hyperopt-Sklearn and FLAML.

We've tried H20 on the wage data. It produced a model that beats OLS with the basic predictor set, which gave a test MSE of 0.229, by producing a test MSE of 0.21. (The difference is not statistically significant.) H20 is similar to the ensemble method that we constructed above. The performance was very impressive because we gave H20 a time budget of just 100 seconds!

## 9.6 When Do Neural Networks Win?

The wage example may give a pessimistic impression on the power of deep learning (and machine learning more generally). A more optimistic impression emerges from examining performance of deep learning in data-rich settings, where large samples and rich features are available.

A recent example comes from Bajari et al. (2021) [19]. Here we are interested in predicting prices of products given their characteristics, which include both text and images. The resulting predictions are called hedonic prices. In this example, neural networks (specifically BERT [20] and ResNet50 [21]) are first used to convert the text and image data into several thousand-dimensional numerical features $X$ (called embeddings). These features extracted from the text and image data are then used as input variables in a deep neural network for predicting product prices. The deep neural network used in the example consists of 3 hidden layers, with the penultimate layer consisting of about 400 neurons.

The features produced in the penultimate layer in a deep neural network are often referred to as embeddings as they encode or "embed" the information from the previous layers that is directly used in producing the final predictions. In the case of hedonic pricing, we may refer to these features as "value embeddings" as the final target is price or value of the product.

The data set used in this example is larger than 10 million observations. The accuracy of prediction for the deep neural network described above, as measured by the $R^2$ on the test sample, is about 90%. In contrast, random forests applied to predict prices using the text and image embeddings as inputs deliver an $R^2$ in the test sample that is in the ballpark of 80%, and a linear model estimated via least squares that uses the text and image embeddings as predictor variables delivers an $R^2$ in the test sample of only around 70%. Ignoring the neural network embeddings of the text and image data and using only simple catalog features, the $R^2$ is lower than 40%.

We will discuss further details of generating embeddings in Chapter 11.

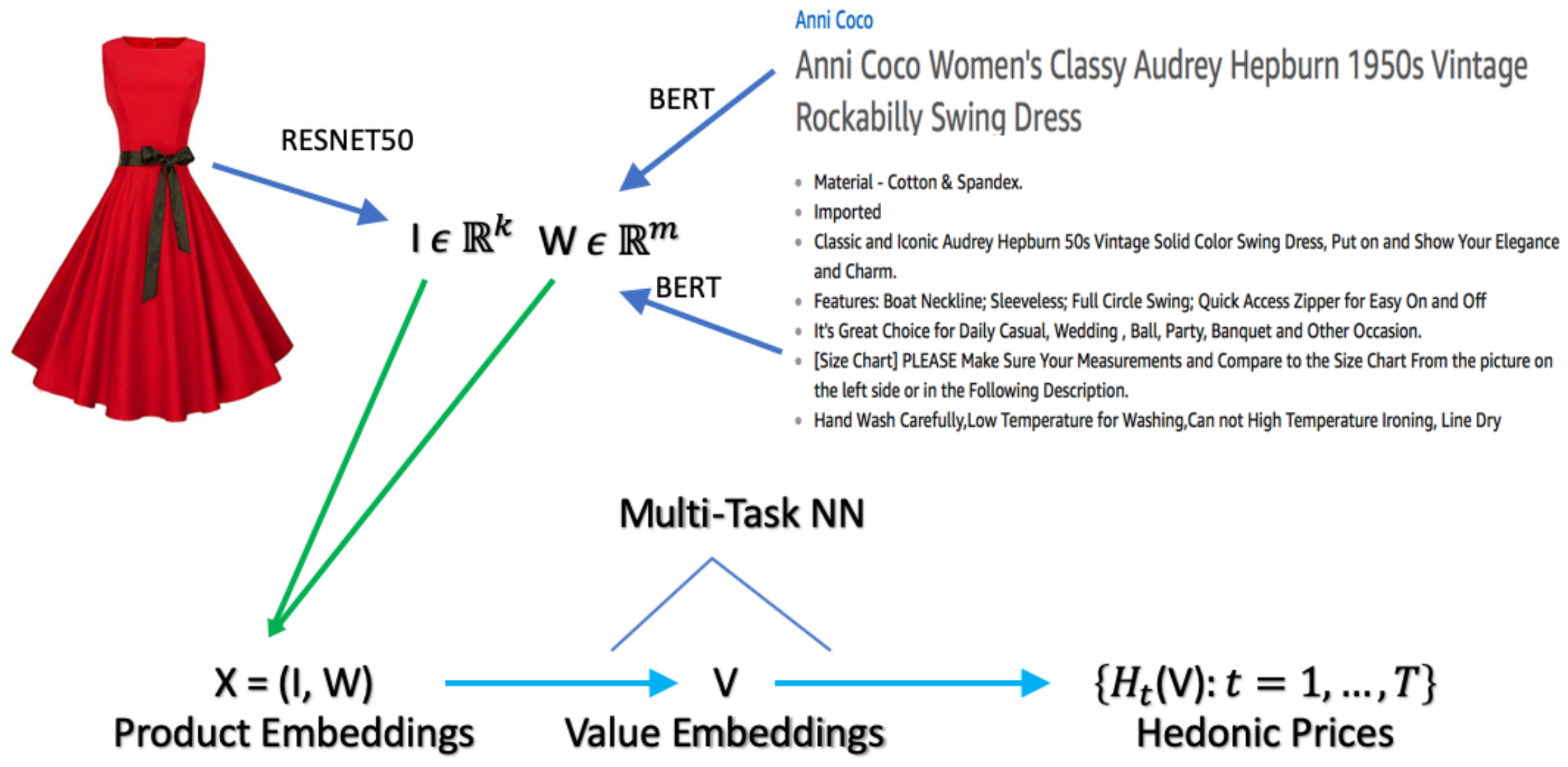

**Figure 9.13:** The structure of the predictive model in Bajari et al. (2021) [19]. The input consists of images and unstructured text data. The first step of the process creates the moderately high-dimensional numerical embeddings $I$ and $W$ for images and text data via state-of-the art deep learning methods, such as ResNet50 and BERT. The second step of the process takes as input $X = (I, W)$ and creates predictions for hedonic prices $H_t(X)$ using deep learning methods with a multi-task structure. The models of the first step are trained on tasks unrelated to predicting prices (e.g., image classification or word prediction), where embeddings are extracted as hidden layers of the neural networks. The models of the second step are trained by price prediction tasks. The multitask price prediction network creates an intermediate lower dimensional embedding $V = V(X)$, called value embedding and then predicts the final prices in all time periods $\{H_t(V), t = 1, \dots, T\}$. Some variations of the method include fine-tuning the embeddings produced by the first step to perform well for price prediction tasks (i.e. optimizing the embedding parameters so as to minimize price prediction loss).

## 9.7 Closing Notes

To sum up, we have discussed assessment of predictive performance of modern linear and non-linear regression methods using splitting of data into training and testing samples. The results could be used to pick the best prediction rule generated by the classical or modern regression methods or to aggregate prediction rules into an ensemble rule, which can result in some improvements. We illustrated these ideas using the wage data from the 2015 Current Population Survey. We finally introduced Auto ML frameworks and commented that Neural Networks perform best in very data-rich settings.

## Notebooks

- Python Notebook on ML-based Prediction of Wages and R Notebook on ML-based Prediction of Wages provide details of implementation of penalized regression, regression trees, random forest, boosted tree and neural network methods, a comparison of various methods and a way to choose the best method or create an ensemble of methods. Moreover, they provide an application of the FLAML (Python) and H2O (R) AutoML framework to the wage prediction problem. With a small time budget, both FLAML and H2O found the model that worked best for predicting wages.

- Python Notebook on Approximation of a Function by Random Forest and Neural Network and R Notebook on Approximation of a Function by Random Forest and Neural Network illustrate the flexibility of these methods in approximating the function $\exp(4x)$.

## Additional resources

- Andrej Karpathy [22] 's Recipe for Training Neural Networks provides a good workflow and practical tips for training good neural network models.

- For practical details of tree-based methods, please see Hastie et al. [23] 's book "Introduction to Statistical Learning".

- For an in-depth treatment of deep learning, see Zhang's et al. [24] 's book "Dive Into Deep Learning", Goodfellow et al. [4] "Deep Learning", and Nielsen [25] "Neural Networks and Deep Learning".

## Notes

Many of the formative developments in modern nonlinear regression were led by the statistics and artificial intelligence communities. The methods were rebranded as machine learning in the 90s, and learning with neural networks was rebranded as deep learning when it was realized that deep network architectures produced phenomenal results in image classification (and later in natural language processing tasks). The success of deep neural networks was a breakthrough associated with advances in both computing power and the ability to collect very large data sets. See the textbooks mentioned above for in-depth treatments of deep learning.

In Chapter 10, we will study the use of the machine learning and deep learning for statistical inference on causal and predictive effects in high-dimensional nonlinear regression settings; and in Chapter 11, we'll be using deep learning for engineering features from text and data (e.g. using images and product descriptions as "regressors").

## Study Problems

1. Use two paragraphs to explain to a friend how one of the tree-based strategies works.

2. Use two paragraphs to explain to a friend how a basic neural network works.

3. Experiment with one of the empirical notebooks provided and summarize your findings. For example, try to see if you can build a better performing neural network in the wage example. One possibility is to use custom models in Keras, where we can construct a partially linear model that borrows the strength of the basic linear model and corrects it slightly with a nonlinear deviation function.

4. Experiment with the last (non-empirical) notebook. See, for example, if you can find a (much) simpler neural network that provides the same quality of fit as the current example in the notebook.

## 9.A Variable Importance via Permutations

There are many ways of assessing variable importance in nonlinear models. A very simple one is the following permutation method.

The importance of variable $j$ in any machine learning algorithm (linear or nonlinear) can be defined by computing the loss in predictive performance that results from replacing the observations of the $j$-th feature $(Z_{ji})_{i=1}^n$ with their random permutation

$$(Z_{j\pi(i)})_{i=1}^n,$$

where $\pi : \{1, ..., n\} \to \{1, ..., n\}$ is a permutation map, generated at random. The loss is averaged over many random permutations, to obtain an average loss measure $L_j$. Then the variables are ranked in terms of $L_j$, from largest to smallest. The top-ranked variables are the most important ones. This idea, that appeared in the original paper by L. Breiman [3], mimics the situation where the permuted regressor is an irrelevant predictor having the same marginal distribution as the observed regressor.

# Statistical Inference on Predictive and Causal Effects in Modern Nonlinear Regression Models

# 10

"Whoever has participated in non-trivial research in any domain of science involving statistical problems must have encountered the difficulty that none of the statistical procedures found in the books fits exactly the practical situation."

– Jerzy Neyman [1].

Here we discuss double/debiased machine learning (DML) methods for performing inference on average predictive or causal effects in two important classes of models: partially linear regression models and interactive regression models. We also present a general DML method for performing inference on a low-dimensional target parameter in the presence of high-dimensional nuisance parameters that are learned using ML methods. Two case studies illustrate the approach.

## 10.1 Introduction

We recall the predictive effect question:

- How does the predicted value of the outcome,

$$\mathrm{E}[Y \mid D, X],$$

  change if a regressor value $D$ increases by a unit, while regressor values $X$ remain unchanged?

This question may have a causal interpretation within any SEM, where conditioning on $X$ is sufficient for identification of the causal effect of $D$ on $Y$. When this condition holds, the question becomes the causal effect question:

- How does the predicted value of the potential outcome,

$$\mathrm{E}[Y(d) \mid X],$$

  change if we intervene and change the treatment value $d$ by a unit, conditional on the observed $X$?

Both questions are interesting and useful to ask, depending on the application. In what follows, we set up double/debiased machine learning (DML) methods for answering these questions with data.[1] These statistical inference methods *do not* distinguish between the two types of questions, so the methods are equally applicable to answering both types.

1: In the book we will use the terms double/debiased machine learning, double machine learning and debiased machine learning interchangeably. It generalizes the double/debiased Lasso approach to generic machine learning methods.

Here we discuss DML methods for performing inference on average predictive or causal effects in two important classes of nonlinear regression models. After presenting these two special cases, we also present a general DML method for performing inference on a low-dimensional target parameter in the presence of high-dimensional nuisance parameters that are learned using ML methods.

The DML method requires a Neyman-orthogonal representation of the target parameters to reduce the spillover of regularization biases inherent in ML methods onto the estimation of the target parameter. The method also makes use of cross-fitting: an efficient form of sample splitting that eliminates biases that may arise from overfitting.

To illustrate the general principles, we provide two case studies. In the first, we perform inference on the effect of gun ownership on homicide rates. In the second, we perform inference on the effect of 401(k) eligibility on financial assets.

## 10.2 DML Inference in the Partially Linear Regression Model (PLM)

We first answer the predictive/causal effect question within the context of the partially linear regression model:

$$Y = \beta D + g(X) + \epsilon, \quad \mathrm{E}[\epsilon \mid D, X] = 0, \tag{10.2.1}$$

where $Y$ is the outcome variable, $D$ is the regressor of interest, and $X$ is a high-dimensional vector of other regressors or features, called "controls." The coefficient $\beta$ answers the predictive effect question. In this segment we discuss estimation and confidence intervals for $\beta$. We also provide a case study, in which we examine the effect of gun ownership on homicide rates.

The model allows a part of the regression function, $g(X)$, to be fully nonlinear, which generalizes the approach from Chapter 4. However, the model is still not fully general, because it imposes additivity in $g(X)$ and $D$. We shall consider a fully unrestricted model in the case of a binary treatment $D$ in Section 10.3. It is worth pointing out before turning to that setting that the partially linear model is not as restrictive as it appears at a first sight since we can consider explicit interactions within the partially linear framework.

**Remark 10.2.1** (Interactions within PLM) Given a raw treatment and a set of controls, $\bar{D}$ and $Z$, we can create the technical treatment $D := \bar{D}T(Z)$, where $T(Z)$ is an $L$−dimensional dictionary of transformations of $Z$. For example, $T(Z)$ could be indicators of various subgroups. Then we can consider the model

$$Y = \sum_{l=1}^{L} \beta_l D_l + g(Z) + \epsilon,$$

where $\mathrm{E}[\epsilon \mid Z, D] = 0$. We can re-write this as

$$Y = \beta_l D_l + g_l(X_l) + \epsilon, \quad \mathrm{E}[\epsilon \mid D_l, X_l] = 0,$$

where $g_l(X_l) := \sum_{k \neq l} \beta_k D_k + g(Z)$ and $X_l := ((D_k)_{k \neq l}, Z)$. We therefore obtain exactly a model of the partially linear form (10.2.1). We can then apply DML methods to learn and perform inference on each element of $(\beta_l)_{l=1}^{L}$ or carry out joint inference (similarly to what we have done in Chapter 4).

In practice and depending on the learner, it may be convenient to treat $g_l(X_l) = h(\{D_k\}_{k \neq l}, Z)$ as a flexible function during estimation rather than impose the structure $g_l(X_l) := \sum_{k \neq l} \beta_k D_k + g(Z)$.

In what follows, we will employ the partialling out $X$ operation of the form that inputs a random variable $V$ and outputs the

residualized form:

$$\tilde{V} := V - \mathrm{E}[V \mid X].$$

Applying this operation to (10.2.1) we obtain

$$\tilde{Y} = \beta \tilde{D} + \epsilon, \quad \mathrm{E}[\epsilon \tilde{D}] = 0, \tag{10.2.2}$$

where $\tilde{Y}$ and $\tilde{D}$ are the residuals left after predicting $Y$ and $D$ using $X$. Specifically, we have that

$$\tilde{Y} := Y - \ell(X) \text{ and } \tilde{D} := D - m(X),$$

where $\ell(X)$ and $m(X)$ are defined as conditional expectations of $Y$ and $D$ given $X$:

$$\ell(X) := \mathrm{E}[Y \mid X] \text{ and } m(X) := \mathrm{E}[D \mid X].$$

Here we recall that the conditional expectations of $Y$ and $D$ given $X$ are the best predictors of $Y$ and $D$ using $X$ .

The equation $\mathrm{E}[\epsilon \tilde{D}] = 0$ above is the Normal Equation for the population regression of $\tilde{Y}$ on $\tilde{D}$. This equation implies the following result:

**Theorem 10.2.1** (FWL Partialling-Out for Partially Linear Model) *Suppose that $Y$, $X$, and $D$ have bounded second moments. Then the population regression coefficient $\beta$ can be recovered from the population linear regression of $\tilde{Y}$ on $\tilde{D}$:*

$$\beta := \{b : \mathrm{E}\left[(\tilde{Y} - b\tilde{D})\tilde{D}\right] = 0\} := (\mathrm{E}[\tilde{D}^2])^{-1}\mathrm{E}[\tilde{D}\tilde{Y}], \tag{10.2.3}$$

*where the second equality and unique definition of $\beta$ follow if $D$ cannot be perfectly predicted by $X$, i.e. if* $\mathrm{E}[\tilde{D}^2] > 0$.

Thus, $\beta$ can be interpreted as a regression coefficient of *residualized* $Y$ on *residualized* $D$, where the residuals are defined by respectively subtracting the conditional expectation of $Y$ given $X$ and $D$ given $X$ from $Y$ and $D$. This result generalizes the FWL from linear models to partially linear models.

Our estimation procedure for $\beta$ in the sample will mimic the partialling out procedure in the population. We also rely on cross-fitting (outlined below) to make sure our estimated residualized quantities are not overfit.

**Double/Orthogonal ML for the Partially Linear Model**

1. Partition data indices into random folds of approximately equal size: $\{1, \ldots, n\} = \cup_{k=1}^{K} I_k$. For each fold $k = 1, \ldots, K$, compute ML estimators $\hat{\ell}_{[k]}$ and $\hat{m}_{[k]}$ of the conditional expectation functions $\ell$ and $m$, leaving out the $k$-th block of data. Obtain the cross-fitted residuals for each $i \in I_k$:

$$\check{Y}_i = Y_i - \hat{\ell}_{[k]}(X_i), \quad \check{D}_i = D_i - \hat{m}_{[k]}(X_i).$$

2. Apply ordinary least squares of $\check{Y}_i$ on $\check{D}_i$. That is, obtain $\hat{\beta}$ as the root in $b$ of the normal equations:

$$\mathbb{E}_n[(\check{Y} - b\check{D})\check{D}] = 0.$$

3. Construct standard errors and confidence intervals as in standard least squares theory.

In what follows it will be convenient to use the notation

$$\|h\|_{L^2} := \sqrt{\mathrm{E}_X h^2(X)},$$

where, as before, $\mathrm{E}_X$ computes the expectation over values of $X$.

**Theorem 10.2.2** (Adaptive Inference on a Target Parameter in PLM [2]) *Consider the PLM model. Suppose that estimators* $\hat{\ell}_{[k]}(X)$ *and* $\hat{m}_{[k]}(X)$ *provide approximations to the best predictors* $\ell(X)$ *and* $m(X)$ *that are of sufficiently high-quality:*

$$n^{1/4}(\|\hat{\ell}_{[k]} - \ell\|_{L^2} + \|\hat{m}_{[k]} - m\|_{L^2}) \approx 0.$$

*Suppose that* $\mathrm{E}[\tilde{D}^2]$ *is bounded away from zero; that is, suppose* $\tilde{D}$ *has non-trival variation left after partialling out. Suppose other regularity conditions listed in [2] hold.*

*Then the estimation error in* $\check{D}_i$ *and* $\check{Y}_i$ *has no first order effect on* $\hat{\beta}$*:*

$$\sqrt{n}(\hat{\beta} - \beta) \approx (\mathbb{E}_n[\tilde{D}^2])^{-1}\sqrt{n}\mathbb{E}_n[\tilde{D}\epsilon].$$

*Consequently,* $\hat{\beta}$ *concentrates in a* $1/\sqrt{n}$ *neighborhood of* $\beta$ *with deviations approximated by the Gaussian law:*

$$\sqrt{n}(\hat{\beta} - \beta) \overset{a}{\sim} N(0, \mathrm{V}),$$

*where*

$$\mathsf{V} = (\mathrm{E}[\tilde{D}^2])^{-1}\mathrm{E}[\tilde{D}^2\epsilon^2](\mathrm{E}[\tilde{D}^2])^{-1}.$$

**Remark 10.2.2** (When PLM fails to hold) Even when the PLM model fails to hold, Theorem 10.2.2 continues to hold when we directly define $\beta$ as in Eq. 10.2.3 of Theorem 10.2.1 for any variable triplet $(X, D, Y)$. That is, $\hat{\beta}$ is in fact an estimate of the BLP of $\tilde{Y}$ as in terms of $\tilde{D}$ regardless of whether the PLM holds. Per Theorem 10.2.1, this coincides with $\beta$ in Eq. (10.2.1) whenever the PLM does hold.

**Confidence Interval** The standard error of $\hat{\beta}$ is $\sqrt{\hat{\mathsf{V}}/n}$, where $\hat{\mathsf{V}}$ is an estimator of $V$. The result implies that the confidence interval

$$\left[\hat{\beta} - 2\sqrt{\hat{\mathsf{V}}/n}, \hat{\beta} + 2\sqrt{\hat{\mathsf{V}}/n}\right]$$

covers $\beta$ in approximately 95% of possible realizations of the sample. In other words, if our sample is not atypical, the interval covers the truth.

**Selecting the Best ML Learners of $\ell$ and $m$.** There may be several methods that satisfy the quality requirements of Theorem 10.2.2, and we may therefore ask what ML methods we should use in practice. Consider a collection of ML methods indexed by $j \in \{1, ..., J\}$. Our goal would be to select the methods that minimize an upper bound on the bias of the DML estimator.

The bias of the DML estimator is controlled by the mean square approximation errors (MSAE):

$$\frac{1}{K}\sum_{k=1}^{K} \|\hat{\ell}_{[k]} - \ell\|^2_{L^2} \text{ and } \frac{1}{K}\sum_{k=1}^{K} \|\hat{m}_{[k]} - m\|^2_{L^2}. \quad (10.2.4)$$

Therefore, we can select the best ML method for estimating $m$ and the best method for estimating $\ell$ to minimize the upper bound on the bias. We will be using mean square prediction errors as proxies for MSAEs.

**Selection of the Best ML Methods for DML to Minimize Bias**. Consider a set of ML methods enumerated by $j \in \{1, ..., J\}$.

- For each method $j$, compute the cross-fitted MSPEs

$$\mathbb{E}_n[\check{Y}_j^2] \text{ and } \mathbb{E}_n[\check{D}_j^2],$$

where the index $j$ reflects the dependency of residuals on the method.

- Select the ML methods $j \in \{1, \dots, J\}$ that give the smallest MSPEs:

$$\hat{j}_\ell = \arg\min_j \mathbb{E}_n[\check{Y}_j^2] \text{ and } \hat{j}_m = \arg\min_j \mathbb{E}_n[\check{D}_j^2].$$

- Use the method $\hat{j}_\ell$ as a learner of $\ell$, and $\hat{j}_m$ as a learner of $m$ in the DML algorithm above.

Two different ML methods may be the best for predicting $Y$ and predicting $D$. By doing MSPE minimization we in fact minimize MSAEs, since MSPEs approximate MSAEs plus terms that do not depend on $j$.

Rather than selecting the single best predictors of $Y$ and $D$, we can also use residuals to form linear ensembles of ML methods that minimize MSPEs.

**Corollary 10.2.3** *The previous inferential result continues to hold if the best or aggregated prediction rules are used as estimators $\hat{m}$ and $\hat{\ell}$ of $m$ and $\ell$ in the DML algorithm. A simple sufficient condition is that the number of ML prediction rules $J$ over which we aggregate or choose from is fixed (meaning small in practice).*

In practical terms, the result of Corollary 10.2.3 means that we should only choose among or aggregate over relatively few ML methods. Otherwise, we may end up overfitting (since we are "cheating" here by using validation data to form the aggregator).

**Remark 10.2.3** (More Technical Condition) A sufficient condition for data dependent selection of which predictor to use when forming residuals to perform well in theory often boils down to requiring $\sqrt{\log J} n^{-1/4} \approx 0$ for choosing the single best method and $\sqrt{J} n^{-1/4} \approx 0$ when using the linear aggregation of methods. However, much work in this area is yet to be formally developed.

## Discussion of DML Construction

The partialling out operation causes the moment equations defining $\beta$ to be Neyman-orthogonal. That is, the moment conditions are insensitive to perturbations of the nuisance parameters $\ell$ and $m$.[2] We discussed Neyman-orthogonality in the context of high-dimensional linear regression models in Chapter 4. We return to and generalize this discussion formally in Section 10.4. This property allows us to get rid of the bias in estimation of $m$ and $\ell$ that arises when ML estimators are applied in high-dimensional settings.

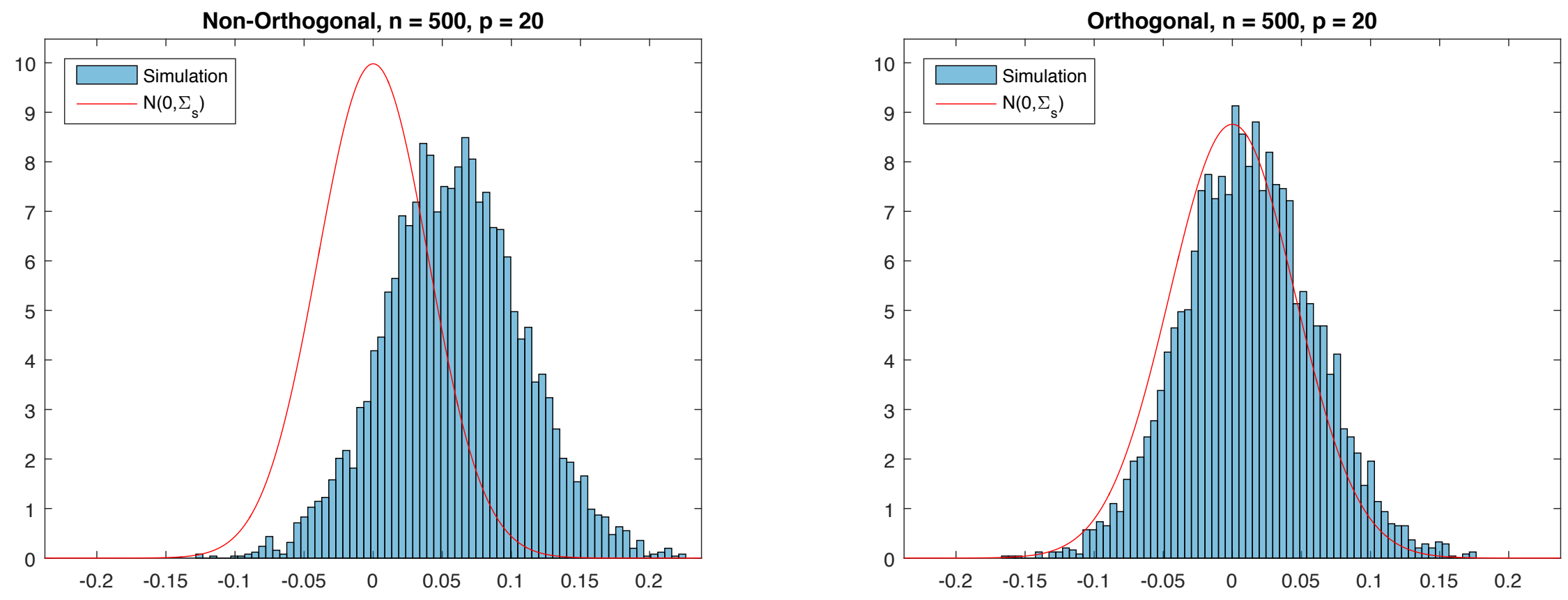

**Figure 10.1:** Left: Behavior of a conventional (non-orthogonal) ML estimator. Right: Behavior of the orthogonal, DML estimator.

2: Generally we use the term nuisance parameters to name parameters that are not the target parameters. Here the target parameter is $\beta$ and $\ell$ and $m$ are nuisance parameters.

Naive application of machine learning methods directly to outcome equations may lead to highly biased estimators, because the resulting strategy is not Neyman-orthogonal. The biases in estimation of $g$, which are unavoidable in high-dimensional estimation, create a non-trivial bias in the estimate of the main effect. This bias is large enough to cause failure of conventional inference.

The left panel of Figure 10.1 illustrates the bias arising due to the use of a non-orthogonal, naive approach for learning $\beta$. Specifically, the figure shows the behavior of a conventional (non-orthogonal) ML estimator, $\tilde{\beta}$, in the partially linear model in a simple simulation experiment where we learn $g$ using a random forest. The $g$ in this experiment is a very smooth function of a small number of variables, so the experiment is seemingly favorable to the use of random forests a priori. The histogram shows the simulated distribution of the centered estimator, $\tilde{\beta} - \beta$. The estimator is badly biased, shifted much to the right relative to the true value $\beta$. Furthermore, the distribution of the estimator (approximated by the blue histogram) is substantively different from a normal approximation (shown by the red curve) derived under the assumption that the bias is negligible.

**Remark 10.2.4** (Bias Transmission) This biased performance of the naive estimator can also be explained analytically. The naive strategy relies on the moment equation:

$$\mathrm{E}[(Y - \beta D - g(X))D] = 0$$

to identify $\beta$ and uses a biased estimate of $g$ in place of $g$. This moment strategy is sensitive to deviations away from the true value. Indeed, let us compute the directional derivative in the direction $\Delta$ away from the true value:

$$\partial_t \mathrm{E}[(Y - \beta D - g(X) + t\Delta(X))D]\Big|_{t=0} = \mathrm{E}[\Delta(X)D] \neq 0.$$

The derivative generally does not vanish, and the biases in estimation of $g$ will transmit to the estimation of $\beta$.

The right panel of Figure 10.1 illustrates the behavior of the (Neyman) orthogonal DML estimator, $\hat{\beta}$, in the partially linear model in a simple experiment where we learn nuisance functions $m$ and $\ell$ using random forests. Note that the simulated data are exactly the same as those underlying the left panel. The simulated distribution of the centered estimator, $\hat{\beta} - \beta$, (given by the blue histogram) illustrates that the estimator is approximately unbiased, concentrates around $\beta$, and is approximately normally distributed. The low bias arises because DML uses the Neyman-orthogonal moment equations.

The DML algorithm uses a form of sample splitting, called cross-fitting, to make sure our estimated residualized quantities are not overfit. Biases arising from overfitting could result from using highly complex fitting methods such as boosting, deep neural networks and random forests. If we don't do sample splitting and the ML estimates overfit, we may end up with very large biases.

Figure 10.2 illustrates how the bias resulting from overfitting in the estimation of nuisance functions can cause the DML (without sample splitting) to be biased and how sample splitting eliminates this problem. In the left panel the histogram shows the finite-sample distribution of the DML estimator in the partially linear model in a simple simulation experiment where nuisance parameters are estimated with overfitting using the full sample, i.e. without sample splitting. The finite-sample distribution is clearly shifted to the left of the true parameter value, demonstrating the substantial bias. In the right panel, the histogram shows the finite-sample distribution of the DML estimator in the same simulation experiment in the partially linear model where nuisance parameters are estimated with sample-splitting using the cross-fitting estimator. Here, we see that the use of sample-splitting has completely eliminated the bias induced by overfitting.

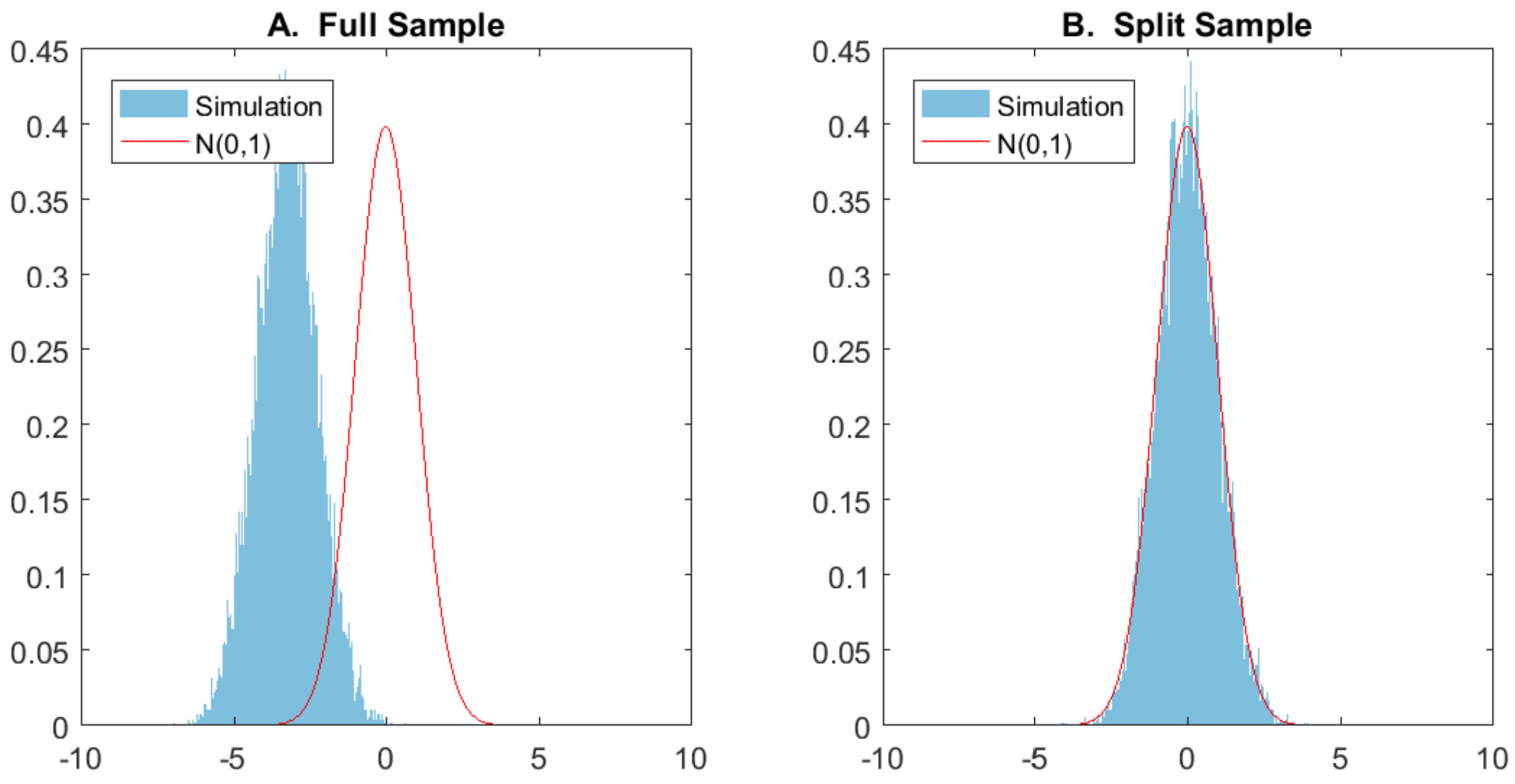

**Figure 10.2:** Left: DML distribution without sample-splitting. Right: DML distribution with cross-fitting.

**Remark 10.2.5** (On overfitting) Note that previously in the context of high-dimensional approximately sparse linear models we were using Lasso with the plug-in choice for the penalty level $\lambda$ which ensures that overfitting is sufficiently well-controlled that we didn't have to use sample splitting. Such refined, theoretically rigorous choices of tuning parameters are not yet available for other machine learning methods. In practice, experienced researchers and machine learning engineers often use intuition, heuristics, and other empirical tools (six packs or witchcraft tables, for example) to set the tuning parameters. While the resulting methods can perform well for prediction purposes, even modest overfitting can result in large biases in DML, as we illustrate in the simulation experiment. Therefore, it is simply safer to rely on sample-splitting in real settings with complicated learners to make sure overfitting of during estimation of our residualized quantities does not contaminate out estimates of the objects of interest.

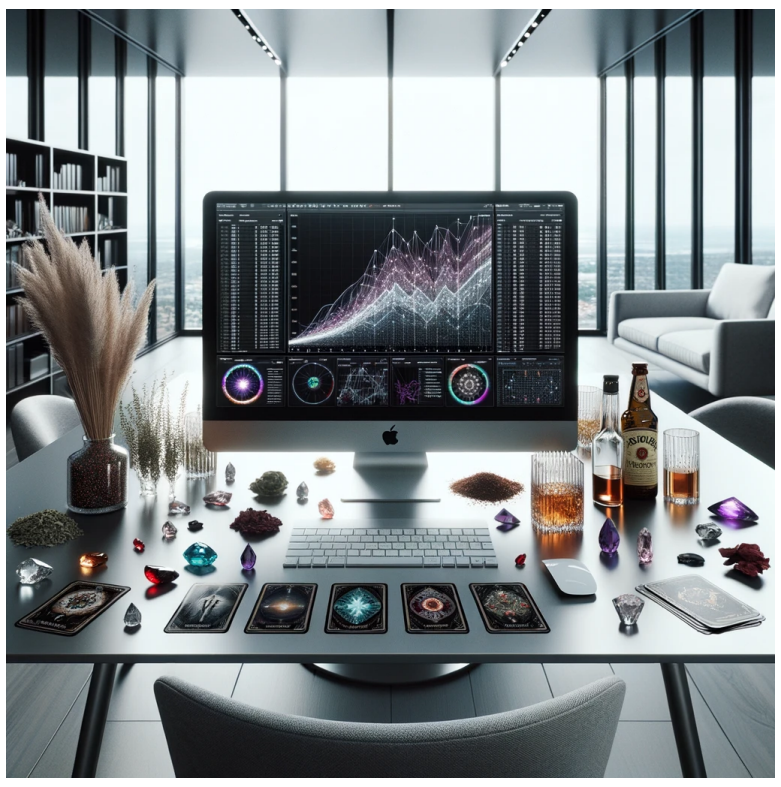

**Figure 10.3:** Witchcraft tables used by some ML practitioners to tune parameters. There are no known theoretical guarantees attached to this tuning method.

## The Effect of Gun Ownership on Gun-Homicide Rates

We consider the problem of estimating the effect of gun ownership on the homicide rate.[3] For this purpose, we estimate the partially linear model:

$$Y_{i,t} = \beta D_{i,(t-1)} + g(X_{i,t}, \bar{X}_i, \bar{X}_t, X_{i,0}, Y_{i,0}, t) + \epsilon_{i,t}.$$

$Y_{i,t}$ is the log homicide rate in county $i$ at time $t$. $D_{i,t-1}$ is the log fraction of suicides committed with a firearm in county $i$ at time $t-1$, which we use as a proxy for gun ownership $G_{i,t}$, which is not observed. $X_{i,t}$ is a set of demographic and economic characteristics of county $i$ at time $t$. We use $\bar{X}_i$ to denote the within county average of $X_{i,t}$ and $\bar{X}_t$ to denote the within time period average of $X_{i,t}$. $X_{i,0}$ and $Y_{i,0}$ denote initial conditions in county $j$. We use $Z_{i,t}$ to denote the set of observed control variables $\{X_{i,t}, \bar{X}_i, \bar{X}_t, X_{i,0}, Y_{i,0}, t\}$. The sample covers 195 large United States counties between the years 1980 through 1999, giving us 3900 observations.[4]

The intent here is that parameter $\beta$ is an approximation of the causal effect of gun ownership $G_{i,t}$ on homicide rates $Y_{i,t}$, controlling for county-level demographic and economic characteristics. We provide further detail about the use of proxy treatments in Section 10.A. To attempt to flexibly account for fixed heterogeneity across counties, common time factors, and deterministic time trends, we include county-level averages, time period averages, initial conditions, and the time index as additional control variables. This strategy is related to strategies for addressing latent sources of heterogeneity via conditioning as in [4]. Finally, for simplicity in this illustration, we assume that all sources of dependence are accounted for by observed variables such that we may take $\epsilon_{i,t}$ as independent across counties, $j$, and over time, $t$.

Raw control variables $X_{i,t}$ are from the U.S. Census Bureau and contain demographic and economic characteristics of the counties such as features of the age distribution, the income distribution, crime rates, federal spending, home ownership rates, house prices, educational attainment, voting patterns, employment statistics, and migration rates.

As a summary statistic we first look at a simple regression of $Y_{i,t}$ on $D_{i,t-1}$ without controls. The point estimate is 0.302 with 95% confidence interval based on the assumption that $\epsilon_{i,t}$ is independent over time and space ranging from 0.277 to 0.333. These results suggest that increases in gun ownership

3: We adapt the basic strategy from Cook and Ludwig [3] who consider using suicide rates as a proxy for gun ownership.

4: Python Notebook on DML for Impact of Gun Ownership on Homicide Rates and R Notebook on DML for Impact of Gun Ownership on Homicide Rates

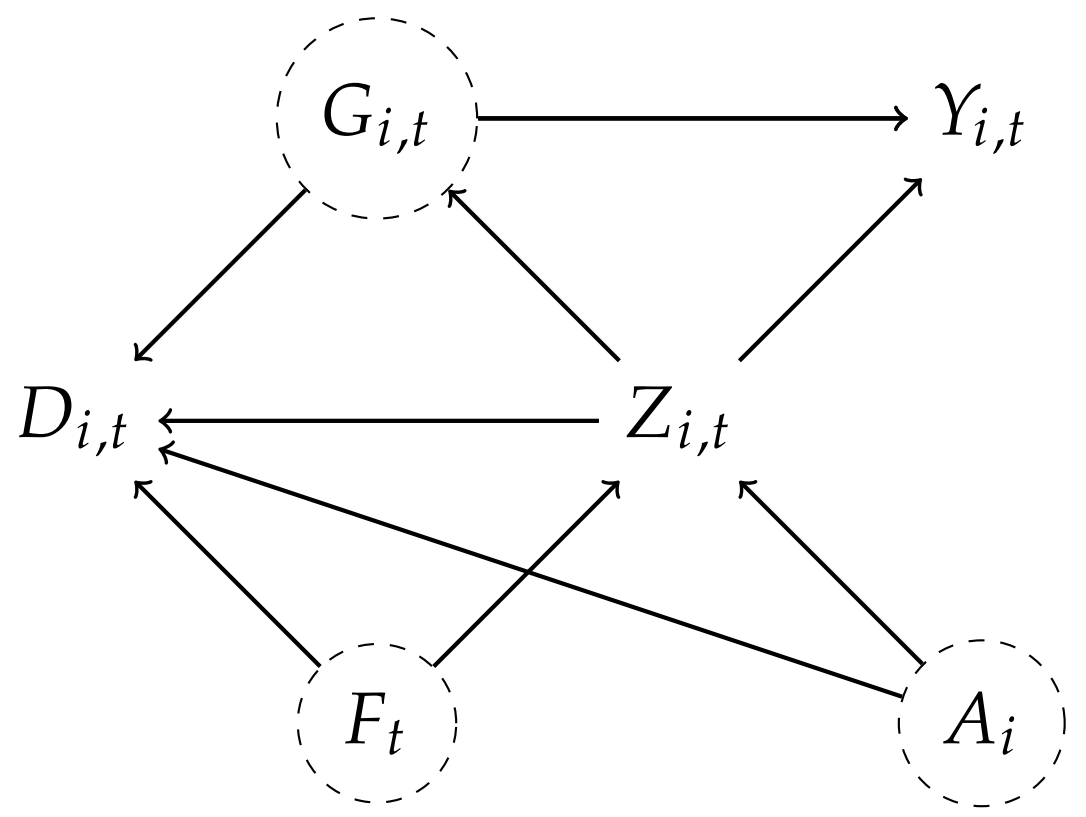

**Figure 10.4:** A Possible DAG Structure for the Gun Ownership Example. Here we approximate the average causal effect $G_{i,t} \rightarrow Y_{i,t}$ only if $G_{i,t} \approx D_{i,t}$. Under the assumption that $D_{i,t}$ is equal to $G_{i,t}$ plus an additive, independent measurement error, the target parameter $\beta$ will be attenuated relative to the true causal effect; see Section 10.A. We also include nodes for latent county specific and time period specific shocks. Often such shocks are accounted for with so-called "fixed effects" which typically leverage strong functional form assumptions. Here, we instead leverage the different, though still strong assumption that flexibly conditioning on observables, including time- and county- specific variables, is sufficient to account for all relevant sources of confounding.

rates are associated with (predict) gun homicide rates – if gun ownership increases by 1% the predicted gun homicide rate goes up by around 0.3% – without controlling for any time factors or county characteristics.

Since our goal is to estimate the effect of gun ownership after controlling for a rich set characteristics, we next include the controls and estimate the model by an array of the modern regression methods that we've learned. Specifically, we consider ten candidate learners for predicting the outcome and for predicting the target variable. We consider linear models estimated with OLS using no control variables (OLS - No Controls), using only the raw control variables (OLS - Basic), and using the raw control variables plus the constructed cross-sectional and time series averages and initial conditions (OLS - All). The remaining methods always take as inputs the complete set of candidate control variables. We use cross-validation to choose tuning parameters for Lasso, Ridge, and Elastic Net. We consider a random forest with default choices and boosted trees constrained to have depth four. Finally, we consider neural nets with two hidden layers of 16 nodes each with early stopping. See R Notebook on DML for Impact of Gun Ownership on Homicide Rates for other training details.

Before turning to estimation results for $\beta$, we look out estimated out-of-sample predictive performance in Table 10.1 which reports cross-fitted root mean square error (RMSE) for the different procedures we consider. The column RMSE Y gives the RMSE

| | RMSE Y | RMSE D |
|---|---|---|
| OLS - No Controls | 1.0964 | 1.2109 |
| OLS - Basic | 0.9540 | 0.4990 |
| OLS - All | 1550360633.3904 | 70320592.0157 |
| Lasso (CV) | 0.4617 | 0.1360 |
| Ridge (CV) | 0.5303 | 0.1450 |
| Elastic Net (.5,CV) | 0.4654 | 0.1346 |
| Random Forest | 0.4021 | 0.1253 |
| Boosted trees - depth 4 | 0.4021 | 0.1224 |
| DNN dropout | 0.6659 | 0.8214 |
| DNN early stopping | 0.5171 | 0.1802 |

**Table 10.1:** Cross-fitted RMSE for predicting outcome (Y) and variable of interest (D) in the gun illustration.

.

for predicting the outcome (log gun homicide rate), and the column RMSE D gives the RMSE for predicting our gun prevalence variable (log of the lagged firearm suicide rate). Here we evidence of the potential relevance of trying several learners rather than just relying on a single, pre-specified choice. There are noticeable differences between performance of most of the learners, with Boosted Trees and Random Forests providing the best performance for predicting both the outcome and policy variable.

Table 10.2 presents the estimated effects of the lagged gun ownership rate on the gun homicide rate as well as the corresponding standard errors. Looking across the results, we see relatively large differences in estimates. These differences suggest that the choice of learner has a material impact in this example. Looking at the measures of predictive performance in Table 10.1, we see that Random Forest and Boosted trees performed best among the considered learners, and we also see that their performance is relatively similar in terms of point estimates of the effect of the lagged gun ownership rate on the gun homicide rate and standard errors. Focusing on the Boosted trees row, the point estimate suggests a 1% increase in the gun proxy is associated with around .1% increase in the gun homicide rate, though the 95% confidence interval is relatively wide: (-0.012,0.211).

The last two rows of the table provide estimates based on using the cross-fit estimates of predictive accuracy of the considered procedures (provided in Table 10.1. The row "Best" uses the method with the lowest MSE as the estimator for $\hat{\ell}(X)$ and $\hat{m}(X)$. In this example, Boosted trees give the best performances in predicting both $Y_{i,t}$ and $D_{i,t-1}$, so the results for "Best" and Boosted trees are identical. The row "Ensemble" uses the linear combination of all ten of the predictors the produces the lowest

| | Estimate | Standard Error |
|---|---|---|
| OLS - No Controls | 0.3020 | 0.0126 |
| OLS - Basic | -0.2568 | 0.0963 |
| OLS - All | -4.1832 | 0.8028 |
| Lasso (CV) | 0.2856 | 0.0568 |
| Ridge (CV) | 0.4624 | 0.0600 |
| Elastic Net (.5,CV) | 0.2888 | 0.0580 |
| Random Forest | 0.0363 | 0.0532 |
| Boosted trees - depth 4 | 0.0997 | 0.0568 |
| DNN dropout | 0.2646 | 0.0110 |
| DNN early stopping | 0.5731 | 0.0496 |
| Best | 0.0997 | 0.0568 |
| Ensemble | 0.0864 | 0.0560 |

**Table 10.2:** Cross-fit estimates for the coefficient on our gun control proxy and standard errors in the gun illustration.

.

MSE for predicting $Y_{i,t}$ or $D_{i,t-1}$ as the estimator $\hat{\ell}(X)$ or $\hat{m}(X)$ respectively. Here the results are similar to the results using only Boosted trees, but differ somewhat due to non-zero linear combination coefficients on the other learners. We also note that the standard error for the ensemble is (slightly) smaller than that of "Best."

## Revisiting the Price Elasticity for Toy Cars

We now revisit again the example from Chapter 0. We are interested in the coefficient $\alpha$ in the PLM:

$$Y = \alpha D + g(W) + \epsilon,$$

where $Y$ is log-reciprocal=sales-rank, $D$ is log-price, and $W$ are product features. In Chapter 4, we let $g(W) = \beta' T(W)$ be a high-dimensional regression using a transformation that included powers and interactions. We now employ flexible nonlinear regression models using DML. We now take $W$ to consist of indicators for brand and subcategory along with physical dimensions interacted with missingness indicators, using no futher transformation, leading to a 2083-dimensional feature vector. We consider inferance on $\alpha$ using DML with different choices of learners applied to both $m(W)$ and $g(W)$: decision trees, gradient boosted trees (with 1000 trees), random forests (with 2000 trees), or a neural network (with two hidden layers of 200 and 20 neurons, respectively, and ReLU activations).

In Table 10.3, we report the cross-validated $R^2$ for predicting $D$ and $Y$ with each of the learners along with the resulting DML

point esimate, standard error estimate, and 95% confidence interval. The first thing we note is that all confidence intervals indicate a substantial negative effect, with a clear indication not only of the direction of the effect but also of its overall magnitude.

Let us first compare these results to the previous ones from when we last revisted this example in Chapter 4. There we saw that OLS with varying number of features failed to exclude 0 from the confidence interval and that Double LASSO lead to an interval [-0.099, -0.029]. We can attribute the latter more negative interval to controlling more confounding, seeing as we expect confounding effects to push the apparent price-sales relationship upward, compared to the theorized downward causal relationship.

Here we see that with more flexible nonlinear methods we obtain an even more negative estimate and confidence interval. This appears to be consistent with the degree to which we are able to control for confounders. LASSO has a cross-validated $R^2$ of 0.09 and 0.32 for predicting $Y$ and $D$, respectively. The $R^2$'s in Table 10.3 are substantially larger. That the corresponding estimates and intervals are also more negative seems to coincide with our theory.

Comparing between the nonlinear methods, this theory appears to remain consistent. Forest and neural net methods have higher $R^2$'s than tree and gradient boosting methods, and the same time have more negative estimates and confidence intervals.

| | $R^2_D$ | $R^2_Y$ | Estimate | Std. Err. | 95% CI |
|---|---|---|---|---|---|
| Tree | 0.40 | 0.19 | -0.109 | 0.018 | [-0.143, -0.074] |
| Boost | 0.41 | 0.17 | -0.102 | 0.019 | [-0.139, -0.064] |
| Forest | 0.49 | 0.26 | -0.134 | 0.019 | [-0.171, -0.096] |
| NNet | 0.47 | 0.21 | -0.132 | 0.020 | [-0.171, -0.093] |

**Table 10.3:** DML estimates of price elasticity based on different learners, along with their $R^2$ for predicting $D$ and $Y$.

Note that just as we can play with transformations in linear models, we can do the same in the PLM. That is, we can modify from partial linearity in the univariate $D$ to partial linearity in a multivariate $T(D)$. We can use this to investigate potentially nonlinear price-sales relationships in this data. Let us transform $D$ using the first $r$ (probabilist's) Hermite polynomials (applied to a location-scale-standardized $D$). We then use DML with neural network learners to learn the coefficients on these polynomial terms.

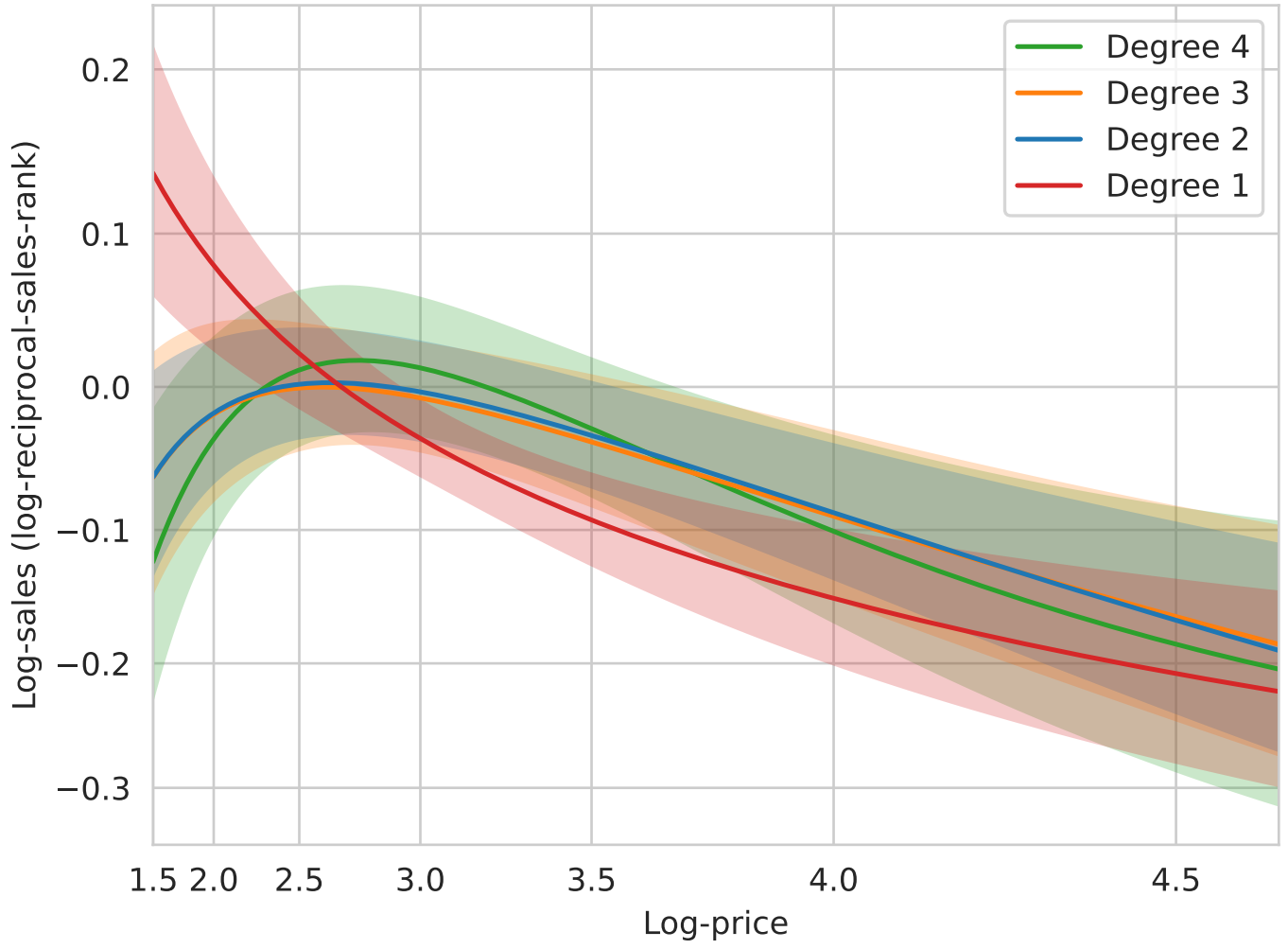

**Figure 10.5:** DML estimates of the price-sales relationship using PLM with higher-order transformations of price. Note the exponential scaling in the axes, which transforms the overall scale back to (non-log) price and sales (reciprocal sales rank).

We plot the resulting estimated functions for $r = 1, \ldots, 4$ in Figure 10.5. As can be seen, the price-sales relationship seems to not be exactly linear, as it stabilizes around a flat-then-decreasing shape for degrees 2, 3, and 4. This shape either suggests that indeed there is less elasticity at lower price points (the mean log-price is 3.06) or that we simply failed to account well for confounding effects at lower price points, which may be idiosyncratic compared to higher-priced toy trucks.

The relationship being not exactly linear does not invalidate using a PLM (in the untransformed univariate $D$). It still corresponds to an average derivative (see Remark 10.3.3, which can still be more interpertable than nonlinear estimates of a causal effect.

# 10.3 DML Inference in the Interactive Regression Model (IRM)

## DML Inference on APEs and ATEs

We consider estimation of average treatment effects when treatment effects are fully heterogeneous and the treatment variable is binary. We consider vectors $W = (Y, D, X)$ and the pair of regression equations:

$$Y = g_0(D, X) + \epsilon, \qquad \mathrm{E}[\epsilon \mid X, D] = 0, \qquad (10.3.1)$$
$$D = m_0(X) + \tilde{D}, \qquad \mathrm{E}[\tilde{D} \mid X] = 0, \qquad (10.3.2)$$

where the second regression equation is presented for convenience. Here $Y$ is an outcome of interest, $D \in \{0, 1\}$ is a binary policy or treatment variable, and $X$ are controls/confounding factors. Since $D$ is not additively separable in the first equation, this model is more general than the partially linear model for the case of binary $D$.

A common target parameter of interest in this model is the average predictive effect (APE),

$$\theta_0 = \mathrm{E}[g_0(1, X) - g_0(0, X)].$$

This quantity is the average predictive effect of switching $D = 0$ to $D = 1$. Under conditional exogeneity discussed in Chapter 5 and Chapter 6, the APE coincides with the average treatment effect (ATE) of the intervention that moves $D = 0$ to $D = 1$.

The confounding factors $X$ affect the policy variable via the propensity score $m_0(X)$ and the outcome variable via the function $g_0(D, X)$. Both of these functions are unknown (except for the case of RCTs, where $m_0(X)$ is known) and potentially complicated, and we can employ ML methods to learn them.

Our construction of the efficient estimator for ATE will be based upon the relation[5]

$$\theta_0 = \mathrm{E}\varphi_0(W), \tag{10.3.3}$$

where

$$\varphi_0(W) = g_0(1, X) - g_0(0, X) + (Y - g_0(D, X))H_0$$

and

$$H_0 = \frac{1(D = 1)}{m_0(X)} - \frac{1(D = 0)}{1 - m_0(X)}$$

is the Horvitz-Thompson transformation.

5: This representation is known as "doubly robust" parameterization, which refers to the fact that $\theta_0$ is recovered whenever the $g$ or $H$ is specified correctly. We don't dwell on this property here – for us, only the Neyman orthogonality property is important.

**Remark 10.3.1** (Regression Adjustment or Propensity Score Reweighting? Use both) We realize that this representation encompasses two equally valid representations of the target parameter: the regression adjusted representation,

$$\theta_0 = \mathrm{E}[g_0(1, X) - g_0(0, X)],$$

and the propensity score reweighting representation,

$$\theta_0 = \mathrm{E}[YH_0].$$

Unfortunately *neither* of these representations is Neyman orthogonal, making them unsuitable for plugging-in machine learning estimators. In sharp contrast, the representation (10.3.3) is Neyman orthogonal, which implies that we can readily deploy ML methods for estimation using the empirical analog of this expression coupled with cross-fitting.

Recall we introduced Neyman orthogonality in Chapter 4. We continue this discussion formally in Section 10.4.

The construction provided in (10.3.1) is equally applicable in cases where the propensity score $\mathrm{P}(D = 1 \mid X)$ is known, as in stratified randomized experiments, and in cases where the propensity score is unknown. When the propensity score is known, the role of regression adjustment in (10.3.1) is to reduce estimation noise.

We will employ the Neyman orthogonal parameterization and cross-fitting to construct a high-quality estimator and perform statistical inference on the target parameter.

**DML for APEs/ATEs in IRM**

1. Partition sample indices into random folds of approximately equal size: $\{1, \ldots, n\} = \cup_{k=1}^{K} I_k$. For each $k = 1, \ldots, K$, compute estimators $\hat{g}_{[k]}$ and $\hat{m}_{[k]}$ of the conditional expectation functions $g_0$ and $m_0$, leaving out the $k$-th block of data, such that $\epsilon \leq \hat{m}_{[k]} \leq 1 - \epsilon$, and for each $i \in I_k$ compute

$$\hat{\varphi}(W_i) = \hat{g}_{[k]}(1, X_i) - \hat{g}_{[k]}(0, X_i) + (Y_i - \hat{g}_{[k]}(D_i, X_i))\hat{H}_i$$

with

$$\hat{H}_i = \frac{1(D_i = 1)}{\hat{m}_{[k]}(X_i)} - \frac{1(D_i = 0)}{1 - \hat{m}_{[k]}(X_i)}.$$

2. Compute the estimator

$$\hat{\theta} = \mathbb{E}_n[\hat{\varphi}(W)]$$

3. Construct standard errors via

$$\sqrt{\hat{\mathrm{V}}/n}, \quad \hat{\mathrm{V}} = \mathbb{E}_n[\hat{\varphi}(W) - \hat{\theta}]^2$$

and use standard normal critical values for inference.

**Remark 10.3.2** (Trimming) An important practical issue is trimming $|\hat{H}_i|$ from taking explosively large values. Large values can occur when estimated propensity scores are near 0 or 1, which may indicate failure of the overlap condition – Assumption 5.2.2 in Chapter 5 and restated in Theorem 10.3.1 below. In the algorithm above, $\hat{H}_i$ can take on the largest absolute value of $\bar{H} = 1/\epsilon$. Therefore, setting $\epsilon = .01$ corresponds to $\bar{H} = 100$. There does not seem to be a good theoretical or practical resolution on how to do trimming.

**Theorem 10.3.1** (Adaptive Inference on ATE with DML) *Suppose conditions specified in [2] hold. In particular, suppose that the*

*overlap condition holds, namely for some $\epsilon > 0$ with probability 1*

$$\epsilon < m_0(X) < 1 - \epsilon.$$

*If estimators $\hat{g}_{[k]}(D, X)$ and $\hat{m}_{[k]}(X)$ are such that $\epsilon \leq \hat{m}_{[k]}(X) \leq 1 - \epsilon$ and provide sufficiently high-quality approximations to the best predictors $g_0(D, X)$ and $m_0(X)$ such that*

$$\|\hat{g}_{[k]} - g_0\|_{L^2} + \|\hat{m}_{[k]} - m_0\|_{L^2} + \sqrt{n}\|\hat{g}_{[k]} - g_0\|_{L^2}\|\hat{m}_{[k]} - m_0\|_{L^2} \approx 0,$$

*then the estimation error in these nuisance parameter has no first order effect on $\hat{\theta}$:*

$$\sqrt{n}(\hat{\theta} - \theta_0) \approx \sqrt{n}\mathbb{E}_n(\varphi_0(W) - \theta_0).$$

*Consequently, the estimator concentrates in $1/\sqrt{n}$ neigborhood of $\theta_0$, with deviations controlled by the Gaussian law:*

$$\sqrt{n}(\hat{\theta} - \theta_0) \overset{a}{\sim} N(0, \mathsf{V})$$

*where*

$$\mathsf{V} = \mathrm{E}(\varphi_0(W) - \theta_0)^2.$$

The condition on the quality of estimators of $g_0$ and $m_0$ provides a possibility of "trading off" the quality of each estimator while retaining the adaptive inference property. The better we estimate the propensity score $m_0$, the worse our estimate of the regression function $g_0$ can be; and vice versa.

## DML Inference for GATEs and ATETs

As discussed in Chapter 5, we may also be interested in average effects for interesting subpopulations such as group ATEs (GATEs) or average treatment effect on the treated (ATET). Recall that a GATE is defined as the average treatment effect within a group:

$$\theta_0 = \mathrm{E}[g_0(1, X) - g_0(0, X) \mid G = 1],$$

where $G$ is a group indicator. For example, we might be interested in the impact of a vaccine on teenagers, in which case we could set $G = 1(13 \leq \text{Age} \leq 19)$, or on older individuals, in which case we might set $G = 1(65 \leq \text{Age})$. DML estimation and inference for GATEs can be carried out similarly to estimation and inference for the ATE by exploiting the relation

$$\theta_0 = \mathrm{E}[\varphi_0(X) \mid G = 1] = \mathrm{E}[\varphi_0(X)G]/\mathrm{P}(G = 1).$$

GATEs are of interest for describing heterogeneity of the average treatment effects across groups. This parameter also has a predictive interpretation in a non-causal sense: It measures the average change in prediction as $D$ switches from 0 to 1, averaging over characteristics of the group $G = 1$.

Another common target parameter ATET:

$$\theta_0 = \mathrm{E}[g_0(1, X) - g_0(0, X) \mid D = 1].$$

In business applications, the ATET is often of the interest for attribution calculations. For example, if the treatment of interest is having experience with a new product, the ATET captures the effect of the new product on those that actually received it.

We provide further detail for DML estimators of GATEs and ATETs in Section 10.4.

**Remark 10.3.3** (Misspecification of PLM as inference on an overlap-weighted APE) In the case of binary treatment $D \in \{0, 1\}$, the IRM (Eqs. 10.3.1 and 10.3.2) generalizes the PLM of Section 10.2 (Eq. 10.2.1) by permitting interaction between the treatment and controls. The PLM, nonetheless, admits a very simple estimator for the treatment coefficient via partialling out: simply regress cross-fitted outcome residuals on cross-fitted treatment residuals, never dividing by propensity scores. What does this get at, however, when the PLM fails to hold? Per Remark 10.2.2, we need only consider the BLP of $\tilde{Y}$ in terms of $\tilde{D}$ in the more general IRM. Writing $g_0(D, X) = g_0(0, X) + D(g_0(0, X) - g_0(1, X))$, we see that $\tilde{Y} = \tilde{D}(g_0(1, X) - g_0(0, X)) + \epsilon$. Since $\mathrm{E}[\tilde{D}^2 \mid X] = m_0(X)(1 - m_0(X))$, we find that the estimand is $\beta = \mathrm{E}[m_0(X)(1 - m_0(X))(g_0(1, X) - g_0(0, X))]/\mathrm{E}[m_0(X)(1 - m_0(X))]$, that is, the APE on the population reweighted by $m_0(X)(1 - m_0(X))/\mathrm{E}[m_0(X)(1 - m_0(X))]$, known as overlap weights as they upweight when $m_0(X)$ is close to $1/2$ and downweight when $m_0(X)$ is close to 0 or 1.

In the case of a continuous univariate treatment on $[0, 1]$, we can leverage the same idea of writing $g_0(D, X)$ as a baseline plus the effect of $D$ using the fundamental theorem of calculus: $g_0(D, X) = g_0(0, X) + \int_0^1 \mathbb{I}[D > t] g_0'(t, X) dt$, where $g_0'$ is the derivative in the first argument. We can then find that $\beta$ identifies some average derivative $\mathrm{E}[w(D, X) g_0'(D, X)]/\mathrm{E}[w(D, X)]$ for some nonnegative weights $w(d, x) = \mathrm{E}[\tilde{D}\mathbb{I}[D > d] \mid X = x]/f(d \mid x) \geq 0$, where $f(d \mid x)$ is the conditional density of $D$ given $X = x$ (see, e.g., Sec. 2.3.1 of [5]). That is, we estimate *some* average causal effect of increasing every value of $D$ by an infinitesimal amount. However, the population over which we average may be highly uninterpertable.

## The Effect of 401(k) Eligibility on Net Financial Assets

Here we re-analyze the impact of 401(k) eligibility on financial assets (Poterba et al., [6] and [7]). The data covers a short period a few years after the introduction of 401(k)'s when they were rapidly increasing in popularity.

R Notebook on DML for Impact of 401(K) Eligibility on Financial Wealth

Python Notebook on DML for Impact of 401(K) Eligibility on Financial Wealth

The key problem in determining the effect of 401(k) eligibility is that working for a firm that offers access to a 401(k) plan is not randomly assigned. To overcome the lack of random assignment, we follow the strategy developed in [6] and [7]. In these papers, the authors use data from the 1991 Survey of Income and Program Participation and argue that eligibility for enrolling in a 401(k) plan in this data can be taken as exogenous after conditioning on a few observables of which the most important for their argument is income.

The basic idea of their argument is that, at least around the time 401(k)'s initially became available, people were unlikely to be basing their employment decisions on whether an employer offered a 401(k) but would instead focus on income and other aspects of the job. Following this argument, whether one is eligible for a 401(k) may then be taken as exogenous after appropriately conditioning on income and other control variables related to job choice.

Compare this argument to the one given below using DAGs.

A key component of the argument underlying the exogeneity of 401(k) eligibility is that eligibility may only be taken as exogenous after conditioning on income and other variables related to job choice that may correlate with whether a firm offers a 401(k). [6] and [7] and many subsequent papers adopt this argument but control for parsimonious, pre-specified functions of what they deem to be relevant characteristics. One might wonder whether such specifications are able to adequately control for income and other related confounders. At the same time, the power to learn about treatment effects decreases as one allows more flexible models. The principled use of flexible ML tools offers one resolution to this tension.

In what follows, we use net financial assets[6] as the outcome variable, $Y$, in the analysis. The treatment variable, $D$, is an indicator for being eligible to enroll in a 401(k) plan. The vector

6: Defined as the sum of IRA balances, 401(k) balances, checking accounts, U.S. saving bonds, other interest-earning accounts in banks and other financial institutions, other interest-earning assets (such as bonds held personally), stocks, and mutual funds less non-mortgage debt.

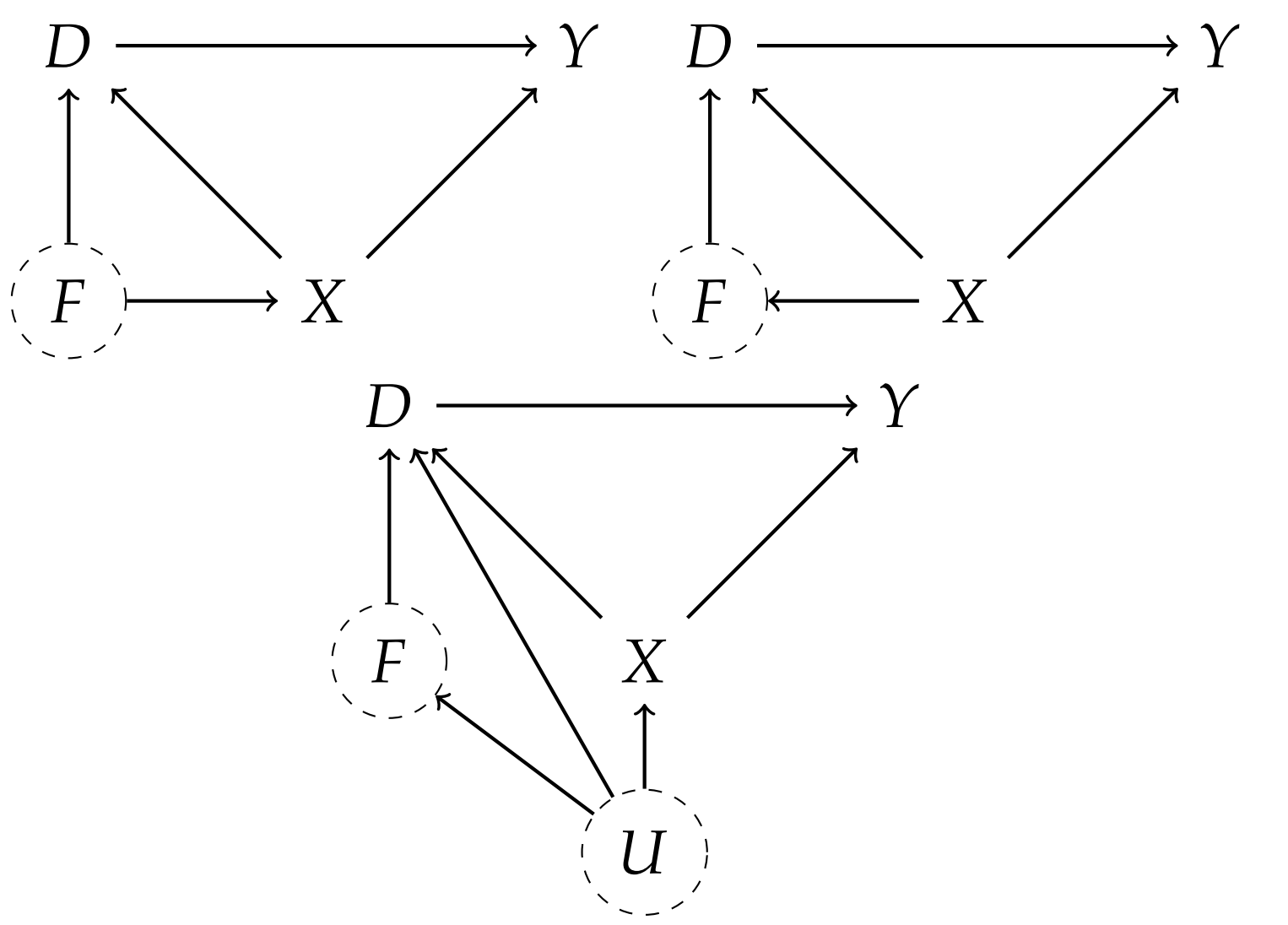

**Figure 10.6:** Three Causal DAGs for analysis of the 401(K) example in which adjusting for $X$ is a valid identification strategy. The bottom figure encompasses the other two as special cases.

of raw covariates, $X$, consists of age, gender, income, family size, years of education, a married indicator, a two-earner status indicator, a defined benefit pension status indicator, an IRA participation indicator, and a home ownership indicator.

It is useful to think about a causal diagram that represents our thinking about identification in this example. In Figure 10.6, we provide three example DAGs for $Y$, the outcome; $D$, the 401(K) eligibility offer which depends on firm characteristics, $F$, which are not observed; and $X$, the worker characteristics. In one structure, $F$ determines the workers characteristics (via the hiring decision), so we have $F \longrightarrow X$. In another structure, workers determine the characteristics of the company they choose to work at, $X \longrightarrow F$. Finally, in the last structure $F$, $X$, and $D$ are jointly determined by a set of latent factors $U$. In any of these cases, $X$ is a valid adjustment set because it is the only parent of $Y$ (other than $D$).

R Notebook on Dagitty-Based Identification in 401(K) Example

It is also useful to consider structures that would break down the identification strategy. We illustrate two such structures in Figures 10.7 and 10.8. In these figures, we introduce a node for the employer match amount, $M$,[7] which could mediate the effect of 401(k) eligibility and have an important effect on financial wealth.

7: Employers often offer a benefit where they will match a proportion of an employee's contribution to their 401k, up to a limit. The limit is referred to as the employer match amount, and averages between 4 and 5% of employee's salaries.

In Figure 10.7, we suppose that $M$ is determined by unobserved firm characteristics, $F$, and worker characteristics, $X$. In this case, adjustment for $X$ is not sufficient as there is a path from latent firm characteristics, which are related to the treatment, to the outcome that is not closed by $X$. However, if $M$ is determined solely by $D$ and $X$ so the red arrow is erased, adjustment for $X$ is sufficient. Therefore, interpreting the target parameter of our estimation strategy as a causal effect is only valid if the match amount is independent of $F$ given $D$ and $X$, that is, if there is no arrow from $F$ to $M$ in the graph. Otherwise, the default interpretation is that we are estimating predictive effects of 401(k) eligibility.

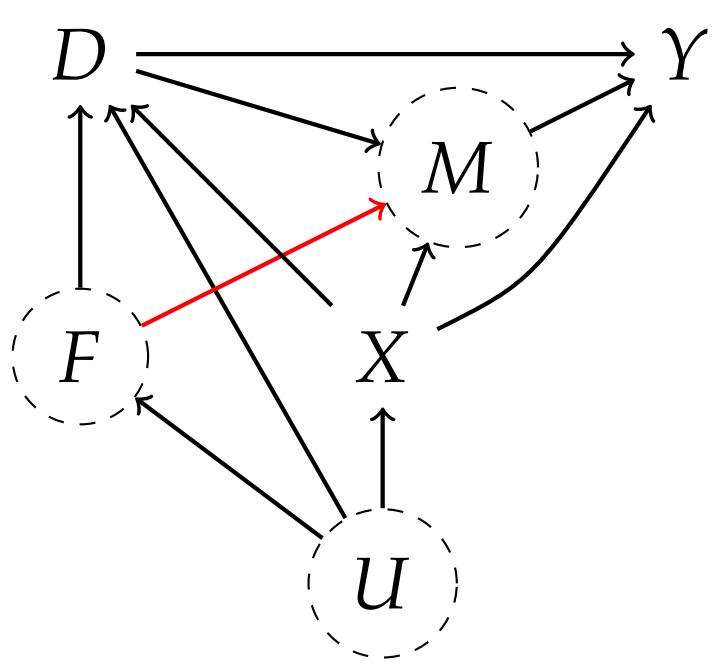

**Figure 10.7:** A DAG Structure where adjusting for $X$ is not sufficient. If there is no arrow from $F$ to $M$, adjusting for $X$ is sufficient.

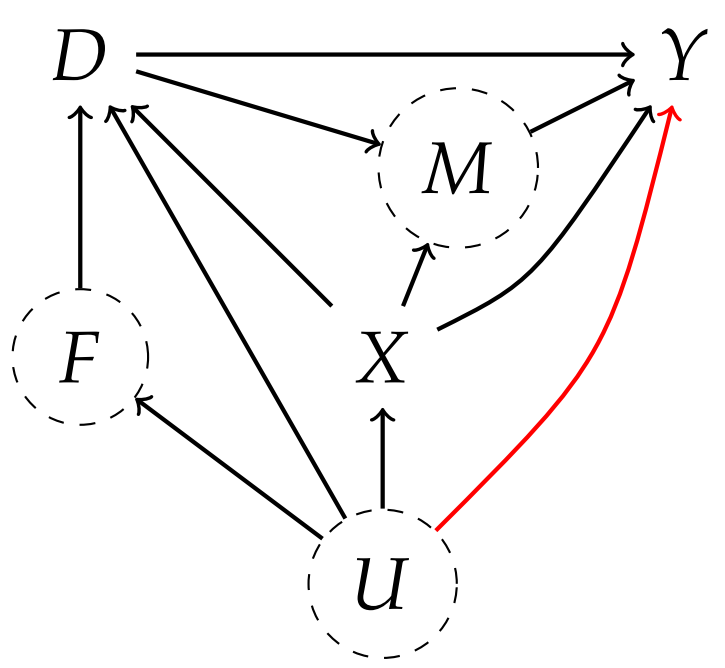

**Figure 10.8:** Another DAG Structure where adjusting for $X$ is not sufficient. Here the latent confounder $U$ affects all variables, so even in the absence of an arrow connecting $F$ to $M$, causal effects cannot be determined after adjusting for $X$. The presence of such latent confounders is always a threat to causal interpretability of any observational study.

In the second example, Figure 10.8, we maintain the assumption that $M$ is independent of $F$ given $D$ and $X$ by eliminating the arrow between nodes $F$ and $M$. However, we now allow for the possibility that latent variables $U$ have a direct effect on $Y$; that is, we have an unobserved confounder or omitted variable. In this example, such a counfounder may be unobserved risk preferences that relate to an individual's preference over jobs, an individual's characteristics, but also have direct effects on savings decisions not channeled purely through observed individual or job characteristics. In general, the possibility of latent confounders always poses a challenge to obtaining estimates of causal effects in non-experimental data. The presence or absence of latent confounders cannot be determined solely from the data in general, and thus their presence must be argued against based on scientific and institutional knowledge in different contexts. See, e.g., discussion in the original papers, [6] and [7], underlying this example. As in the previous example, we must interpret our estimates as predictive effects of 401(k) eligibility if we believe the connection from $U$ to $Y$ exists.

In Table 10.4, we report DML estimates of ATE of 401(k) eligibility on net financial assets both in the partially linear model and

**Table 10.4:** Estimated Effect of 401(k) Eligibility on Net Financial Assets

| | Lasso | Forest | Boost | NNet | Ens | Best |
|---|---|---|---|---|---|---|
| *A. Interactive Regression Model* | | | | | | |
| ATE | 7993 | 8105 | 7713 | 7788 | 7839 | 7753 |
| | [1201] | [1242] | [1155] | [1238] | [1134] | [1237] |
| *B. Partially Linear Regression Model* | | | | | | |
| ATE | 8871 | 9247 | 9110 | 9038 | 9166 | 9215 |
| | [1298] | [1295] | [1314] | [1322] | [1299] | [1294] |

**Note:** Estimated ATE and standard errors from a partially linear model (Panel B) and heterogeneous effect model (Panel A) based on orthogonal estimating equations. Column labels denote the method used to estimate nuisance functions.

the interactive regression model allowing for heterogeneous treatment effects. To reduce the disproportionate impact of extreme propensity score weights in the interactive model, we trim the propensity scores at 0.01 and 0.99.

Turning to the results, it is first worth noting that when no controls are used, the estimated ATE of 401(k) eligibility on net financial assets is $19,559 with an estimated standard error of 1413. Of course, this number is not a valid estimate of the causal effect of 401(k) eligibility on financial assets if there are neglected confounding variables as suggested by [6] and [7]. When we turn to the estimates that flexibly account for confounding reported in Table 10.4, we see that they are substantially attenuated relative to this baseline that does not account for confounding, suggesting much smaller causal effects of 401(k) eligibility on financial asset holdings.

It is interesting and reassuring that the results obtained from the different flexible methods are broadly consistent with each other. This similarity is consistent with the theory that suggests that results obtained through the use of orthogonal estimating equations and any method that provides sufficiently high-quality estimates of the necessary nuisance functions should be similar. Finally, it is interesting that these results are also broadly consistent with those reported in the original work of [6] and [7] which used a simple, intuitively-motivated functional form, suggesting that this intuitive choice was sufficiently flexible to capture much of the confounding variation in this example.

Finally, we can conclude the discussion with a more sobering note that there are credible deviations in the graph structure (e.g. unobserved firm characteristics may affect the match amount) that challenges causal interpretation of the estimates. One approach to dealing with such deviations would be to conduct thorough sensitivity analysis.*

## 10.4 Generic Debiased (or Double) Machine Learning

### Key Ingredients

A general construction upon which DML estimation and inference can be built relies on a method-of-moments estimator for some low-dimensional target parameter $\theta_0$ based upon the empirical analog of the moment condition

$$\mathrm{E}\psi(W;\theta_0,\eta_0) = 0, \tag{10.4.1}$$

where we call $\psi$ the score function, $W$ denotes a data vector, $\theta_0$ denotes the true value of a low-dimensional parameter of interest, and $\eta$ denotes nuisance parameters with true value $\eta_0$.

The first key input of the generic DML procedure is using a score function $\psi(W;\theta,\eta)$ such that

$$\mathsf{M}(\theta,\eta) = \mathrm{E}[\psi(W;\theta,\eta)]$$

identifies $\theta_0$ when $\eta = \eta_0$ – that is,

$$\mathsf{M}(\theta,\eta_0) = 0 \text{ if and only if } \theta = \theta_0 -$$

and the Neyman orthogonality condition is satisfied:

$$\partial_\eta \mathsf{M}(\theta_0,\eta)\Big|_{\eta=\eta_0} = 0. \tag{10.4.2}$$

Here, (10.4.2) ensures that the moment condition (10.4.1) used to identify and estimate $\theta_0$ is insensitive to small perturbations of the nuisance function $\eta$ around $\eta_0$.

* We have done some informal simulations to assess the impact of this threat (using the observation that firms match up to 5% of income), and we estimated the size of the bias to be in the ball park of 10%. Given this, we believe the results reported here are reasonable approximations to the causal effects.

**Remark 10.4.1** The orthogonality condition is named after Neyman [8], because he was the first to propose it in the context of parametric models with nuisance parameters that are estimated at slower than $1/\sqrt{n}$ rates.

Using a Neyman-orthogonal score eliminates the first order biases arising from the replacement of $\eta_0$ with a ML estimator $\hat{\eta}_0$. Eliminating this bias is important because estimators $\hat{\eta}_0$ must be heavily regularized in high-dimensional settings, so these estimators will be biased in general. The Neyman orthogonality property is responsible for the adaptivity of these estimators – namely, their approximate distribution will not depend on the fact that the estimate $\hat{\eta}_0$ contains error as long as the error is sufficiently mild.

**Remark 10.4.2** (Definition of the Derivative) The derivative $\partial_\eta$ denotes the pathwise (Gateaux) derivative operator. Formally it is defined via usual derivatives taken in various directions: Given any "admissible" direction $\Delta = \eta - \eta_0$ and scalar deviation amount $t$, we have that

$$\partial_\eta \mathrm{M}(\theta, \eta)[\Delta] := \partial_t \mathrm{M}(\theta, \eta + t\Delta)\Big|_{t=0}.$$

The statement

$$\partial_\eta \mathrm{M}(\theta_0, \eta_0) = 0$$

means that $\partial_\eta \mathrm{M}(\theta_0, \eta_0)[\Delta] = 0$ for any admissible direction $\Delta$. The direction $\Delta$ is admissible if $\eta_0 + t\Delta$ is in the parameter space for $\eta$ for all small values of $t$.

The second key input is the use of high-quality machine learning estimators of the nuisance parameters. A sufficient condition in the examples given includes the requirement

$$n^{1/4}\|\hat{\eta} - \eta_0\|_{L^2} \approx 0.$$

Different structured assumptions on $\eta_0$ allow us to use different machine-learning tools for estimating $\eta_0$. For instance,

1) approximate sparsity for $\eta_0$ with respect to some dictionary calls for the use of Lasso, post-Lasso, or other sparsity-based techniques;
2) well-approximability of $\eta_0$ by trees calls for the use of regression trees and random forests;

**3)** well-approximability of $\eta_0$ by sparse deep neural nets calls for the use of $\ell_1$-penalized deep neural networks;
**4)** well-approximability of $\eta_0$ by at least one model mentioned in 1)-3) above calls for the use of an ensemble/best choice method over the estimation methods mentioned in 1)-3).

There are performance guarantees for most of these ML methods that make it possible to satisfy the conditions stated above. Ensemble and best choice methods ensure that the performance guarantee is no worse than the performance of the best method.

> The third key input is to use a form of sample splitting at the stage of producing the estimator of the main parameter $\theta_0$, which allows us to avoid *biases* arising from overfitting.

Overfitting can easily occur when using highly complex fitting methods such as boosting, random forests, deep nets, ensembles, and other hybrid machine learning methods. We may heuristically think of overfitting as capturing noise that is particular to the observations used to fit a model in addition to signal. Using overfit estimates of nuisance parameters obtained using the same data as used to estimate the target parameter then heuristically leads to estimation error in these parameters being correlated to outcomes which introduces a type of bias. This bias can be very large, as illustrated in Figure 10.2. We specifically use cross-fitted forms, i.e. sample splitting, of the empirical moments, as detailed below, in estimation of $\theta_0$ to avoid this problem.

## Neyman Orthogonal Scores for Regression Problems

**Scores for Partially Linear Regression Model.** In the PLM, we employ the score function

$$\psi(W;\theta,\eta) := \{Y - \ell(X) - \theta(D - m(X))\}(D - m(X)), \tag{10.4.3}$$

where $W = (Y, D, X)$ is a data vector, and $\eta$ is the nuisance parameter $\eta = (\ell, m)$ with true value $\eta_0 = (\ell_0, m_0)$. Here, $\ell$ and

$m$ are square-integrable functions mapping the support of $X$ to $\mathbb{R}$ whose true values are given by

$$\ell_0(X) = \mathrm{E}[Y \mid X], \quad m_0(X) = \mathrm{E}[D \mid X].$$

The score above is Neyman orthogonal by elementary calculations delegated to Section 10.B. The objects $Y - \ell(X)$ and $D - m(X)$ in the PLM score function (10.4.3) are also clearly the flexible analogs of taking residuals from linear models discussed in Chapter 1.

**Scores for Interactive Regression Model.** For estimation of the ATE parameter in the IRM model, we employ the score

$$\begin{aligned}\psi_1(W;\theta,\eta) := (g(1,X) - g(0,X))& \\ + H(D,X)(Y - g(D,X)) - \theta&,\end{aligned} \tag{10.4.4}$$

where

$$H(D,X) := \frac{D}{m(X)} - \frac{(1-D)}{1-m(X)}, \tag{10.4.5}$$

$W = (Y, D, X)$ is a data vector, and $\eta := (g, m)$ is the nuisance parameter with true value $\eta_0 = (g_0, m_0)$. Here, $g$ is a square-integrable function mapping the support of $(D, X)$ to $\mathbb{R}$, and $m$ is a function mapping the support of $X$ to $(\varepsilon, 1-\varepsilon)$ for some $\varepsilon \in (0, 1/2)$. The true values of $g$ and $m$ are given by

$$g_0(D,X) = \mathrm{E}[Y \mid D, X], \quad m_0(X) = \mathrm{P}[D = 1 \mid X]. \tag{10.4.6}$$

The score above is Neyman orthogonal by elementary calculations delegated to Section 10.B.

For estimation of GATEs we use the score

$$\psi(W;\theta,\eta) := \tfrac{G}{p}\psi_1(W;\theta,\eta); \tag{10.4.7}$$

where $G$ denotes the group membership indicator, the nuisance parameter $\eta$ is $(g, m, p)$ with true value $\eta_0 = (g_0, m_0, p_0)$ for $g_0$ and $m_0$ defined in (10.4.6) and $p_0 = \mathrm{P}(G = 1)$, and $\psi_1$ is the score for the ATE parameter defined in (10.4.4).

For estimation of the ATET parameter, we use the score

$$\psi(W;\theta,\eta) := H(D,X)\frac{m(X)}{p}(Y - g(0,X)) - \frac{D\theta}{p}, \tag{10.4.8}$$

where $H(D, X)$ is given in (10.4.5), and $\eta = (g, m, p)$ is the nuisance parameter with the true value $\eta_0 = (g_0, m_0, p_0)$ for $g_0$ and $m_0$ defined in (10.4.6) and $p_0 = \mathrm{P}(D = 1)$. Note that this score does not require estimating $g_0(1, X)$.

The scores for GATEs and ATET can be shown to be Neyman orthogonal by calculations similar to those in Section 10.B.

## The DML Inference Method

We assume that we have a sample $(W_i)_{i=1}^n$, modeled as i.i.d. copies of data vector $W$, whose law is determined by the probability measure $P$.

**Generic DML**

1. **Inputs:** Provide the data frame $(W_i)_{i=1}^n$, the Neyman-orthogonal score/moment function $\psi(W, \theta, \eta)$ that identifies the statistical parameter of interest, and the name and model for ML estimation method(s) for $\eta$.

2. **Train ML Predictors on Folds:** Take a K-fold random partition $(I_k)_{k=1}^K$ of observation indices $\{1, ..., n\}$ such that the size of each fold is about the same. For each $k \in \{1, \ldots, K\}$, construct a high-quality machine learning estimator $\hat{\eta}_{[k]}$ that depends only on a subset of data $(X_i)_{i \notin I_k}$ that excludes the $k$-th fold.

3. **Estimate Moments:** Letting $k(i) = \{k : i \in I_k\}$, construct the moment equation estimate

$$\hat{\mathsf{M}}(\theta, \hat{\eta}) = \frac{1}{n} \sum_{i=1}^n \psi(W_i; \theta, \hat{\eta}_{[k(i)]})$$

4. **Compute the Estimator:** Set the estimator $\hat{\theta}$ as the solution to the equation.

$$\hat{\mathsf{M}}(\hat{\theta}, \hat{\eta}) = 0. \tag{10.4.9}$$

5. **Estimate Its Variance:** Estimate the asymptotic variance of $\hat{\theta}$ by

$$\hat{\mathsf{V}} = \frac{1}{n} \sum_{i=1}^n [\hat{\varphi}(W_i)\hat{\varphi}(W_i)'] - \frac{1}{n} \sum_{i=1}^n [\hat{\varphi}(W_i)] \frac{1}{n} \sum_{i=1}^n [\hat{\varphi}(W_i)]',$$

where

$$\hat{\varphi}(W_i) = -\hat{J}_0^{-1} \psi(W_i; \hat{\theta}, \hat{\eta}_{[k(i)]})$$

and

$$\hat{J}_0 := \partial_\theta \frac{1}{n} \sum_{i=1}^{n} \psi(W_i; \hat{\theta}, \hat{\eta}_{[k(i)]}).$$

6. **Confidence Intervals:** Form an approximate $(1-\alpha)\%$ confidence interval for any functional $\ell'\theta_0$, where $\ell$ is a vector of constants, as

$$[\ell'\hat{\theta} \pm c\sqrt{\ell'\hat{V}\ell/n}],$$

where $c$ is the $(1-\alpha/2)$ quantile of $N(0,1)$.

7. **Outputs**: Output the results of all steps.

**Remark 10.4.3** (The Case of Linear Scores) The score for most of our examples is linear in $\theta$; that is, the score can be written as

$$\psi(W;\theta,\eta) = \psi^b(W;\eta) - \psi^a(W;\eta)\theta.$$

In such cases the estimator takes the form

$$\hat{\theta} = \hat{J}_0^{-1} \frac{1}{n} \sum_{i=1}^{n} \psi^b(W_i; \hat{\eta}_{[k(i)]}). \tag{10.4.10}$$

where $\hat{J}_0 = \frac{1}{n}\sum_{i=1}^{n} \psi^a(W_i; \hat{\eta}_{[k(i)]})$.

**Remark 10.4.4** (Sample Splitting) In step 2), the estimator $\hat{\eta}_{[k]}$ can be an ensemble or aggregation of several estimators as long as we only use the data $(X_i)_{i \notin I_k}$ outside the $k$-th fold to construct the estimators.

**Remark 10.4.5** (Choosing the number of folds) The choice $K \geq 4-5$ works well based on a variety of empirical examples and in simulations for medium-sized data sets. The choice $K \geq 10$ works well for small data sets.

## Properties of the general DML estimator

We turn now to the properties of the estimator under the assumption of strong identification.

**Definition 10.4.1** (Strong Identification) *We have that* $\mathrm{M}(\theta, \eta_0) =$

0 *if and and only if* $\theta = \theta_0$, *and that*

$$J_0 := \partial_\theta \mathrm{E}[\psi(W;\theta_0,\eta_0)]$$

*has singular values that is bounded away from zero.*

In the context of the PLM, the latter condition is satisfied if $\mathrm{E}[\tilde{D}^2]$ is bounded away from 0, that is, if $\tilde{D}$ has non-trivial variation left after partialing-out controls. In the context of IRM, the latter condition is satisfied if the overlap condition holds.

**Theorem 10.4.1** (Generic Adaptive Inference with DML) *Assume that estimates of nuisance parameters are of sufficiently high-quality, as specified in [2]. Assume strong identification holds.*

*Then, estimation of nuisance parameter does not affect the behavior of the estimator to the first order; namely,*

$$\sqrt{n}(\hat{\theta} - \theta_0) \approx \sqrt{n}\mathbb{E}_n[\varphi_0(W)],$$

*where*

$$\varphi_0(W) = -J_0^{-1}\psi(W;\theta_0,\eta_0), \quad J_0 := \partial_\theta \mathrm{E}[\psi(W;\theta_0,\eta_0)],$$

*and* $J_0 = \mathrm{E}[\psi^a(W;\eta_0)]$ *for linear scores.*

*Consequently,* $\hat{\theta}$ *concentrates in a* $1/\sqrt{n}$*-neighborhood of* $\theta_0$ *and the sampling error* $\sqrt{n}(\hat{\theta} - \theta_0)$ *is approximately normal:*

$$\sqrt{n}(\hat{\theta} - \theta_0) \overset{a}{\sim} N(0, \mathsf{V}), \quad \mathsf{V} := \mathrm{E}[\varphi_0(W)\varphi_0(W)'].$$

**Theorem 10.4.2** *Under the same regularity conditions, the interval* $[\ell'\hat{\theta} \pm c\sqrt{\ell'\hat{\mathsf{V}}\ell/n}]$ *where c is the* $(1-\alpha/2)$ *quantile of a* $N(0,1)$ *contains* $\ell'\theta_0$ *for approximately* $(1-\alpha) \times 100$ *percent of data realizations:*

$$\mathrm{P}\left(\ell'\theta_0 \in [\ell'\hat{\theta} \pm c\sqrt{\ell'\hat{\mathsf{V}}\ell/n}]\right) \approx (1-\alpha).$$

**Selection of the Best ML Methods for DML to Minimize Upper Bounds on Bias**. In many problems the nuisance parameters are regression functions

$$\eta_m = \mathrm{E}[V_m \mid X_m], \quad m \in \{1, \ldots, M\},$$

where $V_m$ are some response variables and $X_m$ are covariate vectors. Consider a set of ML methods enumerated by $j \in \{1, \ldots, J\}$ that produce estimates $\hat{\eta}_{mj[k]}$ when applied to data excluding the $k$-th fold. We have that

$$\check{V}_{i,mj} = V_i - \hat{\eta}_{mj[k(i)]}(X_i), \quad i \in I_k.$$

**Selection of the Best ML Methods for DML to Minimize Bias.**

- For each method $j$, compute the cross-fitted MSPEs
$$\mathbb{E}_n[\check{V}^2_{mj}].$$
- Select the best ML method for predicting $V_m$ via
$$\hat{j}_m = \arg\min_j \mathbb{E}_n[\check{V}^2_{mj}].$$
- Use the method $\hat{j}_m$ as a learner of $\eta_m$ in the Generic DML Algorithm.

**Corollary 10.4.3** *The results of Theorems 10.4.1 and 10.4.2 continue to hold if $J$ is small.*

The precise conditions may depend on the problem at hand. See the Remark 10.2.3 for discussion in the context of the partially linear model.

## Notebooks

- R Notebook on DML for Impact of Gun Ownership on Homicide Rates and Python Notebook on DML for Impact of Gun Ownership on Homicide Rates provide anapplication of DML inference to learn predictive/causal effects of gun ownership on homicide rates across U.S. counties.
- R Notebook on Dagitty-Based Identification in 401(K) Example and Python Notebook on Pgmpy-Based Identification in 401(K) Example analyze graph structures that enable identification of the causal effect of 401(K) eligibility on net financial wealth.
- R Notebook on DML for Impact of 401(K) Eligibility on Financial Wealth and Python Notebook on DML for Impact of 401(K) Eligibility on Financial Wealth provide application of DML inference to learn predictive/causal effects of 401(K) eligibility on net financial wealth. (Note:

The results produced in this notebook and provided in the text are slightly different than those in the original paper [2]. The replication files for [2] are given at the following Github repository. The difference is due to our use of a single split of the sample in producing the results for this text while the results in [2] are based on a method that aggregates results across multiple data splits.)

- R Notebook on DML for Growth Regression Analysis and Python Notebook on DML for Growth Regression Analysis provide an application of DML inference based on ML on predictive/causal effects of countries' initial wealth on the rate of economic growth.

# Notes

For a detailed literature review and technical regularity conditions needed for each of theorems, see [2], which also gives an overview of various analytical methods for generating Neyman-orthogonal scores in a wide variety of problems.

The paper [9] goes further and describes methods for generating higher-order orthogonal scores:

$$\partial_\eta \partial_\eta \mathrm{E}\psi(\theta_0, \eta_0) = 0.$$

The use of higher-order orthogonal scores allows even weaker requirements for the quality of machine learning estimators of the form,

$$n^{1/6} \|\hat{\eta} - \eta_0\|_{L^2} \approx 0,$$

with the caveat that such higher-order orthogonal scores may not always exist for certain subsets of distributions.

The DML method, developed in Chernozhukov, Chetverikov, Demirer, Duflo, Hansen, Newey, and Robins [2], is simply a practical meta-recipe that explicitly incorporates many classical ideas from the parametric and semi-parametric econometrics and statistics literature; see, e.g., Neyman [8]; Bickel, Klassen, Ritov, Wellner [10]; Newey [11]; Robinson [12]; and Robins and Rotnitzky [13]. The intent was to combine ideas from the classical semi-parametric learning literature and prediction methods from the modern machine learning literature to provide immediately practical methods that are ready for rigorous statistical inference on predictive and causal effects. In essence, the approach can be viewed as a modernized version of the "one"-step debiasing correction proposed by Neyman; see, e.g. [14] for a review.

The partialling-out approach has long been employed in classical econometrics. Robinson [12] was the first to employ it in the context of kernel regressions. [2] extended this approach to more modern settings where ML estimators are used for partialling out, with cross-fitting enabling the extension.

For ATE, GATEs and ATET parameters, DML (or "doubly robust" ML) reduces to the use of machine learned "doubly robust scores" with cross-fitting. The idea of using doubly robust scores (also called augmented inverse propensity score weighted scores) is due to Robins and Rotnitzky [13], but also arises as a special case of Newey's [11] fundamental analysis.

Targeted maximum likelihood estimation (TMLE) is another general approach for building orthogonal estimators [15]. This approach relies on doing maximum likelihood estimation for a target parameter, using a least favorable parametric submodel for the parameter of interest as the likelihood function. As with DML, TMLE needs to be combined with cross-fitting in order to deal with general ML estimators to avoid overfitting. The DML and cross-fitted TMLE should generally produce first order equivalent answers under correct specification. However, using TMLE can refine the finite-sample properties.

In the context of ATE, TMLE can be seen as applying a calibrated correction to a nonlinear regression function. We regress $\check{Y}_i = Y_i - \hat{g}(D_i, X_i)$ on $\hat{H}_i$, obtaining

$$\hat{b} = \mathbb{E}_n[\check{Y}\hat{H}]/\mathbb{E}_n[\hat{H}^2].$$

Then we correct the regression function estimate by $\bar{g}(D_i, X_i) = \hat{g}(D_i, X_i) + \hat{b}\hat{H}_i$. This correction was first proposed by Sharfstein, Rotnitzky and Robins [16]. The basic idea is that we know that $Y_i - g(D_i, X_i)$ should be orthogonal to $H_i$. Thus, if our estimate of the regression function does not have this property, we can recalibrate the regression function so the property holds.

For guidance on using DML in empirical studies and on hyper-parameter tuning related to DML we refer to [17].

## Study Problems

1. Experiment with one of the notebooks for the partially linear models (Guns example, Guns with DNNs, or Growth example). For example,

(a) Apply the methods to a different empirical example (e.g., Penn reemployment experiment from CI-1),
(b) or, using the same empirical example, try to use the H20 Auto ML framework as the machine learning tool to estimate $m$ and $\ell$ functions. (See Chapter 9 H20 Auto ML to get started).

Explain what you are doing to a fellow student.

2. Study the 401(K) identification notebook that uses Dagitty. Extend it to another empirical example of your choice. Explain the principles you are using to a fellow student.

3. Study the 401(K) empirical analysis notebook (the part that does not deal with instrumental variables and LATE). Extend it to another empirical example of your choice (Penn reemployment experiment from Chapter 1, for example) or estimate ATE for 401(K) eligibility for a subset of low income (or high-income) workers (Group ATEs).

4. (Theoretical). Explain to a friend the concept of Neyman orthogonality, illustrating it with one of the examples in Appendix B. Extend the calculations in Appendix B to verify Neyman orthogonality for the ATET score specified in (10.4.8).

5. (Theoretical). Explain to a friend the concept of Neyman orthogonality, and explain why the formulations given in Remark 10.3.1 are not Neyman orthogonal.

## 10.A Bias Bounds with Proxy Treatments

Here we explain the measurement error bias in the partially linear structural equation model where treatment is measured with error:

$$\begin{aligned} Y &:= \alpha G + g_Y(X) + \epsilon_Y; \\ D &:= G + g_D(X) + \epsilon_D; \\ G &:= g_G(X) + \epsilon_G; \\ X &:= \epsilon_X; \end{aligned}$$

where $\epsilon$'s are independent and centered. The second equation states that $D$ is generated as a proxy for the actual treatment $G$ using a partially linear structure. In partialed-out form

$$\begin{aligned} \tilde{Y} &:= \alpha \epsilon_G + \epsilon_Y; \\ \tilde{D} &:= \epsilon_G + \epsilon_D; \\ \tilde{G} &:= \epsilon_G. \end{aligned}$$

The projection of $\tilde{Y}$ on $\tilde{D}$ recovers the projection coefficient:

$$\beta = \mathrm{E}[\tilde{Y}\tilde{D}]/\mathrm{E}[\tilde{D}^2] = \alpha \mathrm{E}[\epsilon_G^2]/(\mathrm{E}[\epsilon_G^2] + \mathrm{E}[\epsilon_D^2]).$$

It follows that there is attenuation bias in the estimable quantity $\beta$ relative to the target parameter $\alpha$:

$$|\beta| < |\alpha|.$$

As the proxy error $\mathrm{E}[\epsilon_D^2]$ becomes small, the difference between $\beta$ and $\alpha$ becomes small. Specifically, if $\mathrm{E}[\epsilon_D^2] \to 0$, then $\beta \to \alpha$.

If we somehow knew that

$$R^2_{\tilde{D}\sim\tilde{G}} := \mathrm{E}[\epsilon_G^2]/(\mathrm{E}[\epsilon_G^2] + \mathrm{E}[\epsilon_D^2]) \geq 2/3$$

that is, the true treatment $G$ explains at least two thirds of variance of the proxy treatment $D$ – then we could construct the upper and lower bound on $\alpha$ from $\beta$. E.g. when $\beta > 0$, we would have

$$\beta \leq \alpha \leq \beta / R^2_{D\sim G} = (3/2)\beta.$$

# 10.B Illustrative Neyman Orthogonality Calcuations

**The Score in the Partially Linear Model.** Consider the score for the PLM given in (10.4.3). We have that

$$\mathrm{E}[\psi(W;\beta_0,\eta_0)] = 0$$

by definition of $\beta_0$ of $\eta_0$; recall the 0 indices denote true values. Let $U = (Y - \ell_0(X)) - (D - m_0(X))\beta_0)$. Then, for any $\eta = (m,\ell)$ that are square integrable, the Gateaux derivative in the direction

$$\Delta = \eta - \eta_0 = (m - m_0, \ell - \ell_0)$$

is given by

$$\begin{aligned}
&\partial_\eta \mathrm{E}[\psi(W;\beta_0,\eta_0)][\Delta] \\
&\quad = -\mathrm{E}\Big[U(m(X) - m_0(X))\Big] \\
&\qquad - \mathrm{E}\Big[\Big((m(X) - m_0(X))\beta_0 + (\ell(X) - \ell_0(X))\Big)(D - m_0(X))\Big] \\
&\quad = 0,
\end{aligned}$$

by the law of iterated expectations since $\mathrm{E}[D - m_0(X) \mid X] = 0$ and $\mathrm{E}[U \mid D, X] = 0$.

**The Score for IRM**. Consider the score for the ATE in the IRM given in (10.4.4). We have that

$$\mathrm{E}[\psi(W;\theta_0,\eta_0)] = 0$$

by definition of $\theta_0$ and $\eta_0$. Also, for any $\eta = (g, m)$ that are square integrable with $1/m + 1/(1 - m)$ uniformly bounded, the Gateaux derivative in the direction

$$\Delta = \eta - \eta_0 = (g - g_0, m - m_0)$$

is given by

$$
\begin{aligned}
&\partial_\eta \mathrm{E}[\psi(W;\theta_0,\eta_0)][\Delta] \\
&\quad = \mathrm{E}\Big[g(1,X) - g_0(1,X)\Big] \\
&\qquad - \mathrm{E}\Big[g(0,X) - g_0(0,X)\Big] \\
&\qquad - \mathrm{E}\Big[\frac{D(g(1,X) - g_0(1,X))}{m_0(X)}\Big] \\
&\qquad + \mathrm{E}\Big[\frac{(1-D)(g(0,X) - g_0(0,X))}{1 - m_0(X)}\Big] \\
&\qquad - \mathrm{E}\Big[\frac{D(Y - g_0(1,X))(m(X) - m_0(X))}{m_0^2(X)}\Big] \\
&\qquad - \mathrm{E}\Big[\frac{(1-D)(Y - g_0(0,X))(m(X) - m_0(X))}{(1 - m_0(X))^2}\Big],
\end{aligned}
$$

which is 0 by the law of iterated expectations since $\mathrm{E}[D \mid X] = m_0(X)$, $\mathrm{E}[1 - D \mid X] = 1 - m_0(X)$, $\mathrm{E}[D(Y - g_0(1,X)) \mid X] = 0$, and $\mathrm{E}[(1-D)(Y - g_0(0,X)) \mid X] = 0$.

# 11 Feature Engineering for Causal and Predictive Inference

"It's all about paying attention. [...] Attention is vitality. It connects you with others."

– Susan Sontag [1].

Here we discuss feature engineering as an approach to transform complex objects such as text and images into a collection of relatively low-dimensional numerical features (embeddings) that can be used for standard predictive or causal applications, for example as regressors in a prediction problem. We consider principal components, variational autoencoders and neural networks as general approaches to generate embeddings. We then consider text embeddings in detail, introducing two popular neural network-based Natural Language Processing (NLP) algorithms: ELMo and BERT. We finally consider image embeddings, applying a hedonic price model to apparel data using a neural network algorithm (ResNet50) to generate embeddings.

## 11.1 Introduction

Thus far, we have imposed a significant restriction on the kinds of data on which we can perform inference. While empiricists often consider simple datasets that include variables that have a numeric representation (binary, factor and continuous variables), researchers are increasingly confronted with complex forms of data, such as images and text, that encode a vast amount of information. In this section, we generalize our approach to allow using these types of data.

As a motivating example, we consider the problem of predicting prices of products using the types of characteristics that one might find on a webpage, namely the text in the product description and the product's image. The resulting predicted prices are called hedonic prices, and predictive modeling of this form is motivated by the hedonic price models of economics.

In order to predict prices, we have to convert text and images into relatively low-dimensional numerical features, called "embeddings." The minimal requirement on embeddings is that similar products should have similar embeddings. This requirement guarantees that price predictions for similar products are also similar. The maximal requirement on embeddings is that they should parsimoniously approximate maximal information from text and images that is relevant for price predictions.

The main methods for generating successful embeddings include the following, in order of increasing generality:

- classical principal component analysis,
- auto-encoders, and
- neural networks solving auxiliary prediction tasks.

The auxiliary tasks in the final method may include solving image processing problems, such as object classification and image compression, or natural language processing problems, such as summarization and machine translation.

These auxiliary tasks are not the same as the "main" task. In our price prediction example, the main task is predicting product prices. Before turning to the primary price prediction task, we consider ResNet-50, which is a Residual Network of depth 50, which is designed to perform well on various tasks of object type classification. Consequently, application of ResNet-50 produces embeddings that are useful inputs for solving this auxiliary object classification task. However, because product type is an important determinant of price, the embeddings produced by ResNet-50 that help classify products can also serve as useful inputs to the main task – price prediction.

Analogously, a neural network such as BERT is trained on auxiliary tasks aimed to make it learn word similarity and contextual meaning of words. Consequently, BERT can produce embeddings that provide a useful numerical summary of a product's text description. Because the product description is an important determinant of the price, these embeddings can also serve as useful inputs to the price prediction task.

Embeddings are useful in a variety of predictive and causal inference problems. For example, we can imagine using

- embeddings of product images and descriptions for modeling variety and demand for products;
- embeddings of text resumes for studying the wage offer structure;
- embeddings of countries' characteristics for studying the effect of institutions;
- and please list many of your own here (homework).

There is an emerging literature on the use of embeddings for causal inference; see this repository of papers about using text data in causal inference. See also [2] for a recent review article on the importance and subtleties of using text as data in the social sciences.

## 11.2 From Principal Components to Autoencoders

Principal components are probably the earliest classical example of embeddings. One way to frame principal components is that principal components find unit length orthogonal linear combinations, directions, of a collection of variables that are "best" at reproducing the underlying data. The idea is then that a small number of principal components should capture most of the variability in the original variables and thus may provide a useful low-dimensional summary of the original data.

Specifically, let $(W_1, \ldots, W_n)$ be a sample of $n$ observations of a high-dimensional centered random vector $W_i$ in $\mathbb{R}^d$,[1] and let $\Sigma_n = \mathbb{E}_n[WW'] \in \mathbb{R}^{d \times d}$ denote the empirical covariance matrix. In order to reduce the dimension of $W_i$, suppose we wish to find $K \ll d$ mutually orthogonal rotations

1: Thus, $\mathbb{E}_n[W_j] = 0$ for $j = 1, \ldots, d$.

$$X_{ki} := c_k' W_i, \quad k = 1, \ldots, K,$$

of the original $W_i$'s where

$$c'_\ell c_k = 0 \text{ for } \ell \neq k \text{ and } c'_k c_k = 1 \text{ for each } k$$

such that linear combinations of these variables approximate the original data. These rotations are called principal components of $W_i$. In applications, $W_i$ represent high-dimensional raw features (images, for example), and the principal components

$$X_i^K = (X_{i1}, ...X_{iK})'$$

represent a lower-dimensional encoding or embedding of $W_i$.

More formally, we wish to solve

$$\min_{\{a_j\}_{j=1}^d, \{c_k\}_{k=1}^K} \sum_{j=1}^d \sum_{i=1}^n (W_{ji} - \hat{W}_{ji})^2$$

subject to

$$\begin{aligned} \hat{W}_{ji} := a'_j X_i^K \text{ for } X_i^K = (X_{1i}, ...X_{Ki})' \ j = 1, ..., d \text{ and } i = 1, ..., n, \\ X_{ki} = c'_k W_i \text{ for } i = 1, ..., n \text{ and } k = 1, ..., K, \\ c'_k c_k = 1 \text{ for } \quad k = 1, \dots, K \\ c'_k c_\ell = 0 \text{ for } \ell \neq k \leq K. \end{aligned}$$

The constructed variables resulting from solving this problem,

$$X_i^K = (X_{1i}, ...X_{Ki})'$$

are the first $K$ principal components.

**Remark 11.2.1** The analytical solution to the principal components problem is as follows: The optimal $C_K = [c_1, ..., c_K]$ are the eigenvectors of $\Sigma_n = \mathbb{E}_n[WW']$ corresponding to the $K$ largest eigenvalues $\lambda_1, ..., \lambda_K$ of $\Sigma_n$. That is, $\Sigma_n c_k = \lambda_k c_k$ for each $k$. Furthermore, the optimal $a_j$ is the $j$-th column of $C'_K$.

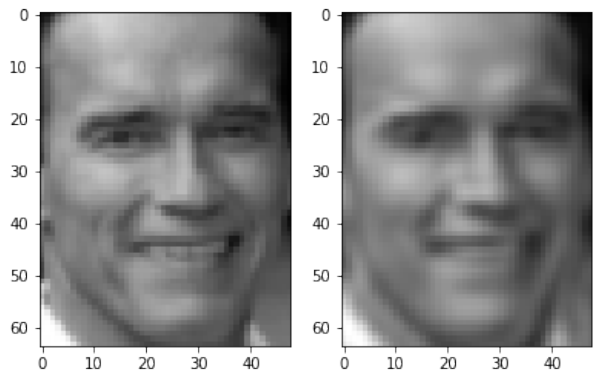

**Figure 11.1:** Featurizing a talented man: The original 3072-dimensional image $W$ and image $\hat{W}$ produced from a 256-dimensional principal component embedding. As a by-product, we've just made an important causal discovery that, surprisingly, doing embedding causes one to be younger ;).

Another interesting feature is of principal components is that they satisfy

$$\mathbb{E}_n[X_k^2] = \lambda_k$$

for $k = 1, ..., K$ and

$$\mathbb{E}_n[X_k X_\ell] = 0$$

for $\ell \neq k$. These properties result from the fact that the $c_k$ are eigenvectors of $\Sigma_n$.

Finding principal components offers one way to produce encod-ings/embeddings of raw inputs. Once we have encodings/embeddings from any method, we can look at how similar the raw inputs $W_k$ and $W_l$ are via the cosine similarity of the embeddings:

$$\mathrm{sim}(W_k, W_l) = X_k' X_l / (\|X_k\| \|X_l\|).$$

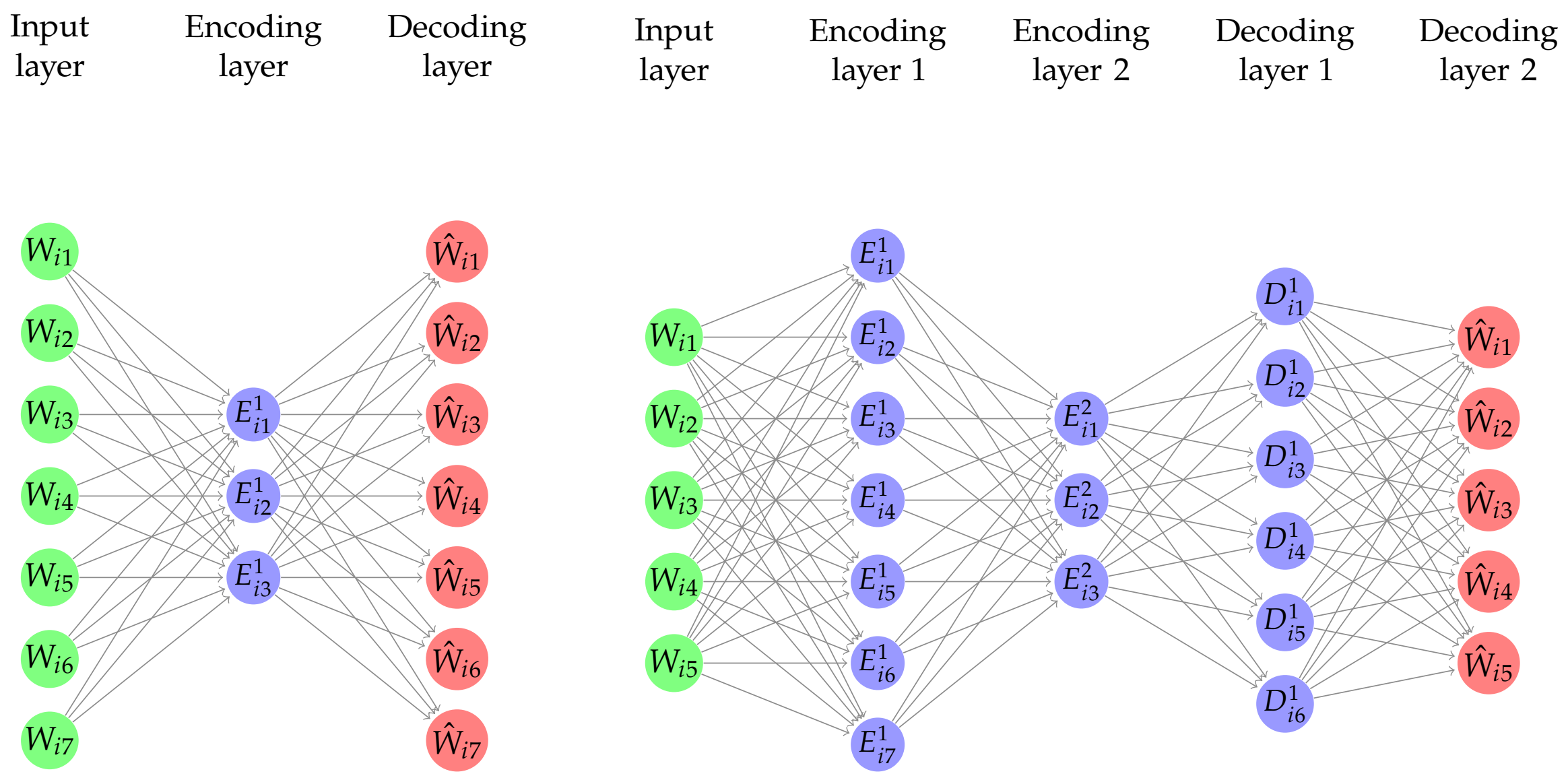

**Figure 11.2:** The left panel shows a linear single layer autoencoder, such as linear principal components. The right panel shows a three layer nonlinear autoencoder; the middle layers can be used as embeddings.

In the context of product embeddings, this approach can be used, for example, to find products that are similar to a given product.

> This predictive exercise underlying principal components can be seen as a linear neural network:
>
> $$\underset{d\times 1}{W_i} \longmapsto C_K' W_i =: \underset{k\times 1}{E} \longmapsto A'E =: \underset{d\times 1}{\hat{W}_i},$$
>
> for $A = [a_1, ...a_d]$. The first step is said to be "encoding" the information in the input, and the second step is said to be "decoding" in the sense of returning the encoded information to the original space. Therefore, principal components are embeddings generated by a linear "encoder-decoder" network (an *autoencoder*, for short). For principal components, the relationship between the encoder and decoder happens to be rather simple, in that $A = C_K'$ (see Remark 11.2.1).

This framing suggests that we can immediately generalize this approach to nonlinearly generated encoders and decoders that

have multiple layers:

$$W_i \overset{g_1}{\longmapsto} E_i^1 \ldots \overset{g_k}{\longmapsto} E_i^k \overset{g_{k+1}}{\longmapsto} D_i^{k+1} \ldots \overset{g_m}{\longmapsto} D_i^m =: \hat{W}_i,$$

where maps $g_\ell$'s are neuron-generating maps. The middle layer or layers of low dimension, represented by the $E_i^k$, are taken to be encoders. The layers of neurons are mnemonically labelled as either "E" or "D," depending on whether they are doing "encoding" or "decoding," though note that there is no strict formal distinction between these types of layers.

Autoencoders are a way of discovering latent, low-dimensional structures in a dataset. In particular, a random data vector $W \in \mathbb{R}^d$ can be said to have low-dimensional structure if we can find some "well-behaved" functions $\mathsf{e} : \mathbb{R}^d \to \mathbb{R}^k$ and $\mathsf{d} : \mathbb{R}^k \to \mathbb{R}^d$, with $k \ll d$, such that

$$(\mathsf{d}(\mathsf{e}(W)) \approx W.$$

In other words, $X = \mathsf{e}(W)$ is a parsimonious, $k$-dimensional representation of $W$ that contains all of the information necessary to approximately reconstruct the full vector $W$. Traditionally, the map $\mathsf{e}(\cdot)$ is called an encoder, and the map $\mathsf{d}(\cdot)$ is called a decoder function. Given this, a general formulation of autoencoders is to minimize the average reconstruction loss,

$$\mathbb{E}_n[\text{loss}(W, \mathsf{d}(\mathsf{e}(W))],$$

over "well-behaved" functions $\mathsf{d} \in \mathsf{D}$ and $\mathsf{e} \in \mathsf{E}$. These classes are often linear, as in principal components, or generated via neural networks.

The qualification of "well-behaved" is important since it is always possible to write down some (completely wild) one-to-one function $\mathsf{e} : \mathbb{R}^d \to \mathbb{R}^1$ such that $\mathsf{e}^{-1}\mathsf{e}(W) = W$.[2]

2: Google "Borel Isomorphism."

**Remark 11.2.2** (Independent Component Analysis) Principal component analysis defines "well-behaved" functions as linear functions whose output $(X_1, \ldots, X_k) = e(W)$ has uncorrelated entries, i.e. $\mathrm{E}[X_i X_j] = 0$. In other words, PCA tries to find latent embedding vectors $(X_1, \ldots, X_k)$ that have the ability to reconstruct the original covariate vectors and are uncorrelated with each other. One could take a step further and require that these latent embeddings are independent of each other, not just uncorrelated, i.e. $X_i \perp\!\!\!\perp X_j$. This leads to the method called linear Independent Component Analysis (ICA) [3]. The intuition of ICA when taken to neural network representations has led to the notion of disentangled representations, i.e. embeddings that encode independent latent dimensions of variation in the data. However, beyond the linear ICA setting, non-linear neural network based versions of ICA have more brittle theoretical foundations in the absence of auxiliary task-related information and task-related outcome variables [4].

**Remark 11.2.3** (Variational Auto-Encoders [5]) The encoding and decoding functions so far in our discussion have been restricted to be deterministic. Implicitly, this assumes that given the observed high-dimensional variables $W$, we can uniquely identify the low-dimensional variables that contain all the information in $W$. Such unique mapping from observed factors to latent factors is not always possible. An important extension of the auto encoder framework is allowing for these mappings to be stochastic. This extension is inspired by Bayesian probabilistic modelling that views the embeddings as latent factors and imagines that the data are drawn by first drawing the latent factors and then drawing the data samples from some distribution that is dependent on the latent factors. The Variational Auto-Encoder is attempting to reverse engineer and learn the latent factor space and the posterior distribution of the latent factors conditional on the observed data using computationally tractable approximate versions of the maximum likelihood method. In the end the method looks very similar to auto-encoders (though stemming from a different reasoning), albeit introducing randomness in the encoding phase. Roughly speaking, we optimize a loss of the form:

$$\mathbb{E}_n[\text{loss}(W, \mathsf{d}(\mathsf{e}(W, Z))] + penalty(\mathsf{e}),$$

where $Z$ is an exogenous jointly independent Gaussian random noise vector and the penalty term forces the encoding function to be non-deterministic and stems from derivations related to the objective of learning the posterior distribution of latent factors. Conditional on an observed $W$, the random variable $\mathsf{e}(W, Z) \mid W$ can be interpreted as a random sample from the posterior distribution of the latent factors that could have generated the observed sample $W$. Moreover, the function $e(W, Z)$, is typically of the form $e(W, Z) = \mu(W) + \Sigma(W)Z$, where the deterministic functions $\mu(W)$ and $\Sigma(W)$ encode the mean and the covariance of the posterior distribution of the latent factors. These deterministic functions $\mu(W), \Sigma(W)$ can be viewed as deterministic embeddings of $W$ and can be used as engineered features in downstream tasks (see e.g. [6]). Alternatively, one can use only $\mu(W)$, which is approximating the mode of the posterior. For a more in-depth introduction to variational auto-encoders see [7].

## 11.3 From Auto-Encoders to General Embeddings

The notion of loss can also be generalized to other loss functions, where the target outcome $A$ may be not $W$, but something else. We can search for embeddings that minimize average prediction loss,

$$\mathbb{E}_n[\text{loss}(A, \mathsf{f}(\mathsf{e}(W))],$$

where the role of $\mathsf{f}(\cdot)$ is no longer just to decode but to predict $A$ (rather than $W$).

For example, in feature engineering from images, $A$ could be a product type or subtype, and $W$ could be the image. In feature engineering from text, $A$ could be a masked word in a sentence and $W$ the sentence containing this word. These alternative approaches could be more useful in relation to the final learning task. For example, to build good hedonic price models, we may be more interested in image or text embeddings that best help to accurately describe the type or subtype of a product (rather than reconstruct the image or text itself).

This approach is generally implemented via neural networks as follows

$$W_i \overset{g_1}{\longmapsto} E_i^1 \ldots \overset{g_k}{\longmapsto} E_i^k \overset{g_{k+1}}{\longmapsto} F_i^{k+1} \ldots \overset{g_m}{\longmapsto} F_i^m =: \hat{A}_i,$$

where $g_\ell$'s are neuron-generating maps. The middle layers $E_i^k$ are taken to be embedding layers, and $F_i^k$'s are predictive layers, aimed to create good predictions of auxiliary targets.

## 11.4 Text Embeddings

### First generation: Word2Vec Embeddings

We first review some basic ideas underlying the Word2Vec algorithm (Mikolov et al., 2013). One way we could encode words that appear in a corpus of documents (e.g. product descriptions) into a vector is to consider a very high-dimensional vector of dimension $d$, where $d$ is the total number of words in the corpus. Then the $j$-th word in the corpus (e.g. in alphabetical order) can be represented as:

$$e_j = (0, .....0, 1, 0, ....0)',$$

with 1 in the j-th position. This encoding has a very high dimension limiting its usefulness. Furthermore, this representation does not capture word similarity – e.g., cosine similarity between two different words $j$ and $k$ is always zero since $e_j{}'e_k = 0$.

Instead we aim to represent words by vectors of much lower dimension, $r$, that are able to capture word similarity. We denote the representation of the $j$-th word by $u_j$, so the dictionary is an $r \times d$ matrix

$$\omega = \{u_1, ..., u_d\},$$

where $r$ is the reduced dimensionality of the dictionary. This dictionary is a linear rotation of the original dictionary $E = \{e_1, ..., e_d\}$, where

$$\omega = \omega E.$$

Therefore, the problem of finding the rotation $\omega$ is analogous to the problem of finding principal components, except that our goal is now to find representations $\omega$ that are able to capture word similarity. Once we are done, each word $t_j$ in a human-readable dictionary can be represented by a new "word" $u_j$. The goal of Word2Vec is to find an effective representation with the dimension $r$ of the embedding being much smaller than the total number of words in the corpus, $d$. We achieve this goal by treating $\omega$ as parameters and estimating them so that the model performs well in some basic natural language processing tasks. These tasks are typically not related to downstream tasks, such as predicting hedonic prices or performing causal inference using text as control features, but are related to language prediction tasks.

Figure 11.3 shows components of embeddings for several words produced by a trained Word2Vec map. The numbers presented in the table are not particularly interpretable in isolation. Each

**Example of Wor2Vec features**

| | | | | | | | | |
|---|---|---|---|---|---|---|---|---|
| womens | 0.387542 | 0.03051 | -0.19703 | 0.179724 | -0.222901 | -0.606905 | 0.306091 | -0.597467 |
| mens | 0.758868 | 0.372418 | 0.370116 | 0.706623 | -0.124954 | 0.5088 | 0.106177 | 0.208935 |
| clothing | 0.149283 | 0.5161 | -0.027684 | 0.218484 | -0.851416 | -0.409885 | 0.386088 | 0.170605 |
| shoes | 1.323812 | -0.358704 | -0.007683 | -0.552144 | 0.011261 | 0.365239 | 0.228273 | -0.565655 |
| women | 0.601477 | -0.045845 | -0.099481 | 0.010576 | -0.096852 | -0.605281 | 0.25606 | -0.550759 |
| girls | 0.417473 | -0.005265 | -0.40939 | -0.531189 | -1.31938 | -0.034746 | -0.940507 | -0.361215 |
| men | 0.778298 | 0.406613 | 0.426292 | 0.534272 | -0.056103 | 0.51756 | 0.107846 | 0.245275 |
| boys | 0.896637 | -0.016821 | -0.001602 | -0.181901 | -1.313441 | 0.449006 | -0.828408 | 0.52121 |
| accessories | 0.86825 | -0.378385 | -1.247708 | 1.541265 | 0.323952 | 0.282909 | -0.491176 | 0.081314 |
| socks | 0.27636 | 0.354296 | 0.185734 | 0.301311 | -0.643142 | -0.021945 | 0.320751 | 0.240676 |
| luggage | 0.796763 | 1.749548 | -2.30671 | -0.559585 | 0.03054 | 0.921458 | 0.417333 | 0.313436 |
| dress | 0.282053 | 0.233192 | 0.043318 | 0.174759 | -0.50114 | -0.381047 | 0.297995 | -0.026033 |
| baby | 0.346065 | -0.550016 | -1.136202 | -0.043899 | -2.004979 | 0.689747 | -1.091575 | 0.009901 |
| jewelry | -0.315784 | 0.347808 | -0.308736 | 0.878713 | -0.766016 | 1.124318 | -0.079883 | -2.039485 |
| black | 0.427496 | 0.030204 | -0.019082 | 0.224096 | -0.162242 | -0.325359 | 0.170407 | -0.172714 |
| boots | 1.009074 | -0.30359 | 0.03197 | -0.334004 | -0.095679 | 0.111328 | 0.11769 | -0.51878 |
| shirts | 0.444152 | 0.452918 | 0.393656 | 0.517929 | -0.531462 | 0.099621 | 0.146202 | 0.204338 |
| shirt | 0.328998 | 0.421561 | 0.226565 | 0.455649 | -0.700352 | 0.067224 | 0.106364 | 0.233862 |
| underwear | 0.230821 | 0.490978 | 0.226338 | 0.202376 | -0.774363 | 0.004693 | 0.228712 | 0.310215 |

**Figure 11.3:** Examples of words converted to numerical features via Word2Vec. Compare embeddings for words "shirt" and "shirts" and for "luggage" and "dress".

column represents a "trait" and the cell entry represents the loading of the word in the row in that trait. The numbers are more useful in comparison with each other across different rows which allows us to understand word similarity. For example, we can see that the very similar words "shirt" and "shirts" have very similar embeddings while the embeddings for the seemingly relatively different words "luggage" and "dress" are quite dissimilar.

In our context, we can think of each word appearing in a datum (e.g. a product description) as a random variable $T$ and denote its corresponding embedding representation by $U$.

> One of the ways to train the word embeddings is to predict the middle word from the words that surround it in word sentences.

Given a subsentence $s$ of $K + 1$ words, we have a central word $T_{c,s}$ whose identity we would like to predict. As predictors, we have the context words $\{T_{o,s}\}$ that surround the central word $T_{c,s}$. One approach for forming the prediction starts by collapsing the embeddings for context words by a sum,[3]

3: Why not? We can try it and see if it works.

$$\bar{U}_o = \frac{1}{K} \sum_o U_{o,s},$$

where $U_{o,s}$ is the element of $\omega$ corresponding to the word $T_{o,s}$. This step imposes a drastically simplifying assumption that the context words are exchangeable – i.e. the position of each word is not important.

The probability of the middle word $T_{c,s}$ being equal to $t$ is modeled via the multinomial logit function:

$$p_s(t; \pi, \omega) := P\Big(T_{c,s} = t \mid \{T_{o,s}\}; \omega\Big) = \frac{\exp(\pi_t' \bar{U}_s(\omega))}{\sum_{\bar{t}} \exp(\pi_{\bar{t}}' \bar{U}_s(\omega))},$$

where $\pi = (\pi_1, ..., \pi_d)$ is an $m \times d$ matrix of parameter vectors defining the choice probabilities. The model constrains the choice probabilities $\pi$ to be $\omega$, and estimates $\omega$ using the maximum quasi-likelihood method:

$$\max_{\omega=\pi} \sum_{s \in \mathcal{S}} \log p_s(T_{i,s}; \pi, \omega),$$

where we sum the log-probabilities over many examples $\mathcal{S}$ of subsentences $s$. Once we are done training, we can generate the embedding for the title or description of product $i$, containing the embedded words $\{U_{j,i}\}_{j=1}^{J}$ by simply averaging them:

$$W_i = \frac{1}{J} \sum_{j=1}^{J} U_{j,i}. \tag{11.4.1}$$

**Remark 11.4.1** In summary, the Word2Vec algorithm transforms text into a vector of numbers that can be used to compactly represent words. The algorithm trains a neural network in a supervised manner such that contextual information is used to predict another part of the text.

For example, let's say that the title description of the item is: "Hiigoo Fashion Women's Multi-pocket Cotton Canvas Handbags Shoulder Bags Totes Purses." The model will be trained using many $n$-word subsentence examples, such that the center word is predicted from the rest. If we just use $n = 3$ subsentence examples, then we train the model using the following examples: (Hiigoo,Women's) $\longrightarrow$ Fashion, (Fashion,Multi-pocket)$\longrightarrow$ Women's, (Women's,Cotton) $\longrightarrow$ Multi-pocket, and so on.

How do we judge whether the text embedding is successful or not? In the hedonic price context, we can check whether Word2Vec features improve the quality of prediction of the price by the hedonic model. We can also check if similar words $T_k$ and $T_l$ have similar embeddings. We can measure the similarity through cosine similarity:

$$\text{sim}(T_k, T_l) = U_k' U_l / (\|U_k\| \|U_l\|) \in [-1, 1].$$

The more similar the words are, according to our human notion of similarity, the higher the value our formal measure of similarity should take. For example, the following are the two words that are most similar to "tie" under the similarity measure: "necktie" and "bowtie." The dense embedding also induces an interesting vector space on the set of words, which seems to encode analogues well. For example, the word "briefcase" is very cosine-similar to the artificial latent word[4]

$$\text{Artificial word} = \text{Word2Vec(men's)} + \text{Word2Vec(handbag)} - \text{Word2Vec(women's)}.$$

This similarity between a real word and our constructed latent word gives some justification for the "averaging" of embeddings to summarize whole sentences or descriptions.

Word2vec embeddings were among the first generation of early successful embedding algorithms. These algorithms have been improved by the next generation of NLP algorithms, such as ELMo and BERT, which are discussed next.

## Second Generation: Sequence Models

A major advance in language modeling has been to represent text as a sequence using recurrent (autoregressive) models. Among the various benefits, this in particular allows for better capturing the context around words. Of note is the Embeddings from Language Models (ELMo) algorithm [10], which uses the idea of the Shannon game where we aim to guess a word in a sentence, $m$, consisting of $n$ total words. Specifically, we consider the problem of predicting word $k+1$ using the preceding $k$ words via

$$p^f_{k,m}(t) = P[T_{k+1,m} = t \mid T_{1,m}, ..., T_{k,m}; \theta]$$

and similarly consider the reverse prediction via

$$p^b_{k,m}(t) = P[T_{k-1,m} = t \mid T_{k,m}, ...T_{n,m}; \theta],$$

where $\theta$ is a parameter vector. ELMo then uses recurrent neural networks (RNNs) to model these probabilities. RNN is a particular architecture for sequence input and output where we use neurons from the previous prediction to make the current prediction.[5] Parameters are estimated by maximimizing parameterized approximate versions of the log-likelihoods of the observed data (aka quasi-likelihoods), typically referred to

4: This example also shows us how word embeddings very easily encode and propagate biases that exist in document corpora that are typically used in machine learning; a realization that has been highlighted by several recent works [8]. One should always be cognizant of such inherent biases in trained embeddings. Recent works in machine learning (e.g. [8]) provide automated approaches that partially correct for these biases, though not completely removing the problem [9].

5: RNNs are essentially the neural network versions of linear autoregressive models, such as ARIMA models, which go back to the early work of statisticians George Box and Gwilym Jenkins [11, 12] and have also been used in economics to model volatility of financial assets in the GARCH model of economist Tim Bollerslev [13].

In its simplest form a RNN parses inputs in a serial manner $T_1, \ldots, T_t, \ldots, T_k$ and at each step $t$ produces a state vector $S_t = \sigma(AT_t + BS_{t-1} + c)$ that is a nonlinear function (a set of neurons) of the current input and the previous state vector. That is, $\sigma$ is an activation function as presented in Section 9.3 applied elementwise to each coordinate. Moreover, a RNN produces an output prediction vector $y_t = \sigma(DS_t + e)$ that is a nonlinear function (a set of neurons) of the current state. The parameters $A, B, c, D, e$ of all these neurons are the same (shared) across steps.

ELMo uses a particular form of recursive neural network called Long Short-Term Memory (LSTM) network. LSTMs improve upon the numerical stability of RNNs by allowing for the "state" to pass through the current step as-is, without any non-linearity applied. Allowing the state to pass through steps without alteration helps in propagating information across distant steps and thus better accommodates long-term memory.

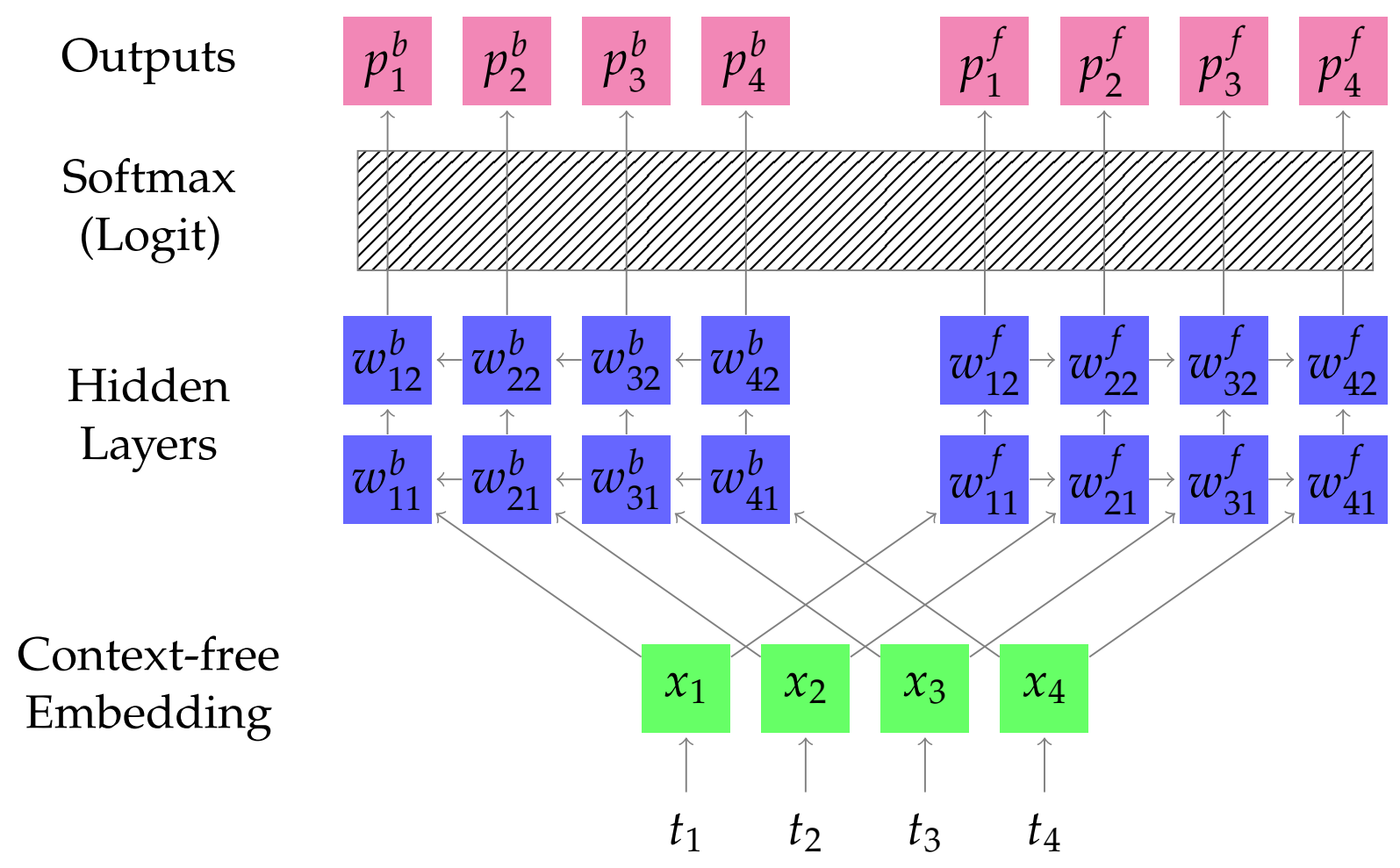

**Figure 11.4:** ELMo Architecture. ELMo network for a string of 4 words, with $L = 2$ hidden layers. Here, the softmax layer (multinomial logit) is a single function mapping each input in $\mathbb{R}^d$ to a probability distribution over the dictionary $\Sigma$.

as quasi-maximum log-likelihood methods, where the forward and backward log quasi-likelihoods are added together.

To give a simple example, suppose we wanted to grasp the positional context better in the previous example. Rather that start by collapsing the embeddings for context words surrounding a target central word via a sum, we could instead keep track of word order and assign individual parameters to each context. For example, we could model the forward predicted probability of word $k$ in sentence $m$ as

$$P(T_{k,m} = t \mid \{T_{j,m}\}_{j=1}^{k-1}) = \frac{e^{\sum_{j=1}^{k-1} \pi'_{t,j} U_{j,m}(\omega)}}{\sum_{\bar{t}} e^{\sum_{j=1}^{k-1} \pi'_{\bar{t},j} U_{j,m}(\omega)}},$$

and similarly model the reverse prediction problem where we recall that $U_{j,m}(\omega)$ is the embedding corresponding to word $j$ in sentence $m$. ELMo uses a more sophisticated (and more parsimonious) recursive nonlinear regression (specifically a recurrent neural network) model to build these probabilities. We illustrate a simple ELMo structure in Figure 11.4.

**The basic structure of ELMo.**
Given a sentence $m$ of $n$ words,

1. Words are mapped to context-free embeddings in $\mathbb{R}^d$

2. A network is trained to predict each word $T_{k,m}$ of a string given (a) words $(T_{1,m}, \ldots, T_{k-1,m})$ (forward prediction) or (b) words $(T_{k+1,m}, \ldots, T_{n,m})$ (backward prediction). The objective is to maximize the average over the sum of the log-likelihoods of the $2n - 2$ words being predicted, where the average is taken over all sentences.

3. The embedding of word $T_{k,m}$ is given by a weighted average of outputs of certain hidden neurons corresponding to this word's entire context. Importantly, a subset of the parameters is coupled across the forward and backward prediction problems (2a) and (2b). In particular, the first layer that goes out of the context-free embedding and the final ("softmax") layer that produces the probabilistic predictions is the same for the two prediction objectives (2a) and (2b). Thus the inputs to this layer, which represent the forward and backward context, are constrained to lie in "the same space."

A softmax layer assigns probabilities to each class in a multi-class problem. It is a multi-class generalization of logistic regression that assumes mutually exclusive classes.

**Training**

In Figure 11.4, the output probability distribution $p_k^f$ is taken as a prediction of $T_{k+1,m}$ using words $(T_{1,m}, \ldots, T_{k,m})$. Similarly, $p_k^b$ is taken as a prediction of $T_{k-1,m}$ using words $(T_{k,m}, \ldots, T_{n,m})$. The parameters of the network, $\theta$, are obtained by maximizing the quasi-log-likelihood:

$$\max_{\theta} \sum_{m \in \mathcal{M}} \left( \sum_{k=1}^{n-1} \log p_{k,m}^f(T_{k+1,m}; \theta) + \sum_{k=2}^{n} \log p_{k,m}^b(T_{k-1,m}; \theta) \right),$$

where $\mathcal{M}$ is a collection of sentences. In our example, $\mathcal{M}$ is the collection of titles and product descriptions taken from product web pages.

**Producing embeddings**

To produce embeddings from the trained network, each word $t_k$ in a sentence $m = (t_1, ..., t_n)$ is mapped to a weighted average of the outputs of the hidden neurons indexed by $k$:

$$t_k \mapsto w_k := \sum_{i=1}^{L} (\gamma_i w_{ki}^f + \bar{\gamma}_i w_{ki}^b).$$

The embedding for the sentence (or an entire product description in our example) is produced by summing the embeddings for each individual word. The weights $\gamma$ and $\bar{\gamma}$ can be tuned by the neural network performing the final task. In principle, the whole network could be plugged in to the network performing the final task and allowed to update. However, the ELMo architecture and methodology is more inline with being used as a feature extractor, with only the final linear layer being trained towards the target task (in sharp contrast to the BERT model that we outline next; see e.g. [14]).

## Third generation: Transformers

A subsequent major advance in language modeling has been the development and use of the transformer architecture. Going beyond backwards and forwards sequences, transformers use a mechanism termed "self-attention" [15] in order to model the importance of different parts of the text in understanding any one other part. Like RNNs, this allows for understanding context, but unlike RNNs, it allows the model to better focus on potentially far away parts of the text that may be relevant. For example, this allows to better understand that "it" in one sentence refers to a particular word from a previous sentence.

An early and prominent example of transformer-based language models is Bidirectional Encoder Representations from Transformers (BERT) [16]. Unlike the language model in ELMo which predicts the next word from previous words, the BERT model is trained on two self-supervised tasks simultaneously:

- Mask Language Model: Randomly mask a certain percentage of the words in a sentence and predict the masked words.
- Next Sentence Prediction: Given a pair of sentences, predict whether one sentence precedes another.

**The basic structure of BERT.**

1. Each word in the input sentence is broken into subwords (tokenized) and each piece is called a "token." Each token is encoded using a context-free embedding called WordPiece. A special token [cls] is added to the beginning of the sequence. x% of the tokens representing individual words are replaced by [mask].

2. For each token, its input representation consists of i) its token embedding from (1), ii) its position embedding indicating the position of the token in the sentence, and iii) its segment embedding indicating whether it belongs to sentence A or B.

3. The input representation of tokens in the sequence is fed into the main model architecture: L layers of Transformer-Encoder blocks. Each block consists of a "multi-head attention layer" (described below), followed by a feed forward layer.

4. The output representation of the mask token [mask] is used to predict the masked word via a softmax layer, and the output representation of the special [cls] token is used for Next Sentence Prediction. The loss function is a combination of the two losses.

We next focus in detail on the main structure step in (3), especially the "multi-head attention" layer.

**Computing the Attention**

We begin with the context-free embeddings $(x_1, x_2, \ldots, x_n)$, for $n$ words, with each $x_k \in \mathbb{R}^d$. Let $X$ denote the matrix whose $k$-th row is the embedding $x_k$. An attention module transforms this matrix of $n$ embeddings $X$ into another matrix of $n$ embeddings, where now each row $k$ contains an embedding of the "information" in a "neighborhood" around token $k$. The notion of "neighborhood" and the notion of "information" are all parameterized by neural network parameters of the attention module and learnable in a data-driven manner as we describe below.

The goal of an attention module is to create weighted neighborhoods (or attention regions) of seemingly distant tokens (in a data-driven manner) and then create embeddings that correspond to linear combinations of the embeddings of the tokens in these neighborhoods (or attention regions). One way to achieve this is to decouple the "neighborhood" representation of a token with the representation of each "meaning" or "value." Thus we will transform each token embedding $x_k$ into a *key embedding* $\kappa_k := x_k' \omega^K$, where $\omega^K \in \mathbb{R}^{d \times d_k}$ is a learnable matrix parameter, and a *value embedding* $v_k := x_k' \omega^V$, where $\omega^V \in \mathbb{R}^{d \times d_v}$ is a learnable matrix parameter. Then a neighborhood can be encoded by a *query* vector $q$ that lies in the same space as the space of keys and such that the weighted neighborhood is defined via a similarity metric between the vector $q$ and the key vectors. Attention mechanisms used in Transformers use a scaled inner product as the similarity, i.e. $s_k := q'\kappa_k/\sqrt{d_k}$. Then this similarity is passed through a soft-max function $\sigma(\cdot)$ to map it to a selection probability in $[0,1]$. Finally, as we alluded to in the beginning the *embedding of the neighborhood* that corresponds to this query $q$ is simply the weighted average of the value embeddings of the tokens, i.e. $a := \sum_{k=1}^{n} \sigma(s_k) v_k$.

Suppose now that we had $n$ neighborhood queries $q_1, \ldots, q_n$, then we could create $n$ such neighborhood embeddings $a_1, \ldots, a_n$. Transformers consider "self-attention" queries, where each of the $n$ queries $q_k$ corresponds to a *query embedding* associated with a particular token and is yet another linear embedding of the form $x_k'\omega^Q$, where $\omega^Q \in \mathbb{R}^{d\times d_k}$ is a learnable matrix parameter. Then for each such query we can calculate the corresponding *neighborhood embedding* $a_k$.

Overall this transformation takes as input a matrix $X \in \mathbb{R}^{n\times d}$, where each row corresponds to an original token embedding and transforms it into a matrix $A$, where each row $k$ corresponds to the neighborhood embedding associated with query $q_k$, which in turn is associated with token $x_k$. We can write this calculation we just described in matrix form; let $Q = X\omega^Q$ denote the matrix with rows corresponding to query embeddings, let $K = X\omega^K$ denote the matrix with rows corresponding to key embeddings, and let $V = X\omega^V$ denote the matrix with rows corresponding to value embeddings. Then the attention embeddings (or neighborhood embeddings) can be written in matrix form as

$$A = \text{Attention}(Q, K, V) := \sigma\left(QK^\top/\sqrt{d_k}\right)V.$$

A Multi-Head Attention mapping, which is the building block of the BERT model, builds many such attention transformations, for different matrix parameters $\{\omega_i^Q, \omega_i^K, \omega_i^V\}_{i=1}^h$, calculates the corresponding attention embedding matrices $A_i \in \mathbb{R}^{n\times d_v}$, then concatenates the results in a big embedding matrix $A = \text{Concatenate}(A_1, \ldots, A_h) \in \mathbb{R}^{n\times h\cdot d_v}$ and applies a linear projection transformation $A\omega^O$, where $\omega^O \in \mathbb{R}^{h\cdot d_v\times d_o}$, to produce the final output encoding. Thus we can define the basic Multi-Head Attention transformation:

$$X \longmapsto \text{MultiHead}(X) := \text{Concatenate}(\text{Head}_1, \ldots, \text{Head}_h)\omega^O,$$

$$\text{Head}_i = \text{Attention}(X\omega_i^Q, X\omega_i^K, X\omega_i^V),$$

Each Transformer building block in BERT consists of a series several repetitions of multi-head attention encodings, followed by a fully connected neural network (applied to each of the $n$ output encodings separately). The input $n$ encodings of each repetition is the output of the previous repetition.

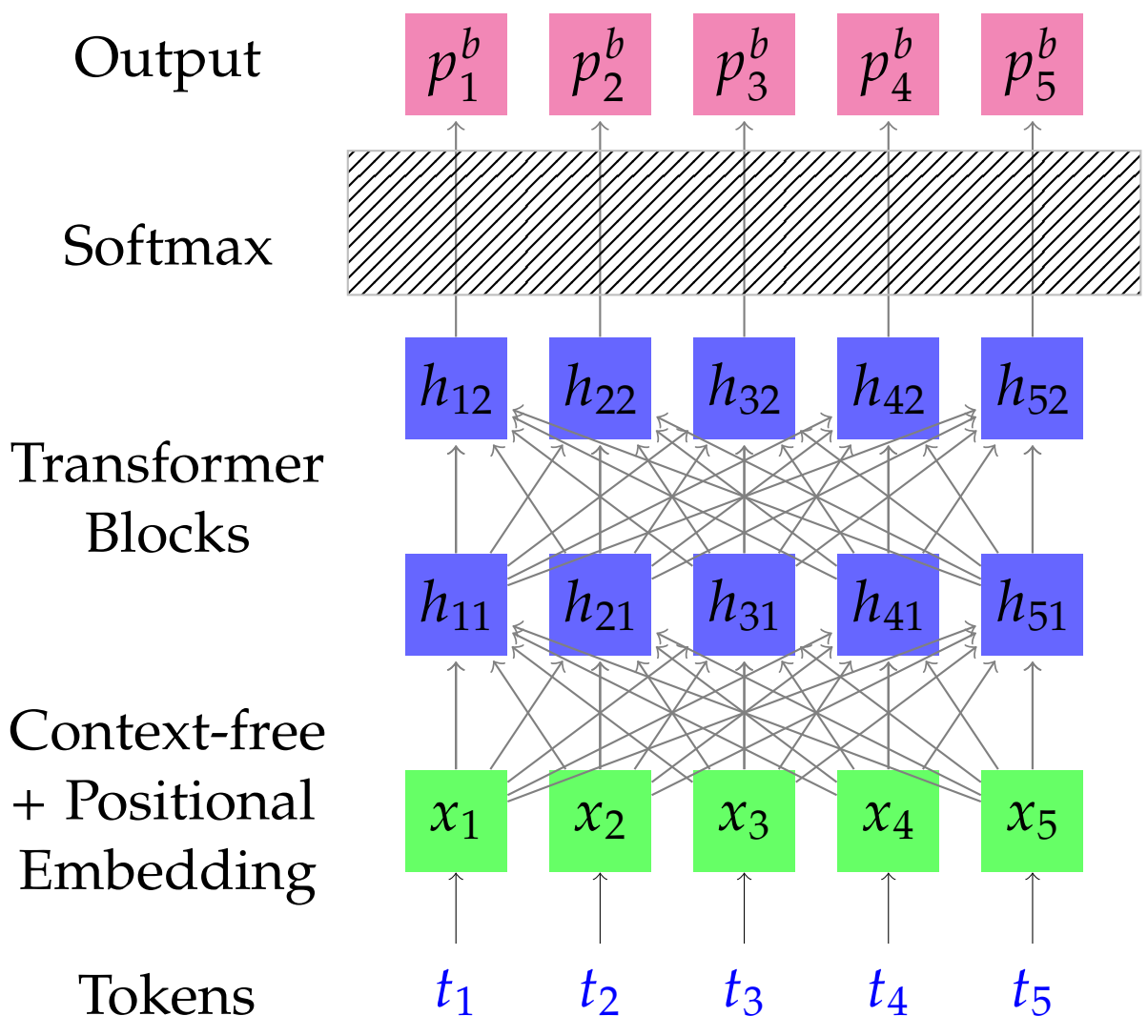

**Figure 11.5:** BERT Architecture

### Generating product embeddings

Depending on specific tasks and resources, Devlin et al. [16] suggested to construct BERT embeddings in various ways:[6]

- Use the last layer, second-to-last layer, or concatenate the last 4 layers of the encoder outputs from the pre-trained BERT model.
- Fine tune the whole BERT model using the downstream task.
- Train the BERT language model from scratch on new data.

6: Tuning only a final linear layer on top of a pre-trained embedding network and freezing all other parameters of the embedding is referred to in the machine learning literature as "linear probing." If one allows for the parameters of the embedding itself to be updated when optimizing for a particular downstream prediction task, then this practice is referred to as "fine-tuning." See also the recent work of [17] for a better way of blending the two modes by first training the final linear layer and then unfreezing the remaining parameters of the embedding and continuing to train. This blending seems to produce substantial gains in generalization ability and accuracy of the resulting predictive model.

In the hedonic price example below, the feature-based approach was chosen, where the second-to-last layer from a pre-trained BERT model was extracted as embeddings. Each product's text embedding is the average of the embeddings of each word/token from the input text field.

## Beyond ELMo and BERT

ELMo and BERT are both important breakthroughs in NLP. The former marked the first contextual word embedding trained from a deep language model, and the latter was the first contextual word embedding using Transformer architecture. The biggest difference lies in the choice of fundamental architectures: ELMo is based on a Recurrent Neural Network (RNN), while BERT is based on the Transformer architecture. RNNs can struggle to capture long-term dependencies, whereas the Transformer architecture is more efficient at capturing long-range dependencies in the text. Furthermore, ELMo creates context by using the left-to-right and right-to-left language model representations, while BERT models the entire context simultaneously.

Large language models are continuously evolving and becoming ever more powerful and sophisticated in their understanding of language and meaning. The latest generation of large language models lie in the Generative Pre-trained Transformer (GPT) family [18–20]. While BERT can be understood to use the transformer architecture as an *encoder*, GPT models use the transformer architecture in a *decoder* for a generative model of the probabilities in an auto-regressive model reminiscent of the one used in ELMo. It combines these modeling ideas from such successful precursors together with pre-training on a large corpus of text. With the rapid development in the space of large language and multi-modal models, the latest and greatest models will certainly advance beyond the descriptions in this book, but the principles of using these models to understand complex data and use it to support robust causal inference may remain the same.

## Revisiting the Price Elasticity for Toy Cars

In Chapter 0, Chapter 4, and Chapter 10, we saw how using increasingly powerful learning methods (OLS, LASSO, nonlinear regression) to control for confounding in the price-sales data for toy cars lead to increasingly more negative estimates and confidence intervals for elasticity. However, these only use the categorical (brand, subcategroy) and numeric (physical dimensions) features, while we actually observe much richer data: all the text on the product page, including the product description. Text embeddings are a great way to leverage these data and include them in a causal analysis of price elasticity. We can take BERT and plug it into neural networks, one to predict price and one for sales: we take the text as input, pass it through BERT initialized at the pre-trained model, add an additional dense layer, and train the whole network. Doing this over 5 folds, we obtain a corss-validated $R^2$ of 0.55 for predicting $Y$ and 0.027 for predicting $D$, improving upon the best nonlinear methods considered in the previous chapter. Applying DML with these new predictors leads to a point estimate for elasticity of -0.174 and 95% confidence interval of [-0.214, -0.135]. This suggests we are able to better control for observed confounders, which generally push apparent elasticity up, leading to more negative estimates.

At the end of the chapter we provide a notebook wherein we repeat the exercise of constructing neural nets using BERT for predicting $Y$ and $D$ and plug them into DML. The results are different as we use a publicly available dataset (which does not have the same range of numeric features and therefore could not be used for comparison to high-dimensional and nonlinear predictive methods in the absence of feature engineering).

## 11.5 Image Embeddings

One of the most successful deep learning models for image classification was the ResNet50 model developed by He et al. [21]. At the time of the release, the paper achieved the best results in image classification, in particular for the ImageNet and COCO datasets.

The central idea of the paper is to exploit "partial linearity": traditional nonlinearly-generated neurons are combined (or added together) with the previous layer of neurons. More specifically, ResNet50 takes a standard feed-forward convolutional neural network and adds skip connections that bypass two (or one or several) convolutional layers at a time. Each skipping step generates a residual block in which the convolution layers predict a residual.

Formally, each $k$-th residual block is a neural network mapping

$$\begin{aligned} v \longmapsto (v, \sigma_k^0(\omega_k^0 v)) &\longmapsto (v, \sigma_k^1 \circ \omega_k^1 \sigma_k^0(\omega_k v)) \\ &\longmapsto v + \sigma_k^1 \circ \omega_k^1 \sigma_k^0(\omega_k^0 v), \end{aligned}$$

where $\omega$'s are matrix-valued parameters or "weights." This structure can be seen as a special case of general neural network architecture, designed so that it is easy to learn the identity sub-maps (entering the composition of the entire network). Putting together many blocks like these sequentially results in the overall architecture depicted in Figure 11.6.

The deep feed-forward convolutional networks developed in prior work suffered from major optimization problems – once the depth was sufficiently high, additional layers often resulted in much higher validation and training error. It was argued that this phenomenon was a result of "vanishing gradients," where in a network of $n$ layers, computation by backpropogation using the chain rule involves multiplying $n$ small numbers (if using traditional activation functions, recent popular activation functions such as RELU do not induce such a small derivative), causing the gradient to "vanish" for early layers and posing a computational challenge. The residual network architecture addresses this by using the residual block architecture: including the residual directly via skip connections reduces the minimizing impact of the activation function. The creation of this architecture has allowed for high quality training even for very deep networks.

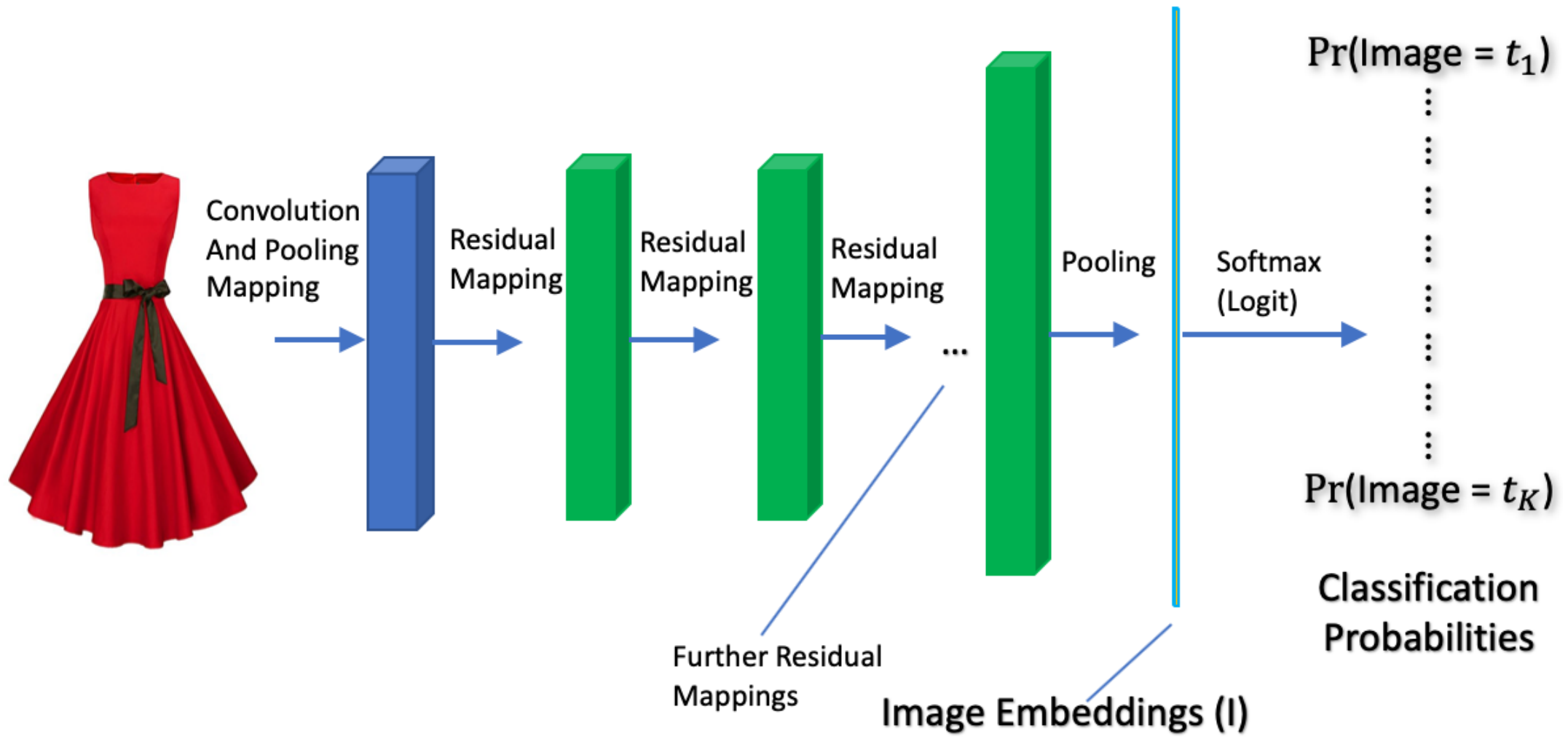

**Figure 11.6:** The ResNet50 operates on numerical 3-dimensional arrays representing images. It first does some pre-processing by applying convolutional and pooling filters, then it applies many L-residual block mappings, producing the arrays shown in green. The penultimate layer produces a high-dimensional vector $I$, the image embedding, which is then used to predict the image type.

Just like with text embeddings, we are not interested in the final predictions of these networks but rather in the last hidden layer, which is taken to be the image embedding. In the next example, one can rely on a publicly trained ResNet50 model to generate the image embeddings.

## Application: Hedonic Prices

Here we apply our new knowledge of embeddings to review an empirical application considered in Bajari et al. [22]. The application is a prediction problem which deals with hedonic price models. An empirical hedonic model is a predictive model for price given a traded object's characteristics.[7] Here, the goal is

7: It can be given structural or causal interpretation using the so-called hedonic price models from economics.

to predict the price of apparel bought and sold on Amazon.com using the product's image and description:

$$P_{it} = H_{it} + \epsilon_{it} = h_t(X_{it}) + \epsilon_{it}, \quad \mathrm{E}[\epsilon_{it} \mid X_{it}] = 0, \qquad (11.5.1)$$

where $P_{it}$ is the price of product $i$ at time $t$ (in months), $X_{it}$ are the product features, and the price function $x \mapsto h_t(x)$ can change from period to period, reflecting the fact that product attributes/features may be valued differently in different periods. [22] use the data from time period $t$ to estimate the function $h_t$ using modern nonlinear regression methods, such as deep neural network methods. The results are contrasted with classical linear regression methods as well as other modern regression methods, such as the random forest.

One of the main uses of hedonic prices is construction of cost of living indices. The use of hedonic prices allows us to "price" the product attributes as well as entire "baskets of attributes" that consumers buy. Then, given a reference "basket of attributes," one can look at the hedonic cost of a basket today compared to its cost in an earlier reference period to determine whether the cost increased or decreased. These types of calculations underlie the construction of commonly used consumer price indices (measuring inflation rates), at least for categories such as apparel products.

A key component of the approach taken in [22] is the use of product features $X_{it}$ generated as neural network embeddings of text and image information about the product. Specifically, $X_{it}$ consists of text embedding features $W_{it}$, constructed by converting the title and product description available on a product's web page into numeric vectors, and image embedding features $I_{it}$ constructed by converting the product image into numeric vectors:

$$X_{it} = (W_{it}', I_{it}')'. \qquad (11.5.2)$$

These embedding features are generated respectively by applying the BERT and ResNet50 mappings.

The model takes high-dimensional text and image features as inputs, converts them into a lower dimensional vector of value embeddings using deep learning methods, and outputs simultaneous predictions of price in all time periods.

The general structure of the model takes the form

$$Z_{it} = \begin{bmatrix} \mathrm{Text}_{it} \\ \mathrm{Image}_{it} \end{bmatrix} \overset{e}{\longmapsto} X_{it}$$

$$\overset{g_1}{\longmapsto} E_{it}^{(1)} \dots \overset{g_m}{\longmapsto} E_{it}^{(m)} =: V_{it} \overset{\theta'}{\longmapsto} \{H_{it}\}_{t=1}^{T} := \{\beta_t' V_{it}\}_{t=1}^{T}.$$

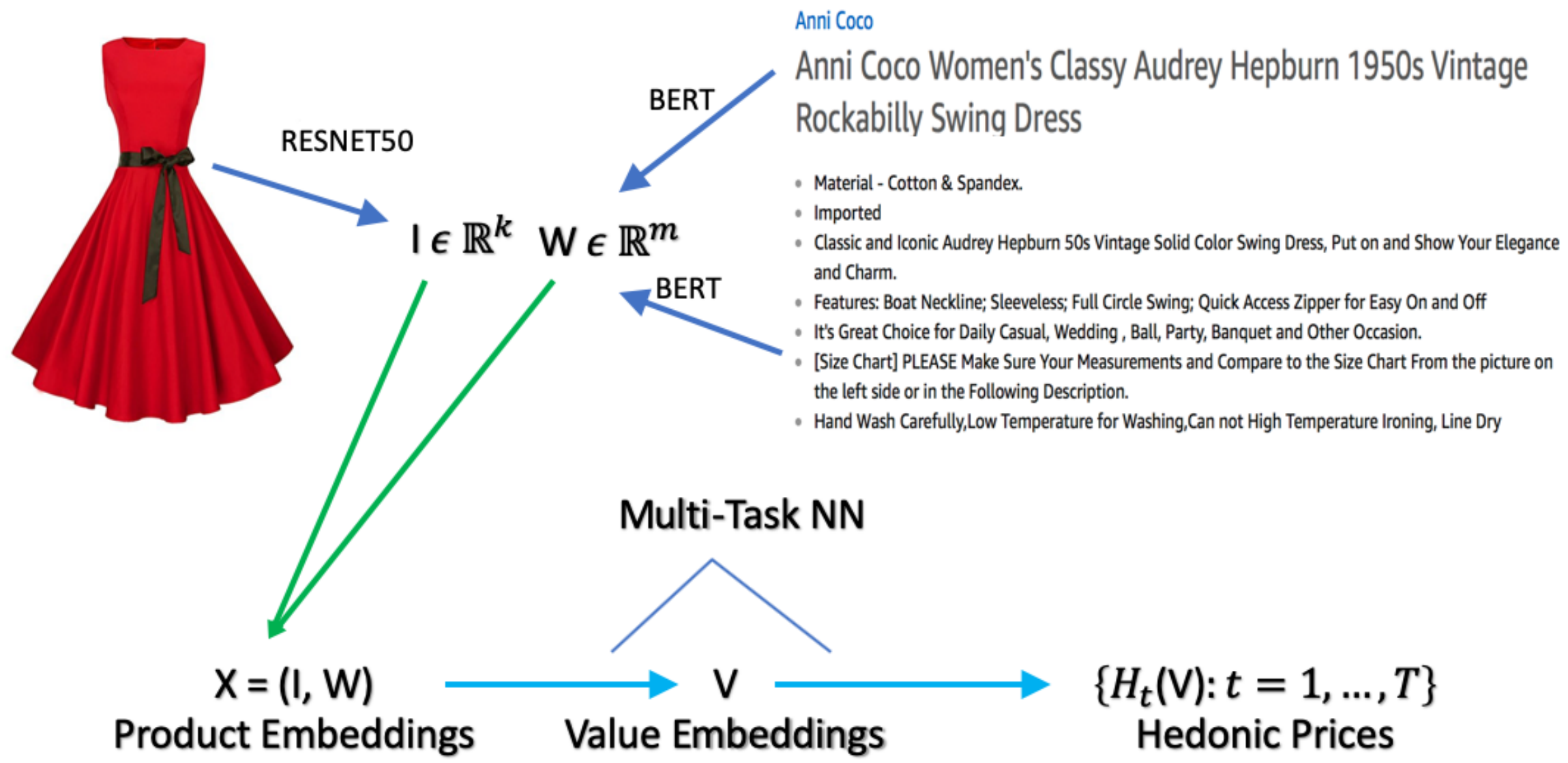

**Figure 11.7:** The structure of the predictive model in Bajari et al. [22]. The input consists of images and unstructured text data. The first step of the process creates numerical embeddings $I$ and $W$ for images and text data via state of the art deep learning methods, such as ResNet50 and BERT. The second step of the process takes as its input $X = (I, W)$ and creates predictions for hedonic prices $H_t(X)$ using deep learning methods with a multi-task structure. The models of the first step are trained on tasks unrelated to predicting prices (e.g., image classification or word prediction), where embeddings are extracted as hidden layers of the neural networks. The models of the second step are trained by price prediction tasks. The multitask price prediction network creates an intermediate lower dimensional embedding $V = V(X)$, called a value embedding, and then predicts the final prices in all time periods $\{H_t(V), t = 1, ..., T\}$. Some variations of the method include fine-tuning the embeddings produced by the first step to perform well for price prediction tasks (i.e. optimizing the embedding parameters so as to minimize price prediction loss).

Here $Z_{it}$,[8] the original input which lies in a very high-dimensional space, is nonlinearly mapped into an embedding vector $X_{it}$ which is of moderately high dimension (up to 5120 dimensions in this example). $X_{it}$ is then further nonlinearly mapped into a lower dimension vector $E^{(1)}_{it}$. This process is repeated to produce the final hidden layer, $V_{it} = E^{(m)}_{it}$, which is then linearly mapped to the final output that consists of hedonic price $H_{it}$ for product $i$ in all time periods $t = 1, ..., T$.

8: As a practical matter, most of the product attributes in [22] are time-invariant - that is, $Z_{it} = Z_i$ has no time variation. We state the model in more generality here.

The last hidden layer $V = E^{(m)}$ is called the *value embedding* in this context – the value embedding represents latent attributes to which dollar values are attached. The embeddings produced in this example are moderately high-dimensional (up to 512 dimensions) summaries of the product, derived from the most common attributes that directly determine the price of the predicted hedonic price of the product. Note that the embeddings $V$ in this example do not depend on time and so may be thought of as representing intrinsic, potentially valuable attributes of the product. However, the predicted price does depend on time $t$ via the coefficient $\beta_t$, reflecting the fact that the different intrinsic attributes are valued differently across time.

The network mapping above comprises a deep neural network with neurons $E_{k,\ell}$ of the form

$$g_\ell : v \longmapsto \{E_{k,\ell}(v)\}_{k=1}^{K_\ell} := \{\sigma_{k,\ell}(v'\alpha_{k,\ell})\}_{k=1}^{K_\ell}. \tag{11.5.3}$$

Here $\sigma_{k,\ell}$ is the activation function that can vary with the layer $\ell$ and can vary with $k$, from one neuron to another.

The model is trained by minimizing the loss function

$$\min_{\eta \in \mathcal{N}, \{\beta_t\}_{t=1}^T} \sum_t \sum_i (P_{it}^c - \beta_t' V_{it}(\eta))^2 Q_{it}, \tag{11.5.4}$$

where $\eta$ denotes all of the parameters of the mapping

$$X_{it} \mapsto V_{it}(\eta)$$

and $\mathcal{N}$ represents the parameter space. Here, we are using a weighted loss where we weight by the quantity of product $i$ sold at time $t$, $Q_{it}$.

Next we review how the initial embedding is generated. A multilingual BERT model is used to convert text information and the ResNet50 model is used to convert images into a subvector of $E_{it}^{(1)}$. These models are trained on auxiliary prediction tasks with auxiliary outputs $A_{T_{it}}$ for text and $A_{I_{it}}$ for images. Introducing these auxiliary tasks can be illustrated diagrammatically as

$$X_{it} = \begin{bmatrix} \text{Text}_{it} \\ \text{Image}_{it} \end{bmatrix} \overset{e}{\longmapsto} \begin{matrix} A_{T_{it}} \\ \uparrow \\ W_i \\ I_i \\ \downarrow \\ A_{I_{it}} \end{matrix} =: E_{it}^{(1)} \dots \longmapsto E_{it}^{(m)} := V_{it} \longmapsto \{\hat{P}_{it}^*\}_{t=1}^T, \tag{11.5.5}$$

The embeddings $W_{it}$ and $X_{it}$ forming $E_{it}^{(1)}$ are obtained by mapping them into auxiliary outputs $A_{Tj}$ and $A_{Ij}$ that are scored on natural language processing tasks and image classification tasks respectively. This step uses data that are not related to prices, as described in detail in the previous sections. The parameters of the mapping generating $E_{it}^{(1)}$ are considered as fixed in our analysis.

The price prediction network we employ in this example contains three hidden layers, with the last hidden layer containing 400 neurons. The network is trained on a large data set with more than 10 million observations. A large enough data set is crucial for training successful neural networks.

The accuracy of prediction as measured by the $R^2$ on the test sample is about

$$90\%.$$

In contrast,

- random forests using embeddings deliver $R^2$ in the ballpark of 80%;
- the linear model using least squares applied to embeddings delivers $R^2$ in the ballpark of 70%;
- the linear model, using simple catalogue features (without embeddings), delivers an $R^2$ lower than 40%.

Thus, embeddings offer a means of making use of complex data for predictions and, at least for large data sets, neural nets can offer predictive improvements relative to competing machine learning approaches.

## Notes

[23] develop "DoubleMLDeep" which allows to use additionally multimodal data, in particular text and images, as confounding variables in the DML framework.

## Notebooks

- Python Auto-Encoders Notebook provides an introduction to auto-encoders, starting from classical principal components.

- Python Toys and Prices Notebook provides an introduction to text embeddings via BERT and provides an application to predicting demand for toys.

## Study Problems

1. Work through the Auto-Encoders notebook. Try to improve the performance of the auto-encoders. Report your findings (even if you don't manage to improve them! :-)).

2. Work through the BERT notebook. Try to experiment with the structure of the neural nets and demand estimation procedure. Report your findings.

# Advanced Topics

# Unobserved Confounders, Instrumental Variables, and Proxy Controls

# 12

"Without Philip Wright
would there have been causal DAGs?
Who can really say?"

– Kei Hirano.*

In this chapter we discuss various models with unobserved confounders, where the adjustment strategies we have discussed no longer work. We start with sensitivity analysis of causal inference to the presence of unobserved confounders. Then we discuss identification of causal effects when instrumental variables or proxy controls are available.

* Sewall Wright, son, and Philip Wright, father, were responsible for some of the greatest ideas in causal inference. Sewall Wright invented causal path diagrams (linear DAGs), and Philip Wright wrote down DAGs for supply-demand equations, proposed IV methods for their identification, and even proposed weather conditions as instruments. Just one of these contributions would probably be enough to get a QJE publication in 1970s and later, but it was not good enough in 1926 or so. Philip Wright is a (causal) parent of Sewall Wright, so he is one of the causes of DAGs (hence the haiku).

## 12.1 The Difficulty of Causal Inference with an Unobserved Confounder

> "All happy statisticians are happy in their own way; but all the unhappy ones are all alike — they all do causal inference with observational data". L. Tolstoy in Anna Karenina (Source: Twitter)

Here we consider models with an unobserved confounding variable. The presence of unobserved confounding variables complicates identification of causal effects. Without further assumptions it is impossible to identify causal effects in a setting with unobserved confounding variables.

For example, consider the following two basic models shown in the margin figure, where we can think of $Y$ as wages, $D$ as education, and $A$ as latent ability.

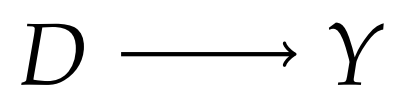

**Figure 12.1:** $D$ causes $Y$

If $A$ is not observed, the two models in Figures 12.1 and 12.2 are statistically indistinguishable from each other. In the first model $D$ has a causal effect on $Y$, and in the second it does not. Even with strong restrictions, as in Gaussian linear SEMs, the observed correlation between $D$ and $Y$ can always be rationalized either as a causal effect of $D$ on $Y$ or the result of a common cause $A$ (homework). This observation applies more generally. While we cannot precisely pin down causal effects in these cases, we can still learn about causal effects by performing sensitivity analysis if we are willing to assume a bound on the strength of unobserved confounders. We discuss a practical and intuitive approach to sensitivity analysis in Section 12.2.

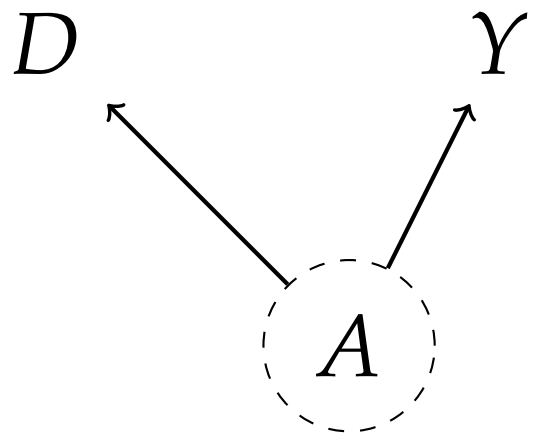

**Figure 12.2:** $D$ and $Y$ are caused by a latent factor $A$

We may also make progress in learning causal effects in the presence of unobserved confounders by considering the use of instrumental variables (IVs) – additional random vectors $Z$ that create exogenous variation in $D$. This approach was introduced by Philip Wright in 1928 [1]. The use of instruments renders many linear ASEMs identifiable, allowing us to perform inference on structural effects $D \to Y$. Some nonlinear ASEMs also become identifiable, though identification still fails for completely unrestricted nonlinear models. We discuss the use of instruments in Sections 12.3-12.4.

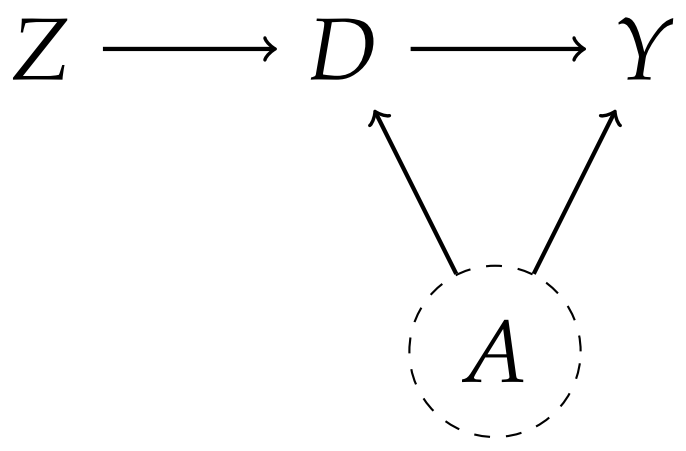

**Figure 12.3:** A DAG with Latent Confounder $A$ and Instrument $Z$.

A related set of problems is when we observe multiple proxy measurements of the latent confounder $A$. For example, we may observe $S$, the SAT score, and $Q$, the ACT score, which may both be proxies for latent confounder, $A$, ability. Note that conditioning on $Q$ and $S$ does not block the backdoor path $Y \leftarrow A \to D$. Hence we cannot use the regression adjustment method for identification of $D \to Y$. However, this problem is related to IVs, because we can effectively use one measurement in place of $A$ and instrument it with another measurement to deal with the measurement error. This process can provide identification of the main effect $D \to Y$. In other words, we can use instrumental variable regression of $Y$ on $D$ and $S$, using $D$ and $Q$ as technical instrumental variables. This approach was introduced by Zvi Griliches in 1977 [2]. This model has also been extensively studied for nonlinear models as well, e.g., Miao et al. [3] and Deaner [4], especially in the recent literature. We discuss proxy approaches in Section 12.6.

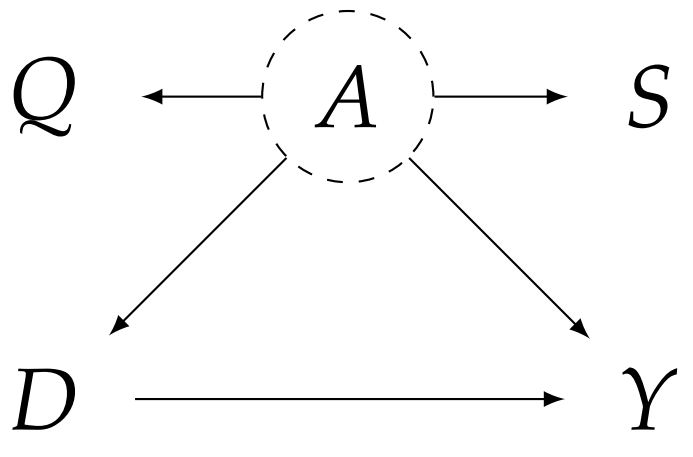

**Figure 12.4:** A DAG with two proxies for latent confounders.

## 12.2 Impact of Confounders on Causal Effect Identification and Sensitivity Analysis

**Example 12.2.1** (Partially Linear SEM) Consider the SEM

$$\begin{aligned} Y &:= \alpha D + \delta A + f_Y(X) + \epsilon_Y, \\ D &:= \gamma A + f_D(X) + \epsilon_D, \\ A &:= f_A(X) + \epsilon_A, \\ X &:= \epsilon_X, \end{aligned}$$

where, conditional on $X$, $\epsilon_Y, \epsilon_D, \epsilon_A$ are both mean zero and mutually uncorrelated. We further normalize

$$\mathrm{E}[\epsilon_A^2] = 1.$$

The key structural parameter is $\alpha$:

$$\alpha = \partial_d Y(d)$$

where

$$Y(d) := (Y : do(D = d)).$$

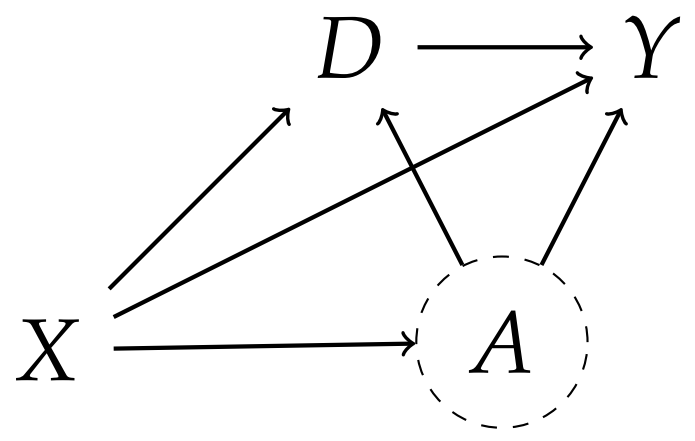

**Figure 12.5:** $X$ are observed confounders, and $A$ are unobserved confounders.

To give context to our example, we can interpret $Y$ as earnings, $D$ as education, $A$ as ability, and $X$ as a set of observed background variables. In this example, we can interpret $\alpha$ as the returns to schooling.

We start by applying the partialling out operator to get rid of the $X$'s in all of the equations. Define the partialling out operation of any random vector $V$ with respect to another random vector $X$ as the residual that is left after subtracting the best predictor

of $V$ given $X$:

$$\tilde{V} = V - \mathrm{E}[V \mid X].$$

If $f$'s are linear, we can replace $\mathrm{E}[V \mid X]$ by linear projection. After partialling out, we have a simplified system:

$$\begin{aligned} \tilde{Y} &:= \alpha\tilde{D} + \delta\tilde{A} + \epsilon_Y, \\ \tilde{D} &:= \gamma\tilde{A} + \epsilon_D, \\ \tilde{A} &:= \epsilon_A, \end{aligned}$$

where $\epsilon_Y$, $\epsilon_D$, and $\epsilon_A$ are uncorrelated.

Then the projection of $\tilde{Y}$ on $\tilde{D}$ recovers

$$\beta = \mathrm{E}[\tilde{Y}\tilde{D}]/\mathrm{E}[\tilde{D}^2] = \alpha + \phi,$$

where

$$\phi = \delta\gamma/\mathrm{E}\left[(\gamma^2 + \epsilon_D^2)\right],$$

is the omitted confounder bias.

Omitted confounder bias is also often referred to as omitted variables bias.

The formula follows from inserting the expression for $\tilde{D}$ into the definition of $\beta$ and then simplifying the resulting expression using the assumptions on the $\epsilon$'s.

We can use this formula to bound $\phi$ directly by making assumptions on the size of $\delta$ and $\gamma$. An alternative approach can be based on the following characterization, based on partial $R^2$'s. This characterization essentially follows from Cinelli and Hazlett [5], with the slight difference that we have adapted the result to the partially linear model.[1]

1: [6] recently obtained a similar result for fully nonlinear models.

**Theorem 12.2.1** (Omitted Confounder Bias in Terms of Partial $R^2$'s) *In the setting given in Example 12.2.1,*

$$\phi^2 = \frac{R^2_{\tilde{Y}\sim\tilde{A}|\tilde{D}} R^2_{\tilde{D}\sim\tilde{A}}}{(1 - R^2_{\tilde{D}\sim\tilde{A}})} \frac{\mathrm{E}\left[(\tilde{Y} - \beta\tilde{D})^2\right]}{\mathrm{E}\left[(\tilde{D})^2\right]},$$

*where $R^2_{V\sim W|X}$ denotes the population $R^2$ in the linear regression of $V$ on $W$, after partialling out linearly $X$ from $V$ and $W$.*

Therefore, if we place bounds on how much of the variation in $\tilde{Y}$ and in $\tilde{D}$ the unobserved confounder $\tilde{A}$ is able to explain, we can bound the omitted confounder bias by

$$\sqrt{\phi^2}.$$

**Example 12.2.2** We consider an empirical example based on data surrounding the Darfur war. Specifically, we are interested in the effect of having experienced direct war violence on attitudes towards peace. The observed controls explain 12-15% of the variance of $Y$, beyond what's explained by the "treatment" variable, and 1% of the variance of treatment $D$. Therefore, suppose we are willing to accept that

$$R^2_{\tilde{Y}\sim\tilde{A}|\tilde{D}} \leq .15, \quad R^2_{\tilde{D}\sim\tilde{A}} \leq .01;$$

that is, we have a latent confounder that is no stronger than the observed controls for predicting $Y$ and for predicting $D$.

Then, the upper/lower bound on $\alpha$ is given by

$$\beta \pm \phi, \quad \phi^2 = \frac{.0015}{.99} \frac{\mathrm{E}\left[(\tilde{Y} - \beta\tilde{D})^2\right]}{\mathrm{E}\left[\tilde{D}^2\right]}.$$

The estimated $\beta$ is about .1. Plugging in estimates of $\mathrm{E}\left[(\tilde{Y} - \beta\tilde{D})^2\right]$ and $\mathrm{E}[(\tilde{D})^2]$ yields an estimated lower bound on $\alpha$ of around .074. In Figure 12.6, we show the combination of all partial $R^2$ such that the bias is less than .026. It shows that our conclusions about causal effects are not very sensitive to the presence of unknown confounders whose power is limited by the stated assumptions.

DML Sensitivity R Notebook carries out sensitivity analysis based on DML and the R package Sensemakr for the analysis of the Darfur wars data.

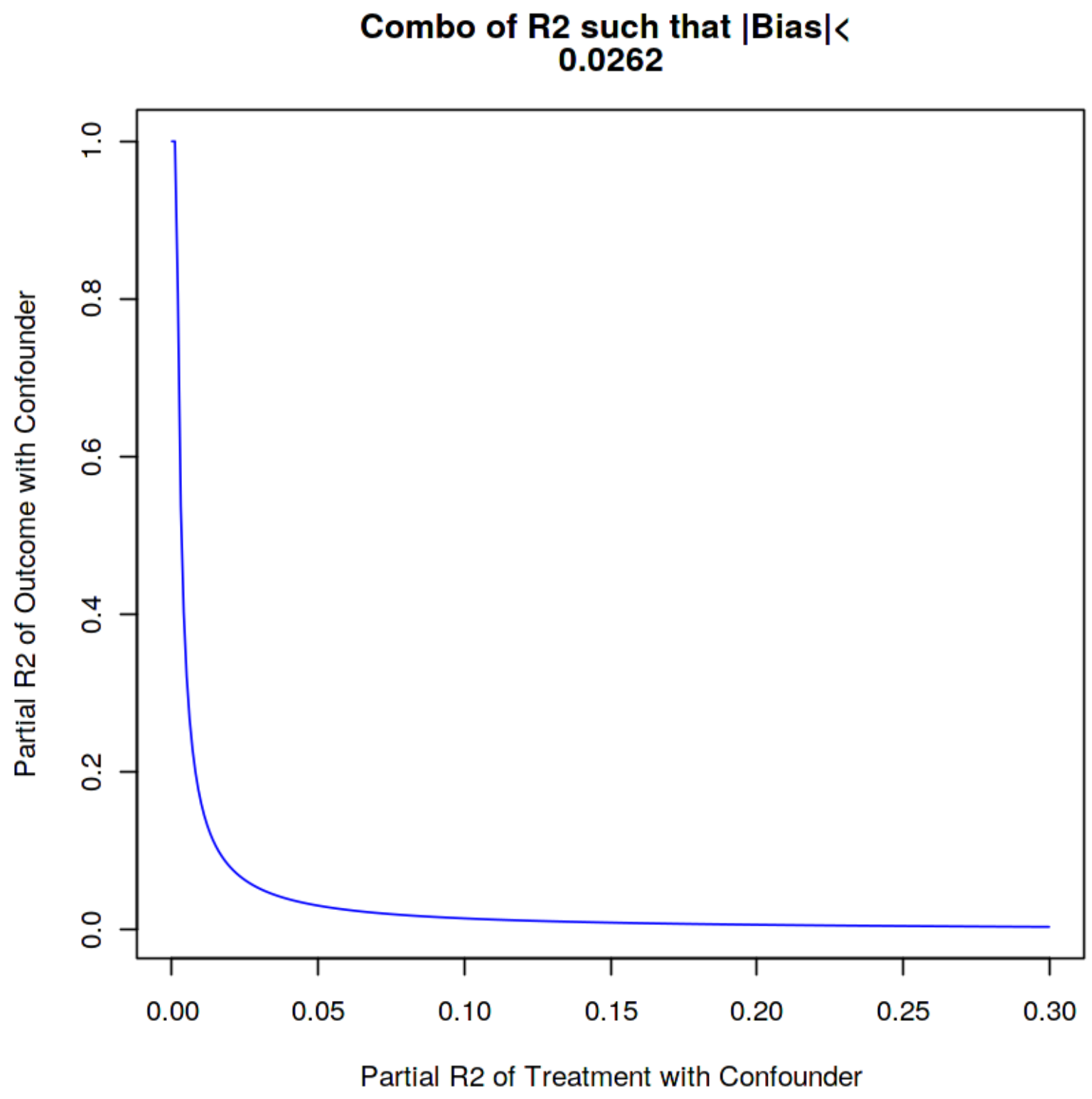

**Figure 12.6:** Sensitivity contour plots: The graph shows values of $R^2_{\tilde{Y}\sim\tilde{D}|\tilde{A}}$ and $R^2_{\tilde{D}\sim\tilde{A}}$ that give a given value of the bias $|\hat{\phi}| = .026$.

## 12.3 Partially Linear IV Models

When instrumental variables are available, it becomes possible to point identify causal effects in partially linear models and certain types of causal effects in nonlinear models. Here we begin with partially linear models.

### A Wage Equation with Unobserved Ability

**Example 12.3.1** (Returns to Education with Omitted Ability; Generalization of Griliches, 1977 [2]) Consider the ASEM

$$
\begin{aligned}
Y &:= \alpha D + \delta A + f_Y(X) + \epsilon_Y, \\
D &:= \beta Z + \gamma A + f_D(X) + \epsilon_D, \\
Z &:= f_Z(X) + \epsilon_Z, \\
A &:= f_A(X) + \epsilon_A, \\
X &:= \epsilon_X,
\end{aligned}
$$

where, conditional on $X$, $\epsilon_Y, \epsilon_D, \epsilon_Z, \epsilon_A$ have mean zero and are mutually uncorrelated.

We can interpret $Y$ as earnings, $D$ as education, $A$ as ability, $Z$

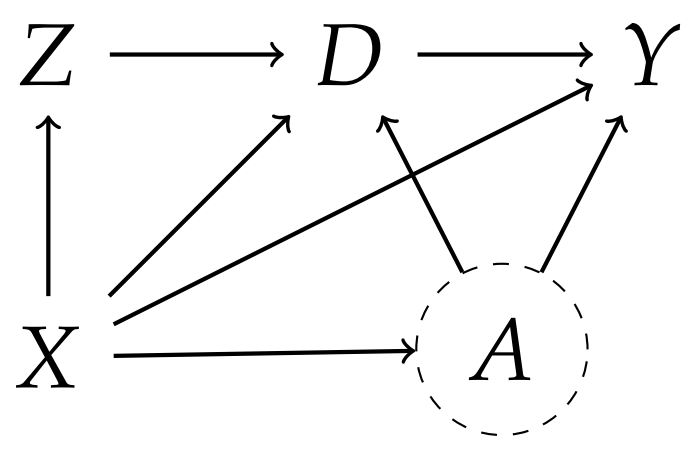

**Figure 12.7:** An IV model with observed and unobserved confounders.

as an observed shifter of education, and $X$ as a set of observed background variables. The key structural parameter is $\alpha$, the returns to schooling, i.e.

$$\alpha = \partial_d Y(d),$$

where

$$Y(d) = Y : do(D = d).$$

Examples of instruments for schooling, $Z$, that have appeared in the literature include

- distance to college (Card [7]),
- compulsory schooling laws (Angrist [8]),
- offer to participate/offer to treat in a training program (many studies), and
- subsidies to finance education (Griliches, Heckman).

We apply the partialling-out operator to get rid of the $X$'s in all of the equations. As before, we define the partialling out operation of any random vector $V$ with respect to another random vector $X$ as the residual that is left after subtracting the best predictor of $V$ given $X$:

$$\tilde{V} = V - \mathrm{E}[V \mid X].$$

If $f$'s are linear, we replace $\mathrm{E}[V \mid X]$ with linear projection.

After partialling-out, we have a simplified system.

$$\begin{aligned}
\tilde{Y} &:= \alpha\tilde{D} + \delta\tilde{A} + \epsilon_Y, \\
\tilde{D} &:= \beta\tilde{Z} + \gamma\tilde{A} + \epsilon_D, \\
\tilde{Z} &:= \epsilon_Z, \\
\tilde{A} &:= \epsilon_A,
\end{aligned}$$

where $\epsilon_Y$, $\epsilon_D$, $\epsilon_Z$, and $\epsilon_A$ are uncorrelated.

We immediately obtain the following result:

**Theorem 12.3.1** *In Example 12.3.1, we can rewrite an econometric measurement model for identification of $\alpha$:*

$$\tilde{Y} := \alpha\tilde{D} + U, \quad U \perp \tilde{Z},$$

*where $U = \delta\tilde{A} + \epsilon_Y$. Alternatively, we can equivalently identify $\alpha$ using the moment restriction*

$$\mathrm{E}\left[(\tilde{Y} - \alpha\tilde{D})\tilde{Z}\right] = 0.$$

*The identification of* $\alpha$ *follows from solving this equation,*

$$\alpha = \mathrm{E}[\tilde{Y}\tilde{Z}]/\mathrm{E}[\tilde{D}\tilde{Z}],$$

*provided the instruments are relevant:* $\mathrm{E}[\tilde{D}\tilde{Z}] \neq 0$ *or* $\beta \neq 0$.

**Remark 12.3.1** (Neyman Orthgonality and DML) The target parameter $\alpha$ is Neyman orthogonal with respect to nuisance parameters – the regression functions $\mathrm{E}[Y \mid X]$, $\mathrm{E}[D \mid X]$, and $\mathrm{E}[Z \mid X]$. Therefore we can use DML for learning and performing statistical inference on the parameter $\alpha$.

## Wright's Causal Path Derivation

Starting from the DAG given in Figure 12.7, we obtain Figure 12.8 after partialling out.

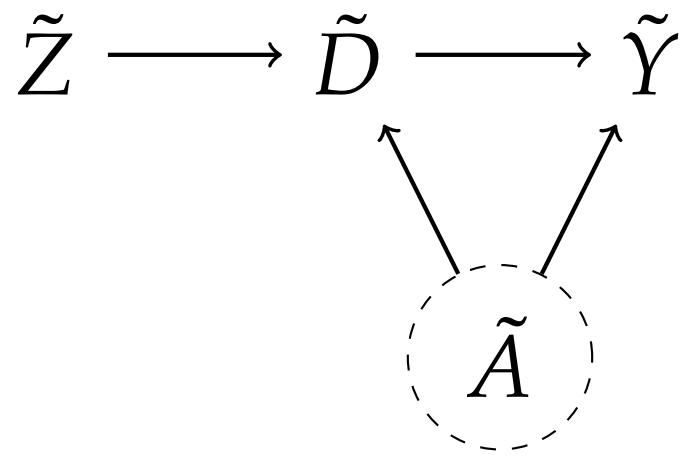

**Figure 12.8:** DAG corresponding to Figure 12.7 after partialling out observed confounder $X$.

Philip Wright (1928) [1] observed that the structural parameter $\beta\alpha$, the effect $\tilde{Z} \to \tilde{Y}$, is identified from the projection of $\tilde{Y} \sim \tilde{Z}$:

$$\beta\alpha = \mathrm{E}[\tilde{Y}\tilde{Z}]/\mathrm{E}[\tilde{Z}^2].$$

The structural parameter $\beta$, the effect of $Z \to D$, is identified from the projection of $\tilde{D} \sim \tilde{Z}$:

$$\beta = \mathrm{E}[\tilde{D}\tilde{Z}]/\mathrm{E}[\tilde{Z}^2].$$

$\alpha$, the effect of $D \to Y$, is then identified by the ratio of the two provided $\beta \neq 0$:

$$\alpha = \frac{\beta\alpha}{\beta} = \mathrm{E}[\tilde{Y}\tilde{Z}]/\mathrm{E}[\tilde{D}\tilde{Z}].$$

## Aggregate Market Demand

Let's apply our approach to a canonical example in economics: the identification of the price elasticity of demand using a supply shifter as an instrument.

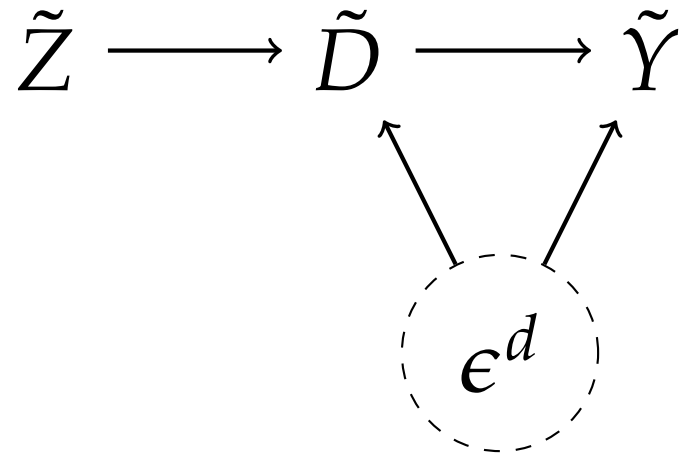

**Figure 12.9:** A DAG for aggregate demand, with the latent node $\epsilon^d$ representing the demand shock

**Example 12.3.2** (Market Demand; Generalization of P. Wright,

1928 [1]) Consider the ASEM

$$\begin{aligned} Y &:= \alpha D + f_Y(X) + \epsilon^d, \\ D &:= \beta Z + f_D(X) + \rho\epsilon^d + \gamma\epsilon^s, \\ Z &:= f_Z(X) + \epsilon_Z \end{aligned}$$

where $\epsilon^d$, $\epsilon^s$ and $\epsilon_Z$ are mean zero and uncorrelated conditional on $X$. In this example, $Y$ is (log) demand, $D$ is (log) price, $Z$ is an observed supply shifter, $X$ is a vector of observed demand shifters, $\epsilon^d$ is a demand shock, and $\epsilon^s$ is a supply shock. The key parameter is $\alpha$, the price elasticity of demand:

$$\alpha = \partial_d Y(d),$$

where $Y(d) := (Y : do(D = d))$. Here we focus on only the demand side of the market and do not attempt to explicitly model the supply side.

In econometrics, the set-up here is sometimes referred to as a *limited information* model or formulation because we are focusing on identifying only a single equation in a more complicated underlying system.

Example 12.3.2 is equivalent to the previous Example 12.3.1 – set $A = \epsilon^d$, $\epsilon_Y = 0$, $\epsilon^s = \epsilon_D$, and so on. Hence, the identification method is the same as before.

## Limits of Average Causal Effect Identification under Partial Linearity

The result in Theorem 12.3.1 extends beyond the partially linear setting presented in Example 12.3.1 to the following non-linear structural equation model:

**Example 12.3.3** (Partially Linear Outcome IV Model) Consider the ASEM

$$\begin{aligned} Y &:= g_Y(\epsilon_Y)D + f_Y(A, X, \epsilon_Y), \\ D &:= f_D(Z, X, A, \epsilon_D), \\ Z &:= f_Z(X, \epsilon_Z), \\ A &:= f_A(X, \epsilon_A), \\ X &:= \epsilon_X, \end{aligned}$$

where, $\epsilon_Y, \epsilon_D, \epsilon_Z, \epsilon_A$ are exogenous and mutually independent. The key structural parameter is:

$$\alpha := \mathrm{E}[\partial_d Y(d)] = \mathrm{E}[g_Y(\epsilon_Y)],$$

where

$$Y(d) = Y : do(D = d).$$

This parameter is typically referred to as the average marginal effect of the treatment.

Theorem 12.3.1 extends almost as is to this more general non-linear structural equation model.

**Theorem 12.3.2** *In Example 12.3.3, we can identify $\alpha$ using the moment restriction*

$$\mathrm{E}\left[(\tilde{Y} - \alpha\tilde{D})\tilde{Z}\right] = 0.$$

*The identification of $\alpha$ follows from solving this equation,*

$$\alpha = \mathrm{E}\left[\tilde{Y}\tilde{Z}\right] / \mathrm{E}\left[\tilde{D}\tilde{Z}\right],$$

*provided the instruments are relevant:* $\mathrm{E}[\tilde{D}\tilde{Z}] \neq 0$.

Note that the non-linear structural equation model in Example 12.3.3 imposes extra assumptions on the structural response function of the outcome $Y$. Thus our identification argument imposes more conditions on the structural equations than the ones that can be encoded via a DAG. Such auxiliary assumptions are required for identification of average treatment effects with instruments.

In particular, the identification argument relies on the fact that the unobserved confounder $A$ enters in an additively separable manner in the outcome equation. If for instance, $A$ was an input to the function $g$, i.e. $Y := g_Y(A, \epsilon_Y)D + f_Y(A, X, \epsilon_Y)$, then the quantity identified by the moment restriction in Theorem 12.3.2 would not correspond to an average treatment effect. In this case, the unobserved confounder creates heterogeneity in the treatment effect and also heterogeneity in the effect of the instrument on the treatment, typically referred to as the "compliance" (i.e., the correlation between $Z$ and $D$ varies with $A$). This property is what renders the ratio quantity $\alpha = \mathrm{E}\left[\tilde{Y}\tilde{Z}\right] / \mathrm{E}\left[\tilde{D}\tilde{Z}\right]$, invalid for the causal estimand of interest.

In fact, it is the joint heterogeneity in both the outcome relationship and the compliance relationship that causes the problem. We show next that we could allow for a much more complex outcome model as long as the effect of the instrument on the treatment (compliance) is not heterogeneous in $A$ or $X$.

**Example 12.3.4** (Partially Linear Compliance IV Model) Con-

sider the ASEM

$$
\begin{aligned}
Y &:= g_Y(A, X, \epsilon_Y)D + f_Y(A, X, \epsilon_Y), \\
D &:= g_D(\epsilon_D)Z + f_D(X, A, \epsilon_D), \\
Z &:= f_Z(X) + \epsilon_Z, \\
A &:= f_A(X, \epsilon_A), \\
X &:= \epsilon_X,
\end{aligned}
$$

where, $\epsilon_Y, \epsilon_D, \epsilon_Z, \epsilon_A$ are exogenous and mutually independent. The key structural parameter is:

$$\alpha := \mathrm{E}[\partial_d Y(d)] = \mathrm{E}[g(A, X, \epsilon_Y)],$$

where

$$Y(d) = Y : do(D = d).$$

**Theorem 12.3.3** *In Example 12.3.4, we can identify $\alpha$ using the moment restriction*

$$\mathrm{E}\left[(\tilde{Y} - \alpha\tilde{D})\tilde{Z}\right] = 0.$$

*The identification of $\alpha$ follows from solving this equation,*

$$\alpha = \mathrm{E}\left[\tilde{Y}\tilde{Z}\right] / \mathrm{E}\left[\tilde{D}\tilde{Z}\right],$$

*provided the instruments are relevant:* $\mathrm{E}[\tilde{D}\tilde{Z}] \neq 0$.

Thus, we see that we need that either the effect of education on wages is not heterogeneous in the unobserved ability variable $A$ or that the effect of the observed education shifter $Z$ (e.g. distance to college) on education $D$ is not heterogeneous in the unobserved ability variable to use the identification strategies presented in this section in the context of our education example. In Section 12.4, we will investigate what causal quantities are identifiable even in non-linear structural equation models, where the unobserved confounder creates heterogeneity in both the treatment effect and in the compliance behavior.

**Remark 12.3.2** (Effect heterogeneity based on observables) We note that allowing for $X$ to enter the $g_Y$ or $g_D$ function in Example 12.3.3 and Example 12.3.4 (i.e. allowing for the treatment effect or compliance, i.e. effect of instrument on treatment, to vary with $X$), is a more benign extension because $X$ is an observed variable. In this case, we can repeat the identification strategies in this section, conditional on $X$, and we can show with similar arguments that

$$\beta(X) := \mathrm{E}[\partial_d Y(d) \mid X] = \frac{\mathrm{E}[\tilde{Y}\tilde{Z} \mid X]}{\mathrm{E}[\tilde{D}\tilde{Z} \mid X]}. \tag{12.3.1}$$

Then we can simply average these conditional estimates to get the average marginal effect:

$$\alpha = \mathrm{E}[\beta(X)]. \tag{12.3.2}$$

Such an identification strategy was initiated in [9, 10] and was also recently used in the context of DML estimators [11–13]. In particular, the following moment condition that identifies $\alpha$,

$$\mathrm{E}\left[\beta(X) + \frac{(\tilde{Y} - \beta(X)\tilde{D})\tilde{Z}}{\mathrm{E}[\tilde{D}\tilde{Z} \mid X]} - \alpha\right] = 0, \tag{12.3.3}$$

is Neyman orthogonal with respect to the nuisance functions $\beta(X)$ and $\gamma(X) := \mathrm{E}[\tilde{D}\tilde{Z} \mid X]$. We note that this identification strategy remains valid even if in Example 12.3.4 the instrument equation is fully non-linear, i.e. $Z := f_Z(X, \epsilon_Z)$.

## 12.4 Nonlinear IV Models

Once we consider nonlinear models, identification becomes a much more delicate matter. We first consider the local average treatment effect (LATE) model, and then we turn to quantile models.

### The LATE Model

An important nonlinear IV model is the local average treatment effect model (LATE), proposed by Imbens and Angrist [14].

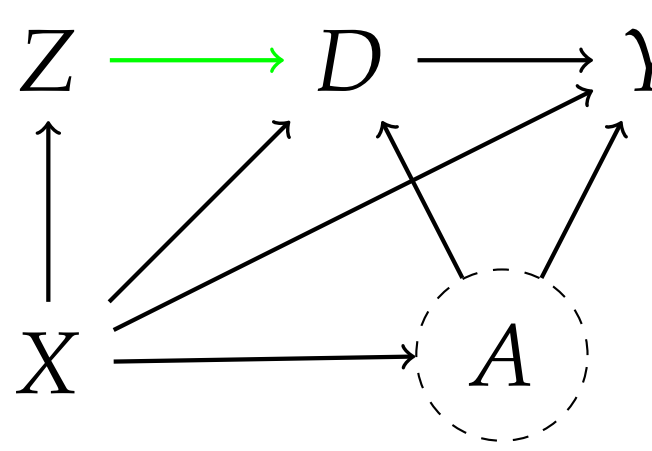

**Figure 12.10:** LATE models. Green arrow denotes a monotone functional relation.

**Example 12.4.1** (LATE) Consider the SEM, where

$$\begin{aligned} Y &:= f_Y(D, X, A, \epsilon_Y) \\ D &:= f_D(Z, X, A, \epsilon_D) \in \{0, 1\}, \\ Z &:= f_Z(X, \epsilon_Z) \in \{0, 1\}, \\ X &:= \epsilon_X, \quad A = \epsilon_A, \end{aligned}$$

where $\epsilon$'s are all independent, and

$z \mapsto f_D(z, A, X, \epsilon_D)$ is weakly increasing (weakly monotone).

Suppose the instrument $Z$ is an offer to participate in a training program and that $D$ is the actual endogenous participation in the training program. Participation in the program may depend on unobservables $A$, such as ability or perseverence, that also affect the eventual outcome $Y$. We can also have background exogenous covariates $X$ in the model.

Define

$$Y(d) := f_Y(d, X, A, \epsilon_Y) \text{ and } D(z) := f_D(z, X, A, \epsilon_D)$$

as the potential outcomes that result from applying fix-interventions in the corresponding equations from Example 12.4.1.

The model allows us to identify the local average treatment effect (LATE), defined as

$$\theta = \mathrm{E}[Y(1) - Y(0) \mid D(1) > D(0)],$$

where $\{D(1) > D(0)\}$ is the compliance event, where switching instrument value from $Z = 0$ to $Z = 1$ induces participation. Therefore LATE measures the average treatment effect conditional on compliance.

**Theorem 12.4.1** *In the LATE model, we have that $\theta$ is identified by the ratio of two statistical parameters,*

$$\theta = \theta_1/\theta_2,$$

*where*

$$\theta_1 := \mathrm{E}\left[\mathrm{E}[Y \mid X, Z = 1] - \mathrm{E}[Y \mid X, Z = 0]\right],$$

*and*

$$\theta_2 := \mathrm{E}\left[\mathrm{E}[D \mid X, Z = 1] - \mathrm{E}[D \mid X, Z = 0]\right],$$

*provided that the instrument $Z$ is relevant, $\theta_2 > 0$, and $Z$ has full conditional support – namely $0 < P(Z = 1 \mid X) < 1$. Moreover, $\theta_2$ identifies the probability of compliance:*

$$\theta_2 = \mathrm{P}[D(1) > D(0)].$$

The result has an intuitive interpretation.[2] In the event of compliance, the instrument moves the treatment as if experimentally, which induces quasi-experimental variation in the outcome. We measure the probability of compliance with $\theta_2$

2: In the model with no $X$ the ratio $\theta_1/\theta_2$ is equivalent to Wright's [1] IV estimand.

and the average induced changes in outcome by $\theta_1$. Taking the ratio is then like conditioning on the compliance event. See the proof in Section 12.A for details.

The ratio can be recognized as the ratio of average treatment effects of $Z$ on $Y$ and $D$,

$$\theta_1 = ATE(Z \to Y),$$

$$\theta_2 = ATE(Z \to D).$$

This assertion follows from the application of the backdoor criterion. Therefore in order to perform inference on LATE, we can simply re-use the tools for performing inference on two ATEs.

**Remark 12.4.1** (DML for $\theta_1/\theta_2$) We can apply DML to obtain $\hat{\theta}_1$ and $\hat{\theta}_2$ and then construct the estimator $\hat{\theta} = \hat{\theta}_1/\hat{\theta}_2$ via the plug-in principle. This approach automatically has the Neyman orthogonality property.

## The IV Quantile Model*

Another nonlinear IV model is the following model that exploits monotonicity in the unobservable shock in the outcome equation to obtain identification.

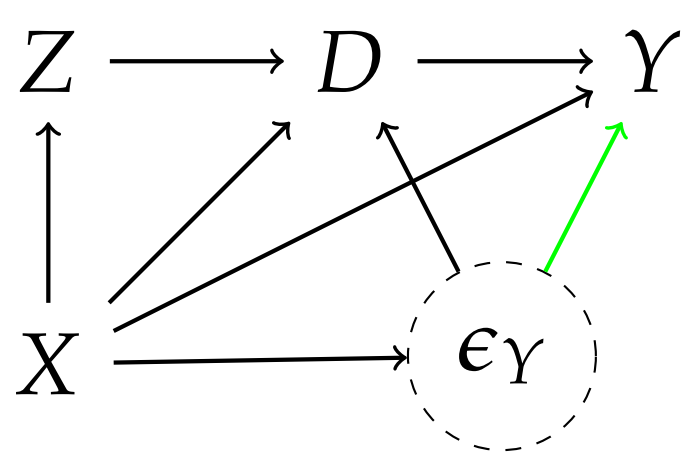

**Figure 12.11:** IV Quantile Model. The green arrow represents a strictly monotonic effect.

**Example 12.4.2** (IV Quantile Model) Consider the SEM

$$\begin{aligned} Y &= f_Y(D, X, \epsilon_Y), \\ D &= f_D(Z, X, \epsilon_Y, \epsilon_D), \\ Z &= f_Z(X, \epsilon_Z), \\ X &= \epsilon_X, \end{aligned}$$

where $\epsilon$'s are all independent,

$$f_Y(D, X, \cdot) : [0, 1] \mapsto \mathbb{R} \text{ is strictly increasing,}$$

and $\epsilon_Y$ is normalized to have uniform distribution on $(0, 1)$. The context could be given from the demand example, where $Y$ is demand, $D$ price, $\epsilon_Y$ a demand shock, $\epsilon_D$ a supply shock; $X$ the set of background variables, and $Z$ a set of instrumental variables. The function $f_Y(d, x, u)$ is the $u$-th quantile of the structural function of $f_Y(d, x, \epsilon_Y)$, which is the demand function in this context. For example, $f_Y(d, x, 1/2)$ is the median structural function.

The testable implication of the IV Quantile Model is the following.

**Theorem 12.4.2** *In the IV Quantile Model, the testable moment restriction is*

$$\mathrm{P}[Y \leq f_Y(D, X, u) \mid Z, X] = u,$$

*for each $u \in (0, 1)$. There exist regularity conditions, analogous to instrument relevance, under which the structural function $f_Y$ is identified from this restriction.*

In practice, linear forms $f_Y(D, X, u) = \alpha(u)'D + \beta(u)'X$ are often used. Adopting a linear functional form leads to method of moments approaches such as the IV quantile regression for performing inference on structural quantile functions.

Code for IV Quantile Models can be found here.

**Remark 12.4.2** (DML for IVQR Models) The problem of constructing DML for IVQR problems is considered open. Neyman-orthogonal approaches for the partially linear IVQR models are sketched out in the review [15] and may be a good place to start.

## 12.5 Partially Linear SEMs with Griliches-Chamberlain Proxy Controls

Suppose we are interested in the causal effect of college education on earnings in the presence of an unobserved confounder – individual ability. Here we show that we can recover the effect of college education on earnings in the presence of latent ability using proxies for ability, but not the effect of ability itself.

**Example 12.5.1** (Earnings with Omitted Ability; Griliches, 1977 [2]; Griliches and Chamberlain, 1977 [16]) Consider the

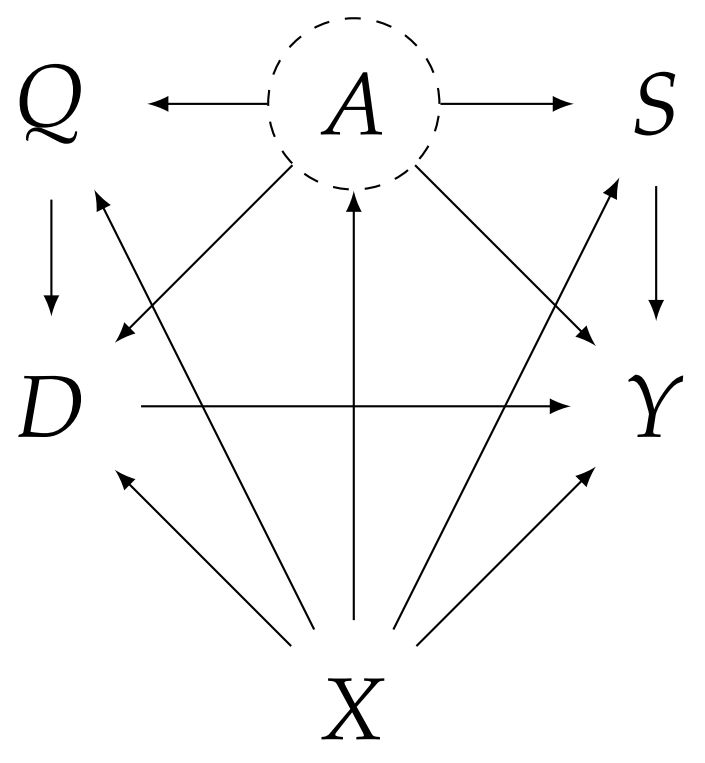

**Figure 12.12:** A DAG with Controls and Proxy Controls

ASEM

$$
\begin{aligned}
Y &:= \alpha D + \delta A + \iota S + f_Y(X) + \epsilon_Y, \\
D &:= \gamma A + \beta Q + f_D(X) + \epsilon_D, \\
Q &:= \eta A + f_Q(X) + \epsilon_Q, \\
S &:= \phi A + f_S(X) + \epsilon_S, \\
A &:= f_A(X) + \epsilon_A, \\
X &:= \epsilon_X,
\end{aligned}
$$

where $\epsilon_Y, \epsilon_D, \epsilon_Q, \epsilon_S, \epsilon_A, \epsilon_X$ have mean zero and are uncorrelated conditional on $X$. Interpret $Y$ as earnings, $D$ as college degree, $A$ as ability, $Q$ and $S$ as proxies of ability, and $X$ as a set of observed background variables. Example proxies $Q$ and $S$ are

- Q is test scores or grades in some period $t_0$ and S is test scores or grades at a later period $t_1$.

The key structural parameter is $\alpha$, the returns to schooling; i.e.

$$\alpha = \partial_d Y(d),$$

where $Y(d) = Y : do(D = d)$.

After partialling out we are left with the DAG in Figure 12.13:

$$
\begin{aligned}
\tilde{Y} &:= \alpha \tilde{D} + \delta \tilde{A} + \iota \tilde{S} + \epsilon_Y, \\
\tilde{D} &:= \gamma \tilde{A} + \beta \tilde{Q} + \epsilon_D, \\
\tilde{Q} &:= \eta \tilde{A} + \epsilon_Q, \\
\tilde{S} &:= \phi \tilde{A} + \epsilon_S, \\
\tilde{A} &:= \epsilon_A,
\end{aligned}
$$

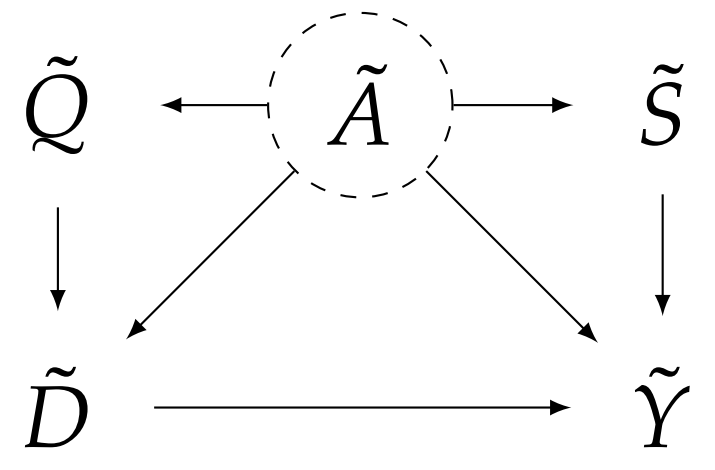

**Figure 12.13:** A DAG with Proxy Controls After Partialling Out

where $\epsilon_Y, \epsilon_D, \epsilon_Q, \epsilon_S, \epsilon_A$ are uncorrelated. The idea now is to replace $\tilde{A}$ in the equation for $\tilde{Y}$ with $\tilde{S}$. Note that because $S$ enters the $Y$ equation directly, we cannot consider using $\tilde{Q}$ to proxy for $\tilde{A}$. We still cannot learn $\alpha$ from the regression of $\tilde{Y}$ on $\tilde{D}$ and $\tilde{S}$ though as $S$ is an imperfect proxy for $A$. The following result, which provides an IV approach to identify $\alpha$, is immediate via substitution.[3]

3: Prove the result as a reading exercise. Substitute $\tilde{A} = (\tilde{S} - \epsilon_S)/\phi$ in the first equation and use the assumptions on the disturbances.

**Theorem 12.5.1** *Assume that all variables in Example 12.5.1 are square-integrable. Then we have the following measurement equation:*

$$\tilde{Y} = \alpha \tilde{D} + \bar{\delta} \tilde{S} + U, \quad \mathrm{E}[U(\tilde{D}, \tilde{Q})] = 0,$$

$$U = -\delta \epsilon_S / \phi + \epsilon_Y; \quad \bar{\delta} = \iota + \delta/\phi.$$

*Here $\alpha$ is identified from the moment condition* $\mathrm{E}[U(\tilde{D}, \tilde{Q})] = 0$, *which is equivalent to using $\tilde{Q}$ as an instrument for $\tilde{S}$, provided that $\tilde{D}$ and the best linear predictor of $\tilde{S}$ using $\tilde{Q}$ and $\tilde{D}$ have non-degenerate covariance matrix.*

Note that $\tilde{Q}$ here plays the role of a *technical* instrument for $\tilde{S}$. This approach recovers $\alpha$, but not $\delta$. For inference, we can employ the DML method for IV models; see also Chapter 13.

**Remark 12.5.1** (Neyman Orthogonality and DML) The formulation of the target parameter given above is Neyman-orthogonal, and high-quality estimation and statistical inference can be carried out using DML. In essence, we just residualize the system, using cross-fitted residuals, and then apply standard instrumental variable methods from econometrics to perform inference on the structural parameter of interest.

## 12.6 Nonlinear Models with Proxy Controls★

An important recent development is "proximal causal inference," which generalizes early work by Griliches and Chamberlain [16].†

**Example 12.6.1** (Miao, Geng, and Tchetgen Tchetgen [3]) We consider the following model encoded in the DAG in Figure 12.14:

$$
\begin{aligned}
Y &:= f_Y(D, S, A, \epsilon_Y), \\
D &:= f_D(A, Q, \epsilon_D), \\
Q &:= f_Q(A, \epsilon_Q), \\
S &:= f_S(A, \epsilon_S), \\
A &:= \epsilon_A,
\end{aligned}
$$

where $\epsilon$'s are mutually independent. We can endow the same context to this model as in Example 12.5.1.

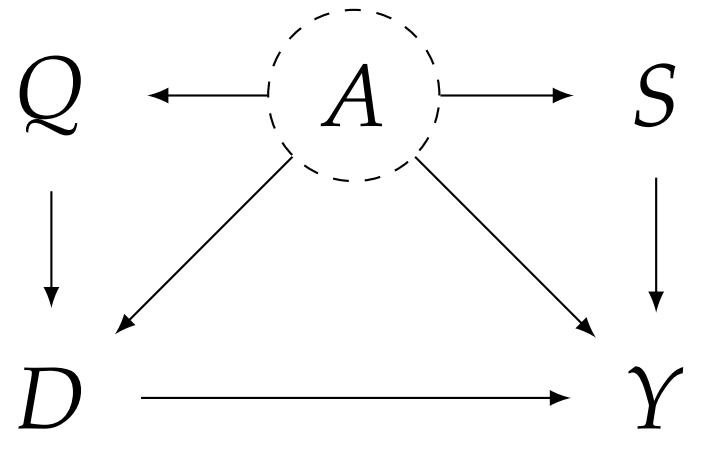

**Figure 12.14:** A SEM with Proxy Controls $Q$ and $S$. Note that conditioning on $Q$ and $S$ does not block the backdoor path $Y \leftarrow A \rightarrow D$, hence we cannot use the regression adjustment method for identification of $D \rightarrow Y$.

† The most relevant papers include, amongst others, the stream of work by Tchetgen Tchetgen and collaborators, as well as the dissertation work of Deaner. Here we describe some results of the first group specialized to the discrete case.

Here we can introduce background exogenous controls $X$ in each of the equations, but we don't do so to save notation. Notice that the model in Example 12.6.1 generalizes the Example 12.5.1 to the nonparametric case.

**Assumption 12.6.1** *In Example 12.6.1, assume*

*(a) Variables $Q$, $S$ and $A$ are finitely discrete and take on the same number of values.*
*(b) The matrix $\Pi(S \mid Q, d)$, whose $s^{\text{th}}$ row and $q^{\text{th}}$ column is $p(s \mid q, d)$, is invertible for each value $d$.*

Condition (b) is analogous to the usual relevance condition in IV and basically says that the two proxies $S$ and $Q$ have sufficient joint variation at any value of $d$ to allow $Q$ to serve as an "instrument" for $S$. The discreteness assumption can be generalized to a more general completeness condition; see Miao et al.[3] and Deaner [4]. As with the usual IV relevance condition, Condition (b) is testable from the data. In contrast, the DAG itself and the other conditions involve an unobserved variable $A$ and are therefore generally untestable. The validity of these untestable conditions must be assessed using contextual knowledge about the empirical problem.

**Theorem 12.6.1** *Under Assumption 12.6.1, $p(y : do(d))$ is identifiable by the proximal formula:*

$$p(y : do(d)) = \Pi(y \mid d, Q)\, \Pi(S \mid Q, d)^{-1}\, \Pi(S)\,, \qquad (12.6.1)$$

*where $\Pi(y \mid d, Q)$ and $\Pi(S)$ are row and column vectors whose entries are of the form $p(y \mid d, q)$ and $p(s)$.*

The mnemonic way to think about the formula above is that we are doing a kind of instrumental regression of $Y$ on $S$, while instrumenting $S$ with $Q$, which is exactly how we dealt with the linear version of this problem in Section 12.6.

**Remark 12.6.1** [17] and and [18] provide moment functions defined in terms of efficient influence functions, which possess the Neyman orthogonality property, for estimating of the average treatment effect within this proxy control setting in the presence of a high-dimensional set of control variables. These moment functions can thus serve as the foundation for the use of DML inference methods for the average treatment effect in such settings.

## Notebooks

- DML Sensitivity R Notebook analyses the sensitivity of the DML estimate in the Darfur wars example to unobserved confounders using the Sensemakr package in R. DML Sensitivity Python Notebook does the same analysis in Python.

- DML for Partially Linear IV R Notebook and DML for Partially Linear IV Python Notebook carry out the DML IV analysis of the Acemoglu-Johnson-Robinson example, which considers the impact of the quality of institutions on economic growth, instrumenting quality of institutions with settler mortality. The notebook explores the partially linear IV model and tests for the presence of weak instruments. See Chapter 14 for further discussion of this example as well as discussion of weak identification/instruments.

- DML for LATE Models R Notebook and DML for LATE Models Python Notebook estimate the Local Average Treatment Effects of 401(K) participation on net financial wealth.

- DML for Linear Proxy Controls Python Notebook provides an application of using proxy controls to estimate the effect of smoking on birth weight.

## Study Problems

1. Explain omitted confounder bias to a fellow student (one paragraph). Explore using sensitivity analysis to aid in understanding robustness of economic conclusions to the presence of unobserved confounders in an empirical example of your choice. The DML Sensitivity R Notebookcan be a helpful starting point but apply the ideas to a different empirical example. (You could use any of the previous examples we have analyzed).

2. Write a brief explanation of the idea of the instrumental variables regression model that would be appropriate for educating a fellow student. Discuss the idea of identifying the causal effect in this setting via path analysis in the spirit of what Philip Wright did. Illustrate your discussion with an empirical example. For example, revisit the analysis in DML for Partially Linear IV R Notebook.

3. (Simulation.) Create a notebook to simulate one of the linear IV or proxy controls models that we've described. Assume there are no $X$'s for simplicity. Demonstrate numerically why using least squares may not be appropriate due to unobserved confounding. Demonstrate numerically how using instrumental variable regression overcomes the issue.

4. (LATE etc.) Explain to a fellow student in writing one of the nonlinear models (e.g. LATE, IV quantile model, or the nonlinear model with proxy controls) and how causal parameters in these models are identified. DML for LATE Models R Notebook could be a starting point for explaining LATE and illustrating your explanation with empirical results. (If you have a good empirical example for proxy controls, please let us know.

## 12.A Proofs

### Latent Confounder Bias Result: Theorem 12.2.1

The proof heavily relies on the Frisch-Waugh-Lovel partialling out theorem (FWL) and the normalization on the variance of the latent confounder:

$$\mathrm{E}[\tilde{A}^2] = 1. \tag{12.A.1}$$

The proof also relies on the properties of $R^2_{U\sim V}$ which measures the proportion of variance of centered random variable $U$ that is linearly explained by another centered random variable $V$:

$$R^2_{U\sim V} = \frac{E\beta^2 V^2}{\mathrm{E}[U^2]} = 1 - \frac{\mathrm{E}[\epsilon^2]}{\mathrm{E}[U^2]} = \frac{(\mathrm{E}[UV])^2}{\mathrm{E}[U^2]\mathrm{E}[V^2]} = \mathrm{Cor}^2(U, V),$$

where $\beta = \mathrm{E}[VU]/\mathrm{E}[V^2]$ is the coefficient of the best linear projection of $U$ onto $V$, $\epsilon = U - \beta V$ is the projection residual, and $\mathrm{Cor}(U, V)$ denotes the correlation between $U$ and $V$. Note that $R^2$ is symmetric in $U$ and $V$: $R^2_{U\sim V} = R^2_{V\sim U}$.

By FWL and the normalization (12.A.1), we have

$$\gamma = \mathrm{E}[\tilde{A}\tilde{D}], \quad \delta = \mathrm{E}[\bar{A}\bar{Y}]/\mathrm{E}[\bar{A}^2],$$

where

$$\bar{Y} = \tilde{Y} - \beta\tilde{D}; \quad \beta = \mathrm{E}[\tilde{Y}\tilde{D}]/\mathrm{E}[\tilde{D}^2];$$

$$\bar{A} = \tilde{A} - \tilde{\beta}\tilde{D}; \quad \tilde{\beta} = \mathrm{E}[\tilde{A}\tilde{D}]/\mathrm{E}[\tilde{D}^2].$$

It follows that

$$\phi^2 = \frac{\gamma^2\delta^2}{(\mathrm{E}[\tilde{D}^2])^2} = \frac{(\mathrm{E}[\tilde{A}\tilde{D}])^2}{(\mathrm{E}[\tilde{D}^2])^2}\frac{(\mathrm{E}[\bar{Y}\bar{A}])^2}{(\mathrm{E}[\bar{A}^2])^2}.$$

Then the result follows from the normalization (12.A.1) and the following relations:

$$(\mathrm{E}[\tilde{D}\tilde{A}])^2 = \mathrm{Cor}^2(\tilde{D}, \tilde{A})\mathrm{E}[\tilde{D}^2] = R^2_{\tilde{D}\sim\tilde{A}}\mathrm{E}[\tilde{D}^2],$$

$$(\mathrm{E}[\bar{Y}\bar{A}])^2 = \mathrm{Cor}^2(\bar{Y}, \bar{A})\mathrm{E}[\bar{Y}^2]\mathrm{E}[\bar{A}^2] = R^2_{\bar{Y}\sim\bar{A}}\mathrm{E}[\bar{Y}^2]\mathrm{E}[\bar{A}^2],$$

$$\mathrm{E}[\bar{A}^2] = 1 - R^2_{\tilde{A}\sim\tilde{D}} = 1 - R^2_{\tilde{D}\sim\tilde{A}}.$$

and noting that by definition $R^2_{\bar{Y}\sim\bar{A}} = R^2_{\tilde{Y}\sim\tilde{A}|\tilde{D}}$.

□

## Partially Linear Outcome IV Model: Theorem 12.3.2

First note that since $\mathrm{E}[\tilde{Z} \mid X] = 0$, we can re-write the moment condition as

$$\mathrm{E}[(Y - \alpha D)\tilde{Z}] = 0.$$

We can use the structural equation for $Y$ to replace $Y$ in the moment equation:

$$\mathrm{E}[(g_Y(\epsilon_Y)D + f_Y(A, X, \epsilon_Y) - \alpha D)\tilde{Z}] = 0.$$

Furthermore, since $\tilde{Z} \perp\!\!\!\perp A, \epsilon_Y \mid X$, we have that

$$\begin{aligned}\mathrm{E}[f_Y(A, X, \epsilon_Y)\tilde{Z}] &= \mathrm{E}[f_Y(A, X, \epsilon_Y)\mathrm{E}[\tilde{Z} \mid X, A, \epsilon_Y]] \\ &= \mathrm{E}[f_Y(A, X, \epsilon_Y)\mathrm{E}[\tilde{Z} \mid X]] = 0.\end{aligned}$$

Thus we can re-write the moment equation as

$$\mathrm{E}[(g_Y(\epsilon_Y)D - \alpha D)\tilde{Z}] = 0.$$

Solving for $\alpha$ and using the fact that $\epsilon_Y \perp\!\!\!\perp \tilde{Z}$, we get

$$\alpha = \frac{\mathrm{E}[g_Y(\epsilon_Y)D\tilde{Z}]}{\mathrm{E}[D\tilde{Z}]} = \frac{\mathrm{E}[g_Y(\epsilon_Y)]\mathrm{E}[D\tilde{Z}]}{\mathrm{E}[D\tilde{Z}]} = \mathrm{E}[g_Y(\epsilon_Y)].$$

□

## Partially Linear Compliance IV Model: Theorem 12.3.3

Using the exact same arguments as in the proof of Theorem 12.3.2, we can deduce that the solution to the moment restriction takes the form

$$\alpha = \frac{\mathrm{E}[g_Y(X, A, \epsilon_Y)D\tilde{Z}]}{\mathrm{E}[D\tilde{Z}]} = \frac{\mathrm{E}[g_Y(X, A, \epsilon_Y)\mathrm{E}[D\tilde{Z} \mid X, A, \epsilon_Y]]}{\mathrm{E}[D\tilde{Z}]}.$$

We now use the assumptions on the structural response functions of $D$ and $Z$ to argue that $\mathrm{E}[D\tilde{Z} \mid X, A, \epsilon_Y] = \mathrm{E}[D\tilde{Z}]$, i.e. the covariance of $D$ and $Z$ (aka compliance) is independent of $X, A, \epsilon_Y$. This independence would then imply the theorem, since we would get that

$$\alpha = \mathrm{E}[g_Y(X, A, \epsilon_Y)].$$

First, we use the assumption on the structural response function of $D$:

$$\mathrm{E}[D\tilde{Z} \mid X, A, \epsilon_Y] = \mathrm{E}[(g_D(\epsilon_D)Z + f_D(X, A, \epsilon_D))\tilde{Z} \mid X, A, \epsilon_Y].$$

Using the fact that $Z \perp\!\!\!\perp A, \epsilon_D, \epsilon_Y \mid X$, and that $\mathrm{E}[\tilde{Z} \mid X] = 0$, we can remove the term $f_D(X, A, \epsilon_D)$ from the above equation:

$$\mathrm{E}[D\tilde{Z} \mid X, A, \epsilon_Y] = \mathrm{E}[g_D(\epsilon_D)Z\tilde{Z} \mid X, A, \epsilon_Y].$$

Using the additively separable assumption on the structural response of $Z$ and the fact that $\epsilon_Z$ is an exogenous independent variable, we have

$$\begin{aligned}\mathrm{E}[D\tilde{Z} \mid X, A, \epsilon_Y] &= \mathrm{E}[g_D(\epsilon_D)Z\epsilon_Z \mid X, A, \epsilon_Y] \\ &= \mathrm{E}[g_D(\epsilon_D)\epsilon_Z^2 \mid X, A, \epsilon_Y] = \mathrm{E}[g_D(\epsilon_D)\epsilon_Z^2]\end{aligned}$$

where we used the fact that all noise variables $\epsilon_D, \epsilon_Y, \epsilon_Z$ are exogenous and mutually independent. □

## Linear Proxy Model: Theorem 12.5.1.

We substitute $\tilde{A} = (\tilde{S} - \epsilon_S)/\phi$ in the equation $\tilde{Y} := \alpha\tilde{D} + \delta\tilde{A} + \iota\tilde{S} + \epsilon_Y$ to obtain

$$\tilde{Y} = \alpha\tilde{D} + \bar{\delta}\tilde{S} + U,$$

$$U = -\delta\epsilon_S/\phi + \epsilon_Y; \quad \bar{\delta} = \iota + \delta/\phi.$$

To verify

$$\mathrm{E}\,[U] = 0$$

we observe using repeated substitutions that:

- $\tilde{D}$ is a linear combination of $(\epsilon_A, \epsilon_Q, \epsilon_D)$,
- $\tilde{Q}$ is a linear combination of $\epsilon_A$ and $\epsilon_Q$.
- $U$ is a linear combination of $(\epsilon_S, \epsilon_Y)$.

The conclusion follows from the assumption that

$$(\epsilon_A, \epsilon_Q, \epsilon_D, \epsilon_S, \epsilon_Y)$$

are all uncorrelated. The conclusion that $\alpha$ is identified provided that $\tilde{D}$ and the best linear predictor of $\tilde{S}$ using $\tilde{Q}$ and $\tilde{D}$ have non-degenerate covariance matrices is left as an exercise.

□

## The LATE Result: Theorem 12.4.1

We can use, for example, the backdoor criterion to conclude that

$$\mathrm{E}[\mathrm{E}[D \mid Z = z, X]] = \mathrm{E}[\mathrm{E}[D(z) \mid X]] = \mathrm{E}D(z).$$

Similarly,

$$\mathrm{E}[\mathrm{E}[Y \mid Z = z, X]] = \mathrm{E}[\mathrm{E}[Y(D(z)) \mid X]] = \mathrm{E}[Y(D(z))].$$

Furthermore, by monotonicity, we have both

$$\theta_2 = \mathrm{E}[D(1) - D(0)] = \mathrm{P}(D(1) > D(0))$$

and

$$\begin{aligned} \theta_1 &= \mathrm{E}[Y(D(1)) - Y(D(0))] \\ &= \mathrm{E}[\{Y(1) - Y(0)\}1\{D(1) > D(0)\}]. \end{aligned}$$

Therefore

$$\theta_1/\theta_2 = \mathrm{E}[Y(1) - Y(0) \mid D(1) > D(0)].$$

□

## Testable Restriction for the IV Quantile Model: Theorem 12.4.2

The result is immediate from (i) the equivalence of the event $Y \leq f_Y(D, X, u)$ and the event $\epsilon_Y \leq u$, which holds under the strict monotoniticity assumption, and (ii) the independence of $\epsilon_Y$ from $Z$ and $X$ which follows from the stated independence conditions. Using (i) and (ii), we have

$$\mathrm{P}[Y \leq f_Y(D, X, u) \mid Z, X] = \mathrm{P}[\epsilon_Y \leq u \mid Z, X]$$

$$= \mathrm{P}[\epsilon_Y \leq u] = \mathrm{P}[U(0,1) \leq u] = u.$$

□

## Identification in the Nonlinear Proxy Variables Model: Theorem 12.6.1

To sketch a proof, the DAG implies that the observed variables $D, Y, Q, S$ and the unobserved variable $A$ obey the two conditional independence relations:

$$(i) \quad S \perp\!\!\!\perp (Q, D) \mid A \quad (ii) \quad Q \perp\!\!\!\perp Y \mid (A, D). \qquad (12.A.2)$$

These in turn imply

$$\begin{aligned} \Pi(S \mid Q, d) &= \Pi(S \mid Q, A, d)\Pi(A \mid Q, d) \\ &= \Pi(S \mid A)\Pi(A \mid Q, d) \end{aligned}$$

and

$$\begin{aligned} \Pi(y \mid Q, d) &= \Pi(y \mid Q, A, d)\Pi(A \mid Q, d) \\ &= \Pi(y \mid A, d)\Pi(A \mid Q, d). \end{aligned}$$

We now want to solve these equations for $\Pi(y \mid A, d)$ in terms of quantities that could be learned in the data.

We will need invertibility of $\Pi(S \mid Q, d)$ which requires invertibility of both $\Pi(S \mid A)$ and $\Pi(A \mid Q, d)$. Under these invertibility conditions, we have

$$\Pi(A \mid Q, d) = \Pi(S \mid A)^{-1}\Pi(S \mid Q, d)$$

and

$$\Pi(y \mid Q, d) = \Pi(y \mid A, d)\Pi(S \mid A)^{-1}\Pi(S \mid Q, d),$$

which yield

$$\Pi(y \mid A, d) = \Pi(y \mid Q, d)\Pi(S \mid Q, d)^{-1}\Pi(S \mid A).$$

Next, because $A$ blocks backdoor paths between $D$ and $Y$, we have that

$$p(y \mid a : do(d)) = p(y \mid a, d) \qquad (12.A.3)$$

or, after integrating out $a$,

$$p(y : do(d)) = \Pi(y \mid A, d)\Pi(A),$$

which can be further expressed as

$$\Pi(y \mid d, Q)\, \Pi(S \mid Q, d)^{-1}\, \Pi(S)\,, \qquad (12.A.4)$$

using the derivations above. □

# 13 DML for IV and Proxy Controls Models and Robust DML Inference under Weak Identification

"Better LATE than nothing."

– Guido Imbens [1].

Here we specialize DML methods to partially linear models with instruments, arising either through endogeneity of the policy variable or through the use of proxy controls as outlined in Chapter 8. We also present DML methods for LATE parameters in the fully nonlinear model with a binary endogenous treatment and binary instrument. We further examine how DML inference method can be modified to cope with weak instruments and weak identification in generic moment problems through the use of Neyman-orthogonal scores and Neyman's $C(\alpha)$ statistic.

## 13.1 DML Inference in Partially Linear IV Models

Here we consider estimation of parameters that obey the following instrumental exclusion restriction:

$$\mathrm{E}[\epsilon \tilde{Z}] = 0,$$

where

$$\epsilon := \tilde{Y} - \theta_0' \tilde{D},$$

and

$$\begin{aligned} \tilde{Y} &= Y - \ell_0(X), \quad \ell_0(X) = \mathrm{E}[Y \mid X], \\ \tilde{D} &= D - r_0(X), \quad r_0(X) = \mathrm{E}[D \mid X], \\ \tilde{Z} &= Z - m_0(X), \quad m_0(X) = \mathrm{E}[Z \mid X]. \end{aligned}$$

Here we take the dimension of $\tilde{Z}$ to be the same as that of $\tilde{D}$ for simplicity.

Two key examples leading to this statistical structure are

- the partially linear instrumental variable model, and
- the partially linear model with proxy controls.

We discussed these examples and showed they fit into this structure in Chapter 8.

To estimate $\theta_0$ and to perform inference on it we can apply the general DML algorithm with the score

$$\psi(W; \theta, \eta) := (Y - \ell(X) - \theta'(D - r(X)))(Z - m(X)), \quad (13.1.1)$$

where $W = (Y, D, X, Z)$ and $\eta = (\ell, m, r)$ with $\ell$, $m$, and $r$ being $P$-square-integrable functions mapping the support of $X$ to $\mathbb{R}$. By definition, we have that

$$\mathrm{E}[\psi(W; \theta_0, \eta_0)] = 0;$$

and it is not difficult to check (via homework) that the Neyman orthogonality condition,

$$\partial_\eta \mathrm{E}[\psi(W; \theta_0, \eta_0)] = 0,$$

holds at the true value $\eta_0 = (\ell_0, m_0, r_0)$ of the nuisance parameters.

**DML for Partially Linear IV and Proxy Models**

1. Partition data indices into $k$ folds of approximately equal size: $\{1,...,n\} = \cup_{k=1}^{K} I_k$. For each fold $k = 1,...,K$, compute ML estimators $\hat{\ell}_{[k]}(X)$, $\hat{m}_{[k]}(X)$, $\hat{r}_{[k]}(X)$ of the best predictors $\ell_0(X)$, $m_0(X)$, $r_0(X)$, leaving out the $k$-th block of data, and obtain the cross-fitted residuals for each $i \in I_k$:

$$\check{Y}_i = Y_i - \hat{\ell}_{[k]}(X_i),$$
$$\check{D}_i = D_i - \hat{r}_{[k]}(X_i),$$
$$\check{Z}_i = Z_i - \hat{m}_{[k]}(X_i).$$

2. Compute the standard IV regression of $\check{Y}_i$ on $\check{D}_i$ using $\check{Z}_i$ as the instrument; that is, obtain $\hat{\theta}$ as the root in $\theta$ of the following equation:

$$\mathbb{E}_n[(\check{Y} - \theta'\check{D})\check{Z}] = 0.$$

3. Construct standard errors and confidence intervals as in the standard linear instrumental variables regression theory.

In what follows it will be convenient to use the following notation

$$\|h\|_{L^2} := \sqrt{\mathrm{E}_X[h^2(X)]},$$

where, as before, $\mathrm{E}_X$ computes the expectation over values of $X$.

**Theorem 13.1.1** (Adaptive Inference in the Partially Linear IV Model) *Impose technical regularity conditions as in [2] which include the following key conditions: (1) the instruments are strong – namely, the singular values of* $\mathrm{E}[\tilde{D}\tilde{Z}]$ *are well-separated from zero – and (2) the estimators* $\hat{\ell}_{[k]}(X)$, $\hat{m}_{[k]}(X)$, *and* $\hat{r}_{[k]}(X)$ *provide high-quality approximations to the best predictors* $\ell_0(X)$, $m_0(X)$, *and* $r_0(X)$ *– namely,*

$$n^{1/4}\|\hat{\ell}_{[k]} - \ell_0\|_{L^2} \approx 0, \quad n^{1/4}\|\hat{m}_{[k]} - m_0\|_{L^2} \approx 0,$$

*and*

$$n^{1/4}\|\hat{r}_{[k]} - r_0\|_{L^2} \approx 0.$$

*Then the estimation error in* $\check{D}_i$ *and* $\check{Y}_i$ *has no first order effect on the behavior of* $\hat{\theta}$*:*

$$\sqrt{n}(\hat{\theta} - \theta_0) \approx (\mathbb{E}_n[\tilde{D}\tilde{Z}])^{-1}\sqrt{n}\mathbb{E}_n[\tilde{Z}\epsilon],$$

*and, as a result, $\hat{\theta}$ concentrates in a $1/\sqrt{n}$ neighborhood of $\theta$ with deviations approximated by the Gaussian law:*

$$\sqrt{n}(\hat{\theta} - \theta_0) \overset{a}{\sim} N(0, \mathsf{V}),$$

*where*

$$\mathsf{V} = (\mathrm{E}[\tilde{D}\tilde{Z}'])^{-1}\mathrm{E}[\tilde{Z}\tilde{Z}'\epsilon^2](\mathrm{E}[\tilde{Z}\tilde{D}])^{-1}.$$

The standard error of $\hat{\theta}$ is estimated as $\sqrt{\hat{\mathsf{V}}/n}$, where $\hat{\mathsf{V}}$ is an estimator of $V$ based on the plug-in principle. The result implies that the confidence interval

$$[\hat{\theta} - 2\sqrt{\hat{\mathsf{V}}/n}, \hat{\theta} + 2\sqrt{\hat{\mathsf{V}}/n}]$$

covers $\theta$ for approximately 95% of the realizations of the sample. In other words, if our sample is not atypical, the interval covers the truth.

## The Effect of Institutions on Economic Growth

To demonstrate DML estimation of partially linear structural equation models with instrumental variables, we consider estimating the effect of institutions on aggregate output following the work of [3] (AJR).

We use the same set of 64 country-level observations as AJR. The data set contains measurements of GDP, settler mortality, an index which measures protection against expropriation risk and geographic information. The outcome variable, $Y$, is the logarithm of GDP per capita and the endogenous explanatory variable, $D$, is a measure of the strength of individual property rights that is used as a proxy for the strength of institutions. To deal with endogeneity, we use an instrumental variable $Z$, which is mortality rates for early European settlers. Our raw set of control variables, $X$, include distance from the equator and dummy variables for Africa, Asia, North America, and South America.

Estimating the effect of institutions on output is complicated by the clear potential for simultaneity between institutions and output: Better institutions may generate higher incomes, but higher incomes may also lead to the development of better institutions. To help overcome this simultaneity, AJR use mortality rates for early European settlers as an instrument for institution quality. The validity of this instrument hinges on the argument that settlers set up better institutions in places where they were more likely to establish long-term settlements, that where they were likely to settle for the long term is related to settler mortality at the time of initial colonization, and that institutions are highly persistent. The exclusion restriction for the instrumental variable is then motivated by the argument that GDP, while persistent, is unlikely to be strongly influenced by mortality in the previous century, or earlier, except through institutions.

In their paper, AJR note that their instrumental variable strategy will be invalidated if other factors are also highly persistent and related to the development of institutions within a country and to the country's GDP. A leading candidate for such a factor, as they discuss, is geography. AJR address this by assuming that the confounding effect of geography is adequately captured by a linear term in distance from the equator and a set of continent dummy variables. Using DML allows us to relax this assumption and replace it by a weaker assumption that geography can be sufficiently controlled by an unknown function of distance from the equator and continent dummies which can be learned by ML methods.

We present the verbal identification argument above in the form of a DAG in Figure 13.1. In the DAG, $Y$ is wealth, $O$ the quality of early institutions, $D$ the quality of modern institutions, $X$ observed measures of geography, $Z$ early settler mortality, $A$ the present day latent factors jointly determining modern institutions and wealth, and $L$ early latent factors affecting early settler mortality. Applying the IV method here requires the identification of the causal effect of $Z \to D$ and $Z \to Y$. From the DAG, we see that $X$ blocks the backdoor paths from $Y$ to $Z$ and from $D \to Z$. This means that the instrument satisfies the required exogeneity condition conditional on $X$.

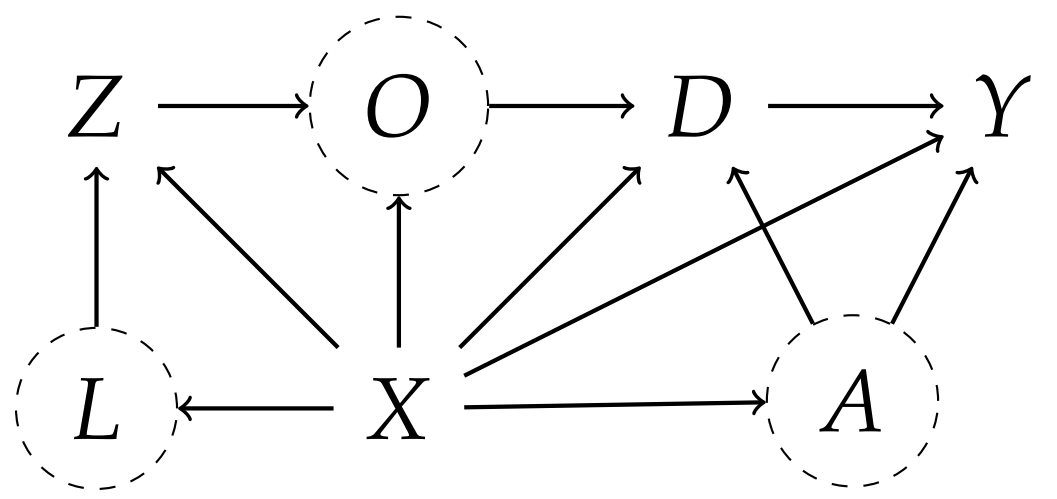

**Figure 13.1:** DAG for the Effect of Quality of Institutions on Wealth.

We think the story sounds plausible, but it is always important to consider threats to identification. The direct threat to identification would be if $L$ directly affected $Z$ and either $O$, $D$, or $Y$, or, in words, if early latent factors directly affected early settler mortality and either present-day quality of institutions or present day wealth. In such cases we would need to include $L$ as additional controls. $L$ could represent many different latent

| Lasso | Forest | Best |
|---|---|---|
| 0.77 | 0.88 | 0.88 |
| (0.17) | (0.32) | (0.32) |

**Note:** Estimated coefficient from DML based estimation of a linear instrumental variables model based on orthogonal estimating equations in the AJR example. Column labels denote the method used to estimate nuisance functions. We used $K = 20$ folds for cross-fitting.

**Table 13.1:** DML Estimates of the Effect of Institutions on Output

factors. For example, one might conjecture that the religion of early European settlers (e.g., Catholic vs Protestant) is related to the type of institutions they would establish and to their mortality rates upon colonization. In their original study, AJR did examine this threat by checking robustness of their result with respect to the inclusion of religion variables. They also examined the use of other additional controls to assess robustness to other potential sources of confounding.[1]

1: It is good to revisit their analysis using ML tools. See their Data archive to get started.

We report results from applying DML following the procedure outlined in Section 10.4 in Table 13.1. For cross-fitting, we use 20 folds given how small the data set is. Here we just tried two successful methods, Lasso and Random Forests, for learning the nuisance functions $\eta$. As predictors in the Lasso estimates, we used a dictionary formed by taking latitude and latitude$^2$ interacted with continent dummies as technical controls. For the Random Forest estimates, we simply include latitude and continent dummies as raw controls. The Random Forest predicts outcomes $Y$, $D$, and $Z$ better than Lasso. The resulting best DML estimate is therefore based on DML with Random Forest used in all ML steps.

In this example, we see uniformly large and positive point estimates across all procedures considered, and estimated effects are statistically significant at the 5% level in all cases. We note the estimates are somewhat smaller than the baseline estimates reported in AJR – an estimated coefficient of 1.10 with estimated standard error of 0.46 ([3], Table 4, Panel A, column 7) – but are qualitatively similar, indicating a strong and positive effect of institutions on output.

## 13.2 DML Inference in the Interactive IV Regression Model (IRM)

### DML Inference on LATE

In this section, we consider estimation of local average treatment effects (LATE) with a binary treatment variable, $D \in \{0,1\}$, and a binary instrument, $Z \in \{0,1\}$. As before, $Y$ denotes the outcome variable, and $X$ is the vector of covariates. Consider the following statistical parameter:

$$\theta_0 = \frac{\mathrm{E}[\mathrm{E}[Y \mid Z=1,X] - \mathrm{E}[Y \mid Z=1,X]]}{\mathrm{E}[\mathrm{E}[D \mid Z=1,X] - \mathrm{E}[D \mid Z=0,X]]}.$$

This parameter is the ratio of the average predictive effects of $Z$ on $Y$ and of $D$ on $Y$. Under the assumptions laid out in Chapter 8, this statistical parameter is a causal parameter – the average treatment effect for compliers (LATE).

To set up estimation, define the regression functions:

$$\begin{aligned} \mu_0(Z,X) &= \mathrm{E}[Y \mid Z,X] \\ m_0(Z,X) &= \mathrm{E}[D \mid Z,X] \\ p_0(X) &= \mathrm{E}[Z \mid X]. \end{aligned}$$

Define the nuisance parameter $\eta = (\mu, m, p)$ to denote square-integrable functions $\mu$, $m$, and $p$, with $\mu$ mapping the support of $(Z,X)$ to $\mathbb{R}$ and $m$ and $p$ respectively mapping the support of $(Z,X)$ and $X$ to $(\varepsilon, 1-\varepsilon)$ for some $\varepsilon \in (0,1/2)$. The true value of the nuisance parameter is $\eta_0 = (\mu_0, m_0, p_0)$, the regression functions defined above.

The DML estimator of $\theta_0$ employs the orthogonal score

$$\begin{aligned} \psi(W;\theta,\eta) := {} & \mu(1,X) - \mu(0,X) + H(p)(Y - \mu(Z,X)) \\ & - \Big(m(1,X) - m(0,X) + H(p)(D - m(Z,X)\Big)\theta, \end{aligned}$$

for $W = (Y,D,X,Z)$ and

$$H(p) := \frac{Z}{p(X)} - \frac{(1-Z)}{1-p(X)}.$$

It is easy to verify (for homework) that this score satisfies the moment condition

$$\mathrm{E}[\psi(W;\theta_0,\eta_0)] = 0$$

and also the Neyman orthogonality condition

$$\partial_\eta \mathrm{E}\psi(W; \theta_0, \eta_0) = 0$$

at the true value $\eta_0 = (\mu_0, m_0, p_0)$ of the nuisance parameter.

Therefore we can apply the generic ML algorithm to this problem, including the selection of the best ML methods to estimate the nuisance parameters.

**Theorem 13.2.1** (DML for LATE) *Suppose conditions specified in [2] hold. In particular, suppose that the overlap condition holds; namely, for some $\epsilon > 0$ with probability 1,*

$$\epsilon < p_0(X) < 1 - \epsilon.$$

*Further, suppose $\epsilon < \hat{p}_{[k]}(X) < 1 - \epsilon$ and that estimators $\hat{p}_{[k]}$, $\hat{m}_{[k]}$, $\hat{\mu}_{[k]}$ provide high-quality approximation to $p_0$, $m_0$, and $\mu_0$ in the sense that*

$$n^{1/2} \|\hat{p}_0 - p_0\|_{L^2} \times \left( \|\hat{\mu}_0 - \mu_0\|_{L^2} + \|\hat{m}_0 - m_0\|_{L^2} \right) \approx 0.$$

*Then estimation of the nuisance parameters does not affect the behavior of the estimator to the first order; namely,*

$$\sqrt{n}(\hat{\theta} - \theta_0) \approx \sqrt{n}\mathbb{E}_n[\varphi_0(W)],$$

*where*

$$\varphi_0(W) = -J_0^{-1}\psi(W; \theta_0, \eta_0), \quad J_0 := \mathrm{E}\Big[m_0(1, X) - m_0(0, X)\Big].$$

*Consequently, $\hat{\theta}$ concentrates in a $1/\sqrt{n}$-neighborhood of $\theta_0$ and the sampling error $\sqrt{n}(\hat{\theta} - \theta_0)$ is approximately normal*

$$\sqrt{n}(\hat{\theta} - \theta_0) \overset{a}{\sim} N(0, \mathsf{V}), \quad \mathsf{V} := \mathrm{E}[\varphi_0(W)\varphi_0(W)'].$$

Variance estimation and confidence intervals are constructed as in the generic DML algorithm.

## The Effect of 401(k) Participation on Net Financial Assets

Here we continue to re-analyze the effects of 401(k)'s on household financial assets, picking up from Section 10.3. In this section, we report the LATE in this example where we take the endogenous treatment variable to be *participating* in a 401(k) plan using 401(k) *eligibility* as instrument. Even after controlling for features related to job choice, it seems likely that the actual choice of whether to participate in an offered plan would be endogenous. Of course, we can use eligibility for a 401(k) plan as an instrument for participation in a 401(k) plan under the conditions that were used to justify the exogeneity of eligibility for a 401(k) plan outlined in Section 10.3.

R Notebook on DML for Impact of 401(K) Participation on Financial Wealth

We report DML results of estimating the LATE of 401(k) participation using 401(k) eligibility as an instrument in Table 13.2. We employ the procedure outlined in Section 13.2 using the same ML estimators to estimate the quantities used to form the orthogonal estimating equation as we employed to estimate the ATE of 401(k) eligibility in Section 10.3, so we omit the details for brevity. Looking at the results, we see that the estimated causal effect of 401(k) participation on net financial assets is uniformly positive and statistically significant across all of the considered methods. As when looking at the ATE of 401(k) eligibility, it is reassuring that the results obtained from the different flexible methods are broadly consistent with each other. It is also interesting that the results based on flexible ML methods are broadly consistent with, though somewhat attenuated relative to, those obtained by applying the same specification for controls as used in [4] and [5] and using a linear IV model which returns an estimated effect of participation of \$13,102 with estimated standard error of (1922). The attenuation may suggest that the simple intuitive control specification used in the original baseline specification is not sufficiently flexible.

| Lasso | Forest | Boosting | Neural Net. | Ensemble | Best |
|---|---|---|---|---|---|
| 8944 | 11764 | 11133 | 11186 | 11173 | 11113 |
| (2259) | (1788) | (1661) | (1795) | (1641) | (1645) |

**Note:** Estimated LATE under a fully interactive IV model. Column labels denote the method used to estimate nuisance functions.

**Table 13.2:** DML Estimates of LATE on 401(k) Participation on Net Financial Assets

## 13.3 DML Inference with Weak Instruments

### Motivation

As a simple motivating example, consider a statistical model with instrumental moment conditions and one-dimensional endogenous variable $D$ when there are either no controls or we are able to partial them out perfectly. In this case, the IV estimator takes the form

$$\hat{\theta} = \mathbb{E}_n[\tilde{Z}\tilde{Y}]/\mathbb{E}_n[\tilde{Z}\tilde{D}],$$

and we have that

$$\sqrt{n}(\hat{\theta} - \theta) = \sqrt{n}\mathbb{E}_n[\tilde{Z}\epsilon]/\mathbb{E}_n[\tilde{Z}\tilde{D}].$$

When $\mathbb{E}_n[\tilde{Z}\tilde{D}]$ is well-separated away from zero, we invoke the approximation

$$\sqrt{n}\mathbb{E}_n[\tilde{Z}\epsilon]/\mathbb{E}_n[\tilde{Z}\tilde{D}] \overset{a}{\sim} N(0, \mathrm{E}[\tilde{Z}^2\epsilon^2])/\mathrm{E}[\tilde{Z}\tilde{D}]. \qquad (13.3.1)$$

However, this approximation is not reliable when instruments are "weak" – when $\mathbb{E}_n[\tilde{Z}\tilde{D}]$ appears close to zero. Intuitively, we may worry that small changes in a sample that result in relatively small changes in $\mathbb{E}_n[\tilde{Z}\tilde{D}]$ may still have large impacts on the estimator $\hat{\theta}$ when $\mathbb{E}_n[\tilde{Z}\tilde{D}]$ is near zero because $\mathbb{E}_n[\tilde{Z}\tilde{D}]$ shows up in the denominator. That is, (13.3.1), which essentially ignores sampling variation in $\mathbb{E}_n[\tilde{Z}\tilde{D}]$, may provide a very poor approximation to the actual finite sample sampling behavior of the IV estimator.

"Weak identification" (or "weak instruments" in IV models) refers to settings in which we cannot confidently conclude a testable identifying assumption holds in our data. In our simple IV model, the parameter $\theta$ is not identified when $\mathrm{E}[\tilde{Z}\tilde{D}] = 0$ as solving the population moment condition requires solving $\mathrm{E}[\tilde{Z}\tilde{D}]\theta = \mathrm{E}[\tilde{Z}\tilde{Y}]$.

We illustrate the potential poor performance of the usual aymptotic approximation (13.3.1) in Figure 13.2 which reports results from a simulation experiment in which $\mathrm{E}[\tilde{Z}\tilde{D}]$ is close to zero. Here we see the sampling distribution (given by the blue curve) of the IV estimator deviates strongly from the normal approximation (given by the red curve). Note that by varying how close $\mathrm{E}[\tilde{Z}\tilde{D}]$ is to zero, one can make the differences more or less pronounced.

In principle, we can detect the weak instrument problem by testing whether $\beta = 0$ in the projection equation

$$\tilde{D} = \beta\tilde{Z} + U, \quad \mathrm{E}[\tilde{Z}\tilde{D}].$$

Econometricians have found that the normal approximation above is adequate for inferential properties if the t-statistic for testing the null $\beta = 0$ is bigger than 4:[2]

$$|\hat{\beta} - \beta|/\mathrm{se}(\hat{\beta}) > 4.$$

2: These are called "rules of thumb" and are based on simulation experiments.

If this happens, then we are said to have a "strong" instrument. If this test for the strong instrument is passed, then it is relatively safe to use the normal approximation for inference with the IV estimator. If not, using the usual asymptotic approximation is not safe, but is there anything else that we can do?

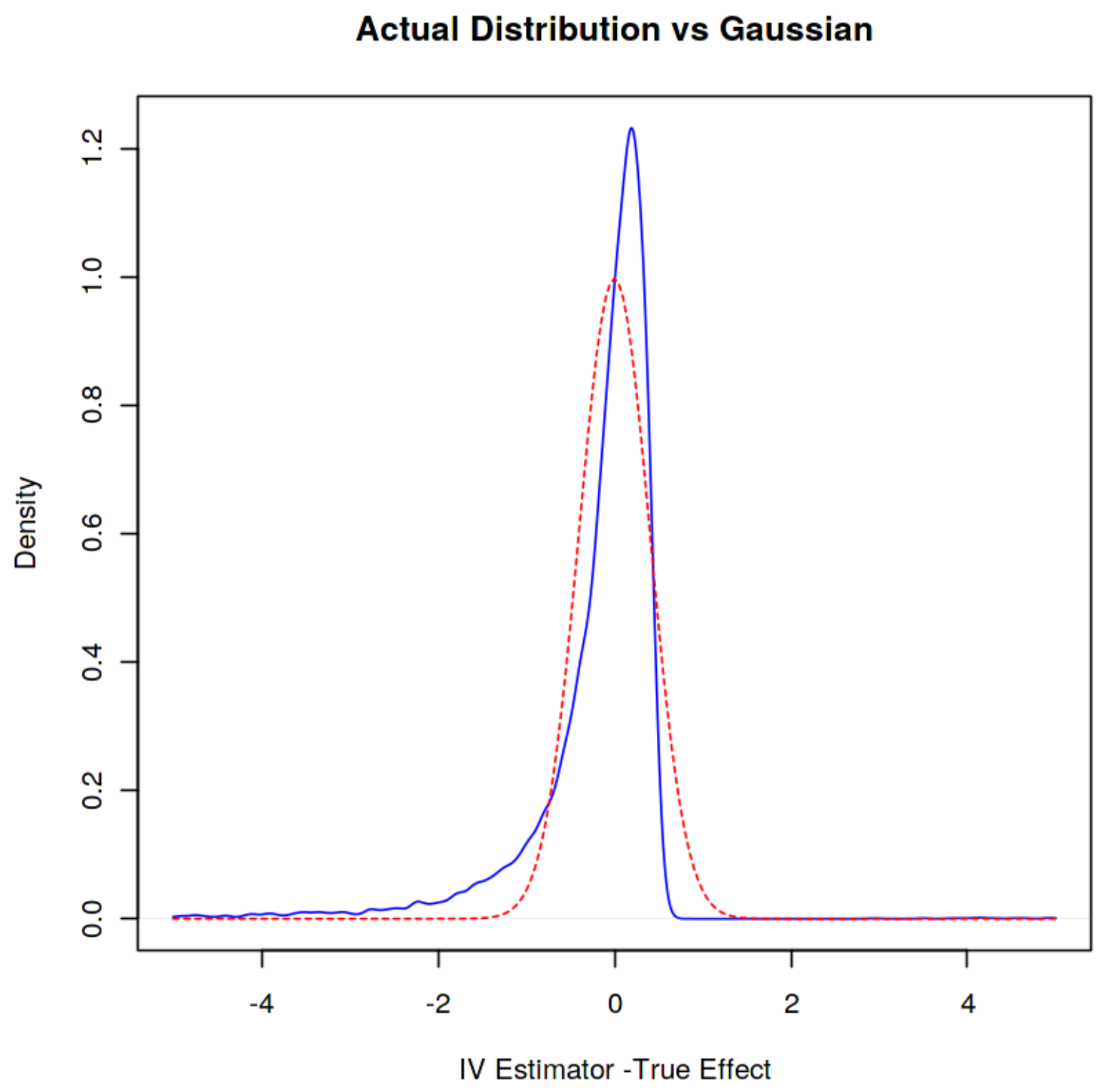

**Figure 13.2:** Actual sampling distribution of the IV estimator in a simulation experiment vs the normal approximation of the IV Estimator using weak instrument.

Of course there is. There are a variety of alternative inferential procedures whose behavior does not hinge on the well-separation of $\mathbb{E}_n[\tilde{Z}\tilde{D}]$ from zero. Here, we consider one specific approach based upon the statistic[3]

$$C(\theta) = \frac{|\mathbb{E}_n[(\tilde{Y} - \theta\tilde{D})\tilde{Z}]|^2}{\mathbb{V}_n[(\tilde{Y} - \theta\tilde{D})\tilde{Z}]/n}.$$

3: The empirical variance $\mathbb{V}_n$ is defined as:

$$\mathbb{V}_n[g(W_i)] := \mathbb{E}_n g(W_i)g(W_i)' - \mathbb{E}_n[g(W_i)]\mathbb{E}_n[g(W_i)]'.$$

If $\theta_0 = \theta$, then $C(\theta) \overset{a}{\sim} N(0,1)^2 = \chi^2(1)$. Therefore, we can reject the hypothesis $\theta_0 = \theta$ at level $a$ (for example $a = .05$ for a 5% level test) if $C(\theta) > c(1-a)$ where $c(1-a)$ is the $(1-a)$-quantile of a $\chi^2(1)$ variable. The probability of falsely rejecting the true hypothesis is approximately $a \times 100\%$. To construct a $(1-a) \times 100\%$ confidence region for $\theta$, we can then simply invert the test by collecting all parameter values that are not rejected at the $a$ level:

$$CR(\theta) = \{\theta \in \Theta : C(\theta) \leq c(1-a)\}.$$

In more complex settings with many controls or controls that enter with unknown functional form, we can simply replace the residuals $\tilde{Y}$, $\tilde{D}$, and $\tilde{Z}$ by machine learned cross-fitted residuals $\check{Y}$, $\check{D}$, and $\check{Z}$. Thanks to the orthogonality of the IV moment condition underlying the formulation outlined above, we can formally assert that the properties of $C(\theta)$ and the subsequent testing procedure and confidence region for $\theta$ continue to hold when using cross-fitted residuals. We will further be able to apply the general procedure to cases where $D$ is a vector, with a suitable adjustment of the statistic $C(\theta)$.

## DML Inference Robust to Weak-IV in PLMs

Here, we present a more general version of weak identification robust inference, including implementation and theoretical details, in settings where we want to use machine learning to aid in controlling for confounding variables $X$.

**DML Weak-IV-Robust Inference for PLIV Model**

1. **Initialize:** Let $\Theta$ be a known parameter space that contains the true value $\theta_0$. Using the DML-PLIV algorithm, produce the cross-fitted residuals: $\check{Y}_i$, $\check{D}_i$, and $\check{Z}_i$. Using the cross-fitted residuals and for $\theta \in \Theta$, compute the moment function

$$\check{\mathsf{M}}(\theta) := \mathbb{E}_n[(\check{Y}_i - \theta'\check{D}_i)\check{Z}_i],$$

the empirical covariance function

$$\check{\Omega}(\theta) := \mathbb{V}_n[(\check{Y} - \theta'\check{D})\check{Z}],$$

and the score statistic

$$C(\theta) := n\check{\mathsf{M}}(\theta)'\check{\Omega}^{-1}(\theta)\check{\mathsf{M}}(\theta).$$

2. **Robust Confidence Region:** Construct the approximate $(1 - a) \times 100\%$ confidence region as

$$CR(\theta_0) = \{\theta \in \Theta : C(\theta) \leq c(1 - a)\},$$

where $c(1 - a) := (1 - a)$-quantile of a $\chi^2(m)$ variable, where $m = \dim(Z_i)$.

In order to state the next result, define the oracle version of the moment and covariance functions given in Step 1 of the DML Weak-IV-Robust Inference algorithm,

$$\hat{\mathsf{M}}(\theta) = \mathbb{E}_n[(\tilde{Y} - \theta'\tilde{D})\tilde{Z}]$$

and

$$\hat{\Omega}(\theta) = \mathbb{V}_n[(\tilde{Y} - \theta'\tilde{D})\tilde{Z}],$$

which are defined in terms of the true residuals $\tilde{Y}_i$, $\tilde{D}_i$, and $\tilde{Z}_i$.

**Theorem 13.3.1** *Under regularity conditions, estimation of the nuisance parameters does not affect the behavior of the C statistic in the sense that*

$$C(\theta_0) \approx n\hat{\mathrm{M}}(\theta_0)'\hat{\Omega}^{-1}(\theta_0)\hat{\mathrm{M}}(\theta_0) \overset{a}{\sim} \chi^2(m).$$

*Consequently, the test rejects the true value with approximate probability $a$,*

$$\mathrm{P}(C(\theta) \geq c(1-a)) \approx a,$$

*and the confidence region $CR(\theta_0)$ contains $\theta_0$ with approximate probability $(1-a)$,*

$$\mathrm{P}(\theta_0 \in CR(\theta_0)) \approx (1-a).$$

## The Effect of Institutions on Economic Growth Revisited

We illustrate the use of DML weak identification robust inference by revisiting the AJR example from Section 13.1. Recall that Random Forests performed best in all auxiliary predictive steps in our original exercise in this example, so we only consider the use of Random Forests to form residuals in this section.

After partialling out controls using Random Forests, we run the regression of $\check{D}$ on $\check{Z}$ to assess the strength of the instruments. The resulting t-statistic is approximately 2, much lower than the "safety" threshold of 4. As such, we conclude that we have a weak instrument and proceed with weak identification robust inference.

We implement the robust inferential approach from the previous subsection considering $\Theta = [-2, 2]$ as our parameter space for the causal effect of institutions on wealth. We note that, because the outcome we consider is the logarithm of GDP per capita, the range [-2,2] includes extremely (likely implausibly) large negative and positive effects, so restricting attention to this range *a priori* seems reasonable. We illustrate the procedure in Figure 13.3 which plots the value of the test statistic $C(\theta)$ for $\theta \in [-2, 2]$.

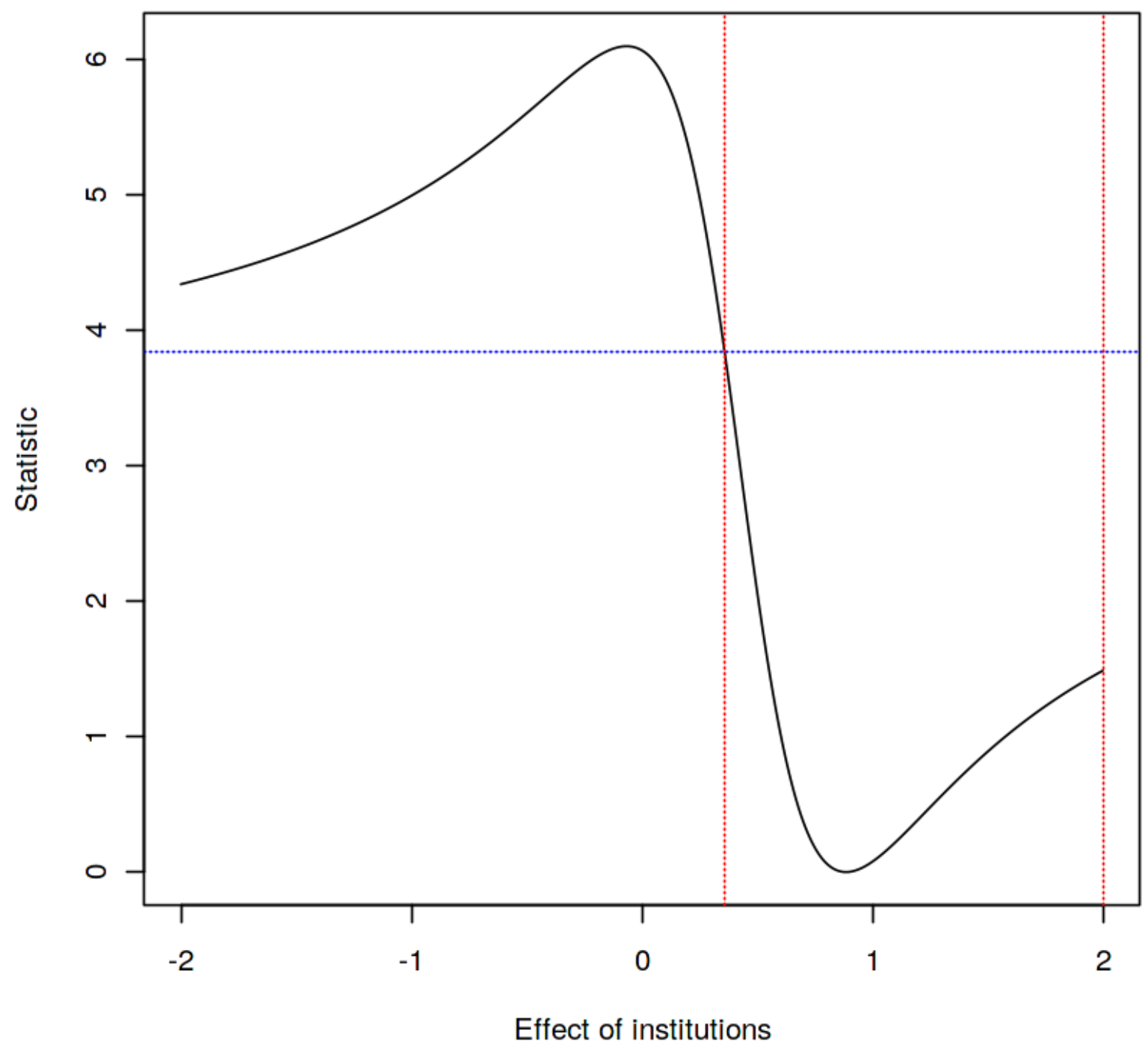

**Figure 13.3:** Construction of weak IV robust confidence regions for the effect of institutions on output using DML. Values of the $C(\theta)$ statistic are shown on the vertical axis; values of $\theta$ tested on the horizontal axis. The 90% confidence region is given by the red vertical bars.

The resulting 95% confidence region is

$$[.35, 2].$$

We can compare this region to the confidence region produced by the usual Gaussian asymptotic approximation which is not robust to weak identification:

$$[.88 \pm 2 \cdot 0.32] = [.24, 1.52].$$

Both the usual and robust confidence regions are consistent with relatively large positive effects of institutions on wealth. However, it is interesting that the lower end of the robust confidence region is larger than the lower end of the usual region and that this difference is economically meaningful. That is, we could not rule out that a one unit increase in quality of institutions causes an approximately a 27% increase in GDP per capita looking at the usual interval, while we could rule out all effect sizes smaller than 42% with the robust interval. The difference between a 27% and 42% increase in GDP per capita is certainly economically relevant. Given that the instruments are weak, we should, of course, rely on the robust confidence interval.

## 13.4 Generic DML Inference under Weak Identification

We now present a generally applicable formulation of weak identification robust inference. This formulation covers the problem of weak instruments in the context of LATE estimation as well as other problems where Neyman-orthogonal scores are available.

The initialization and first two steps to our approach to weak identification robust inference are the same as in the Generic DML Algorithm: We then use these estimates of the nuisance parameters in conjunction with the score function at a fixed value of $\theta$ to construct a score test statistic analogous to $C(\theta)$ from the previous section which can be used to test the hypothesis that $\theta_0 = \theta$ and to form confidence regions. We collect this procedure in the following algorithm:

**Generic DML Robust to Weak Identification**

1. **Initialize:** Provide the data frame $(W_i)_{i=1}^n$, the Neyman-orthogonal score/moment function $\psi(W, \theta, \eta)$ and the name and model for ML estimation method(s) for learning nuisance parameters $\eta$. Specify $\Theta$ to be a known parameter space that contains the true value $\theta_0$. We then take a K-fold random partition $(I_k)_{k=1}^K$ of observation indices $\{1, ..., n\}$ such that the size of each fold is about the same, and for each $k \in \{1, \ldots, K\}$, we construct a machine learning estimator $\hat{\eta}_{[k]}$ using data $(W_i)_{i \notin I_k}$, that is, all the data *except* the data from the $k^{\text{th}}$ fold.

2. **Estimate Moments and Their Variance:** Letting $k(i) = \{k : i \in I_k\}$, construct the moment function

$$\check{\mathsf{M}}(\theta) = \frac{1}{n} \sum_{i=1}^{n} \psi(W_i; \theta, \hat{\eta}_{[k(i)]})$$

covariance function,

$$\check{\Omega}(\theta) = \frac{1}{n} \sum_{i=1}^{n} [\psi(W_i; \theta, \hat{\eta}_{[k(i)]})\psi(W_i; \theta, \hat{\eta}_{[k(i)]})'] - \frac{1}{n} \sum_{i=1}^{n} [\hat{\psi}(W_i; \theta, \hat{\eta}_{[k(i)]})] \frac{1}{n} \sum_{i=1}^{n} [\psi(W_i; \theta, \hat{\eta}_{[k(i)]})]',$$

and score statistic

$$C(\theta) = n\check{\mathrm{M}}(\theta)'\check{\Omega}^{-1}(\theta)\check{\mathrm{M}}(\theta).$$

3. **Confidence Region:** Construct the approximate $(1 - a) \times 100\%$ confidence region as

$$CR(\theta_0) = \{\theta \in \Theta : C(\theta) \leq c(1 - a)\}$$

where $c(1 - a)$ is the $(1 - a)$–quantile of a $\chi^2(m)$ variable where $m = \dim(\check{\mathrm{M}}(\theta))$.

Note that this confidence region simply collects all values $\theta \in \Theta$ that are not rejected by testing $\theta_0 = \theta$ using test statistic $C(\theta)$ at the $a$-level.

As in the previous section, we define oracle versions of the moment and covariance functions from the preceding algorithm for use in stating formal results:

$$\hat{\mathrm{M}}(\theta) = \mathbb{E}_n[\psi(W; \theta, \eta_0)],$$

$$\hat{\Omega}(\theta) = \mathbb{V}_n[\psi(W; \theta, \eta_0)].$$

**Theorem 13.4.1** *Under regularity conditions, estimation of nuisance parameters does not affect the behavior of the $C(\theta)$ statistic in the sense that*

$$C(\theta_0) \approx n\hat{\mathrm{M}}(\theta_0)\hat{\Omega}^{-1}(\theta_0)\hat{\mathrm{M}}(\theta_0) \overset{a}{\sim} \chi^2(m).$$

*Consequently, a test that rejects when $C(\theta) \geq c(1-a)$, for $c(1-a)$ the $(1-a)$–quantile of a $\chi^2(m)$ variable, rejects the true value with approximate probability $a$:*

$$\mathrm{P}(C(\theta_0) \geq c(1 - a)) \approx a.$$

*Similarly, the confidence region corresponding to this test, $CR(\theta_0)$, contains $\theta_0$ with approximate probability $(1 - a)$:*

$$\mathrm{P}(\theta_0 \in CR(\theta_0)) \approx (1 - a).$$

# Notebooks

- The R Notebook on Weak IV provides a simulation experiment illustrating the weak instrument problem with IV

estimators.

- R Notebook on DML for Impact of Institutions on Output provides DML analysis of the impact of institutions on a country's wealth following AJR. The notebook first proceeds with the analysis assuming strong identification. It then notes the weak instrument problem and performs DML analysis that is robust to weak identification.
- R Notebook on DML for Impact of 401(K) Eligibility on Financial Wealth provides application of DML inference to learn predictive/causal effects of 401(K) eligibility on net financial wealth. (Note: The results produced in this notebook and provided in the text are slightly different than those in the original paper [2]. The replication files for [2] are given at the following Github repository. The difference is due to our use of a single split of the sample in producing the results for this text while the results in [2] are based on a method that aggregates results across multiple data splits.)

## Notes

The statistic $C(\theta)$ is Neyman's $C(\alpha)$ statistic.

## Study Problems

1. Experiment with The R Notebook on Weak IV, varying the strength of the instrument. How strong should the instrument be in order for the conventional normal approximation based on strong identification to provide accurate inference? Based on your experiments, provide a brief explanation of the weak IV problem to a friend.
2. Experiment with R Notebook on DML for Impact of 401(K) Eligibility on Financial Wealth. Apply the analysis to another data-set. For example, try the JTPA data from Joshua Angrist's data archive. Don't forget to draw your DAGs!
3. Experiment with R Notebook on DML for Impact of Institutions on Output. Try to extend the analysis by including other control variables (e.g. religion, other measures of geography, or measures of natural resources) or consider another empirical application to another IV example. (See some potential applications at the the Angrist data archive). In the case of a new application, don't forget to draw your DAGs!

4. (Theoretical). Verify that the scores for the partially linear IV methods are Neyman orthogonal.

# Statistical Inference on Heterogeneous Treatment Effects

# 14

"Never cross a river that is on average four feet deep."

– Nassim Nicholas Taleb [1].

We study estimation and inference on heterogeneous treatment effects. We introduce DML for inference on heterogenous treatment effects. We first review conditional and group average treatment effects as methods to analyse differences in the impact of treatment arising from the value of covariates. We show how these effects can be estimated using OLS. We illustrate the approach using the 401(k) example. We then consider more flexible inference on heterogeneous effects using adaptations of Random Forest methods, known as Causal Forests and illustrate the approach with an application on a large social science experiment studying the effect of the use of the word "welfare" in policy documents, on public perception.

## 14.1 CATEs under Conditional Exogeneity

We consider the standard setup for analyzing the effect of a binary treatment in the presence of a high-dimensional set of controls $Z$. Specifically, we have potential outcomes $Y(0)$ and $Y(1)$ and assigned treatment $D$ that obey the conditional exogeneity condition:

We focus on the binary treatment case, but note that the approach readily extends to more general settings.

$$D \perp\!\!\!\perp Y(d) \mid Z.$$

We observe the outcome $Y := Y(D)$, the treatment assignment $D$, and the high-dimensional set of controls $Z$.

Our main interest in this section is the Conditional Average Treatment Effect (CATE) defined as

$$\tau_0(X) = \mathrm{E}[Y(1) - Y(0) \mid X],$$

where $X$ is (typically) a low-dimensional subset of covariates $Z$. We have already seen in prior sections that under conditional exogeneity, the conditional average treatment effect is identified by the conditional average predictive effect (c.f. Theorem 5.2.1), which leads to the simple identification equation:

$$\tau_0(X) = \mathrm{E}[\mathrm{E}[Y \mid D = 1, Z] - \mathrm{E}[Y \mid D = 0, Z] \mid X] \qquad (14.1.1)$$

**The value of CATE estimation** So far in our analysis we have primarily been focusing on average causal effects. However, average effects are not informative of *whom to treat*. At best they can inform uniform policies, where we decide whether to roll out or not a new treatment on the whole population. Such uniform policies can have two major drawbacks. If the average effect is significantly positive and we decide to uniformly deploy the treatment, then there could potentially exist sub-groups in the population for which the treatment can have severe adverse effects. Analogously, if the average effect is significantly negative or a statistical null, then we might choose not to deploy a new policy or treatment. However, there could exist *responder sub-groups* in the population, for which the new treatment can have a significant positive impact. In both cases, by focusing on average causal effects, we are causing harm on sub-groups of the population, either by depriving of or forcing a new treatment.

Conditional average treatment effects allow us to identify such heterogeneities of the effect and discover in a data-driven manner the sub-groups of the population for which the treatment can be harmful or beneficial. Good estimates of the CATE, allows us to deploy personalized policies; personalizing the offered treatment based on observable characteristics of each unit. For this reason, the study of CATE estimation has become increasingly more widespread, especially in settings where we have rich datasets, with many informative covariates and in many application domains; with a frontrunner domain being digital experimentation, where datasets are rich and personalization is easily implementable and deployable.

**The hardness of CATE estimation** From a statistical viewpoint, estimation and inference on the CATE is inherently harder than estimation and inference of average effects. So far, most of the policy relevant target parameters that we have been interested in, take the form of some low-dimensional vector valued parameter. This is the first time, where our target causal parameter of interest is actually a function or the value of a function at a particular point. The closest estimation problem to the CATE is that of estimating a Best Prediction rule or a Conditional Expectation Function. Note that even if we had access to both counterfactuals $Y(1), Y(0)$, then estimation of the CATE is as hard as estimating a regression function corresponding to the outcome $Y(1) - Y(0)$. For such problems, thus far we were content at estimating them with respect to the mean-squared-error metric, and at a reasonable rate that decays to zero, potentially slower than the parametric rate of $n^{-1/2}$. On the contrary for most causal effects of interest, we were not really content with simply a mean-squared error rate; we typically sought the ability to construct confidence intervals and were striving for very accurate estimation, most of the times at parametric rates.

For this reason, when it comes to CATE estimation, we will need to re-calibrate our expectations and potentially relax our goals. In this and the next chapter, we will consider four such avenues:

- Target the estimation of the best linear approximation (BLA) of the CATE function, with a set of predefined low-dimensional engineered features. In this case, we can essentially recover all the desiderata of target causal quantities: estimation at parametric rates, confidence intervals for the BLA at a particular point and even simultaneous confidence bands for the BLA at a set of target evaluation points.
- Target inference on other summarizations of CATE such as its tail expectations, the value of a covariate-based

treatment policy, and the value of the optimal such policy. Again, we recover the desiderata of target causal quantities.

- Construct non-parametric confidence intervals for CATE predictions at a particular point, using novel methods (such as Causal Forests), which are practically powerful and marry machine learning techniques with uncertainty quantification, but which are theoretically valid only when $X$ is low-dimensional, and which in practice can be more brittle and are not as *assumption-lean* as inference based on OLS.
- Drop our desire to produce confidence intervals on the CATE function and only require good accuracy of the learned CATE function as captured by the mean-squared-error metric. In this case, we will be essentially treating the CATE problem as a best prediction problem and we will need to develop analogous methods for model selection, ensembling and out-of-sample evaluation. To compensate for the lack of confidence intervals for the CATE predictions, we will develop hypothesis tests that can be performed out-of-sample, that act as validation metrics that measure the quality of the CATE model as whole, as summarized in particular dimensions. For instance, we can test out of sample, whether the model picked up any statistically significant signal of heterogeneity, or if we use the model to prioritize treatment among the population, then will it lead to statistically significant policy gains.
- Drop the emphasis on learning the effect heterogeneity and focus only on the value of personalized policies that come out of our estimation process. In this case, we view CATE only as a means to our goal of designing personalized policies and in that respect we might want to measure the quality of our process, solely based on the personalized policy gains over some baseline, and not on the accuracy of the magnitude of the effect. Note that to learn a good policy, we are primarily interested in learning the sign of the effect and not necessarily its magnitude and appropriately partitioning the population such that the sign of the effect is relatively homogeneous within each sub-group. From this perspective, learning a good policy is more akin to a *classification* problem (classifying for which parts of the population the effect is positive/negative) as opposed to a regression problem and we will investigate such a formal equivalence.

## 14.2 Inference on Best Linear Approximations

Our main goal is summarizing the potentially complex and high-dimensional treatment effect function, which may depend on the entire vector $Z$, in terms of a lower-dimensional object $X$. We may be interested in such summaries for aiding interpretation or for policy reasons where we are interested in effects among particular recipients defined by observable characteristics.

For example, in the context of the 401(K) analysis from previous chapters, we have that $Y$ is a household's total net financial assets, $D$ is 401(k) eligibility status, and $Z$ is the entire set of household characteristics. We might then take $X$ to be income in which case the CATE $\tau_0(X)$ shows the expected effect of 401(k) eligibility on total financial assets for a subject whose income level is $X$.

The key to adaptively estimating and potentially performing inference for the CATE is expressing it as a conditional expectation of an unbiased signal:

$$\tau_0(X) = \mathrm{E}[Y(\eta_0) \mid X],$$

where the signal takes the form

$$Y(\eta) = H(\mu)\ (Y - g(D,Z)) + g(1,Z) - g(0,Z),$$

with nuisance parameters $\eta := (\mu, g)$ and

$$H(\mu) := \frac{D}{\mu(Z)} - \frac{1-D}{1-\mu(Z)}.$$

Here, $g(D,Z)$ and $\mu(Z)$ are square integrable functions with $\mu(Z)$ taking on values in $[\epsilon, 1-\epsilon]$ for some $\epsilon > 0$. The true values of these nuisance parameters are $\eta_0 := (\mu_0, g_0)$ defined as

$$\mu_0(Z) := \mathrm{P}(D = 1 \mid Z), \quad g_0(D,Z) := \mathrm{E}[Y \mid Z, D].$$

Importantly, the signal has the Neyman orthogonality property:

$$\partial_\eta \mathrm{E}[Y(\eta_0) \mid X] = 0.$$

Making use of the representation of the CATE as the conditional expectation of $Y(\eta_0)$, we then estimate the CATE using the following steps:

**Generic DML for CATE**

1. Partition data indices into $k$ folds of approximately equal size: $\{1, ..., n\} = \cup_{k=1}^{K} I_k$. For each fold $k = 1, ..., K$, compute ML estimators $\hat{g}_{[k]}(D, Z)$ and $\hat{\mu}_{[k]}(Z)$ of the best predictors $g_0(D, Z)$ and $\mu_0(Z)$ leaving out the $k$-th block of data. For any observation $i \in I_k$, define

$$\begin{aligned} Y_i(\widehat{\eta}) &= Y_i(\widehat{\eta}_k) \\ &= H_i(Y_i - \hat{g}_{[k]}(D_i, Z_i)) + \hat{g}_{[k]}(1, Z_i) - \hat{g}_{[k]}(0, Z_i) \end{aligned}$$

where $H_i = \dfrac{D_i}{\hat{\mu}_{[k]}(Z_i)} - \dfrac{1 - D_i}{1 - \hat{\mu}_{[k]}(Z_i)}$.

2. Use low-dimensional or high-dimensional regression methods to regress $Y_i(\hat{\eta})$ on covariate features $X_i$. If low-dimensional methods are used, inference on CATE can proceed using standard results for low-dimensional methods.

Under regularity conditions, the second step is adaptive, meaning all the learning guarantees and confidence intervals are approximately the same as if we knew the nuisance parameters $\eta_0$. This adaptation holds true because of the conditional Neyman orthogonality of $Y(\eta)$. We note that this adaptivity *does not imply* that inferential objects, e.g. confidence intervals, can readily be obtained if high-dimensional methods are used in Step 2. We discuss implementation and inferential issues in more detail in the following sections.

## Least Squares Methods for Learning CATEs

Here we focus on using least squares in the second step of the general approach given above.

Consider approximating or summarizing the function $t(x)$ by a linear combination of basis functions:

$$p(x)'\beta_0,$$

where $p(x)$ is $d$-dimensional dictionary with

$$d \ll n.$$

For example, $p(x)$ could be a vector of group indicators or a vector of orthogonal polynomials or splines.

The parameter $\beta_0$ is chosen to minimize the approximation error to the CATE:

$$\min_{\beta} \mathrm{E}(\tau_0(X) - p(X)'\beta)^2.$$

$p(x)'\beta_0$ is thus the best linear predictor for the CATE; that is,

$$\beta_0 = \left(\mathrm{E}\left[p(X)p(X)\right]\right)^{-1} \mathrm{E}\left[p(X)Y(\eta_0)\right].$$

An important, easily interpretable special case is when we choose to use group indicators in forming the basis functions $p(x)$. Specifically, we define group indicators as

$$G_k(X) = 1(X \in R_k),$$

where $R_k's$ are mutually exclusive regions in the covariate space. For example, in the 401(k) example, we may be interested in average treatment effects for observations with household income less than \$10,000, observations with income between \$10,000 and \$20,000, etc. which we could capture by defining $G_1(X) = 1(\text{Income} < \$10,000)$, $G_2(X) = 1(\$10,000 \leq \text{Income} < \$20,000)$, etc. With the group indicators defined, we then set

$$p(X) = (G_1(X), \ldots, G_K(X))'.$$

In this case, the Best Linear Predictor $\beta_0$ recovers the GATEs (group average treatment effects).

More generally, $p(x) \in \mathbb{R}^d$ represents a $d$-dimensional dictionary of series/sieve basis functions – e.g., polynomials or splines – and $p(x)'\beta_0$ corresponds to the best linear approximation to the target function $\tau_0(x)$ in the given dictionary. Under some smoothness conditions, $\pi(x) = p(x)'\beta_0$ will approximate $\tau_0(X)$ as the dimension of the dictionary becomes large, and our inference will target this function.

Taking the approach motivated above to a sample of data, we have that the natural estimator of the best linear predictor of the CATE is

$$p(x)'\widehat{\beta},$$

where $\widehat{\beta}$ is the ordinary least squares estimate of $\beta_0$ defined on the random sample $(X_i, D_i, Y_i)_{i=1}^N$:

$$\widehat{\beta} := \left( \frac{1}{N} \sum_{i=1}^{N} p(X_i)p(X_i)' \right)^{-1} \frac{1}{N} \sum_{i=1}^{N} p(X_i)Y_i(\widehat{\eta}).$$

Semenova et al. [2] derive a complete set of results for the properties of $p(x)'\widehat{\beta}$ as an estimator of the best linear predictor curve $x \mapsto p(x)'\beta_0$. Importantly, these results establish an asymptotic approximation that allows simultaneous inference on all parameters of the best linear predictor curve. The key result verifies that the large sample properties of $\widehat{\beta}$ are the same as those of

$$\bar{\beta} := \left( \frac{1}{N} \sum_{i=1}^{N} p(X_i)p(X_i)' \right)^{-1} \frac{1}{N} \sum_{i=1}^{N} p(X_i)Y_i(\eta_0),$$

when ML tools are used to estimate the nuisance parameter $\eta_0$ so long as the ML tools perform sufficiently well. Thus, we can employ standard methods for inference about $\beta_0$ and the best linear predictor curve functional $x \mapsto p(x)'\beta_0$.

Specifically, leveraging that $\widehat{\beta}$ and $\bar{\beta}$ have the same large sample properties, we have

$$\hat{\beta} - \beta_0 \sim_a N(0, \hat{\Omega}/N),$$

where

$$\widehat{\Omega} := \widehat{Q}^{-1} \left[ \mathbb{E}_n p(X_i)p(X_i)'(Y_i(\widehat{\eta}) - p(X_i)'\widehat{\beta})^2 \right] \widehat{Q}^{-1} \qquad (14.2.1)$$

for $\widehat{Q} = \mathbb{E}_n p(X_i)p(X_i)'$.

This result can be used to construct uniform confidence bands for

$$x \mapsto p(x)'\beta_0,$$

which can be interpreted as confidence intervals for CATE $x \mapsto \tau_0(x)$ if the approximation error is small.

## Application to 401(k) Example

R Notebook for DML on CATE analyzes the ATE of 401(K) conditional on income.

We illustrate estimation of CATEs and GATEs by revisiting the 401(k) example. Here, we consider the effect of 401(k) eligibility on net total financial assets controlling for household characteristics. We consider heterogeneity of this effect as a function of income. We consider two different ways to summarize these heterogeneous effects: GATEs based on coarse income categories

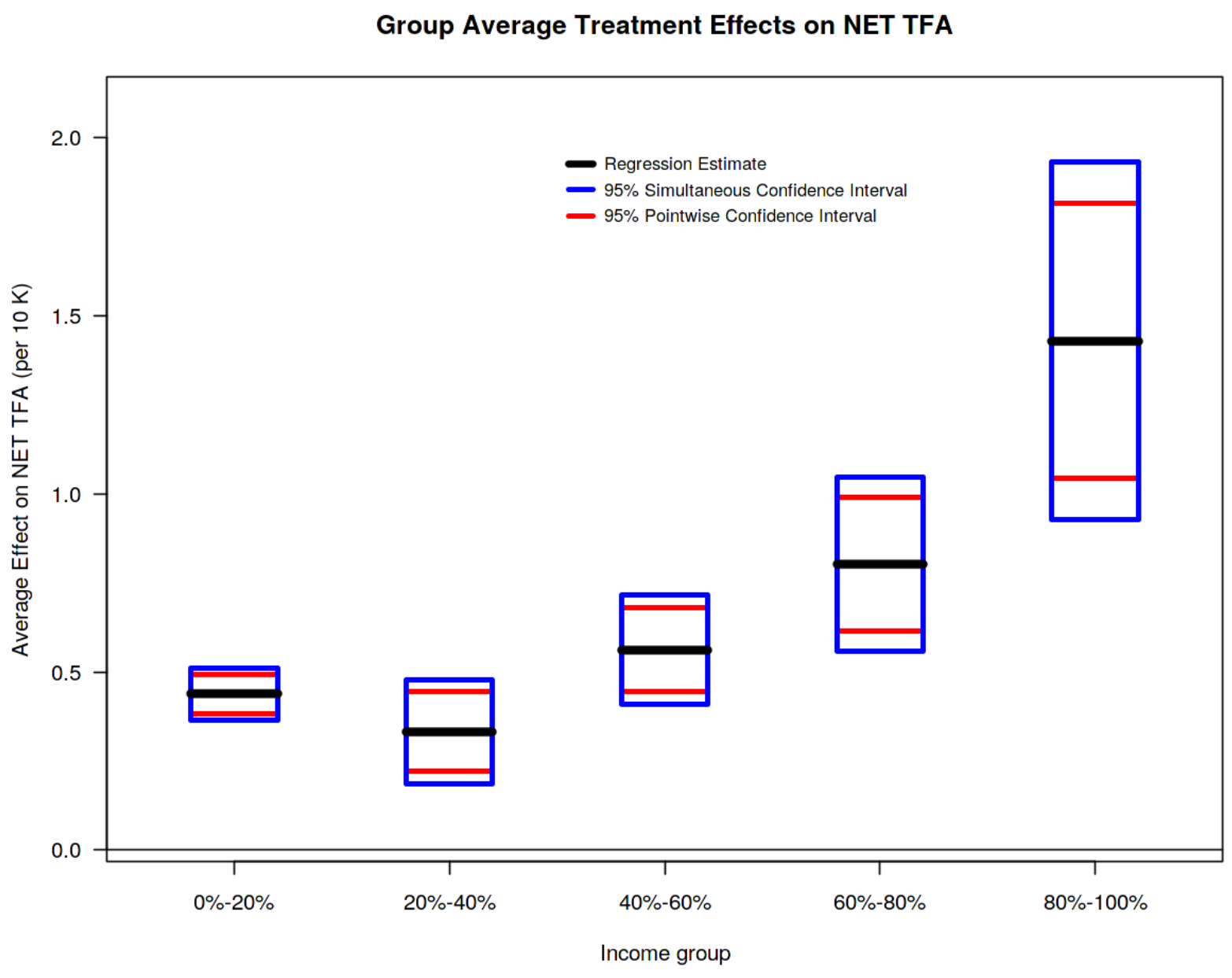

**Figure 14.1:** Inference on ATE of 401(k) Eligibility by Income Group

and a summary of the CATE given income based on a collection of polynomial terms in log(Income).

We show estimates and confidence bands on GATEs by income groups in Figure 14.1. Here, groups correspond to income quintiles; e.g the first group has households with income smaller than the $20^{\text{th}}$ percentile, the second group has households with income between the $20^{\text{th}}$ and $40^{\text{th}}$ percentile, and so on. Point estimates are provided by the central solid black bands. We represent pointwise confidence bands with the red lines in the interior of the box for each GATE. These bands would be appropriate for inference if one were interested *ex ante* in a single, pre-specified GATE. For example, one might be specifically interested in the eligibility effect among low income individuals and thus focus on the pointwise intervals over the first GATE. Finally, uniform confidence bands are given by the upper and lower bounds of the box for each GATE. These uniform bands provide a coverage guarantee for all five reported GATEs and would be appropriate for inference in settings where one was interested in all five effects and did not *ex ante* have a single specific GATE of interest.

We illustrate using a polynomial in log income to approximate the CATE in Figure 14.2. Point estimates are given by the central black line while the blue lines provide confidence bands. The narrower – dashed – confidence bands are pointwise and would be appropriate for a scenario in which one had a single, pre-specified value of income of interest. The wider confidence bands are uniform, providing a coverage guarantee for the *entire* best linear predictor curve $x \mapsto p(x)\beta_0$. That is, any path for the entire curve that would not be rejected will lie entirely within the uniform confidence band. Finally, note that the coverage guarantee extends to the true CATE function $x \mapsto \tau_0(x)$ if the approximation error of the polynomial to the true CATE is small.

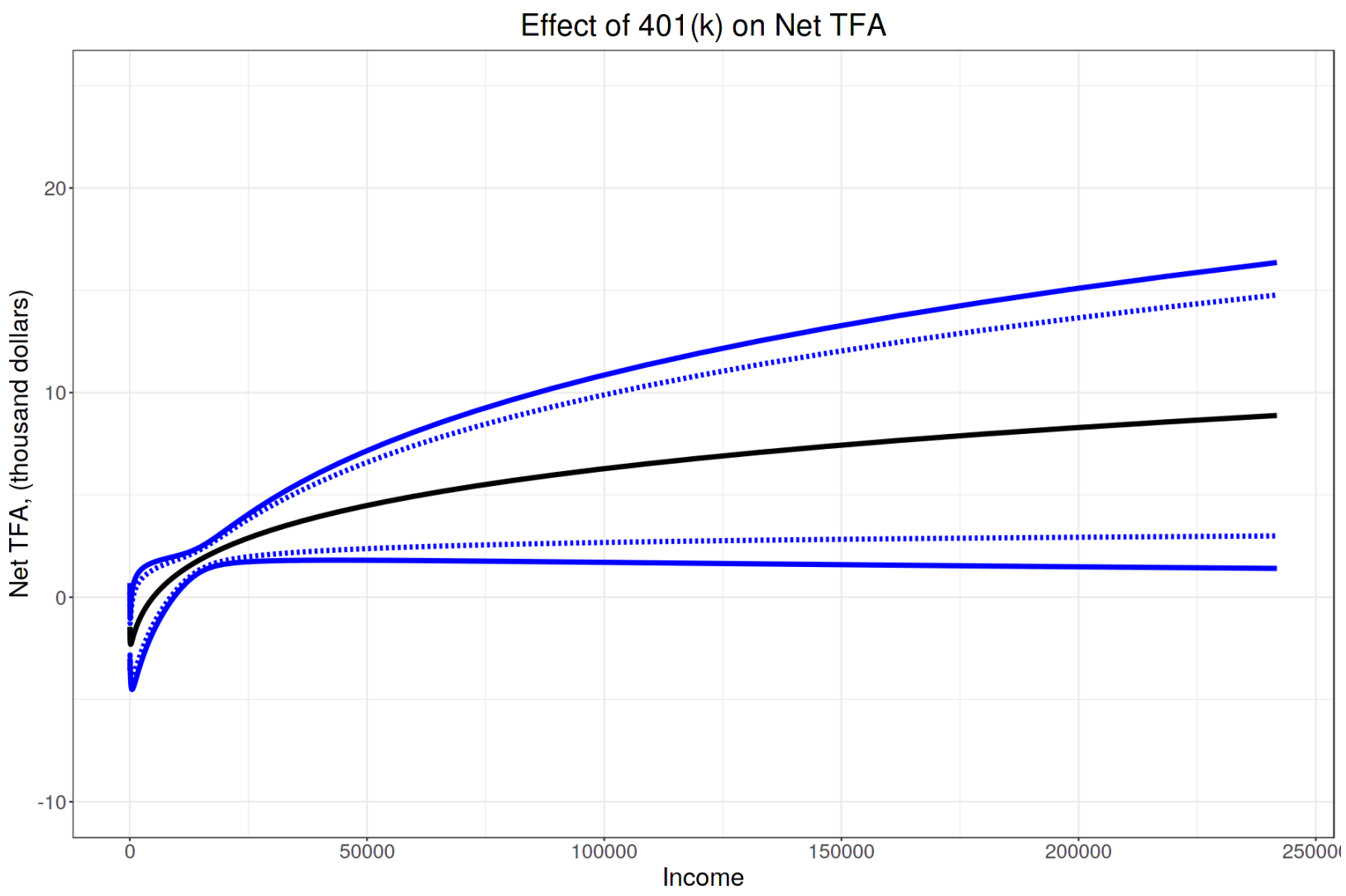

**Figure 14.2:** Inference on CATE of 401(k) Eligibility Conditional on Log-Income

## 14.3 Personalized Policies and Inference on Their Values

At the opening of this section we hinted that one reason why one might want to estimate a CATE model is so as to deploy a more personalized or contextual policy or to stratify and prioritize the treatment assignment, so as to maximize the outcome of interest. We formalize a personalized treatment policy $\pi$ as function that given any instance of the variable $X$ returns a probability $\pi(X) \in [0, 1]$ with which we to give treatment or not. Note that if the probability is 1 or 0 it is a deterministic assignment to treat or not treat. We are interested in conducting inference on the value of any policy $\pi$ and in particular the maximum such possible value.

Given any policy $\pi$, we define its value as its gain in the average outcome if we were to follow $\pi$'s treatment recommendation for everyone in the population compared to treating no one:

$$\begin{aligned} V(\pi) &:= \mathrm{E}[\pi(X)\,Y(1) + (1 - \pi(X))\,Y(0)] - \mathrm{E}[Y(0)] \\ &= \mathrm{E}[\pi(X)\,(Y(1) - Y(0))] \\ &= \mathrm{E}[\pi(X)\,\tau_0(X)]. \end{aligned} \tag{14.3.1}$$

Since we have already seen that the CATE $\tau_0(X)$ can be identified as $\mathrm{E}[Y(\eta_0) \mid X]$, we derive that the policy gains of any candidate policy can be identified as:

$$V(\pi) := \mathrm{E}\left[\pi(X)\,\mathrm{E}[Y(\eta_0) \mid X]\right] = \mathrm{E}[\pi(X)Y(\eta_0)].$$

For a given fixed $\pi$, this quantity is akin to the ATE considered in Section 10.3 weighted by a known function $\pi(X)$. In particular, when $\pi(X) = 1$ treats everyone, $V(\pi)$ is the ATE. The policy value is also akin to the GATE considered in the same section: if we scale it up by $1/\mathrm{E}[\pi(X)]$ then it is the GATE among those treated by $\pi$. Correspondingly, we can conduct inference on it by following the same recipe as in Section 10.4. This corresponds to the estimate

$$\widehat{V}(\pi) = \frac{1}{n}\sum_{i=1}^{n} \pi(X_i)Y(\hat{\eta}_{[k(i)]}),$$

and the statement of Theorem 10.3.1 still applies to this weighted ATE, providing for inference on $V(\pi)$.

One measure of the heterogeneity of treatment effects is how much we can deviate from the average effect by carefully tailoring who gets assigned which treatment, that is, how large we can make $V(\pi)$:

$$V^* = \max_{\pi} V(\pi) = \mathrm{E}[\max_{p\in[0,1]} p\,\tau_0(X)] = V(\mathbb{1}\{\tau_0(X) \geq 0\}).$$

The last equality shows that $\pi^*(X) = \mathbb{1}\{\tau_0(X) \geq 0\}$ is an optimal policy (there may be multiple if $\mathrm{P}(\tau_0(X) = 0) > 0$). This suggests we can estimate $V^*$ by following the same recipe as before and treating $\tau_0$ as one more nuisance to estimate and plug into the policy we are evaluating. This, in fact, works well whenever $\pi^*$ is uniquely optimal because then first order conditions for the optimization problem defining $V^*$ will automatically ensure a zero derivative in $\tau_0$, i.e., Neyman orthogonality as in Section 10.4. Namely, write $V^* = M(\tau_0, \eta_0) = \max_\tau M(\tau, \eta_0)$, where $M(\tau, \eta) = \mathrm{E}[\mathbb{1}\{\tau(X) \geq 0\}Y(\eta)]$. We already know that $\partial_\eta M(\tau_0, \eta_0) = 0$ from the case of evaluating any given policy, seen as a weighted ATE. For the derivative in $\tau$ we have

$$\begin{aligned}
&\frac{1}{t}|M(\tau_0 + t\xi, \eta_0) - M(\tau_0, \eta_0)| \\
&\quad = \frac{1}{t}\mathrm{E}[\tau_0(X)(\mathbb{1}\{-t\xi(X) \leq \tau_0(X) < 0 \ \vee \ 0 \leq \tau_0(X) < -t\xi(X)\})] \\
&\quad \leq \mathrm{E}[|\xi(X)|\mathbb{1}\{|\tau_0(X)| \leq t|\xi(X)|\}] \\
&\quad \leq \|\xi\|_2\sqrt{\mathrm{P}(|\tau_0(X)| \leq t|\xi(X)|)},
\end{aligned}$$

which has limit 0 as $t \to 0$ by the continuity of probability and $\mathrm{P}(\tau_0(X) = 0) = 0$ (since otherwise $\mathbb{1}\{\tau_0(X) > 0\}$ would not be almost surely the same as $\pi^*(X)$, contradicting its uniqueness). Thus, it suffices we learn $\tau_0$ at non-parametric rates and plug it into a guess for $\pi^*$, which we then evaluate the same as any fixed policy.[1]

1: Just uniqueness of $\pi^*$ may not be enough satisfy the additional regularity assumptions of Theorem 10.4.1 beyond Neyman orthogonality. We may need to assume not only that $\tau_0(X)$ has no atom at 0 but that it in fact has a bounded density in a neighborhood of 0. Additionally, the norm in which estimates of $\tau_0$ converge is important and interacts with the allowable rate.

While $V^*$ provides insight into how much we can get out of a treatment and covariates $X$ if we are careful about personalizing, it need not give a full picture of the heterogeneity of treatment effects across $X$. For example, it may well be 0 if treatment effects are all negative, even if they are very heterogeneous. Another lens into heterogeneity may be the value of the optimal policy, when constrained to treat *exactly* $q$-fraction of the population:

$$\begin{aligned} V_q^* &= \max_{\pi:\mathrm{E}[\pi(X)]=q} \mathrm{E}[\pi(X)\tau_0(X)] \\ &= \max_{\pi} \min_{\lambda} \mathrm{E}[\pi(X)\tau_0(X) + \lambda(q - \pi(X))] \\ &= \min_{\lambda} \max_{\pi} \mathrm{E}[\pi(X)\tau_0(X) + \lambda(q - \pi(X))] \\ &= \min_{\lambda} q\lambda + \mathrm{E}[0 \vee (\tau_0(X) - \lambda)]. \end{aligned} \tag{14.3.2}$$

When, on the other hand, $\pi^*$ is not at all unique, inference on $V^*$ can be tricky. A solution to this challenging problem is given in [3].

Exchanging the order of max and min in the penultimate line of Eq. (14.3.2) is justified by a result known as Sion's minimax theorem.

We recognize the minimizer of the check loss in Eq. (14.3.2) as the quantile. So the latter minimization is attained at $\lambda$ equal to the $(1 - q)$-th quantile $\mu(\tau_0, q) = \inf\{t : \mathrm{P}(\tau_0(X) > t) \le q\}$. Thus, any optimal constrained policy has $\pi_q^*(X) = 1$ when $\tau_0(X) > \mu(\tau_0, q)$ and $\pi_q^*(X) = 0$ when $\tau_0(X) < \mu(\tau_0, q)$. The quantity $V_q^*/q$ is exactly the average treatment effect among the $q$-fraction of a subpopulation with largest values of $\tau_0(X)$, also known as the superquantile or the conditional value at risk. When $\mathrm{P}(\tau_0(X) = \mu(\tau_0, q)) > 0$ this subpopulation may not be unique and there can be different ways of splitting the group with $\tau_0(X) = \mu(\tau_0, q)$ to obtain a subpopulation of fraction exactly $q$ (assuming either an infinite population or a finite population of infinitely divisible units). Notice the quantity $V_q^*$ is still well-defined even in this non-unique case, and that as we vary $q$, we obtain a full characterization of the distribution of $\tau_0(X)$.

Now, suppose $\mathrm{P}(\tau_0(X) = \mu(\tau_0, q)) = 0$. Then, the constrained optimal policy is uniquely given by $\pi_q^*(X) = \mathbb{1}\{\tau_0(X) \ge \mu(\tau_0, q)\}$ and we have that $V_q^*/q = \mathrm{E}[Y(1) - Y(0) \mid \tau_0(X) \ge \mu(\tau_0, q)]$ is exactly the GATE among those with CATE above the $(1 - q)$-th quantile. Moreover, for $q' > q$ with $\mathrm{P}(\tau_0(X) = \mu(\tau_0, q)) = \mathrm{P}(\tau_0(X) = \mu(\tau_0, q')) = 0$, we have that $(V_{q'}^* - V_q^*)/(q' - q)$ is the GATE among those with CATES between the $(1 - q')$-th and $(1 - q)$-th quantiles.

In the latter unique case, we may be tempted to follow the same recipe as before to also estimate $V_q^*$: plug in estimates of $\tau_0, \mu(\tau_0, q)$ to form a guess of $\pi_q^*$ and evaluate it as we would any fixed policy. This worked for $V^*$ because $\pi^*$ maximized value, automatically inducing Neyman orthogonality. However, $\pi_q^*$ does not globally maximize value, only subject to constraints. In other words, we do have $V^* = M(\tau_0 - \mu(\tau_0, q), \eta_0)$, but generally $\partial M(\tau_0 - \mu(\tau_0, q), \eta_0) \neq 0$.

To recover the same orthogonality-via-optimality as before we need to introduce a cost of violating the constraint. This is exactly what Eq. (14.3.2) does. Namely, writing $V_q^* = M_q(\tau_0, \mu(\tau_0, q), \eta_0) = \min_\lambda \max_\tau M_q(\tau, \lambda, \eta_0)$, where $M_q(\tau, \lambda, \eta) = q\lambda + \mathrm{E}[\mathbb{1}\{\tau(X) \geq \lambda\}(Y(\eta) - \lambda)]$, we will again find that $\partial_\tau M_q(\tau_0, \mu(\tau_0, q), \eta_0) = 0$ and $\partial_\lambda M_q(\tau_0, \mu(\tau_0, q), \eta_0) = 0$ because $\tau_0, \mu(\tau_0, q)$ by first order conditions, and $\partial_\eta M_q(\tau_0, \mu(\tau_0, q), \eta_0) = 0$ as always for weighted ATE estimation. This provides a recipe for estimating $V_q^*$. The minimax formulation also bestows a special property that if we get $\tau_0$ wrong (or, estimate it too slowly), we will still get a lower bound on $V_q^*$ as long as we estimate a corresponding best-response $\lambda$ well.[2]

2: Complete details on how to do inference on $V_q^*$ and guarantees thereon are given in [4].

## 14.4 Non-Parametric Inference for CATEs with Causal Forests

An inherently harder task is performing inference on the value $\tau_0(x)$ at a given point $x$. Statistically this problem can be even harder than performing inference on the value of a regression function at a particular point $x$. In fact, one solution casts the problem as such. Note that we already argued that:

$$\tau_0(x) = \mathrm{E}\left[Y(\eta_0) \mid X = x\right] \tag{14.4.1}$$

Thus one approach is to estimate the nuisance parameters $\eta_0$ in a cross-fitting manner and then use any flexible regression method that supports prediction intervals and apply it to the regression problem $Y(\hat{\eta}) \sim X$. In low dimensions, many classical approaches, such as kernel regression are applicable and can be invoked. In high-dimensions these methods will struggle to provide any meaningful insight.

An alternative is to use Random Forest based methods that will perform much better in practice. Standard Random Forest approaches typically equally balance bias and variance and hence do not allow for confidence interval construction. Recent work of [5, 6] proposes adaptations of Random Forests that, in low-dimensions, provably produce asymptotically normal and un-biased predictions and provide theoretically justified construction of confidence intervals. The key ingredients in these adaptations are:

i *honesty*: a separate sample is used to construct the structure of the tree and a separate sample is used to calculate the estimates at the leaf nodes of the tree, ii *balancedness*: every split should leave at least a $\rho \geq 0.2$ fraction of the samples on each side, iii *random feature split*: every feature should have a probability of at least $\pi/d$ to be chosen on each split, where $d$ is the number of features (e.g. this can be achieved by choosing a random feature to split on with probability $\pi$), iv *fully grown*: the tree should be grown fully, such that the number of samples that fall in every leaf should be at most some small constant, ii *sub-sampling*: unlike typical random forest methods that use bootstrap sub-samples to build each tree (i.e. of the same size as the original samples and drawn with replacement), these adapted forests use sub-samples without replacement and of a smaller size $s \ll n$ than the original sample on each tree (the size of the sub-sample needs to be chosen carefully for the validity of the confidence intervals and should be of the order of $n^{\frac{\alpha d}{\alpha d+1}}$, where $\alpha = \frac{\log(1/\rho)}{\pi \log(1/(1-\rho))}$).

We will refer to any forest construction process that satisfies these properties as an *Honest Random Forest*.

We will refer to Honest Random Regression Forests that are trained on the doubly robust proxy labels $Y(\hat{\eta})$, with cross-fitted estimates of the nuisance functions, as *Doubly Robust Forests*. Based on the results in [6, 7], one can show that the validity of the confidence intervals of Honest Random Regression Forests is maintained even when the labels are biased due to the estimation error of the nuisance parameters $\hat{\eta}$, so long as:

$$\sqrt{n/s}\mathrm{E}[(H(\hat{\mu} - H(\mu_0))\,(\hat{g}(D,Z) - g_0(D,Z)) \mid X = x] \approx 0$$

Note that this requires accuracy of the nuisance estimates with respect to a confidional mean squared error. [7] shows how this can be achieved even in settings where $Z$ is high-dimensional, albeit $X$ remains low-dimensional. In particular, one should expect the size of the confidence interval or the error of the estimate to decay to zero at a rate of $\approx n^{-\frac{1}{2(\alpha d+1)}}$, where $d$ is the dimension of the covariates in $X$. This result is based on a conditional variant of the Neyman orthogonality property that is satisfied by the doubly robust proxy labels, in conjunction with the asymptotic normality of predictions that stem from Honest Random Forests.

Finally, under stronger assumptions and for binary covariates, the recent work of [8] also shows that confidence intervals of Honest Regression Random Forests trained with the squared loss criterion and *without the balancedness or random feature split* property, are asympotitcally valid even in high-dimensions, as long as the true regression function $\mathrm{E}[Y \mid X = x]$ is sparse (i.e. only a small constant number of variables are relevant for predicting $Y$). However, an upper bound on the degree of sparsity (i.e. the number of relevant features) needs to be known. Extending this inference result to high-dimensional continuous features remains and active area of investigation. Similarly, extending this result to simultaneous confidence bands and not just pointwise confidence intervals is another active area of investigation.

While adaptive estimation of the CATE can be obtained fairly generally, it is important to note that $X$ should be low dimensional if we want to obtain confidence intervals or perform hypothesis tests. Genovese and Wasserman (Annals of Stats, 2008) [9] show that there do not exist adaptive confidence bands for estimation of the curve $\tau_0(X)$ except under very restrictive assumptions more generally. They suggest instead to construct adaptive bands that cover a surrogate function $\pi$ which is close to, but simpler than, $\tau_0$.

In the previous section, where we discuss the use of OLS with low-dimensional $X$, the surrogate $\pi$ represents either GATEs or the best linear approximation of the CATE. Inferential guarantees are also available for the case where $X$ is low-dimensional and Random Forests are used. Inferential results for low-dimensional surrogates $\pi$ based on other methods should also be possible, though we note that GATEs and best linear predictors more generally are readily interpretable and will likely be useful in many settings.

Despite these theoretical limitations, forest based approaches are empirically powerful as they tend to identify the most relevant factors that drive treatment effect heterogeneity, while at the same time providing some signal of uncertainty of the prediction. Even though this uncertainty quantification is more brittle than for instance the confidence intervals of an OLS regression, as it depends on many more assumptions and holds only under particular choices of the hyperparameters of the method (which are typically violated in practice; especially when data-driven hyperparameter tuning is invoked via cross-validation, which is the typical case), honest random forest based approaches still provide a meaningful signal of how uncertain the model is about its CATE predictions, at different regions of the covariate space.

**Generalized Random Forests (GRF)** An alternative approach is to formulate the CATE problem as solving a local or conditional version of a moment restriction [6, 7]:

$$E[m(W; \tau_0(x), \eta_0) \mid X = x] = 0 \tag{14.4.2}$$

where $m(W; \tau, \eta)$ is a vector of moment restrictions of the same dimension as $\tau$.

Such Generalized Random Forests are trained so as to maximize the induced heterogeneity in $\hat{\tau}(x)$ with every split. For every node $P$ in some tree of the forest, let $\hat{\tau}_P$ denote our estimate of $E[m(W; \tau_P, \eta_0) \mid X \in P] = 0$. Such an estimate can be constructed by solving the moment restriction with respect to $\tau_P$ using only the samples that fall in node $P$ and using an estimate $\hat{\eta}$ of $\eta_0$ based on auxiliary data or in a cross-fitting manner. Let $C_1, C_2$ denote the child nodes that will be created from some candidate split, with sample sizes $n_1$ and $n_2$ correspondingly. Then a proxy criterion that targets maximizing heterogeneity is maximizing $n_1\,\tau_{C_1}^2 + n_2\,\tau_{C_2}^2$. This is one of the criteria typically used in Generalized Random Forests. Moreover, to avoid the computational burden of resolving the moment equation for every candidate split, typically an approximation of the quantities $\tau_{C_1}$ and $\tau_{C_2}$ is used. In particular, a local linear approximation around the estimate of the parent node is being used and locally updated, i.e. $\tau_{C_1} \approx \tau_P - \frac{1}{n_1} \sum_{i \in C_1} J_P m(W_i; \tau_P, \hat{\eta})$, with $J_P = \frac{1}{n_P} \sum_{i \in P} \partial_\tau m(W; \tau_P, \hat{\eta})$, with $n_P$ being the number of samples in the parent node.

Moreover, the final estimate $\hat{\tau}(x)$ is derived in a manner slightly different than regression forests (albeit it coincides for the case of a regression moment, i.e. $m(W; \tau(x)) = y - \tau(x)$). For more general moments, for every target point $x$ for which we want to predict the CATE, the Random Forest structure is used to construct weights for every other sample $i \in \{1, \ldots, n\}$, that capture the degree of "similarity" of $x$ to $X_i$. These weights roughly correspond to the fraction of trees in the forest, for which $X_i$ falls in the same leaf node as $x$, downweighting leafs of larger size. Thus if we have trained a forest with $B$ trees and we let $L_b(x)$ denote the leaf node that a sample with covariates $x$ falls in at tree $b$ and let $|L_b(x)|$ the number of samples in that

leaf, then we have:

$$K(x, X_i) = \frac{1}{B}\sum_{b=1}^{B} \frac{1\{L_b(x) = L_b(X_i)\}}{|L_b(x)|}$$

Then to calculate $\hat{\tau}(x)$, we solve with respect to $\hat{\tau}(x)$, a weighted empirical average version of the moment condition:

$$\sum_{i=1}^{n} K(x, X_i)\, m(W_i; \hat{\tau}(x), \hat{\eta}) = 0 \qquad (14.4.3)$$

using the weights that are derived based on the similarity metric induced by the forest structure.

When the forest construction process satisfies the criteria we defined earlier in the section of *honesty*, *balancedness*, *random feature splitting*, *fully grown trees* and *sub-sampling without replacement*, then under similar conditions as in the case of Regression Forests, the prediction of a Generalized Random Forest (and its extension, the Orthogonal Random Forest) can be shown to be asymptotically normal and an asymptotically valid confidence interval construction can be employed. Albeit the same limitations as we described in the regression case, carry over to the confidence intervals produced by these methods.

**Causal Forests: a GRF for CATE** We describe here an empirically popular variant of causal forests that uses the Generalized Random Forest formulation. Albeit, unlike the Doubly Robust Forest approach, this approach is valid only if $X = Z$ or if we make the stronger further assumption that the high-dimensional CATE function $\delta_0(Z) = \mathrm{E}[Y(1) - Y(0) \mid Z]$, is only a function of the variables $X$, i.e. $\delta_0(Z) = \tau_0(X)$ and $\tau_0$.

In this case, for a binary treatment, we can write without loss of generality

$$\mathrm{E}[Y \mid D, Z] = \pi_0(Z)\, D + g_0(Z)$$

where $\pi_0(Z) = \mathrm{E}[Y \mid D = 1, Z] - \mathrm{E}[Y \mid D = 0, Z]$ is the conditional average predictive effect (CAPE). Moreover, by conditional exogeneity, the CAPE function $\pi_0$ is equal to the high-dimensional CATE function $\delta_0$. Thus, for a binary treatment we can always write the regression equation:

$$Y = \delta_0(Z)\, D + g_0(Z) + \epsilon, \qquad \mathrm{E}[\epsilon \mid D, Z] = 0$$

From this, we can derive that $\mathrm{E}[Y \mid Z] = \delta_0(Z)\mathrm{E}[D \mid Z] + g_0(Z)$.

Subsequently, we can write:

$$Y - \mathrm{E}[Y \mid Z] = \delta_0(X)\,(D - \mathrm{E}[D \mid Z]) + \epsilon$$

Letting $\hat{Y} = Y - \mathrm{E}[Y \mid Z]$ and $\hat{D} = D - \mathrm{E}[D \mid Z]$ and since $Z, \hat{D}$ is uniquely determined by $D, Z$, we can conclude that the following regression equation holds:

$$\hat{Y} = \delta_0(Z)\,\hat{D} + \epsilon, \qquad \mathrm{E}[\epsilon \mid \hat{D}, X] = 0$$

If we now further assume that $\delta_0(Z) = \tau_0(X)$, and since $X, \hat{D}$ is a subset of $Z, \hat{D}$, we can write the regression equation:

$$\hat{Y} = \tau_0(X)\,\hat{D} + \epsilon, \qquad \mathrm{E}[\epsilon \mid \hat{D}, X] = 0 \tag{14.4.4}$$

From this regression equation we can derive the moment constraint:

$$\mathrm{E}\left[(\hat{Y} - \tau_0(x)\,\hat{D})\,\hat{D} \mid X = x\right] = \mathrm{E}\left[\epsilon\,\hat{D} \mid X = x\right] = 0$$

Note that this moment equation is a conditional analogue of the Normal Equation that we used in the PLR model, where we used the equation

$$\mathrm{E}(\hat{Y} - \theta_0\,\hat{D})\,\hat{D} = 0$$

to estimate the constant treatment effect under a partially linear model $\mathrm{E}(Y \mid D, X) = \theta_0 D + g(Z)$. Now that the coefficient associated with $D$ is allowed to vary with $X$, we can estimate the heterogeneous coefficient by solving the same moment but conditional on $X$, i.e.

$$\mathrm{E}\left[(\hat{Y} - \tau_0(x)\,\hat{D})\,\hat{D} \mid X = x\right] = 0 \tag{14.4.5}$$

Note that the above method falls in the general framework that can be handled by Generalized Random Forests and their extension, the Orthogonal Random Forests. We can estimate $\hat{\tau}(x)$ by estimating the nuisance function $\eta_0 = (p_0, q_0)$, where $p_0(Z) = \mathrm{E}(D \mid Z)$ and $q_0(Z) = \mathrm{E}(Y \mid Z)$ in a cross-fitting manner, letting $\check{Y} = Y - \hat{p}(Z)$, $\check{D} = D - \hat{q}(Z)$ and then applying the Generalized Random Forest method with moment equation:

$$m(W; \tau(x), \hat{\eta}) = (\check{Y} - \tau(x)\,\check{D})\,\check{D}$$

The formal analysis of the validity of the confidence intervals of this approach can be found in [6] for the case when $X = Z$ and is low-dimensional, in which case, one does not need to account for the errors in $\hat{\eta}$, as long as a constant offset is also added to the moment equation, solving the vector of moments:

$$m(W;\tau(x),\beta(x),\hat{\eta}) = (\check{Y} - \tau(x)\,\check{D} - \beta(x)) \begin{pmatrix} \check{D} \\ 1 \end{pmatrix}$$

A formal analysis of the case when $X \subseteq Z$ and $Z$ can be high-dimensional (or when the constant term is not added to the moment equation) can be found in [7], which also accounts for the impact of the nuisance estimation errors.

**Example 14.4.1** (Forests in the 401k Example) We revisit the 401(k) example that we used in the previous section and apply the forest based methods for CATE estimation. In this case, we used all the variables for heterogeneity (i.e. $X = Z$) and let the forest methods identify the relevant dimensions of heterogeneity in a more data-driven manner. We applied both the Doubly Robut Forest and the Causal Forest approach. For the nuisance estimates, in all cases, we used gradient boosted forests and estimated the nuisances in a cross-fitting manner with 5-fold cross-fitting.

In Figure 14.3 we depict the predictions and confidence intervals of the Doubly Robust Forest method, where the x-axis corresponds to income (while other co-variates are fixed to their overall median values). In Figure 14.4 we depict the analogous plot for the Causal Forest method. We find that both methods identify a similar CATE and that this CATE is inline with the intuitive property that the effect of 401(k) eligibility on net financial assets is larger for larger incomes. Moreover, unlike the BLP estimates, we find that the forest based estimates, behave more reasonably at the extreme ends of the income distribution as they do not extrapolate linearly and identify, in a data-driven manner, a more sigmoid effect curve, between $\approx$ \$5$k$ and $\approx$ \$22$k$. The results are almost identical for the two methods. Moreover, the confidence intervals are informative that the CATE prediction is quite uncertain at the upper extreme part of the income distribution were samples are much more spread out and there is a long tail. Finally, when looking at measures of feature importance for Random Forests, income was identified as the most important feature.

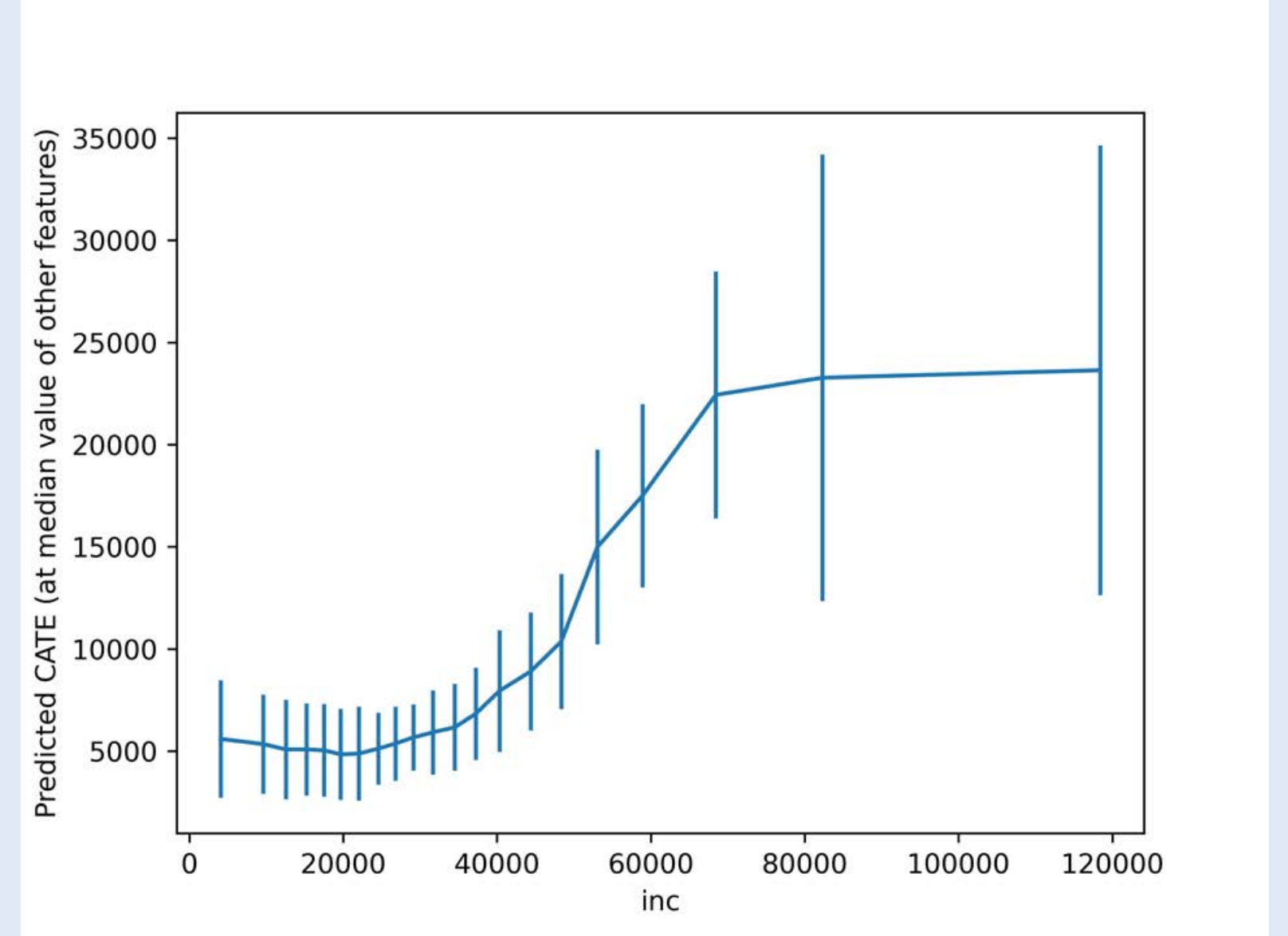

**Figure 14.3:** Doubly Robust Forest in the 401k example.

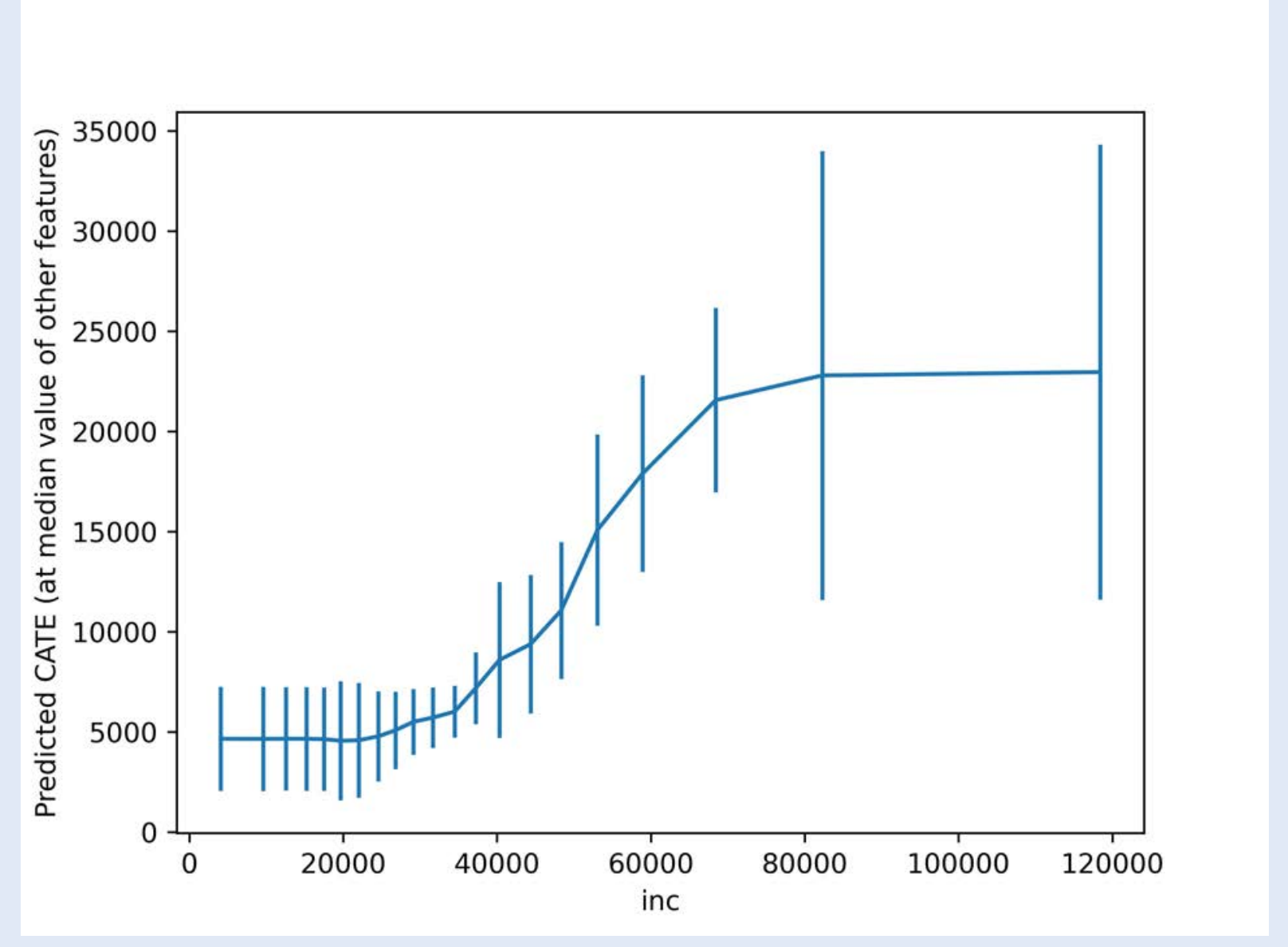

**Figure 14.4:** Causal Forest in the 401k example.

## Empirical Example: The "Welfare" Experiment

We illustrate the estimation of CATEs with forests, with an empirical application on studying the effects of the word “welfare” on the support of people for government programs. Starting in the 1980s, the General Social Survey (GSS) started including a question around satisfaction with public spending. What is more important, the GSS conducted a randomized controlled trial where the respondents where assigned one of two variations of the same question at random. Both variations had the same meaning and introduction, albeit one was asking about satisfaction of the respondent with respect to government spending for “welfare programs”, while the other variation was phrased as government spending for "assistance to the poor". This small variation has been found to have substantial average effect on the response and several studies have attempted to parse out treatment effect heterogeneity.

In this section we applied Causal Forests and Doubly Robust Forests on a dataset collected from such GSS surveys from 1986 to 2010, as described in [10]. The dataset consists of 12907 samples containing i) the variant of the question that was assigned to the participant (with $D = 1$ corresponding to "welfare programs" and $D = 0$ corresponding to "assistance to the poor), ii) their numerical level of satisfaction response to the question ($Y$) and 42 features ($X$) that contain many characteristics of the respondent related to gender, income, education, family size and marital status, race, political views and occupation sector. The average treatment effect based on a simple two-means estimate is $-0.3681$ as reported in Figure 14.5.

| | **coef** | **std err** | **P> \|z\|** | **[0.025** | **0.975]** |
|---|---|---|---|---|---|
| **const** | 0.4798 | 0.006 | 0.000 | 0.467 | 0.492 |
| $D$ | -0.3681 | 0.007 | 0.000 | -0.383 | -0.354 |

**Figure 14.5:** Average treatment effect in welfare experiment.

We constructed a Causal Forest and a Doubly Robust Forest using all the 42 variables for treatment effect heterogeneity and as controls. We used gradient boosting regression with cross-fitting to calculate the nuisance functions required for each of the forests. The hyperparameters of the nuisance estimators were selected based on cross validation. Subsequently, we looked at the most important feature in the forest, as measured by a feature importance criterion that roughly corresponds to the average reduction in the splitting criterion, every time that the feature was used for splitting. The most important feature came out to be political views, both in the Causal Forest and in the Doubly Robust Forest. Subsequently, in Figure 14.6 and Figure 14.7 we report the heterogeneous effect for each value of the polviews covariate and imputing all other covariates at their median value. The point estimate and the corresponding 5%-95% confidence intervals that are provided by the forest methods are depicted.

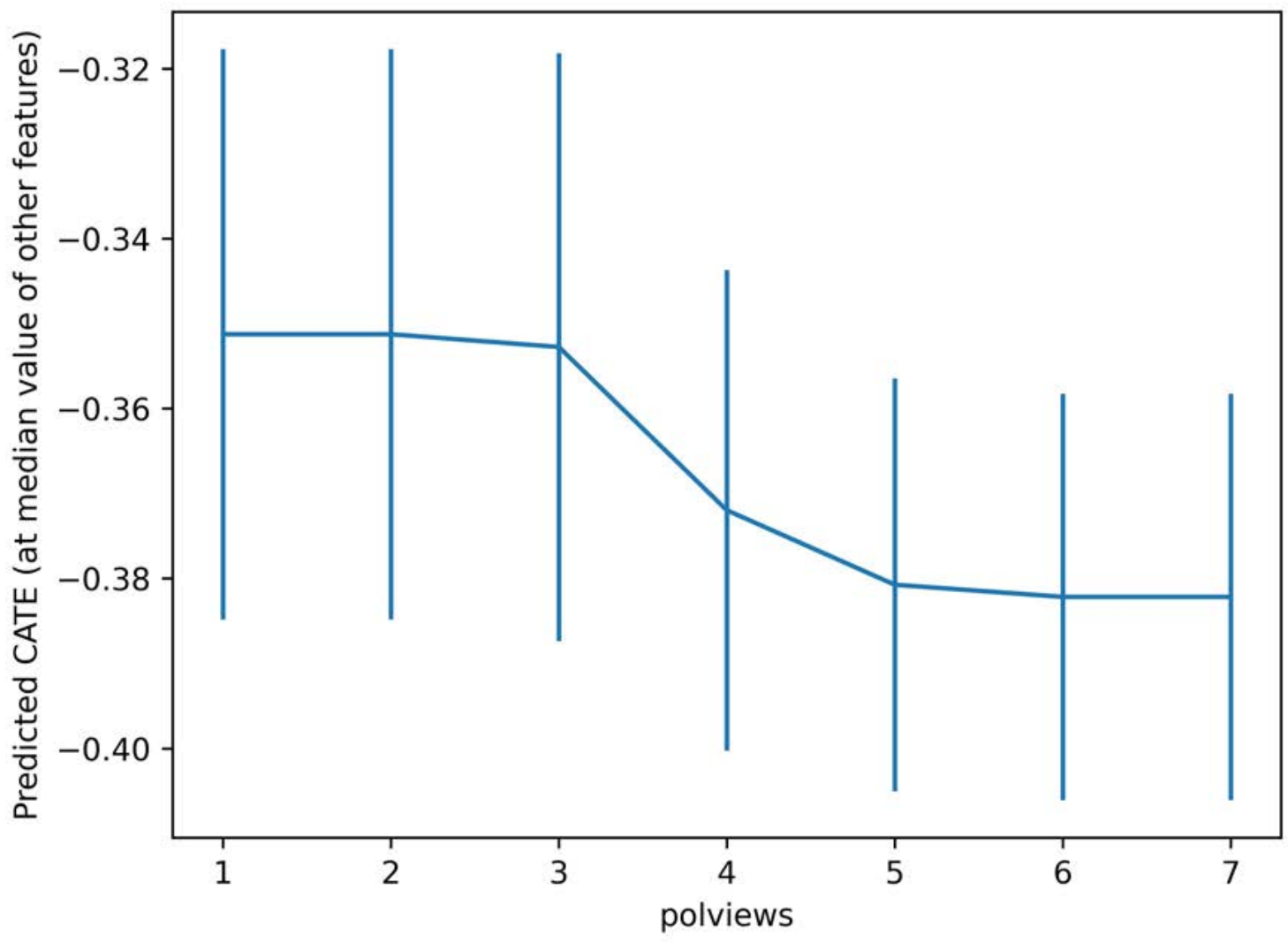

**Figure 14.6:** R-Learner based causal forest in the welfare example.

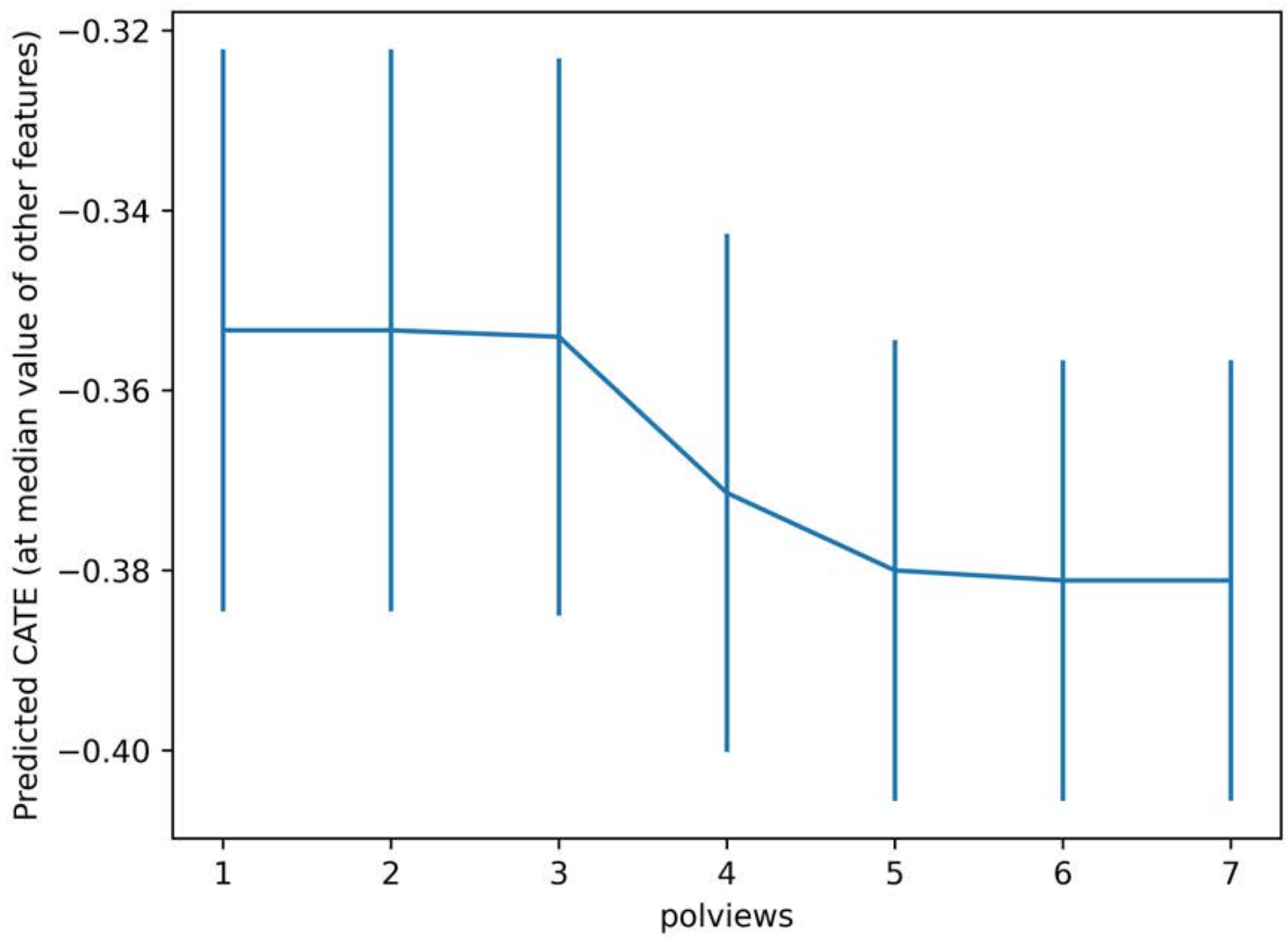

**Figure 14.7:** Doubly Robust honest regression forest in the welfare example.

# Notebooks

- R Notebook for DML on CATE analyzes ATE of 401(K) conditional on income.
- Python Notebook for CATE Inference analyzes CATE of welfare experiment and for 401k experiment with Best Linear Predictors of CATE and with Random Forest and Causal Forest based methods.

# Estimation and Validation of Heterogeneous Treatment Effects

# 15

"You can, for example, never foretell what any one man will do, but you can say with precision what an average number will be up to."

– Sherlock Holmes [1].

We study flexible estimation of heterogeneous treatment effects. We target the construction of an estimate of the true CATE function and not its projection on a simpler function space, with as small root-mean-squared-error as possible. We consider flexible estimation using generic ML techniques and discuss how one can perform model selection and out-of-sample validation of the quality of the learned model of heterogeneity. We conclude with the topic of policy learning, i.e. constructing optimal personalized policies.

## 15.1 ML Methods for CATE Estimation

We consider the same setting as in Chapter 14 of analyzing the heterogeneous effect of a binary treatment in the presence of a high-dimensional set of observed controls $Z$, under conditional exogeneity. In this section, we target the construction of an estimate $\hat{\tau}(X)$ of the true CATE function $\tau_0(X)$ and not its best linear approximation, using generic ML techniques, in a manner such that the mean squared error $\mathrm{E}_X(\tau_0(X) - \hat{\tau}(X))^2$, which is also what we used in Chapter 10 to measure the quality of non-linear predictive ML models, is minimized. We will also be interested in the mean squared error of the estimate with respect to the best approximation of the CATE over some flexible, potentially non-linear function space $T$, i.e. the function $\tau_*$ defined as

$$\tau_* = \arg\min_{\tau \in T} \mathrm{E}_X(\tau_0(X) - \tau(X))^2. \quad (15.1.1)$$

As in the previous section, the key is to decompose the estimation of the CATE into a sequence of regression problems. Then generic ML techniques can be used to address each of these regression problems. This approach has been coined *meta-learning* in the literature on CATE estimation, since we are trying to treat ML techniques as a black-box oracle that solves any regression problem and we are trying to build on top of that oracle to learn the CATE. Motivated by the ability to construct confidence intervals, in the previous section, we provided one such choice of a reduction, as we will explain later. However, when one is primarily interested in mean squared error, other decompositions could potentially have better finite sample performance. We present here the multitude of such meta-learning approaches that have been proposed in the literature and we will conclude with a comparative analysis of each of them.

### Meta-Learning Strategies for CATE Estimation

To simplify the exposition, and emphasizing the meta-learning aspect of these methods, we will introduce a notation for a regression estimate oracle. We denote with $O_H(\{X_i, Y_i, W_i\}_{i=1}^n)$ an oracle algorithm that takes as input a dataset of $n$ i.i.d. samples, consisting of covariatex $X_i$, regression labels $Y_i$ and sample weights $W_i$ (where weights are assumed to be independent of $Y_i$ given $X_i$) and produces an estimate $\hat{h}$ of the function that minimizes the sample-weighted square loss:

$$\begin{aligned} h_0 &= \arg\min_{h \in H} \mathrm{E} W (Y - h(X))^2 \\ &= \arg\min_{h \in H} \mathrm{E} W (\mathrm{E}(Y \mid X) - h(X))^2 \end{aligned} \tag{15.1.2}$$

over some function space $H$. When sample weights $W_i$ are omitted, they will be assumed to be equal to 1. This oracle will typically correspond to some ML approach to solving this weighted regression problem and we will be assuming that such an oracle provides an estimate $\hat{h}$ that converges to $h_0$ at some rate, with respect to the mean-squared-error metric, i.e.

$$\|\hat{h} - h_0\|_{L^2(X)} = r_n \to 0.$$

**Single (S)-Learner** Starting from the very simple identification formula for the CATE in Equation (14.1.1), we can learn the CATE by first invoking an ML regression method to construct an estimate $\hat{g}$ of the conditional expectation function $g_0(D, Z) := \mathrm{E}[Y \mid D, Z]$, assuming that $g_0$ lies in some function space $G$. Then we can construct a model $\hat{\tau}$ of the CATE by invoking an ML regression method to construct an estimate of the conditional expecation function $\mathrm{E}[\hat{g}(1, Z) - \hat{g}(0, Z) \mid X]$, over some function space $T$. Overall we arrive at the following meta-learning algorithm:

**Single Learner (S-Learner)**

$$\begin{aligned} \hat{g} &:= O_G(\{(D_i, Z_i), Y_i\}_{i=1}^n) \\ \hat{\tau} &:= O_T\left(\{X_i, \hat{g}(1, Z_i) - \hat{g}(0, Z_i)\}_{i=1}^n\right) \end{aligned} \tag{15.1.3}$$

Even if $T$ does not contain $\tau_0$, as long as $g_0 \in G$ and $\|\hat{g} - g_0\|_{L^2(D,Z)} \to 0$, the S-Learner estimate will be converging to the best approximation of the CATE within the space $T$, i.e. $\|\hat{\tau} - \tau_*\|_{L^2(X)} \to 0$.

**Two (T)-Learner** Estimating a single regression model that predicts the outcome $Y$ from the treatment $D$ and the controls $Z$, can overly regularize the treatment variable. Especially in settings where the treatment has a small effect, many ML algorithms will most probably shrink the treatment effect to zero and prioritize the inclusion of other informative covariates in the selected model. For this reason, it seems natural to weaken this regularization bias on the treatment. This can be achieved by fitting two separate models, one model $\hat{g}^T$ that estimates the relationship between the outcome $Y$ and the covariates $Z$ within the treated group, i.e. $g_0^T := \mathrm{E}[Y \mid Z, D = 1]$ and one model $\hat{g}^C$ that learns the same relationship within the control group, i.e. $g_0^C - := \mathrm{E}[Y \mid Z, D = 0]$. Then the CATE can be estimated by invoking an ML regression method to construct an estimate of the CEF $\mathrm{E}[\hat{g}^T(Z) - \hat{g}^C(Z) \mid X]$. Overall, we arrive at the following meta-learning algorithm:

**Two Learner (T-Learner)**

$$\begin{aligned} \hat{g}^C &:= O_{G^C}(\{Z_i, Y_i\}_{i\in\{1,\ldots,n\}:D_i=0}) \\ \hat{g}^T &:= O_{G^T}(\{Z_i, Y_i\}_{i\in\{1,\ldots,n\}:D_i=1}) \\ \hat{\tau} &:= O_T(\{X_i, \hat{g}^T(Z_i) - \hat{g}^C(Z_i)\}_{i=1}^n) \end{aligned} \tag{15.1.4}$$

Similar to the S-Learner, as long as $g_0^C \in G^C$ and $g_0^T \in G^T$, then the result of the T-Learner will always be converging to $\tau_*$, i.e. the best approximation of the CATE within $T$.

**Doubly Robust (DR)-Learner** The above approaches rely fully on accurate outcome modelling. If we face settings where the conditional counterfactual outcomes $\mathrm{E}[Y \mid D = 1, Z]$ are complicated functions that are hard to model and estimate, but the CATE function $\tau(X)$ is relatively simple, the aforementioned two meta-learners will suffer from large estimation errors in $\hat{g}, \hat{g}^T, \hat{g}^C$. If for instance, we are in a randomized controlled trial and we know the propensity $\mu_0$, then we also know that the random variable $YH(\mu_0)$ satisfies

$$\begin{aligned} \mathrm{E}[YH(\mu_0) \mid X] &= \mathrm{E}[\mathrm{E}[YH(\mu_0) \mid Z] \mid X] \\ &= \mathrm{E}[\mathrm{E}[Y \mid D = 1, Z] - \mathrm{E}[Y \mid D = 0, Z] \mid X] \\ &= \tau(X). \end{aligned}$$

Thus when solving this regression problem we only need to be accurately approximating the potentially simpler CATE function, as opposed to the response functions under treatment or control.[1]

1: An estimation strategy based on running a regression of $YH(\hat{\mu})$ on $X$, is referred to in the literature as the Inverse Propensity Score (IPS)-Learner, but we will omit more details on it.

Beyond randomized control trials, the above approach is too heavily dependent on constructing a good estimate $\hat{\mu}$ of the propensity score. Moreover, even for randomized control trials, the latter method can have very large variance, due to dividing the outcome $Y$ by the inverse propensity. For this reasons, it might be beneficial even when we care solely about mean squared error, to use the doubly robust approach, which combines propensity and regression modelling and can reduce both bias due to errors in estimating the propensity and variance by dividing only the un-explained variation in the outcome by the propensity. This leads to the doubly robust meta-learner (we will describe its two-learner variant, which is advisable in practice):

**Doubly Robust Learner (DR-Learner)**

$$
\begin{aligned}
\hat{g}^C &:= O_{G^C}(\{Z_i, Y_i\}_{i\in\{1,\dots,n\}:D_i=0}) \\
\hat{g}^T &:= O_{G^T}(\{Z_i, Y_i\}_{i\in\{1,\dots,n\}:D_i=1}) \\
\hat{\mu} &:= O_M(\{Z_i, D_i\}_{i\in\{1,\dots,n\}}) \\
\hat{g}(D, Z) &:= \hat{g}^T(Z)\, D + \hat{g}^C(Z)\,(1-D) \\
Y_i(\hat{\eta}) &:= H_i(\hat{\mu})\,(Y_i - \hat{g}(D_i, Z_i)) + \hat{g}^T(Z_i) - \hat{g}^C(Z_i) \\
\hat{\tau} &:= O_T\left(\{X_i, Y_i(\hat{\eta})\}_{i=1}^n\right)
\end{aligned}
\tag{15.1.5}
$$

The previous section can be seen as a special of this meta-learner, where we use OLS as our regression oracle in the final step and were we use cross-fitting when we estimate the first three steps and calculate the proxy labels $\hat{Y}_i$.

The DR-Learner inherits certain type of double robustness properties that we have, even when analyzing the mean squared error. In particular, suppose that we knew the true regression functions $g_0^C, g_0^T, \mu_0$ and based on some appropriate argument, we could show that the regression oracle in the last step, when ran with the ideal labels $Y(\eta_0)$, achieves a mean squared error rate of the order of $r_n^2$. Then, assuming $g_0^C \in G^C$, $g_0^T \in G^T$ and $\mu_0 \in M$, one can argue, under benign regularity conditions, that the mean squared error of the DR-Learner, can be upper bounded as:

$$
\mathrm{E}(\hat{\tau}(X) - \tau_*(X))^2 \lesssim r_n^2 + \mathrm{Error}(\hat{g})^2 \cdot \mathrm{Error}(H(\hat{\mu}))^2
$$

where the error of these nuisances can always be taken to be the fourth moment of the prediction error[2] and under further regularity conditions on the function space $T$ used in the estimation for $\tau$, it can be taken to be the root-mean-squared-error.[3] Thus as long as the product of the errors in modelling the regression and the propensity function are small, then the mean squared error for $\hat{\tau}$ will not be significantly impacted by these first stage estimation errors. For formal versions of variants of such results, we defer the reader to the following papers [2–6].

2: For any estimate $\hat{f}$ of a function $f_0$, that takes as input some random variable $W$, the fourth moment of the prediction error $\|\hat{f} - f_0\|_{L^4(X)}$ is defined as

$$
\left(\mathrm{E}_W(\hat{f}(W) - f_0(W))^4\right)^{1/4}
$$

and is a slightly strong measure of performance that the root-mean-squared-error, i.e.

$$
\sqrt{\mathrm{E}_W(\hat{f}(W) - f_0(W))^2}
$$

.

3: In fact, one can always use the slightly better error metric:

$$
\left(\mathrm{E}_X\left[\mathrm{E}_W[(\hat{f}(W) - \hat{f}(W))^2 \mid X]^2\right]\right)^{1/4}
$$

where $X$ is the set of variables that enter the CATE function $\tau$. When $X$ is the empty set, as in the case of average causal effects this boils down to the mean squared error and otherwise this requires better control, on average, of the conditional mean squared error of the nuisance functions, conditional on the variables $X$ that enter the CATE function.

**Residual (R)-Learner** If we know that the true CATE model lies in a simple function space and even if we knew the true nuisance parameters $\eta_0$, the labels that are used in the final stage of the DR-Learner can still have a large magnitude, due to the division by the propensity. In settings were the overlap assumption is almost violated at particular regions of the covariate space, the regression labels $Y_i(\eta_0)$ will be taking very large values in absolute magnitude. This can lead to a high-variance estimate. For instance, if we knew that the treatment effect is constant, then we are essentially assuming the partially linear regression model and we shouldn't be using the doubly robust method, but rather the residual-on-residual method, which minimizes the loss $\mathrm{E}(\hat{Y}_i - \tau\hat{D}_i)^2$, where $\hat{Y} = Y - \mathrm{E}[Y \mid Z]$ and $\hat{D} = D - \mathrm{E}[D \mid Z]$. Similarly, if we are willing to assume that the CATE function is linear in some engineered features of only the variables $X$, i.e. $\delta(Z) = \tau(X) = \beta' p(X)$, then we should instead be estimating a linear interactive model, where we interact the treatment with the engineered features and apply the residualization apporach to arrive at the loss function

$$\mathrm{E}(\hat{Y}_i - \beta' p(X)\,\hat{D}_i)^2$$

since $p(X)D - \mathrm{E}[p(X)D \mid Z] = p(X)\,(D - \mathrm{E}[D \mid Z]) = p(X)\,\hat{D}$.

Analogously, if we know that the high-dimensional CATE function $\delta_0(Z) = \mathrm{E}[Y(1) - Y(0) \mid Z]$, is only a function of the variables $X$, i.e. $\delta_0(Z) = \tau_0(X)$ and $\tau_0$ lies in some simple space $T$, a lower variance loss function, than the doubly robust loss function would be:

$$\min_{\tau \in T} \mathrm{E}(\hat{Y}_i - \tau(X)\hat{D}_i)^2$$

As we already showed in Equation (14.4.4) in Section 14.4, under the aforementioned assumptions we can write the regression equation:

$$Y = \tau_0(X)\,D + g_0(Z) + \epsilon, \qquad \mathrm{E}[\epsilon \mid D, Z] = 0$$

Thus, we are faced with a non-linear regression equation, regressing $\hat{Y}$ on $\hat{D}, X$, where we know that the CEF is of the form $\mathrm{E}[\hat{Y} \mid \hat{D}, X] = \tau(X)\,\hat{D}$, for some function $\tau$ in some simple function space $T$. To estimate this regression problem, we should thus minimize the square loss, over the space of such CEFs, i.e.

$$\min_{\tau \in T} \mathrm{E}(\hat{Y} - \tau(X)\hat{D})^2, \tag{15.1.6}$$

which is exactly the R-Learner loss.

When taken to estimation, the residuals $\hat{Y}, \hat{D}$ will be replaced by the estimated residuals $\check{Y}, \check{D}$, where $\check{Y} = Y - \hat{h}(Z)$ and $\check{D} = D - \hat{\mu}(Z)$, with $\hat{h}$ being an estimate of the CEF $\mathrm{E}[Y \mid Z]$ (e.g. one could use the two-learner based estimate

$$\hat{h}(Z) := \hat{g}^T(Z)\,\hat{\mu}(Z) + \hat{g}^C(Z)\,(1 - \hat{\mu}(Z)).$$

or a direct regression, regressing $Y$ on $Z$. Moreover, note that minimizing the R-Learner loss, is equivalent to minimizing a sample-weighted square loss, where the covariates are $X$, the labels are $\hat{Y}/\hat{D}$ and the weights are $\hat{D}^2$:

$$\mathrm{E}(\hat{Y} - \tau(X)\hat{D})^2 = \mathrm{E}\hat{D}^2\,(\hat{Y}/\hat{D} - \tau(X))^2,$$

Thus the final step in the R-Learner also corresponds to a sample-weighted regression oracle problem. This leads to the following meta-learner algorithm:

**Residual Learner (R-Learner)**

$$\begin{aligned}
\hat{h} &:= O_H(\{Z_i, Y_i\}_{i\in\{1,\ldots,n\}}) \\
\hat{\mu} &:= O_M(\{Z_i, D_i\}_{i\in\{1,\ldots,n\}}) \\
\check{Y}_i &:= Y_i - \hat{h}(Z_i) \\
\check{D}_i &:= D_i - \hat{\mu}(Z_i) \\
\hat{\tau} &:= O_T\left(\{X_i, \check{Y}_i/\check{D}_i, \check{D}_i^2)\}_{i=1}^n\right)
\end{aligned} \tag{15.1.7}$$

Under the assumption that $\delta_0 = \tau_0 \in T$, and that $h_0 \in H$, $\mu_0 \in M$, then the R-Learner converges to the true CATE $\tau_0$. Moreover, this approach inherits similar robustness properties as the partialling out approach for the case of estimating average causal effects. In particular, if we let $r_n^2$ denote the mean squared error that the final regression oracle would have achieved had we known the true nuisance parameters $h_0, \mu_0$, then under regularity conditions, one can show that:

$$\mathrm{E}_X(\hat{\tau}(X) - \tau_0(X))^2 \lesssim r_n^2 + \mathrm{Error}(\hat{\mu})^4 + \mathrm{Error}(\hat{\mu})^2\,\mathrm{Error}(\hat{h})^2$$

Unlike the DR-Learner, we see here that accurate estimation of the propensity is more important and cannot be compensated by more accurate estimation of the outcome regression problem. Similar to the DR-Learner, the error function in the above claim can always be taken to be the fourth moment of the prediction error and under further restrictions on the function space $T$, it can be taken to be the root-mean-squared-error. For formal versions of this claim see [4, 6, 7].

One may also wonder what does the R-Learner estimate when the assumption that $\delta_0 = \tau_0$ or that $\tau_0 \in T$ is violated. Unlike all prior meta-learners, the R-Learner does not converge necessarily to the best approximation $\tau_*$ of the CATE within $T$. For instance, consider the extreme case where $T$ contains only constant functions. Then we are estimating an average treatment effect based on a partialling out approach, while the partial linear response function does not hold and there exists treatment effect heterogeneity. In this case, the partialling out approach will not be converging to the average causal effect and similarly for any $T$, the R-Learner will not be converging to $\tau_*$.

To understand the limit point of the R-Learner, let us examine the R-Learner loss as defined in Equation (15.1.6). By construction, $\hat{\tau}$ will be converging to the solution to that minimization problem. As we have already argued, under conditional exogeneity, we can always write $\hat{Y} = \delta_0(Z)\hat{D} + \epsilon$, with $\mathrm{E}[\epsilon \mid \hat{D}, Z] = 0$. Thus we can re-write the R-Learner loss as:

$$\begin{aligned} \mathrm{E}(\hat{Y} - \tau(X)\hat{D})^2 &= \mathrm{E}(\delta_0(Z)\hat{D} - \tau(X)\hat{D})^2 + \mathrm{E}\epsilon^2 \\ &= \mathrm{E}\left[(\delta_0(Z) - \tau(X))^2 \,\mathrm{Var}(D \mid Z)\right] + \mathrm{E}\epsilon^2 \end{aligned}$$

where we used the fact that $\mathrm{E}[\hat{D}^2 \mid Z] = \mathrm{Var}(D \mid Z)$. Thus minimizing the R-Learner loss is equivalent to minimizing a treatment-variance-weighted square loss and the estimate will be converging to the best treatment-variance-weighted approximation of the high-dimensional CATE function, i.e.

$$\tilde{\tau} = \arg\min_{\tau \in T} \mathrm{E}\left[(\delta_0(Z) - \tau(X))^2 \,\mathrm{Var}(D \mid Z)\right] \tag{15.1.8}$$

This solution is essentially putting more weight on regions of the covariate space $Z$, where the treatment was more randomly assigned. If for instance parts of the population were almost always treated or almost always not treated, then these parts of the population will not be considered when constructing the best approximation. We will refer to this solution as the *best overlap-weighted approximation*, since it assigns weights to parts of the population, dependent on the degree of "overlap" (i.e. whether both treatments were observed for this part of the population). For instance, suppose that $T$ is the space of constant functions and that the treatment is randomly assigned for some parts of the population and is essentially deterministic for other parts. Then $\tilde{\tau}$ will recover the average treatment effect of the subset of the population for which treatment was randomly assigned. On the contrary, in this case the doubly robust estimate will try to recover the average causal effect of the overall population, but because of that it will inadvertently be very high variance and unstable, since for some parts of the population it barely ever sees one of the two treatments.

**Cross (X)-Learner** The Cross Learner tries to combine propensity to improve on outcome modelling in a manner qualitatively very different from the DR- or R-learner and not with the target of reducing the sensitivity to errors in the nuisance models. Rather it does so primarily motivated from an *accuracy and covariate-shift consideration*. Moreover, it begins with a very different starting point and idea. As a first one realizes that the high-dimensional CATE $\delta_0(Z)$ is the same, whether we measure it on the treated or on the control! In other words, the Conditional Average Treatment Effect on the Treated (CATT) is equal to the Conditional Average Treatment Effect on the Control (CATC), unlike the average treatment effect, which can be different due to different distributions of $Z$ in treatment and control. This can be easily seen as, by conditional exogeneity:

$$\begin{aligned}\delta_0^T(Z) &:= \mathrm{E}[Y(1) - Y(0) \mid Z, D = 1] \\ &= \mathrm{E}[Y(1) - Y(0) \mid Z] = \delta_0(Z)\end{aligned}$$

and similarly for $\tau^0$.[4]

4: Note that the same wouldn't be true if we condition a subset $X$ of $Z$:

$$\begin{aligned}\tau_0^T(X) &:= \mathrm{E}[Y(1) - Y(0) \mid X, D = 1] \\ &= \mathrm{E}[\mathrm{E}[Y(1) - Y(0) \mid Z] \mid X, D = 1] \\ &= \mathrm{E}[\delta(Z) \mid X, D = 1] \\ &\neq \mathrm{E}[\delta(Z) \mid X] =: \tau_0(X)\end{aligned}$$

Moreover, when we try to measure the CATT, then we actually observe the counterfactual under treatment and therefore we do not need to impute this counterfactual outcome (e.g. by learning $g_0^1$). Similarly for the CATC. Thus we can identify the CATT and CATC as:

$$\begin{aligned}\delta_0^T(Z) &= \mathrm{E}[Y - \mathrm{E}[Y \mid Z, D = 0] \mid Z, D = 1] \\ \delta_0^C(Z) &= \mathrm{E}[\mathrm{E}[Y \mid Z, D = 1] - Y \mid Z, D = 0]\end{aligned}$$

This yields two ways of identifying the CATE $\delta_0(Z)$ and any convex combination of these two solutions, would also be a valid identification strategy for the CATE. This approach, allows us to avoid having to model both response models well for all regions of the covariate space (which would be the case for the S-, T-, or DR-Learners). This can be powerfull if we know that the CATE is a much simpler function to learn than a mean counterfactual response model.

If we believe that the hard part is modelling the mean counterfactual response under some treatment but not the treatment effect, then we can use the following strategy: for parts of the covariate space $Z$, where we have more control data (i.e. $\mu_0(Z)$ is small), we can use the CATT strategy, which only requires estimating the mean counterfacutal response under control, i.e.

$E(Y \mid Z, D = 0)$, but not under treatment. Of course, we still have to learn the effect function using only the treated data, which we don't have that many in this part of $Z$, but since we believe that the effect function is simple, this is a more benign problem. Similarly, if for parts of the covariate space $Z$, we have more treated data (i.e. $\mu_0(Z)$ is large), we can use the CATC strategy, which only requires estimating the mean counterfactual response under treatment, i.e. $E(Y \mid Z, D = 1)$, but not under control. This motivates using the following convex combination as our final identification formula for the CATE:

$$\delta_0(Z) = \delta_0^T(Z)\,(1 - \mu_0(Z)) + \delta_0^C(Z)\,\mu_0(Z)$$

Subsequently, for any subset $X$ of $Z$, we can use the fact that $\tau_0(X) = E[\delta_0(Z) \mid X]$. This identification strategy leads to the following meta-learning estimation strategy:

**Cross Learner (X-Learner)**

$$\begin{aligned}
\hat{g}^C &:= O_{G^C}(\{Z_i, Y_i\}_{i \in \{1,\dots,n\}: D_i = 0}) \\
\hat{g}^T &:= O_{G^T}(\{Z_i, Y_i\}_{i \in \{1,\dots,n\}: D_i = 1}) \\
\hat{\mu} &:= O_M(\{Z_i, D_i\}_{i \in \{1,\dots,n\}}) \\
\hat{\delta}^C &:= O_\Delta(\{Z_i, \hat{g}^T(Z_i) - Y_i\}_{i \in \{1,\dots,n\}: D_i = 0}) \\
\hat{\delta}^T &:= O_\Delta(\{Z_i, Y_i - \hat{g}^C(Z_i)\}_{i \in \{1,\dots,n\}: D_i = 1}) \\
\hat{\delta}^X(Z) &:= \hat{\delta}^T(Z)\,(1 - \hat{\mu}(Z)) + \hat{\delta}^C(Z)\,\hat{\mu}(Z) \\
\hat{\tau} &:= O_T\left(\{X_i, \hat{\delta}^X(Z_i)\}_{i=1}^n\right)
\end{aligned} \tag{15.1.9}$$

Assuming that the function spaces used in the nuisance oracles contain the true functions, the final step of the X-learner will converge to the best approximation of the CATE $\tau_*$, within the space $T$. Moreover, this estimation strategy can have substantial benefits when the CATE function $\delta_0$ is much simpler than the response functions $g_0^C, g_0^T$ and when there are substantial imbalances in the treatment across the population (i.e. the propensities substantially deviate from 1/2). The latter many times arises in digital experimentation, where only a small fraction of the population receives the treatment. In this case, the response under control can be much more accurately estimated. In fact, in many such settings we have a lot of historical data, prior to running an experiment, where the treatment was un-available and which can be used as auxiliary datasets for learning the baseline response; with the small treated data from the experiment only being used to estimate the heterogeneous effect function $\delta_0^T$.

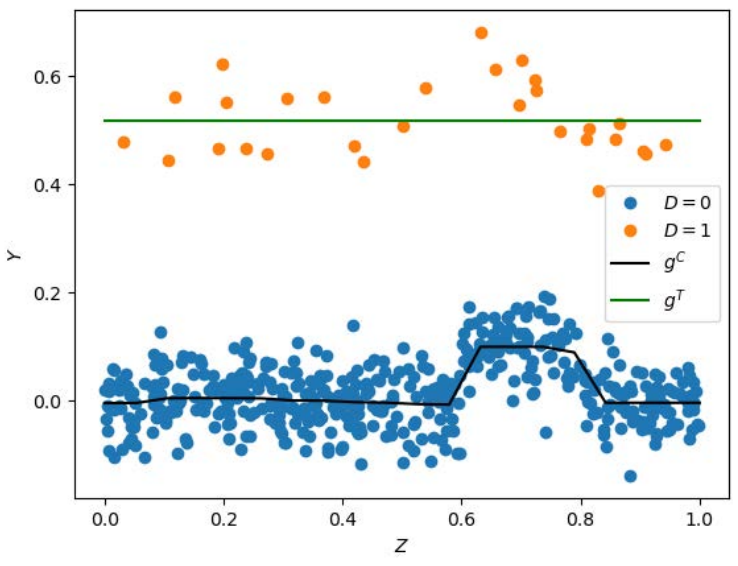

**Figure 15.1:** DGP 1. Imbalanced dataset were baseline response is more complex than heterogeneous effect.

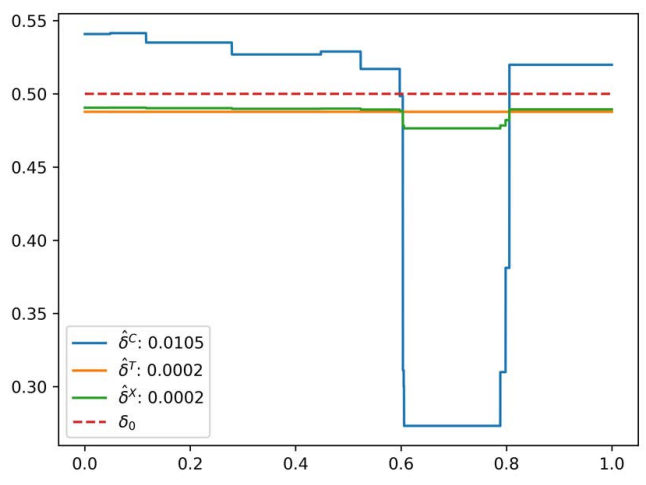

**Figure 15.2:** DGP 1. Different CATE estimates in the X-Learner. The legend displays the mean squared error of each estimate.

**Example 15.1.1** (Imbalanced Dataset with Hard Baseline Response and Simple CATE) As a stark example, consider the case when $Z = X \sim U[0, 1]$, treatment is very rare, i.e. $\mu(Z) = .05$, the treatment effect is constant, i.e. $\delta_0(Z) = .5$ and the baseline response is complex and contains a discontinuity:

$$\begin{aligned} Y &= .5\, D + .3\, \mathbb{1}\{Z \in [.6, .8]\} + N(0, \sigma = .05), \\ D &= \text{Bernoulli}(\mu(Z) = .05) \end{aligned} \qquad \text{(DGP 1)}$$

In this case, the data that we collect, for $n = 500$, are depicted in Figure 15.7. If we use gradient boosted forest regression to estimate the two response functions under treatment and under control, we find that the $\hat{g}^T$ response function is substantially more regularized and the discontinuity is not learned, due to the small sample size. On the other hand $\hat{g}^C$ is much more accurate and the discontinuity is learned due to the large sample size. Subsequently, we see in Figure 15.2 that the estimate based on the CATC identification strategy is much less accurate than the one based on the CATT identification strategy. Moreover, the X-Learner is putting almost all the weight on the CATT estimate $\hat{\delta}^T$ and is highly accurate compared to $\hat{\delta}^C$. However, in this setting, we also find that other strategies that also use propensity modelling (e.g the R- or DR-Learners) also manage to correct the error in the T-Learner regression models and achieve similar accuracy to the X-Learner.

On the other hand, if the inductive bias that the CATE is simpler than the response functions under either treatment or control does not hold, then the superiority of the X-Learner strategy as compared for instance to the T-learner strategy for outcome modelling vanishes. For instance, if we instead have an outcome model of:

$$\begin{aligned} Y &= .5\, \mathbb{1}\{Z \in [.6, .8]\}\, D + .1 + N(0, \sigma = .05), \\ D &= \text{Bernoulli}(\mu(Z) = .05) \end{aligned} \qquad \text{(DGP 2)}$$

then all methods that only rely on outcome modelling fail and methods that also combine propensity based identification start to outperform (see Figure 15.4). Even more vivid is the flip in performance if we further make the treatment more prevalent than the baseline:

$$\begin{aligned} Y &= .5\, \mathbb{1}\{Z \in [.6, .8]\}\, D + .1 + N(0, \sigma = .05), \\ D &= \text{Bernoulli}(\mu(Z) = .95) \end{aligned} \qquad \text{(DGP 3)}$$

In this case it is more important to use the large amount of treated data, since $\mu(Z) = .95$, not to learn the response function, but rather to learn the CATE function (see Figure 15.5). In this case a T-Learner outcome modelling strategy and a T-Learner based DR-Learner is a better option.

We conclude by noting that the reasoning in the cross learner strategy can actually be used as a sub-process to improve outcome modelling in all other learners. In particular, note that the key advance of the cross learner is to observe than when the treatment is very rare, then we should be estimating the response $\hat{g}^C$ under control and then estimating only the effect $\hat{\delta}^T$ using the treatment data. In this case, we can also use $\hat{g}^T = \hat{g}^C + \hat{\delta}^T$ as our estimate of the response under treatment. Similarly, if the control group is very rare, then we should be estimating the response $\hat{g}^T$ under treatment and then estimating only the effect $\hat{\delta}^C$ using the control data. In this case, again we can also use $\hat{g}^C = \hat{g}^T - \hat{\delta}^C$ as our estimate of the response under control. Moreover, we can locally blend these two estimation strategies by weighting both estimates of the two response functions using the propensity, i.e. putting a weight of $(1 - \hat{\mu}(Z))$ to the first estimation strategy and a weight of $\hat{\mu}(Z)$ to the second estimation strategy. This approach is an alterantive outcome modelling process that can be used instead of the $S$ or $T$ learner approaches for learning the response functions under the different treatments. In that respect, the X-Learner outcome modelling strategy can be used in conjunction with the DR- or the R-Learner approaches, if one wants to introduce some robustness with respect to outcome modelling by incorprorating identification by propensity approaches. For instance, in Figure 15.3, we also depict the CATE learned if we combine the X-learner approach to outcome modelling with the doubly robust correction (coined the DRX-Learner).

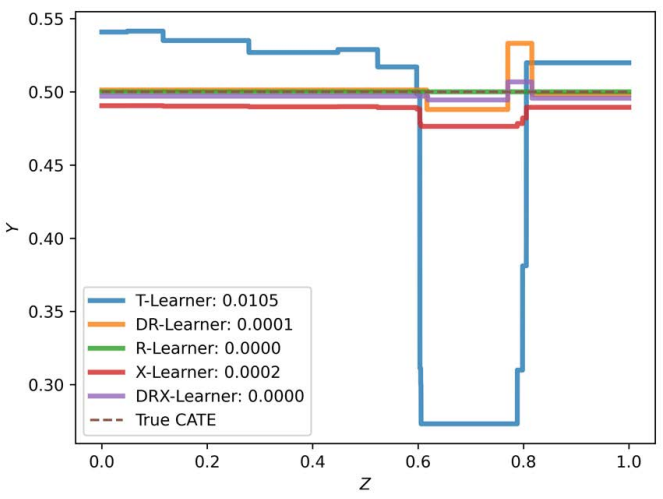

**Figure 15.3:** DGP 1. CATE estimates ($n = 500$) from other meta-learners. The legend displays the mean squared error of each learner.

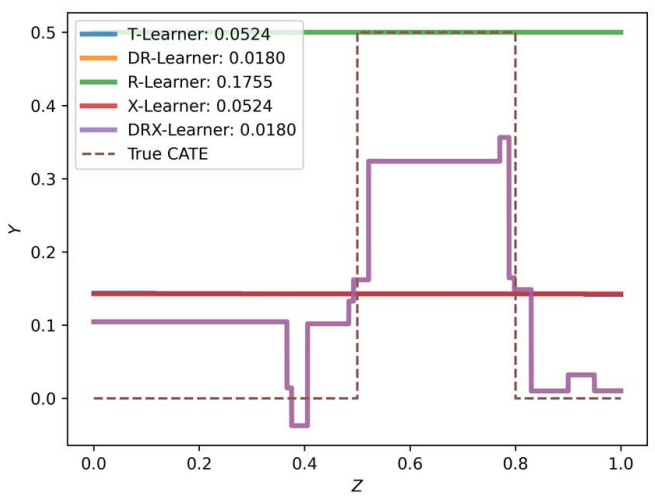

**Figure 15.4:** DGP 2. CATE estimates ($n = 500$) from meta-learners, when CATE is complex and baseline simple. The legend displays the mean squared error of each learner.

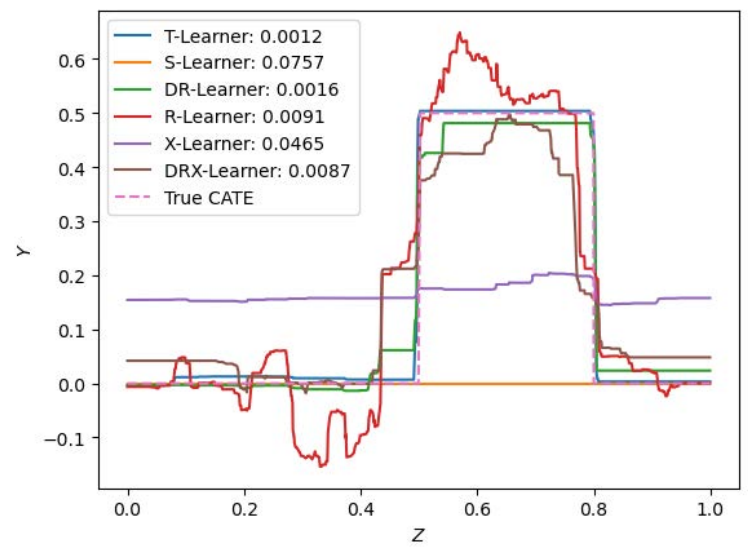

**Figure 15.5:** DGP 3. CATE estimates ($n = 500$) from meta-learners, when CATE is complex and baseline simple and treatment very prevalent. The legend displays the mean squared error of each learner.

## Qualitative Comparison and Guidelines

We present here a set of bullet points that can guide the reader through the choosing among the different meta-learner strategies, dependent on inductive biases about their setting:

- S/T-Learner: they heavily rely on correct outcome modelling, trying to learn how the outcome relates to the control co-variates $Z$. If this estimation problem is hard to learn, then they will have poor performance, especially when the effect is a simple function and $X$ is much lower dimensional than $Z$. However, they can have very low variance and be more stable as they only depend on simple regression strategies. If one has to choose among the two, then the T-Learner should be preferable, if both treatments are sufficiently represented, since it avoids overly regularizing the treatment, which reduces the bias of the treatment effect estimate. If one of the two treatments is rare, then the X-Learner should be more preferable than the T-Learner.

- X-Learner: even if the X learner estimates a propensity, the propensity is primarily used to select locally, which outcome model is better and is not used to identify the effect. Thus the X learner is essentially also only performing outcome modelling. If we believe that the CATE function is simpler than the response functions under treatment or control, this outcome modelling estimation strategy should be preferred. Otherwise, if we believe that the CATE function is equally or more complex than the response functions, then a T-Learner approach can outperform an X-Learner approach. If we further believe that learning any of these outcome processes could potentially be a substantially harder task than learning the propensity, then this method can be heavily biased. In that case the DR-Learner or the R-Learner should be prefered. However, the X learner reasoning can still be useful in improving the outcome modelling part of the DR or R learners.
- DR-Learner: possesses doubly robust properties, in that the mean squared error of the CATE is small if either the outcome model is learned accurately or the propensity model. It is particularly useful in learning projections of the CATE on simpler function spaces or on small subsets $X$ of the control variables $Z$. If $X$ is very small compared to $Z$, then even the X-Learner needs to accurately learn the complex effect function $\delta_0(Z)$ accurately. However, the DR-Learner can learn the simpler CATE $\tau_0(X)$, if the propensity model is accurately learned. For instance, we saw in DGP 3 in Example 15.1.1, that when the CATE function was complex, then methods such as the DR-Learner that incorporate propensity modelling are more accurate. However, contrary to the S/T/X-learners, when the true data generating process has extreme propensities in parts of the covariate space (i.e. parts of the population are almost deterministically either treated or not treated), then the DR-Learner can have high variance and become unstable. On the other hand the R-Learner will be extrapolating the CATE from nearby regions where there is more overlap and assuming the CATE model space is smooth enough, such model-based extrapolation can be quite accurate.

- R-Learner: posseses insensitivity properties related to Neyman orthogonality, in that the impact of errors in the propensity model or the outcome model impact the CATE model only in a second order manner. In particular, if the outcome model is wrong, but the propensity model is very accurate, the CATE will be highly accurate. However, it is more heavily relying on moderately accurate propensity modelling, unlike the DR-Learner. If for instance the outcome model is perfect, but the propensity model is very wrong, then the DR-Learner will be highly accurate but the R-Learner will not be. On the positive side, the R-Learner is much more stable than the DR-Learner in the presence of extreme propensities, as it does not divide by the propensity score when constructing the regression labels. The reason that it can bypass that is that it inherently estimates only an overlap-weighted projection of the CATE and not the true projection of the high-dimensional CATE $\delta_0$, when the CATE model is either mis-specified or does not solely depend on $X$ and not on the larger set of covariates $Z$. In both cases the DR-Learner converges to the true CATE, while the R-Learner can potentially ignore large parts of the population to reduce its variance; introducing bias and extrapolating the CATE from nearby highly overlapping regions. For instance, we show that in the case of DGP 1 in Example 15.1.1 the R-Learner was out-performing the DR-Learner, since the treatment effect was constant and the propensity very small. The R-Learner should be prefered to the DR-Learner when such overlap weighted projections are acceptable within the application context and when we believe we have a relatively accurate propensity model. In principle, a similar variance reduction can also be performed for the DR-Learner, by multiplying the DR-Learner loss with sample weights $W = \mathrm{Var}(D \mid Z)^2$, which would then avoid dividing by the propensity and would converge to an overlap weighted projection of the CATE $\delta_0$, with the aforementioned sample weights, while preserving the double robust nature of the estimation (i.e. that errors in propensity can be compensated by more accurate estimation in the outcome model and vice versa) (see e.g. [8]).

All-in-all, one should note that there is no clear winner among the X-, R- and DR-Learner methods and each can potentially be the best performer in different contexts. The above discussion gives a high-level strategy of which method to use dependent on which types of phenomena one should be expecting to arise in their data. In the next section we give a more data-driven selection among these methods using out-of-sample scoring and ensembling.

**Remark 15.1.1** (Guarding for Overfitting with Cross-fitting) To avoid having to worry about overfitted estimators, all the first stage nuisance models across all the meta-learners should preferably be estimated in a cross-fitting manner (i.e., the models $\hat{g}, \hat{g}^C, \hat{g}^T, \hat{\mu}, \hat{h}$) while the CATE models (i.e., $\hat{\tau}, \hat{\delta}^C, \hat{\delta}^T$) should be estimated using all the samples.

**Remark 15.1.2** (Explainability and Interpretability) An important side benefit of the meta-learning approach to CATE estimation is that in the end we end up with an ML regression model that represents our estimate of the CATE function. Even though this regression model can be quite complicated (e.g. a random forest, a gradient boosted forest or a neural network), we can apply the multitude of interpretability approaches in machine learning to interpret the learned model. For instance, we can summarize how different features change the value of the CATE model via the widely used SHAP values [9] or approximate locally the CATE model with simpler linear models based on the widely used LIME framework. Finally, we can invoke *distillation methods* that fit simpler and easy to visualize models using the learned model predictions as labels. For instance, we can train a shallow binary tree regression model that approximates the CATE model predictions and then visualize the learned tree. For more elaborate treatment of interpretability methods in machine learning see [10].

## Guarding for Covariate Shift

When machine learning models are evaluated on a different population of covariates than the one that they were trained on, then an important finite sample consideration is deterioration of their performance due to the covariate shift. Such population mis-match between training and evaluation typically arises when we employ ML algorithms within a CATE estimation. For instance, in the T-Learner we train an ML model on the treated datapoints and then we evaluate it on all the datapoints.

Similarly, in the X-Learner we train a CATE model on the treated points and then we evaluate it on all the datapoints.

In such settings, we should only expect the oracle ML model to have small mean squared error with respect to the distribution of its training data and with respect to the best approximation of the CEF, where the approximation error is calculated with respect to the training data. For instance, suppose that we estimate a regression model $\hat{h}$ that takes as input random variables $X$ and predicts a variable $Y$, with sample weights $W$, by invoking an ML regression oracle as defined in Equation (15.1.2). Assuming that the CEF $h_* := \mathrm{E}(Y \mid X)$ does not change between train and evaluation data and letting $D_t$ denote the distribution of $X$ in the training data and $D_e$ in the evaluation data, then our regression estimate satisfies that:

$$\mathrm{E}_{X\sim D_t} W(\hat{h}(X) - h_0(X))^2 \leq r_n$$

where $h_0 = \arg\min_{h\in H} \mathrm{E}_{X\sim D_t} W(h_*(X) - h(X))^2$. Since we evaluate this regression model on a different population, we would typically care about the following mean squared error:

$$\mathrm{E}_{X\sim D_e} W_e(\hat{h}(X) - h_*(X))^2$$

with some set of weights $W_e$ that depend on some downstream use of the model.

**Example 15.1.2** (Covariate Shift in X-Learner) In the context of the X-Learner, we train a model $\hat{\delta}^T$ on the treated data and then we use it to calculate $\hat{\delta}(Z) = \hat{\delta}^T(Z)\,(1 - \mu(Z)) + \hat{\delta}^C(Z)\,\mu(Z)$ on all the data points. Thus in this case, when measuring the quality of the downstream CATE estimate $\hat{\tau}$ in the final step of the X-Learner, we care about the quality of $\hat{\delta}^T$ as measured by the metric:

$$\mathrm{E}_Z(1 - \mu(Z))^2(\hat{\delta}^T(Z) - \delta_0(Z))^2$$

On the contrary, the oracle for $\hat{\delta}^T$ would be guaranteeing:

$$\mathrm{E}_{Z|D=1}(\hat{\delta}^T(Z) - \tilde{\delta}_0(Z))^2$$

where $\tilde{\delta}_0 = \arg\min_{\delta\in\Delta} \mathrm{E}_{Z|D=1}(\delta_0(Z) - \delta(Z))^2$.

There are two sources of discrepancy: *first* the approximation error can be substantially different if we use the best approximation with respect to a different distribution and *second* the mean squared error is measured with respect to the wrong distribution. If the true CEF $h_0$ lies in the function space $H$, then the first problem vanishes (though in finite samples and with some growing sieve space, we should always expect some finite sample approximation bias). Simiarly, if we denote with $p_t$ the density of $X$ under $D_t$ and $p_e$ under $D_e$, then if the density ratio $p_e(X)/p_t(X)$ is upper and lower bounded by some constants $[c, C]$, then we always have that:

$$\begin{aligned} \mathrm{E}_{X\sim D_e} W_e(h(X) - h_0(X))^2 &= \mathrm{E}_{X\sim D_t} \frac{p_e(X)}{p_t(X)} W_e(h(X) - h_0(X))^2 \\ &\in [c, C] \cdot \mathrm{E}_{X\sim D_t} W_e(h(X) - h_0(X))^2 \end{aligned}$$

Thus even if we don't take any measures to address the co-variate shift, by minimizing the squared error under the training distribution, we are approximately minimizing the error under the evaluation distribution. However, these constants can be quite large in practice and the magnitude of the discrepancy can be comparable to the sample size.

For these reasons a large literature in machine learning has focused on addressing such co-variate shift problems by changing how we train the model, when we know what the target evaluation distribution or metric will be. In its simplest form, one can instead optimize for the density ratio weighted error, i.e.:

$$\mathrm{E}_{X\sim D_t} \frac{p_e(X)}{p_t(X)} W_e(h(X) - h_0(X))^2$$

Noting also that $\frac{p_e(X)}{p_t(X)} = \frac{p(X|e)}{p(X|t)} = \frac{p(e|X)p(t)}{p(t|X)p(t)}$, the above is equivalent to minimizing:

$$\mathrm{E}_{X\sim D_t} \frac{p(e \mid X)}{p(t \mid X)} W_e(h(X) - h_0(X))^2$$

which requires solving two classification problems (i.e. predicting the probability that a sample is in population $e$ given $X$ and predicting whether the sample is in population $t$ given $X$, using the union of the populations).

**Example 15.1.3** (Covariate Shift in X-Learner (continued)) Going back to our X-Learner example, we have $p(e \mid Z) = 1$ (since we evaluate on all the population) and $p(t \mid Z) = \mu_0(Z)$ (since we train only on the training population). Moreover, we care about evaluation weights $W_e = (1 - \hat{\mu}(Z))^2$. Thus it would potentially be better in finite samples if one optimizes:

$$\mathrm{E}_{Z|D=1}\frac{1}{\hat{\mu}(Z)}(1-\hat{\mu}(Z))^2(\hat{\delta}^T(Z)-\tilde{\delta}_0(Z))^2$$

In other words, calling the ML oracle when training $\hat{\delta}^T$ with sample weights $W = \frac{1}{\hat{\mu}(Z)}(1-\hat{\mu}(Z))^2$.

For instance, if we employ such co-variate shift techniques in DGP 3 from Example 15.1.1, then we find that the performance of the Domain Adapted X-Learner (DAX-Learner) is restored (see Figure 15.6).

Analogous finite sample corrections can be taken throughout the meta-learner algorithms by first working out what is the target evaluation population and metric we care about and changing the training of the ML model appropriately.

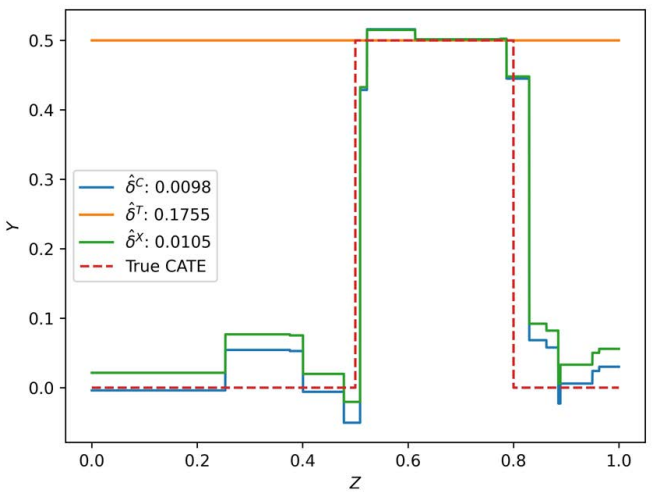

**Figure 15.6:** DGP 3. CATE estimates ($n = 500$) from meta-learners in adapted X-Learner, with co-variate shift corrections.

**Covariate shift techniques when overlap fails.** Beyond this simple approach of density weighting, many other ML methods have been developed in the literature to guard against covariate shift. One advantage of many of these alternative methods, is that they are applicable even when there is lack of overlap (i.e. when the density ratio can be unbounded or zero). For instance, one large class of covariate shift approaches within the context of neural network training, makes the assumption that overlap holds on some latent representation space $\phi(X)$ and not on the observed covariate space $X$ and that the conditional expectation function can be written as a function of these latent variables, i.e. $\mathrm{E}(Y \mid X) \approx \mathrm{E}(Y \mid \phi(X))$. In this case, one can train a neural network architecture where the first few layers of the neural network are used to construct the mapping $\phi(X)$ and the subsequent layers are used to construct $\mathrm{E}(Y \mid \phi(X))$. Subsequently a distribution distance measure is introduced as a regularizer, that measures the distribution distance of $\phi(X)$ between samples that stem from the training and evaluation population. A popular metric is a variant of the Wasserstein distance. In this manner, we are trying to construct a latent representation that has approximately the same distribution under the two populations and which predicts well the target $Y$.

**Shared representation learning with neural networks.** In the context of CATE estimation, the latter approach was utilized by [11, 12] within the T-Learner framework for outcome modelling. In particular, the first few layers of the network are used to represent $\phi(Z)$, which then is used to represent both $g_0^T$ and $g_0^C$. Moreover, a wasserstein penalty is introduced so that the distribution of $\phi(Z)$ is similar between the treated and control population. The resulting method is typically referred to as the CFR-Net. If one believes that their setting satisfies this inductive bias, i.e. that there exists a latent representation that is sufficient for the CEF of the outcome and in which overlap holds, then this approach can be used for better outcome modelling within the context of any meta-learner. For instance, one can use the CFR-Net together with the DR- or R- learners for estimating $g_0^T, g_0^C$ and hence also $g_0$ and $\hat{h}$ (potentially using the same shared representation, when estimating the propensity; to avoid extreme propensities). See also [13] for the empirical evaluation of variants of such neural network approaches, combined with doubly robust learning.

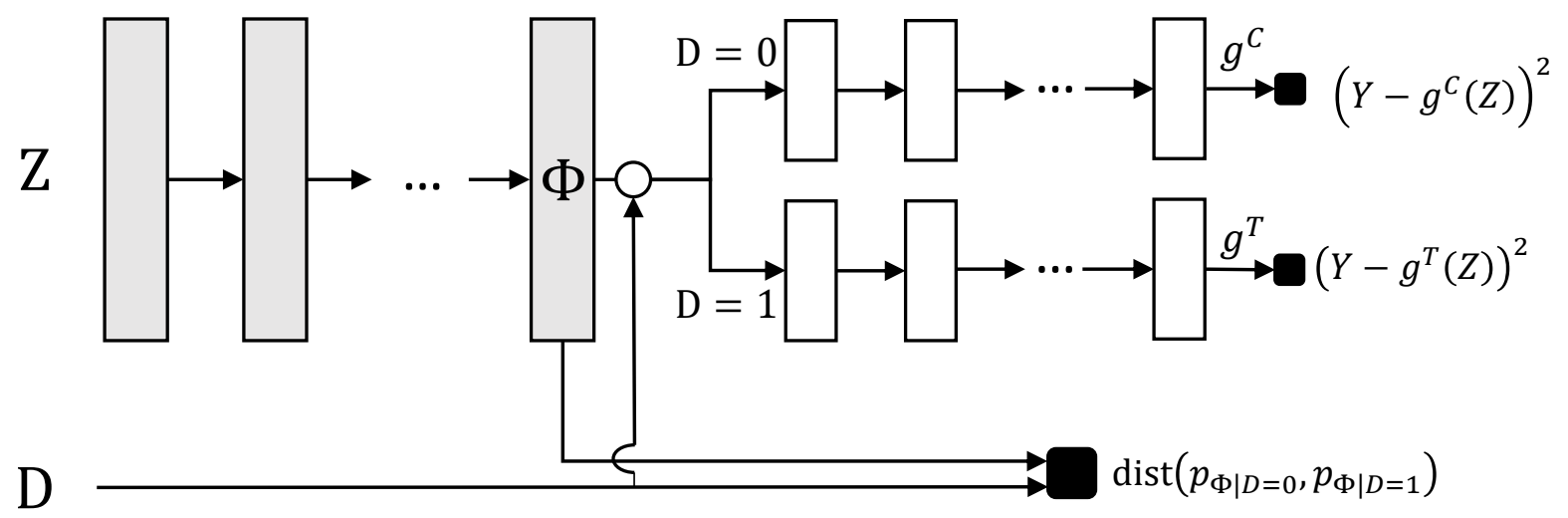

**Figure 15.7:** Counterfactual regret network of [11, 12], to guard against covariate shift in the T-Learner.

**Example 15.1.4** (Meta-Learners in the 401(k) Example) We applied each of the meta-learner models to estimate the CATE in the 401(k) example. We estimated a CATE model that uses all the available variables for heterogeneity (i.e. $X = Z$) and used gradient boosted forests (based on the xgboost library) as oracle regression models for each step of each meta-learner. We depict below the CATE predictions of each of the meta-learner models as a function of income (x-axis), when all other features are fixed to their overall median value.

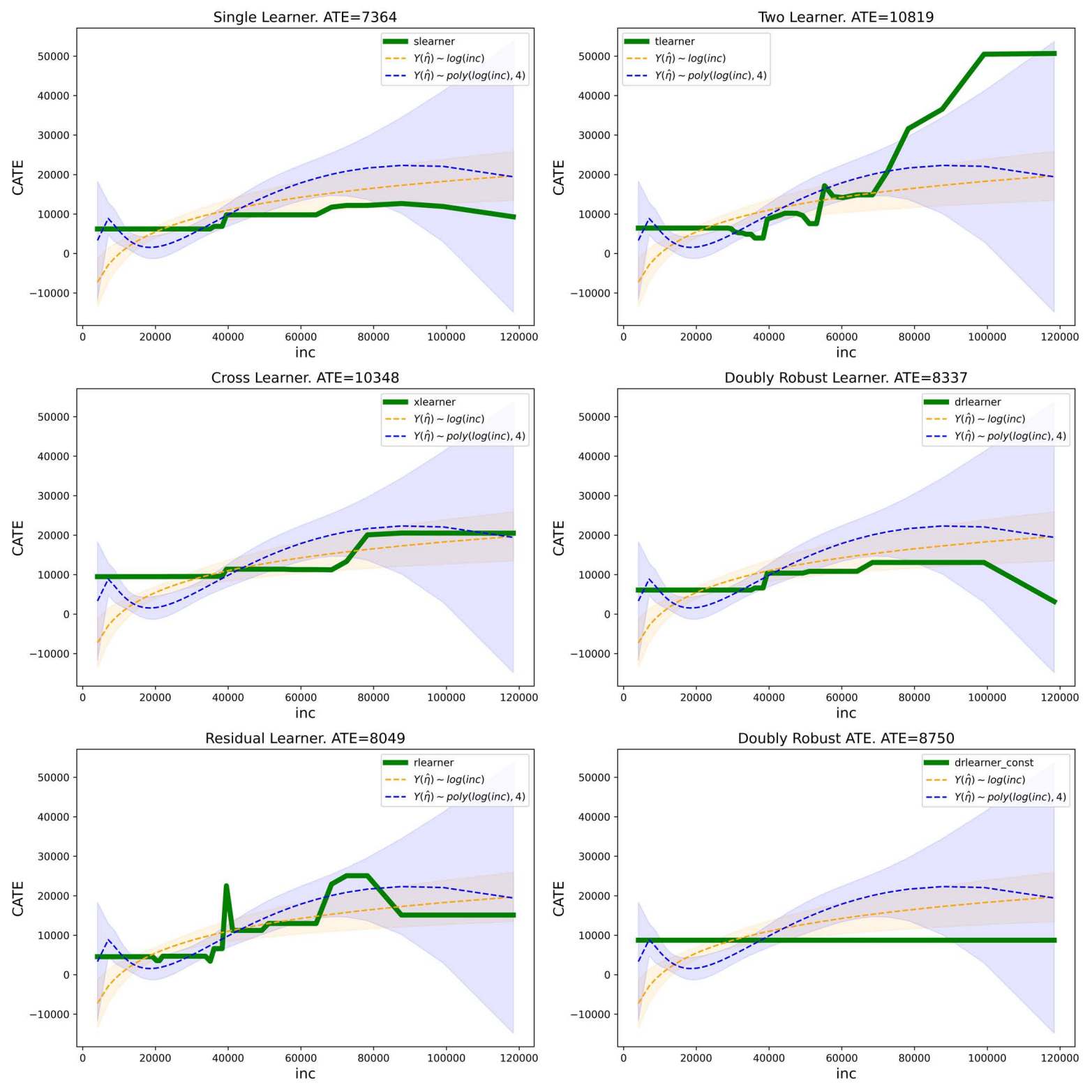

**Figure 15.8:** CATE predictions of different meta-learners in the 401k example. Gradient boosted forests (via the xgboost library) were used as ML oracles for regression and classification. The CATE is predicted on a grid of income points, corresponding to equally spaced income quantiles. All other covariates were imputed at their median values. For comparison, each plot also displays the doubly robust best linear predictor of the CATE with 5-95% confidence intervals on a simple linear form of engineered features of the income.

Subsequently, we investigate for interpretability reasons, the main factors that are driving the predictions of the DR-Learner model. We do this by fitting a simple shallow binary regression tree on the predictions of the model. We find that the model's CATE predictions are primarily driven by income and age factors. In particular, the model finds that 401(k) eligibility has the lowest effect ($\approx \$6k$) in net financial assets for low income ($\lesssim \$39k$) and younger people ($< 59$ years), while it has the highest effect ($\approx \$13k$) for high income people ($\gtrsim \$68k$).

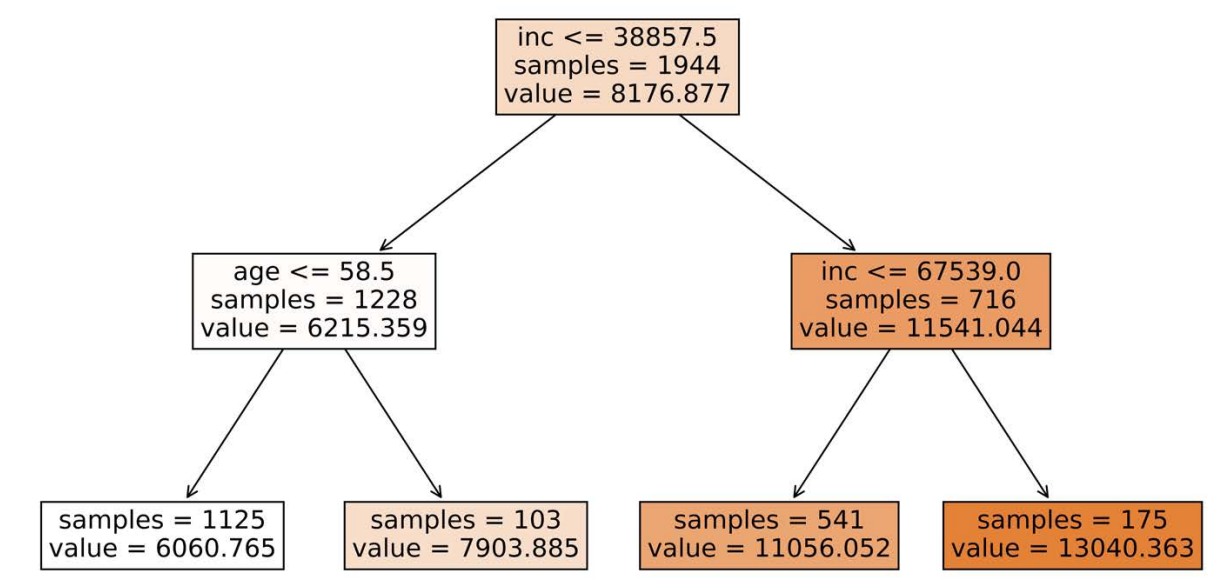

We also depict the SHAP values for each feature in the CATE model, that identifies the directionality and magnitude of the change that each feature drives in the model's output. We again identify that the main factors that drive variation in the output of the DR-Learner CATE model are income and age.

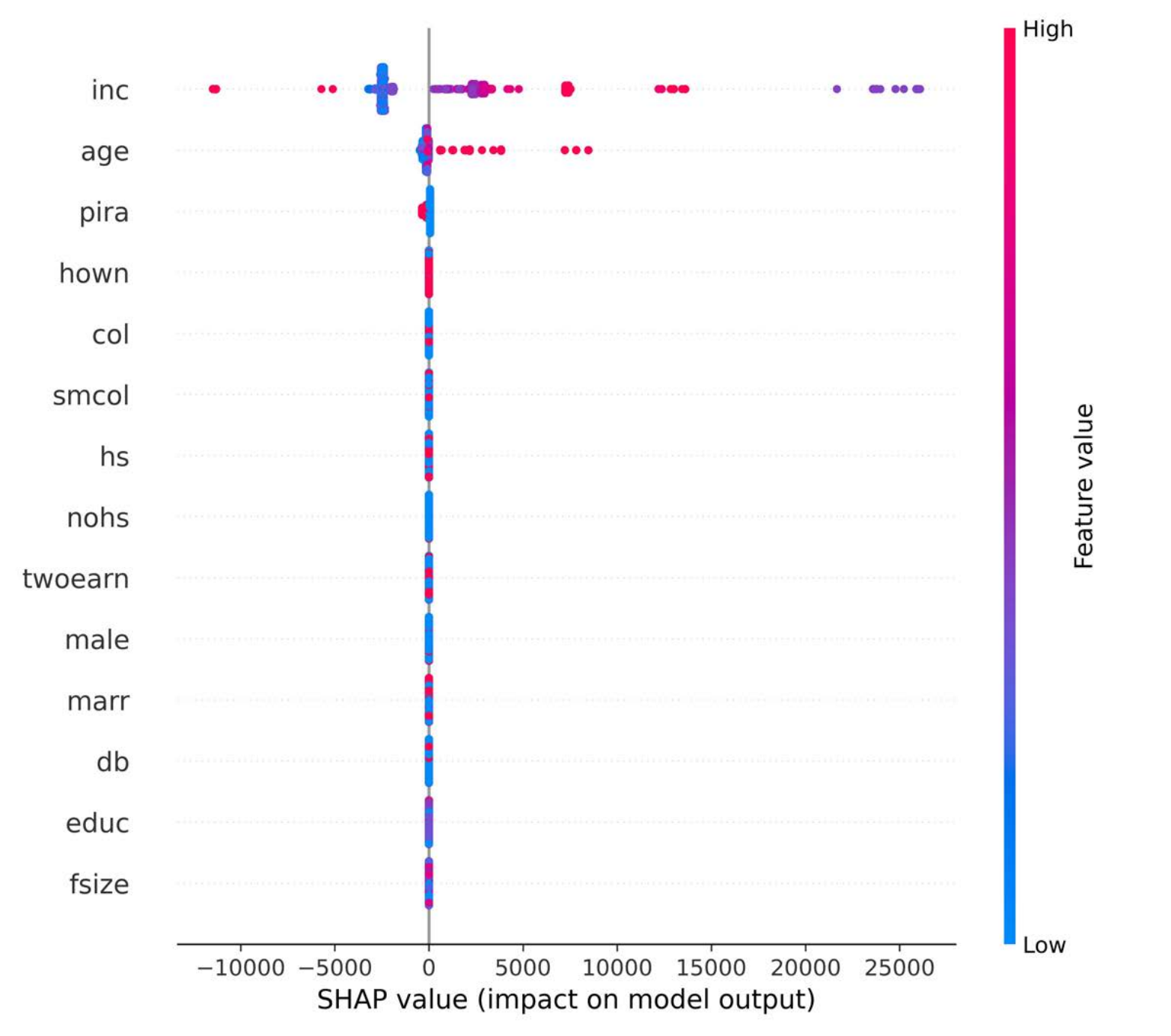

We find that even though we did not hard-code income or age as factors of effect heterogeneity, the generic ML approach identified these two factors as the key drivers; a conclusion that is inline with domain knowledge.

## 15.2 Scoring for CATE Model Selection and Ensembling

The previous section gave an overview of how to qualitatively select among the different meta-learning strategies. Here we discuss how one can automate the process of selection using out-of-sample scoring and moreover how to potentially ensemble the models that come out of different estimation strategies into a single CATE model. In this section, we envision that the user has split their data into a training and scoring set and based on the training set they have fitted a set of candidate CATE models $T := \{\tau_1, \ldots, \tau_M\}$. These models could correspond to the result of the different meta-learning strategies and with different regression style oracles. For instance, $\tau_1$ could be the result of an X-Learner with random forest regression oracles, $\tau_2$ the result of an X-Learner with gradient boosted regression oracles, $\tau_3$ the result of the DR-Learner with linear/logistic model oracles.

Let $n$ denote the size of the scoring set. Our goal is to be able to use the scoring set in order to evaluate which of these $M$ models is more accurate (with confidence) and to device approaches to select a single model $\tau_*$ that could either correspond to one of the models $T$ or to an ensemble of these models with weights $(w_1, \ldots, w_M)$, such that $\tau_*(X) = \sum_{i=1}^{M} w_i \tau_i(X)$. Such a model $\tau_*$ should ideally be competing with the best model in $T$, i.e., with high probability:

$$\mathrm{E}(\tau_*(X) - \tau_0(X))^2 \leq \min_{j=1}^{M} \mathrm{E}(\tau_j(X) - \tau_0(X))^2 + \epsilon(n, M) \tag{15.2.1}$$

for some error function $\epsilon(n, M)$ that should decay fast to zero as a function of $n$ and should grow slowly with the number of candidate models $M$. We can use again the doubly robust outcome approach, viewing the problem as a regression problem with the doubly robust proxy outcomes $Y_i(\hat{\eta})$ as the labels and utilize techniques from model scoring, ensembling, model selection and stacking for regression problems.

## Comparing Models with Confidence

We can use the doubly robust loss:

$$\hat{L}_{DR}(\tau; \hat{\eta}) := \mathbb{E}_n(Y(\hat{\eta}) - \tau(X))^2 \tag{15.2.2}$$

as a quality score for each of the candidate models. Since we care about selecting among the models in $T$, we primarily care about choosing a score function that orders the models accurately. Hence, we primarily care that *differences* in the score between two models, i.e.:

$$\hat{\delta}_{i,j}(\hat{\eta}) = \hat{L}_{DR}(\tau_i; \hat{\eta}) - \hat{L}_{DR}(\tau_j; \hat{\eta}), \tag{15.2.3}$$

approximate well *differences* in mean squared error, i.e.:

$$\delta^*_{i,j} = \mathrm{E}(\tau_i(X) - \tau_0(X))^2 - \mathrm{E}(\tau_j(X) - \tau_0(X))^2 \tag{15.2.4}$$

Consider the population analogues of the score and differences in the score, i.e.:

$$L_{DR}(\tau;\eta) := \mathrm{E}(Y(\eta) - \tau(X))^2 \quad (15.2.5)$$

$$\delta_{i,j}(\eta) := L_{DR}(\tau_i;\eta) - L_{DR}(\tau_j;\eta) \quad (15.2.6)$$

Since $\mathrm{E}[Y(\eta_0) \mid X] = \tau_0(X)$, we have that:[5]

5: Prove this as an exercise.

$$\delta_{i,j}(\eta_0) = \delta^*_{i,j} \quad (15.2.7)$$

Moreover, note that the population difference at the estimate $\hat{\eta}$ satisfies (by simply expanding the squares):

$$\begin{aligned}\delta_{i,j}(\hat{\eta}) &= \mathrm{E}[\tau_i(X)^2 - \tau_j(X)^2 - 2\,Y(\hat{\eta})\,(\tau_i(X) - \tau_j(X))] \\ &= \mathrm{E}[\tau_i(X)^2 - \tau_j(X)^2 - 2\,\mathrm{E}[Y(\hat{\eta}) \mid X]\,(\tau_i(X) - \tau_j(X))]\end{aligned}$$

Thus the error in the model comparison due to the error in the estimate $\hat{\eta}$ is:

$$\delta_{i,j}(\hat{\eta}) - \delta_{i,j}(\eta_0) = 2\,\mathrm{E}[\mathrm{E}[Y(\eta_0) - Y(\hat{\eta}) \mid X]\,(\tau_i(X) - \tau_j(X))]$$

We see that a good scoring rule, should be using proxy labels that have small bias, i.e.:

$$\mathrm{bias}(X;\hat{\eta}) := \mathrm{E}[Y(\eta_0) - Y(\hat{\eta}) \mid X] \quad (15.2.8)$$

The doubly robust proxy labels exactly achieve this property. In particular, we can show based on results in prior sections:[6]

6: Prove this as an exercise.

$$\mathrm{bias}(X;\hat{\eta}) = (H(\mu_0) - H(\hat{\mu}))\,(g_0(D,Z) - \hat{g}(D,Z)) \quad (15.2.9)$$

Thus we derived that the error in the comparison between model $\tau_i$ and model $\tau_j$, due to the estimation error in $\hat{\eta}$ is:

$$2\,\mathrm{E}[(H(\mu_0) - H(\hat{\mu}))\,(g_0(D,Z) - \hat{g}(D,Z))\,(\tau_i(X) - \tau_j(X))]$$

This has doubly robust properties, i.e. if either $H(\hat{\mu})$ is accurate or $\hat{g}$ is accurate, then the comparison between the two models will be accurate. Moreover, the difference in scores also satisfies the Neyman orthogonality property:[7]

7: Prove this as an exercise.

$$\partial_\eta \delta_{i,j}(\eta_0) = 0 \quad (15.2.10)$$

and since $\hat{\delta}_{i,j}(\eta)$ is the empirical analogue of $\delta_{i,j}(\eta)$, we can apply the general framework of Neyman orthogonality to deduce that the score difference estimate $\hat{\delta}_{i,j}(\hat{\eta})$ is root-$n$ asymptotically normal:[8]

8: A similar theorem also holds for the case of cross-fitted estimates. In practice, one can either use the nuisance estimates that were constructed on the training set, which was also used to construct the functions $\{\tau_1, \ldots, \tau_M\}$ or perform cross-fitting within the scoring set.

**Theorem 15.2.1** *Let* $\nu_{ij}(X) = \tau_i(X) - \tau_j(X)$ *and suppose that* $\mathbb{E}_n \nu_{ij}(X)^2 \geq c$ *for some constant* $c > 0$ *and let* $n$ *grow to infinity. As long as* $\hat{\mu}$ *and* $\hat{g}$ *are estimated on a separate sample and satisfy that:*

$$\sqrt{n}\mathrm{E}[(H(\mu_0) - H(\hat{\mu}))\,(g_0(D,Z) - \hat{g}(D,Z))\,\nu_{ij}(X)] \approx 0$$

*and both nuisance functions are consistent, i.e.:*

$$\|\hat{\mu} - \mu_0\|_{L^2} + \|\hat{g} - g_0\|_{L^2} \approx 0 \tag{15.2.11}$$

*Then the estimation error in the nuisance functions* $\hat{\mu}, \hat{\eta}$, *does not have a first order effect in the estimation error of the score difference between two models:*

$$\sqrt{n}(\hat{\delta}_{i,j}(\hat{\eta}) - \delta^*_{i,j}) \approx \sqrt{n}\mathbb{E}_n(Y(\eta_0) - \tau_i(X))^2 - (Y(\eta_0) - \tau_j(X))^2$$

*Consequently, the estimate* $\hat{\delta}_{i,j}$ *concentrates in a* $1/\sqrt{n}$ *neigborhood of* $\delta^*_{i,j}$ *with deviations controlled by the Gaussian law:*

$$\sqrt{n}(\hat{\delta}_{i,j}(\hat{\eta}) - \delta^*_{i,j}) \overset{a}{\sim} N(0, \mathrm{V}) \tag{15.2.12}$$

*where:*

$$\mathrm{V} := \mathrm{E}\left((Y(\eta_0) - \tau_i(X))^2 - (Y(\eta_0) - \tau_j(X))^2 - \delta^*_{i,j}\right)^2$$

*Moreover, confidence intervals on the performance difference between two models can be constructed as:*

$$\mathrm{P}\left(\delta^*_{i,j} \in \left[\hat{\delta}_{i,j}(\hat{\eta}) \pm c\sqrt{\hat{\mathrm{V}}/n}\right]\right) \approx 1 - \alpha \tag{15.2.13}$$

*where* $c$ *is the* $(1-\alpha/2)$*-quantile of the standard normal distribution and*

$$\hat{\mathrm{V}} := \mathbb{E}_n\left((Y(\hat{\eta}) - \tau_i(X))^2 - (Y(\hat{\eta}) - \tau_j(X))^2 - \hat{\delta}_{i,j}(\hat{\eta})\right)^2$$

The above theorem can also directly be used to construct

**Remark 15.2.1** (Sample-dependent base models) The assumption that $\mathbb{E}_n \nu_{ij}(X)^2$ is some non-zero constant $c$ independent of $n$ required so that the variance $V$ is non-zero. In practice, the candidate models will also be changing with $n$ as we will be growing the sample size of the training set together with the scoring set. As the size of the training set converges to infinity it is highly probable that $c$ will be converging to zero, in which case $V$ will also be converging to zero. The above theorem allows one to compare models that are distinct in their average predictions by at least some constant. If we want to be comparing models whose distinctness (i.e. $\mathrm{E}\nu_{ij}(X)^2$) shrinks with the sample size, then we need to be more careful in the asymptotic normal approximation. In this case, it is more appropriate to consider the asymptotic properties of the self-normalized quanity:

$$\sqrt{\frac{n}{V_n}}(\hat{\delta}_{i,j}(\hat{\eta}) - \delta^*_{i,j}),$$

where $V_n$ is now allowed to depend on $n$, since $\tau_i, \tau_j$ are allowed to depend on $n$. In this case, to ignore the error due to $\hat{\eta}$ we would need that:

$$\sqrt{\frac{n}{V_n}}\mathrm{E}[(H(\mu_0) - H(\hat{\mu}))\,(g_0(D,Z) - \hat{g}(D,Z))\,\nu_{ij}(X)] \approx 0$$

As we show in Appendix 15.A:

$$V_n \geq 4\mathrm{E}\nu_{ij}(X)^2\,\mathrm{Var}(Y(\eta_0) \mid X)$$

Thus if we assume that $\mathrm{Var}(Y(\eta_0) \mid X) \geq c > 0$, then $V_n \geq 4c\|\nu_{ij}\|_{L^2}^2$ and it suffices that:

$$\sqrt{n}\mathrm{E}\left[(H(\mu_0) - H(\hat{\mu}))\,(g_0(D,Z) - \hat{g}(D,Z))\,\frac{\nu_{ij}(X)}{\|\nu_{ij}\|_{L^2}}\right] \approx 0$$

If $|\nu_{ij}(X)| \leq C\|\nu_{ij}\|_{L^2}$ almost surely, then the above would hold under the standard condition that:

$$\sqrt{n}\|H(\mu_0) - H(\hat{\mu})\|_{L^2}\,\|g_0 - \hat{g}\|_{L^2} \approx 0$$

Even when this condition does not hold, by an application of the Cauchy-Schwarz inequality it suffices that:

$$\sqrt{n}\sqrt{\mathrm{E}\left[(H(\mu_0) - H(\hat{\mu}))^2\,(g_0(D,Z) - \hat{g}(D,Z))^2\right]} \approx 0$$

which would hold if:

$$\sqrt{n}\|H(\mu_0) - H(\hat{\mu})\|_{L^4}\,\|g_0 - \hat{g}\|_{L^4} \approx 0 \tag{15.2.14}$$

Moreover, for the confidence interval to be valid, we would also need that:

$$\frac{|\mathsf{V} - \hat{\mathsf{V}}|}{\hat{\mathsf{V}}} \approx 0 \tag{15.2.15}$$

If the denominator is lower bounded by a constant, the above holds under benign regularity conditions. However, if we consider models whose separation shrinks at some rate $\sigma_n^2$, we can enforce the same rate of shrinkage on our estimate, of the variance, in which case we would need the estimation error in the variance to shrink faster than $\sigma_n^2$. Thus we can only consider comparison of models that are separated by at least some amount that dominates the error we expect in our variance estimate. This separation would always be of larger order than $1/n$. However, how small we can take this rate also depends on rates of convergence of our nuisance estimates individually and not just their product.

**Remark 15.2.2** (Normalized Interpretable DR-Score) The loss $\hat{L}_{DR}(\tau;\hat{\eta})$ might not be very interpretable in practice as the result depends on the unobserved heterogeneity of the outcome and on the units of the outcome. As a more interpretable performance metric we can consider comparing the loss of any candidate model as compared to the loss of the best constant effect model fitted on the training sample. Let $\hat{\tau}_c$ denote the constant effect model that always outputs the estimate $\hat{\theta}$ of the average treatment effect, estimated on the training data. Then we can define the normalized score:

$$\hat{S}(\tau;\hat{\eta}) = \frac{\hat{L}_{DR}(\hat{\tau}_c;\hat{\eta}) - \hat{L}_{DR}(\tau;\hat{\eta})}{\hat{L}_{DR}(\hat{\tau}_c;\hat{\eta})} \tag{15.2.16}$$

This can be interpreted as a relative improvement in performance over a constant model and is a number in $[-\infty, 1]$. A larger score hints at a better CATE model. Moreover, for any reasonable model model this score will be a non-negative number in $[0, 1]$.

## Competing with the Best Model

The doubly robust loss can also be used for constructin an ensemble $\tau_*$ that competes with the best model in $T$. The simplest approach would be to choose the model with the best score, i.e.:

$$\tau_* = \arg\min_{\tau\in T} \hat{L}_{DR}(\tau;\hat{\eta}) \tag{15.2.17}$$

Such a model satisfies the oracle performance guarantee in Equation (15.2.1) with (see e.g. [4])

$$\epsilon(n, M) \lesssim \sqrt{\frac{\log(M)}{n}} + \|H(\mu_0) - H(\hat{\mu})\|_{L^2} \, \|g_0 - \hat{g}\|_{L^2}$$

The leading term in this result is unfortunate, since it does not decay fast with the sample size $n$, i.e. as $1/n$. For instance, for parametric base models, we would expect the base models to have RMSE performance of $\lesssim 1/n$, in which case the above $1/\sqrt{n}$ rate becomes a dominant term.

One problem with this approach is the non-convexity of the space of models over which we are optimizing (i.e. optimizing over singleton models). This non-convexity can be alleviate by stacking approaches that convexify the optimization space over which we optimize and minimize the doubly robust loss over linear combinations of the base cate models, i.e.:

$$\tau_* := \sum_{i=1}^{M} w_i^* \tau_i, \quad w^* := \arg\min_{w \in W} \hat{L}_{DR}\left(\sum_{i=1}^{M} w_i \tau_i ; \hat{\eta}\right) \tag{15.2.18}$$

where $W$ could either be $\mathbb{R}^M$, in which case this is simply OLS regression with covariates $\tau_1(X), \ldots, \tau_M(X)$ and target outcome $Y(\hat{\eta})$,[9] or $W$ could be the $M$-dimensional simplex, i.e.

$$W := \left\{ w \in \mathbb{R}^M : w_i \geq 0, \sum_{i=1}^{M} w_i = 1 \right\},$$

9: This is mathematically equivalent to the BLP approach we described in the first section of this chapter, albeit using the predictions of the base CATE models as the engineered features.

in which case this corresponds to a convex regression with the same covariates and outcome as in the OLS case. In the absence of further assumptions on the quality of the base models $\tau_i$, the above yield a model $\tau_*$ that satisfies the oracle performance guarantee in Equation (15.2.1) with (see e.g. [4, 14])

$$\epsilon(n, M) \lesssim \min\left\{ \frac{M}{n} + \|H(\mu_0) - H(\hat{\mu})\|_{L^4} \, \|g_0 - \hat{g}\|_{L^4}, \right.$$
$$\left. \sqrt{\frac{\log(M)}{n}} + \|H(\mu_0) - H(\hat{\mu})\|_{L^2} \, \|g_0 - \hat{g}\|_{L^2} \right\}$$

The above approach yields a fast rate guarantee with respect to the sample size, but suffers from a large set of base models $M$. The reason being that the convexification of the optimization space introduced $M$ parameters that correspond to the weights for each model and no penalty to encourage sparsity of the solution.

One can achieve the ideal leading rate of $\log(M)/n$, that is both fast with respect to the sample size $n$ and grows only logarith-mically with the number of base models $M$, by a penalized stacking approach called Q-aggregation [15], which penalizes different models based on their individual performance:

$$w = \arg\min_{w \in W} \hat{L}_{DR}\left(\sum_{i=1}^{M} w_i \tau_i; \hat{\eta}\right) + \sum_{i=1}^{M} w_i \hat{L}_{DR}(\tau_i; \hat{\eta}) \quad (15.2.19)$$

where $W$ is the $M$-dimensional simplex. This is an $M$-dimensional convex optimization program that can be solved very fast with modern convex optimization solvers. The resulting ensemble model competes with the best model at the statistically optimal leading rate of (see [16]):

$$\epsilon(n, M) \lesssim \frac{\log(M)}{n} + \|H(\mu_0) - H(\hat{\mu})\|_{L^4} \|g_0 - \hat{g}\|_{L^4}$$

**Remark 15.2.3** (ATE and Intercept of Stacked Model) In practice, one might also include an intercept in the stacking model, i.e.

$$w = \arg\min_{w \in W, c \in \mathbb{R}} \hat{L}_{DR}\left(c + \sum_{i=1}^{M} w_i \tilde{\tau}_i; \hat{\eta}\right) \quad (15.2.20)$$

where $\tilde{\tau}_i$ are de-meaned versions of the CATE models, i.e. $\tilde{\tau}_i(X) = \tau_i(X) - \mathbb{E}_n \tau_i(X)$. In this case, the constant $c$ can be interpreted as the final estimate of the Average Treatment Effect. Given that typically the scoring dataset will be smaller than the training dataset, it might be more advisable to use an estimate of the ATE that comes from the trainin dataset. For instance, we can use as $c$ the doubly robust estimate of the ATE from the training dataset, denoted as $\hat{\theta}_{DR}^{\text{train}}$ and not optimize over it in the scoring dataset, i.e.

$$w = \arg\min_{w \in W} \hat{L}_{DR}\left(\hat{\theta}_{DR}^{\text{train}} + \sum_{i=1}^{M} w_i \tilde{\tau}_i; \hat{\eta}\right) \quad (15.2.21)$$

**Remark 15.2.4** (Generic Stacking for CATE) Another approach that is typically employed in regression stacking is using more flexible stacking regressors. In the case of stacking for the CATE we can treat the ensemble problem as yet another regression problem of predicting $Y(\hat{\eta})$ from the covariates

$$\tau \circ X := (\tau_1(X), \ldots, \tau_M(X)),$$

using the scoring dataset. Thus we can call a generic ML regression oracle to get $\tau_*$:

$$\tau_* := O_{T^*}\left(\{\tau \circ X_i, Y_i(\hat{\eta})\}_{i=1}^n\right)$$

However, in this case one should worry about variance and overfitting, as typically the scoring dataset will be of smaller size than the training dataset. Thus, very flexible models can deteriorate the performance. The benefits of more general stacking regressors are not clear if one wants to solely compete with the best base model. However, more general regressors can potentially find models that perform better than the best base model. In practice, the most commonly used oracle models are penalized linear models, such as Lasso or Ridge regression, potentially with postivity constraints on the coefficients.

**Remark 15.2.5** (Stability) The final CATE ensemble model that we selected based on the aforementioned process (i.e. training generic meta-learner models on a training set and scoring and stacking on a test set) unfortunately does not come with confidence intervals. Even though this is a process that can lead to a model with small mean squared error, the exact model can be quite sensitive to small variants of the data analysis pipeline (e.g. the randomness in the train/score split, the randomness in the estimators, the removal of a few samples). Even though we cannot construct valid confidence intervals for the predictions of the model or the findings in its structure (i.e. which are the important features), it is still advisable to perform some form of sensitivity or stability check of the model to such variants. For instance, one can run the same pipeline with different random seeds or remove random small fractions of the data and see how stable the different aspects of the model are. In the next section, we will also discuss more formal statistical tests that one can perform on a separate validation set (i.e. if one splits their data into train/scoring/validation sets), which validate aspects of the chosen CATE model.

**Example 15.2.1** (Model Selection in Simple DGPs) We revisit the three data generating processes from Example 15.1.1. We depict below the performance of the Q-aggregation ensemble in a random sample of each of the data generating processes ($n = 500$).

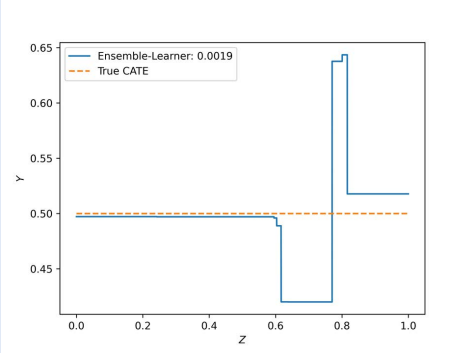

**(a)** DGP 1

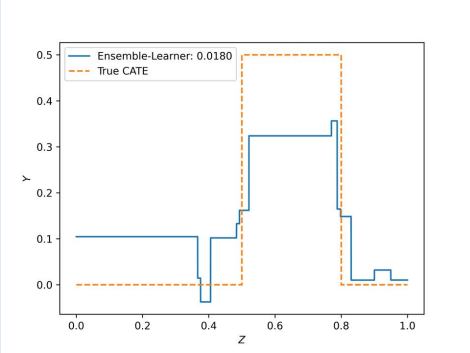

**(b)** DGP 1

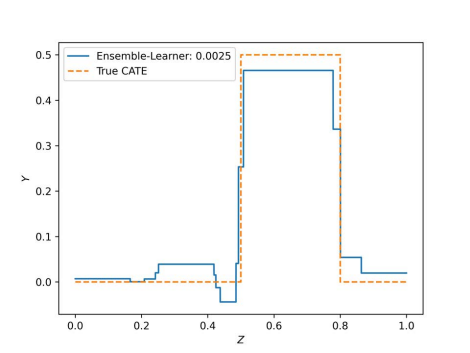

**(c)** DGP 1

Moreover, in Table 15.10, we depict the average performance of each meta-learning method and of three variants of the ensemble methods (based on Q-aggregation, convex regression and simply choosing the single best score model) in terms of CATE RMSE over 100 experiments. We find that even though different learners are optimal in each of the DGPs, the ensemble learners are consistently close to the best performer across the board, while each of the other learners fails by a large margin in at least one DGP.

| | DGP 1 | DGP 2 | DGP 3 |
|---|---|---|---|
| DR | [0.016 ± 0.012]<br>0.015 (0.036) | [0.193 ± 0.036]<br>0.195 (0.243) | [0.049 ± 0.011]<br>0.047 (0.070) |
| DRX | [0.012 ± 0.009]<br>0.010 (0.030) | [0.193 ± 0.037]<br>0.195 (0.243) | [0.178 ± 0.034]<br>0.181 (0.230) |
| R | [0.002 ± 0.007]<br>0.000 (0.020) | [0.374 ± 0.080]<br>0.418 (0.421) | [0.374 ± 0.079]<br>0.418 (0.421) |
| T | [0.038 ± 0.005]<br>0.037 (0.047) | [0.235 ± 0.007]<br>0.232 (0.245) | [0.036 ± 0.011]<br>0.035 (0.056) |
| X | [0.010 ± 0.008]<br>0.008 (0.028) | [0.235 ± 0.007]<br>0.232 (0.245) | [0.223 ± 0.006]<br>0.221 (0.235) |
| DAX | [0.013 ± 0.008]<br>0.010 (0.030) | [0.156 ± 0.053]<br>0.151 (0.243) | [0.155 ± 0.045]<br>0.149 (0.232) |
| Q | [0.017 ± 0.014]<br>0.014 (0.038) | [0.165 ± 0.049]<br>0.161 (0.243) | [0.037 ± 0.012]<br>0.035 (0.056) |
| Convex | [0.019 ± 0.013]<br>0.018 (0.037) | [0.163 ± 0.042]<br>0.164 (0.236) | [0.038 ± 0.011]<br>0.037 (0.056) |
| Best | [0.017 ± 0.015]<br>0.011 (0.042) | [0.171 ± 0.055]<br>0.164 (0.269) | [0.037 ± 0.013]<br>0.036 (0.061) |

**Figure 15.10:** CATE RMSE performance of each of the meta-learning and ensemble methods in the three simple DGPs across 100 experiments. Each cell displays [mean ± standard deviation], median (95%) of the RMSE across the 100 experiments.

**Example 15.2.2** (Model Selection in the 401(k) Example) We also revisit the 401(k) example from the perspective of model comparison and ensembling. We first investigate the comparison of each of the performance of each of the meta-learning models compared to the constant model in Figure 15.11. We find that we cannot statistically conclude that the RMSE performance of any of these models is better than the constant effect model.

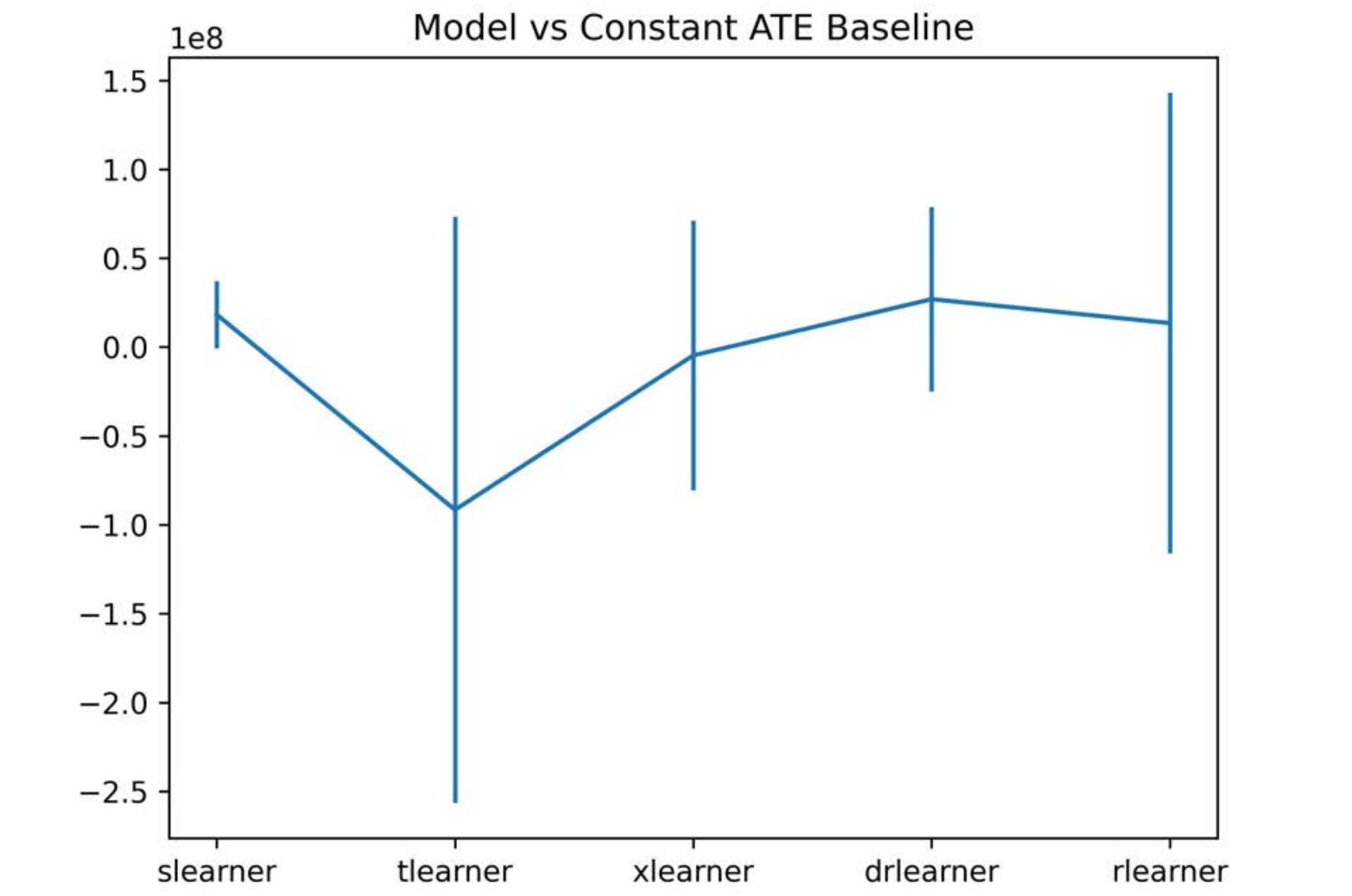

**Figure 15.11:** Score difference and confidence interval for each of the meta-learner models as compared to a constant treatment effect baseline. We find that among all models, only for the s-learner we can barely find that it has better CATE evaluation accuracy as compared to a constant effect model with statistical significance.

Subsequently, we investigate the ensemble models that are selected by each stacking method. We find that very flexible methods such as OLS, or Ridge can be quite un-stable, while methods that either constrain the weights to be positive or lie in the simplex, or induce sparse ensembles are qualitatively more reasonable.

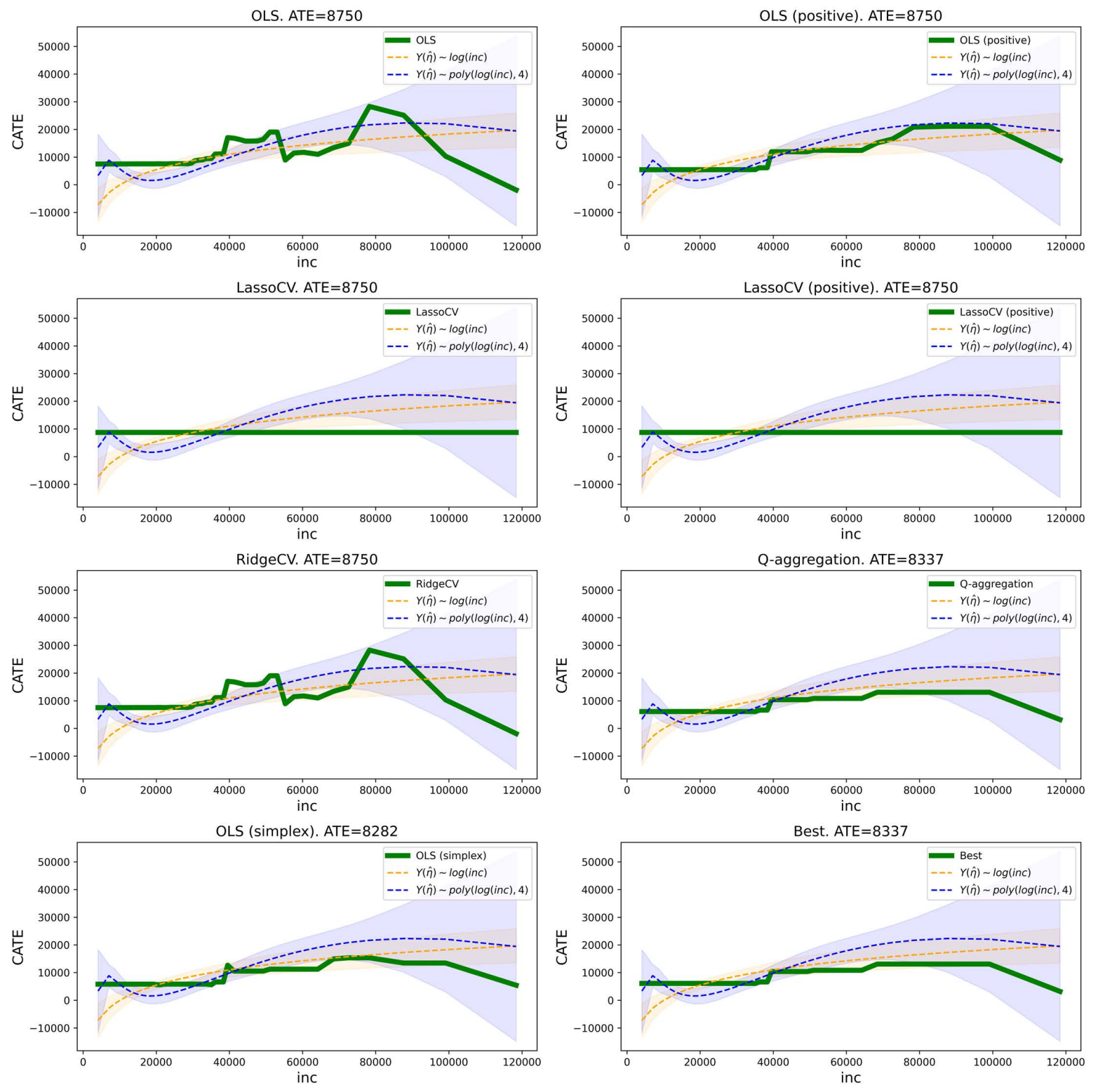

**Figure 15.12:** CATE predictions of different stacked ensemble models in the 401k example. Gradient boosted forests (via the xgboost library) were used as ML oracles for regression and classification. The CATE is predicted on a grid of income points, corresponding to equally spaced income quantiles. All other covariates were imputed at their median values. For comparison, each plot also displays the doubly robust best linear predictor of the CATE with 5-95% confidence intervals on a simple linear form of engineered features of the income.

Subsequently, we investigate for interpretability reasons, the main factors that are driving the predictions of the ensemble model chosen by Q-aggregation. We do this by fitting a simple shallow binary regression tree on the predictions of the model. Given that the ensemble chooses to put weight primarily on the DR-Learner model, the insights are similar to those in Example 15.1.4.

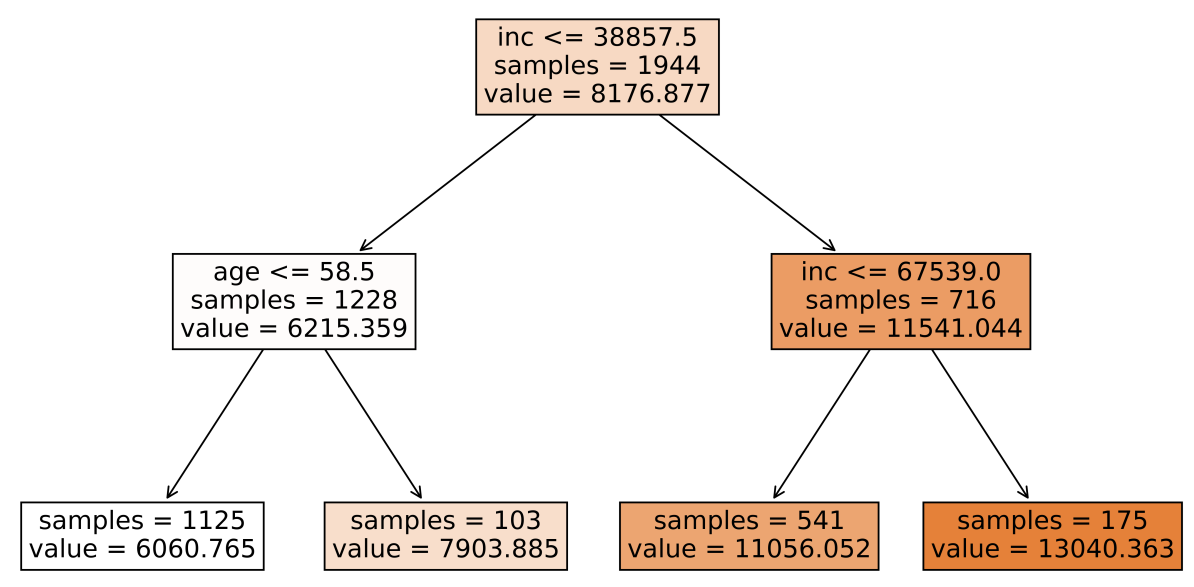

**Figure 15.13:** Single binary regression tree distillation of the Q-aggregation based stacked ensemble.

## 15.3 CATE Model Validation

Now that we have a selected a winning CATE model or ensemble (e.g., the ensemble that comes out of Q-aggregation on the scoring data), we want to run formal statistical tests that validate whether the model that we chose contains any signal of treatment effect heterogeneity, or whether it is a confident model on average, or whether it is a useful model to drive personalized policy decisions as compared to simple benchmarks. We will refer to all these methodologies as CATE model validation and all the techniques can be thought as diagnostics that one should run on their CATE model before deployment or before using it to drive personalized decisions. As a side benefit, many of these diagnostics can also be used as a formal statistical test of the presence of treatment effect heterogeneity.

Throughout this section we assume that one has held out yet another dataset, called the test set (e.g., by splitting their data into train, validation and test) and that one has selected a CATE model $\tau_*$ without using the test set (e.g., by running some ensemble pipeline on on the train and validation set).

## Heterogeneity Test Based on Doubly Robust BLP

If we calculate the doubly robust pseudo outcomes $Y^{DR}(\hat{\eta})$ on the test set (using cross-fitting within the test set to estimate the models $\hat{\eta}$ or using the union of the training and scoring data to estimate $\hat{\eta}$). Then we know that if the model of the CATE $\tau_*$ is good, the best linear predictor of the true CATE using $(1, \tau_*(X))$ as features should yield a statistically significant coefficient on the feature associated with the CATE model. In fact, in an ideal world this coefficient should be 1.

Thus we can run such a significance test to measure whether the CATE model $\tau_*$ has picked up any signal that is correlated with the true CATE. Note that if $\tau_0(X)$ is the true CATE $\mathrm{E}[Y(1)-Y(0) \mid X]$, then the coefficient associated with $\tau_*$ in the OLS regression $Y(\hat{\eta}) \sim (1, \tau_*(X))$ is converging in the population limit to the quantity:

$$\beta_1 := \frac{\mathrm{Cov}(\tau_0(X), \tau_*(X))}{\mathrm{Var}(\tau_*(X))} = \frac{\mathrm{Cov}(Y(1) - Y(0), \tau_*(X))}{\mathrm{Var}(\tau_*(X))} \tag{15.3.1}$$

Thus, the statistical test of whether $\beta_1$ is non-zero is a statistical test on the correlation of the individual treatment effect $Y(1) - Y(0)$ and the learned model $\tau_*(X)$. Note that if this test comes up as significant, then this also implies that there exists treatment effect heterogeneity, as a function of the observed features $X$. Moreover, the theory from the first section of this chapter applies here to show that the statistical test based on OLS regression is a valid test, as long as the product of the regression and propensity estimation errors, converges faster than $n^{-1/2}$.

**Example 15.3.1** (Heterogeneity Test in 401(k) Example) Returning to our 401k example, we can split the data in three sets (train, score, test) and employ the heterogeneity test on the score set using the ensemble model chosen based on the Q-aggregation method of stacking. Out of the 9716 samples used in this analysis, 60% were used for training and 20% for scoring and 20% for testing. For the training of the nuisance parameters $\hat{\eta}$ that are used in the doubly robust proxy labels in the test set, we used the union of the (train, score) datasets, due to the relatively small size of the test set. Moreover, for better interpretability of the results we ran OLS on the centered CATE predictions, i.e. on features $(1, \tau_*(X) - \mathbb{E}_n \tau_*(X))$. This does not change the intepretation of $\beta_1$, but it changes the interpretation of the constant term as the ATE (which wouldn't have been the case without centering).

The results are depicted in Figure 15.14. We find that the ensemble model does indeed have a statistically significant correlation with the individual treatment effect $Y(1) - Y(0)$ and that the confidence interval on that coefficient includes the ideal coefficient of 1. We did find, however, large variation of the specific numbers reported in the table, dependent on the random split that was chosen in (train, score, test) sets from the original data.

| | **coef** | **std err** | **P> \|z\|** | **[0.025** | **0.975]** |
|---|---|---|---|---|---|
| **const** | 7634.4018 | 1805.185 | 0.000 | 4096.303 | 1.12e+04 |
| $\tau_*(X)$ | 2.2406 | 0.718 | 0.002 | 0.833 | 3.648 |

**Figure 15.14:** OLS statistical test regression $Y^{DR}(\hat{\eta})$ on the features $(1, \tau_*(X) - \mathbb{E}_n\tau_*(X))$ in the 401(k) example. Standard Errors are heteroscedasticity robust (HC1). $\tau_*$ corresponds to the stacked ensemble based on Q-aggregation.

## Validation Based on Calibration

A good CATE model should also be well calibrated. In the context of regression, a regression model $g(X)$ that predicts some outcome $Y$, is calibrated if the expected value of the outcome, conditional on the model returning a value of $\gamma$, should be equal to $\gamma$, i.e.

$$\mathrm{E}[Y \mid g(X) = \gamma] = \gamma. \tag{15.3.2}$$

Similarly, we can say that a CATE model $\tau_*$ is calibrated if the expected value of the treatment effect, conditional on the model returning a value of $t$, should be equal to $t$, i.e.

$$\gamma(\tau_*, t) := \mathrm{E}[Y(1) - Y(0) \mid \tau_*(X) = t] = t \tag{15.3.3}$$

Moreover, if we let $\mathrm{P}_{\tau_*}$ denote the distribution of treatment effects returned by model $\tau_*$, then we can define average calibration scores across different values of $t$. Some popular measures defined in the literature [17–19] are either the $\ell_2$ or the $\ell_1$-expected calibration error:

$$\mathrm{CAL}_1(\tau_*) := \int |\gamma(\tau_*, t) - t|\, d\mathrm{P}_{\tau_*}(t) \tag{15.3.4}$$

$$\mathrm{CAL}_2(\tau_*) := \int (\gamma(\tau_*, t) - t)^2 d\mathrm{P}_{\tau_*}(t) \tag{15.3.5}$$

In fact, an interesting property of the $\ell_2$-calibration error, is that the MSE of a CATE model $\tau_*$ satisfies a calibration-distortion decomposition (analogous to the bias-variance decomposition):

$$\|\tau_* - \tau_0\|_{L^2} = \mathrm{CAL}_2(\tau_*) + \mathrm{DIS}(\tau_*) \tag{15.3.6}$$

where $\text{DIS}(\tau_*) = \text{E}\left[\text{Var}(\tau_0(X) \mid \tau_*(X))\right]$. Thus any consistent $\tau_*$ model will eventually also be calibrated. However, calibration is a self-consistency guarantee that should be desireable for many models and should not account for the majority of the MSE. Moreover, even if a model is far from $\tau_0$, it is still desirable from a "steakholder experience" perspective that it should be calibrated.

The aforementioned desiderata can be taken to data by invoking again the proxy outcome regression approach. In particular, note that by the properties of the doubly robust proxy labels:

$$\gamma(\tau_*, t) = \text{E}[Y(\eta_0) \mid \tau_*(X) = t] \tag{15.3.7}$$

we can use observed data and out-of-sample estimates $\hat{\eta}$ of the nuisance functions $\eta_0$, to measure the calibration properties of a candidate CATE model.

To avoid having to run a non-parametric regression of $Y(\eta_0)$ on $\tau_*(X)$, in order to estimate the function $\gamma(\tau_*, t)$, a typical way that calbiration is evaluated is by looking at quantile bins of the distribution of CATE. For instance, if we let $q_1 \leq \ldots \leq q_K$ denote a set of $K$ equally spaced quantiles of the distribution $\text{P}_{\tau_*}$, then a well-calibrated model should satisfy that:

$$\text{E}[Y(\eta_0) \mid \tau_*(X) \in [q_t, q_{t+1}]] = \text{E}[\tau_*(X) \mid \tau_*(X) \in [q_t, q_{t+1}]]$$

In other words, consider any group $G_t$, defined defined by some quantile interval $[q_t, q_{t+1}]$ of the predictions of the model $\tau_*$. Then the group average treatment effect (GATE) for the group $G_t$, should be the same, whether we calculate it by using the doubly robust GATE, i.e., $\text{E}[Y(\eta_0) \mid X \in G_t]$ or whether we calculate it by using the average CATE value of the model $\tau_*$ within that group, i.e., $\text{E}[\tau_*(X) \mid X \in G_t]$.

We can now easily take the latter approach to data. For some small $K$ (e.g. $K = 4$), we can consider a set of thresholds $q_1 \leq \ldots \leq q_{K+1}$ that roughly approximate equally spaced quantiles of the CATE distribution $\text{P}_{\tau_*}$ and which are calculate without looking at the test sample (e.g. this can be calculated as the empirical quantiles of the empirical distribution of values of the model $\tau_*$ on the union of the training and scoring samples). These now define a set of $K$ groups, $G_1, \ldots, G_K$ as described in the previous paragraph. Subsequently, we can estimate the GATE for each group, using the doubly robust approach on the test data, i.e.

$$\hat{\theta}_k^{DR} = \frac{1}{|\{i \in [n] : X_i \in G_k\}|} \sum_{i \in [n]: X_i \in G_k} Y_i(\hat{\eta}) \tag{15.3.8}$$

where $\hat{\eta}$ is either estimated in a cross-fitting manner on the test set or using the union of training and scoring samples. Equivalently, we can simulaneoulsy estimated all these parameters by running OLS of $Y(e\hat{t}a)$ on the one-hot-encodings of the group membership indicator functions, as in the first section of the chapter. Moreover, confidence intervals can be directly obtained for these values (e.g. based on the OLS heteroskedasticity robust confidence intervals or based on the simple formula for the standard error of an average of i.i.d. observations; in this case we have the average of the $|\{i \in [n] : X_i \in G_k\}|$ observations $\{Y_i(\hat{\eta}) : i \in [n], X_i \in G_k\}$). These confidence intervals can also be used to test whether these different groups have statistically signficant different average treatment effects, i.e. whether the groups are separated statistically.

Moreover, these estimates can then also be used to construct approximate analogues of the $\ell_2$ and $\ell_2$-average calibration scores. For each group $G_k$, we can also calculate the average value of the model $\tau_*$, i.e.,

$$\hat{\theta}_k^{*} = \frac{1}{|\{i \in [n] : X_i \in G_k\}|} \sum_{i \in [n]: X_i \in G_k} \tau_*(X_i) \tag{15.3.9}$$

Ideally, if the model was reasonable, $\hat{\theta}_k^*$ should be very close to $\hat{\theta}_k^{DR}$. The average difference can be considered as a quality metric of $\tau_*$, i.e.,

$$\widetilde{\mathrm{CAL}}_1(\tau_*) := \sum_{k=1}^{K} \left|\hat{\theta}_k^{DR} - \hat{\theta}_k^*\right| \cdot |\{i \in [n] : X_i \in G_k\}| \tag{15.3.10}$$

$$\widetilde{\mathrm{CAL}}_2(\tau_*) := \sum_{k=1}^{K} \left(\hat{\theta}_k^{DR} - \hat{\theta}_k^*\right)^2 \cdot |\{i \in [n] : X_i \in G_k\}| \tag{15.3.11}$$

These can be viewed as binning approximations to the $\ell_1$- and $\ell_2$-average calibration scores (the first one was recommended as a calibration score in the context of randomized trials by [20]).

**Example 15.3.2** (Calibration in the 401(k) example) We revisit the 401(k) example from the perspective of calibration. Following the same data analysis pipeline as in Example 15.3.1, we now also calculate the doubly robust GATEs for groups defined by quartiles of the CATE distribution of the ensemble $\tau_*$ constructed based on Q-aggregation stacking. The bottom group corresponds to the bottom 25% of predicted CATEs, the next group to the 25%-50% of predicted CATEs, etc. In Figure 15.15, we depict on the x-axis the average CATEs, as calculated based on $\tau_*$, within each group and on the y-axis and the doubly robust estimate and 5-95% confidence interval for the GATE as calculated based on the doubly robust proxy labels $Y(\hat{\eta})$ on the test set.

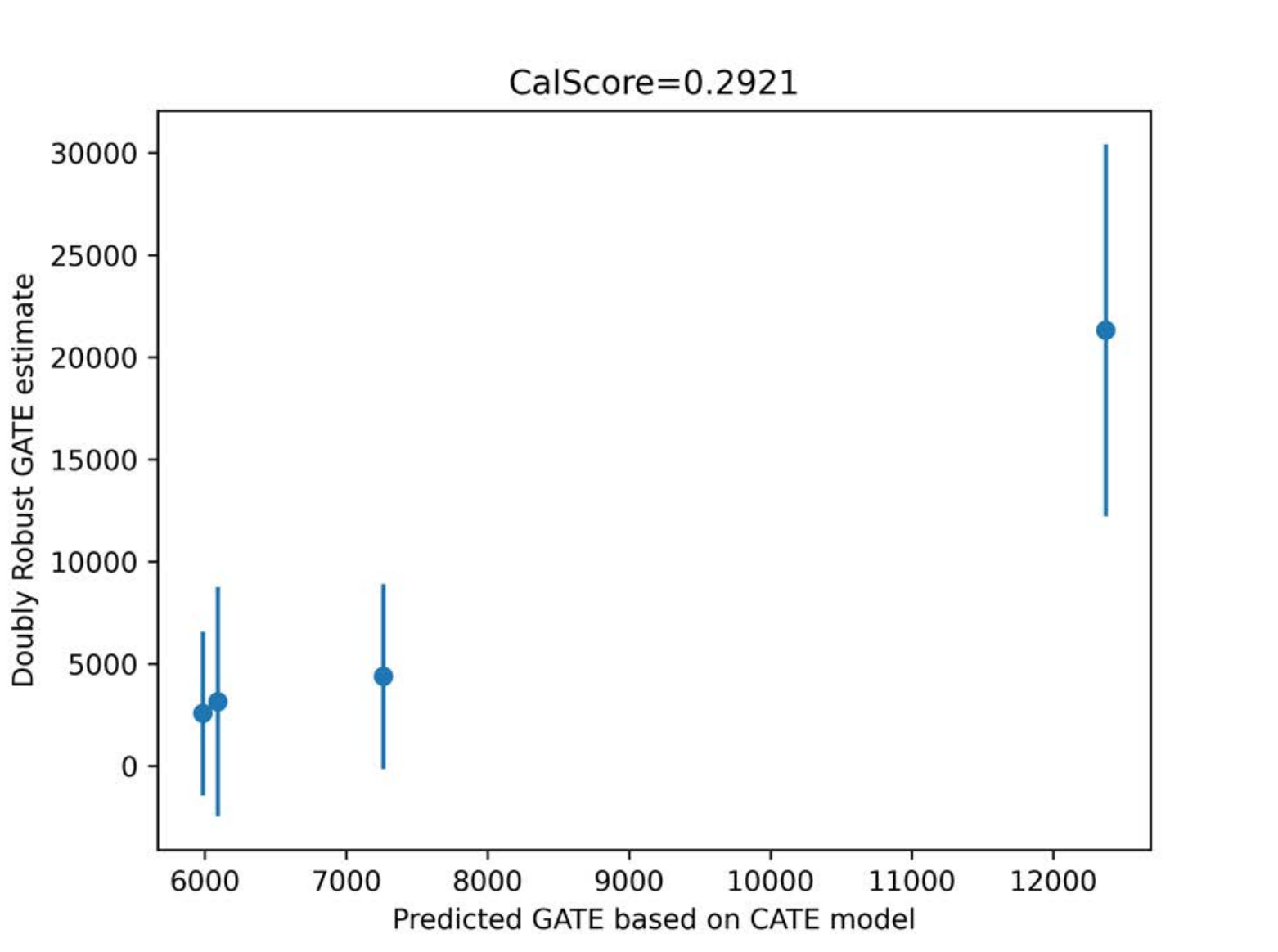

**Figure 15.15:** Calibration check for chosen ensemble model $\tau_*$ in the 401(k) example. Test samples are splitted in four groups based on CATE predictions and CATE quantiles (e.g. bottom group contains samples whose CATE predictions lie in the bottoms 25% of predictions). The x-axis depicts the average predicted CATE within each group based on $\tau_*$, while the y-axis depicts the GATE as calculated based on the doubly robust pseudo-outcomes calculated on the test set.

**Interpretation via Distillation and Group Differences.** We can also try to interpret what are the differences of characteristics between the top and bottom CATE groups; if we find that they have statistically significantly different GATEs. We can do that by either reporting the mean values of the covariates in the two groups or building some interpretable classification model that distinguishes between the two groups.

| | group1<br>mean ± s.e. | group2<br>mean ± s.e. | group1 - group2<br>mean ± s.e. |
|---|---|---|---|
| age | 40.01 ± 0.27 | 42.56 ± 0.45 | -2.56 ± 0.72 |
| inc | 26898 ± 346 | 65771 ± 760 | -38873 ± 1106 |
| fsize | 2.82 ± 0.04 | 3.12 ± 0.07 | -0.30 ± 0.11 |
| educ | 12.77 ± 0.07 | 14.74 ± 0.11 | -1.97 ± 0.18 |
| db | 0.24 ± 0.01 | 0.37 ± 0.02 | -0.14 ± 0.03 |
| marr | 0.52 ± 0.01 | 0.84 ± 0.02 | -0.32 ± 0.03 |
| male | 0.22 ± 0.01 | 0.20 ± 0.02 | 0.02 ± 0.03 |
| twoearn | 0.29 ± 0.01 | 0.66 ± 0.02 | -0.37 ± 0.03 |
| pira | 0.16 ± 0.01 | 0.42 ± 0.02 | -0.26 ± 0.03 |
| nohs | 0.16 ± 0.01 | 0.02 ± 0.01 | 0.13 ± 0.02 |
| hs | 0.41 ± 0.01 | 0.26 ± 0.02 | 0.15 ± 0.03 |
| smcol | 0.24 ± 0.01 | 0.26 ± 0.02 | -0.02 ± 0.03 |
| col | 0.19 ± 0.01 | 0.46 ± 0.02 | -0.27 ± 0.03 |
| hown | 0.57 ± 0.01 | 0.84 ± 0.02 | -0.27 ± 0.03 |

**Figure 15.16:** Group differences between the top 25% predicted CATE group (group2) and the bottom 75% predicted CATE group (group1) in the 401k example.

For instance, we can train a shallow binary classification tree that tries to predict whether a sample comes from the bottom or the top group, based on $X$, using the union of samples from the two groups.

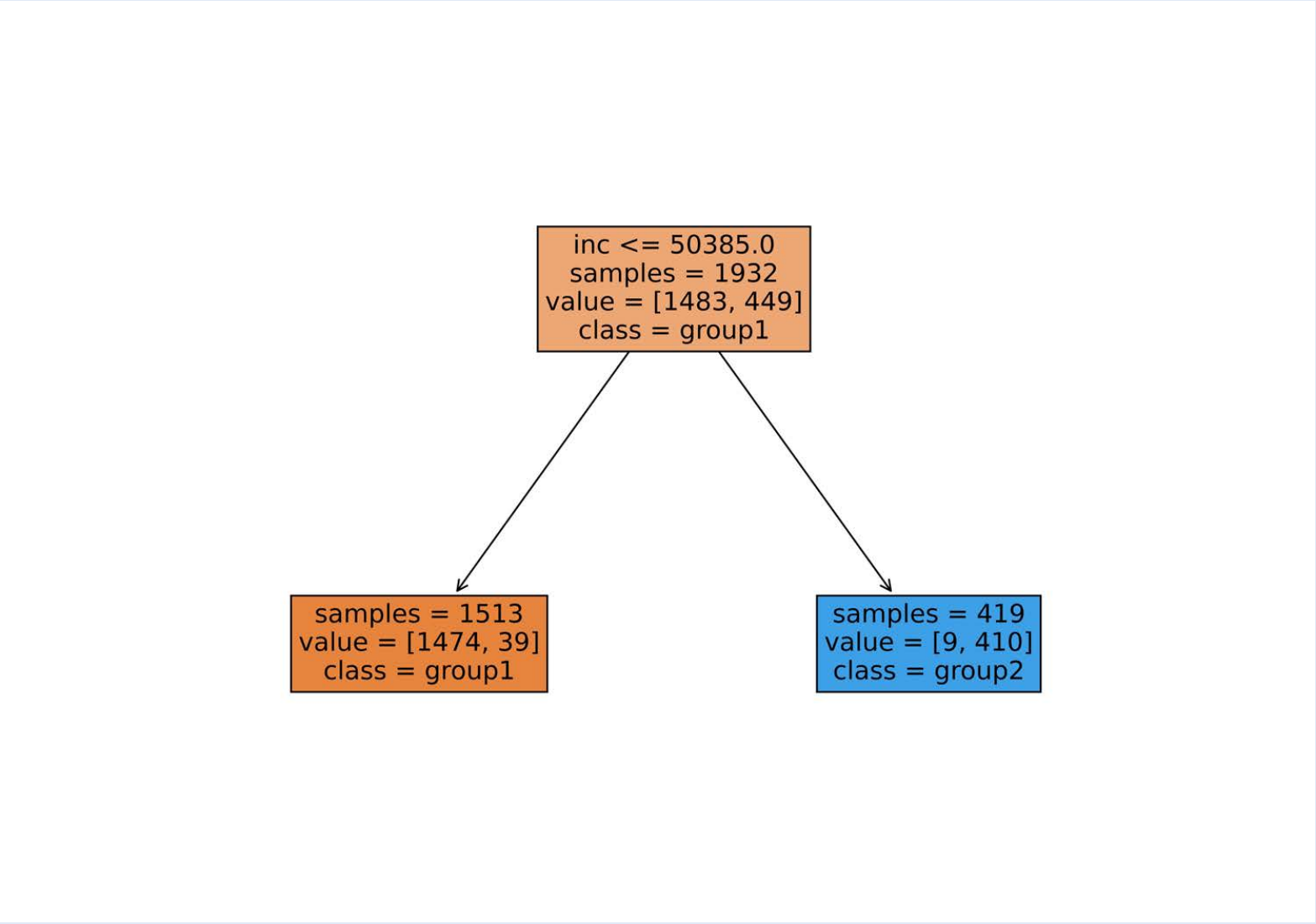

**Figure 15.17:** Decision tree that distills the main differences between group1 and group2 as defined in Figure 15.16.

## Validation Based on Uplift Curves

Another way that we can judge the quality of a CATE model is by testing its ability to help us prioritize or stratify which part of the population we should be treating. In Section 14.3, we studied the value of the optimal policy subject to treating exactly a $q$-fraction of the overall population. In a sense, how much this value varies or how different it is from the ATE is a measure of the amount of uplift offered by personalizing optimally based on $X$ using the true CATE $\tau_0$. We can study the same question from the lens of $\tau_*$, which can give a lesser or equal level of uplift, the higher the uplift the better the model. Unlike Section 14.3, where $\tau_0$ was a nuisance to be estimated and plugged into the optimal constrained or unconstrained policy, here $\tau_*$ is fixed and given (as it is based on a separate data set). In particular, our evaluation procedures would work even if $\tau_0$ were hard to learn or had discontinuities in its distribution, since we focus on a fixed $\tau_*$ instead.

Let $\mu(\tau_*, q)$ denote an estimate based on non-test data of the $1-q$ quantile of the distribution $\mathrm{P}_{\tau_*}$ of CATEs produced by the model $\tau_*$. Then the group $\{X : \tau_*(X) \geq \mu(\tau_*, q)\}$ is fixed in terms of the test data. The corresponding GATE is

$$\mathrm{GATE}(q) := \mathrm{E}[Y(1) - Y(0) \mid \tau_*(X) \geq \mu(\tau_*, q)] \tag{15.3.12}$$

The improvement in the average effect of the treated, induced by the prioritization rule based on $\tau_*$, as compared to treating a random $q$ fraction of the population, would be:

$$\mathrm{TOC}(q) := \mathrm{GATE}(q) - \mathrm{ATE} \tag{15.3.13}$$

and the improvement in the total effect would be:

$$\mathrm{QINI}(q) := \mathrm{TOC}(q)\,\mathrm{P}(\tau_*(X) \geq \mu(\tau_*, q)). \tag{15.3.14}$$

For any fixed CATE model $\tau_*$, the function $\mathrm{TOC}(\tau_*, \cdot)$ is referred to in the literature as the *Treatment Operating Characteristic* curve, while the function $\mathrm{QINI}(\tau_*, \cdot)$ is referred to as the *QINI* curve (analogous to the Gini curve for classification models).[10]

10: These terminologies primarily stem from the uplift modelling literature in Computer Science [21–23].

These curves also have interesting interpretations as covariances of the individual treatment effect $Y(1) - Y(0)$ with non-linear functions of the CATE model $\tau_*$ (see proofs in Appendix 15.B):

$$\begin{aligned}
\mathrm{TOC}(q) &= \mathrm{Cov}\left(Y(1) - Y(0), \frac{1\{\tau_*(X) \geq \mu(\tau_*, q)\}}{\mathrm{P}(\tau_*(X) \geq \mu(\tau_*, q))}\right) \\
\mathrm{QINI}(q) &= \mathrm{Cov}\left(Y(1) - Y(0), 1\{\tau_*(X) \geq \mu(\tau_*, q)\}\right)
\end{aligned}$$

Since the second term in each covariance is a function of $X$ alone and $\mathrm{E}[Y(1) - Y(0) \mid X] = \mathrm{E}[Y(\eta_0) \mid X]$, these quantities are identified by replacing the individual effects with the doubly robust pseudo-outcomes:

$$\begin{aligned}
\mathrm{TOC}(q) &= \mathrm{Cov}\left(Y(\eta_0), \frac{1\{\tau_*(X) \geq \mu(\tau_*, q)\}}{\mathrm{P}(\tau_*(X) \geq \mu(\tau_*, q))}\right) \\
\mathrm{QINI}(q) &= \mathrm{Cov}\left(Y(\eta_0), 1\{\tau_*(X) \geq \mu(\tau_*, q)\}\right)
\end{aligned}$$

**Area Under the Curve (AUC)** Viewing the above two quantities as functions of the target fraction $q$, we can calculate the areas under these two curves as scalar measures of quality of the model $\tau_*$ in its ability to correctly target sub-parts of the population at different levels of treatment population size targets, i.e.

$$AUTOC := \int_0^1 \mathrm{TOC}(q)dq \tag{15.3.15}$$

$$AUQC := \int_0^1 \mathrm{QINI}(q)dq \tag{15.3.16}$$

The larger the Area Under the Curve, the better the CATE model is at treatment prioritization or stratification.

Moreover, these measures are signals of treatment effect heterogeneity. If any of the two measures are statistically non-zero, then treatment effect heterogeneity was detected with statistical significance. In fact, if we detect that any of these curves lies above zero at any point $q$, with statistical significance, then that also serves as a test for treatment effect heterogeneity. For this reason, we will now develop confidence intervals and simultaneous confidence bands for these two curves, when they are estimated from samples.

**Remark 15.3.1** (Tie-Breaking) If our CATE model returns a constant effect for a large segment of the population, then the quantile estimate function $\mu(\tau_*, q)$ will contain many flat regions and ties cannot be ignored. In this case, if we want to approximately target a $q$-fraction of the population, then we should be treating deterministically everyone with $\tau_*(X) > \mu(\tau_*, q)$ and units with $\tau_*(X) = \mu(\tau_*, q)$, we should be treating with probability:

$$\frac{q - \mathrm{P}(\tau_*(X) > \mu(\tau_*, q))}{\mathrm{P}(\tau_*(X) = \mu(\tau_*, q))}$$

which corresponds to the probability mass that remains after treating everyone above $\mu(\tau_*, q)$ (i.e. $q - \mathrm{P}(\tau_*(X) > \mu(\tau_*, q))$), divided by the probability mass in the group of units that have predicted CATE equal to $\mu(\tau_*, q)$. We can then use an estimate of this quantity using the training and scoring datasets, e.g.

$$\lambda = \frac{q - \mathrm{P}_n(\tau_*(X) > \mu(\tau_*, q))}{\mathrm{P}_n(\tau_*(X) = \mu(\tau_*, q))}$$

and consider the policy that treats deterministically for units with $\tau_*(X) > \mu(\tau_*, q)$ and treats with probability $\lambda$ for units with $\tau_*(X) = \mu(\tau_*, q)$. In this case, the TOC and QINI curves will take a slightly more complex form:

$$\mathrm{Cov}\left(Y(\eta_0), \frac{1\{\tau_*(X) > \mu(\tau_*, q)\} + \lambda 1\{\tau_*(X) = \mu(\tau_*, q)\}}{\mathrm{P}(\tau_*(X) \geq \mu(\tau_*, q)) + \lambda \mathrm{P}(\tau_*(X) = \mu(\tau_*, q))}\right)$$
$$\mathrm{Cov}\left(Y(\eta_0), 1\{\tau_*(X) \geq \mu(\tau_*, q)\} + \lambda 1\{\tau_*(X) = \mu(\tau_*, q)\}\right)$$

All the conclusions and intution in what follows, directly extends to account for tie-breaking, so we will omit tie-breaking for simplicity of exposition.

**Estimation and inference.** To estimate the TOC and the QINI curves, we will use the doubly robust proxy outcome approach. We will train nuisance models $\hat{\eta}$ without using the test sample and then construct estimates of the TOC and QINI curves as:

$$\widehat{\mathrm{TOC}}(q) = \mathrm{Cov}_n\left(Y(\hat{\eta}), \frac{1\{\tau_*(X) \geq \mu(\tau_*, q)\}}{\hat{\pi}(q)}\right) \tag{15.3.17}$$

$$\widehat{\mathrm{QINI}}(q) = \mathrm{Cov}_n\left(Y(\hat{\eta}), 1\{\tau_*(X) \geq \mu(\tau_*, q)\}\right) \tag{15.3.18}$$

where $\hat{\pi}(q) = \mathbb{E}_n 1\{\tau_*(X) \geq \mu(\tau_*, q)\}$ and we used the shorthand notation:

$$\mathrm{Cov}_n(A, B) = \mathbb{E}_n\left[(A - \mathbb{E}_n(A))(B - \mathbb{E}_n(B))\right]$$

Both of these estimates are of the general estimation form that can be handled by the Neyman orthogonality framework. For each $q$, we can view each of the estimates as an estimate of the form:

$$\hat{\theta}(q; \nu) = \mathbb{E}_n[\psi(W; \nu)]$$

for some appropriate defined function $\psi$ and with $\nu$ being a vector of nuisance quantities, which contain $\eta$, $\pi$ and $\theta_0 = \mathrm{E}[Y(\eta_0)]$ and which satisfies Neyman orthogonality with respect to all of these nuisance quantities. Thus these estimates will be asymptotically Gaussian with the effect of the nuisances being negligible. Moreover, even if we evaluate these curves at many points, as long as the number of points $q$ that we use does not grow exponentially with the sample size, then these estimates will be jointly Gaussian and we can construct simultaneous confidence bands as in Section 4.4.

**Theorem 15.3.1** *Let $q \in Q := \{q_1, \ldots, q_p\}$ denote a grid of quantiles. Let $\alpha = (\alpha_1, \ldots, \alpha_p)$ denote the $p$-dimensional vector* *whose t-th coordinate is TOC(q) and $\hat{\alpha}$ the corresponding vector of estimates $\widehat{TOC}(q)$. Let $\mathbb{I}(q) := 1\{\tau_*(X) \geq \mu(\tau_*, q)\}$ and:*

$$\psi_\ell(W) = (Y(\eta_0) - \theta_0)\left(\frac{\mathbb{I}(q_\ell)}{\pi_0(q_\ell)} - 1\right) - \alpha_\ell$$

*with $\theta_0 = \mathrm{E}Y(\eta_0)$ and $\pi_0(q) = \mathrm{E}\mathbb{I}(q)$. Suppose that the nuisance estimates $\hat{\eta}$ are trained on a separate sample and satisfy,*

$$\sqrt{n}\|H(\hat{\mu}) - H(\mu_0)\|_{L^2}\|\hat{g} - g_0\|_{L^2} \approx 0$$
$$\|H(\hat{\mu}) - H(\mu_0)\|_{L^2} + \|\hat{g} - g_0\|_{L^2} \approx 0$$

*Provided that $\log(p)^5/n$ is small and the estimates satisfy the adaptivity property:*

$$\sqrt{\log(p)}\max_{\ell=1}^{p}\left|\sqrt{n}(\hat{\alpha}_\ell - \alpha_\ell) - \mathbb{E}_n\phi_\ell(W)\right| \approx 0$$

*the following Gaussian approximation holds:*

$$\sqrt{n}(\hat{\alpha} - \alpha_0) \overset{a}{\sim} N(0, V),$$

*where*

$$V_{\ell k} = \mathrm{E}\psi_\ell(W)\,\psi_k(W)$$

Analogous theorem also applies to the QINI curve estimates.

This result can be used to construct simultaneous confidence bands for the value of the TOC curve at many quantiles $q$ as described in Remark 4.4.1. We can consider the estimate of the variance:

$$\hat{V}_{\ell k} = \mathbb{E}_n\hat{\psi}_\ell(W)\hat{\psi}_k(W) \quad \hat{\psi}_\ell(W) = (Y(\hat{\eta}) - \hat{\theta})\left(\frac{\mathbb{I}(q_\ell)}{\hat{\pi}(q_\ell)} - 1\right) - \hat{\alpha}_\ell$$

and construct a confidence band at confidence level $\alpha$:

$$CR = \times_{\ell=1}^{p}[\hat{\alpha}_\ell \pm c\sqrt{\hat{V}_{\ell\ell}/n}]$$

where $c$ is the $1-\alpha$ quantile of the distribution of $\|Z\|_\infty$ for a random variable $Z \sim N(0, \hat{D}^{-1/2}\hat{V}\hat{D}^{-1/2})$, where $\hat{D} = \mathrm{diag}(\hat{V})$ is the matrix with diagonal entries $\hat{V}_{\ell\ell}$ and zero off-diagonal entries.

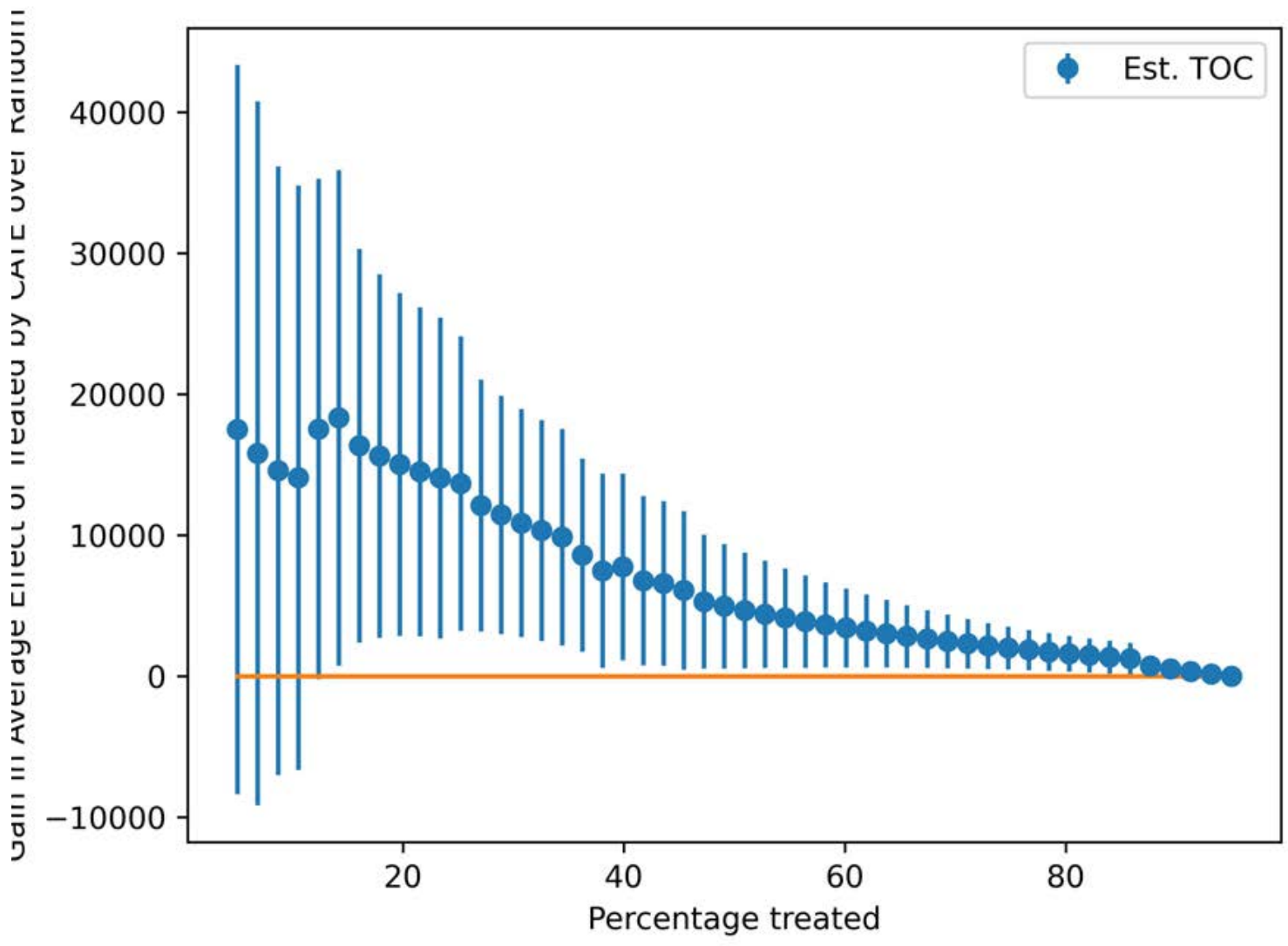

**Figure 15.18:** Point estimates and uniform confidence band of the TOC curve for the Q-aggregation ensemble $\tau_*$ in the 401(k) example.

Note that if there is any point that is above the zero line, with confidence, in this curve, then the CATE model $\hat{\tau}$ has identified heterogeneity in the effect in a statistically significant manner. For such a test we can calculate a one-sided confidence interval, as we only care that the quantities are larger than some value with high confidence. Using the Gaussian approximation, a one-sided confidence band, at confidence level $\alpha$, can be calculated as:

$$CR = \times_{\ell=1}^{p} \left[ \hat{\alpha}_\ell - c\sqrt{\hat{V}_{\ell\ell}/n}, \infty \right)$$

where $c$ is the $1 - \alpha/2$ quantile of the distribution of $\|Z\|_\infty$ for a random variable $Z$ as defined in the previous paragraph.

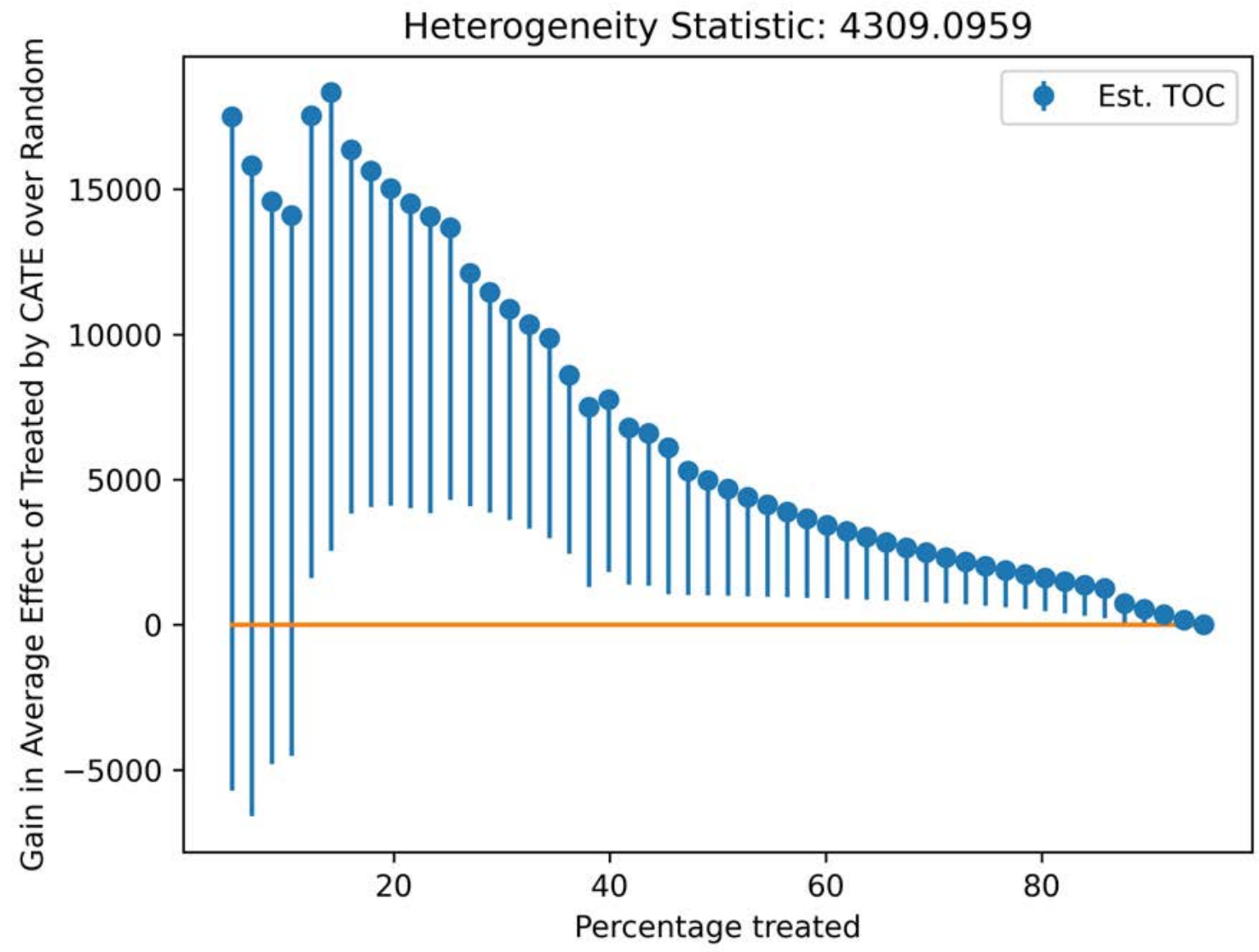

**Figure 15.19:** Point estimates and one-sided uniform confidence band of the TOC curve for the Q-aggregation ensemble $\tau_*$ in the 401(k) example. The heterogeneity statistic depicted in the title corresponds to the largest lower bound of the confidence band across all quantile points and is a statistical signal for the presence treatment effect heterogeneity.

We can also calculate the area under the curve using the discrete difference approximation:

$$\widehat{AUTOC} = \sum_{\ell=1}^{p} \widehat{\text{TOC}}(q_\ell)\,(q_{\ell+1} - q_\ell) \tag{15.3.19}$$

Note that since this is a linear combination of the estimates at each $q_\ell$, under the assumptions of Theorem 15.3.1, the estimate of the area under the curve will be asympotically normal and centered around the quantity:

$$AUTOC = \sum_{\ell=1}^{p} \text{TOC}(q_\ell)\,(q_{\ell+1} - q_\ell)$$

and we can calculate a one-sided confidence interval as:

$$AUTOC \in \left[\widehat{AUTOC} - \sqrt{\hat{V}/n}, \infty\right)$$

where the estimate of the variance is:

$$\hat{V} = \mathbb{E}_n \hat{\psi}(W)^2 \qquad \hat{\psi}(W) = \sum_{\ell=1}^{p} \hat{\psi}_\ell(W)\,(q_{\ell+1} - q_\ell)$$

If the confidence interval does not contain zero, then we have again detected heterogeneity.

| AUTOC | s.e. | One-Sided 95% CI |
|---|---|---|
| 5228.9137 | 1471.8731 | [2807.8980, Infty] |

**Figure 15.20:** AUTOC point estimate and one-sided confidence interval for the Q-aggregation ensemble $\tau_*$ in the 401(k) example.

The exact same analysis can be conducted for the QINI curve, constructing doubly robust point estimates and a simultaneous one-sided confidence band, as well as a one-sided confidence interval for the discretized quantile approximation of the AUQC, i.e.

$$AUQC = \sum_{\ell=1}^{p} \text{QINI}(q_\ell)\,(q_{\ell+1} - q_\ell)$$

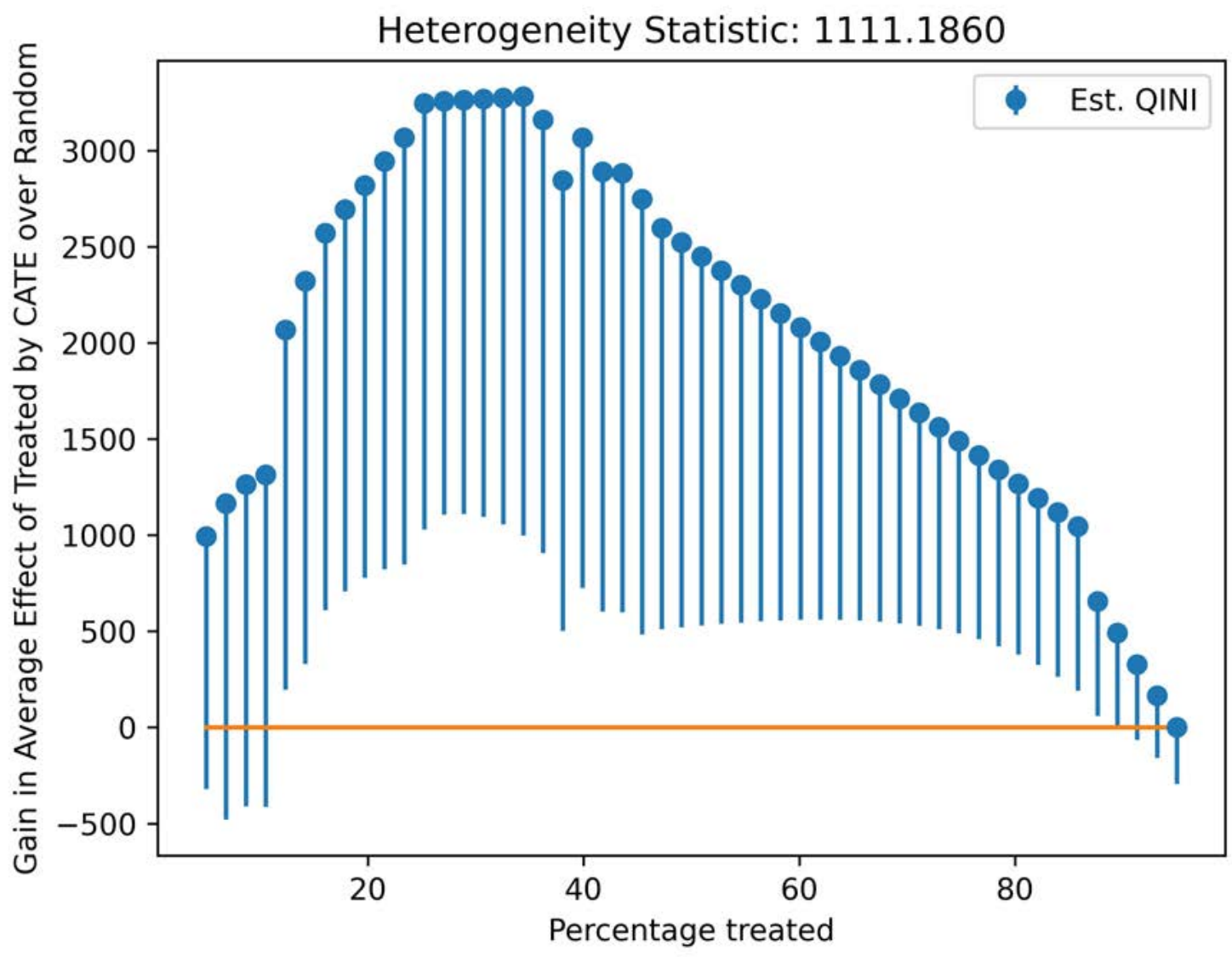

**Figure 15.21:** Point estimates and one-sided uniform confidence band of the QINI curve for the Q-aggregation ensemble $\tau_*$ in the 401(k) example. The heterogeneity statistic depicted in the title corresponds to the largest lower bound of the confidence band across all quantile points and is a statistical signal for the presence treatment effect heterogeneity.

| AUQC | s.e. | One-Sided 95% CI |
|---|---|---|
| 1542.4581 | 385.2292 | [908.8125, Infty] |

**Figure 15.22:** AUQC point estimate and one-sided confidence interval for the Q-aggregation ensemble $\tau_*$ in the 401(k) example.

**Remark 15.3.2** (Rank-Average Weighted Treatment Effects) The analysis in this section viewed the quantile function $\mu(\tau_*, q)$ as fixed and considered a variant of the uplift curves based on the targeting policy $\pi_q(X) := 1\{\tau_*(X) \geq \mu(\tau_*, q)\}$. If this was the policy that was deployed in the population, then we define the TOC curve at value $q$, as the average effect of the treated population of policy $\pi_q$ and we perform inference on this quantity. An alternative view of the TOC curve is to consider the ranking viewpoint that at deployment time will rank the population based on the prediction models predictions and will treat exactly the top $q$ fraction of the population. From this viewpoint, the accuracy of the quantile estimate matters a lot and should be incorporated into the uncertainty estimates. Quantifying the uncertainty that stems from estimation errors in the quantiles $\mu(\tau_*, q)$ is a more involved topic. The recent work of [24] takes this view and performs inference on the ranking interpretation of the TOC and QINI curves, correctly accounting for the uncertainty in the estimation of the quantiles of the CATE distribution and offers procedures for asymptotically correct confidence intervals (albeit not confidence bands).

## 15.4 Personalized Policy Learning

In Section 14.3 we studied evaluation of personalized policies, in particular optimal ones. However, we did not delve into the learning of optimal policies, just as we discussed inference on CATE inChapter 14 but did not delve into learning it using flexible non-parametric methods. While any CATE model learned as in the present chapter can be used to prioritize treatment, a CATE model would only be a means to an end and not the object of interest itself, which may be learned more directly. The primary object of interest is a good personalized treatment policy $\pi$ that given any instance of the variable $X$ returns a treatment assignment $\pi(X) \in \{0, 1\}$.

Note that learning a good policy is an inherently different statistical task than learning a good CATE model. For a good unconstrained policy, it suffices that we learn whether the CATE $\tau(X) = \mathrm{E}[Y(1) - Y(0) \mid X]$ is positive or negative. As seen in Section 14.3, the optimal policy is given by looking at the sign of CATE: $\pi^*(X) = \mathbb{1}\{\tau_0(X) \geq 0\}$. Thus policy learning is more akin to a classification problem that tries to predict the sign of the CATE as opposed to a regression problem that tries to learn the magnitude of CATE too. Of course, mistakes in predicting the sign are more detrimental when the magnitude of the CATE is larger and therefore should be weighed differently. Thus policy learning is more accurately described as a classification problem with sample dependent mis-classification costs, known in the machine learning literature as *cost-sensitive classification*.

Recall from Section 14.3 that we define in Eq. 14.3.1 the gains of policy over no treatment as $V(\pi) = \mathrm{E}[\pi(X)Y(1) + (1 - \pi(X))Y(0)] - \mathrm{E}[Y(0)] = \mathrm{E}[\pi(X)Y(\eta_0)]$. Optimizing $V(\pi)$ over $\pi$ is equivalent to a sample-weighted classification problem, where the goal of $\pi$ is to match the sign of $Y(\eta_0)$, with sample weights $|Y(\eta_0)|$. More formally, note that:

$$\mathrm{argmax}_\pi V(\pi) = \mathrm{argmax}_\pi \mathrm{E}\left[(2\pi(X) - 1)\, Y(\eta_0)\right]$$

and we can simplify the latter as:

$$\begin{aligned}\mathrm{E}\left[(2\pi(X) - 1)\, Y(\eta_0)\right] &= \mathrm{E}\left[(2\pi(X) - 1)\ \mathrm{sign}\left(Y(\eta_0)\right)\ |Y(\eta_0)|\right]\\ &= \mathrm{E}\left[\mathbb{1}\left\{2\pi(X) - 1 = \mathrm{sign}\left(Y(\eta_0)\right)\right\} |Y(\eta_0)|\right]\end{aligned}$$

Thus we can treat the sign of $Y(\eta_0)$ as the "label" of the sample in a classification problem and $|Y(\eta_0)|$ as the weight of the sample, and our centered treatment policy $2\pi(Z) - 1$ is trying to predict the label. We can therefore invoke any machine learning classification approach in a meta-learning manner, so as to solve this weighted classification problem. One popular approach is to use a decision tree classifier, since it will lead to an interpretable policy that is easy to visualize.

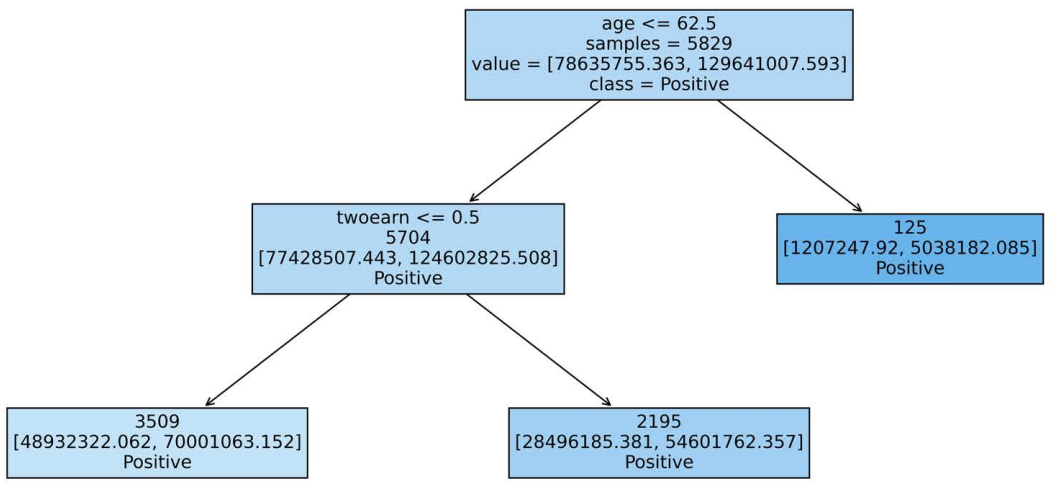

**Figure 15.23:** The details that are displayed on each node are also useful in understanding the group average treatment effect for each node. In particular, the information 'samples=N', gives us the size of each node $N$, and the information 'value=[A, B]', then 'A' is the sum of the $|Y(\hat{\eta})|$ for the samples where $Y(\hat{\eta}) < 0$ and similarly, 'B' is the sum of $|Y(\hat{\eta})|$ for the samples where $Y(\hat{\eta}) > 0$. Thus to get the GATE for each node, we simply do '(B-A)/N', which would correspond to $\frac{1}{N}\sum_{i\in\text{node}} Y(\hat{\eta})$, which is the doubly robust estimate of the GATE for the node.

In finite samples, we would also need to construct estimates $\hat{\eta}$ of the nuisance parameters $\eta_0$ in a cross-fitting manner using arbitrary ML regression methods, as discussed in prior sections and then solve a sample weighted classification problem with samples $\{(X_i, \text{sign}(Y_i(\hat{\eta})), W_i = |Y_i(\hat{\eta})|\}_{i=1}^n$. The results in [25] show that the regret of the returned policy $\hat{\pi}$, as compared to the optimal policy within some policy space $\Pi$, i.e.:

$$R(\hat{\pi}) = \max_{\pi_*\in\Pi} V(\pi_*) - V(\hat{\pi}) \tag{15.4.1}$$

inherit the double robustness property and decay at the order of

$$\approx \sqrt{\frac{V_* \, VC(\Pi)}{n}} + \|H(\hat{\mu}) - H(\mu_0))\|_{L^2} \, \|\hat{g} - g_0\|_{L^2}$$

where $VC(\Pi)$ is a measure of statistical complexity of the policy space $\Pi$ (e.g. a small constant for shallow binary decision trees) and $V_*$ is a constant that in many practical scenarios can be thought as some constant multiple of the variance of the value of the optimal policy in the class $\pi_* = \text{argmax}_{\pi\in\Pi} V(\pi)$, i.e.

$$V_* \approx \text{Var}(\pi_*(X)Y(\eta_0))$$

See also [4, 26] for generalizations and variations of this result.

**Remark 15.4.1** (Probabilistic Policies) The aforementioned analysis also applies if we allow our policy space to output probabilistic choices, i.e. $\pi(X) \in [0,1]$ denotes the probability of treatment. In this case, our objective can equivalently be thought as optimizing a weighted classiciation problem of the form:

$$\mathrm{E}\left[\mathrm{P}_{D\sim\pi}[(2D-1) = \mathrm{sign}(Y(\eta_0))]\,|Y(\eta_0)|\right] \tag{15.4.2}$$

**Remark 15.4.2** (Variance Penalization Methods) One caveat of treating the policy optimization problem as a weighted classification problem and calling a classification oracle is that we might be artificially favoring policies that have high variance. In particular, suppose that some policy $\pi$ assigns a large probability to a treatment at some region of $X$ in which the observed data that has a very low probability. In this case, the variance of this policy is very large, due to the fact that we are dividing by the propensity in the observed data. In this case, one would expect that $V_*$ in the aforementioned regret rate will be a very large multiple of the variance of the optimal policy $\mathrm{Var}(\pi_*(X)\,Y(\eta_0))$. To avoid dependence on the worst case such overlap ratio between any policy in $\Pi$ and the observed policy, i.e. $\sup_{x\in X}\frac{\pi(x)}{\mu_0(x)}$, one needs to amend the objective function that we are optimizing to penalize policies that are expected to have large variance (equiv. small overlap with the policy that was deployed in the observational data). In it's simplest form, one can invoke explicit variance penalization in the empirical objective:

$$\max_{\pi\in\Pi} \mathbb{E}_n\left[\pi(X)\,Y(\hat{\eta})\right] - \lambda\sqrt{\mathrm{Var}_n(\pi(X)\,Y(\hat{\eta}))} \tag{15.4.3}$$

where $\lambda$ is a hyper-parameter that for policy spaces with bounded VC dimension should be set to some constant multiple of $\sqrt{\frac{VC(\Pi)\log(n)}{n}}$.

The one caveat of this approach is that the optimization problem is no longer a simple classification problem and one cannot invoke an out-of-the-box ML classification oracle. Several other approaches have been proposed in the literature that have benefits either on the computational side or on the statistical side, such as distributionally robust optimization [27] (i.e. optimizing the worst case policy value over a ball of distributions that are close to the empirical distribuiton), pessimism [28] (i.e. optimizing a proxy of the lower bound of a confidence interval for the value of a policy), out-of-sample regularization [26] (i.e. optimizing policies that also achieve small error on a held-out sample). This is an active area of research, especially in the area of reinforcement learning and is referred to in the literature as offline policy learning or offline reinforcement learning. Even more complex is the problem of adaptively collecting data and randomizing in an adaptive manner, so as to find an optimal policy, which is referred to in the literature as online policy learning or online reinforcement learning. See [29] for a recent survey.

## 15.5 Empirical Example: The "Welfare" Experiment

We revisit the welfare experiment dataset that we analzed in Chapter 14 and deploy all the methods described in this section. We remind that this dataset corresponds to an experiment that was run as part of the General Social Survey (GSS)[11] , where some respondents received a questionaire about their willingness to support a “Welfare Program” (which will be viewed as the treatment, i.e. $D = 1$, in our analysis), while others received the same questionaire but the program was referred to as “Assistance to the Poor”(which will be viewed as the control, i.e. $D = 0$, in our analysis).

11: See e.g. https://gssdataexplorer.norc.org/variables/vfilter for a full description of the variables in the survey.

After some preprocessing, the dataset contains 12907 individuals and 42 covariates. Instead of simply estimating the projection of the CATE onto a simple model that is linear in the political views variable or its one-hot-encoding, we instead train generic ML models based on all the methods outlined in this chapter. We then score each of the models and construct an ensemble CATE model using Q-aggregation.

In Figure 15.24 we depict the predictions of the Q-aggregation ensemble, as a function of the political views variable, fixing all other covariates to their median values. We find that the fully data-driven model did pick up political views as a relevant variable, but the degree of variation is much smaller than the one that is identified using the doubly robust method for the projection of the CATE on political views. Potentially, this demonstrates that other variables are also relevant and some of the variation picked up by the CATE projection models should have been attributed to other covariates that co-vary with political views.

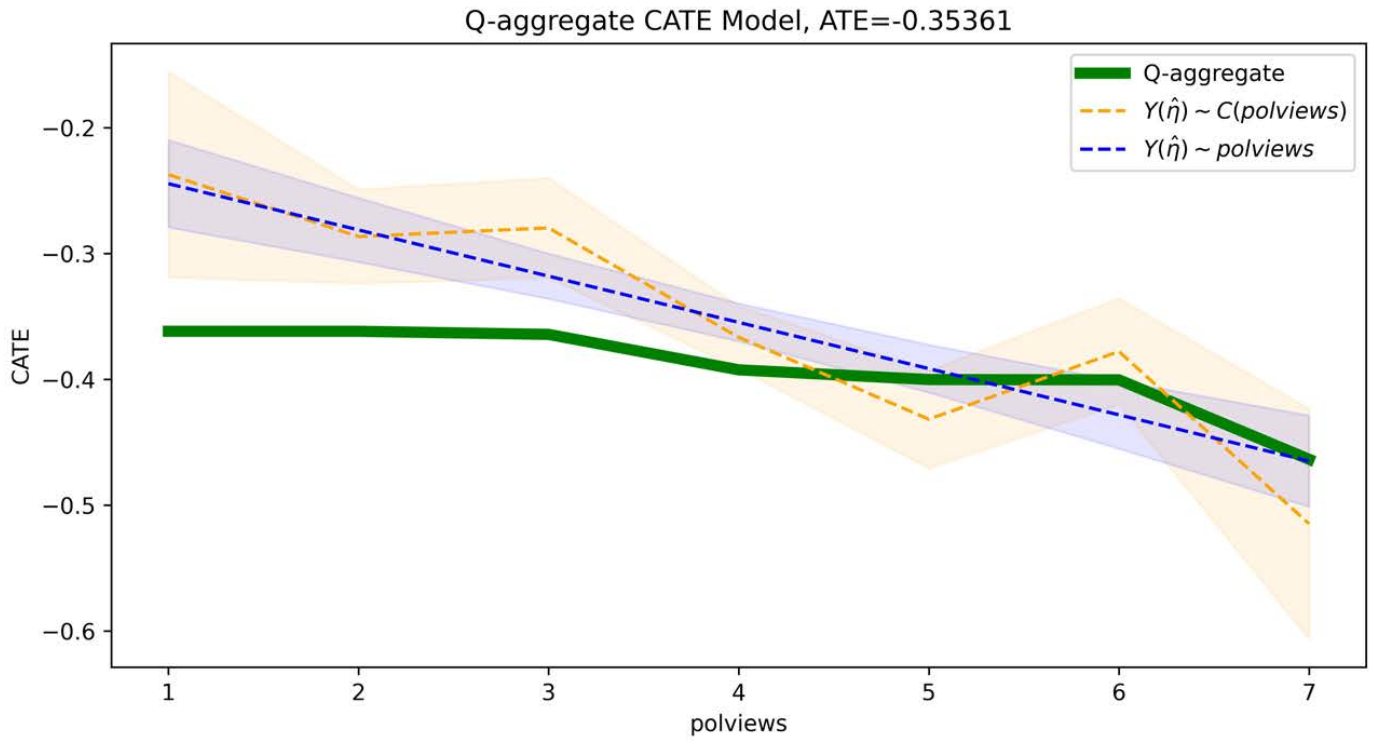

**Figure 15.24:** CATE predictions of the Q-aggregation stacked ensemble. Gradient boosted forests (via the xgboost library) were used as ML oracles for regression and classification. The CATE is predicted on a grid of income points, corresponding to equally spaced income quantiles. All other covariates were imputed at their median values. For comparison, each plot also displays the doubly robust best linear predictor of the CATE with 5-95% confidence intervals as a linear function of the covariate 'polviews' and as a linear function of the one-hot-encoding of the covariate 'polviews'.

To understand the heterogeneity patterns that were identified by the ensemble CATE model, we fit a single shallow binary decision tree to the predictions of the CATE ensemble model. We depict the learned tree in Figure 15.25. We see that political views is indeed the single most important factor that discriminates the predictions of the learned model, however we also see that the ensemble model also learned that education and race also creates heterogeneity in the reaction to programs labeled as "welfare" (as opposed to "assistance to the poor"). In particular, the model identified that people with more left-wing political views and more than 15 years of education (i.e., 4-year college educated individuals) have the least adverse reaction to the word "welfare", while more right wing individuals who did not identify as black (i.e., race2=0) have the most adverse reaction to the term "welfare". Moreover, we see that political views alone does not create a large variation, but it is the combination of political views and college education that creates the largest degree of heterogeneity in the effect.

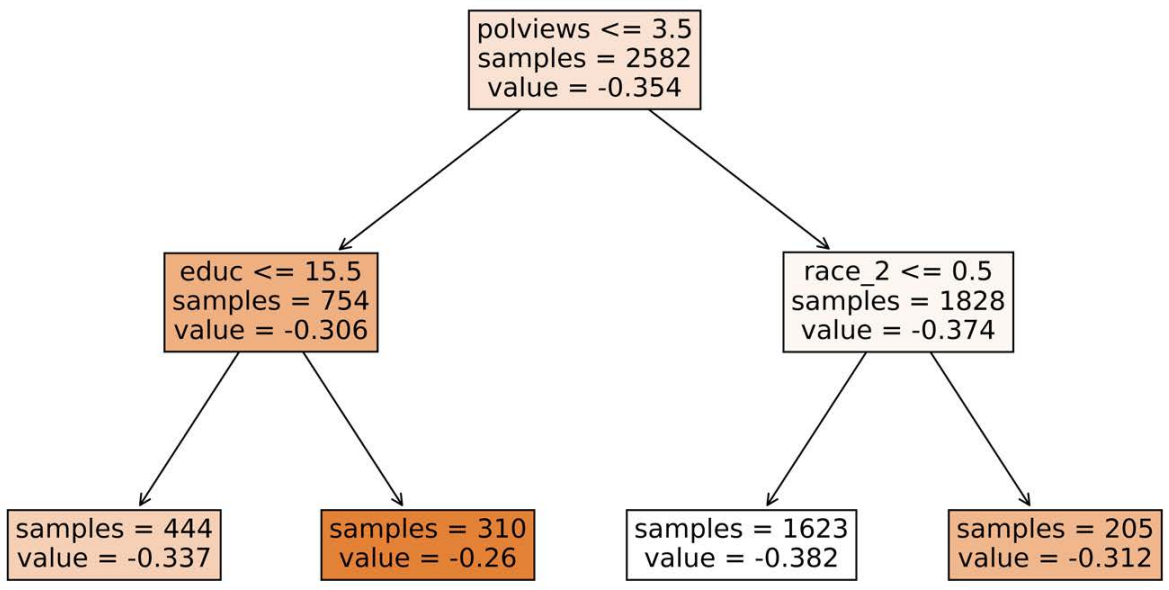

**Figure 15.25:** Single binary regression tree distillation of the Q-aggregation based stacked ensemble.

An alternative way to visualize the importance of the different variables in changing the output of the CATE ensemble is by using the SHAP values. In Figure 15.26. These values identify how each individual variable contributes to changes in the output of the ensemble model. We again identify that political views and education create the largest variation in the output, though here we see that other variables can also be attributed changes in the prediction, such as the number of hours worked last week (hrs1). We see here that having worked less hours last week increases the output of the model, i.e., leads to less adverse reaction to the word "welfare". So people that worked more hours were less eager to contribute to a program termed "welfare".

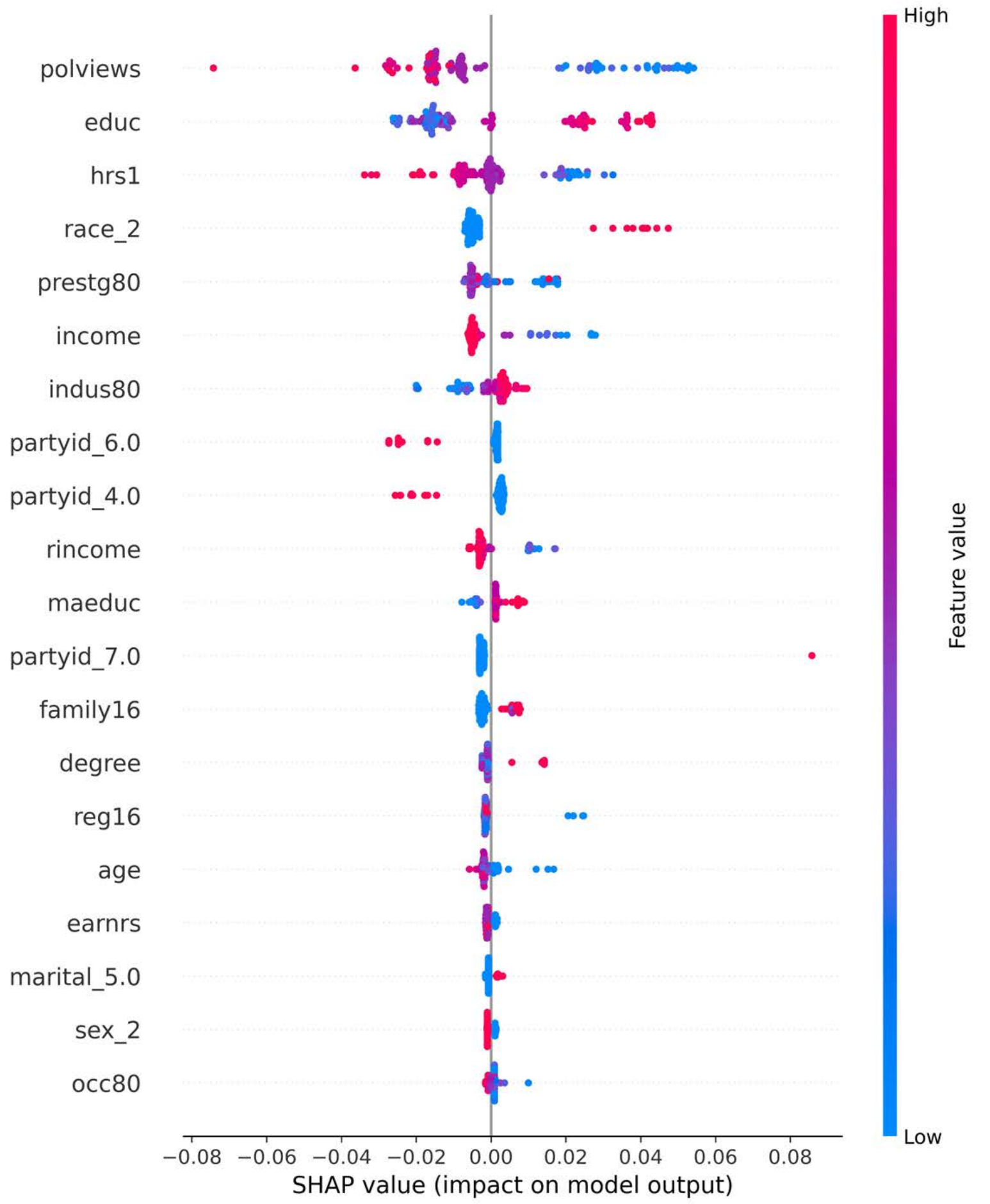

**Figure 15.26:** SHAP values for the Q-aggreagation based stacked ensemble in the welfare experiment dataset.

We can also validate the learned model by running several statistical tests on a held-out sample. For instance, in Figure 15.27 we run an OLS regression of the doubly robust outcome $Y(\hat{\eta})$ on $(1, \tau_*(X))$. We find that the coefficient associated with the stacked ensemble was statistically significant and the confidence interval included the value 1. Hence, this validates that the model carries significant information on the heterogeneity of the effect.

| | coef | std err | P> \|z\| | [0.025 | 0.975] |
|---|---|---|---|---|---|
| **const** | -0.3839 | 0.016 | 0.000 | -0.416 | -0.352 |
| $\tau_*(X)$ | 1.4655 | 0.267 | 0.000 | 0.943 | 1.988 |

**Figure 15.27:** OLS statistical test regression $Y(\hat{\eta})$ on $(1, \tau_*(X))$ in the Criteo example. Standard Errors are heteroscedasticity robust (HC1). $\tau_*$ corresponds to the stacked ensemble based on Q-aggregation.

We also evaluate how calibrated the model is by depicting the group average treatment effects for each quartile of the predicted CATE distribution. The GATE was estimated using the doubly robust approach on the held-out sample. We see that the bottom and top quartiles are separated in a statistically significant manner, while also the calibration score of the model is quite high (0.4461).

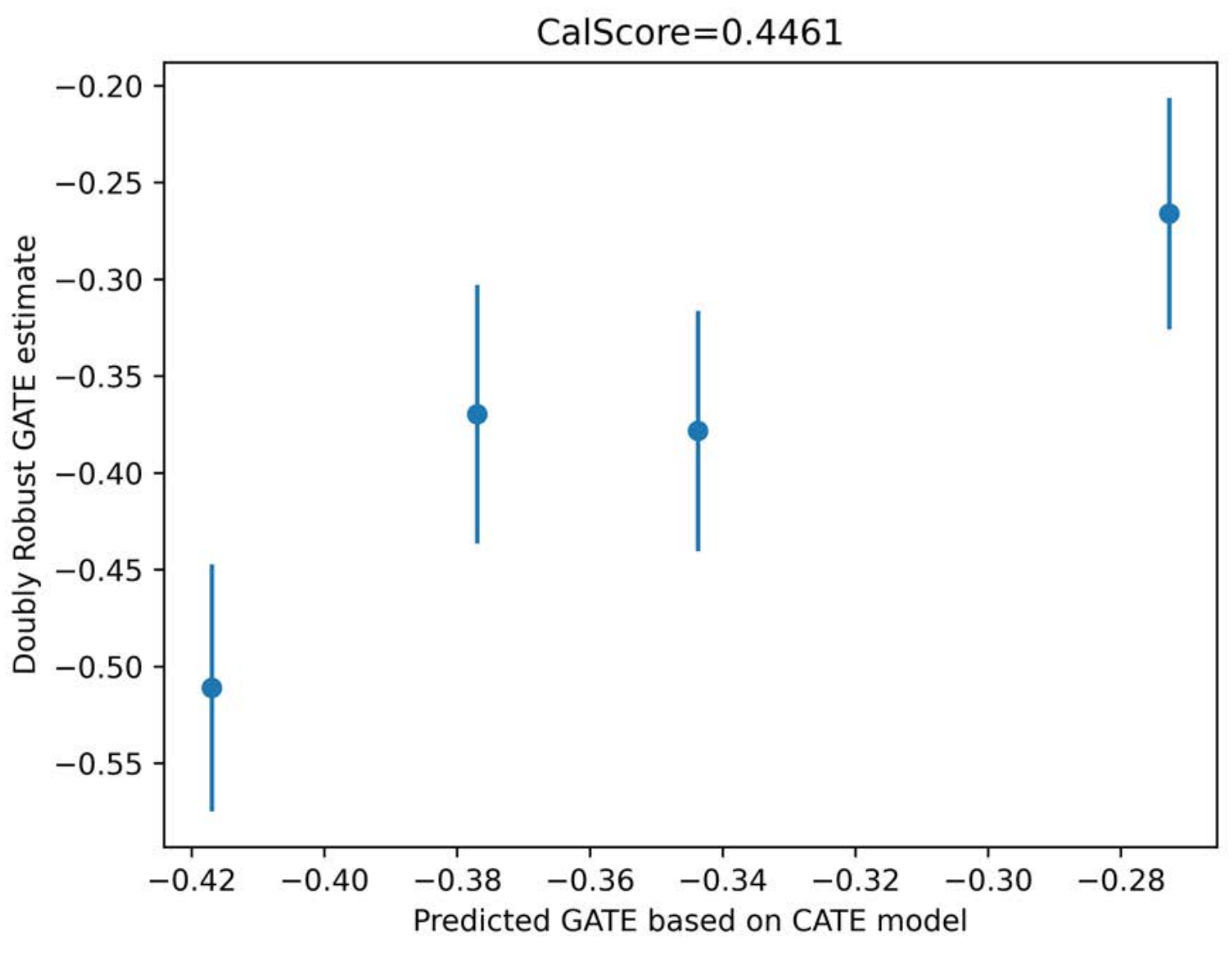

**Figure 15.28:** Calibration check for chosen ensemble model $\tau_*$ in the welfare example. Test samples are splitted in four groups based on CATE predictions and CATE quantiles (e.g. bottom group contains samples whose CATE predictions lie in the bottoms 25% of predictions). The x-axis depicts the average predicted CATE within each group based on $\tau_*$, while the y-axis depicts the GATE as calculated based on the doubly robust pseudo-outcomes calculated on the test set.

Given that we identified that the bottom and top quartile are different in a statistically significant manner, we can also visualize the differences of these two groups, by simply depicting the difference in means of each of the covariates in the two groups. We see for instance, that hours worked last week was significantly different in the two groups, as well as income, age, political views and education, reinforcing our prior findings.

| | group1<br>mean ± s.e. | group2<br>mean ± s.e. | group1 - group2<br>mean ± s.e. |
|---|---|---|---|
| hrs1 | 44.16 ± 0.30 | 36.74 ± 0.58 | 7.42 ± 0.88 |
| income | 11.52 ± 0.03 | 10.61 ± 0.10 | 0.92 ± 0.12 |
| rincome | 10.58 ± 0.05 | 9.23 ± 0.14 | 1.35 ± 0.19 |
| age | 41.75 ± 0.27 | 37.52 ± 0.49 | 4.23 ± 0.76 |
| polviews | 4.39 ± 0.03 | 3.07 ± 0.05 | 1.32 ± 0.08 |
| educ | 13.58 ± 0.06 | 15.41 ± 0.11 | -1.82 ± 0.17 |
| earnrs | 1.81 ± 0.02 | 1.60 ± 0.03 | 0.21 ± 0.05 |
| sibs | 3.40 ± 0.06 | 3.42 ± 0.14 | -0.02 ± 0.20 |
| childs | 1.68 ± 0.03 | 1.24 ± 0.06 | 0.44 ± 0.09 |
| occ80 | 351.52 ± 5.66 | 280.24 ± 8.53 | 71.28 ± 14.19 |

**Figure 15.29:** Group differences between the top 25% predicted CATE group (group2) and the bottom 25% predicted CATE group (group1) in the welfare example.

We can also visualize the differences between the two groups by fitting a shallow binary classification tree to predict membership in the top quartile vs bottom quartile groups. We see again that political views, education and race are the most important distinguishing factors for membership in the two groups. For instance, as we see in Figure 15.30, in the held-out dataset, among the 309 college-educated and left-wing individuals, only 5 were in the bottom quartile group (which had a statistically signficant more adverse reaction to welfare), while 304 were in the top quartile group. Similarly, among the 1622 right-wing and not black individuals, 1541 were in the bottom quartile group vs. 81 in the top quartile group.

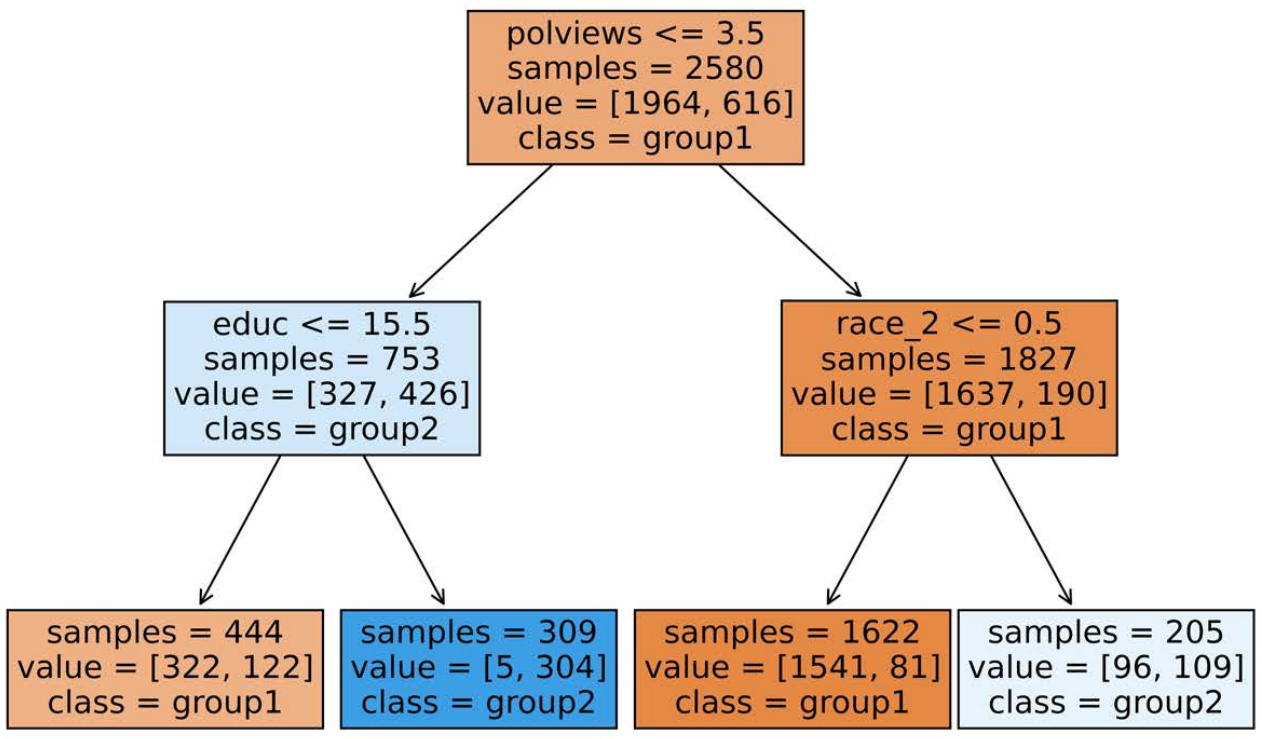

**Figure 15.30:** Decision tree that distills the main differences between group1 and group2 as defined in Figure 15.29.

Finally, we can verify that we detected a statistically significant heterogeneity of effect by looking at the uplift curves, i.e. the TOC (c.f. Figure 15.31) and QINI (c.f. Figure 15.32) curves. We find that both curves lie above the zero line, even when we incorporate one-sided confidence bands. The largest lower point of this confidence band is depicted as a heterogeneity statistic in the title. For instance, we see that the largest lower point is 0.1039 in the TOC curve, which occurs at roughly 5%. This means that, with 95% confidence level, if we look at the group that corresponds to the top 5% of the CATE predictions, then we expect to see an average effect within that group that is at least 0.1039 larger than the average effect in the overall population.

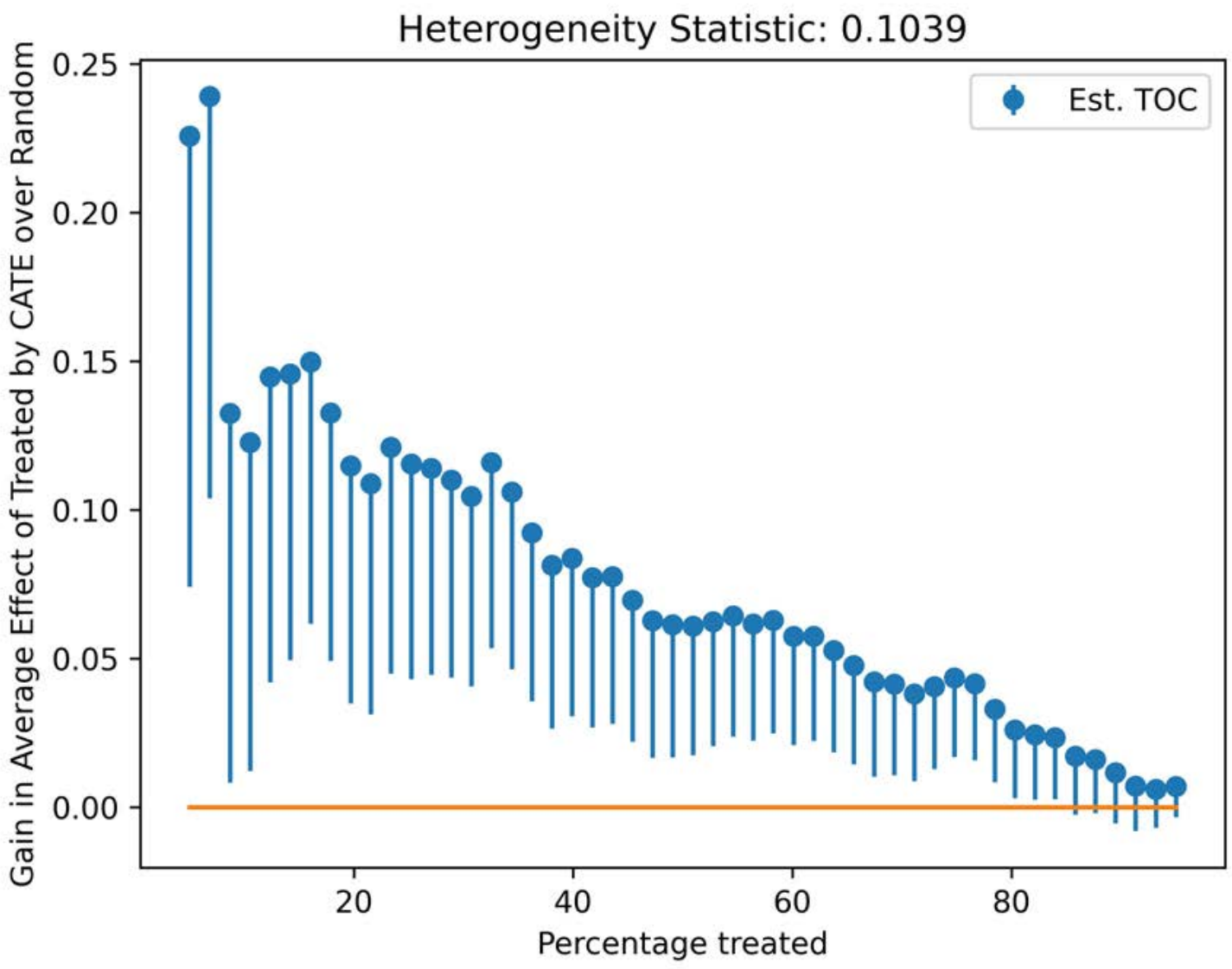

**Figure 15.31:** TOC curve with a 95% one-sided confidence band for the welfare experiment dataset.

Similarly, in the QINI curve we find that this heterogeneity statistic is 0.0164 and occurs at roughly 30%, which means that if were to treat the group of people that corresponds to the top 30% of CATE predictions, then we would expect to get a total effect in the population that is 0.0164 larger than if we were to treat a random 30% fraction of the population.

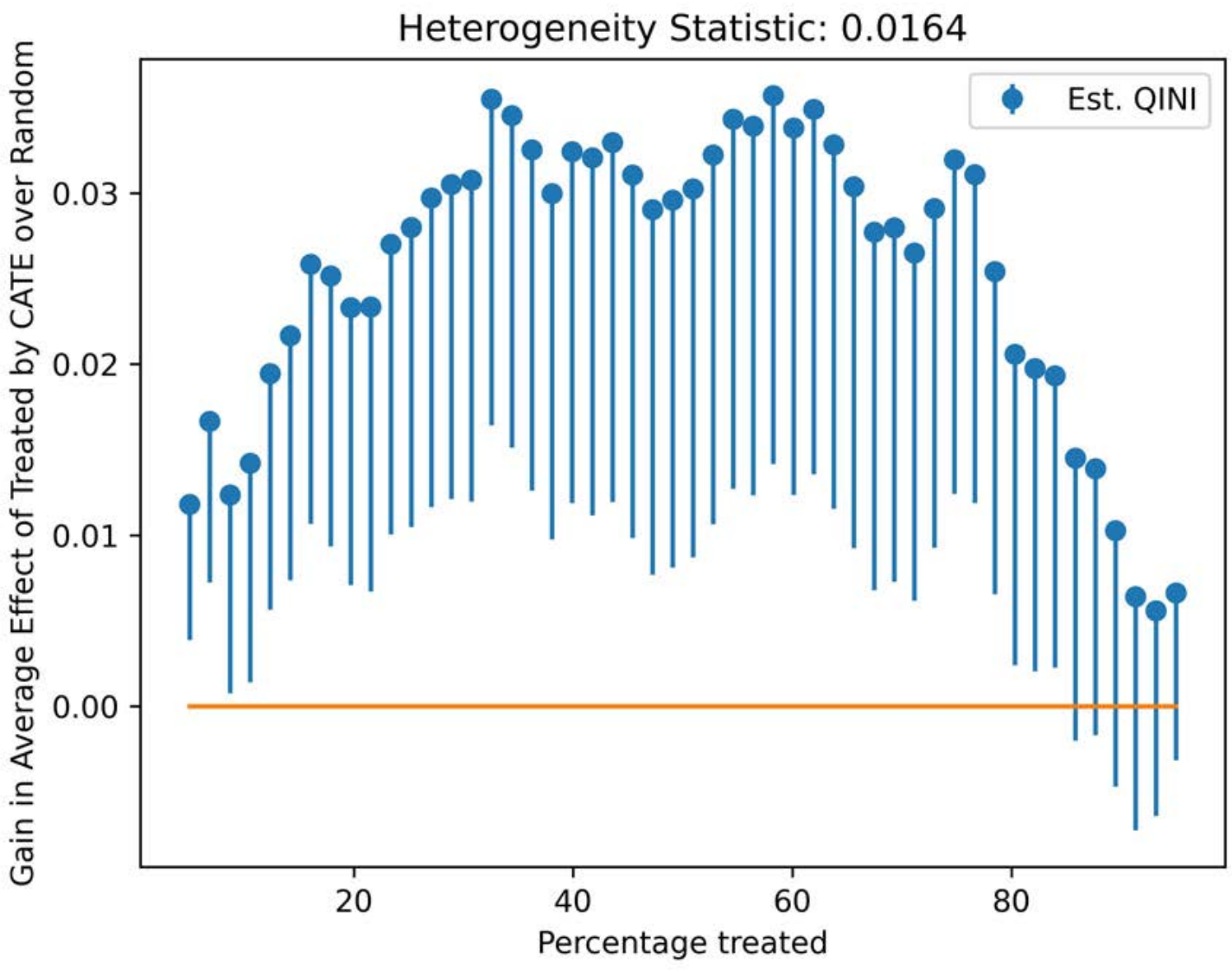

**Figure 15.32:** QINI curve with a 95% one-sided confidence band for the welfare experiment dataset.

We can also calcualte the area under these curves and the confidence interval for that area. If the confidence interval does not contain zero, then we have again detected heterogeneity with statistical significance.

| AUTOC | s.e. | One-Sided 95% CI |
|---|---|---|
| 0.0667 | 0.0128 | [0.0457, Infty] |

| AUT Qini | s.e. | One-Sided 95% CI |
|---|---|---|
| 0.0232 | 0.0046 | [0.0156, Infty] |

## 15.6 Empirical Example: Digital Advertising A/B Test

We now consider an application of CATE estimation in the context of estimating the effects of digital advertising. We will be using a publicly available dataset released by the digital advertising company Criteo®[30]. The dataset consists of approximately 14 million samples (each corresponding to an online visitor) and 10 (anonymized) features that describe the user and the context of the visit. This dataset is the combination of several incrementality tests ran by the company. In each incrementality test a random subset of the population is prevented from being targeted by digital advertising. Subsequently the company tracks whether the user visited or not the webpage that corresponded to the ad that was being shown in some period after the advertising campaign. The latter will be the outcome of interest. Thus we will be measuring the effects of digital advertising on drawing traffic to a particular webpage (the dataset also contains other relevant outcomes such as "conversion", i.e. whether the visitor purchased an associated product).

We applied the CATE estimation pipeline outlined in this section. In Figure 15.33 we depict the performance of each of the meta-learners, compared to the performance of a baseline model that fits a constant treatment effect, as measured by the doubly robust score (see Theorem 15.2.1). Note that since here we are in a randomized trial, the propensity is known and hence the rate requirements in that theorem are satisfied. We see that all meta-learners perform better than a constant effect, indicating statistically significant heterogeneity. Moreover, we see that all learners except the S-learner have comparable performance.

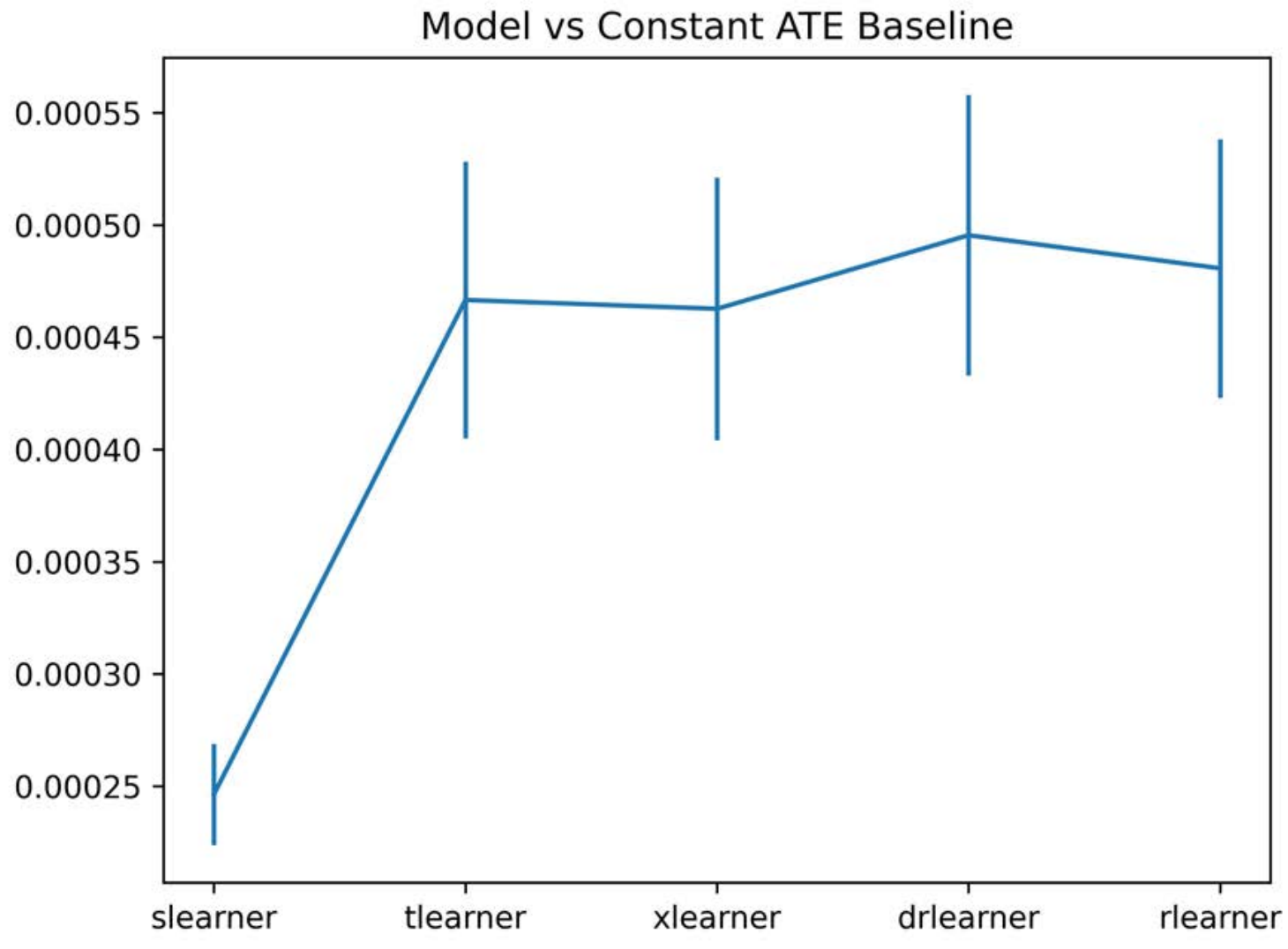

**Figure 15.33:** Performance (and 95% confidence intervals) of meta-learner models in the Criteo example compared to a constant effect model, as measured by the Doubly Robust score.

We also compared the meta-learning approach to the Best-Linear-Predictor approach presented in the prior chapter on heterogeneous treatment effects. Instead of learning a CATE model using all the features, we fitted the best linear CATE when using only feature 'f3' or a second degree polynomial of that feature. We find that this BLP approach in this setting is quite un-stable due to poor extrapolation behavior. In particular, the feature 'f3' has very heavy negative tails (see Figure 15.35). The different parametric models overfit the parametric curve to the region of high density and extrapolate very poorly in the heavy negative tail. On the contrary the Q-aggregation ensemble of the meta-learning models is more stable and regularizes appropriately in this regime.

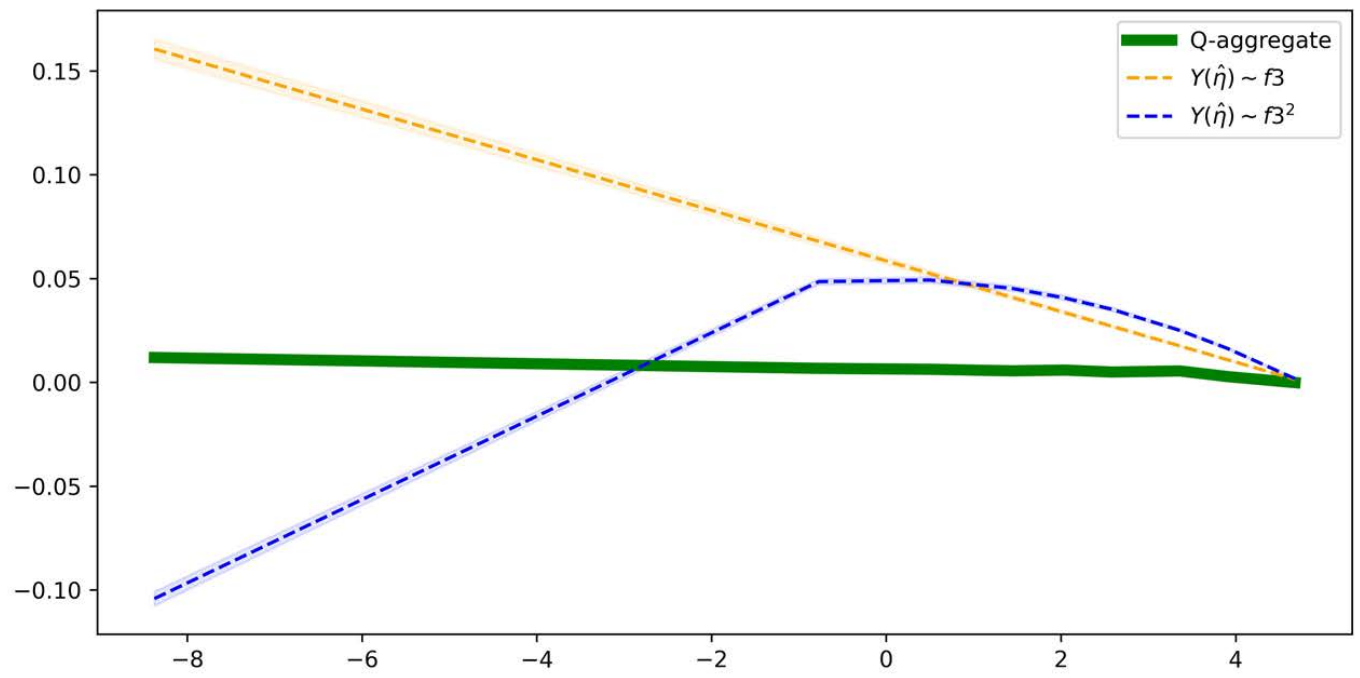

**Figure 15.34:** Predictions of the Q-aggregation stacked ensemble and of the Doubly Robust BLP of CATE as a linear or quadratic function of feature 'f3' in Criteo example.

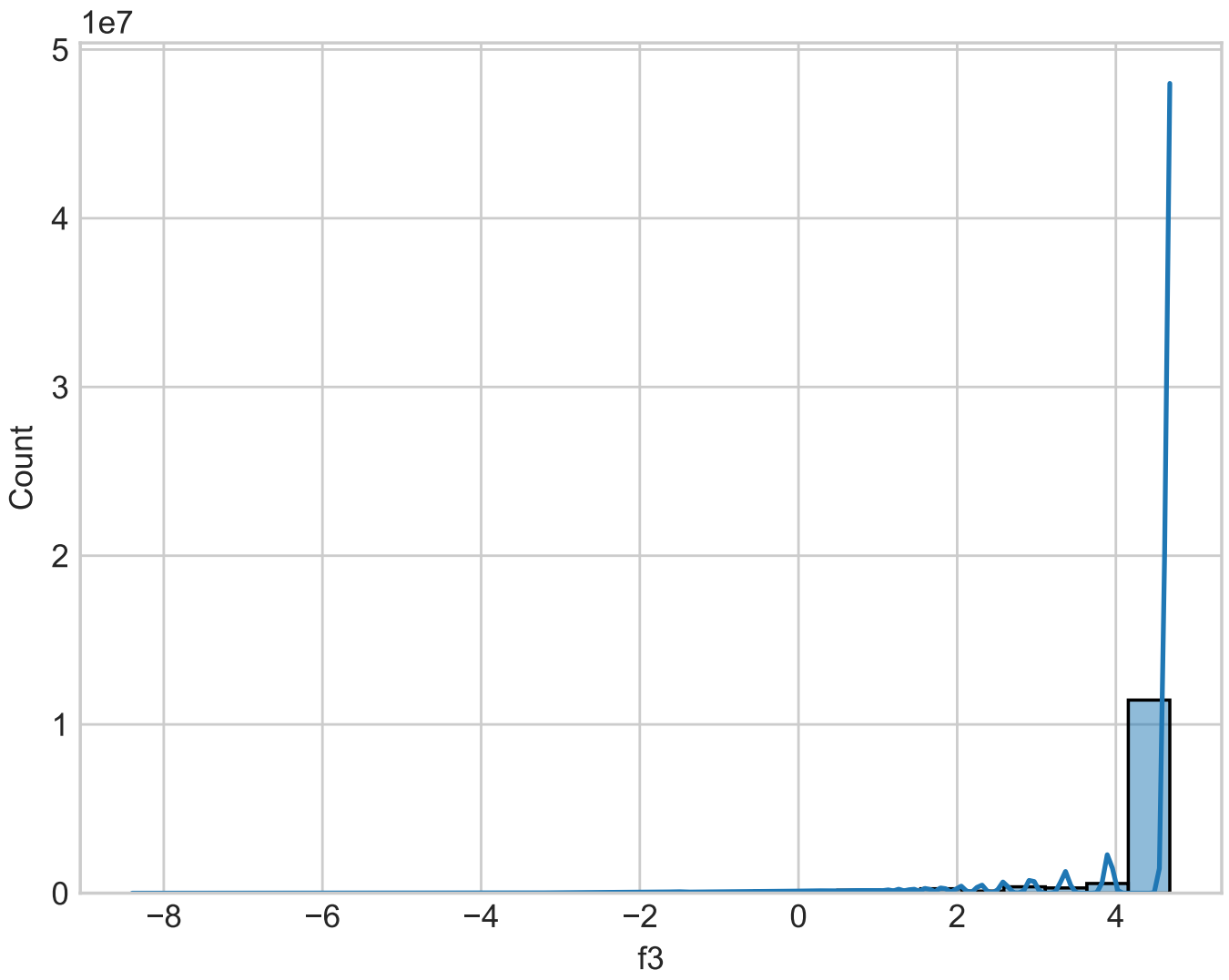

**Figure 15.35:** Histogram of distribution of feature 'f3'

We then validate the Q-aggregation ensemble using all the validation methods presented in this chapter. In Table 15.36 we run an OLS regression of the doubly robust pseudo-outcome on the CATE predictions and an intercept. We find that the coefficient of the CATE predictor is very accurately estimated to be 1.

| | **coef** | **std err** | **P> \|z\|** | **[0.025** | **0.975]** |
|---|---|---|---|---|---|
| **const** | 0.0074 | 0.000 | 0.000 | 0.007 | 0.008 |
| $\tau_*(X)$ | 1.0096 | 0.036 | 0.000 | 0.940 | 1.079 |

**Figure 15.36:** OLS statistical test regression $Y(\hat{\eta})$ on $(1, \tau_*(X))$ in the Criteo example. Standard Errors are heteroscedasticity robust (HC1). $\tau_*$ corresponds to the stacked ensemble based on Q-aggregation.

We also see that the predictions of the CATE model are very well calibrated and the doubly robust GATE estimates for each quartile of CATE prediction groups lies almost on the 45 degree line, with a calibration score that is very close to 1.

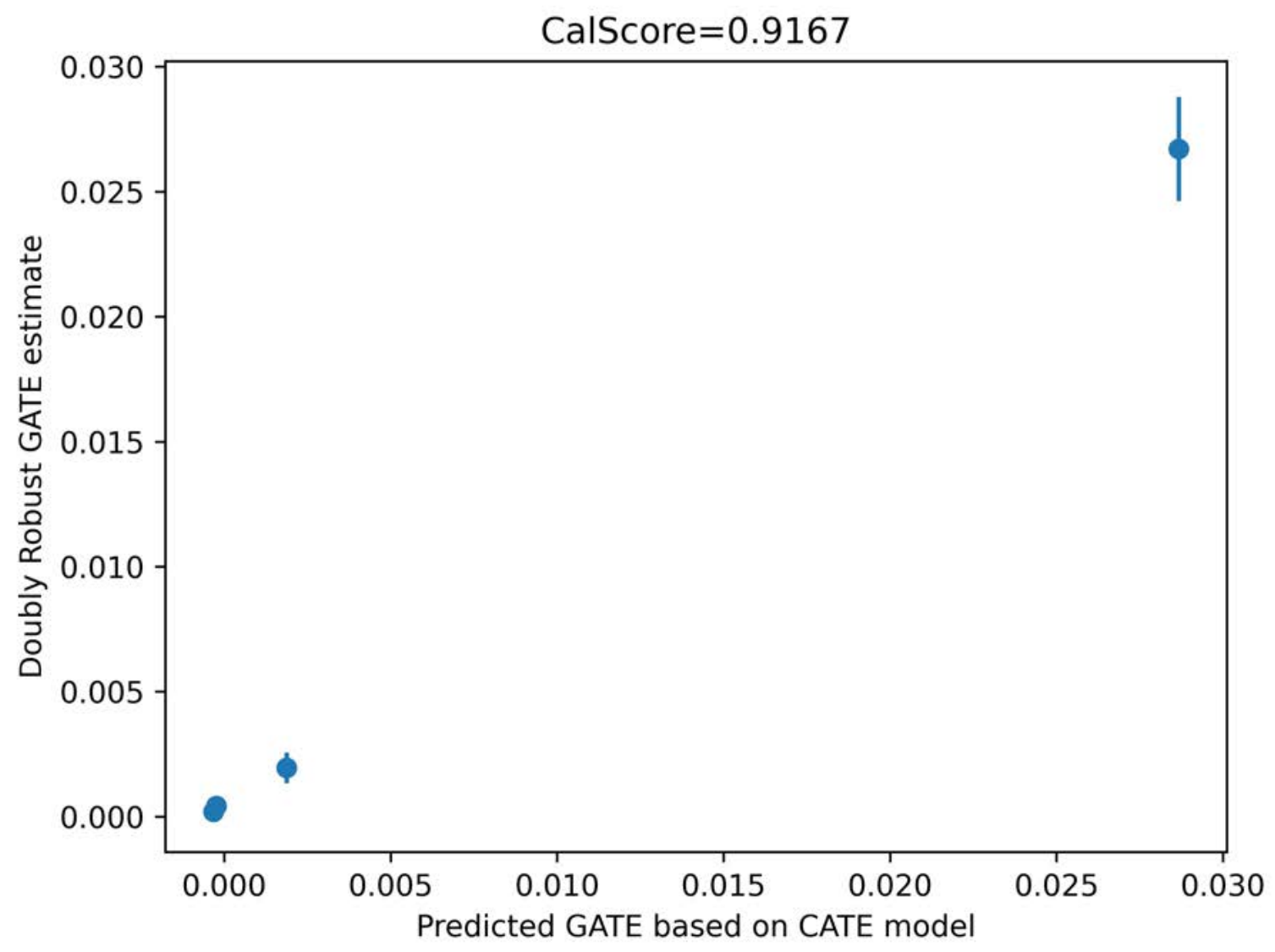

**Figure 15.37:** Calibration check for chosen ensemble model $\tau_*$ in the welfare example. Test samples are splitted in four groups based on CATE predictions and CATE quantiles (e.g. bottom group contains samples whose CATE predictions lie in the bottoms 25% of predictions). The x-axis depicts the average predicted CATE within each group based on $\tau_*$, while the y-axis depicts the GATE as calculated based on the doubly robust pseudo-outcomes calculated on the test set.

The TOC and Qini Curves are depicted in Figure 15.38 and Figure 15.39, with one-sided 95% uniform confidence bands. We see that there is statistically significant heterogeneity as these curves lie well above the zero line. For instance, the TOC curve tells as that if we treat roughly the group that corresponds to roughly the top 5% of CATE predictions then we should expect that the average treatment effect of that group to be approximately 0.07 larger than the average treatment effect. Moreover, the Qini curve tells us that if we treat approximately the group that corresponds to the top 15% of CATE predictions, then we should expect the total effect of such a treatment policy to be $\approx$ 0.005 larger than the total effect if we were to treat a random 15% subset of the population. Thus our CATE model carries significant information that is valuable for better ad targeting.

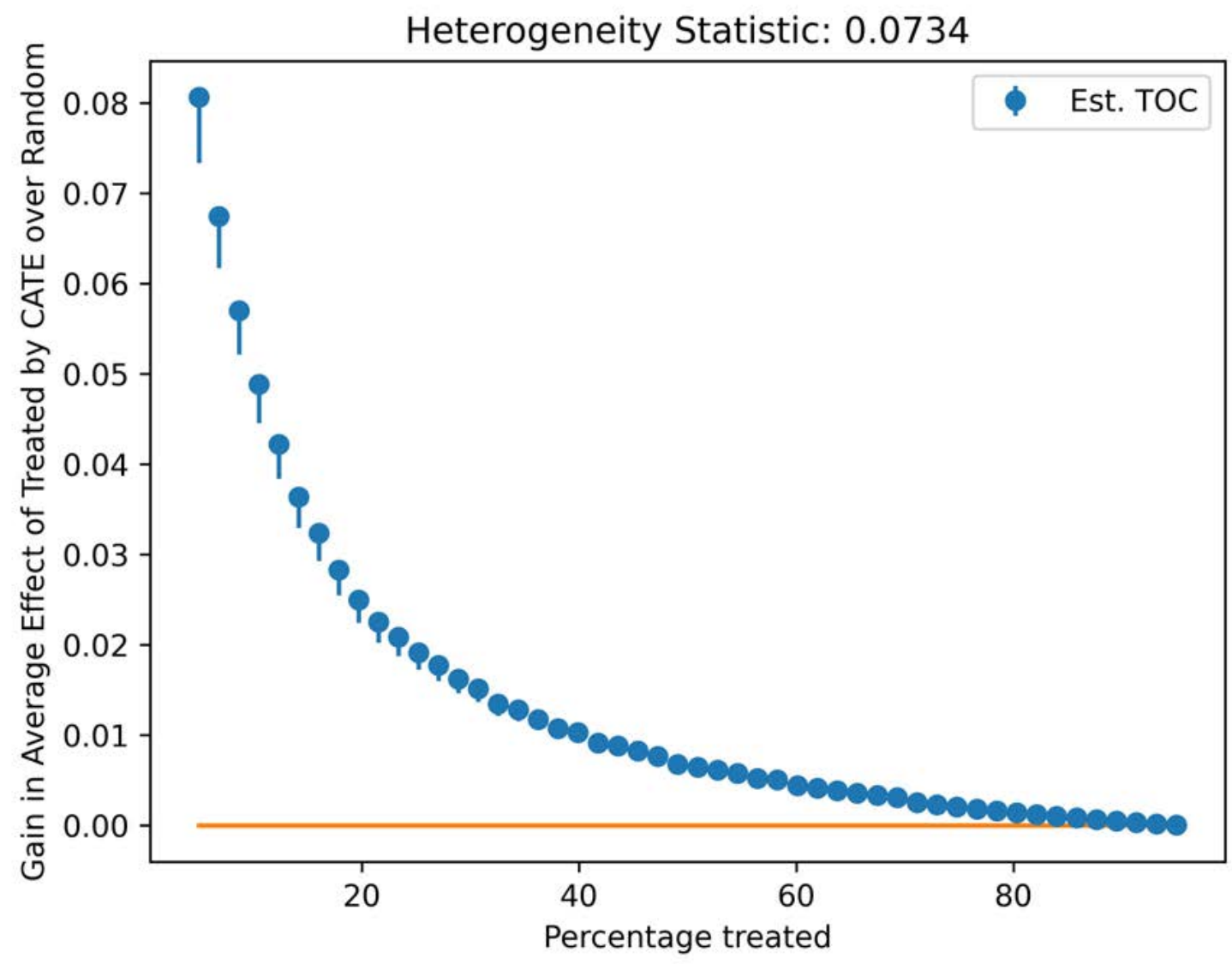

**Figure 15.38:** TOC curve with a 95% one-sided confidence band for the digital advertising dataset.

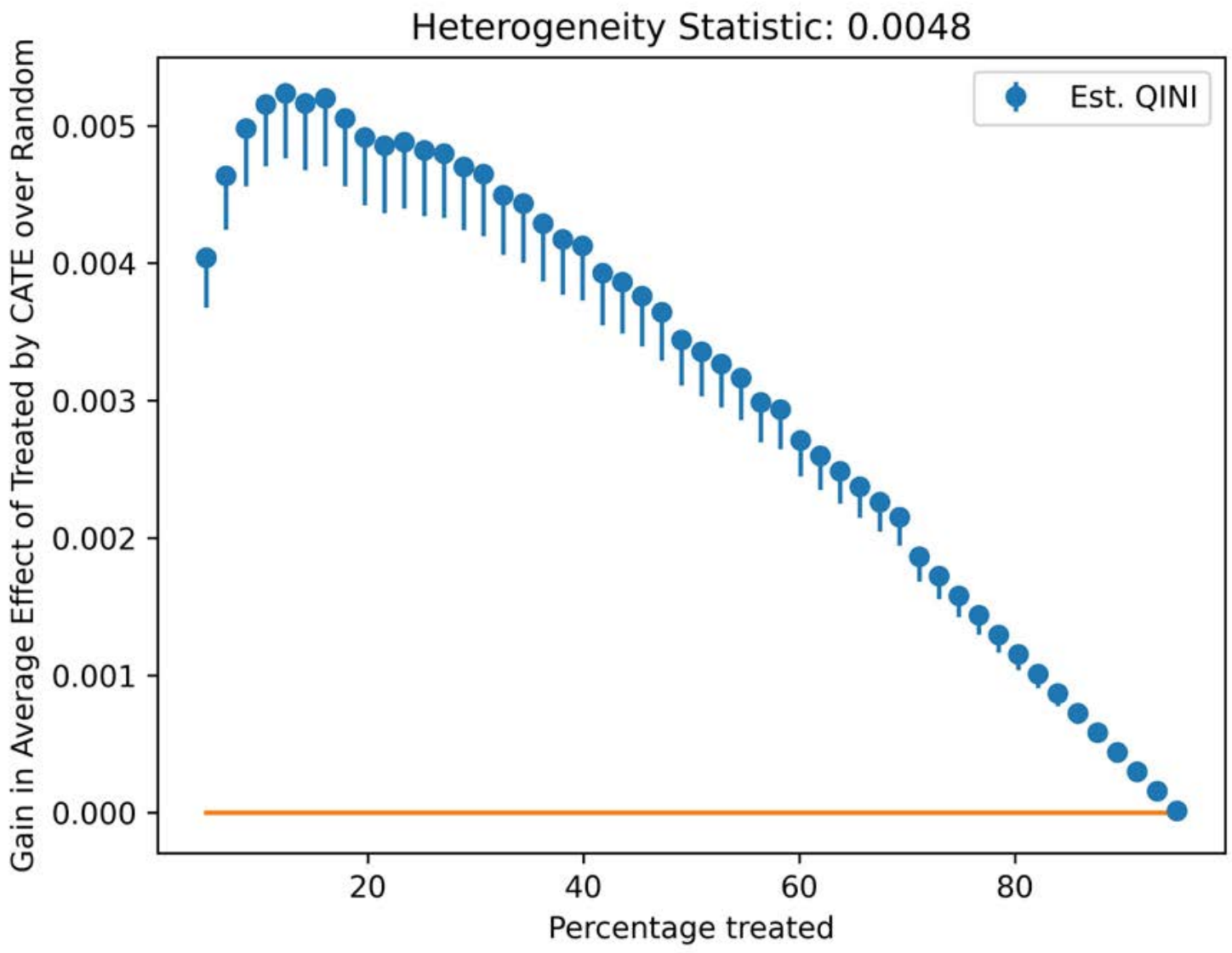

**Figure 15.39:** QINI curve with a 95% one-sided confidence band for the digital advertising dataset.

Given that in this setting we primarily care about personalized ad targeting, we can also apply the direct policy learning methodologies and learn an interpretable decision tree policy of who to target. We assume here that the cost of treatment is 0.04 (which can be interpreted as the cost of ad display divided by the average value of a webpage visit) and learn a binary decision tree that dictates who should be treated. The learned policy tree is depicted in Figure 15.40. We see that the method learned that we should not be treating visitors with a high value of the feature 'f8' and we should definitely be treating visitors with a low value for both the features 'f8' and 'f3'. For visitors with a small 'f8' and large 'f3' the model is rather indifferent

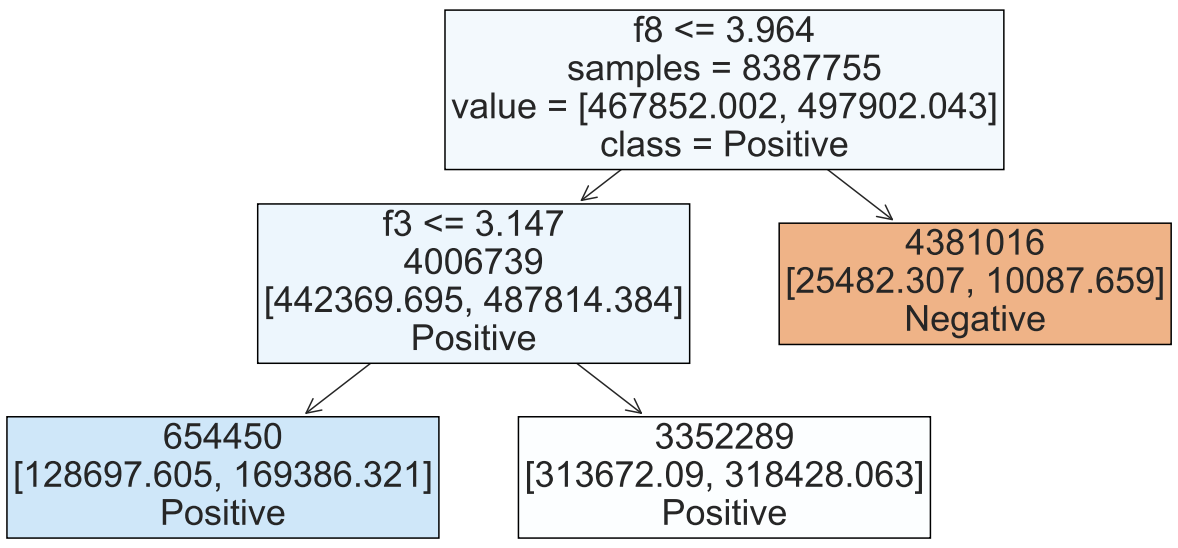

**Figure 15.40:** The details that are displayed on each node are also useful in understanding the group average treatment effect for each node. In particular, the information 'samples=N', gives us the size of each node $N$, and the information 'value=[A, B]', then 'A' is the sum of the $|Y^{DR}(g,p)|$ for the samples where $Y^{DR}(g,p) < 0$ and similarly, 'B' is the sum of $|Y^{DR}(g,p)|$ for the samples where $Y^{DR}(g,p) > 0$. Thus to get the GATE for each node, we simply do '(B-A)/N', which would correspond to $\frac{1}{N}\sum_{i\in\text{node}} Y^{DR}(g,p)$, which is the doubly robust estimate of the GATE for the node.

and weakly recommends treatment.

We can evaluate the performance of the learned policy out of sample using the policy evaluation method presented in Section 14.3. We find that the value of the learned policy is 0.00669 with standard error 0.00029 and 95% confidence interval $[0.00613, 0.00725]$. On the contrary if we were to treat everyone in the population, i.e. display an ad to everyone, then we would get a policy value of 0.00304 with standard error 0.00029 and 95% confidence interval $[0.00248, 0.00360]$. This showcases again the large benefits of personalized policy learning, which yields almost a double net profit.

## Notebooks

- Python Notebook for CATE analyzes CATE of welfare experiment and criteo experiment and for the 401k dataset with generic machine learning.

## 15.A Appendix: Lower Bound on Variance in Model Comparison

First we observe that:

$$\begin{aligned}\Delta_{i,j} &:= (Y(\eta_0) - \tau_i(X))^2 - (Y(\eta_0) - \tau_j(X))^2 \\ &= \tau_i(X)^2 - \tau_j(X)^2 - 2Y(\eta_0)(\tau_i(X) - \tau_j(X)) \\ &= (\tau_i(X) - \tau_j(X))\,(\tau_i(X) + \tau_j(X) - 2Y(\eta_0))\end{aligned}$$

and note that:

$$\mathsf{V}_n = \mathrm{E}\Delta_{i,j}^2 - \left(\mathrm{E}\Delta_{i,j}\right)^2$$

Moreover,

$$\mathrm{E}\Delta_{i,j}^2 = \mathrm{E}\left[(\tau_i(X) - \tau_j(X))^2\,\mathrm{E}\left[(\tau_i(X) + \tau_j(X) - 2Y(\eta_0))^2 \mid X\right]\right]$$

By a variance decomposition argument and since $\tau_0(X) = \mathrm{E}(Y(\eta_0) \mid X)$:

$$\begin{aligned}\mathrm{E}&\left[(\tau_i(X) + \tau_j(X) - 2Y(\eta_0))^2 \mid X\right]\\ &= 4\,\mathrm{Var}(Y(\eta_0) \mid X) + \mathrm{E}\left[(\tau_i(X) + \tau_j(X) - 2\tau_0(X))^2 \mid X\right]\end{aligned}$$

Thus we have derived that:

$$\begin{aligned}\mathrm{E}\Delta_{i,j}^2 &= 4\mathrm{E}\left[(\tau_i(X) - \tau_j(X))^2\ \mathrm{Var}(Y(\eta_0) \mid X)\right]\\ &\quad + \mathrm{E}\left[(\tau_i(X) - \tau_j(X))^2\,(\tau_i(X) + \tau_j(X) - 2\tau_0(X))^2\right]\end{aligned}$$

Moreover, note that by Jensen's inequality:

$$\begin{aligned}\left(\mathrm{E}\Delta_{i,j}\right)^2 &= \left(\mathrm{E}(\tau_i(X) - \tau_j(X))\,(\tau_i(X) + \tau_j(X) - 2\tau_0(X))\right)^2\\ &\leq \mathrm{E}(\tau_i(X) - \tau_j(X))^2\,(\tau_i(X) + \tau_j(X) - 2\tau_0(X))^2\end{aligned}$$

Thus we can conclude that:

$$\mathsf{V}_n \geq 4\mathrm{E}\left[(\tau_i(X) - \tau_j(X))^2\ \mathrm{Var}(Y(\eta_0) \mid X)\right] \tag{15.A.1}$$

## 15.B Appendix: Interpretation of Uplift curves

We derive first the covariance interpretation of the TOC uplift curve.

$$\begin{aligned}\mathrm{TOC}(q) &= \mathrm{E}[Y(1) - Y(0) \mid \hat{\tau}(X) \geq \mu(q)] - \mathrm{E}[Y(1) - Y(0)]\\ &= \mathrm{E}\left[(Y(1) - Y(0))\frac{1\{\hat{\tau}(X) \geq \mu(q)\}}{\mathrm{P}(\hat{\tau}(X) \geq \mu(q))}\right] - \mathrm{E}[Y(1) - Y(0)]\end{aligned}$$

Let $A = Y(1) - Y(0)$ and $B = \frac{1\{\hat{\tau}(X) \geq \mu(q)\}}{\mathrm{P}(\hat{\tau}(X) \geq \mu(q))}$ and note that $\mathrm{E}[B] = 1$. Thus we have:

$$\mathrm{TOC}(q) = \mathrm{E}\,[A\,B] - \mathrm{E}[A] = \mathrm{E}\,[A\,B] - \mathrm{E}[A]\,\mathrm{E}[B] = \mathrm{Cov}(A, B)$$

Next, we derive the covariance interpretation of the QINI uplift curve. Let $A = Y(1) - Y(0)$ and $B = \hat{\tau}(Z) \geq \mu(q)$. Then by the definition of the QINI curve:

$$\begin{aligned}
\tau_{\text{QINI}}(q) &:= \tau(q)\,\mathrm{P}(\hat{\tau}(Z) \geq \hat{\mu}(q)) \\
&= \left(\mathrm{E}[A \mid B] - \mathrm{E}[A]\right)\,\mathrm{P}(B) \\
&= \left(\mathrm{E}[A\frac{\mathbb{1}\{B\}}{\mathrm{P}(B)}] - \mathrm{E}[A]\right)\,\mathrm{P}(B) \\
&= \mathrm{E}[A\mathbb{1}\{B\}] - \mathrm{E}[A]\mathrm{E}[\mathbb{1}\{B\}] \;=\; \mathrm{Cov}\left(A, \mathbb{1}\{B\}\right)
\end{aligned}$$

# 16 Difference-in-Differences

"Corpus omne perseverare in statu suo quiescendi vel movendi uniformiter in directum, nisi quatenus a viribus impressis cogitur statum illum mutare." ("An object remains in its state of rest or of moving uniformly in a straight direction, unless forced to change that state by impressed forces.")

– Isaac Newton [1].

Here we discuss debiased machine learning (DML) methods for performing inference on average causal effects in panel (or longitudinal) or repeated cross-section data in the difference-in-differences (DiD) framework. We present and discuss the key identifying assumption for the average treatment effect on the treated based on DiD – the so-called "parallel trends" assumption – allowing for high-dimensional observed confounding variables. This assumption suggests a natural estimation strategy that directly applies DML to estimate average treatment effects on the treated using differenced outcomes.

## 16.1 Introduction

We now consider estimation of causal effects in panel (longitudinal) data where we observe individual units in multiple time periods or repeated cross-section data. While there are many potential approaches for analyzing data with both a cross-sectional and temporal component, we specifically look at difference-in-differences (DiD) and closely related approaches.

DiD and related methods are widely used in empirical work in the social sciences and in policy analysis. The basic DiD structure relies on having two groups of observations – a treatment group and a control group – for two time periods – a pre-treatment and a post-treatment period. Canonical DiD analysis then proceeds by comparing changes in the average pre- and post-treatment outcomes in the treatment group to changes in the average pre- and post-treatment outcomes in the control group. Attaching a causal interpretation to this comparison relies on an assumption that imposes that changes in the treatment group *in the absence of treatment* would have been the same as changes in the control group. This assumption captures the intuition that the treatment group would have evolved along the same path as the control group in the absence of treatment – i.e., the two groups share "parallel trends." Under the parallel trends assumption, the difference between the treatment and control differences between the pre- and post-treatment averages identifies the average treatment effect on the treated (ATET).

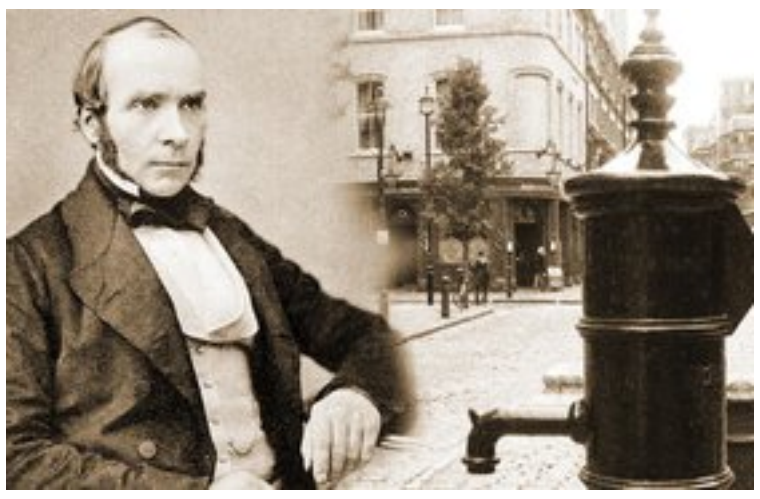

**Figure 16.1:** DiD is perhaps the oldest quasi-experimental research design. John Snow was a London doctor and is often considered the father of modern epidemiology. [2] is essentially an effort to provide convincing evidence that water is the causal agent for cholera transmission. It presents and discusses multiple pieces of evidence – including a DiD. *Source:* `https://www.micropia.nl/en/discover/microbiology/john-snow/`, accessed 6/7/23.

In this chapter, we review the basic DiD framework. We then focus on DiD in a setting where a researcher wishes to impose *conditional parallel trends*. That is, we consider settings where there are observed variables that are thought to be related to the evolution of the outcome of interest such that parallel trends holds only after conditioning on these variables. After suitably defining the conditional parallel trends assumption, we illustrate that the DML approach to estimating ATET from Chapter 10 can be readily applied within the DiD context.

## 16.2 The Basic Difference-in-Differences Framework: Parallel Worlds

The basic DiD structure has many appealing features. It is intuitive. It allows for essentially unrestricted differences in baseline outcomes for the treatment and control groups and allows for treatment to depend on those baseline differences. Estimation and inference are also relatively straightforward. Here we review the DiD structure using potential outcomes notation and highlight the key identifying assumptions.

The canonical DiD structure relies on existence of two time periods, denoted $t = 1$ and $t = 2$, and maintains that all observations are in the control state at $t = 1$. As such, we introduce potential outcomes

$$Y_t(d)$$

where $d \in \{0, 1\}$ denotes the treatment state in period $t = 2$. For example, $Y_1(1)$ denotes the period one outcome under treatment – that is, the outcome in the period before treatment is received – and $Y_2(1)$ denotes the period two outcome under treatment. Let $D \in \{0, 1\}$ be the treatment group indicator with $D = 1$ indicating that treatment is received at $t = 2$ and $D = 0$ indicating no treatment in either time period. Observed outcomes in period $t$ may then be represented as $Y_t = DY_t(1) + (1 - D)Y_t(0)$. As in other causal inference contexts, we are left with missing data as we are unable to observe observations simultaneously in the treatment and control state.

One can also define four potential outcomes $(Y_t(0,0), Y_t(0,1), Y_t(1,0), Y_t(1,1))$ for each time period. The DiD structure imposes that $(Y_t(1,0), Y_t(1,1))$ can never be observed so it is impossible to learn about the effects of treatment paths that have treatment occur at t = 1. We choose the simpler representation with a single argument in the potential outcomes for notational clarity. Keeping explicit track of potential outcomes for different treatment paths is important in more complicated settings with more potential treatment paths as may arise with many time periods or more complex treatment variables.

DiD proceeds under the following key assumption:

**Assumption 16.2.1** (Parallel Trends and No Anticipation) *Potential outcomes satisfy*

$$E[Y_2(0) - Y_1(0) \mid D = 1] = E[Y_2(0) - Y_1(0) \mid D = 0] \quad (16.2.1)$$

*and*

$$E[Y_1(0) \mid D = 1] = E[Y_1(1) \mid D = 1]. \quad (16.2.2)$$

Condition (16.2.1) is the *parallel trends* assumption. It requires that, in expectation, the change in control potential outcomes among the treatment group is the same as the change in the control potential outcomes among the control group. Condition (16.2.2) imposes that receipt of treatment at $t = 2$ does not impact average period 1 outcomes. Here, we are effectively ruling out anticipation effects. Importantly, (16.2.2) allows for systematic differences between average potential outcomes among treated and control observations in the pre-treatment period. That is, it does not impose that $E[Y_1(0) \mid D = 1] = E[Y_1(0) \mid D = 0]$. Thus, we can accommodate, for example, scenarios where we believe that period two treatment assignment is related to period one outcomes.

Often, the no anticipation assumption is left implicit or ignored. We state it for clarity and because it allows clean definition of the causal effect of interest.

It is worth explicitly noting that the parallel trends assumption is typically functional form dependent. That is, if $\mathrm{E}[Y_2(0) - Y_1(0) \mid D = 1] = \mathrm{E}[Y_2(0) - Y_1(0) \mid D = 0]$, it will generally not be the case that $\mathrm{E}[g(Y_2(0)) - g(Y_1(0)) \mid D = 1] = \mathrm{E}[g(Y_2(0)) - g(Y_1(0)) \mid D = 0]$. For example, suppose the outcome of interest is wages. Parallel trends holding for wage does not imply that parallel trends holds for log(wage), and the DiD estimator based on log(wage) need not recover a causal effect. Intuitively, this functional form dependence arises because parallel trends relies on latent sources of confounding being additively separable so that they are eliminated by the differencing operation.

One situation where parallel trends hold regardless of the outcome transformation is when $D$ is randomly assigned (i.e., when $(Y_1(0), Y_1(1), Y_2(0), Y_2(1) \perp D)$. See [3] for further discussion.

A related framework to DiD that is independent to monotone transformations (at the cost of other restrictions, of course) is the changes-in-changes model of [4].

It is straightforward to verify that ATET is identified under Assumption 16.2.1. Note that the right-hand-side of Eq. (16.2.1) is an observable quantity while the left-hand-side corresponds to the unobservable change in the control potential outcomes of treated units. Parallel trends allows us to impute this latent change from the observed change in the control units. Effectively, we are assuming that the treated observations would have changed in the same way as the control observations in the absence of treatment. Similarly, the right-hand-side of Eq. (16.2.2) is an observable quantity while the left-hand-side is the unobserved average of control potential outcomes in period one for the treated group. Eq. (16.2.2) allows us to impute this baseline average from the observed baseline average in the treatment group. We can then reconstruct the counterfactual average of the control potential outcome in the post-treatment period by adjusting this baseline average by the observed change in average outcomes between the two periods in the control group. Figure 16.2 presents a graphical illustration of the identification argument.

More formally, we can put this together to write the ATET as

$$\begin{aligned}
\alpha &= \mathrm{E}[Y_2(1) - Y_2(0) \mid D = 1] \\
&= \mathrm{E}[Y_2(1) \mid D = 1] - \mathrm{E}[Y_2(0) \mid D = 1] \\
&= \mathrm{E}[Y_2(1) \mid D = 1] \\
&\qquad - (\mathrm{E}[Y_1(1) \mid D = 1] + \mathrm{E}[Y_2(0) - Y_1(0) \mid D = 0]) \\
&= \mathrm{E}[Y_2(1) - Y_1(1) \mid D = 1] - \mathrm{E}[Y_2(0) - Y_1(0) \mid D = 0]
\end{aligned} \tag{16.2.3}$$

where the second equality follows from a direct application of Assumption 16.2.1. The expression in the last line is exactly the difference between the difference between post- and pre-treatment period average outcomes in the treatment group and the difference between post- and pre-treatment period average outcomes in the control group – hence, difference-in-differences.

One could identify the ATE by augmenting Assumption 16.2.1 with $\mathrm{E}[Y_2(1) - Y_1(1) \mid D = 0] = \mathrm{E}[Y_2(1) - Y_1(1) \mid D = 1]$ and $\mathrm{E}[Y_1(0) \mid D = 0] = \mathrm{E}[Y_1(1) \mid D = 0]$. The first condition is a restriction not on evolution in the untreated state but treatment effects themselves which seems hard to motivate in realistic settings. As such, we follow the majority of the DiD literature in focusing on estimation of ATET.

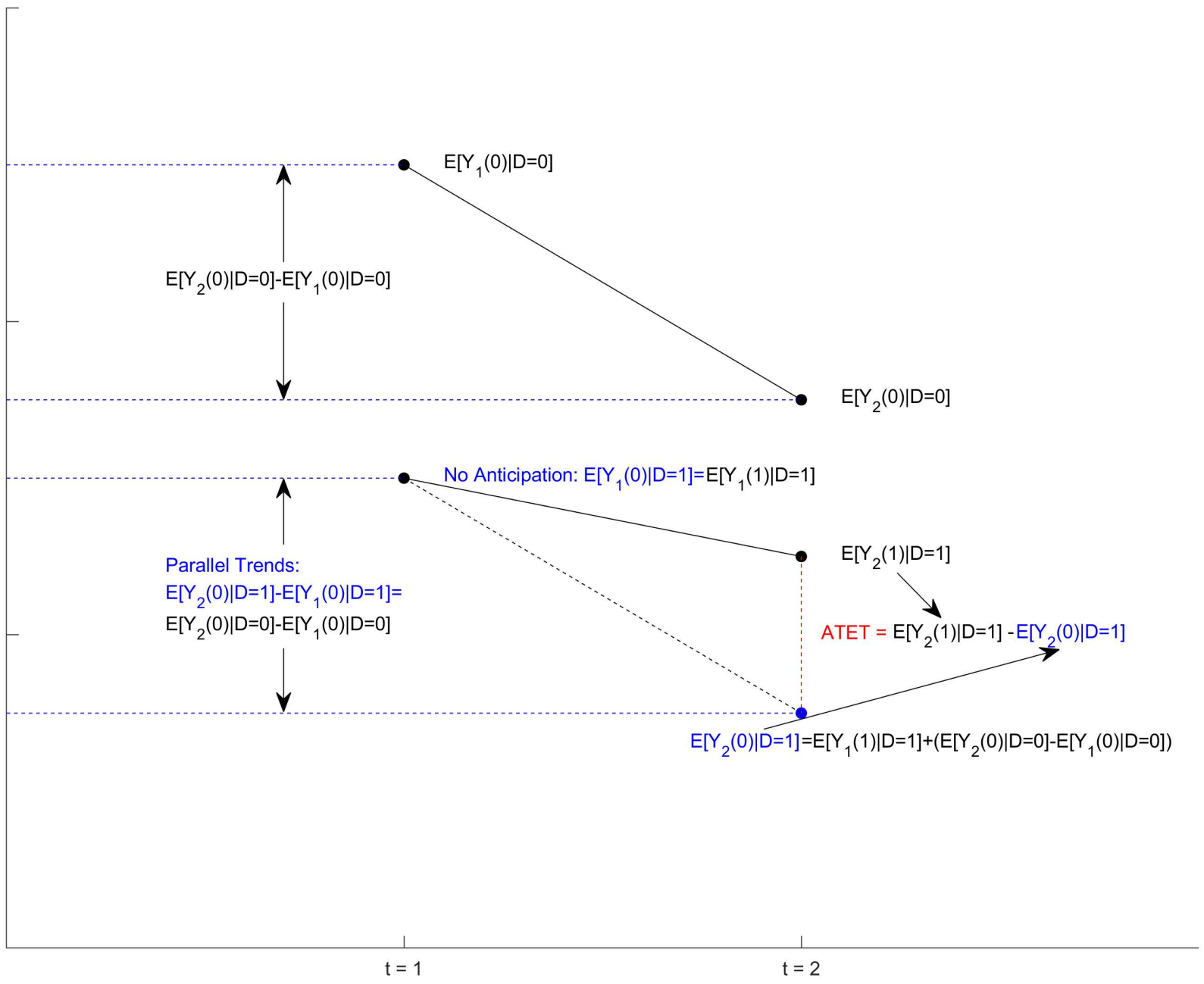

**Figure 16.2: DiD Identification.** This figure illustrates identification of the ATET in the canonical DiD framework. Objects represented in black are observable. Objects in blue are unobserved and identified via Assumption 16.2.1. Visually we impute the unobserved $E[Y_2(0) \mid D = 1]$ by extrapolating from the observed $E[Y_1(1) \mid D = 1]$ using the observed "trend" between $E[Y_1(0) \mid D = 0]$ and $E[Y_2(0) \mid D = 0]$. The ATET is then the difference between the observed $E[Y_2(1) \mid D = 1]$ and the imputed $E[Y_2(0) \mid D = 1]$.

Estimation of the ATET in canonical DiD in a finite sample is straightforward by considering four group means:

$$\hat{\theta}_s(d) = \frac{\mathbb{E}_n[Y1(D = d, t = s)]}{\mathbb{E}_n[1(D = d, t = s)]}.$$

Defining the estimator of the ATET as $\widehat{\alpha}$, we have

$$\widehat{\alpha} = (\hat{\theta}_2(1) - \hat{\theta}_1(1)) - (\hat{\theta}_2(0) - \hat{\theta}_1(0)). \tag{16.2.4}$$

Asymptotic properties under independence follow in a fashion similar to difference-in-mean estimators for the ATE outlined in Chapter 2.

We can also obtain a numerically equivalent estimator of the ATET via regression. Specifically, the ordinary least squares estimator of the parameter $\alpha$ in the linear model

$$Y = \beta_0 + \beta_1 D + \beta_2 P + \alpha DP + U, \tag{16.2.5}$$

where $P$ is a binary variable with $P = 1$ indicating the post-treatment time period (t = 2), is numerically equivalent to $\widehat{\alpha}$ in (16.2.4). The regression formulation is especially convenient for obtaining standard errors under different dependence assumptions.

**Table 16.1:** DiD Estimation of the Effect of the Mariel Boatlift on Unemployment

| | 1979 | 1981 | Difference |
|---|---|---|---|
| Miami Unemployment | 5.1 | 3.9 | -1.2 |
| | (1.1) | (0.9) | (1.4) |
| Comparison Unemployment | 4.4 | 4.3 | -0.1 |
| | (0.3) | (0.3) | (0.4) |
| Difference (Miami - Comparison) | 0.7 | -0.4 | -1.1 |
| | (1.1) | (0.9) | (1.5) |

**Note:** Unemployment rates among white individuals in Miami and four comparison cities – Atlanta, Los Angeles, Houston, and Tampa-St. Petersburg – reproduced from [5]. Standard errors assuming independence are in parentheses. The DiD estimate is provided in the entry in the last row and column.

## The Mariel Boatlift

Card's analysis of the impact of the Mariel Boatlift on the Miami labor market, [5], provides a prototypical application of DiD. For example, it is the example of DiD in Angrist and Krueger's *Handbook of Labor Economics* chapter on methods [6]. The basic idea of the study was to use the Mariel Boatlift – a sudden and arguably unexpected inflow of immigrants that increased the Miami labor force by about 7% between May and September of 1980 – to understand the impact of immigration on low-skilled labor market outcomes.

A key component of the analysis was arguing that Atlanta, Los Angeles, Houston, and Tampa-St. Petersburg provide valid control cities in the sense that we might plausibly believe that the change in labor market outcomes in these cities from the late 1970's to the early 1980's is useful for inferring how the Miami labor market would have changed in the absence of the Mariel immigration. Part of the argument in [5] relies on evidence that the cities had similar characteristics in the pre-treatment period. Effectively, this argument relies on parallel trends holding *conditional* on these pre-treatment characteristics. We consider using DML to flexibly control for rich covariates in Section 16.3.

We illustrate canonical DiD in the Mariel Boatlift example in Table 16.1 which uses numbers taken from Table 4 in [5]. Here, we see the DiD estimate of the ATET on unemployment is -1.1 with standard error 1.5, which does not provide strong evidence of a large impact of the Mariel immigration on unemployment.

## 16.3 DML and Conditional Difference-in-Differences

In many empirical applications, researchers deviate from the canonical DiD framework by including additional control variables. The fundamental motivation is similar to that for including control variables in other causal contexts, e.g., as motivated in Chapter 5: it is easier to believe that parallel trends holds among units that are identical in terms of observed characteristics. In this section, we explore flexibly including control variables in a DiD framework leveraging DML methods.

We restate the canonical DiD assumptions so that they hold after conditioning on pre-treatment/strictly exogenous characteristics $X$.

**Assumption 16.3.1** (Conditional DiD Assumptions) *Potential outcomes satisfy conditional parallel trends*

$$\mathrm{E}[Y_2(0) - Y_1(0) \mid D = 1, X] = \mathrm{E}[Y_2(0) - Y_1(0) \mid D = 0, X] \ a.s. \tag{16.3.1}$$

*and no anticipation*

$$\mathrm{E}[Y_1(0) \mid D = 1, X] = \mathrm{E}[Y_1(1) \mid D = 1, X] \ a.s. \tag{16.3.2}$$

*In addition, there is a treatment group and its characteristics overlap with the control group*

$$\exists\, \varepsilon : \mathrm{P}(D = 1) \geq \varepsilon \text{ and } \mathrm{P}(D = 1 \mid X) \leq 1 - \varepsilon \ a.s. \tag{16.3.3}$$

The intuition for (16.3.1) and (16.3.2) is essentially identical to the intuition for Assumption 16.2.1 discussed in the previous section. The only difference is that these conditions are now imposed within observationally identical groups as defined by $X$. Condition (16.3.3) is a standard overlap condition for identifying ATET which essentially imposes that there are control observations available for every value of $X$. Under Assumption 16.3.1, it is straightforward to verify that the ATET is identified by repeating the argument in (16.2.3) conditional on $X$ and averaging over the distribution of $X$ in the $D = 1$ group.

We leave verification of identification of the ATET in the conditional DiD framework as an exercise.

Similar to estimating parameters in the partially linear model or average treatment effects under confounding as discussed in Chapter 10, obtaining estimates of the ATET in the conditional DiD setting will require estimating high-dimensional nuisance objects. We thus exploit DML methods to accommodate the use of flexible methods in estimating these objects.

A key input into DML estimation is a Neyman-orthogonal score. In the conditional DiD framework with panel data,[1]

$$\psi(W;\alpha,\eta) = \frac{D - m(X)}{p(1 - m(X))}(\Delta Y - g(0,X)) - \frac{D}{p}\alpha \qquad (16.3.4)$$

provides an orthogonal score for the ATET, $\alpha$, where $W = (Y_1, Y_2, D, X)$ denotes the observable variables; $\Delta Y = Y_2 - Y_1$; $\eta = (p, m, g)$ denotes nuisance parameters with true values $p_0 = \mathrm{E}[D]$, $m_0(X) = \mathrm{E}[D \mid X]$, and $g_0(0, X) = \mathrm{E}[\Delta Y \mid D = 0, X]$. See also [7], [8], [9]. Comparing to the score for the ATET provided in Chapter 10, we see that this score is identical to that for learning the ATET under conditional ignorability where the outcome variable is simply defined as $\Delta Y$.

1: We provide the Neyman-orthogonal score and discuss DML estimation with repeated cross-sections in Section 16.A.

Given the Neyman-orthogonal score (16.3.4), it is then straightforward to implement DML to estimate the ATET. $\sqrt{n}$-asymptotic normality of $\widehat{\alpha}$, the DML estimator of the ATET, follows from Theorem 10.4.1 in Chapter 10.

**DML for ATET in Conditional DiD**

Let $(W_i)_{i=1}^n = (Y_{1i}, Y_{2i}, D_i, X_i)_{i=1}^n$ be the observed data.

1. Partition sample indices into random folds of approximately equal size: $\{1, \dots, n\} = \cup_{k=1}^K I_k$. For each $k = 1, \dots, K$, compute estimators $\hat{p}_{[k]}$, $\hat{g}_{[k]}$, and $\hat{m}_{[k]}$ of $\mathrm{E}[D]$ and the conditional expectation functions $g_0(0, X) = \mathrm{E}[\Delta Y \mid D = 0, X]$ and $m_0(X) = \mathrm{E}[D \mid X]$ leaving out the $k^{\text{th}}$ block of data and enforcing $\hat{m}_{[k]} \le 1 - \epsilon$.
2. For each $i \in I_k$, let

$$\hat{\psi}(W_i;\alpha) = \frac{D_i - \hat{m}_{[k]}(X_i)}{\hat{p}_{[k]}(1 - \hat{m}_{[k]}(X_i))}(\Delta Y_i - \hat{g}_{[k]}(0, X_i)) - \frac{D_i}{\hat{p}_{[k]}}\alpha.$$

Compute the estimator $\widehat{\alpha}$ as the solution to $\mathbb{E}_n[\hat{\psi}(W_i;\alpha)] = 0$ which yields

$$\widehat{\alpha} = \frac{\mathbb{E}_n\left[\frac{D_i - \hat{m}_{[k]}(X_i)}{\hat{p}_{[k]}(1-\hat{m}_{[k]}(X_i))}(\Delta Y_i - \hat{g}_{[k]}(0, X_i))\right]}{\mathbb{E}_n\left[\frac{D_i}{\hat{p}_{[k]}}\right]}.$$

3. Let

$$\hat{\varphi}(W_i) = \frac{\hat{\psi}(W_i;\hat{\alpha})}{\mathbb{E}_n\left[\frac{D_i}{\hat{p}_{[k]}}\right]}.$$

Construct standard errors via

$$\sqrt{\hat{\mathrm{V}}/n}, \quad \hat{\mathrm{V}} = \mathbb{E}_n[\hat{\varphi}(W_i)^2]$$

and use standard normal critical values for inference.

### Comparison to Adding Regression Controls

The equivalence between the ATET estimator obtained by directly looking to the difference between the treatment and control differences in means and the ordinary least squares estimator of the coefficient $\alpha$ in the linear model (16.2.5) in the canonical DiD setting suggests a simple approach to incorporating control variables by augmenting the regression model to include controls linearly. That is, add $\beta' X$ to the model in (16.2.5). However, the coefficient on the $DP$-interaction term is not equivalent to the ATET and need not uncover any sensible causal effect without very strong functional form restrictions and restrictions on treatment effect heterogeneity. See, e.g., [10] for further discussion. In contrast, the DML estimator always targets the ATET under Assumption 16.3.1 and is relatively simple to implement.

Minimum Wage R Notebook and Minimum Wage Python Notebook contain the code for the minimum wage example.

## 16.4 Example: Minimum Wage

In this section, we use DML for DiD to estimate the effect of minimum wage increases on teen employment. We use data from and roughly follow the approach of [11]. The data are annual county level data from the United States covering 2001 to 2007. The outcome variable is log county-level teen employment, and the treatment variable is an indicator for whether the county has a minimum wage above the federal minimum wage.[2] Note that this definition of the treatment variable makes the analysis straightforward but ignores the nuances of the exact value of the minimum wage in each county and how far those values are from the federal minimum.[3] The data also includes county population and county average annual pay. We follow [11] by removing observations with missing entries which produces a balanced panel with data from counties in 42 states. See [11], [12], and [13] for further details regarding the data.

2: The federal minimum wage over 2001-2007 was constant at $5.15.

3: Under these definitions, this example is an example of *staggered adoption*. Staggered adoption refers to a setting with a binary, absorbing treatment variable. That is, once an observation becomes treated it remains treated thereafter. This setting is straightforward to analyze as treatment paths are completely characterized by the treatment date and controls can be constructed from observations that are not treated during the sample period (the never treated) or observations that are not treated prior to the treatment date and remain untreated in the period in which one wants to estimate the ATET (the as-yet not treated).

We focus our analysis exclusively on the set of counties that had wage increases away from the federal minimum wage in 2004. That is, we treat 2003 and earlier as the pre-treatment period and the period 2004-2007 as the post-treatment period. We assume that parallel trends holds after conditioning on three pre-treatment variables – 2001 population, 2001 average pay, and 2001 teen employment – and the region to which each county belongs.[4]

4: We follow [11] and categorize each observation as belonging to one of four U.S. census regions.

We estimate dynamic effects by estimating the ATET in 2004-2007 corresponding to the effect in the year of treatment and one, two, and three years after the treatment. For control observations, we use the set of observations that still have minimum wage equal to the federal minimum in each year – the "as-yet not treated" – so the control group changes from period to period. For example, we use all observations that had minimum wage equal to the federal minimum in 2004 as control observations when estimating the ATET in 2004, but we use all observations that had minimum wage equal to the federal minimum in 2005 as control observations to estimate the 2005 ATET. These definitions yield 102 treatment observations for estimating each ATET and 2389, 2327, 2080, and 1417 control observations for 2004, 2005, 2006, and 2007 respectively.

Since our goal is to estimate the ATET of the county level minimum wage being larger than the federal minimum imposing that parallel trends holds after flexibly controlling for region and our pre-treatment variables, we employ DML using the algorithm from Section 16.3 using an array of methods including several of the modern regression methods that we discussed in previous chapters. Specifically, we consider ten candidate learners for the high-dimensional nuisance functions $g_0(0, X) = \mathrm{E}[\Delta Y \mid D = 0, X]$ and $m_0(X) = \mathrm{E}[D \mid X]$. We consider using no control variables (No Controls) which corresponds to maintaining unconditional parallel trends. We consider linear index models using only the raw control variables (Basic) – the four region dummies and log of 2001 population, log of 2001 average pay, and log of 2001 employment – and using a full cubic expansion of the raw control variables including all third order interactions (Expansion).[5] We consider Lasso and Ridge with the cubic expansion of the raw variables and penalty parameter chosen by cross-validation (Lasso (CV) and Ridge (CV)). We consider a random forest with no randomization over input variables and 1000 trees (Random Forest). Additionally, we consider three different tree models: a tree with depth 15 (Deep Tree), a tree with depth 3 (Shallow Tree), and a tree tuned using cross-validation (Tree (CV)). For random forest

5: We use a linear model estimated by OLS for $g_0(0, X)$ and a logistic model with linear index in the stated variables for $m_0(X)$.

| | 2004 | 2005 | 2006 | 2007 |
|---|---|---|---|---|
| | A. $E[\Delta Y \mid D = 0, X]$ | | | |
| No Controls | 0.1633 | 0.1882 | 0.2235 | 0.2302 |
| Basic | 0.1634 | 0.1854 | 0.2191 | 0.2216 |
| Expansion | 0.1887 | 0.2122 | 0.2445 | 0.2710 |
| Lasso (CV) | 0.1631 | 0.1851 | 0.2193 | 0.2214 |
| Ridge (CV) | 0.1631 | 0.1851 | 0.2191 | 0.2213 |
| Random Forest | 0.1716 | 0.1982 | 0.2330 | 0.2388 |
| Deep Tree | 0.1922 | 0.2250 | 0.2599 | 0.2708 |
| Shallow Tree | 0.1678 | 0.1924 | 0.2279 | 0.2290 |
| Tree (CV) | 0.1633 | 0.1889 | 0.2178 | 0.2227 |
| | B. $E[D \mid X]$ | | | |
| No Controls | 0.1983 | 0.2006 | 0.2111 | 0.2503 |
| Basic | 0.1986 | 0.2009 | 0.2113 | 0.2217 |
| Expansion | 0.1988 | 0.2007 | 0.2113 | 0.2217 |
| Lasso (CV) | 0.1968 | 0.1986 | 0.2083 | 0.2197 |
| Ridge (CV) | 0.1971 | 0.1989 | 0.2086 | 0.2198 |
| Random Forest | 0.2005 | 0.2051 | 0.2128 | 0.2355 |
| Deep Tree | 0.2207 | 0.2364 | 0.2303 | 0.2744 |
| Shallow Tree | 0.1921 | 0.1944 | 0.2029 | 0.2301 |
| Tree (CV) | 0.1937 | 0.1955 | 0.2039 | 0.2311 |

**Note:** Cross-fit RMSE for predicting $\Delta Y$ and treatment status $D$ in the minimum wage example. Row labels denote the method used to estimate the nuisance function, and column labels indicate the year for which we are calculating the ATET, with 2004, 2005, 2006, and 2007 respectively corresponding to the year of the treatment, one year after treatment, two years after treatment, and three years after treatment.

**Table 16.2:** RMSE for Learners in Minimum Wage example

and the tree models, we use region, log of 2001 population, log of 2001 average pay, and log of 2001 employment as input variables. Finally, we consider estimation using the learner for $E[\Delta Y \mid D = 0, X]$ and for $E[D \mid X]$ that produce the lowest RMSE during cross-fitting (Best) allowing for a different learner to be selected for each task.[6]

We start by reporting the RMSE obtained during cross-fitting for each learner in each period in Table 16.2. Here we see that the Deep Tree systematically performs substantially worse in terms of cross-fit predictions than the other learners for both tasks and that Expansion performs similarly poorly for the outcome prediction. It also appears there is some signal in the regressors, especially for the propensity score, as all methods outside of Deep Tree and Expansion produce smaller RMSEs than the No Controls baseline. The other methods all produce similar RMSEs, with a small edge going to Ridge and Lasso. While it would be hard to reliably conclude which of the relatively good performing methods is statistically best here, one could exclude Expansion and Deep Tree from further consideration on the basis of out-of-sample performance suggesting they

6: For any observation with estimated propensity score larger than 0.95, we replace the propensity score with 0.95. Applying this trimming, we replace 12, 10, 13, and 21 observations for the deep tree in 2004-2007 respectively and replace 2, 2, and 1 observation for Basic, Expansion, and Lasso (CV) in 2007.

| | 2004 | 2005 | 2006 | 2007 |
|---|---|---|---|---|
| No Controls | -0.039 | -0.076 | -0.117 | -0.131 |
| | (0.019) | (0.021) | (0.023) | (0.026) |
| Basic | -0.037 | -0.066 | -0.088 | -0.041 |
| | (0.018) | (0.020) | (0.021) | (0.033) |
| Expansion | -0.022 | -0.046 | -0.061 | 0.303 |
| | (0.025) | (0.030) | (0.033) | (0.227) |
| Lasso (CV) | -0.035 | -0.062 | -0.082 | -0.049 |
| | (0.018) | (0.020) | (0.021) | (0.031) |
| Ridge (CV) | -0.035 | -0.062 | -0.083 | -0.061 |
| | (0.018) | (0.020) | (0.021) | (0.025) |
| Random Forest | 0.013 | -0.056 | -0.039 | -0.071 |
| | (0.029) | (0.024) | (0.028) | (0.038) |
| Deep Tree | 0.077 | 0.007 | 0.100 | -0.470 |
| | (0.079) | (0.172) | (0.080) | (0.178) |
| Shallow Tree | -0.028 | -0.040 | -0.058 | -0.065 |
| | (0.019) | (0.021) | (0.021) | (0.026) |
| Tree (CV) | -0.027 | -0.045 | -0.060 | -0.069 |
| | (0.019) | (0.021) | (0.021) | (0.025) |
| Best | -0.028 | -0.051 | -0.055 | -0.055 |
| | (0.019) | (0.021) | (0.021) | (0.031) |

**Note:** Estimated ATET and standard errors (in parentheses) in the minimum wage example. Row labels denote the method used to estimate the nuisance function, and column labels indicate the year for which we are calculating the ATET, with 2004, 2005, 2006, and 2007 respectively corresponding to the year of the treatment, one year after treatment, two years after treatment, and three years after treatment.

**Table 16.3:** Estimated ATET in Minimum Wage example

are doing a poor job approximating the nuisance functions. Best (or a different ensemble) provides a good baseline that is principled in the sense that one could pre-commit to using the best learners without having first looked at the subsequent estimation results.

We report estimates of the ATET in each period in Table 16.3. Here, we see that the majority of methods provide point estimates that suggest the minimum wage increase leads to decreases in youth employment with small effects in the initial period that become larger in the years following the treatment. This pattern seems economically plausible as it may take time for firms to adjust employment and other input choices in response to the minimum wage change. The methods that produce estimates that are not consistent with this pattern are Deep Tree and Expansion which are both suspect as they systematically underperform in terms of having poor cross-fit prediction performance. In terms of point estimates, the other pattern that emerges is that all estimates that use the covariates produce ATET estimates that are systematically smaller in magnitude than the No Controls baseline, suggesting that failing to include the controls may lead to overstatement of treatment effects in this example.

Turning to inference, we would reject the hypothesis of no minimum wage effect two or more years after the change at the 5% level, even after multiple testing correction, if we were to focus on the row "Best" (or many of the other individual rows). Focusing on "Best" is a reasonable ex ante strategy that could be committed to prior to conducting any analysis. It is, of course, reassuring that this broad conclusion is also obtained using many of the individual learners suggesting some robustness to the exact choice of learner made.

Because we have data for the period 2001-2007, we can perform a so-called *placebo* or *pre-trends* test to provide some evidence about the plausibility of the conditional DiD assumptions, Assumption 16.3.1. Specifically, we can continue to use 2003 as the reference period but now consider 2002 to be the treatment period. Sensible economic mechanisms underlying Assumption 16.3.1 would then typically suggest that the ATET in 2002 – before the 2004 minimum wage change we are considering – should be zero. Finding evidence that the ATET in 2002 is non-zero then calls into question the validity of Assumption 16.3.1.

We repeat the exercise for obtaining our ATET estimates and standard error for 2004-2007 and report the results in Table 16.4. Here we see broad agreement across all methods in the sense of returning point estimates that are small in magnitude and small relative to standard errors. In no case would we reject the hypothesis that the pre-event effect in 2002 is different from zero at usual levels of significance. We note that failing to reject the hypothesis of no pre-event effects certainly does not imply that Assumption 16.3.1 is in fact satisfied. For example, confidence intervals include values that would be consistent with relatively large pre-event effects. Conditioning inference on the results of such an assessment is also generally a bad idea; see, e.g. [14] and [15] for a discussion specifically in the context of DiD. However, it is reassuring to see that there is not strong evidence of a violation of the underlying identifying assumption.

**Table 16.4:** Pre-trends Assessment

| | RMSE Y | RMSE D | ATET | s.e. |
|---|---|---|---|---|
| No Controls | 0.1543 | 0.1945 | -0.0037 | (0.0131) |
| Basic | 0.1541 | 0.1949 | -0.0044 | (0.0130) |
| Expansion | 0.1577 | 0.1949 | 0.0046 | (0.0140) |
| Lasso (CV) | 0.1544 | 0.1932 | -0.0039 | (0.0131) |
| Ridge (CV) | 0.1544 | 0.1935 | -0.0053 | (0.0131) |
| Random Forest | 0.1635 | 0.2265 | 0.0230 | (0.0265) |
| Deep Tree | 0.1822 | 0.2234 | 0.0080 | (0.0276) |
| Shallow Tree | 0.1620 | 0.1884 | -0.0037 | (0.0134) |
| Tree (CV) | 0.1550 | 0.1905 | -0.0056 | (0.0133) |
| Best | 0.1541 | 0.1884 | -0.0031 | (0.0134) |

**Note:** Estimated pre-event (2002) ATET and standard errors (in parentheses) in the minimum wage example. Row labels denote the method used to estimate the nuisance function. RMSE Y and RMSE D give cross-fit RMSE for the outcome and treatment respectively. ATET provides the point estimate of the ATET based on the method in the row label with standard error given in column s.e.

## Notebooks

- Minimum Wage R Notebook and Minimum Wage Python Notebook contain the analysis of minimum wage example.

## Notes

There is a relatively large literature focusing on flexibly estimating ATET in DiD contexts. Much of this work has focused on potential failure of the usual practice of estimating homogeneous coefficient linear models with additive fixed effects for groups and time periods under heterogeneous treatment effects. Specifically, much of the work has noted that coefficients on a treatment variable in a homogeneous linear model with fixed effects need not be proper weighted averages of heterogeneous treatment effects but may place negative weights on some effects. The possibility of negative weights then leaves open the possibility of, for example, having uniformly positive treatment effects but obtaining negative and significant estimates of the coefficient on a treatment variable in a linear model. The DML approach we present in this chapter offers one solution to this problem that allows for flexibly accommodating control variables that can account for heterogeneity. See the excellent review papers [11], [16], [17] for more discussion.

## Study Problems

1. Verify that the ATET is identified under Assumption 16.3.1. Provide a short explanation of the intuition for the identification result. Give an intuitive example where the ATET would be identified after conditioning on covariates but where identification of the ATET would fail in the canonical DiD framework (i.e. without conditioning on additional covariates).

2. Study the minimum wage empirical analysis notebook. Estimate the ATET for observations treated in a year different than 2004 – e.g. repeat the analysis doing the exercise for observations treated in 2005.

3. Study the minimum wage empirical analysis notebook. Estimate the ATET using the never treated as opposed to the not-yet treated as the control group.

## 16.A Conditional Difference-in-Differences with Repeated Cross-Sections

Here we provide the Neyman-orthogonal score for the ATET in the conditional DiD context with repeated cross-section data. For additional development including formal statement of additional assumptions for DiD with repeated cross sections, see [7], [8], [9].

The chief difference in this setting relative to when one has panel data is that we can not directly construct the difference between outcomes in the first and second period as we do not see the same individuals across time periods. Rather, we revert to the analog of the canonical DiD estimator by directly working with the four conditional means defined by grouping the treated and control observations pre- and post-treatment. Specifically, we make use of the score function

$$
\begin{aligned}
\psi(W, \alpha, \eta) = & \left( \frac{DT}{p\lambda}(Y - g(1,2,X)) \right. \\
& \left. - \frac{D(1-T)}{p(1-\lambda}(Y - g(1,1,X)) \right) \\
& - \left( \frac{m(X)(1-D)T}{p\lambda(1-m(X))}(Y - g(0,2,X)) \right. \\
& \left. - \frac{m(X)(1-D)(1-T)}{p(1-\lambda)(1-m(X))}(Y - g(0,1,X)) \right) \\
& + \frac{D}{p}\left(g(1,2,X) - g(1,1,X)\right) \\
& - \frac{D}{p}\left(g(0,2,X) - g(0,1,X)\right) - \frac{D}{p}\alpha
\end{aligned}
\tag{16.A.1}
$$

where $W = (Y, T, D, X)$ denotes the observable variables for each observation with $T$ an indicator which equals one if the observation is in the post-treatment period (period 2) and $\eta = (p, \lambda, m, g)$ denotes nuisance parameters with true values $p_0 = \mathrm{E}[D]$, $\lambda_0 = \mathrm{E}[T]$, $m_0(X) = \mathrm{E}[D \mid X]$, and $g_0(d, t, X) = \mathrm{E}[Y \mid D = d, T = t, X]$.

Under iid sampling, we can directly apply the generic cross-fitting approach to DML as in Section 10.4. In many DiD settings, researchers wish to allow for unmodeled dependence between observations corresponding to different groups such as cities or counties. As long as there are many such groups, it is straightforward to modify the DML algorithm to accommodate this dependence. The algorithm needs to be adjusted by forming the cross-fitting folds such that all observations within groups are included together in the same fold. Similarly, it is straightforward to adjust inference to account for this dependence by applying clustered standard errors with clustering done at the group level.

# 17 Regression Discontinuity Designs

"la nature ne fait jamais des sauts."
("nature never makes jumps.")

– Gottfried Leibniz [1].

In this chapter we discuss Regression Discontinuity Design (RDD). First, we introduce the basic idea of Regression Discontinuity (RD). RDDs, when they exist, offer a highly credible way to identify causal effects. However, leveraging RDDs without covariates can fall short in practice, whether due to lack of observations near the RD or the lack of generalizability away from the RD. We show how modern machine learning methods can be utilized for estimation in RDDs with very many covariates.

## 17.1 Introduction

Like many other methods presented in the Advanced Materials – IV, proxy controls, and DiD – RDDs are also widely used in empirical work for measuring causal effects in non-experimental settings where we cannot reliably measure all confounders.

The basic RDD structure relies on a so-called running variable or score which determines treatment: units whose score is above a cutoff value are assigned to the treatment, while units with score below the cutoff are assigned to control. Examples are reward of a scholarship if a student's grade average exceeds a certain threshold, bestowing of license to practice (say, medicine or law) if one's exam score exceeds a threshold, assignment of a particular medical treatment if a biomarker is above a cutoff, or getting social benefits if the income is below some income threshold.

We can always negate the running variable or rename the treatment if the relationship is the other way.

The intuition for identification is that units marginally above and below the threshold are comparable in terms of potential outcomes, since they are the same in all ways except the assignment to treatment, assuming of course that there are no other discontinuities at the cutoff that would also render them different in other ways. The latter continuity in potential outcomes is the identifying assumption in RDDs. For example, suppose we are interested in the causal effect of a student receiving a scholarship on their future academic success. While the future academic success of students with low grade averages is very different from those with high averages, with or without a scholarship, the students right at the cutoff essentially have the same averages and are comparable, but those just above have a scholar and those just below do not.

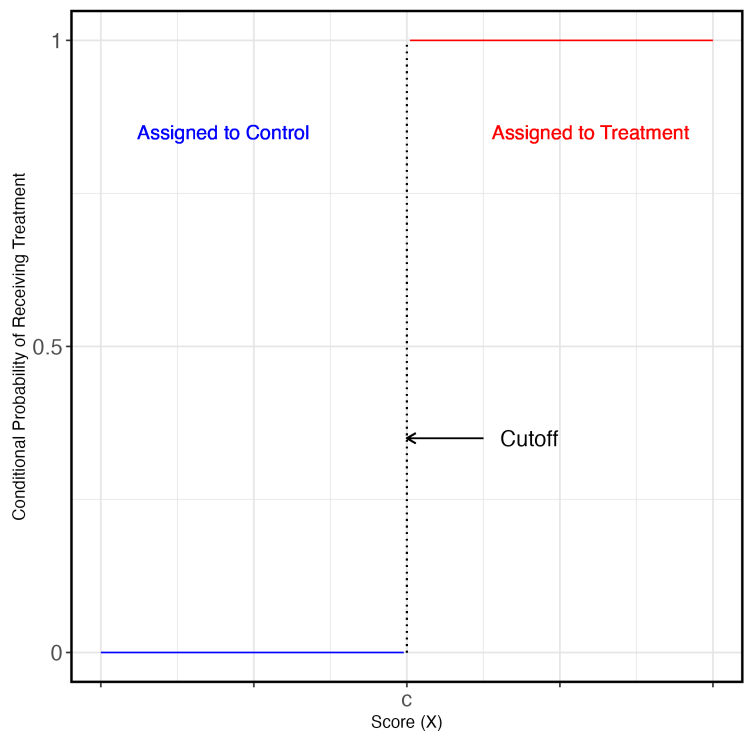

**Figure 17.1:** In the sharp RDD the assignment of treatment depends in a deterministic way on the underlying . Units with values of the running variable below a cutoff are not treated, while units above the threshold are treated.

**Figure 17.2:** In the sharp RDD the assignment of treatment depends in a deterministic way on the underlying running variable. Units with values of the running variable below a cutoff are not treated, while units above the threshold are treated.

We can also conceive of being above or below as random "luck," i.e., exogenous variation. E.g., getting just one more question right on the exam is a random event that has nothing to do with the academic preparedness of the student – it can happen to any one. This is an alternative approach to identification in RDDs based on local randomization [2].

## 17.2 The Basic RDD Framework

### Setting

In the sharp RDD the binary treatment variable $D_i \in \{0, 1\}$ for individual $i$ is assigned on basis of a running variable $X_i$

in a deterministic ("sharp") way: $D_i = 1(X_i \geq c)$, where 1 denotes the indicator function and $c$ the cutoff value. An unit is treated ($D_i = 1$) if the value of the running variable is above the threshold and in the control group ($D_i = 0$) otherwise. For each individual we observe additionally the outcome $Y_i$ and potentially some pre-treatment variables $Z_i \in \mathbb{R}^p$. The observed data $\{W_i\}_{i=1}^n = \{(Y_i, X_i, Z_i)\}_{i=1}^n$ are an i.i.d. sample of size $n$ from the distribution of $W = (Y, X, Z)$.

The parameter of interest in RDD is the ATE at the cutoff value $c$:

$$\tau_{\mathrm{RD}} = \mathbb{E}\left[Y_i(1) - Y_i(0) \mid X_i = c\right].$$

For identification of this causal effect it is required, that i) the conditional mean functions of the potential outcome $\mathbb{E}\left(Y_i(t) \mid X_i = x\right)$ are continuous at the cutoff level for $t \in \{0, 1\}$ and ii) that the density of the running variable near the cutoff is positive.

Under these conditions we have

$$\tau_{\mathrm{RD}} = \lim_{x \downarrow c} \mathbb{E}\left(Y_i \mid X_i = x\right) - \lim_{x \uparrow c} \mathbb{E}\left(Y_i \mid X_i = x\right).$$

$\lim_{x \downarrow c}$ and $\lim_{x \uparrow c}$ denote the right-sided and left-sided limit.

Hence, the jump in the conditional expectation functions $\mathbb{E}\left(Y_i \mid X_i = x\right)$ of the observed outcome at the threshold determines the causal effect of interest.

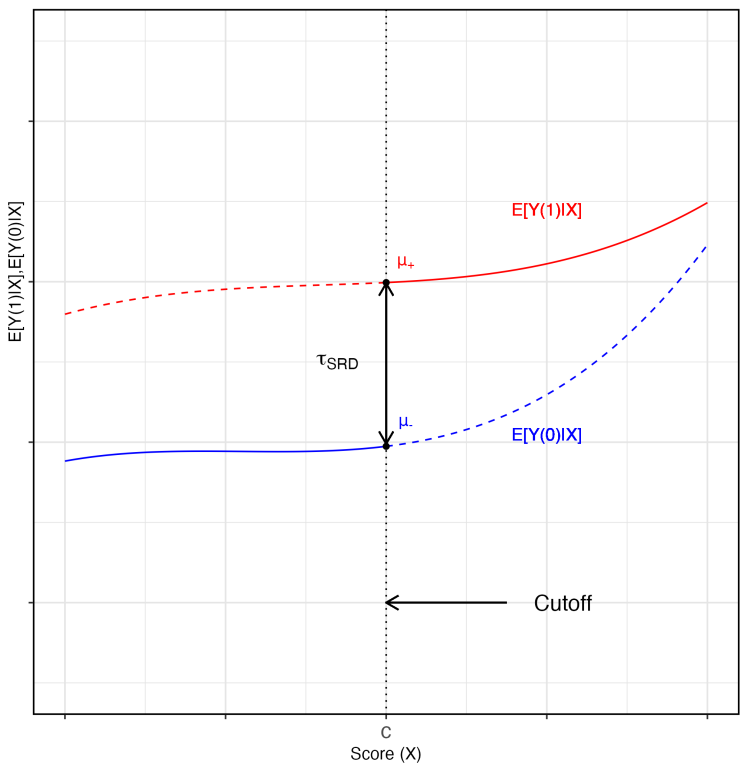

**Figure 17.3:** Identifictation and estimation in the sharp RDD.

## Estimation

In the sharp RDD we are faced with the problem of estimating the jump in the conditional mean functions at the cutoff value which boils down to estimation of the conditional mean functions at the left and right of the cutoff value. For this nonparametric methods, like sieves, kernel, and local polynomials can be used. Local polynomial estimation has become the default method for this, and therefore we will focus on this method following the notation and exposition in [3] closely.

*Standard RD Estimator:* Without covariates, a weighted linear regression of $Y_i$ on $X_i$ is estimated locally around the cutoff to estimate the parameter of interest:

$$\hat{\tau}_{h,\mathrm{base}} = e_2^\top \operatorname*{argmin}_{\theta \in \mathbb{R}^4} \sum_i^n K_h\left(X_i\right)\left(Y_i - V_i^\top \theta\right)^2.$$

$K$ denotes a kernel function, $h > 0$ the bandwidth, $K_h(x) = K(x/h)/h, V_i = (1, D_i, X_i/h, D_iX_i/h)^\top$ a vector of appropriate transformations of the running variable, and $e_2 = (0,1,0,0)^\top$ the unit vector to select the coefficient of $D_i$, which is the target parameter.

In a setting, where standard conditions are met, such as the continuity of the running variable and the bandwidth $h$ approaching zero at a suitable rate, the estimator $\widehat{\tau}_{\text{base}}(h)$ demonstrates an approximate normal distribution in large samples with a bias of the order $h^2$ and a variance of the order of $(nh)^{-1}$:

$$\widehat{\tau}_{\text{base}}(h) \overset{a}{\sim} N\left(\tau + h^2 B_{\text{base}}, (nh)^{-1} V_{\text{base}}\right).$$

Bias and variance are given by

$$\begin{aligned} B_{\text{base}} &= \frac{\bar{\nu}}{2}\left(\partial_x^2 \mathbb{E}\left[Y_i \mid X_i = x\right]\big|_{x=0^+} - \partial_x^2 \mathbb{E}\left[Y_i \mid X_i = x\right]\big|_{x=0^-}\right) \text{ and} \\ V_{\text{base}} &= \frac{\bar{\kappa}}{f_X(0)}\left(\mathbb{V}\left[Y_i \mid X_i = 0^+\right] + \mathbb{V}\left[Y_i \mid X_i = 0^-\right]\right). \end{aligned}$$

Here $\bar{\nu}$ and $\bar{\kappa}$ are kernel constants, defined as
$\bar{\nu} = \left(\bar{\nu}_2^2 - \bar{\nu}_1\bar{\nu}_3\right)/\left(\bar{\nu}_2\bar{\nu}_0 - \bar{\nu}_1^2\right)$ for $\bar{\nu}_j = \int_0^\infty v^j K(v)dv$
and $\bar{\kappa} = \int_0^\infty \left(K(v)\left(\bar{\nu}_1 v - \bar{\nu}_2\right)\right)^2 dv / \left(\bar{\nu}_2\bar{\nu}_0 - \bar{\nu}_1^2\right)^2$, and $f_X$ denotes the density of $X_i$.

*RDD with Covariates:* In empirical work, covariates (pretreatment variables) $Z_i$ are often available that could also be included in the analysis. This is analogous to randomized control trials, where additional covariates can reduce the variance of the estimator and usually do not effect the point estimate. There are several ways how to adjust the RD estimator for covariates. [4] analyse in detail the use of additional regressors in RDD. The standard approach is simply to take up the regressors in the weighted least squares regression. The modified estimator is given by:

$$\hat{\tau}_{h,\text{adj}} = e_2^\top \underset{(\theta,\gamma)\in\mathbb{R}^{4+p}}{\operatorname{argmin}} \sum_i^n K_h(X_i)\left(Y_i - V_i^\top\theta - Z_i^\top\gamma\right)^2. \quad (17.2.1)$$

$Z_i$ denotes the vector of covariates and $\gamma$ the corresponding coefficient vector.

An important insight is, that the estimator can be equivalently written as a RD estimator without covariates, but with a covariate-adjusted outcome, $Y_i - Z_i^\top \widehat{\gamma}_h$, where $\widehat{\gamma}_h$ is the vector of linear projection coefficients. The adjusted estimator is then given by:

$$\widehat{\tau}_{\text{lin}}(h) = \sum_{i=1}^{n} w_i(h)\left(Y_i - Z_i^\top \widehat{\gamma}_h\right),$$

with data-dependent weights $w_i(h)$ which depend only on the realizations of the running variable.

[4] show that $\widehat{\tau}_{\text{lin}}(h)$ is consistent for the RD parameter if the conditional distribution of the regressors given the running variable varies smoothly around the cutoff. The surprising part is that no functional form assumptions on the underlying conditional expectations are required.

Specifically, if $\mathbb{E}[Z_i \mid X_i = x]$ is twice continuously differentiable around the cutoff, then

$$\widehat{\tau}_{\text{lin}}(h) \overset{a}{\sim} N\left(\tau + h^2 B_{\text{base}}, (nh)^{-1} V_{\text{lin}}\right)$$

under regularity conditions similar to those for the estimator without covariates, where the bias term $B_{\text{base}}$ is as above and the new variance term is

$$V_{\text{lin}} = \frac{\bar{\kappa}}{f_X(0)}\left(\mathbb{V}\left[Y_i - Z_i^\top \gamma_0 \mid X_i = 0^+\right] + \mathbb{V}\left[Y_i - Z_i^\top \gamma_0 \mid X_i = 0^-\right]\right)$$

with $\gamma_0$, a non-random vector of projection coefficients, the probability limit of $\widehat{\gamma}_h$ (see also [5]).

The linear adjustment estimator generally has smaller asymptotic variance than the estimator without covariates, i.e. $V_{\text{lin}} \leq V_{\text{base}}$ which was shown in [5]. See also the discussions in [4].

## 17.3 RDD with (Many) Covariates

### Motivation for Using Covariates

For the identification and estimation of the average treatment effect at the cut-off value no covariate information is required except the running variable, but nevertheless in many applications additional covariates are collected, which might be exploited for the analysis. Following [6], the use of covariates is beneficial for:

1. *Efficiency and power improvements:* Similar as in randomized control trials, using covariates can increase efficiency and improve power, as we discussed in the previous section. [7] show that the inclusion of covariates in a local polynomial analysis (additional to the score) can lead to asymptotic efficiency gains, if carefully implemented.
2. *Auxiliary information:* In RDD the score determines the assignment of the treatment and measurement errors in the running variable can distort the results. Additional covariates can be exploited to overcome these issues or deal with missing data problems.
3. *Treatment effect heterogeneity:* Covariates can be used to define subgroups in which the treatment effects differ.
4. *Other parameters of interest and extrapolation:* As the identified treatment effect in RDD is local at the cutoff, additional covariates might help for extrapolation of the treatment effects or identify other causal parameters.

For an extensive discussion of the use of covariates in RDDs we refer to [6].

## High-Dimensional Covariates

### RDD with LASSO estimation

In the case where many covariates are potentially included in the local polynomial regression of the RDD, Lasso can be used for variable selection. This has been analyzed by [8] and [5]. Here we follow [5] closely. The idea is that in a first step the relevant variables are selected with a localized / weighted Lasso regression. In the second step, the local linear RDD estimation with the selected covariates from the first step is conducted. In detail, the procedure is given by:

1. Using a preliminary bandwidth $b$ and a penalty parameter $\lambda$, one solves the following Lasso version of the weighted least squares problem by adding a penalty term:

$$\left(\widetilde{\theta}, \widetilde{\gamma}\right) = \underset{(\theta,\gamma)\in\mathbb{R}^{4+p}}{\operatorname{argmin}} \sum_{i=1}^{n} K_b\left(X_i\right)\left(Y_i - V_i^\top\theta - \left(Z_i - \hat{\mu}_Z\right)^\top\gamma\right)^2 + \lambda \sum_{k=1}^{p} \hat{w}_k \left|\gamma_k\right|,$$

where

$$\hat{\mu}_Z = \frac{1}{n}\sum_{i=1}^{n} Z_i K_b(X_i) \text{ and } \hat{\varpi}_k^2 = \frac{b}{n}\sum_{i=1}^{n}\left(K_b(X_i) Z_i^{(k)} - \mu_Z^{(k)}\right)^2$$

are the local sample mean and variance, respectively, of the covariates.

2. Using a final bandwidth $h$, one computes the restricted post-Lasso estimate of $\tau_{\mathrm{RD}}$ as $\hat{\tau}_h\left(\hat{J}\right)$ as in 17.2.1, where $\hat{J} = \left\{k \in \{1,\ldots,p\} : \widetilde{\gamma}^{(k)} \neq 0\right\}$ is the set of the indices of those covariates selected in the first step.

*Results:*

A key assumption, which is widely used for studying Lasso, is an approximate sparsity condition, which has been already discussed earlier in this book. To state and adapt this assumption more formally, the following population regression coefficients and corresponding residuals for any $J \subset \{1,\ldots,p\}$ and bandwidth $h$ are defined:

$$(\theta_0(J,h), \gamma_0(J,h)) = \underset{(\theta,\gamma)}{\operatorname{argmin}} \mathbb{E}\left[K_h(X_i)\left(Y_i - V_i^\top\theta - Z_i(J)^\top\gamma\right)^2\right],$$
$$r_i(J,h) = Y_i - V_i^\top\theta_0(J,h) - Z_i(J)^\top\gamma_0(J,h).$$

Approximate sparsity then means that there exist covariate sets $J \subset \{1,\ldots,p\}$ that contains a "small" number $s \equiv |J| \ll p$ of regressors. For these, the local correlation between the corresponding regression errors $r_i(J,h)$ and each component of $Z_i$ is small relative to the estimation error:

$$\max_{j=1,\ldots,p}\left|\mathbb{E}\left[K_h(X_i) Z_i^{(j)} r_i(J,h)\right]\right| = O\left(\sqrt{\frac{\log p}{nh}}\right).$$

Moreover, this condition needs to be satisfied for an appropriate range of bandwidths, so that the sequence $J$ does not depend on the exact choice of $h$.

Under this and other regularity conditions, [5] show that the post-Lasso estimator $\hat{\tau}_h\left(\hat{J}\right)$ has the same first-order asymptotic properties as an infeasible estimator $\hat{\tau}_h(J)$ that uses the true target set, and then prove an asymptotic normality result for the latter. Taken together, this yields the main result of [5], which is that the post-Lasso estimator $\hat{\tau}_h\left(\hat{J}\right)$ of $\tau_{\mathrm{RD}}$ satisfies

$$\frac{\sqrt{nh}\left(\hat{\tau}_h\left(\hat{J}\right) - \tau_{\mathrm{RD}} - h^2\mathscr{B}_n\right)}{\mathcal{S}_n} \xrightarrow{d} \mathcal{N}(0,1),$$

with asymptotic bias and variance, respectively, such that

$$\mathscr{B}_n \approx \frac{C_{\mathscr{B}}}{2}\left(\mu''_{\widetilde{Y}+} - \mu''_{\widetilde{Y}-}\right) \quad \text{and} \quad \mathcal{S}_n^2 \approx \frac{C_{\mathcal{S}}}{f_X(0)}\left(\sigma^2_{\widetilde{Y}+} + \sigma^2_{\widetilde{Y}-}\right).$$

Here $C_{\mathscr{B}}$ and $C_{\mathcal{S}}$ are constants that depend on the kernel function $K$ only, and

$$\widetilde{Y}_i = Y_i - Z_i\left(J_n\right)^{\top}\gamma_n, \text{ with } \gamma_n = \left(\sigma^2_{Z(J_n)-} + \sigma^2_{Z(J_n)+}\right)^{-1}\left(\sigma^2_{YZ(J_n)-} + \sigma^2_{YZ(J_n)+}\right),$$

is a covariate-adjusted version of the outcome variable that uses a vector $\gamma_n$ that can be thought of as an approximation of $\gamma_0\left(J, h\right)$ that is independent of the bandwidth. The estimator is thus first-order asymptotically equivalent to a basic sharp RD estimator with the covariate-adjusted outcome $\widetilde{Y}_i$ replacing the original outcome $Y_i$

Here, we used the following notation: For generic random vectors $A$ and $B$, we use the notation that $\mu_A(x) = \mathbb{E}(A \mid X = x)$, $\mu_{AB}(x) = \mathbb{E}\left(AB^{\top} \mid X = x\right)$, $\sigma^2_{AB}(x) = \mu_{AB}(x) - \mu_A(x)\mu_B(x)^{\top}$; and write $\sigma^2_A(x) = \sigma^2_{AA}(x)$ for simplicity. For a generic function $f$, we also write $f_+ = \lim_{x\downarrow 0} f(x)$ and $f_- = \lim_{x\uparrow 0} f(x)$ for its right and left limit at zero, respectively, so that $\tau_{\mathrm{RD}} = \mu_{Y+} - \mu_{Y-}$.

**RDD with generic ML Methods**

As mentioned above, instead of using covariates in the weighted least squares regression (linear adjustment estimator), [4] show that it is asymptotically equivalent to run a local linear RD regression with a modified outcome variable $Y_i - Z_i'\gamma$ with a projection coefficient $\gamma$. [3] argue that this approach can be extended to allow for more general modifications of the form $Y_i - \eta_0(Z_i)$ for any function $\eta_0$. Different choices of $\eta_0$ give the same estimand, since treatment has no effect on $Z$, but may change the performance of an estimator based on such a modified centered outcome variable. The optimal choice of $\eta_0$ with regard to the asymptotic variance is the average of the conditional expectation functions of the outcome given the running variables and covariates just to the right and left of the cutoff value. In fact, that we get the same estimand for any $\eta_0$ also means that we must be insensitive to any errors in $\eta_0$, that is, we have Neyman-orthogonality. Thanks to this, by using a DML procedure, modern machine learning methods can then be used to estimate the function $\eta_0$ (especially, the optimal one) in a first step and then the modified outcome is used in a local RDD regression as second step, all with cross-fitting to ensure independence between the steps.

[3] extend the approach of [4] to allow for flexible covariate adjustment in high-dimensional settings using modern machine learning methods. The estimator they propose employs cross-fitting and consists of two steps:

1. Randomly split the data $\{W_i\}_{i\in[n]}$ into $S$ folds of equal size, collecting the corresponding indices in the sets $I_s$, for $s \in [S]$. In practice, $S = 5$ or $S = 10$ are common choices for the number of cross-fitting folds. Let $\widehat{\eta}(z) = \widehat{\eta}\left(z; \{W_i\}_{i\in[n]}\right)$ be the researcher's preferred estimator of $\eta_0$, calculated on the full sample; and let $\widehat{\eta}_s(z) = \widehat{\eta}\left(z; \{W_i\}_{i\in I_s^c}\right)$, for $s \in [S]$, be a version of this estimator that only uses data outside the $s$ th fold.

2. Estimate $\tau$ by computing a local linear "no covariates" RD estimator that uses the adjusted outcome $M_i\left(\widehat{\eta}_{s(i)}\right) = Y_i - \widehat{\eta}_{s(i)}\left(Z_i\right)$ as the dependent variable, where $s(i)$ denotes the fold that contains observation $i$ :

$$\widehat{\tau}(h; \widehat{\eta}) = \sum_{i=1}^{n} w_i(h) M_i\left(\widehat{\eta}_{s(i)}\right).$$

[3] establish that the estimator $\widehat{\tau}(h; \widehat{\eta})$ is asymptotically equivalent to the infeasible estimator $\widehat{\tau}(h; \bar{\eta}) = \sum_{i=1}^{n} w_i(h) M_i(\bar{\eta})$ that uses the variable $M_i(\bar{\eta})$ as the outcome, where $\bar{\eta}$ is a deterministic approximation of $\widehat{\eta}$ whose error vanishes in large samples in some appropriate sense. It then holds that

$$\widehat{\tau}(h; \widehat{\eta}) \stackrel{a}{\sim} N\left(\tau + h^2 B_{\text{base}}, (nh)^{-1} V(\bar{\eta})\right)$$

The asymptotic variance in the above expression is minimized if $\widehat{\eta}$ is consistent for $\eta_0$, in the sense that $\bar{\eta} = \eta_0$. However, the distributional approximation is valid even if $\bar{\eta} \neq \eta_0$ because the moment condition (3.2) holds for (essentially) all adjustment functions, and not just the optimal one. In that sense, the procedure allows for misspecification in the first stage. Moreover, even under misspecification $V(\bar{\eta})$ is typically smaller than $V_{\text{base}}$. Valid confidence intervals can easily be constructed for $\tau$ by applying standard methods developed for settings without covariates to a data set with running variable $X_i$ and outcome $M_i\left(\widehat{\eta}_{s(i)}\right)$, ignoring sampling uncertainty about the estimated adjustment function.

## Heterogeneous Treatment Effects and Adjustments for Heterogeneity

So far we have used covariates in order to increase efficiency for the same estimand, $\tau_{\mathrm{RD}}$ that was defined in their absence. Covariates, however, can also help us understand and and control for heterogeneity. In particular, at a conceptual level, we can repeat the setup in Section 17.2 for (almost) every stratum $Z = z$, leading to the *CATE at the cutoff*:

$$\begin{aligned}\tau_{\mathrm{C-RD}}(Z) &= \mathrm{E}[Y(1) - Y(0) \mid Z, X = c] \\ &= \lim_{x \downarrow c} g(x, Z) - \lim_{x \uparrow c} g_0(x, Z),\end{aligned}$$

where $g_0(X, Z) = \mathrm{E}[Y \mid X, Z]$.

A potentially policy-relevant summary of $\tau_{\mathrm{C-RD}}(Z)$ is its average,

$$\tau_{\mathrm{A-C-RD}} = \mathrm{E}\tau_{\mathrm{C-RD}}(Z) = \mathrm{E}[\mathrm{E}[Y(1) - Y(0) \mid Z, X = c]].$$

For example, if we were to assume that $Z$ accounts for all treatment effect heterogeneity across values of the running variable, that is, $Y(1) - Y(0) \perp\!\!\!\perp X \mid Z$, then we would conclude that $\tau_{\mathrm{A-C-RD}} = \mathrm{E}[Y(1) - Y(0)]$ is the marginal ATE in the population, not just at the cutoff. More generally, we can say that $\tau_{\mathrm{A-C-RD}}$ controls for the heterogeneity modulated by $Z$, whether it is all of the heterogeneity or not.

The weaker conditional mean-independence of $Y(1) - Y(0)$ and $\mathbb{1}[X = c]$, given $Z$, suffices, but is perhaps harder to reason about.

Luckily, we can leverage DML to estimate $\tau_{\mathrm{A-C-RD}}$. For $h > 0$, consider a smoothed version of the same parameter:

$$\tilde{\tau}_h = \int_{-\infty}^{\infty} (4\mathbb{1}[x > c] - 2) K_h(x - c) \mathrm{E}[g_0(x, Z)] \mathrm{d}x,$$

where $K_h(x) = K(x/h)/h$ for a kernel $K$. Note that under appropriate continuity of $g_0(x, W)$ near but not at $x = c$, for almost every $W$, we have that $\lim_{h \to 0} \tilde{\tau}_h = \tau_{\mathrm{A-C-RD}}$. The quantity $\theta_0 = \tilde{\tau}_h$ is a simple linear summary of $g_0$, similar to those we studied in Chapter 10. We can then apply DML to estimate it using the Neyman orthogonal-score

$$\psi(W;\theta,\eta) = \int_{-\infty}^{\infty} (4\mathbb{1}[x > c] - 2)K_h(x-c)g(x,Z)\mathrm{d}x + \frac{(4\mathbb{1}[x > c] - 2)K_h(X-c)}{f(X \mid Z)}(Y - g(X,Z)) - \theta,$$

where $\eta = (g, f)$ are the nuisances, with the true value for the latter nuisance being $f_0$, the conditional density of $X$ given $Z$.

Note we can do this for every $h$, with our MSE to $\tau_{\text{A-C-RD}}$ (up to $o_p(1/n)$) consisting of the variance $\mathrm{E}[\psi(W;\tilde{\tau}_h,\eta_0)^2]/n$ and the squared bias $(\tau_{\text{A-C-RD}} - \tilde{\tau}_h)^2$. What remains is to choose $h$ to balance the two. Depending on the smoothness of $g_0$, we can further reduce the bias by using higher-order kernels (see [9]) or leveraging higher-order local-polynomial regression (instead of the local-constant regression used to define $\tilde{\tau}_h$ above). Depending on how much we can drive the bias down, we can achieve a better MSE rate.

## 17.4 Empirical Example

In this section, the effect of the antipoverty program Progresa/Opportunidades on the consumption behavior of families in Mexico in the early 2000s is analyzed. The analysis is accompanied by two notebooks.

The program was intended for families in extreme poverty and included financial incentives for participation in measures that improved the family's health, nutrition and children's education. The effect of this program is a widely studied problem in social and economic sciences and, according to the WHO, was a very successful measure in terms of reducing extreme poverty in Mexico.

Eligibility for the program was determined based on a pre-intervention household poverty-index. Individuals above a certain threshold received treatment (participation in the program), while individuals below the threshold were excluded and recorded as a control group. All observations above the threshold participated in the program, which makes the analysis fall into the standard (sharp) regression discontinuity design.

Data for this application are provided by [10] and in the presentation of the results we follow [3].*

* Links to notebooks for a replication are provided in the Notebook section.

Outcome variables are food and non-food consumption, one year and two years after the implementation of the program. The treatment variables is defined as eligibility for the cash transfer (intention-to-treat analysis). The data set contains 1,944 observations and 85 socio-economic pre-treatment variables like household size, gender, years of education and information on the house. Without considering pre-treatment variables participation in the program reduced food consumption by 22.1 units in the year following the intervention. With including additional pre-treatment variables and using ML methods for estimation, the point estimates for the effect of the program remain almost unchanged, but the confidence intervals are different. For an in-depth discussion of the results, we refer to the notebooks.

Without any covariate adjustments the effect of the cash transfer on food consumption one year after the program was introduced is estimated with −18.6 (s.e. 16.6). Utilizing linear adjustments for the covariates leads to an estimate of −14.8 and a reduced variance of 13.7. Using machine learning methods for the adjustment leads to to estimates of the effect between −16.0 and −21.5 and to a reduction of the standard errors compared to the baseline model (standard errors between 14 and 16). Notably, zero is contained in all confidence intervals (95% confidence level).

## Notebooks

- Python notebook for RDD provides an analysis of the effect of the antipoverty program Progresa/ Opportunidades on the consumption behavior of families in Mexico in the early 2000s.
- R notebook version for RDD

## Notes

The ideas behind RDDs and IVs come together in *fuzzy RDDs*. Whereas in sharp RDDs the treatment assignment is deterministic depending on being above or below the cutoff, in fuzzy RDDs the assignment mechanism is assigned at random with a assignment probability that need not be 0 or 1. Nonetheless, as in the sharp case, there is a discontinuity at the cutoff level. Then, for the units in an infinitesimal neighborhood of the cutoff, being just above or just below can be understood as an *instrument* for the treatment, with the assignment probability reflecting the compliance and the size of the discontinuity therein being the strength of the instrument. Almost the same tools for IV can be used once we localize to the cutoff.

Excellent introductions and surveys for RDD are the "classics" [11] and [12]. Updates including recent results are [13], [14], [15] and the monographs [16] and [2].

## Study Problems

1. Derive the moment conditions which identify the target parameter in RDD and show that it is orthogonal with regard to covariates.

2. In Israel, there is a strict restriction on the maximum size of public-school classrooms. For several decades in the previous century, the maximum was 40, such that, say, having 81 enrolled in a single grade meant a school has to open three parallel classrooms for that grade so that no one classroom has more than 40 students. Discuss why does this induces an RDD for the study of the impact of class size on academic performance? Assuming we have the school id, class id, and test scores of each individual student in, say the 5th grade in 1991, how would you construct an RDD: what would be the unit of analysis, the running variable, and the cutoff? How should we interpret the ATE and to what kind of student population might it not be relevant for and why? (Once you have thought about this study question, you can read about the study that famously leveraged this RDD in [17].)

# Index